\documentclass[12pt,twoside]{mitthesis}
\usepackage{amsmath,graphicx,bm,amsbsy,amssymb,aas_macros,color,enumerate,lgrind,cmap,rotating,url,array,hyperref,appendix,ctable,capt-of,setspace,subfig,float}
\usepackage[subfigure]{tocloft}
\usepackage{pdflscape}
\usepackage[numbers]{natbib}
\usepackage[T1]{fontenc} 

\newcommand{\x}{\mathbf{x}}
\newcommand{\y}{\mathbf{y}}

\newcommand{\C}{\mathbf{C}}
\newcommand{\R}{\mathbf{R}}
\newcommand{\U}{\mathbf{U}}
\newcommand{\G}{\mathbf{G}}
\newcommand{\N}{\mathbf{N}}
\newcommand{\F}{\mathbf{F}}
\newcommand{\Q}{\mathbf{Q}}
\newcommand{\I}{\mathbf{I}}

\newcommand{\ev}{\mathbf{v}}

\newcommand{\BigO}{\mathcal{O}}
\newcommand{\Pre}{\mathbf{P}}
\newcommand{\trans}{\mathsf{T}}
\newcommand{\Proj}{\mathbf{\Pi}}
\newcommand{\rv}{\mathbf{r}}
\newcommand{\kv}{\mathbf{k}}
\newcommand{\D}{\mathbf{D}}

\newcommand{\Ubar}{\overline{\U}}

\newcommand{\lbar}{\overline{\lambda}}
\newcommand{\Gam}{\mathbf{\Gamma}}
\newcommand{\GammaBar}{\overline{\Gam}}
\newcommand{\beq}{\begin{equation}}
\newcommand{\eeq}{\end{equation}}
\newcommand{\T}{\mathbf{T}}
\newcommand{\perpInt}{\int_{k_{\perp}^\alpha -\Delta k_{\perp}/2}^{k_{\perp}^\alpha +\Delta k_{\perp}/2}}
\newcommand{\paraIntPos}{\int_{k_{\|}^\alpha -\Delta k_{\|}/2}^{k_{\|}^\alpha +\Delta k_{\|}/2}}
\newcommand{\paraIntNeg}{\int_{-k_{\|}^\alpha +\Delta k_{\|}/2}^{-k_{\|}^\alpha -\Delta k_{\|}/2}}
\newcommand{\M}{\mathbf{M}}

\newcommand{\Eye}{\mathbf{I}}
\newcommand{\kvec}{\mathbf{k}}
\newcommand{\n}{\mathbf{n}}

\newcommand{\mean}{\boldsymbol\mu}
\newcommand{\base}{\mathbf{b}}

\newcommand{\A}{\mathbf{A}}
\newcommand{\p}{\mathbf{p}}

\newcommand{\PSF}{\mathbf{P}} 
\newcommand{\K}{\mathbf{K}}
\newcommand{\J}{\mathbf{J}}
\newcommand{\E}{\mathbf{E}}

\newcommand{\rhat}{\hat{\mathbf{r}}}

\def\ie{{\frenchspacing\it i.e.}}

\def\spose#1{\hbox to 0pt{#1\hss}}
\def\simlt{\mathrel{\spose{\lower 3pt\hbox{$\mathchar"218$}}
   \raise 2.0pt\hbox{$\mathchar"13C$}}}
\def\simgt{\mathrel{\spose{\lower 3pt\hbox{$\mathchar"218$}}
     \raise 2.0pt\hbox{$\mathchar"13E$}}}
 \def\simpropto{\mathrel{\spose{\lower 3pt\hbox{$\mathchar"218$}}
     \raise 2.0pt\hbox{$\propto$}}}

\newcommand{\xhat}{\hat{\mathbf{x}}}

\newcommand{\CHat}{\widehat{\C}}

\newcommand{\dtb}{T_b}
\newcommand{\Mpci}{\text{ Mpc}^{-1}}
\newcommand{\Mpc}{\text{ Mpc}}
\newcommand{\MHz}{\text{ MHz}}
\newcommand{\GHz}{\text{ GHz}}
\newcommand{\mJy}{\text{ mJy}}
\newcommand{\muJy}{ \text{ } \mu \text{Jy}}
\newcommand{\nJy}{\text{ nJy}}
\newcommand{\Msun}{ M_{\odot}}

\newcommand{\kperp}{k_\perp}
\newcommand{\kpara}{k_\parallel}

\newcommand{\be}{\begin{equation}}
\newcommand{\ee}{\end{equation}}
\newcommand{\mK}{\text{ mK}}
\newcommand{\meter}{\text{ m}}
\newcommand{\km}{\text{km}}
\def\Let@{\def\\{\notag\math@cr}}

\newcolumntype{L}[1]{>{\raggedright\let\newline\\\arraybackslash\hspace{0pt}}m{#1}}
\newcolumntype{C}[1]{>{\centering\let\newline\\\arraybackslash\hspace{0pt}}m{#1}}
\newcolumntype{R}[1]{>{\raggedleft\let\newline\\\arraybackslash\hspace{0pt}}m{#1}}

\newlength\mylen

\renewcommand\cftpartpresnum{Part~}
\begin{document} 

%
%
%
%
%
%
%
%
%
%
%

\title{It's Always Darkest Before the Cosmic Dawn: \\ Early Results from Novel Tools and Telescopes for \\ 21 cm Cosmology} 

\author{Joshua Shane Dillon}
       \prevdegrees{B.S., Stanford University (2009)}
\department{Department of Physics}

\degree{Doctor of Philosophy}

\degreemonth{June}
\degreeyear{2015}
\thesisdate{April 17, 2015}


\supervisor{Max Tegmark}{Professor of Physics}

\chairman{Nergis Mavalvala}{Professor of Physics \\ Associate Department Head for Education}

\maketitle



\cleardoublepage
\setcounter{savepage}{\thepage}
\begin{abstractpage}
%
%
%

21\,cm cosmology, the statistical observation of the high redshift universe using the hyperfine transition of neutral hydrogen, has the potential to revolutionize our understanding of cosmology and the astrophysical processes that underlie the formation of the first stars, galaxies, and black holes during the ``Cosmic Dawn.'' By making tomographic maps with low frequency radio interferometers, we can study the evolution of the 21\,cm signal with time and spatial scale and use it to understand the density, temperature, and ionization evolution of the intergalactic medium over this dramatic period in the history of the universe. 

For my Ph.D.~thesis, I explore a number of advancements toward detecting and characterizing the 21\,cm signal from the Cosmic Dawn, especially during its final stage, the epoch of reionization. In seven different previously published or currently submitted papers, I explore new techniques for the statistical analysis of interferometric measurements, apply them to data from current generation telescopes like the Murchison Widefield Array, and look forward to what we might measure with the next generation of 21\,cm observatories. I focus in particular on estimating the power spectrum of 21\,cm brightness temperature fluctuations in the presence enormous astrophysical foregrounds and how those measurements may constrain the physics of the Cosmic Dawn.
\end{abstractpage}


%
%
\newfloat{openingQuote}{p}

\pagebreak[4]

\begin{openingQuote}
\emph{Thus the explorations of space end on a note of uncertainty. And necessarily so. We are, by definition, at the very center of the observable region. We know our immediate neighborhood rather intimately. With increasing distance, our knowledge fades, and fades rapidly. Eventually, we reach the dim boundary---the utmost limits of our telescopes. There, we measure shadows, and we search among ghostly errors of measurement for landmarks that are scarcely more substantial.
\\ \\
The search will continue. Not until the empirical resources are exhausted, need we pass onto the dreamy realms of speculation.}\\ \\

\begin{centering}
\textsc{Edwin Hubble} \\
\emph{The Realm of the Nebulae}, 1936

\end{centering}
\end{openingQuote}

\pagebreak[4]

\cleardoublepage


\pagestyle{plain}

\tableofcontents
\newpage
\listoffigures
\newpage
\listoftables

\chapter*{Acknowledgments}
\addcontentsline{toc}{chapter}{Acknowledgments}

These past six months have put me in a reflective mood. Applying for postdocs and then writing this thesis have both been exercises in looking back, summarizing my graduate career, and then trying to project that momentum forward. As I come into my own as a physicist and move into this next stage of my career, I see that my time at MIT has given me a real running start. It's been a exhilarating and joyous six years and I have a lot of people to thank for that.

First, I have to thank my advisor, Max Tegmark. Max is responsible for guiding me on the difficult road from being interested in science to being a scientist. Of course, I learned a tremendous amount from him about how to tackle problems in cosmology and data analysis by dividing them into manageable parts with simple sense-checks. Max extended to me considerable freedom in working on what I found interesting while never letting me lose sight of the big picture. But more importantly, Max was, and I'm sure will continue to be, an invaluable source of sage career advice. I learned from Max how to pick interesting problems and good collaborators to work on them with and the essential importance of being nice. Max knew how important it was for me to work on HERA and I owe my postdoc in large part to his prescience. 

Next I'd like to thank Jackie Hewitt. While not technically my advisor, I've worked closely enough with her over the last few years that she's basically been a co-advisor. Working with Jackie has afforded me the exciting and edifying opportunity to see how my theoretical techniques stand up to real data. Like Max, she always kept me focused on the big picture, though her hard-nosed realism always felt like a excellent counterpoint to Max's ebullience. Jackie is a model for leadership in academia and I'm grateful to have witnessed and benefited from her direction of both her research group and the MKI.

The nexus of 21\,cm cosmology that Max and Jackie have built at MIT has led to the incredible opportunity to collaborate with seven other graduate students all working toward the same goals. First and foremost, I have to thank Adrian Liu for serving as a mentor, role model, and mini-advisor. His perspective from a few years down basically the same road I'm traveling has been incredibly helpful. Adrian's selflessness as a collaborator is the model I have tried mimic (however imperfectly) in my own relationships with the younger 21\,cm grad students, Aaron Ewall-Wice, Abraham Neben, and Jeff Zheng. I want to thank them for their tireless work and for the trust they placed in me to act as a mentor and sounding board, even if I sometimes led them down blind alleys. And of course, Chris Williams, Mike Matejek, and Andy Lutomirski have been patient tutors in all things radio astronomy and generous collaborators. The science I've done together with all seven friends and colleagues has been far more interesting, enjoyable, and fruitful than anything I could have done alone. 

I have also had the great fortune to work with a number of outstanding collaborators in both the MWA and HERA. In particular, I am grateful for my recommenders, Aaron Parsons and Miguel Morales. I've already learned so much from both and them and I have no doubt that will continue as we build HERA togther. I also want to thank Jonnie Pober for his wonderful pessimism (and for letting me use his postdoc applications as a model), Danny Jacobs and Bryna Hazelton for reminding me that experimental reality is always more complicated than I think, Cath Trott for making me think deeply about estimators, and Adam Beardsley for the postdoc application commiseration. 

Beyond my immediate research area, I am immensely grateful for the amazing intellectual community at MIT, especially in the astrophysics division. I found great friends, role models, and tutors among the older graduate students, including Ben Cain, Robyn Sanderson, Phil Zukin, Leo Stein, Leslie Rodgers, Scott Hertel, Nick Smith, John Rutherford, Uchupol Ruangsri, and Becky Levinson. I am honored and grateful to have shared the long road with my astrograd peers and dear friends---Kat Deck, Roberto Sanchis Ojeda, Lu Feng, and David Hernandez---through courses and exams, late nights at the office and at the pub, and triumphs and setbacks. And I feel privileged to have overlapped with a fantastic group of younger graduate students, including Adam Anderson, Reed Essick, Peter Sullivan, Jessamyn Allen, Greg Dooley, Tom Cooper, and Alex Ji.

I am also grateful for the role faculty and postdocs played in building the intellectual environment at MIT. In particular, I want to thank Rob Simcoe for his excellent teaching, fair Part III questions, and for agreeing to serve on my committee and read this monstrously long thesis. I also want to thank my academic advisor, Nergis Mavalvala, and the rest of the amazing astro faculty, especially Ed Bertschinger, Bernie Burke, John Belcher, Deepto Chakrabarty, Anna Frebel, Alan Guth, Scott Hughes, Saul Rappaport, Paul Schechter, Nevin Weinberg, and Josh Winn, with whom I've enjoyed classes, journal clubs, colloquia, seminars, coffees, lunches, dinners, and random hallway encounters. I am especially grateful to Al Levine, a font of wisdom and a selfless aide to all Part III examinees. I am grateful for the morning coffee crew of postdocs and research scientists, including Kevin Schlaufman, Mike McDonald, Dan Castro, Laura Lopez (especially for her postdoc application advice), Zach Berta-Thompson, Bryce Croll, Simon Albrecht, Kathy Cooksey who got morning coffee started, and Eric Miller who sustained it. They have been the source of stimulating intellectual discourse and for timely career advice.

I'm grateful for all the really cool people I've met in the department, friends with whom I've commiserated and celebrated. I'm sure I haven't seen the last of Robin Chisnell, Christina Ignarra, Minde Lekaveckas, Dan Pilon, Laura Popa, Axel Schmidt, and Eugene Yurtsev. I'd especially like to thank my fantastic roommates over the last several years, Chris Aakre, Ognjen Ilic, Ethan Dyer, Sam Ocko, and Cosmin Deaconu. Having such a group of fun, interesting, and hard-working roommates was a source both of great happiness and of considerable motivation to commit myself to my research. It has also been an immense pleasure to have so many Stanford friends in the Boston area, especially my fellow Phi Psis: Andy Lutomirski, John and Te Rutherford, Naveen Sinha, A.J. Kumar, Phil Aguilar, Tony Pan, Harvey Xiao, Fah Sathirapongsasuti, Walter Vulej, Jon Kass Ian Counts, and Frank Wang. The Phi Psis really showed me around Boston and made me feel at home here and I am very grateful to have had an instant group of close friends when I moved here. Lastly, I am immeasurably thankful to have maintained such a close relationship with my high school friends (and, in a strange twist of fate, business partners) and I have to thank Daniel Dranove, Ben Hantoot, Eli Halpern, David Munk, David Pinsof, Max Temkin, and Eliot Weinstein for their years of true and loyal friendship that significantly predate this thesis and I'm sure will endure for many years beyond it.

Finally I need to thank my family. I am proud to come from a family that holds academic and intellectual pursuits in such high esteem. My parents and grandparents furnished me with every possible opportunity to discover my passions, pursue them, and excel in them. Even when there wasn't always money for the newest toy or a fancy vacation, there was always money for books. They taught both Molly and I the virtue of intellectual work in the service of humanity and I am grateful that we have encouraged each other in our individual pursuits of that goal. I can draw a straight line from those values to the completion of this thesis and my Ph.D. Their support, their pride in my accomplishments, and their love sustained this effort and for that I will always be grateful.


\chapter*{Preface}
\addcontentsline{toc}{chapter}{Preface}

\section*{The Golden Age of Cosmology}

I've often heard it said that we are living in a ``golden age of cosmology.'' Perhaps just as often, I hear the equally sweeping claim that cosmology just finished its golden age and that all the exciting discoveries have probably been made. As I step back to survey the scientific landscape that I am graduating into---one that I hope to shape with my work---I have to ask myself: which is it?

The late University of Chicago cosmologist David Schramm is credited with first declaring the end of the twentieth century a golden age. In a meeting report on dark matter \cite{SchrammShadows} he began:

\begin{singlespacing} \vspace{-12pt}
\begin{quotation}
\noindent Let me open by noting that we're in the golden age of cosmology... Now cosmologists finally have the technology that allows experiments that tell us about the universe as a whole. We have been able to study it in a truly quantitative way, and we've been able to establish that the early universe was hot and dense. 
\end{quotation}
\end{singlespacing}
By that he meant that the surprising discovery of the recession of almost all observed galaxies, coupled with the discovery for the cosmic microwave background and the precise measurement of the cosmic abundance of light elements, all upheld the remarkable theory that the universe began with a hot big bang. The weight of evidence had just reached the point where a basic framework could be worked out and (more or less) agreed upon---now it was time to fill in the gaps. 

His sentiment was met with a mix of bemusement and skepticism. In one anecdote:

\begin{singlespacing}  \vspace{-12pt}
\begin{quotation}
\noindent He kept proclaiming that cosmology was in a ``golden age.'' His chamber of commerce enthusiasm seemed to grate on some of his colleagues; after all, one does not become a cosmologist to fill in the details left by pioneers. After Schramm's umpteeth ``golden age'' proclamation, one physicist snapped that you cannot know an age is golden when you are \emph{in} that age but only in retrospect. Schramm jokes proliferated. One colleague speculated that the stocky physicist might represent the solution to the dark matter problem. Another proposed that Schramm be employed as a plug to prevent our universe from being sucked down a wormhole.
\end{quotation}
\end{singlespacing}
That story comes from John Horgan's provocatively titled \emph{The End of Science} \cite{HorganEndOfScience}. In it he argues that scientists across virtually all disciplines are already beginning to sense an end, a butting up against the limits of knowledge that comes with the extraordinary successes of fundamental science over the last few centuries. He worries that the ``great revelations or revolutions'' are behind us that that the ultimate \emph{telos} of science---the ``primordial quest to understand the universe and our place in it''---has been mostly accomplished. Writing on cosmology specifically, he asks:

\begin{singlespacing}  \vspace{-12pt}
\begin{quotation}
\noindent What if Schramm was right? What if cosmologists had, in the big bang theory, the major answer to the puzzle of the universe? What if all that remained was tying up loose ends, those that could be tied up?
\end{quotation}
\end{singlespacing}

I think Horgan misses the point. Schramm didn't think he lived in the golden age of cosmology because the biggest discoveries had just been made. It was a golden age because the recent triumph of the big bang model had opened up whole new lines of inquiry. Rather suddenly, cosmologists realized that they were solving an entirely different puzzle than they had been before. That doesn't mean that all pieces were in hand, or that all the pieces they'd find would fit in so neatly.

Thomas Kuhn, the philosopher and historian of science, famously wrote about this process in \emph{The Structure of Scientific Revolutions} \cite{KuhnSSR}. He describes the typical progress of an scientific discipline as puzzle solving or ``normal science.'' Working within a shared framework, a community of scientists has common set of values and theories---a paradigm---which makes sensible a new set of questions about nature, a new set of puzzles to solve. When enough puzzles arise that linger unsolved as anomalies, a need arises for a new paradigm. Ideally it is more accurate, more predictive, of greater scope, and simpler than previous theories. Rarely is it so clear. In time, the better theory wins the consensus, if perhaps not unanimous support. As Max Planck put it, ``science advances one funeral at a time.''

Our stories about science invariably romanticize the revolutionaries. We aspiring scientists all want to be the revolutionaries, but only a lucky few get the privilege. The more I study science, both its past and its present, the more I love the puzzle solving. I know that scientific revolution is impossible without puzzles that defy resolution, without the hard work that extracts from the full complexity of nature a slow trickle of anomalies.

This thesis is about the development of a new technique for exploring one of the last unobserved epochs in the history of the cosmos. We are looking for the faint radio signature of the impact of the first stars, galaxies, and black holes on the intergalactic hydrogen gas that pervades the universe. We call this period that spans from the first stars through the eventual heating and reionization of the intergalactic gas the ``Cosmic Dawn.'' We haven't seen it yet. 

It's easy to despair at the challenge of detecting that faint signal amidst contaminants orders of magnitude stronger. And it's easy to despair that the golden age is over and that all we're doing is filling in the gaps. In a sense, that's literally true. There's a blank space in our cosmic timeline and we're trying to fill it in. But in the way that really matters, I don't believe any of that. I've titled this thesis \emph{It's Always Darkest Before the Cosmic Dawn} because, despite any occasional doubt or despair, I think what we're doing is important and stands a good chance of being something really big. We're not done yet. I'm not done yet. The search will continue.

I really believe that we're still in a golden age of cosmology. The golden age continues because the advance of our technology continues. Bigger and faster computers let us store and analyze more data. For radio astronomy, better computers leads directly to bigger and more sensitive telescopes. It's a golden age of statistics and of ``big data'' (whatever that means) and cosmology is fundamentally a statistical discipline. If we are, as Hubble put it, to ``measure shadows'' and ``search among ghostly errors of measurement for landmarks that are scarcely more substantial,'' it sure helps to make a lot of measurements. 

Big discoveries don't end golden ages---they help us to see that we're in one. Big discoveries lead to new puzzles. We need to solve puzzles to find anomalies. We need anomalies before we can have revolutions. We need revolutions for new paradigms with new puzzles to solve. To do science, one must remember that testing theories by solving puzzles and advancing revolutions by finding anomalies go hand-in-hand.
It also helps to remember this. \emph{The End of Science} was published on May 12th, 1996. David Schramm died in a tragic plane crash in December 19, 1997. Three months later, the High-z Supernova Search Team announced the discovery of dark energy. 

We're not out of puzzles yet.


\chapter{Introduction} \label{ch:Introduction}
   
\section{The Cosmic Dawn}

Over its 13.8 billion year history, our universe has undergone a dramatic transformation. Just 380,000 years after the big bang, when electrons and nuclei combined for the first time and the sea of cosmic microwave background (CMB) photons decoupled from them, the universe was nearly homogenous and isotropic. Fluctuations in density and temperature were a mere part in 100,000. This exotic early universe bears almost no resemblance to today's universe, with its incredible complexity and diversity of phenomena. From the sparsest intergalactic gas to the densest cores of neutron stars, modern densities range by more than a factor of $10^{44}$. 

Part of that transformation was driven by the expansion of the universe, the history of which we now know very precisely. Our standard cosmological model, $\Lambda$CDM, describes a universe that is today is only 5\% ordinary matter with the rest, 26\% dark matter and 69\% dark energy \cite{PlanckCosmoParams2015}, made out of stuff (for lack of a better word) that we know very little about. $\Lambda$ represents dark energy that acts a ``cosmological constant;'' it has an energy density that doesn't change as the universe expands and leads to accelerated expansion. CDM stands for cold dark matter, stuff which does not interact electromagnetically but which is massive enough and slow enough to get trapped gravitationally into halos which host modern galaxies.  Along with a handful of other parameters, this cosmological model describes the expansion history of our universe very precisely and fits all available data. 

If it weren't for those initial seed fluctuations in density, our universe would be far bigger and colder than it was 13.8 billion years ago, but just as boring and lifeless. Those tiny fluctuations in the density of both dark and ordinary matter evolved into stars and galaxies and planets and people. The source of those fluctuations is a great mystery, one potentially solved by invoking cosmic inflation, an early period of exponential expansion in a tiny fraction of a second. 

\begin{figure*}[] 
	\centering 
	\includegraphics[width=\textwidth]{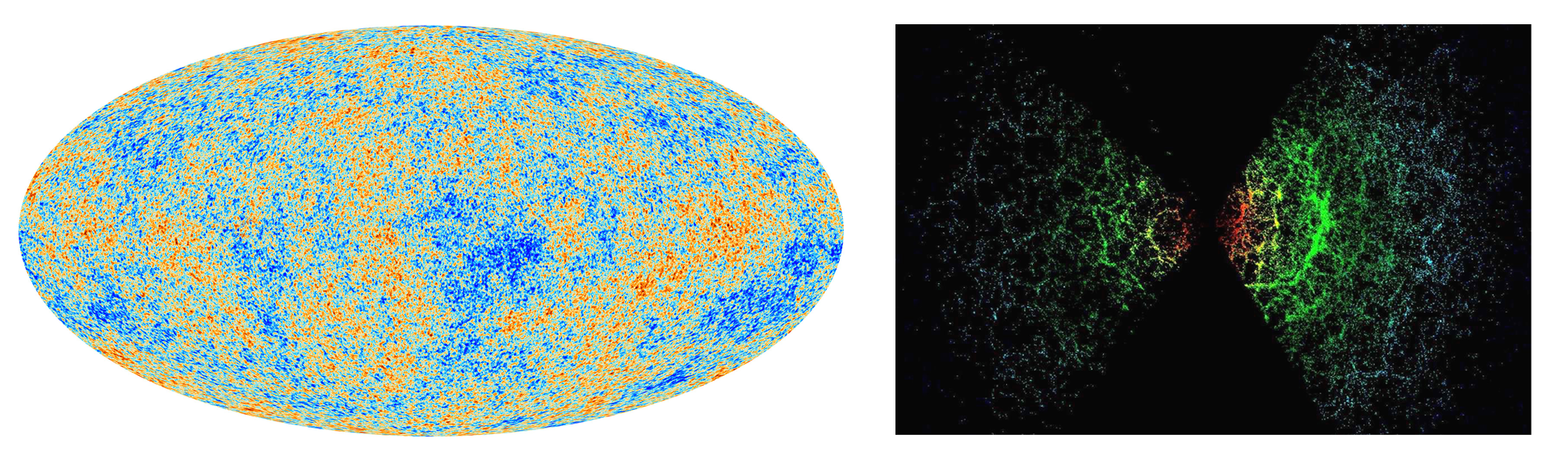}
	\caption[Two probes of our universe: the CMB and galaxy surveys.]{Two probes of the distribution of matter in our universe. The cosmic microwave background (left) gives us a snapshot of a nearly homogenous universe only 380,000 years after the big bang. Galaxy surveys (right) give us information about the relatively local universe, having been transformed dramatically over its 13.8 billion year evolution. \emph{Image credits: the Planck Collaboration and the Sloan Digital Sky Survey Collaboration, respectively.}}
	\label{fig:CMBvsGal}
\end{figure*}

Another daunting challenge is to explain that evolution over large time, mass, and spatial scales. Our success so far speaks volumes of the incredible progress of modern cosmology. Our understanding of the growth of structure in the universe is anchored at both ends by observation. Our record of the earliest times comes from the CMB, that thermal relic of the big bang, which we observed highly redshifted ($z\approx 1100$) by the expansion of the universe. It arrives at our telescopes today largely unperturbed by the intervening structure. In the local universe, we can probe the distribution of matter by cataloging the brightest tracers of it---namely galaxies and the supermassive blackholes they host---among other techniques (see Figure \ref{fig:CMBvsGal}). 

\begin{figure*}[] 
	\centering 
	\includegraphics[width=.6\textwidth]{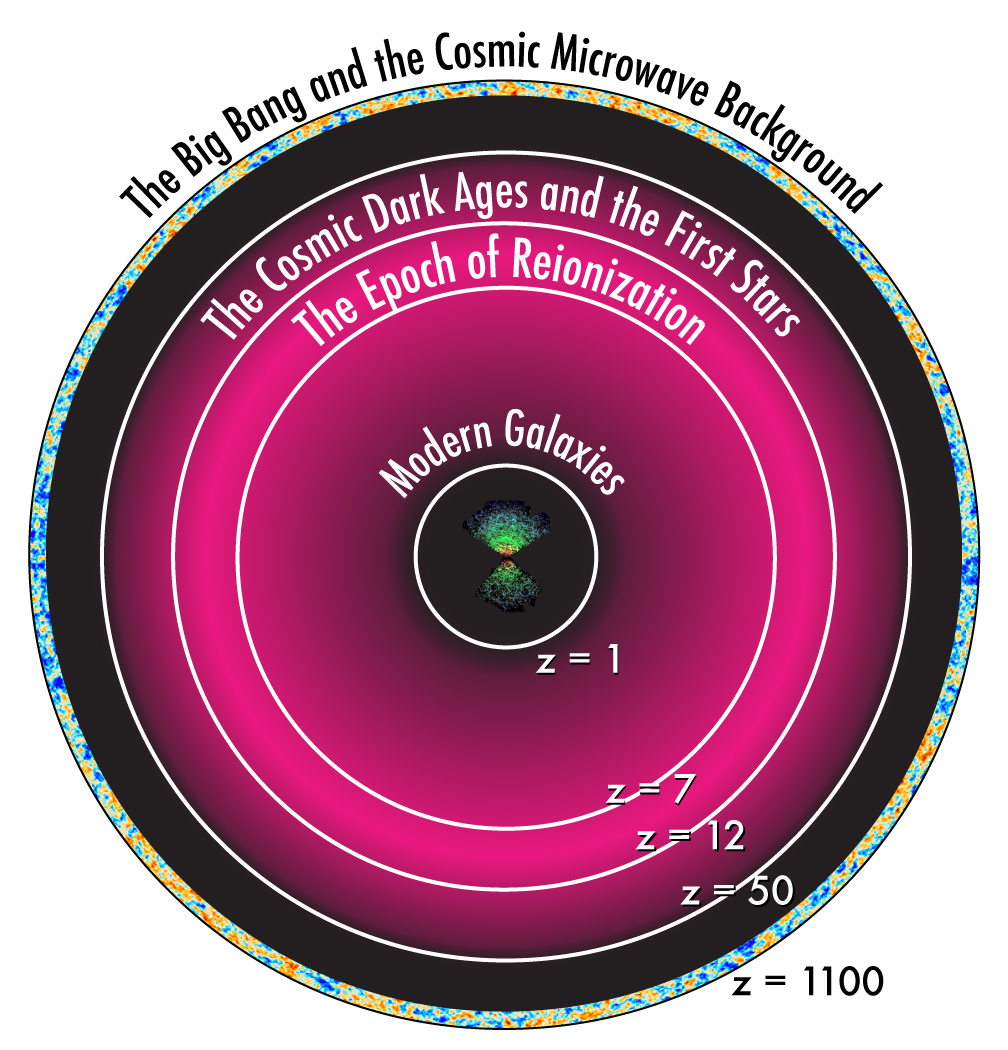}
	\caption[Our Hubble volume and how we can observe it.]{The most complete maps of the distribution of matter in the universe---CMB measurements and galaxy surveys---only probe a small fraction of the volume of the observable universe. In fact, due to its expansion, 80\% of the volume of the observable universe today corresponds to regions we can see from when the universe was less than a billion years old ($z \gtrsim 6$). 21\,cm cosmology, the probe that this thesis focuses on developing, may one day make the entire pink region accessible to direct observation.  \emph{Adapted from \cite{FFTT}.}}
	\label{fig:hubblevolume}
\end{figure*}

In between, our knowledge gets sparser, especially as we look further back in cosmic time. We're limited to observing only the brightest galaxies and active galactic nuclei (AGN) and, with some hard work, the structure along the lines of sight to those bright objects. As Figure \ref{fig:hubblevolume} shows, an incredible fraction of the volume of the universe is unexplored, especially during the first billion years after the big bang. 

\begin{figure*}[] 
	\centering 
	\includegraphics[width=\textwidth]{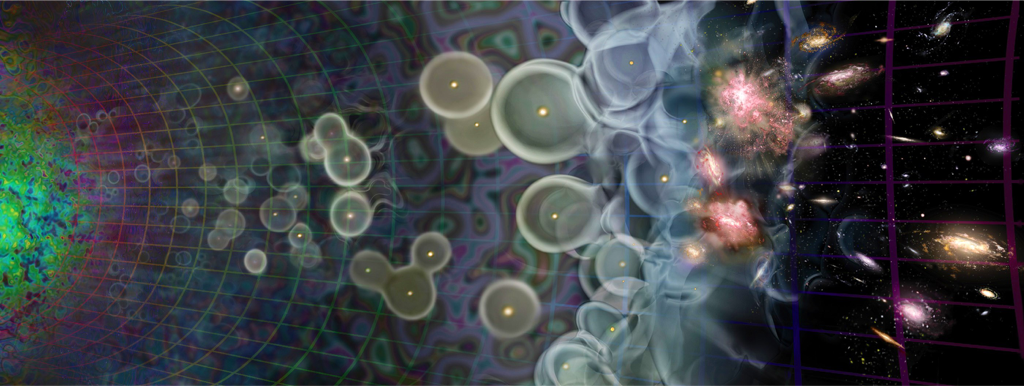}
	\caption[A schematic picture of the Cosmic Dawn.]{The ``Cosmic Dawn'' is a period in the early history of the universe between the cosmic microwave background (on the left) and modern stars and galaxies (on the right). During this time, the first stars and galaxies form and eventually heat up the intergalactic gas. They also begin to ionize the gas around them, inhomogenously filling the universe with merging bubbles of ionized hydrogen. \emph{Image credit: Abraham Loeb and Scientific American.}}
	\label{fig:CosmicDawn}
\end{figure*}

As we look backwards to earlier and earlier times, we also see evidence for another dramatic transition. As dark matter halos condensed gravitationally, the ordinary matter they host cooled and collapsed to form the first generation of stars and galaxies. The formation of these first luminous objects, including the first black holes that grow by accreting matter and shining brightly in X-rays, had a dramatic effect on the rest of the universe. They heated the gas between galaxies, the intergalactic medium (IGM), and eventually reionized it. This process,  starting with the first luminous objects and going through the reionization of the intergalactic gas (depicted schematically in Figure \ref{fig:CosmicDawn}) is known as the ``Cosmic Dawn.''

Precious little is known about exactly how this process proceeded. This thesis is devoted to advancing a new probe of the Cosmic Dawn known as ``21\,cm Cosmology.'' Before I explain the theory behind 21\,cm cosmology in Section \ref{sec:PromiseOf21cm} and the ongoing observational efforts in Section \ref{sec:ObsChallenges}, I will briefly review what we do and don't know about the Cosmic Dawn.

\subsection{What Do We Know?}
Since we can observe the universe before and after the Cosmic Dawn, it's fair to say that we know its basic story from $\Lambda$CDM. Dark matter halos collapsed, starting with the smallest overdensities and growing hierarchically. They played host to the collapse and cooling of gas into the first generation of stars and galaxies which eventually heated the IGM and reionized the universe. To try to confirm the basic story that we see play out in our simulations, we do have a few indirect observations.

First, we know that the universe finished reionization when it was about a billion years old (redshift $z\approx 6$). That information comes from observing the absorption of the first electronic transition of hydrogen, the Lyman-alpha line, in the spectra of high redshift AGN. Even a small amount of neutral hydrogen along the line of sight completely saturates the absorption signature, creating a spectral feature known as the Gunn-Peterson trough \cite{BeckerGunnPeterson}. While these observations fix the endpoint of reionization and the Cosmic Dawn, they can't tell us much about the process itself.

From the cosmic microwave background, we can get a rough sense of the midpoint of reionization. The amplitude of the fluctuations (see Figure \ref{fig:CMBvsGal}) is affected by the scattering of CMB photons off free electrons along the line of sight. The longer the universe has been ionized, the larger the damping of fluctuations in the CMB. Combining those measurements with large-scale fluctuations in the polarization of the CMB, the total level of scattering corresponds to a midpoint redshift of $z=8.8$, assuming instantaneous reionization \cite{PlanckCosmoParams2015}. Of course, reionization didn't happen instantaneously across the entire universe, but the integrated constraint from the CMB is a useful starting point.

Lastly, we can make some inferences about the few galaxies we can see at very high redshift. Observations of the Hubble Ultra Deep Field, tell us about the abundance of the absolute brightest end of the distribution of galaxy luminosities up to about redshift $z=10$ \cite{MadauDickinsonSFR2014}. Since young, massive, and short-lived stars dominate the production of ionizing photons, the star formation rate in these galaxies should hold an important clue as to how the universe reionized. Extrapolating from a relationship that relates infrared and ultraviolet brightness in the local universe to measured star formation rates, \citet{RobertsonReionization2015} find that, with reasonable modeling assumptions and an extrapolation down to very faint galaxies, those observations are consistent with the reionization inferred from the CMB. All of that is still very uncertain and model-dependent, but it confirms that the basic story is plausible.

\subsection{What Don't We Know?}

Though we have a plausible picture of reionization, the exact set of astrophysical processes that drove it are still largely unconstrained. We would like to know:
\begin{itemize}
\item When exactly did reionization happen and how long did it take?
\item How did early stars and galaxies affect the IGM before reionization?
\item When did the first stars form? 
\item How were they different from later generations of stars? 
\item What role did early black holes in X-ray binaries play in the thermal and ionization history of the IGM?
\item What role did the accretion of matter onto growing supermassive black holes play?
\item Which galaxies were responsible for reionization?
\item How did reionization affect any subsequent formation of stars, especially in dwarf galaxies?
\end{itemize}
All of these, and many more, are important unresolved questions that we want to answer to verify our general picture of the formation of structure in the universe. We have a general outline and it is essential to see if all the details fit together.

If we find inconsistencies between our observations and our theories of the astrophysical processes that describe the formation of early stars and galaxies and their interaction with the IGM, that would be very interesting. But what makes cosmology so exciting is the potential for surprising new discoveries that would completely change our understanding of the cosmos. By exploring huge fraction of the volume of the universe only accessible by looking back to the Cosmic Dawn, we can perform very sensitive tests of our standard model of cosmology. For example:
\begin{itemize}
\item Measuring the statistical distribution of matter during the Cosmic Dawn via the matter power spectrum could provide extremely precise constraints on $\Lambda$CDM parameters, or reveal inconsistencies between the model and observations \cite{Yi}. Of particular interest are the possible constraints---both from the power spectrum and from higher order statistics---on the simplest models of inflation, the source of those primordial density perturbations.
\item Another test of $\Lambda$CDM picture and inflation would be to look for the effect of the relative velocities of dark matter and ordinary matter, which should show up in sufficiently sensitive statistical probes of the Cosmic Dawn \cite{DarkMatterRelativeVelocity2010}.
\item The strength of primordial magnetic fields, also a prediction of inflation, could be constrained by their effect on the thermal history of the IGM \cite{SKACosmicDawn2015}.
\item ``Warm'' dark matter models, where the dark matter particle was light enough to remain relativistic  for much of the history of the universe, tend to wash out structure on small scales. If small fluctuations are responsible for early heating, this would affect the thermal and ionization history of the IGM \cite{Mesinger:2013c}. Warm dark matter is currently a topic of great interest because it would help explain some potential discrepancies in the comparison of simulations observations of galaxy formation and because of a reported X-ray spectral line consistent with decaying warm dark matter \cite{BulbulXRay2014}.
\item Likewise, there's also interest in recent reports of a gamma ray signature from the galactic center consistent with cold but annihilating dark matter \cite{FermiGammaCaseForDarkMatter2014}. The annihilating dark matter may also alter the thermal history of the IGM \cite{Mesinger:2013c}, meaning that observations of the Cosmic Dawn may prove a sensitive test of these theories.
\end{itemize}
To begin to answer all these questions and test these ideas with precision measurements, we need new ways to directly probe the universe during the Cosmic Dawn.

\section{The Promise of 21\,cm Cosmology} \label{sec:PromiseOf21cm}
While it is exceedingly difficult to observe the first stars and galaxies to form in our universe directly, that doesn't mean that the Cosmic Dawn is unobservable. Instead of studying the brightest early objects, we can study the gas between them, the IGM. The IGM plays a fundamental role in the development of structure, since it is the source of fuel for early star-forming galaxies and it is dramatically impacted by their evolution. It is not surprising therefore that the density, temperature, and ionization evolution of the IGM across cosmic time encodes considerable information about our universe and the Cosmic Dawn.

Our best hope to directly observe the IGM during the Cosmic Dawn is to look for the radio signature of neutral hydrogen. Its ground state is very slightly split into two energy levels related to the relative spins of the proton and the electron. The incredibly precisely measured transition between the states two corresponds the emission or absorption of a photon with a frequency of $\nu_0 = 1420.40575177$\,MHz, or a wavelength of about 21\,cm.

Since we understand the expansion of the universe very well, we can directly relate radio maps at different frequencies to different redshifts and thus different distances from us. Multi-frequency maps---which are comparatively easy to produce with low-frequency radio telescopes---thus represent large 3-D volumes of the universe. In this way, we can build up enormous tomographic maps, one frequency at a time. An enormous volume of the universe may be observable with these techniques (see Figure \ref{fig:hubblevolume}). 

The scientific potential of these maps is tremendous. As I will discuss in Section \ref{sec:21SigCosmicTime}, the huge volume of the universe accessible will enable precise tests of $\Lambda$CDM. At the same time, they will also provide the first direct observations of the astrophysical processes that drove the Cosmic Dawn. 

In this section, I will review the physical processes that create the 21\,cm signal and make it visible against the backdrop of the CMB (Section \ref{sec:21cmPhysics}). Then I will review how the 21\,cm signal  is expected to vary across cosmic time and how that will translate into statistical probes of neutral hydrogen in the high-redshift IGM (Section \ref{sec:21SigCosmicTime}).

\subsection{The Astrophysics of Neutral Hydrogen Cosmology} \label{sec:21cmPhysics}

The 21\,cm transition has been astrophysically useful since it was first observed in 1951 by Ewen and Purcell \cite{EwenPurcell1951}. It can be used to trace neutral gas in nearby galaxies and to measure their rotation curves.  In the local universe, 21\,cm emission can only be seen in galaxies where gas can cool enough to form neutral hydrogen and where the gas is dense enough that it is effectively shielded from the ionizing background. Before and during reionization, the IGM can be observed in 21\,cm emission or absorption relative to the CMB. In this section, I will explain the reasons why the 21\,cm signal is visible relative the CMB and how it traces ionization, temperature, and density fluctuations in the IGM.

In radio astronomy, we typically measure the specific intensity of emission at the frequency $\nu$, $I_\nu$. At frequencies much lower than the peak of the CMB, we can use the Rayleigh-Jeans limit of the blackbody spectrum to represent observed intensities as brightness temperatures $T_b$, where
\beq
I_\nu \equiv \frac{2 k_B T_b \nu^2}{c^2}.
\eeq
In this limit the equation of radiative transfer through a cloud of hydrogen backlit by the CMB can be written \cite{FurlanettoReview} as 
\beq
T_b(\nu) = T_{S}\left(1-e^{-\tau_\nu}\right) + T_\gamma(z)e^{-\tau_\nu}.
\eeq
Here $T_\gamma(z)$ is the temperature of the CMB at the epoch considered, $\tau_\nu$ is the optical depth of the cloud due to the 21\,cm transition, and $T_S$ is the spin temperature of the gas. The spin temperature, which is the excitation temperature of the hyperfine transition, is defined in terms of the Boltzmann factor for the spin-singlet and spin-triplet hyperfine levels of the ground state of hydrogen,
\beq
\frac{n_\text{triplet}}{n_\text{singlet}} = 3 e^{-h \nu_0 / k_B T_S}. \label{eq:SpinTempDef}
\eeq
The factor of 3 comes from three-fold degeneracy of the triplet state (hence the name). 

The 21\,cm transition is highly ``forbidden'' quantum mechanically, leading to a calculated lifetime for spontaneous emission of about $3\times 10^7$ years \cite{FurlanettoReview}, making $\tau_\nu$ small and the entire IGM optically thin. It follows then that contrast in the 21\,cm signal observed today relative to the CMB, $\delta T_b^\text{obs}$ is given by 
\begin{align}
\delta T_b^\text{obs} &= \frac{T_b(z)}{1+z} - T_\gamma(z=0)  \nonumber \\
&= \frac{\left(T_S - T_\gamma(z)\right)\left(1-e^{-\tau_{\nu_0}}\right)}{1+z} \nonumber \\
&\approx \frac{T_S - T_\gamma(z)}{1+z} \tau_{\nu_0}.
\end{align}
I omit here the detailed calculation of the optical depth integrated over frequency to get $\tau_{\nu_0}$ and, following \citet{FurlanettoReview} and \citet{PritchardLoebReview}, simply state the final result:
\beq
\delta T_b^\text{obs}(\rhat,z) \approx (\text{27\,mK}) x_\text{HI} \left(\frac{T_S-T_\gamma(z)}{T_S}\right) (1+\delta_b)\sqrt{\frac{1+z}{10}} \left[\frac{(1+z)H(z)}{\partial v_\| / \partial r_\|} \right]. \label{eq:deltaTbObs}
\eeq
Of course, as we observe in different directions $\rhat$ or at different redshifts, we see different values of $\delta T_b^\text{obs}$. These fluctuations are sourced in three principal ways. First, ionization can drive the neutral fraction, $x_\text{HI}$, to from 1, fully neutral, to 0, fully ionized with no 21\,cm signal at all. The second is due to spin temperature fluctuations relative to the CMB temperature as a function of time and position. When $T_S \gg T_\gamma$, this term saturates. However, when $T_S$ is very cold, this can drive the signal into strong absorption relative to the CMB.   Third, baryon over-densities $\delta_b$ lead to stronger signals. The last factor in Equation \ref{eq:deltaTbObs} comes from the Doppler broadening of the 21\,cm line, which depends on the Hubble factor, $H(z)$, and gradient of the proper velocity along the line of sight, $\partial v_\| / \partial r_\|$, which includes both the Hubble expansion and the peculiar velocity of the gas cloud \cite{FurlanettoReview}.

It is clear from Equation \ref{eq:deltaTbObs} that the spin temperature plays a key role in determining the observability of the 21\,cm signal. If $T_S$ is in equilibrium with $T_\gamma$, then $\delta T_b = 0$. If $T_S \ll T_\gamma$, then the 21\,cm signal shows up very strongly in absorption. $T_S$ is determined \cite{FurlanettoReview,PritchardLoebReview} by the interplay of three processes:
\begin{itemize}
\item CMB photons at or near the 21\,cm transition can be absorbed or lead to stimulated emission. This couples $T_S$ to $T_\gamma$.
\item Collisions between neutral hydrogen atoms and other particles may induce exchanges of angular momentum, causing a spin flip. This effect is dominated by hydrogen-hydrogen collisions, hydrogen-electron collisions, and hydrogen-proton collisions, all of which couple $T_S$ to the kinetic gas temperature, $T_K$.
\item Absorption and remission of Lyman-alpha photons allows an indirect path to changing the hyperfine state of hydrogen, since transitions from the 1S state of hydrogen to some of the 2P states and back allow a net spin flip. This couples $T_S$ to $T_\alpha$, the color temperature of the Lyman-alpha transition, defined analogously to Equation \ref{eq:SpinTempDef}. This pathway for hyperfine transitions is known as the \emph{Wouthuysen-Field effect} \cite{Wouthuysen:1952,Field:1958}.
\end{itemize}
In equilibrium, the spin temperature is given by 
\beq
T_S^{-1} = \frac{T_\gamma^{-1} + x_c T_K^{-1} + x_\alpha T_\alpha^{-1}}{1 + x_c+ x_\alpha},
\eeq
where $x_c$ and $x_\alpha$, the collisional and Lyman-alpha coupling coefficients depend on the subtle atomic processes that govern these effects, which themselves have complicated temperature and density dependences \cite{FurlanettoReview,PritchardLoebReview}.

\subsection{The 21\,cm Signal Across Cosmic Time} \label{sec:21SigCosmicTime}

The physical processes that drive ionization, spin temperature, and density changes that create $\delta T_b^\text{obs}$ in Equation \ref{eq:deltaTbObs} are both inhomogenous and time-dependent. Across cosmic time, the 21\,cm signal and its underlying statistics are expected to change dramatically, though the precise evolution depends on the poorly understand processes that drove the Cosmic Dawn.

\begin{figure*}[] 
	\centering 
	\includegraphics[width=1\textwidth]{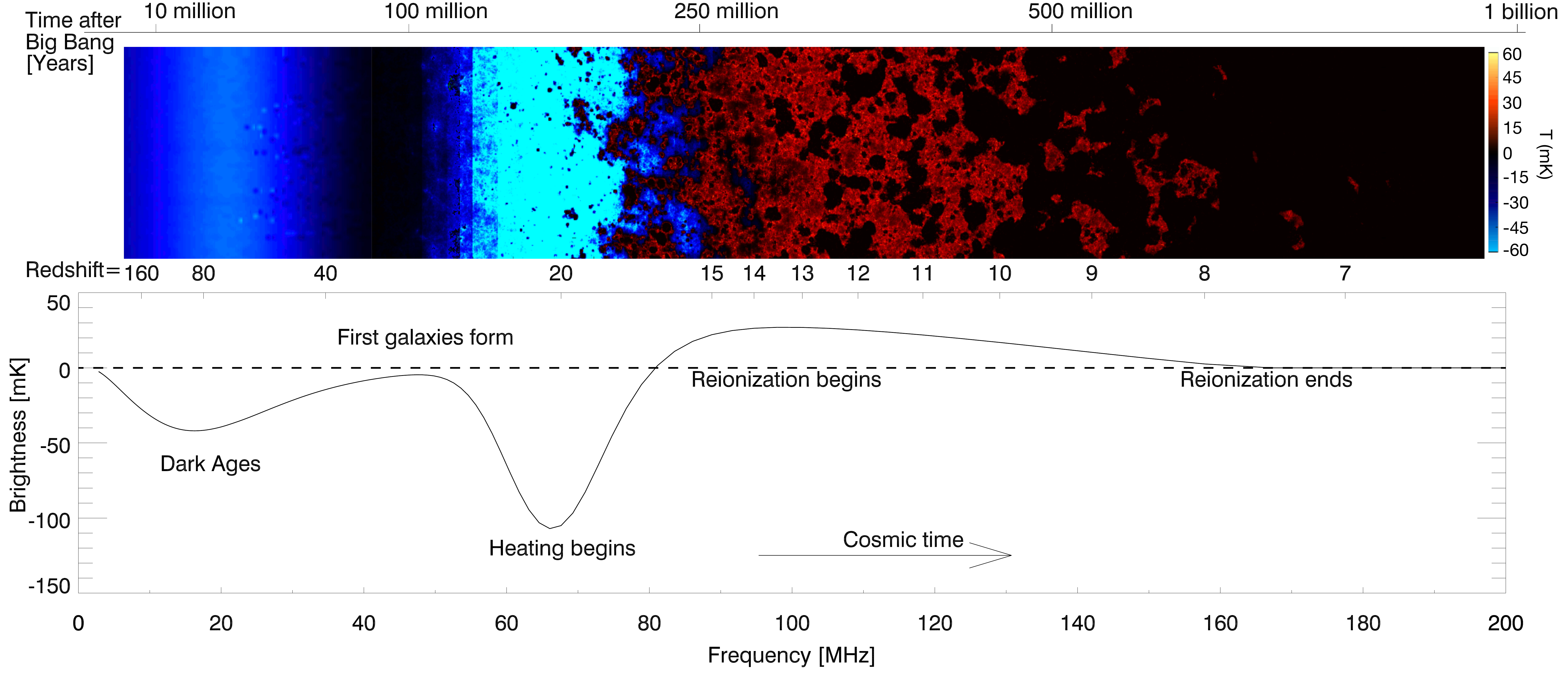}
	\caption[The evolution of the global 21\,cm signal.]{The 21\,cm brightness temperature evolves in complex ways over the course of the dark ages and the Cosmic Dawn. Different physical processes at different times cause it to appear either in absorption or in emission in constrast to the CMB, sometimes globally and sometimes inhomogenously. \emph{Top panel:} one slice through a simulation shows the evolution of the brightness temperature of the signal and the patchy heating and ionization caused by the first generation stars and galaxies. \emph{Bottom panel:} the sky-averaged global 21\,cm signal, which largely traces the evolution of the spin temperature and neutral fraction of hydrogen before and during the Cosmic Dawn. \emph{Reproduced from \cite{PritchardLoebReview}.}}
	\label{fig:GlobalSignal} 
\end{figure*}

In the top panel of Figure \ref{fig:GlobalSignal}, I show one possible history of $\delta T_b^\text{obs}$, reproduced from \citet{PritchardLoebReview}. We can see readily that the evolution of the brightness temperature is complicated and markedly different during different epochs. Fully extracting cosmological and astrophysical information from this process requires large, detailed maps across many redshifts. 

Such maps are very difficult to produce and interpret, as I will discuss in Section \ref{sec:ObsChallenges}, so it is useful to consider reduced data products that take advantage of the approximate statistical isotropy of the signal. The simplest statistical description of the evolution of $\delta T_b^\text{obs}$ is the sky-averaged global signal. The global signal, plotted in the bottom panel of Figure \ref{fig:GlobalSignal}, is expected to go through peaks and troughs as the spin temperature and ionization fraction evolve before and during the Cosmic Dawn.

Another useful way to statistically probe the 21\,cm signal would be to look for correlations on particular length scales. During reionization, for example, we expect correlations on the characteristic length scale associated with growing ionized bubbles around early galaxies. This quantity is most conveniently represented in Fourier space as the power spectrum, $P(\mathbf{k})$, where 
\beq
\langle \widetilde{\delta T}_b^*(\mathbf{k})\widetilde{\delta T}_b(\mathbf{k}') \rangle \equiv (2\pi)^3 \delta(\mathbf{k} - \mathbf{k}')P(\mathbf{k}),
\eeq
where angle brackets denotes an ensemble average, $\widetilde{\delta T}_b(\mathbf{k})$ is the Fourier transform of $\delta T_b(\mathbf{r})$, and $\delta(\mathbf{k} - \mathbf{k}')$ is the Dirac delta function. If the 21\,cm signal is statistically isotropic---which should be a good approximation---then $P(\mathbf{k})$ reduces $P(k)$. Often the power spectrum is reported as a ``dimensionless'' power spectrum\footnote{For the brightness temperature power spectra we measure in 21\,cm cosmology, it actually has units of temperature squared.} $\Delta_{21}^2(k)$ where 
\beq
\Delta_{21}^2(k) \equiv \frac{k^3}{2\pi^2} P(k).
\eeq

Because the 21\,cm signal is not a Gaussian random field, the power spectrum does not contain all of the cosmological information in the maps themselves. But by measuring just a few values of the power spectrum as a function of $k$ and $z$, we can extract much of the available information while significantly reducing the noise on our final measurements. Most of this thesis is concerned with the estimation of the 21\,cm power spectrum, both in theory and in practice, and how it can be used to constrain the physics behind the Cosmic Dawn.

In the remainder of this section, I will briefly summarize the theorized stages in the evolution of the 21\,cm signal and their observable statistical properties. Further information on these processes can be found in \citet{PritchardLoebReview}.

\subsubsection{High Redshifts}
The 21\,cm signal first becomes distinguishable from the CMB around $z\approx 200$. Before that redshift, residual free electrons couple the gas kinetic temperature to the CMB temperature, setting both $T_K$ and $T_S$ to $T_\gamma$. Around $z=200$, this process is no longer effective and the gas begins to cool adiabatically. Therefore, while the temperature of the CMB goes as $T_\gamma \propto (1+z)$, the gas cools like $T_K \propto (1+z)^2$. As long as collisional coupling is effective, which it is thought to be until $z\approx 40$, this sets $T_S < T_\gamma$ and makes the signal appear in absorption. This process accounts for the first dip in the global signal in Figure \ref{fig:GlobalSignal}. Since $T_S$ is fairly uniform during this period and $x_\text{HI} \approx 1$, spatial fluctuations in the 21\,cm signal are sourced by density fluctuations alone. Being able to observe these fluctuations would provide a spectacularly clean probe of the matter power spectrum and a precise test of $\Lambda$CDM, though observations at this redshift are well beyond the limits of current technology.

The second dip in the global signal is caused by the combination of two processes. As the first stars in the universe form, they produce enough Lyman-alpha photons to couple $T_S$ to $T_\alpha$ via the Wouthuysen-Field effect. Since the universe is mostly neutral and the optical depth to Lyman-alpha in the IGM is very large, $T_\alpha$ is driven toward $T_K$, which is less than $T_\gamma$. That causes the 21\,cm signal to be visible again in absorption. Fluctuations in the 21\,cm field are caused by variations in Lyman-alpha field corresponding to the first dark matter halos to collapse and form stars. Eventually, heating of the IGM by X-ray sources, like the first X-ray binaries and micro-quasars, drives $T_K$ above $T_\gamma$ and the 21\,cm signal into emission. Since this process happens inhomogeneously, it is expect that that the signal will be visible in emission in parts of the sky and absorption in other parts of the sky simultaneously, potentially leading to observable effects in the 21\,cm power spectrum (see Chapter \ref{ch:Forest}). This ``Epoch of X-ray Heating'' drives $T_K \gg T_\gamma$, saturating the $T_S$ term in Equation \ref{eq:deltaTbObs}.

\subsubsection{The Epoch of Reionization}

\begin{figure*}[] 
	\centering 
	\includegraphics[width=1\textwidth]{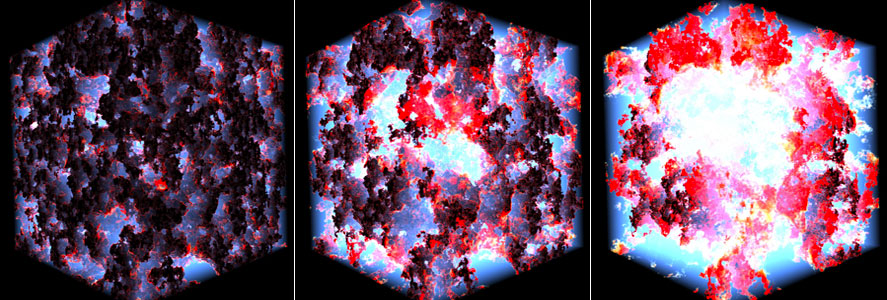}
	\caption[Stages of a reionization simulation.]{A simulation of reionization in a cube approximately 300\,Mpc on a side. As reionization proceeds small bubbles of ionized hydrogen (bright areas) grow to eventually dominate the neutral gas (dark areas) and completely ionize the IGM. \emph{Image credit: Marcelo Alvarez, Ralf Kaehler, and Tom Abel}.}
	\label{fig:ReionizationCube}
\end{figure*}

Around that time, reionization of the intergalactic medium by ultraviolet photons from young, high-mass stars is expected to begin, leading to growing bubbles of ionized gas around early galaxies. As the simulation in Figure \ref{fig:ReionizationCube} shows, ionized bubbles eventually grow and coalesce. This reduces the fraction of neutral hydrogen and thus the strength of the 21\,cm global signal. 

\begin{figure*}[] 
	\centering 
	\includegraphics[width=.8\textwidth]{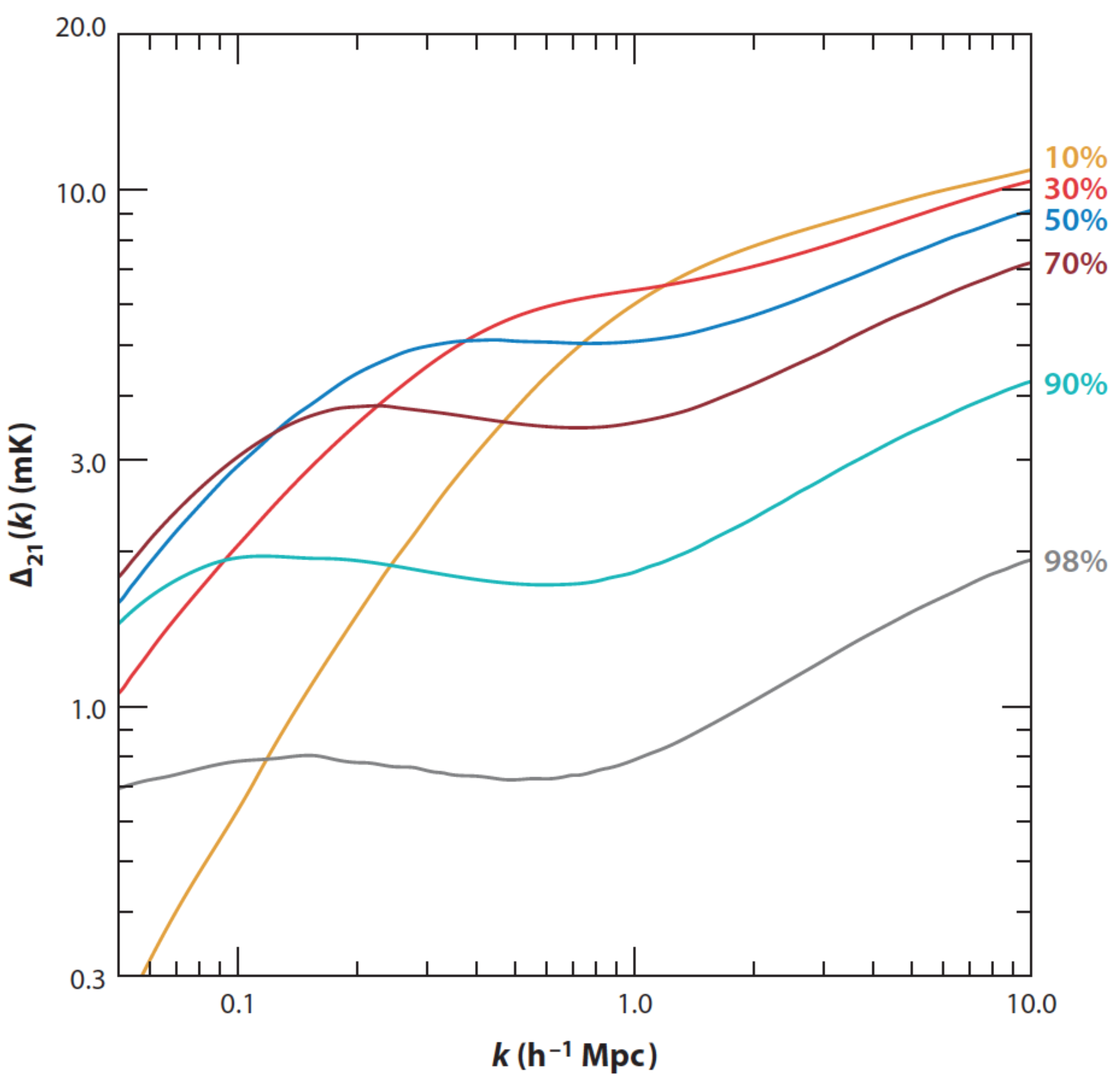}
	\caption[Fiducial reionization power specrtrum.]{The evolution of the 21\,cm power spectrum with ionized fraction is expected to reveal a tremendous amount of information about the processes that drove reionization and the physics of the first stars and galaxies. As the universe goes from mostly neutral (yellow) to mostly ionized (gray), the overall amplitude of the power spectrum is expected to decrease, since $x_\text{HI}$ normalizes $\delta T_b$. However, the growth of ionized bubbles creates correlations on the characteristic size scale of those bubbles, increasing low $k$ power during the early stages of reionization. \emph{Reproduced from \cite{miguelreview}.}}
	\label{fig:FiducialPS}
\end{figure*}

If the spin temperature at reionization is far larger than the temperature of the CMB, then variations in $\delta T_b^\text{obs}$ are created by density and ionization fluctuations, the later of which evolved dramatically over the course of the EoR. At the beginning of reionization, density fluctuations determine the 21\,cm power spectrum, leading to higher power at high $k$ in $\Delta_{21}(k)$. As the ionized bubbles grow, they erase very small scale (high $k$) fluctuations but create correlations on large scales (low $k$). This is reflected in the expected evolution of the 21\,cm power spectrum in Figure \ref{fig:FiducialPS}. As reionization proceeds, the overall amplitude of the power spectrum decreases because it is proportional to $x_\text{HI}^2$. But we also see the formation of the ``knee'' in the power spectrum that moves to lower $k$ as the characteristic bubble size increases.

Simulations of the 21\,cm power spectrum \cite{McQuinn06} have found that it depends more strongly on $x_\text{HI}$ than on the redshift of reionization. It follows them that the power spectrum will be a sensitive probe of the ionization history of the universe, which is still largely unknown. 

The exact shape of the power spectrum and its evolution with $x_\text{HI}$ or $z$ depends sensitively on the astrophysics of reionization. In Chapter \ref{ch:NextGen} my collaborators and I examine the qualitative differences between power spectra when varying parameters of a relatively simplistic reionization model. Because the 21\,cm power spectrum varies so dramatically over cosmic time as a function of $k$, it can be used to sensitively probe the physics that drove it. Specifically, we found that a next-generation telescope could constrain these parameters at roughly the 5\% level using the power spectrum. Though we know that reionization was over by  $z\approx 6$, we don't know exactly when it began or how long it took. Thus, observations aimed at this signal usually observe at $13 \lesssim z \lesssim 6$, corresponding to a frequency between 100 and 200\,MHz.

Of course, part of the promise of 21\,cm cosmology is that it makes an enormous volume of the universe accessible to observation, providing an exquisite test of $\Lambda$CDM and possible extensions to it. If $P(\mathbf{k})$ is decomposed into powers of $\mu$ where $\mu \equiv \mathbf{k}\cdot \rhat$, it can be shown from linear perturbation theory that the $\mu^4$ term depends only on density fluctuations \cite{Barkana1,Barkana2}. With a large enough telescope optimized for 21\,cm cosmology, \citet{Yi} showed that 21\,cm power spectra measured over a fairly large range of redshifts can reduce the errors on cosmological parameters like $\Omega_\Lambda$, $\Omega_m$, $\Omega_b$, $n_s$, $\Omega_k$, and $\sum m_\nu$ by an order of magnitude or more compared to what's possible with current CMB observations. While these measurements are still rather futuristic, they serve as a shining example of what's possible with 21\,cm tomography.

\subsubsection{Low Redshifts}
Though hydrogen in the IGM was completely ionized by $z\approx 6$, galactic halos can still host residual neutral hydrogen where densities are high enough that recombination rates exceed ionization rates, shielding the neutral gas. While it will be very difficult to observe individual galaxies, low resolution images that average together emission from many galaxies may enable a measurement of the underlying matter power spectrum. However, this requires modeling the bias factor that relates dark matter halos to the amount of neutral hydrogen that they host, which may vary as a function of galaxy mass, size, and age in non-trivial ways. 

More promising is the ongoing effort to measure baryon acoustic oscillations in the power spectrum \cite{wyithe2008,ChangDE,CHIMEpathfinder}. Since the baryon acoustic scale serves as a standard ruler, measuring it in the 21\,cm power spectrum as a function of $z$ can constrain the expansion history of the universe and thus the dark energy equation of state. Since the acoustic scale at 150\,Mpc is much larger than individual galaxies, the difficulty of measuring the signal from individual galaxies is less important than ease of building sensitive telescopes with wide fields of view and precise redshift information. Unlike with optical and infrared surveys that have measured the baryon acoustic signal, 21\,cm ``intensity mapping'' experiments get redshift information basically for free, potentially making cosmic-variance-limited measurements relatively inexpensive.

\section{Observational Challenges of 21\,cm Cosmology} \label{sec:ObsChallenges}
Though a detection and characterization of the 21\,cm signal from the epoch of reionization would be an invaluable tool for understanding our Cosmic Dawn, actually making the measurement has proven extremely difficult. In fact, Parts I and II of this thesis are devoted to exploring and overcoming both the theoretical and real-world challenges of making a detection. In this section I will review the basics of interferometry (Section \ref{sec:radioInterferometers}), how we plan to separate out astrophysical foregrounds that are many orders of magnitude stronger than the cosmological signal (Section \ref{sec:theproblemofforegrounds}), and the current (Section \ref{sec:firstgentelescopes}) and next generation (Section \ref{sec:nextgentelescopes}) efforts to detect the 21\,cm signal.

\subsection{Low Frequency Radio Interferometers} \label{sec:radioInterferometers}
\begin{figure*}[] 
	\centering 
	\includegraphics[width=1\textwidth]{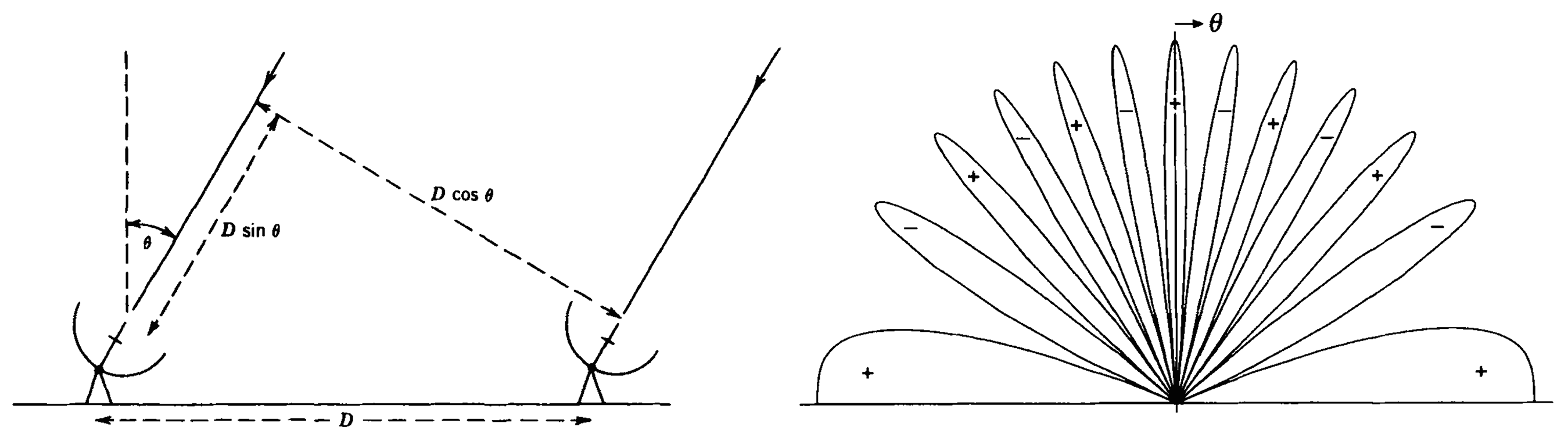}
	\caption[Schematic diagram of the basics of radio interferometry.]{By correlating the signals from two spatially seprated antennas, radio interferometers are sensitive to the delay between when light from a source arrives at one antenna and when it arrives at the other (see the lefthand panel). These correlations, called ``visibilities,'' are weighted measurements of Fourier modes on the sky (see the righthand panel). With many such measurements with a variety of antenna separations and relative orientations, images of the sky can be reconstructed with high sensitivity. \emph{Reproduced from \cite{ThompsonMoranSwenson}.}}
	\label{fig:interferometry}
\end{figure*}

Unlike traditional telescopes that measure energy deposited in a focal plane, radio telescopes measure incident electric fields from the sky directly. If we make the generally very accurate approximation that radio emission from different sources on the sky is incoherent, then it follows that the correlation of measurements from different antennas can tell us about what's on the sky. We call this time-averaged correlation between signals measured at antenna $i$ and antenna $j$ the ``visibility,'' $V_{ij}$. It's given by\footnote{For this discussion, I ignore the complications that arise when measuring a polarized signal. A more complete treatment can be found in \citet{ThompsonMoranSwenson}.}
\beq
V_{ij}(\nu) = \int B_{ij}\left(\rhat,\nu \right)
I(\rhat,\nu) \exp\left[-2 \pi i  \frac{\nu}{c} \base_{ij} \cdot \rhat \right] d\Omega. \label{eq:visibilitydefinition}
\eeq  
This equation can be interpreted as saying that a pair of antennas displaced by vector $\base_{ij}$ are sensitive to the sky, $I(\rhat,\nu)$, weighted by the product of the sensitivities of the antennas, $B_{ij}\left(\rhat,\nu \right)$, also known as the ``primary beam.'' However, the correlation between the signals from two antennas is only observed with an extra time-delay corresponding to the separation between antennas along the line of sight to a source (see the lefthand panel of Figure \ref{fig:interferometry}). This extra time delay introduces the phase factor in Equation \ref{eq:visibilitydefinition}.

As a result of that phase factor, visibilities really measure Fourier modes of the beam-weighted sky. Parts of the sky interfere constructively, other parts destructively, as the righthand panel of Figure \ref{fig:interferometry} illustrates. A pair of antennas can be very sensitive to changes in position perpendicular to their orientation, since that can rapidly change the phase factor. If the antennas are nearby, or if position changes are perpendicular to their separation, the phase changes slowly. 

This can be generalized. With $N$ antennas, we can measure $N(N-1)/2$ different visibilities. As the Earth rotates, $\base_{ij} \cdot \rhat$ changes, allowing for the measurement of new Fourier modes. With enough independently measured Fourier components of the sky, an image can be reconstructed via ``aperture synthesis.'' These measure the sky convolved with a point spread function (PSF) or ``synthesized beam'' related to the observed antenna separations or ``baselines.''

Typically, astronomers build interferometric telescopes because they are interested in making measurements with very high angular resolution. Roughly speaking, the angular resolution of an interferometer is set by $\lambda / b_\text{max}$, the ratio of the wavelength observed to the longest baseline. For 21\,cm cosmology, our aim is not angular resolution---we get most of our sensitivity to small spatial scales from spectral resolution---but high sensitivity and large fields of view. The cost of large, single-dish radio telescopes usually scales with the collecting area as $A^{1.35}$ \cite{FFTT}. The physical hardware cost of building a radio interferometers scales only linearly with the collecting area, since more antennas yield more sensitivity. The computing cost of performing the correlation between antennas to calculate visibilities usually scales as $N^2$ and for large enough $N$, it can be a limiting factor. This is not true for all interferometers, as I will explain in Chapter \ref{ch:MITEOR}.

High sensitivity is extremely important for 21\,cm cosmology precisely because the 21\,cm signal is so weak compared to the astrophysical foregrounds, as I will discuss in Section \ref{sec:theproblemofforegrounds}. Since most of the signal measured by a radio antenna comes from incoherent sky signals, the noise in a visibility is set by $T_\text{sky}$, which is roughly the average sky brightness temperature. $T_\text{sky}$ sets the system temperature, $T_\text{sys}$ because it is usually hundreds of Kelvin at EoR frequencies and thus dominates over the electronic noise in the receiver. The relationship between noise in a visibility and noise in the power spectrum is discussed in Chapters \ref{ch:FastPowerSpectrum} and \ref{ch:Mapmaking}. Suffice it to say that first generation instruments (which I will discuss in greater detail in Section \ref{sec:firstgentelescopes}) likely need a thousand or more hours of observation to make a confident detection of the EoR signal \cite{MiguelNoise,Judd06,LidzRiseFall,LOFAR2,AaronSensitivity,MWAsensitivity,PoberNextGen}. Thus, the need for large collecting areas, combined with the relative inexpensiveness of individual antenna elements designed to operate at low frequencies, has driven the field toward interferometers.

\subsection{The Problem of Foregrounds} \label{sec:theproblemofforegrounds}
Astrophysical foregrounds remain the most daunting challenge for 21\,cm cosmology. The brightness temperature we measure on the sky inevitably contains both the 21\,cm signal from the Cosmic Dawn and relatively nearby, radio-bright objects that fill the entire sky at the angular resolution of our instruments. Our hope of separating the astrophysical foregrounds relies on their spectral smoothness. Measurements of CMB anisotropies faced a similar problem; they also contained smooth spectrum foregrounds much brighter than the signal they sought. In the case of the CMB, measurements at different frequencies have the same thermal blackbody signal and the same foreground contaminants. The strategy for the CMB was to look at different frequencies to differentiate the two based on their frequency dependence. In the case of 21\,cm cosmology, each frequency probes an entirely new cosmological signal. That's the whole point. The ability for tomography to explore a vast volume is also the reason why the problem of foregrounds is so difficult. We need new approaches which take their cues from previous work on the CMB but must be adapted to the thornier problem at hand. In this section, I will explain what the foregrounds are, how they appear in our measurements, and what we can do about them.

\subsubsection{What Are the Foregrounds?}

\begin{figure*}[] 
	\centering 
	\includegraphics[width=.8\textwidth]{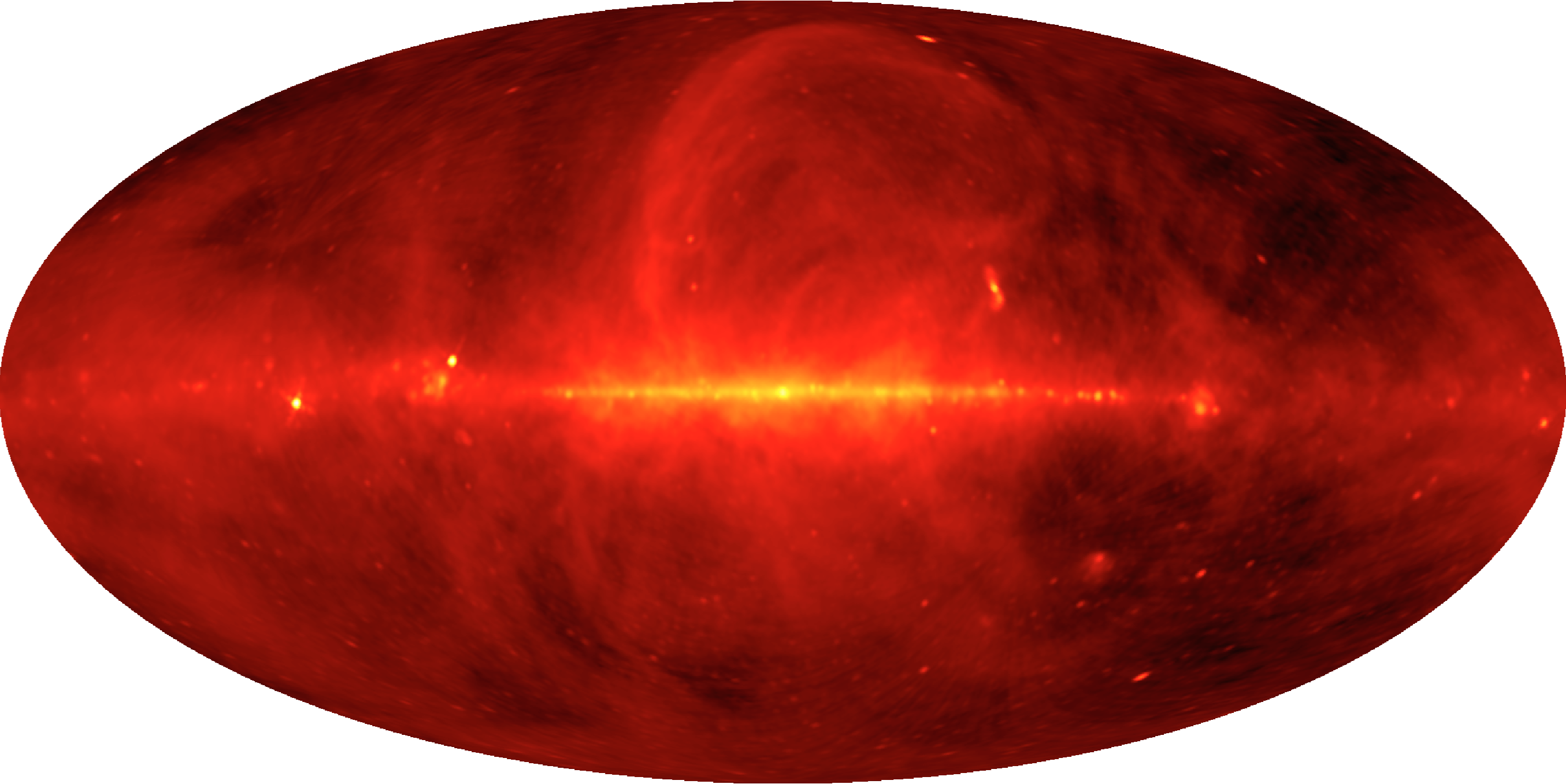}
	\caption[The global sky model of diffuse radio emission at 150\,MHz.]{A map of diffuse radio emission, mostly from our Galaxy, at 150\,MHz. At EoR frequencies, the galaxy is hundreds of Kelvin, compared to the cosmological 21\,cm signal which is likely less than 10\,mK. \emph{Produced using the results of \cite{GSM}.}}
	\label{fig:GalaxyAsForeground}
\end{figure*}

At the frequencies of interest, the dominant foregrounds are synchrotron emission from our Galaxy and other radio galaxies. Synchrotron emission from our Galaxy---the result of ultrarelativistic charged particles bending in the Galaxy's magnetic fields---has some spatial structure, but is highly spatially correlated, as I show in Figure \ref{fig:GalaxyAsForeground}. Free-free emission also contributes, albeit at a much lower level \cite{FurlanettoReview}. Both sources produce very spectrally smooth foregrounds because of the physical mechanism behind synchrotron and free-free emission. 

Additionally, bright radio galaxies, which are usually unresolved by our instruments, contribute considerable flux. They are generally sourced by the interaction of jets from active galactic nuclei with the surrounding IGM. They too are synchrotron dominated and are therefore spectrally smooth. Many sources contaminate every pixel of our maps and create a confusion-limited sea of unresolved point sources. 

Because the dominant foregrounds are driven by processes that create inherently spectrally smooth emission, they can be well-characterized using maps at just a few frequencies. When we make maps at hundreds of frequencies, as we often do in 21\,cm tomography, we can expect only a small fraction of the total information about the cosmological signal to be completely lost due to foreground uncertainty \cite{AdrianForegrounds}.

There are also foregrounds that are not so spectrally smooth. Man-made radio frequency interference (RFI) can be even brighter than the astrophysical foregrounds, but can usually be isolated in time and frequency and mitigated by building arrays at remote sites. Polarized foregrounds, if they leak into maps of unpolarized emission, may also acquire spectral structure due to Faraday rotation. So far, this effect appears to be small \cite{MoorePolarization}.

\subsubsection{How Foregrounds Interact with an Interferometer}

An interferometer is an inherently chromatic instrument. The phase term in Equation \ref{eq:visibilitydefinition} depends frequency, so we should expect that the PSF or synthesized beam should also depend on frequency in non-trivial ways. The primary beam is also frequency dependent, so PSFs can vary spatially as well. Taking this into account properly is the subject of Chapter \ref{ch:Mapmaking}. 

The spatial and spectra dependence of PSFs complicates the simple story of how foregrounds can be separated from the 21\,cm signal. While intrinsic foregrounds are very spectrally smooth, observed foregrounds can have complex spectral structure. With a sufficiently precise understanding of the operation of the instrument---including exquisite calibration---the complex spectral structure can be modeled with just a few foreground parameters per line of sight. But actually understanding our instrumental calibration and primary beams to the roughly 0.01\% level necessary is very difficult.

\begin{figure*}[] 
	\centering 
	\includegraphics[width=1\textwidth]{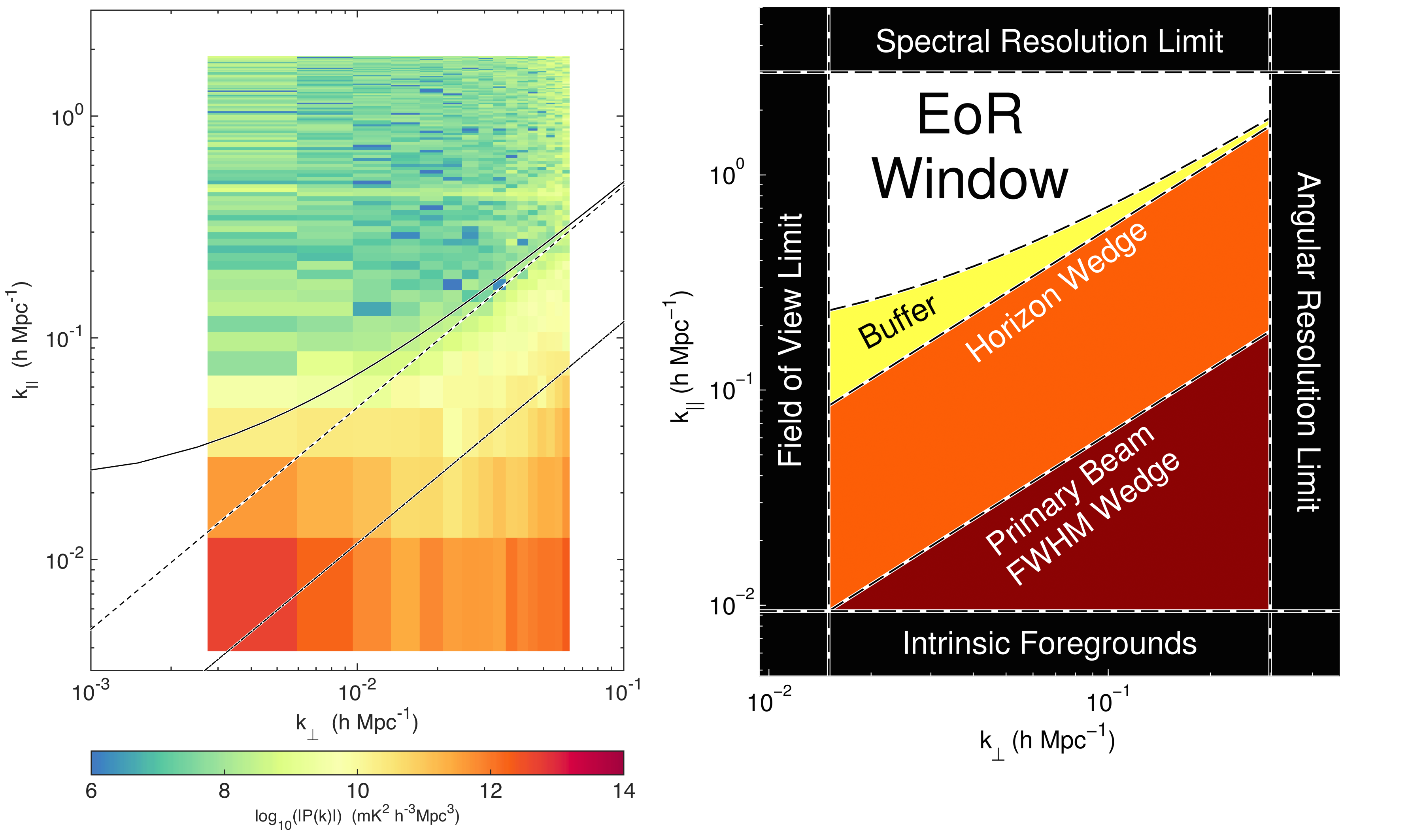}
	\caption[The wedge both schematically and in data.]{When bright but spectrally smooth astrophysical foregrounds are observed with an interferometer, an inherently chromatic insrument, they take on spectral structure. In 2-D Fourier space, where $k_\|$ measures Fourier modes along the line of sight and $k_\perp$ measures Fourier modes perpendicular to the line of sight, the foregrounds show up as a ``wedge.'' That's because finer spatial scales---probed by longer baselines---have more spectral structure. The safest way to detect the 21\,cm power spectrum is to work outside the wedge, in the ``EoR window'' (righthand panel). The EoR window has thus far proven relatively foreground free (see the lefthand panel), though working only in the window comes at the cost of sensitivity, as I will discuss further in Chapter \ref{ch:NextGen}. \emph{Figures reproduced from Chapters \ref{ch:Mapmaking} and \ref{ch:EmpiricalCovariance}, respectively.}}
	\label{fig:WedgeTheoryData}
\end{figure*}

So, when we make 3-D maps we expect foreground contamination at every frequency, which is a proxy for distance. The signal we'd ultimately like to measure depends only on $|\kvec|\equiv k$. However, to separate foregrounds, which behave differently along the line of sight than perpendicular to it, we form power spectra in cylindrically-averaged 2-D Fourier space, parametrized by $k_\|$ and $k_\perp$. Were it not for the chromaticity of the instrument, we would expect foregrounds to only contaminate the lowest $k_\|$ modes. But, as we can see in the 2-D power spectrum plotted in the lefthand panel of Figure \ref{fig:WedgeTheoryData}, the brightest, most foreground-dominated region depends both on $k_\|$ and $k_\perp$. 

Thankfully, the smallest scale of spectral structure the instrument can impart on a given baseline corresponds to the geometric delay associated with sources at the horizon \cite{AaronDelay}. There the phase term in Equation \ref{eq:visibilitydefinition} is maximized. Baseline length determines angular resolution and thus spatial resolution. Therefore longer baselines probe higher $k_\perp$ modes of the 21\,cm power spectrum. Likewise, since delay is a Fourier dual to frequency which is a substitute for distance in 21\,cm tomography, longer delays correspond to higher $k_\|$ modes. This explains the structure we see in Figure \ref{fig:WedgeTheoryData} called ``the wedge,'' which has also been seen in simulations \cite{Dattapowerspec,AaronDelay,VedanthamWedge,MoralesPSShapes,Hazelton2013,CathWedge,ThyagarajanWedge,EoRWindow1,EoRWindow2} and in observations \cite{PoberWedge,X13,NithyaPitchfork}.

Outside the wedge lies the so-called ``EoR window'' which should be free of foreground contamination. Exactly where the wedge-window divide occurs depends on the instrument and the foregrounds. While most interferometers are designed to have little sensitivity near the horizon, a large fraction of the total solid angle of the observable celestial sphere is near the horizon \cite{NithyaPitchfork}. While most observed foreground emission may fall within the main lobe of the primary beam, enough foregrounds to swamp the cosmological signal may still be present in the sidelobes. If the foregrounds have some spectral structure, they are expected to leak into a buffer just beyond the wedge \cite{PoberWedge,EoRWindow1}, as I show in the schematic illustration of the EoR window in the righthand panel of Figure \ref{fig:WedgeTheoryData}.

\subsubsection{Two Strategies for Foreground Removal}

The current leading strategy for detecting the 21\,cm EoR signal relies on avoiding foregrounds by working only within the window. The current best limits in \citet{PAPER64Limits} and the strategy my collaborators and I employed in Chapters \ref{ch:MWAX13} and \ref{ch:EmpiricalCovariance} used only data from inside the the window. As I will discuss in Sections \ref{sec:firstgentelescopes} and \ref{sec:nextgentelescopes}, some telescopes are being designed to take advantage of this strategy and eschew imaging fidelity and angular resolution in favor of many short, redundant baselines that probe low $k_\perp$ modes less contaminated by the wedge. 

The downside to foreground avoidance is that it sacrifices sensitivity. As my coauthors and I found in Chapter \ref{ch:NextGen}, giving up on Fourier modes near the edge of the wedge results in a roughly 70\% drop in sensitivity even for a highly compact array. Using the yellow and perhaps even the orange modes in the righthand panel of Figure \ref{fig:WedgeTheoryData} can mean the difference between an upper limit on the 21\,cm power spectrum with current generation interferometers and a solid detection.

To work in those regions, we must find a way to subtract foregrounds from our data. Foreground subtraction is very difficult and has been the subject of many papers over the last several years (e.g.~\cite{Miguelstatistical,Judd08,paper2,LT11,Chapman1,DillonFast,Chapman2}). We need to subtract foregrounds orders of magnitude stronger than the cosmological signal which have been convolved with an instrument whose effect is only imperfectly understood. We need precise models of both foregrounds and our instrument. And most importantly, we must take our own uncertainty about these models into account. If we do not, we risk mistakenly claiming a detection. Much of this thesis (Chapters \ref{ch:FastPowerSpectrum}, \ref{ch:Mapmaking}, \ref{ch:MWAX13}, and \ref{ch:EmpiricalCovariance}) is concerned with precisely this question: what do we need to know to subtract foregrounds and how do we translate our uncertainty about their subtraction into errors on our power spectrum measurements? The goal is to claw back as much of the EoR window as we are justified in doing, and no more.

Even if we are simply seeking to avoid foregrounds by excising the wedge region, the techniques my collaborators and I have developed are important because they can minimize the leakage of foreground power into the EoR window (see Chapters \ref{ch:MWAX13} and \ref{ch:EmpiricalCovariance}). Regardless of whether or not we work within the wedge, we need to know the errors on our measurements, the correlations between those errors, and the relationship of our measurements to the true cosmological $P(k)$. 

Whether or not we will ever understand the foregrounds and our instruments well enough to work within the wedge is an open question. Perhaps the most important message of this thesis is that we should try to achieve the marked increase in sensitivity possible with foreground subtraction and that, even if we fail, as long as we understand our uncertainties, we'll make the best measurements that we can.

\subsection{First Generation Interferometers and Results} \label{sec:firstgentelescopes}

The quest to detect the 21\,cm signal from the epoch of reionization is well underway and a number of telescopes have set limits on the power spectrum. In this section, I'll discuss several of them, review their progress thus far, and compare their strategies for detecting the 21\,cm signal from the EoR.

\subsubsection{Giant Metrewave Radio Telescope}
\begin{figure*}[] 
	\centering 
	\includegraphics[width=1\textwidth]{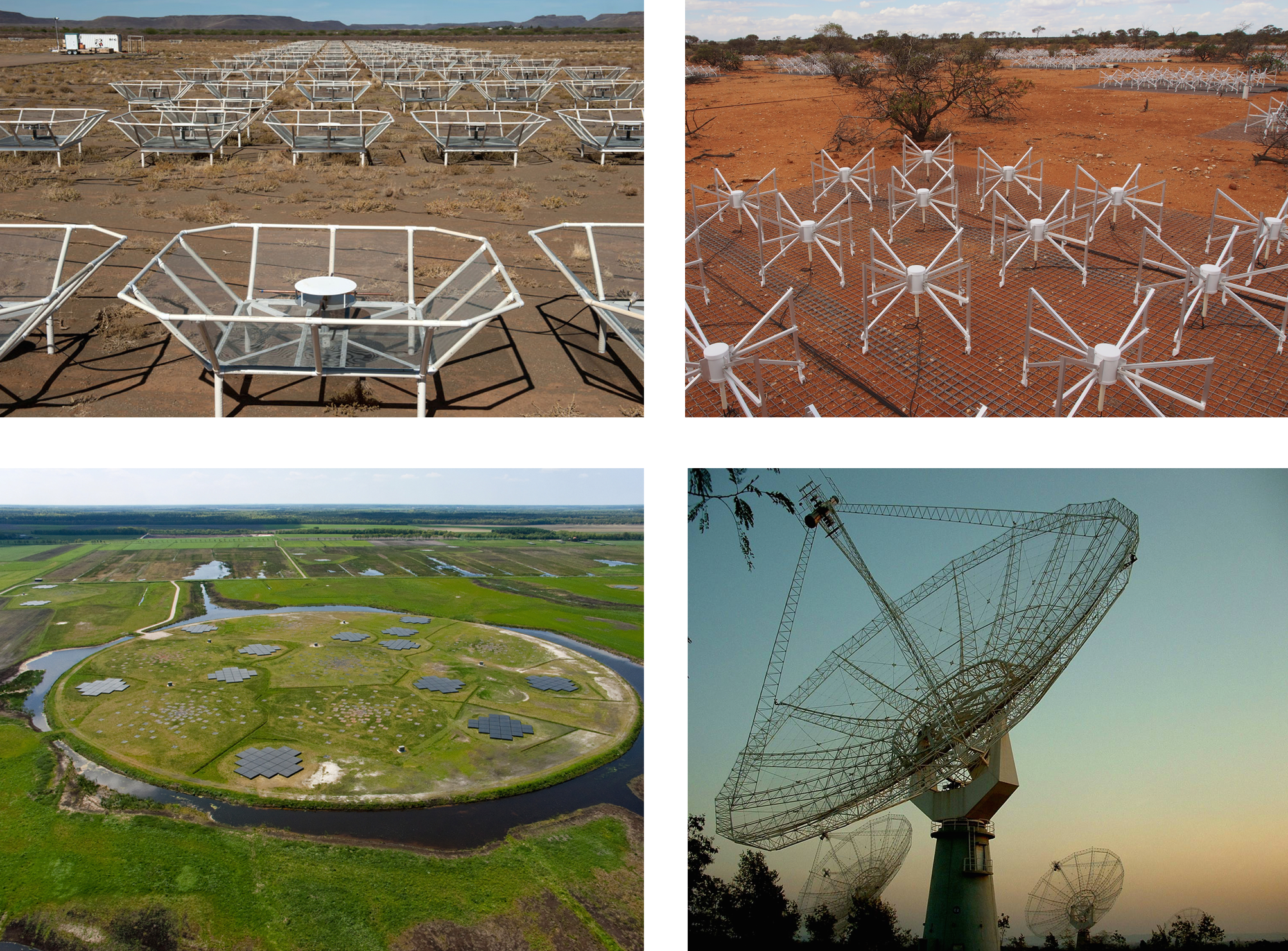}
	\caption[First generation interferometers: PAPER, MWA, LOFAR, and GMRT.]{First generation interferometers trying to detect the 21\,cm signal from the EoR. \emph{Top left:} the Donald C. Backer Precision Array for Probing the Epoch of Reionization (PAPER). \emph{Top right:} the Murchison Widefield Array (MWA). \emph{Bottom left:} the Low Frequency Array (LOFAR). \emph{Bottom right: } the Giant Metrewave Radio Telescope (GMRT). \emph{Image credits: SKA South Africa, the SKA Collaboration, LOFAR/ASTRON, and Tzu-Ching Chang, respectively.}}
	\label{fig:firstgen}
\end{figure*}

The Giant Metrewave Radio Telescope (GMRT) is the oldest of the 21\,cm observatories and consists of 30 steerable dishes, each 45\,m in diameter (see the lower right panel of Figure \ref{fig:firstgen}). At the frequencies of interest, this yields a field of view roughly $3^\circ$ across. It is a multi-purpose observatory located 80\,km North of Pune, India. 

The first upper limit on the 21\,cm power spectrum was set with the GMRT \cite{GMRT}, though it was later revised when it was discovered that the analysis technique for removing foregrounds also removed 21\,cm signal \cite{newGMRT}. The current best limit from GMRT is $\Delta^2(k) < 6.2\times 10^4$\,mK$^{2}$ at $z=8.6$ and $k=0.50$\,h\,Mpc$^{-1}$.

\subsubsection{The Murchison Widefield Array}
The Murchison Widefield Array (MWA), like the GMRT, is a multi-purpose observatory. However, is more focused on 21\,cm cosmology than previous telescopes. It consists of 128 tiles, each made of 16 dual-polarization dipole antennas (see the upper right panel of Figure \ref{fig:firstgen}). The signals from the dipoles are added with an appropriate set of delays by an analog beamformer to focus the sensitivity of the array on particular parts of the sky. This allows the MWA to form a discrete set of primary beams on the sky, each with a full-width at half-maximum of roughly $30^\circ$. For EoR observations, this allows observers to adapt a ``drift and shift'' strategy, where the primary beam changes roughly once every half hour. The MWA is located in the Murchison Radio-astronomy Observatory in a remote part of Western Australia, 600\,km north of Perth.

Chapter \ref{ch:MWAX13} of this thesis contains an analysis of 32-tile MWA prototype data. Chapter \ref{ch:EmpiricalCovariance} updates that analysis with new foreground residual covariance modeling and applies it to 128-tile data, yielding a best (though as-yet-unpublished) upper limit of $\Delta^2(k) < 3.7\times 10^4$\,mK$^2$ at $z=6.8$ and $k=0.18$\,h\,Mpc$^{-1}$. Both chapters contain significantly more detail about the design and operation of the instrument. The MWA has over 1000 hours of total observation already on disk (split across two fields and two frequency bands) and analysis of deeper observations is ongoing.

\subsubsection{The Precision Array for Probing the Epoch of Reionization}
The Donald C. Backer Precision Array for Probing the Epoch of Reionization (PAPER) differs from competing telescopes in that it is a focused experiment designed exclusively for EoR observations. It is located in the Karoo Radio Astronomy Reserve in the Karoo Desert of South Africa. Its 128 dipoles sit atop relatively small frustrum-shaped ground screens arranged in a highly redundant configuration (see the top left panel of Figure \ref{fig:firstgen}). The redundant configuration simplifies calibration (see Chapter \ref{ch:MITEOR}) and focuses the maximum sensitivity on a small number of baselines \cite{AaronDelay}. However, by foregoing imaging fidelity, it makes foreground avoidance the only feasible strategy.

Despite having the least collecting area, its focused design and observing strategy have helped PAPER produce the world's best upper limits on the 21\,cm power spectrum. Using only the previous 64 element configuration, PAPER set an upper limit of $\Delta^2(k) < 500$\,mK$^{2}$ at $z=8.4$ between $k=0.15$ and $k=0.50$\,h\,Mpc$^{-1}$ \cite{PAPER64Limits}. This limit allowed the PAPER team to determine that the IGM was heated above roughly 7\,K at $z=8.4$, otherwise $T_S$ would be so far below $T_\gamma$ that the 21\,cm signal would show up brightly in absorption \cite{PoberPAPER64Heating}. Under a wide range of assumptions, achieving that level of heating requires inferring a population of high-redshift galaxies dimmer than those currently directly observed. This result is not surprising, but it one of the first constraints on the Cosmic Dawn from 21\,cm observations.

\subsubsection{The Low Frequency Array}
The Low Frequency Array (LOFAR) is actually two interferometers, the High Band Array, which observes at EoR frequencies, and the Low Band Array, which was designed for other science. The High Band Array bears many similarities to the MWA in that each element of the interferometer is a analog phased array of 16 dipoles. In the LOFAR core, which is located near Exloo in the Netherlands, 24 such tiles are arranged into each of 40 ``fields'' (22 of which are visible in the lower left panel of Figure \ref{fig:firstgen}). Though LOFAR has a much larger collecting area than the MWA, it cannot correlate every tile with every other tile and instead generally forms beams digitally on a per-field basis, each about $10^\circ$ across. Because beams are formed digitally, multiple simultaneous beams can be formed within the tile beam, though this process is limited by the tradeoff between simultaneous bandwidth and the the computing power required for correlation of what amounts to multiple interferometers simultaneously. Correlation is generally more costly for LOFAR than for PAPER or the MWA because the high level of RFI in the Netherlands necessitates very fine frequency resolution.

Thus far, LOFAR has not published any upper limits on the 21\,cm power spectrum, though they have published some initial calibration, mapmaking, and source-finding results \cite{InitialLOFAR1}. By utilizing far-flung LOFAR stations across Northern Europe, LOFAR can achieve far higher angular resolution than other telescopes. They are attempting to use the angular resolution to subtract individual sources down past the level of the EoR signal using a number of different subtraction techniques \cite{Chapman1,Chapman2}; their baselines are mostly so long that foreground avoidance is too costly. The LOFAR team is also trying to measure the ``variance statistic,'' which is effectively a power spectrum averaged over all $k$ bins, in order to probe the redshift evolution of the cosmological signal with maximum sensitivity \cite{VarianceStatistic}. Interpreting that result will be more difficult than interpreting a power spectrum and it's not clear whether a measurement of the variance statistic will prove a convincing detection of the EoR.

\subsubsection{MITEoR} 
Though it was not designed to have enough sensitivity to detect the EoR, the MIT EoR experiment (or ``MITEoR'' for short) was a small interferometer constructed over a series of expeditions to The Forks, Maine. By our last expedition in the summer of 2013, we deployed 64 dual-polarization MWA dipoles, all fully correlated. The purpose of the experiment was to demonstrate technology for highly scalable interferometers that use redundant calibration \cite{redundant} which makes Fast Fourier Transform correlation possible \cite{FFTT}. More details on the design, deployment, and initial results from MITEoR can be found in Chapter \ref{ch:MITEOR}.

\subsection{Next Generation 21\,cm Interferometers} \label{sec:nextgentelescopes}
While the first generation of 21\,cm observatories is still taking and analyzing data, hoping to make a detection of the 21\,cm signal, none can do much better than that. To not just detect but also characterize the power spectrum during the epoch of reionization, much larger telescopes are needed. Two are planned, the Hydrogen Epoch of Reionization Array and the Square Kilometre Array, each with different technological heritages and design philosophies.

\subsubsection{The Hydrogen Epoch of Reionization Array}
\begin{figure*}[] 
	\centering 
	\includegraphics[width=1\textwidth]{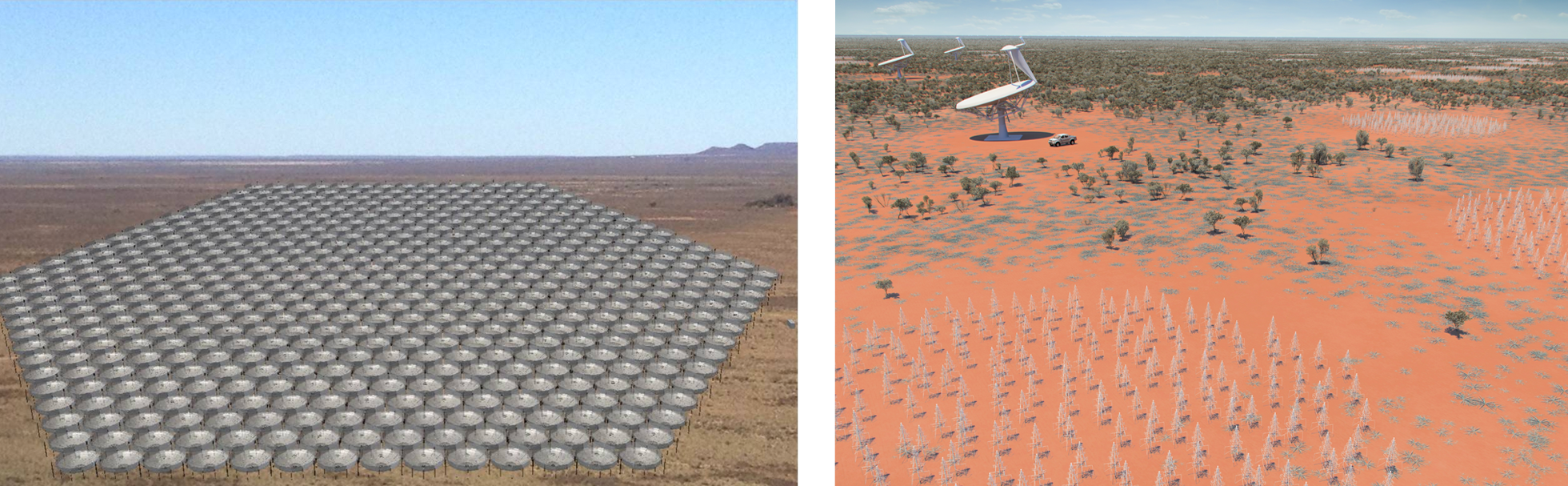}
	\caption[Next generation interferometers: HERA and the SKA.]{Renderings of next generation interferometers: the Hydrogen Epoch of Reionization Array (HERA, left) and the Square Kilometre Array (SKA, right). \emph{Image credits: David DeBoer and the SKA Collaboration, respectively.}}
	\label{fig:HERAandSKA}
\end{figure*}

The Hydrogen Epoch of Reionization Array (HERA) is a planned focused EoR experiment. It will contain 352 crossed dipoles suspended at prime focus over fixed 14\,m parabolic dishes (see the left panel of Figure \ref{fig:HERAandSKA}). HERA is thus a pure drift-scanning instrument. The inner 331 dishes, which are constructed from telephone poles, wire mesh, and PVC pipe, are in a maximally packed hexagonal configuration.\footnote{The hexagonal packing was my first and certainly my most visible contribution to the HERA design.} HERA is funded under the NSF's Mid-Scale Innovations Program to begin construction with 37 dishes. Observations with the first 19 elements are scheduled to begin later this year.

HERA is the spiritual successor to PAPER; it has a densely packed, highly redundant configuration, a simple element design, and is being constructed on the PAPER site in South Africa. It maximizes the collecting area inexpensively by sacrificing sky coverage and the ability to point. Unlike PAPER, its Fourier sampling is dense enough that low-resolution, high-sensitivity imaging should be possible. While HERA is optimized for foreground avoidance, it may be possible to improve its performance with foreground subtraction. HERA's simple design will make this easier, though by no means easy.

Chapter \ref{ch:Mapmaking} of this thesis was written with HERA in mind and uses HERA as a fiducial array. Chapter \ref{ch:NextGen} is an analysis of the ability of HERA detect the EoR and constrain reionization parameters, though it was performed with an earlier design of HERA that called for 547 dishes in the hexagonal core. Regardless, a single observation season with HERA can definitely yield a robust EoR detection and scientifically novel constraints on the physics behind the EoR, even in the foreground avoidance regime.

\subsubsection{The Square Kilometre Array}
By contrast, the Square Kilometre Array (SKA) is the spiritual successor to LOFAR and, to a lesser extent, the MWA. The first phase (SKA1) of the long-planned telescope will actually be two telescopes, the SKA1-LOW near the MWA site in Australia and the SKA1-MID near the PAPER site in South Africa. The SKA1-LOW, the telescope relevant to 21\,cm cosmology during the Cosmic Dawn will consist of 130,000 ``christmas tree'' dipoles (see the righthand panel of Figure \ref{fig:HERAandSKA}) arranged into approximately 500 stations for a total collecting area of about 0.4\,km$^2$. Each dipole will be individually digitized and station dipoles will be added together to form 30 simultaneous beams, each roughly 1 square degree.  Construction of SKA1 is projected to begin in 2018 and be finished by 2023.

Unlike HERA, the SKA is a general purpose observatory with many different scientific objectives. Still, exploring the Cosmic Dawn via a number of probes is one key science drivers \cite{SKACosmologyOverview2015,SKACosmicDawn2015,SKA21cmIGM2015}. Like LOFAR, the SKA will have many fewer short baselines and much less redundancy than HERA, making redundant calibration and foreground avoidance more difficult. For that reason, despite its much larger collecting area the SKA's sensitivity will only be marginally better than HERA's if foregrounds can't be subtracted (see \citet{21CMMC} or Chapter \ref{ch:NextGen} for estimates). On the other hand, with its increased collecting area and resolution, the SKA should be able to easily image the ionized bubbles \cite{SKAImagingHII}, making it a more capable instrument for moving beyond the 21\,cm power spectrum toward other statistical measurements of the EoR \cite{SKAEoRCrossCorr}.

\subsubsection{Omniscopes}
The cost to build very large radio interferometers is eventually dominated by the cost of the correlator. Correlating every element with every other element usually requires computing resources that scale as $N^2$, where $N$ is the number of antenna elements. The MWA, LOFAR, and the SKA avoid this problem using phased arrays, thereby not correlating every antenna with every other antenna, but rather tiles or groups of tiles together. GMRT and HERA get their sensitivity from large individual elements, instead of large $N$, at the cost of field of view. Eventually, if we want to very precisely test $\Lambda$CDM with 21\,cm tomography, we'll want telescopes with both large fields of view and large collecting area \cite{Yi}. The only way I know to achieve that is to build an interferometer that uses fast Fourier transform correlation. 

All telescopes are Fourier transformers. Optical telescopes convert incoming photon momenta into positions in the focal plane. Interferometers sample the incoming radiation field in Fourier space using correlators to compare antenna signals at different baseline separations, effectively performing a discrete Fourier transform. It is not so surprising that, as \citet{FFTT} prove, any regular grid of antennas can be correlated with the fast Fourier transform (FFT). In fact, \citet{FFTT2} showed that any hierarchically regular arrangement of elements can correlated with only $\BigO(N\log N)$ calculations and called that class of telescopes ``omniscopes'' for their broad spectral coverage and wide field of view.

Building such an array will be a major challenge. By design, they only save data from unique baselines, meaning that they must be calibrated in real time. Part of the motivation for Chapter \ref{ch:MITEOR} was to show that the technical advances necessary for this sort of telescope are within reach. HERA, with its highly redundant configuration, will also be an interesting testbed for FFT correlation. I believe that these designs are the future for 21\,cm interferometers with truly massive collecting areas and I'm excited for what that future holds.

\section{A Roadmap for this Thesis}

The work that constitutes this thesis was originally written as seven different papers. The papers appear here as Chapters \ref{ch:FastPowerSpectrum} through \ref{ch:NextGen} and are reproduced verbatim with the permission of their primary co-authors. I played a significant role in the development and writing of all seven papers and served as the first author on four of them---in this thesis, Chapters \ref{ch:FastPowerSpectrum}, \ref{ch:Mapmaking}, \ref{ch:MWAX13}, and \ref{ch:EmpiricalCovariance}. Six of them have already been published in peer-reviewed journals; Chapter \ref{ch:EmpiricalCovariance} has been submitted and is still under review.

Instead of presenting the papers chronologically, I have organized this thesis into three thematic parts. In Part I, \emph{Novel Data Analysis Tools}, I begin with two chapters devoted to rigorous but fast techniques for data analysis for 21\,cm tomography.
\begin{itemize}

\item Chapter \ref{ch:FastPowerSpectrum} reproduces the published paper \emph{A fast method for power spectrum and foreground analysis for 21\,cm cosmology} \cite{DillonFast}, written in collaboration with Adrian Liu and Max Tegmark. It presents a method for fast power spectrum estimation that extends and accelerates the method developed by \citet{LT11}. It also serves as a starting point for the rest of this thesis, much of which focuses on applying and refining these analysis techniques. The work in this chapter was conducted under the supervision of Max Tegmark in close consultation with Adrian Liu, but the project was lead and carried out largely by me.

\item Chapter \ref{ch:Mapmaking} reproduces the published paper \emph{Mapmaking for precision 21\,cm cosmology} \cite{Mapmaking}, written in collaboration with Max Tegmark, Adrian Liu, Aaron Ewall-Wice, Jackie Hewitt, Miguel Morales, Abraham Neben, Aaron Parsons, and Jeff Zheng. It focuses on relaxing a key assumption in Chapter \ref{ch:FastPowerSpectrum} that the PSF is not direction dependent. Understanding the precise statistics of interferometric maps is essential to the separation of Fourier space into the ``EoR window'' and the ``wedge.'' Relaxing this assumption presents a number of computational difficulties, which the second half of the paper focuses on overcoming with a few well-controlled approximations. The work in this chapter was conducted under the supervision of Max Tegmark, whose appendix in \citet{FFTT2} served as the original inspiration for the paper, but was lead and carried out largely by me.
\end{itemize}
 
In Part II, \emph{Early Results from New Telescopes}, I turn from the theoretical development of new data analysis techniques to the application of those methods (and related techniques developed in the Tegmark group) to real data from new radio telescopes---the Murchison Widefield Array and MITEoR. All three chapters in Part II refine previously published analysis techniques to help them meet the challenges of real data. Likewise, all three present the results of those analyses on early data from those telescopes. 
\begin{itemize}

\item Chapter \ref{ch:MWAX13} reproduces the published paper \emph{Overcoming real-world obstacles in 21\,cm power spectrum estimation: A method demonstration and results from early Murchison Widefield Array data} \cite{X13}, co-authored with Adrian Liu and written in collaboration with Chris Williams, Jackie Hewitt, Max Tegmark, and a number of other MWA members. It discusses numerous challenges presented by real-world data that the idealized analyses of \citet{LT11} and Chapter \ref{ch:FastPowerSpectrum} ignored or glossed-over and found ways to consistently deal with them in order to produce the MWA's first limit on the 21\,cm power spectrum. The paper was an equal effort by Adrian Liu and myself. Adrian developed the majority of the methods detailed in Section \ref{sec:Methods} and wrote most of that section. The data was prepared by Chris Williams and I performed the method demonstration and power spectrum analysis that constituted Section \ref{sec:WorkedExample}, most of which I wrote.

\item Chapter \ref{ch:EmpiricalCovariance} reproduces the paper \emph{Empirical covariance modeling for 21\,cm power spectrum estimation: A method demonstration and new limits from early Murchison Widefield Array 128-tile data} \cite{EmpiricalCovariance} which is currently being reviewed by \emph{Physical Review D}. It was written in collaboration with Abraham Neben and under the supervision of Jackie Hewitt and Max Tegmark; the MWA EoR collaboration and Builder's List are also co-authors. The paper is a follow-up to Chapter \ref{ch:MWAX13} and similarly presents new limits on the 21\,cm power spectrum with a few hours of MWA observations. These limits demonstrate the efficacy of the method developed to estimate realistic foreground residual covariance models that are empirically motivated but constrained by our prior beliefs about the frequency structure of the foregrounds. Abraham prepared the maps for power spectrum analysis, provided some of the original ideas for covariance estimation in Fourier space, and wrote Sections \ref{sec:observation} and \ref{sec:processing}. I developed the empirical covariance estimation method, performed the power spectrum analysis, and wrote the rest of the paper.

\item Chapter \ref{ch:MITEOR} reproduces the published paper \emph{MITEoR: a scalable interferometer for precision 21 cm cosmology} \cite{MITEoR}, authored by Jeff Zheng under the supervision of Max Tegmark. Jeff performed the plurality of the work bringing the MITEoR projection to fruition, though it was the culmination of years of effort in the Tegmark group to build and demonstrate an interferometer capable of real-time FFT correlation. I am the fourth author on the paper. My role in the project varied over the years and included data analysis for the first expedition, deployment of several later expeditions, satellite tracking software, and visibility simulations. While Jeff performed the final data analysis and wrote the majority of this paper, I served in a consulting role during the development of the techniques discussed and, along with Max and Adrian Liu, as the primary editor of the paper. Many undergraduate researchers, graduate students, postdocs, and other scientists contributed to the MITEoR project and are also authors on the paper.
\end{itemize}  

Finally, in Part III, \emph{The Cosmic Dawn on the Horizon}, I look forward to what we might be able to measure with next generation 21\,cm interferometers. This part includes two chapters based on previously published forecasts that examine the potential for astrophysical constraints on the first stars, galaxies, black holes and their effect on the IGM.
\begin{itemize}
\item Chapter \ref{ch:Forest} reproduces the published paper \emph{Detecting the 21\,cm forest in the 21\,cm power spectrum} \cite{AaronForest}, written with Aaron Ewall-Wice, Andrei Mesinger, and Jackie Hewitt. The paper investigates the effect of 21\,cm absorption along lines of sight to high-redshift, radio-loud quasars on the the 21\,cm power spectrum. While the effect depends on the relatively unconstrained population of high-redshift quasars and on the thermal history of the IGM, it potentially has a detectible and distinguishable impact on future measurements. This project was lead by Aaron, who performed most of the analysis and wrote most of the paper. Andrei performed the IGM simulations and Jackie supervised the project. As second author, I performed the detailed detectability calculations in Section \ref{sec:det} and served as the primary editor.
 
\item Chapter \ref{ch:NextGen} reproduces the published paper \emph{What Next-Generation 21\,cm Power Spectrum Measurements Can Teach Us About the Epoch of Reionization} \cite{PoberNextGen}, written by Jonnie Pober, Adrian Liu, and myself in collaboration with several other members of the HERA team. The work began with a detailed sensitivity calculation comparison between Jonnie and myself, which eventually led the calculation of the errors HERA should expect on a measurement of the power spectrum for a variety of reionization models and foreground mitigation strategies (Section \ref{sec:results}). Adrian followed up that work with a detailed Fisher matrix analysis of the potential constraints on a parameterized model of reionization in Section \ref{sec:adrian}.

\end{itemize}

It is my hope that this thesis presents a broad picture of how we might eventually overcome the difficulties of detecting the 21\,cm signal, the progress we have already made with the first generation of telescopes, and the exciting science we'll be able to do with those measurements.

\part{Novel Data Analysis Tools}
\chapter{A Fast Method for Power Spectrum and Foreground Analysis for 21\,cm Cosmology} \label{ch:FastPowerSpectrum}

\emph{The content of this chapter was submitted to \emph{Physical Review D} on November 27, 2012 and published \cite{DillonFast} as \emph{A fast method for power spectrum and foreground analysis for 21\,cm cosmology} on February 12, 2013.}

\section{Introduction}
Neutral hydrogen tomography with the 21\,cm line promises to shed light on vast and unexplored epoch of the early universe.  As a cosmological probe, it offers the opportunity to directly learn about the evolution of structure in our universe during the cosmological dark ages and the subsequent Epoch of Reionization (EoR) \cite{BLreview,FurlanettoReview,miguelreview,PritchardLoebReview}.  More importantly, the huge volume of space and wide range of cosmological scales probed makes 21\,cm tomography uniquely suited for precise statistical determination of the parameters that govern modern cosmological and astrophysical models for how our universe transitioned from hot and smooth to cool and clumpy \citep{Matt3,Santos2, juddjackiemiguel1,Whitepaper2,wyithe2008,ChangDE,Rees, Tozzi2, Tozzi, Iliev,furlanetto1,Loeb1,furlanetto2,Barkana1,Mack,Yi,ClesseBackgroundReionizationOmniscopes}.  It has the potential to surpass even the Cosmic Microwave Background (CMB) in its sensitivity as a cosmological probe \citep{Yi}. 

The central idea behind 21\,cm tomography is that images produced by low frequency radio interferometers at different frequencies can create a series of images at different redshifts, forming a three dimensional map of the 21\,cm brightness temperature.  Yet we expect that our images will be dominated by synchrotron emission from our galaxies and others.  In fact, we expect those foreground signals to dominate over the elusive cosmological signal by about four orders of magnitude \citep{Angelica,bernardi}.  

One major challenge for 21\,cm cosmology is the extraction of the brightness temperature power spectrum, a key prediction of theoretical models of the dark ages and the EoR, out from underneath a mountain of foregrounds and instrumental noise.  Liu \& Tegmark (\citep{LT11}, hereafter ``LT'') presented a method for power spectrum estimation that has many advantages over previous approaches (on which we will elaborate in Section \ref{bruteForce}).  It has, however, one unfortunate drawback: it is very slow.  The LT method relies on multiplying and inverting very large matrices, operations that scale as $\BigO(N^3)$, where $N$ is the number of voxels of data to analyze.  

The goal of the present paper is to develop and demonstrate a way of achieving the results of the LT method that scales only as $\BigO(N\log N)$.  Along the way, we will also show how LT can be extended to take advantage of additional information about the brightest point sources in the map while maintaining a reasonable algorithmic scaling with $N$.  Current generation interferometers, including the Low Frequency Array (LOFAR, \citep{LOFARinstrument}), the Giant Metrewave Radio Telescope (GMRT, \citep{GMRT}), the Murchinson Widefield Array (MWA, \citep{TingaySummary}), and the Precision Array for Probing the Epoch of Reionization (PAPER, \citep{PAPER}) are already producing massive data sets at or near the megavoxel scale (e.g. \citep{ChrisMWA}).  These data sets are simply too large to be tackled by the LT method.  We expect next generation observational efforts, like the Hydrogen Epoch of Reionization Array \cite{HERA}, a massive Omniscope \citep{FFTT2}, or the Square Kilometer Array \cite{SKAspecifications}, to produce even larger volumes of data.  Moreover, as computer processing speed continues to grow exponentially, the ability to observe with increasingly fine frequency resolution will enable the investigation of the higher Fourier modes of the power spectrum at the cost of yet larger data sets.  The need for an acceleration of the LT method is pressing and becoming more urgent.

Our paper has a similar objective to \citep{UeLiFast}, which also seeks to speed up algorithms for power spectrum estimation with iterative and Monte Carlo techniques.  The major differences between the paper arise from the our specialization to the problem of 21\,cm cosmology and the added complications presented by foregrounds, especially with regard to the basis in which various covariance matrices are easiest to manipulate.   Our paper also shares similarities to \citep{GibbsPSE}.  Like \citep{UeLiFast}, \citep{GibbsPSE} does not extend its analysis to include foregrounds.  It differs also from this paper in spirit because that it seeks to go from interferometric visibilities to a power spectrum within a Bayesian framework rather than from a map to a power spectrum and because it considers one frequency channel at a time.  In this paper, we take advantage of many frequency channels simultaneously in order to address the problem of foregrounds.

This paper is organized as follows.  We begin with Section \ref{bruteForce} wherein we review the motivation for and details of the LT method.  In Section \ref{Fast} we present the novel aspects of our technique for measuring the 21\,cm brightness temperature power spectrum.  We discuss the extension of the method to bright point sources and the assumptions we must make to accelerate our analysis.  In Section \ref{results} we demonstrate end-to-end tests of the algorithm and show some of its first predictions for the ability of the upcoming 128-tile deployment of the MWA to detect the statistical signal of the Epoch of Reionization.
 

\section{The Brute Force Method}\label{bruteForce} 

The solution to the problem of power spectrum estimation in the presence of foregrounds put forward by LT offers a number of improvements over previous proposals that rely primarily on line of sight foreground information \citep{xiaomin,nusserforegrounds,Judd08,paper1,LOFAR,Harker,paper2,LOFAR2,ChoForegrounds}.  The problem of 21\,cm power spectrum estimation shares essential qualities with both CMB and galaxy survey power spectrum estimation efforts.  Like with galaxy surveys, we are interested in measuring a three dimensional power spectrum.  On the other hand, our noise and foreground contaminants bear more similarity to the problems faced by CMB studies---though the foregrounds we face are orders of magnitude larger.

The LT method therefore builds on the literature of both CMB and galaxy surveys, providing a unified framework for the treatment of geometric and foreground effects by employing the quadratic estimator formalism for inverse variance foreground subtraction and power spectrum estimation.  In Section \ref{21cmStatistics} we will review precisely how it is implemented.  

The LT formalism has a number of important advantages over its predecessors.  By treating foregrounds as a form of correlated noise, both foregrounds and noise can be downweighted in a way that is unbiased and lossless in the sense that it maintains all the cosmological information in the data.  Furthermore, the method allows for the simultaneous estimation of both the errors on power spectrum estimates and the window functions or ``horizontal'' error bars.  

Unfortunately, the LT method suffers from computational difficulties.  Because it involves inverting and multiplying very large matrices, it cannot be accomplished faster than in $\BigO(N^3)$ steps, where $N$ is the number of voxels in the data to be analyzed.  This makes analyzing large data sets with this method infeasible.  The primary goal of this paper is to demonstrate an adaptation of the method that can be run much faster.  But first, we need to review the essential elements of the method to put our adaptations and improvements into proper context.  In Sections \ref{dataOrganization} and \ref{PSdiscretization}, we describe our conventions and notation and explain the relationship between the measured quantities and those we seek to estimate.  In Section \ref{21cmStatistics}, we review the LT statistical estimators and how the Fisher information matrix is used to calculate statistical errors on power spectrum measurements.  Then in \ref{AdrianModels} we explain the LT model of noise and foregrounds in order to motivate and justify our refinements that will greatly speed up the algorithm in Section \ref{Fast}.

\subsection{Data Organization and Conventions} \label{dataOrganization}
We begin with a grid of data that represents the brightness temperatures at different positions on the sky as a function of frequency from which we wish to estimate the 21\,cm brightness temperature power spectrum.  We summarize that information using a data vector $\mathbf{x}$ which can be thought of as a one dimensional object of length $n_{x} n_{y} n_{z}\equiv N$,\footnote{While it is helpful to think of $\mathbf{x}$ as a vector in the matrix operations below, it is important to remember that the index $i$ in $x_i$, which refers to the different components of $\mathbf{x}$, actually runs over different values of the spatial coordinates $x$, $y$, and $z$.} the number of voxels in the data cube.

\begin{figure}
  \centering 
    \includegraphics[width=.6\textwidth]{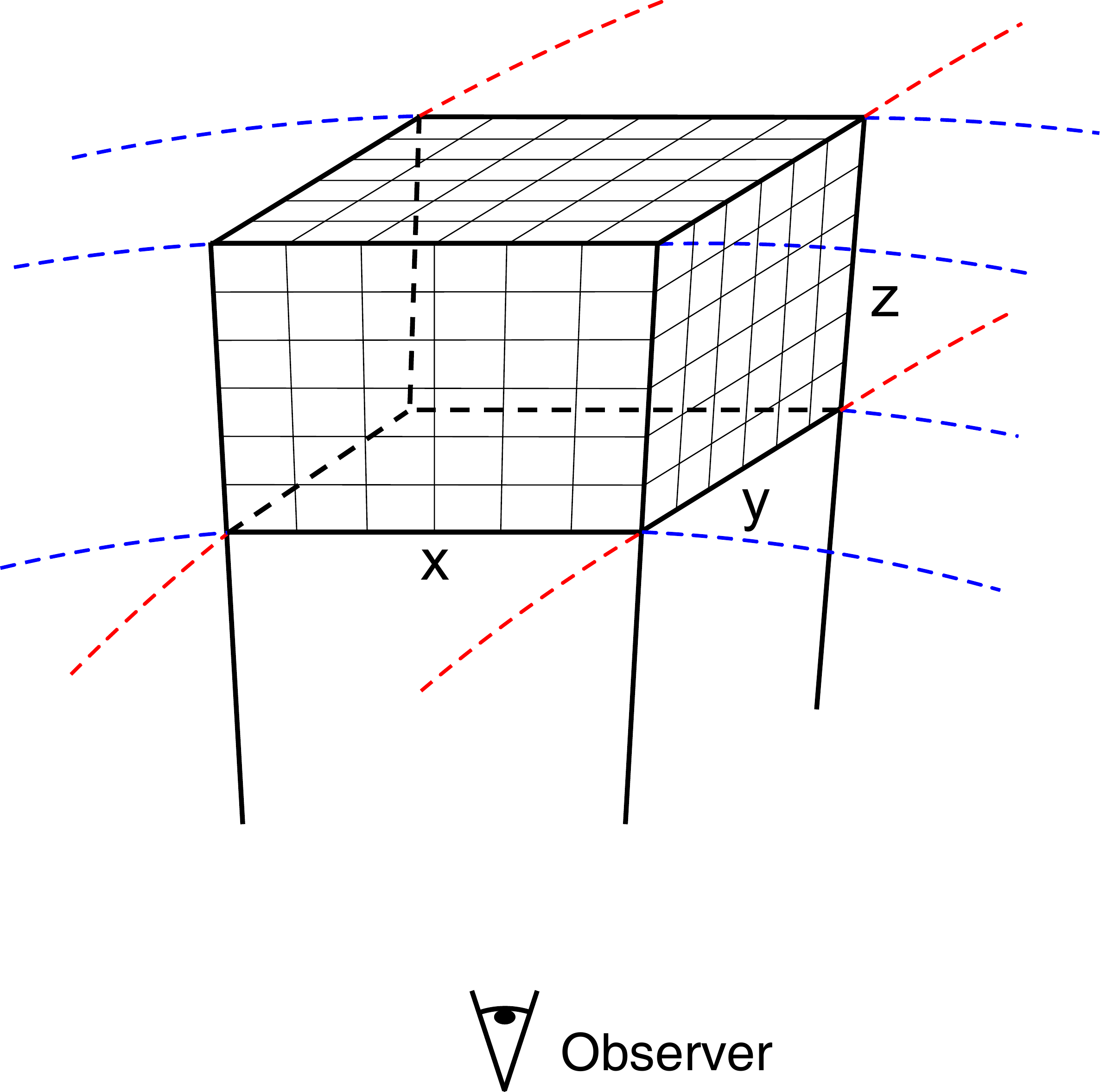}
  \caption[Illustration of the flat sky approxmiation.]{This exaggerated schematic illustrates the flat sky approximation.  It shows great circles (colored and dashed) approximated linearly in the region considered, with lines tracing back to the observer treated as if they were parallel.  Our data cube contains the measured brightness temperatures for every small voxel.}
  \label{flatsky}
\end{figure}

Although the LT technique works for arbitrary survey geometries, we restrict ourselves to the simpler case of a data ``cube'' that corresponds to a relatively small rectilinear section of our universe of size $\ell_x \times \ell_y \times \ell_z$ in comoving coordinates.\footnote{This restriction and its attendant approximations lie at the heart of our strategy for speeding up these calculations, as we explain in Section \ref{Fast}.}  We pick our box to be a subset of the total 21\,cm brightness temperature 3D map that a large interferometric observatory would produce.  Unlike the LT method, our fast method requires that the range of positions on the sky must be small enough for the flat sky approximation to hold (Figure \ref{flatsky}).
Similarly, our range of frequencies (and thus redshifts) in the data cube must correspond to an epoch short enough so that $P(k,z)$ might be approximated as constant in time. Following simulations by \cite{McQuinnLyman}, \cite{Yi} argued that we can conservatively extend the redshift ranges of our data cubes to about $\Delta z \lesssim 0.5$.  At typical EoR redshifts, such a small range in $\Delta z$ allows a very nearly linear mapping between the frequencies measured by an interferometer to a regularly spaced range of comoving distances, $d_C(z)$, although in general $d_C(z)$ is not a linear function of $z$ or $\nu$.  This also justifies the approximation that our data cube corresponds to an evenly partitioned volume of our universe.

If the measured brightness temperatures, $x_i$, were only the result of redshifted 21\,cm radiation, then each measurement would represent the average value in some small box of volume $\Delta x \Delta y \Delta z$ centered on $\rv_i$ of a continuous brightness temperature field $x(\rv)$ \citep{Maxgalaxysurvey1}: 
\beq
x_i \equiv \int \psi_i(\rv)x(\rv) d^3r, \label{voxAvg}
\eeq
where our discretization function $\psi_i$ is defined as $\psi_i(\rv) \equiv \psi_0(\rv -\rv_i)$, where
\beq
\psi_0(\rv) \equiv \frac{\Pi(\frac{x}{\Delta x})\Pi(\frac{y}{\Delta y})\Pi(\frac{z}{\Delta z})}{\Delta x \Delta y \Delta z} \label{tophat}
\eeq
and where $\Pi(x)$ is the normalized, symmetric boxcar function ($\Pi(x)= 1$ if $|x|< \frac{1}{2}$ and $0$ otherwise).  This choice of pixelization encapsulates the idea that each measured brightness temperature is the average over a continuous temperature field inside each voxel.  In this paper, we improve on the LT method by including the effect of finite pixelation.  This will manifest itself as an extra $\Phi(\kv)$ term that will we define in Equation \ref{FTpsi} and that will reappear throughout this paper.

\subsection{The Discretized 21\,cm Power Spectrum} \label{PSdiscretization}
Ultimately, the goal of this paper is to estimate the 21\,cm power spectrum $P(\kv)$, defined via
\beq
\langle \widetilde{x}^*(\kv) \widetilde{x}(\kv') \rangle \equiv (2\pi)^3 \delta(\kv - \kv') P(\kv),
\eeq
where $\widetilde{x}(\kv)$ is the Fourier transformed brightness temperature field and where angle brackets denote the ensemble average of all possible universes obeying the same statistics.  

Our choice of pixelization determines the relationship between the continuous power spectrum, $P(\kv)$, and the 21\,cm signal covariance matrix, which we call $\mathbf{S}$ for \textbf{S}ignal.  It is fairly straightforward to show, given Equation \ref{voxAvg} and the definition of the power spectrum \citep{Maxgalaxysurvey1}, that:
\beq
S_{ij} \equiv \langle x_i x_j\rangle - \langle x_i\rangle\langle x_j\rangle = \int \widetilde{\psi}_{i}(\kv)\widetilde{\psi}_{j}^{*}(\kv)P(\kv)\frac{d^3k}{(2\pi)^{3}}, \label{CDef}
\eeq
where $\widetilde{\psi}_{i}(\kv)$ is the Fourier transform of $\psi_i(\rv)$:
\beq
\widetilde{\psi}_{i}(\kv) \equiv  \int e^{-i\kv\cdot\rv}\psi_{i}(\rv)d^{3}\rv.
\eeq
Separating this integral into each of the three Cartesian coordinates and integrating yields
\begin{align}
\widetilde{\psi}_{i}(\kv) &= e^{i \kv \cdot \rv_i} \Phi(\kv), \text{ where} \nonumber \\
\Phi(\kv) &\equiv j_0\left(\frac{k_x \Delta x}{2}\right) j_0\left(\frac{k_y \Delta y}{2}\right) j_0\left(\frac{k_z \Delta z}{2}\right), \label{FTpsi}
\end{align}
where $j_0(x) = \sin x / x$ is the zeroth spherical Bessel function.  Because we can only make a finite number of measurements of the power spectrum, we parametrize and discretize $P(\kv)$ by approximating it as a piecewise constant function:
\beq
P(\kv) \approx \sum_\alpha p^\alpha \chi^\alpha(\kv), \label{QSpectral}
\eeq
where the ``band power'' $p^\alpha$ gives the power in region $\alpha$ of Fourier space,\footnote{In contrast to lowered Latin indices, which we use to pick out voxels in a real space or Fourier space data cube, we will use raised Greek indices to pick out power spectrum bins, which will generally each run a range in $k_\|$ and in $k_\perp$.} specified by the characteristic function $\chi^\alpha(\kv)$ which equals 1 inside the region and vanishes elsewhere.

Combining Equations \ref{CDef} and \ref{QSpectral} we can write down $S_{ij}$:
\begin{align}
S_{ij} &= \sum_\alpha p^\alpha Q^\alpha_{ij}, \text{ where} \nonumber \\
Q^\alpha_{ij} &\equiv \int \widetilde{\psi}_{i}(\kv)\widetilde{\psi}_{j}^{*}(\kv)\chi^\alpha(\kv)\frac{d^3k}{(2\pi)^{3}}.
\end{align}
We choose these $\chi^\alpha(\kv)$ to produce band powers that reflect the symmetries of the observation.  Our universe is isotropic in three dimensions, but due to redshift space distortions, foregrounds, and other effects, our measurements will be isotropic only perpendicular to the line of sight \citep{Yi,Barkana2,AliAP, NusserAP,BarkanaAP}.  This suggests cylindrical binning of the power spectrum; in the directions perpendicular to the line of sight, we bin $k_x$ and $k_y$ together radially to get a region in k-space extending from from $k_\perp^\alpha -\Delta k_\perp /2$ to $k_\perp^\alpha +\Delta k_\perp /2$ where $k_\perp^2 \equiv k_x^2 + k_y^2$.  Likewise, in the direction parallel to the line of sight, we integrate over a region of k-space both from $k_\|^\alpha -\Delta k_\| /2$ to $k_\|^\alpha +\Delta k_\| /2$ and, because the power spectrum only depends on $k_\| \equiv |k_z|$, from $-k_\|^\alpha +\Delta k_\| /2$ to $-k_\|^\alpha -\Delta k_\| /2$.  Therefore, we have
\begin{align}
Q_{ij}^{\alpha} = \frac{1}{(2\pi)^{3}} \left[ \paraIntPos - \paraIntNeg \right] \perpInt |\Phi(\kv)|^2 e^{i\mathbf{k}\cdot(\mathbf{r}_{i}-\mathbf{r}_{j})} k_{\bot}d\theta dk_\perp dk_{\|}. \label{Qderivation}
\end{align}
Without the factor of $|\Phi(\kv)|^2$, the LT method was able to evaluate this integral analytically.  With it, the integral must be evaluated numerically if it is to be evaluated at all.  This is of no consequence; we will return to this formula in Section \ref{fastPSE} to show how the matrix $\Q^\alpha$ naturally lends itself to approximate multiplication by vectors using fast Fourier techniques.

\subsection{21\,cm Power Spectrum Statistics} \label{21cmStatistics}
In order to interpret the data from any experiment, we need to be able to estimate both the 21\,cm brightness temperature power spectrum and the correlated errors induced by the survey parameters, the instrument, and the foregrounds.  The LT method does both at the same time; with it, the calculation of the error bars immediately enables power spectrum estimation. 
 
\subsubsection{Inverse Variance Weighted Power Spectrum Estimation}
The LT method adapts the inverse variance weighted quadratic estimator formalism \citep{Maxpowerspeclossless, BJK} for calculating 21\,cm power spectrum statistics.  The first step towards constructing the estimator $\widehat{p}^\alpha$ for $p^\alpha$ is to compute a quadratic quantity, called $\widehat{q}^\alpha$ whose relationship\footnote{Unlike the notation in LT, we do not include the bias term in $\widehat{q}^\alpha$ but will later include it in our power spectrum estimator.  The result is the same.} to $\widehat{p}^\alpha$ we will explain shortly:
\begin{equation}
\widehat{q}^\alpha \equiv \frac{1}{2}(\mathbf{x} - \langle \x \rangle)^{\trans}\mathbf{C}^{-1}\mathbf{Q}^{\alpha}\mathbf{C}^{-1}(\mathbf{x} - \langle \x \rangle). \label{estimator}
\end{equation}
Here $\mathbf{C}$ is the covariance matrix of $\mathbf{x}$, so
\begin{equation}
\mathbf{C} \equiv \langle\mathbf{x}\mathbf{x}^{\trans}\rangle - \langle\mathbf{x}\rangle\langle\mathbf{x}\rangle^{\trans}.
\end{equation}
For any given value of $\alpha$, the right-hand side of Equation (\ref{estimator}) yields a scalar. Were both our signal and foregrounds Gaussian, this estimator would be optimal in the sense that it preserves all the cosmological information contained in the data.  Of course, with a non-Gaussian signal, the power spectrum cannot contain all of the information, though it still can be very useful \citep{Maxpowerspeclossless}.
 
Our interest in the quadratic estimators $\widehat{q}^\alpha$ lies in their simple relationship to the underlying band powers.  In \cite{Maxpowerspeclossless}, it is shown that:
\beq
\left< \widehat{\mathbf{q}} \right> = \mathbf{F}\mathbf{p} + \mathbf{b} \label{expectqhat}
\eeq
where each $b^\alpha$ is the bias in the estimator and $\F$ is the Fisher information matrix, which is related to the probability of having measured our data given a particular set of band powers, $f(\mathbf{x}|p^{\alpha})$.  The matrix is defined \citep{fisher} as:
\begin{equation}
F^{\alpha\beta} \equiv -\left<\frac{\partial^{2}\mbox{ln}f(\mathbf{x}|p^{\alpha})}{\partial p^{\alpha}\partial p^{\beta}}\right>.
\end{equation}
The LT method employs the estimators by calculating both $\F$ and $\mathbf{b}$ using relationships derived in \citep{Maxpowerspeclossless}:
\begin{align}
F^{\alpha\beta} &= \frac{1}{2}\mbox{tr}[\mathbf{C}^{-1}\mathbf{Q}^{\alpha}\mathbf{C}^{-1}\mathbf{Q}^{\beta}] \text{ and} \label{fisherTrace} \\
b^\alpha &= \frac{1}{2}\mbox{tr}[(\C - \mathbf{S})\mathbf{C}^{-1}\mathbf{Q}^{\alpha}\mathbf{C}^{-1}]. \label{biasTrace}
\end{align}

We want our $\widehat{p}^\alpha$ to be unbiased estimators of the true underlying band powers, which means that we will have to take care to remove the biases for each band power, $b^\alpha$.  We construct our estimators\footnote{Here were differ slightly from the LT method in the normalization, which does not have the property from Equation \ref{weightedAverage}.  We instead follow \citep{THX2df}. } as linear combinations of the quadratic estimators $\widehat{q}^\alpha$ that have been corrected for bias:
\beq
\widehat{\mathbf{p}} = \mathbf{M}(\widehat{\mathbf{q}} - \mathbf{b}), \label{phatdef}
\eeq
where $\mathbf{M}$ is a matrix which correctly normalizes the power spectrum estimates; the form of $\mathbf{M}$ represents a choice in the trade-off between small error bars and narrow window functions, as we will explain shortly.

How do we expect this estimator to behave statistically?  The only random variable on the right hand side of Equation \ref{phatdef} is $\mathbf{\widehat{q}}$, so we can combine Equations \ref{expectqhat} and \ref{phatdef} to see that our choice of $\widehat{\mathbf{p}}$ indeed removes the bias term:
\beq
\left< \widehat{\mathbf{p}} \right> = \mathbf{M}\F \mathbf{p} +\mathbf{Mb} - \mathbf{Mb} = \mathbf{M}\F \mathbf{p} = \mathbf{W}\mathbf{p}. \label{expectphat} 
\eeq
We have defined the matrix of ``window functions'' $\mathbf{W} \equiv \mathbf{M}\F$ because Equation \ref{expectphat} tells us that we can expect our band power spectrum estimator, $\widehat{\mathbf{p}}$, be be a weighted average of the true, underlying band powers, $\mathbf{p}$.  That definition imposes the condition on $\mathbf{W}$ that
\beq
\sum_\beta W^{\alpha\beta} = 1 \label{weightedAverage}
\eeq
which is equivalent to the statement that the weights in a weighted average must add up to one.  The condition on $\mathbf{W}$ constrains our choice of $\mathbf{M}$, though as long as $\mathbf{M}$ is an invertible matrix,\footnote{None of the choices of $\mathbf{M}$ involve anything more computationally intensive than inverting $\F$. This is fine, since $\F$ is a much smaller matrix than $\C$.} the choice of $\mathbf{M}$ does not change the information content of our power spectrum estimate, only the way we choose to represent our result.

\subsubsection{Window Functions and Error Bars}

In this paper, we choose a form of $\widehat{\mathbf{p}}$ where $\mathbf{M} \propto \F^{-1/2}$.  Two other choices for $\mathbf{M}$ are presented in \cite{THX2df}: one where $\mathbf{M} \propto \I$ and another where $\mathbf{M} \propto \F^{-1}$.  The former produces the smallest possible error bars, but at the cost of wide window functions and correlated measurement errors.  The latter produces $\delta$-function windows, but large and anticorrelated measurement errors.  This choice of $\mathbf{M} \propto \F^{-1/2}$ has proven to be a happy medium between those other two choices for $\mathbf{M}$.  It produces reasonably narrow window functions and reasonably small error bars which have the added advantage of being completely uncorrelated, so that each measurement contains a statistically independent piece of information.  Because $\mathbf{W} \equiv \mathbf{M} \F$ and because of the condition on $\mathbf{W}$ in Equation \ref{weightedAverage}, there is only one such $\mathbf{M}$:
\beq
M^{\alpha\beta} \equiv \frac{\left(\mathbf{F}^{-1/2}\right)^{\alpha\beta}}{\sum_\gamma (\F^{1/2})^{\alpha\gamma}}.
\eeq
With this choice of $\mathbf{M}$ we get window functions of the form
\beq
W^{\alpha\beta} = \frac{(\F^{1/2})^{\alpha\beta}}{\sum_\gamma (\F^{1/2})^{\alpha\gamma}}
\eeq
which we can use to put ``horizontal error bars'' on our power spectrum estimates.

Using Equation \ref{phatdef} and the fact derived in \citep{Maxpowerspeclossless} that an equivalent formula for $\F$ is given by
\beq
\mathbf{F} = \langle\widehat{\mathbf{q}}\widehat{\mathbf{q}}^{\trans}\rangle - \langle\widehat{\mathbf{q}}\rangle\langle\widehat{\mathbf{q}}\rangle^{\trans}, \label{covq}
\eeq
we can see that the covariance of $\widehat{\mathbf{p}}$ takes on a simple form:
\beq
\langle\widehat{\mathbf{p}}\widehat{\mathbf{p}}^{\trans}\rangle - \langle\widehat{\mathbf{p}}\rangle\langle\widehat{\mathbf{p}}\rangle^{\trans} = \mathbf{M}\F \mathbf{M}^\trans.
\eeq
This allows us to write down the ``vertical error bars'' on our individual power spectrum estimates:
\beq
\Delta \widehat{p}^\alpha = \left[\left(\mathbf{M}\F \mathbf{M}^\trans\right)^{\alpha\alpha}\right]^{1/2} = \frac{1}{\sum_\gamma (\F^{1/2})^{\alpha\gamma}}.
\eeq
As in LT, we can transform our power spectrum estimates and our vertical error bars into temperature units:
\beq
\widehat{T}^\alpha \equiv \left[\frac{(k^\alpha_\perp)^2 k^\alpha_\|}{2\pi^2} p^\alpha \right]^{1/2}
\eeq
and likewise,
\beq
\Delta \widehat{T}^\alpha = \left[\frac{(k^\alpha_\perp)^2 k^\alpha_\|}{2\pi^2 \left(\sum_\gamma (\F^{1/2})^{\alpha\gamma}\right)} \right]^{1/2}. \label{TemperatureUnits}
\eeq
This makes it easier to compare to theoretical predictions, which are often quoted in units of K or mK.

\subsection{Foreground and Noise Models}\label{AdrianModels}
The structure of the matrix $\C$ that goes into our inverse variance weighted estimator depends on the way we model our foregrounds, noise, and signal.  We assume that those contributions are the sum of five uncorrelated components:
\begin{align}
\mathbf{C} &= \sum_{c \text{ } \in \text{ components}}\langle\mathbf{x}_c\mathbf{x}_c^{\trans}\rangle - \langle\mathbf{x}_c\rangle\langle\mathbf{x}_c\rangle^{\trans} \nonumber \\ &\equiv \mathbf{S} +  \mathbf{R} + \mathbf{U} + \mathbf{G} + \mathbf{N}.
\end{align}
These are the covariance matrices due to 21\,cm \textbf{S}ignal, bright point sources \textbf{R}esolved from one another, \textbf{U}nresolved point sources, the \textbf{G}alactic synchrotron, and detector \textbf{N}oise, respectively.  This deconstruction of $\mathbf{C}$ is both physically motivated and will ultimately let us approximate $\mathbf{C}^{-1}(\mathbf{x}-\langle \x \rangle)$ much more quickly than by just inverting the matrix.   

Following LT, we neglect the small cosmological $\mathbf{S}$ because it is only important for taking cosmic variance into account.  It is straightforward to include the $\mathbf{S}$ matrix in our method, especially because we expect it to have a very simple form, but this will only be necessary once the experimental field moves from upper limits to detection and characterization of the 21\,cm brightness temperature power spectrum.  

In this paper, we will develop an accelerated version of the LT method using the models delineated in LT.  That speed-up relies on the fact that all of these covariance matrices can be multiplied by vectors $\mathcal{O}(N\mbox{log}N)$ time.  However, our techniques for acceleration will work on a large class of models for $\C$ as long as certain assumptions about translation invariance and spectral structure are respected.  In this section, we review the three contaminant matrices from LT: $\U$, $\G$, and $\N$.  When we discuss methods to incorporate these matrices into a faster technique in Section \ref{FastCov}, we will also expand the discussion of foregrounds to include $\R$, which is a natural extension of $\U$.
\begin{figure*} 	
	\centering 
	\includegraphics[width=.8\textwidth]{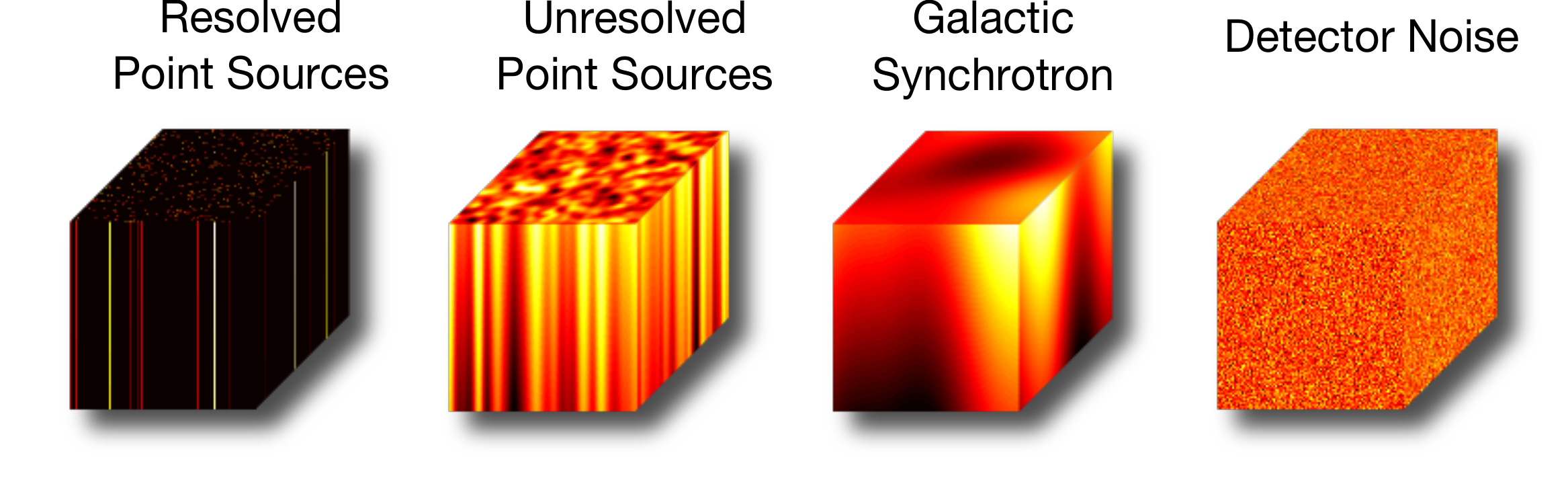}
	\caption[Example contaminant data cubes.]{These example data cubes (with the line of sight drawn vertically) illustrate the strong or weak correlations between different voxels in the same cube.  In Section \ref{random} we explain how these simulated data cubes are generated quickly. The addition of resolved point sources, which is not included in LT, is discussed in Section \ref{RFast}.  To best exemplify the detailed structure of the models, the color scales are different for each of the cubes.}
	\label{DataCubes}
\end{figure*}

\subsubsection{Unresolved Point Sources} \label{AdrianU}
For a typical current generation or near future experiment, the pixels perpendicular to the line of sight are so large that every one is virtually guaranteed to have a point source in it bright enough to be an important foreground to our 21\,cm signal.  These confusion limited point sources are taken into account using their strong correlations parallel to the line of sight and weaker correlations perpendicular to the line of sight, both of which are easily discerned in Figure \ref{DataCubes}.

Following LT we split $\U$ into the tensor product of two parts, one representing correlations perpendicular to the line of sight and the other parallel to the line of sight:
\beq
\U \equiv  \U_\perp \otimes \U_\|
\eeq
Covariance perpendicular to the line of sight is modeled as an unnormalized Gaussian:
\beq
(U_\perp)_{ij} \equiv \exp\left[\ \frac{\left((\mathbf{r}_\perp)_{i} - (\mathbf{r}_\perp)_{j}\right)^2}{2 \sigma^2_\perp} \right] \label{Uperp}
\eeq
where $\sigma_\perp$ represents the correlation length perpendicular to the line of sight.  Following LT, we take this to be a comoving distance corresponding to 7 arcminutes, representing the weak clustering of point sources. 

The covariance along the line of sight assumes a Poisson distributed number of point sources below some flux cut, $S_{\text{cut}}$, which we take to be 0.1 Jy, each with a spectral index drawn from a Gaussian distribution with mean $\bar{\kappa}$ and standard deviation $\sigma_\kappa$.  Given a differential source count \citep{dimatteo1} of
\begin{align}
\frac{dn}{dS} = (4000 \text{ Jy}^{-1} \text{sr}^{-1}) \times \begin{cases} \left(\frac{S}{0.880 \text{ Jy}}\right)^{-2.51} & \text{for S $>$ 0.880 Jy} \\ \left(\frac{S}{0.880 \text{ Jy}}\right)^{-1.75} & \text{for S $\leq$ 0.880 Jy,} \end{cases} \label{sourceCounts}
\end{align}
we get a covariance parallel to the line of sight of 
\begin{align}
(U_\|)_{ij} = (1.4\times10^{-3}\text{ K})^2\ (\eta_i \eta_j)^{-2-\overline{\kappa}}\left(\frac{\Omega_{pix}}{\text{1 sr}}\right)^{-1}
& \text{exp}\left[\frac{\sigma^2_\kappa}{2}(\text{ln}(\eta_i \eta_j))^2\right] I_2(S_{cut}). \label{Uexact}
\end{align}
where we have assumed a power law spectrum for the point sources where $\eta_i \equiv \nu_i / \nu_*$, $\nu_* = $ 150 MHz, and $\overline{\kappa}$ and $\sigma_\kappa$ are the average value and standard deviation of the distribution of spectral indices of the point sources.  We define $I_2(S_{cut})$ as
\beq
I_2(S_{cut}) \equiv \int_0^{S_{\text{cut}}}S^2\frac{dn}{dS}dS \label{I2fluxintegral}
\eeq
Following LT, we take $\overline{\kappa}=0.5$ and $\sigma_\kappa = 0.5$, both of which are consistent with the results of \citep{ChrisMWA}. In Section \ref{Ufast}, we will return to Equation \ref{Uexact} and show how it can be put into an approximate form that can be quickly multiplied by a vector.

\subsubsection{Galactic Synchrotron Radiation} \label{AdrianG}

Following LT, we model Galactic synchrotron emission in the same way that we model unresolved point sources.  Fundamentally, both are spatially correlated synchrotron signals contributing to the brightness temperature of every pixel in our data cube.  However the galactic synchrotron is much more highly correlated spatially, which can be clearly seen in the sample data cube in Figure \ref{DataCubes}. This leads to our adoption of a much larger value of $\sigma_\perp$; we take $\sigma_\perp$ to be a comoving distance corresponding to $30^\circ$ on the sky.  Following LT, we take $\overline{\kappa}=0.8$ and $\sigma_\kappa = 0.4$.

This is an admittedly crude model for the galactic synchrotron, in part because it fails to take into account the roughly planar spatial distribution of the Galactic synchrotron.  A more sophisticated model for $\G$ that incorporates a more informative map of the Galactic synchrotron can only produce smaller error bars and narrower window functions.  However, such a model might involve breaking the assumption of the translational invariance of correlations, which could be problematic for the technique we use in Section \ref{Fast} to speed up this algorithm.  In practice, we expect very little benefit from an improved spatial model of the Galactic synchrotron due to the restriction imposed by the flat sky approximation that our map encompass a relatively small solid angle.

\subsubsection{Instrumental Noise} \label{AdrianN}

Here we diverge from LT to adopt a form of the noise power spectrum from \citep{FFTT} that is more readily adaptable to the pixelization scheme we will introduce:
\beq
P^N(\kv,\lambda) = \frac{\lambda^2 T_\text{sys}^2 y d_M^2}{A f^\text{cover} \tau} B^{-2}(\kv,\lambda). \label{NPowerSpectrum}
\eeq
Here $T_\text{sys}$ is the system temperature (which is sky noise dominated in our case), $A$ is the total effective collecting area of the array, and $\tau$ is the is the total observing time.  $B(\kv,\lambda)$ is a function representing the $uv$-coverage, normalized to peak at unity, which changes with wavelength. Lastly,  $y$ is the conversion from bandwidth to the comoving length of the box parallel to the line of sight and $d_M$ is the transverse comoving distance\footnote{The transverse comoving distance, $d_M(z)$, is the ratio of an object's comoving size the angle it subtends, as opposed to the angular diameter distance, $d_A(z)$, which is the ratio of its physical size to the angle it subtends.  It is sometimes called the ``comoving angular diameter distance'' and it is even sometimes written as $d_A(z)$.  See \cite{HoggDistance} for a helpful summary of these often confusingly named quantities.}, so $y d_M^2 \Omega_\text{pix} \Delta \nu = \Delta x \Delta y \Delta z$ with $\Omega_\text{pix}$ being the angular size of our pixels and $\Delta \nu$ being the frequency channel width. This form of the noise power spectrum assumes that the entire map is observed for the same time $\tau$, which is why the ratio of the angular size of the map to the field of view does not appear.

We use Equation \ref{CDef} to discretize the power spectrum and get $\N$:
\beq
N_{ij} = \int e^{i \kv \cdot \rv_i} e^{-i \kv \cdot \rv_j} |\Phi(\kv)|^2 P^N(\kv,\lambda) \frac{d^3k}{(2\pi)^{3}} \label{Ndef}
\eeq
Instead of evaluating this integral, we will show in Section \ref{Nfast} that it can be approximated using the discrete Fourier transform.

\subsection{Computational Challenges to the Brute Force Method}
For a large data cube, the LT method requires the application of large matrices that are memory-intensive to store and computationally infeasible to invert.  However, we need to be able to multiply by and often invert these large matrices to calculate our quadratic estimators (Equations \ref{Qderivation} and \ref{estimator}), the Fisher matrix (Equation \ref{fisherTrace}), and the bias (Equation \ref{biasTrace}).  A $10^6$ voxel data cube, for example, would take $\BigO(10^{18})$ computational steps to analyze.  This is simply infeasible for next-generation radio interferometers and we have therefore endeavored to find a faster way to compute 21 power spectrum statistics.


\section{Our Fast Method} \label{Fast}
To avoid the computational challenges of the LT method, we seek to exploit symmetries and simpler forms in certain bases of the various matrices out of which we construct our estimate of the 21\,cm power spectrum and its attendant errors and window functions.  In this section, we describe the mathematical and computational techniques we employ to create a fast and scalable algorithm.  

Our fast method combines the following six separate ideas:
\begin{enumerate}
\item A Monte Carlo technique for computing the Fisher information matrix and the bias (Section \ref{fisherMC}).
\item An FFT-based technique for computing band powers using the $\mathbf{Q}^\alpha$ matrices (Section \ref{fastPSE}).
\item An application of the conjugate gradient method that eliminates the need to invert $\C$ (Section \ref{CGPSE}).
\item A Toeplitz matrix technique for multiplying vectors quickly by the constituent matrices of $\C$ (Section \ref{FastCov}).  
\item A combined FFT and spectral technique for preconditioning $\C$ to improve converge of the conjugate gradient method (Section \ref{FastPrecon})
\item A technique using spectral decomposition and Toeplitz matrices for rapid simulation of data cubes for our Monte Carlo (Section \ref{random}).
\end{enumerate}
In this Section, we explain how all six are realized and how they fit into our fast method for power spectrum estimation.  Finally, in Section \ref{fisherConverge}, we verify the algorithm in an end-to-end test.

\subsection{Monte Carlo Calculation of the Fisher Information Matrix} \label{fisherMC}
In order to turn the results of our quadratic estimator into estimates of the power spectrum with proper vertical and horizontal error bars, we need to be able to calculate the Fisher information matrix and the bias term.  Instead of using the form of $\F$ in Equation \ref{fisherTrace} that the LT method employs, we take advantage of the relationship between $\F$ and $\widehat{\mathbf{q}}$ in Equation \ref{covq} that $\F = \langle\widehat{\mathbf{q}}\widehat{\mathbf{q}}^{\trans}\rangle - \langle\widehat{\mathbf{q}}\rangle\langle\widehat{\mathbf{q}}\rangle^{\trans}$.  If we can generate a large number of simulated data sets $\mathbf{x}$ drawn from the same covariance $\C$ and then compute $\widehat{\mathbf{q}}$ from each one, then we can iteratively approximate $\F$ with a Monte Carlo. In other words, a solution to the problem of quickly calculating $\widehat{\mathbf{q}}$ also provides us with a way to estimate $\F$.  What's more, the solution is trivially parallelizable; creating artificial data cubes and analyzing them can be done by many CPUs simultaneously.

In calculating $\F$, we can get $b^\alpha$ out essentially for free.  If we take the average of all our $\widehat{\mathbf{q}}$ vectors, we expect to that 
\begin{align}
\langle \widehat{q}^\alpha \rangle &= \left< \frac{1}{2}(\mathbf{x} - \langle \x \rangle)^{\trans}\mathbf{C}^{-1}\mathbf{Q}^{\alpha}\mathbf{C}^{-1}(\mathbf{x} - \langle \x \rangle) \right> \nonumber \\ &= \mbox{tr}\left[ \left< (\mathbf{x} - \langle \x \rangle)(\mathbf{x} - \langle \x \rangle)^{\trans} \right> \mathbf{C}^{-1}\mathbf{Q}^{\alpha}\mathbf{C}^{-1}\right] \nonumber \\ &= \mbox{tr}\left[ \mathbf{Q}^\alpha \C^{-1} \right] = b^\alpha
\end{align}
in the limit where $\mathbf{S}$ is negligibly small.  This implies that $\widehat{\mathbf{p}}$ can be written in an even simpler way:
\beq
\widehat{p}^\alpha = \frac{1}{\sum_\gamma F^{\alpha\gamma}}\left(\widehat{q}^\alpha - \left< \widehat{q}^\alpha \right>\right)
\eeq
where, recall, $\mathbf{F}$ is calculated as the sample covariance of our $\widehat{\mathbf{q}}$ vectors.  We therefore can calculate all the components of our power spectrum estimate and its error bars using a Monte Carlo.

In Section \ref{fisherConverge} we will return to assess how well the Monte Carlo technique works and its convergence properties.  But first, we need to tackle the three impediments to computing $\widehat{q}^\alpha$ in Equation \ref{estimator} quickly: generating a random $\x$ drawn from $\C$, computing $\C^{-1}(\x-\langle \x \rangle)$, and applying $\mathbf{Q}^\alpha$.

\subsection{Fast Power Spectrum Estimation Without Noise or Foregrounds}\label{fastPSE}
If we make the definition that
\beq
\mathbf{y} \equiv \mathbf{C}^{-1}(\mathbf{x} - \langle \x \rangle) 
\eeq
to simplify Equation \ref{estimator} to
\beq
q^\alpha = \mathbf{y}^\trans \mathbf{Q}^\alpha \mathbf{y} \label{yQy},
\eeq
we can see that even if we have managed to calculate $\mathbf{y}$ quickly, we still need to multiply it by a $N\times N$ element $\mathbf{Q}^\alpha$ matrix for each band power $\alpha$.   Though each $\Q^\alpha$ respects translation invariance that could make multiplying by vectors faster, there exists an even faster technique that can calculate every entry of $\widehat{\mathbf{p}}$ simultaneously using fast Fourier transforms.

To see that this is the case, we substitute Equation \ref{Qderivation} into Equation \ref{yQy}, reversing the order of summation and integration and factoring the integrand:
\begin{align}
\widehat{q}^{\alpha} = \text{ }& \frac{1}{2} \left[ \paraIntPos - \paraIntNeg \right]  \perpInt \nonumber \\ &  \bigg(\sum_i y_i e^{i\mathbf{k}\cdot\mathbf{r}_i}\bigg) \bigg(\sum_j y_j e^{-i\mathbf{k}\cdot\mathbf{r}_j}\bigg) |\Phi(\kv)|^2 \frac{k_{\perp} d\theta dk_\perp dk_{\|}}{(2\pi)^{3}}. \label{integralEstimator}
\end{align}
The two sums inside the integral are very nearly discrete, 3D Fourier transforms.  All that remains is to discretize the Fourier space conjugate variable $\kv$ as we have already discretized the real space variable $\rv$.

In order to evaluate the outer integrals, we approximate them as a sum over grid points in Fourier space. The most natural choice for discretization in $\mathbf{k}$ is one that follows naturally from the FFT of $\mathbf{y}$ in real space.  If our box is of size $\ell_x \ell_y  \ell_z$ and broken into $n_x \times n_y \times n_z$ voxels\footnote{For simplicity and consistency we assume that $n_x$, $n_y$, and $n_z$ are all even and we take the origin the to be the second of the two center bins.} we have that 
\begin{align}
\mathbf{r}_j =& \left( \frac{j_x\ell_x}{n_x}, \frac{j_y\ell_y}{n_y},\frac{j_z\ell_z}{n_z} \right)  \label{rvector} \\ &\mbox{where } j_x,j_y,j_z \in \left\{-\frac{n_{x,y,z}}{2},...,0,...,\frac{n_{x,y,z}}{2}-1\right\}.\nonumber 
\end{align}
The natural 3D Fourier space discretization is
\begin{align}
\mathbf{k}_m =& \left( \frac{2 \pi m_x }{\ell_x}, \frac{2 \pi m_y }{\ell_y},\frac{2 \pi m_z }{\ell_z} \right) \\ & \mbox{where } m_x,m_y,m_z \in \left\{-\frac{n_{x,y,z}}{2},...,0,...,\frac{n_{x,y,z}}{2}-1\right\} \nonumber 
\end{align}
with a Fourier space voxel volume 
\beq
(\Delta k)^3 = \frac{2 \pi}{\ell_x} \times \frac{2 \pi}{\ell_y} \times \frac{2 \pi}{\ell_z}.
\eeq

With this choice of discretization, we will simplify our integrals by sampling Fourier space with delta functions, applying the approximation in the integrand of Equation \ref{integralEstimator} that 
\beq
1 \approx \sum_m \frac{(2 \pi)^3 \delta^3(\mathbf{k} - \mathbf{k_m})}{\ell_x \ell_y \ell_z}. \label{FourierApprox}
\eeq
This simplifies Equation \ref{integralEstimator} considerably:
\begin{align}
\widehat{q}^{\alpha} = \text{ } \frac{1}{2} \sum_m \bigg(\sum_i y_i e^{i\mathbf{k}_m\cdot\mathbf{r}_i}\bigg) \bigg(\sum_j y_j e^{-i\mathbf{k}_m\cdot\mathbf{r}_j}\bigg) \frac{\chi^\alpha(\mathbf{k}_m) |\Phi(\kv_m)|^2}{\ell_x \ell_y \ell_z}. 
\end{align}
If we define $\widetilde{y}_m \equiv \sum_j y_i  e^{-i \mathbf{k}_m \cdot \mathbf{r}_j}$, then we can write $\widehat{\mathbf{q}}$ as:
\begin{align}
\widehat{q}^\alpha &\approx \frac{1}{2 \ell_x \ell_y \ell_z} \sum_m \widetilde{y}_m^* \widetilde{y}_m \chi^\alpha(\mathbf{k}_m) |\Phi(\kv_m)|^2 \nonumber \\&= \frac{1}{2 \ell_x \ell_y \ell_z} \sum_m |\widetilde{y}_m|^2 \chi^\alpha(\mathbf{k}_m) |\Phi(\kv_m)|^2. \label{FFTPSE}
\end{align}
This result makes a lot of sense: after all, the power spectrum is---very roughly speaking---the data Fourier transformed, squared, and binned with an appropriate convolution kernel.  

This is a very quick way to calculate $\widehat{\mathbf{q}}$ because we can compute $\widetilde{\y}$ in $\mathcal{O}(N\log N)$ time (if we already have $\y$) and then we simply need to add $|\widetilde{y}_m|^2$ for every $m$, weighted by the value of the analytic function $|\Phi(\kv_m)|^2$ to the appropriate band power $\alpha$.\footnote{For simplicty, we choose band power spectrum bins with the same width as our Fourier space bins (before zero padding).  This linear binning scheme makes plotting, which is typically logarithmic in the literature, more challenging.  On the other hand, it better spreads out the number of Fourier space data cube bins assigned to each band power.}  Each value of $|\widetilde{y}_m|^2$ gets mapped uniquely to one value of $\alpha$, so there are only $N$ steps involved in performing the binning.  Unlike in the LT method, we perform the calculation of $\widehat{q}^\alpha$ for all values of $\alpha$ simultaneously.

\begin{figure} 
	\centering 
	\includegraphics[width=.6\textwidth]{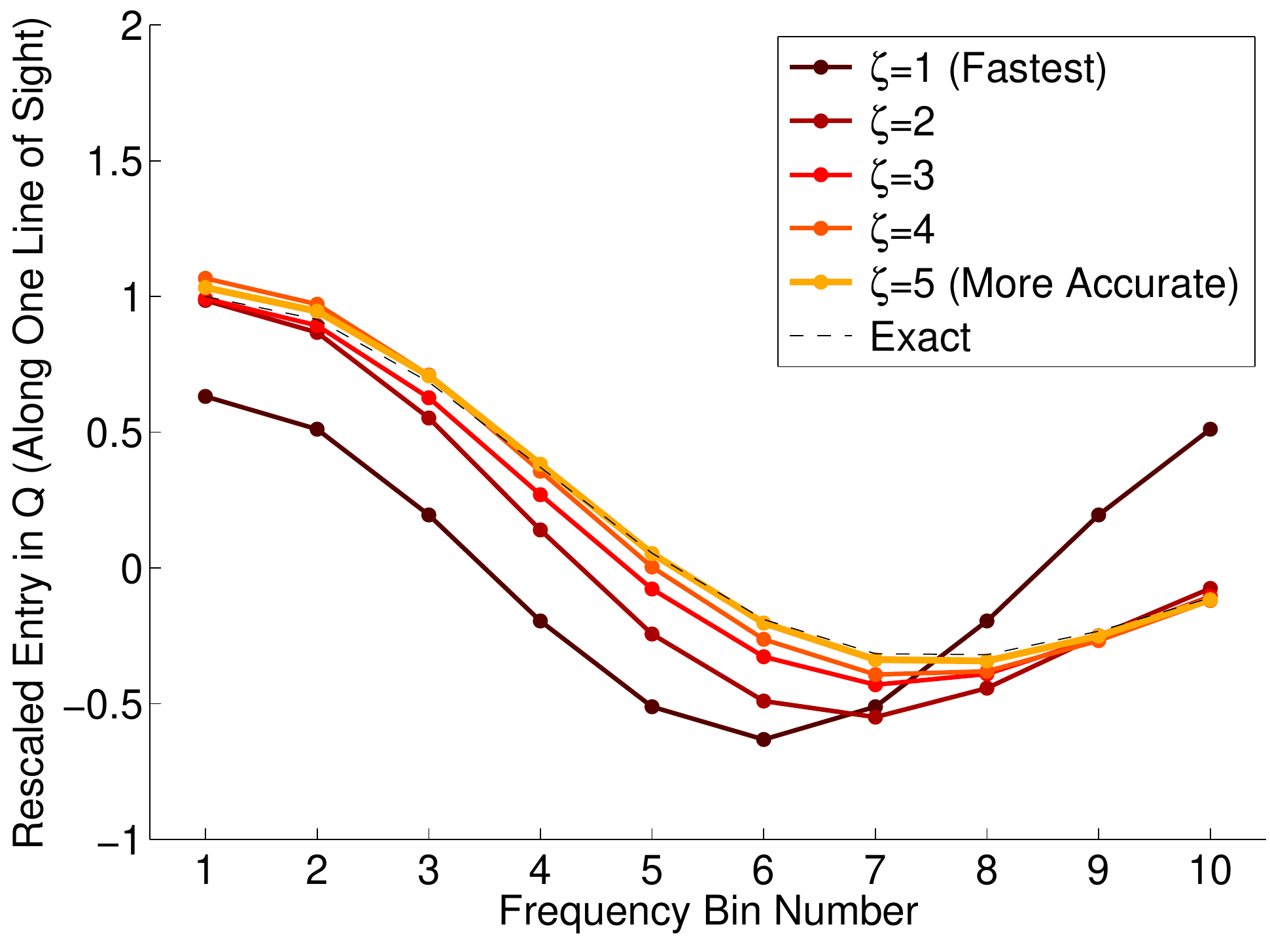}
	\caption[Approximation of $\Q^\alpha$ with the FFT.]{We use an FFT-based technique to approximate the action of the matrix $\Q^\alpha$ that encodes the Fourier transforming, binning, and pixelization factors.  In this Figure, we show how the approximation improves with different factors of the zero padding parameter, $\zeta$, while varying a single coordinate of one of the $\mathbf{Q}^\alpha$ matrices.  For a fairly small value of $\zeta$, the approximation is quite good, meaning that the binning and Fourier transforming step contributes subdomninantly to the complexity of the overall algorithm.}
	\label{FFTapproximationToQ}
\end{figure}

However, the FFT approximation to $Q_{ij}^\alpha$ from Equation \ref{FourierApprox} does not work very well at large values of $(\mathbf{r}_i - \mathbf{r}_j)$ because the discrete version of $\mathbf{Q}^\alpha$ does not sample the continuous version of $\mathbf{Q}^\alpha$ very finely.  This can be improved by zero padding the input vector $y_i$, embedding it inside of a data cube of all zeros a factor of $\zeta^3$ larger.  For simplicity, we restrict $\zeta$ to integer values where $\zeta = 1$ represents no zero padding.  By increasing our box size, we decrease the step size in Fourier space and thus the distance between each grid point in Fourier space where we sample $k$ with delta functions.  Repeating the derivation from Equations \ref{rvector} through \ref{FFTPSE} yields: 
\beq
\widehat{q}^\alpha \approx \frac{1}{2\ell_x \ell_y \ell_z \zeta^3} \sum_m |\widetilde{y}_m|^2 \chi^\alpha(\mathbf{k}_m) |\Phi(\kv_m)|^2,
\eeq
where $\widetilde{\y}$ has been zero padded and then Fourier transformed.  This technique of power spectrum estimation scales as $\mathcal{O}(\zeta^3 N\log N)$, which is fine as long as $\zeta$ is small,\footnote{Though not the computational bottleneck, this step is the most memory intensive; it involves writing down an array of $\zeta^3 N$ double-precision complex numbers.  This can reach into the gigabytes for very large data cubes.}  In Figure \ref{FFTapproximationToQ} we see how increasing $\zeta$ from 1 to 5 greatly improves accuracy.

\subsection{Inverse Variance Weighting with the Conjugate Gradient Method} \label{CGPSE}
We now know how to calculate $\widehat{q}^\alpha$ quickly provided that we can also calculate $\y \equiv \C^{-1}(\x - \langle \x \rangle)$ quickly.  The latter turns out to be the most challenging part of the problem; we will address the various difficulties that it presents in this Section through Section \ref{FastPrecon}.  We take our inspiration for a solution from a similar problem that the WMAP team faced in making their maps.  They employed the preconditioned conjugate gradient method to great success \citep{OhCMBConjGrad,WMAPconjugategrad}.  

The conjugate gradient method \cite{ConjGradOriginal} is an iterative technique for solving a system of linear equations such as $\C\y = (\x - \langle \x \rangle) $.  Although directly solving this system involves inverting the matrix $\C$, the conjugate gradient method can approximate the solution to arbitrary precision with only a limited number of multiplications of vectors by $\C$.  If we can figure out a way to quickly multiply vectors by $\C$ by investigating the structure of its constituent matrices, then we can fairly quickly approximate $\y$.  We will not spell out the entire algorithm here but rather refer the reader to the helpful and comprehensive description of it in \citep{ConjGrad}.

Whenever iterative algorithms are employed, it is important to understand how quickly they converge and what their rates of convergence depend upon.  If we are trying to achieve an error $\varepsilon$ on our approximation $\y_\text{CGM}$ to $\y$ where
\beq
\varepsilon \equiv \frac{|\C\y_\text{CGM} - (\x - \langle \x \rangle)|}{|\x - \langle \x \rangle|}.
\eeq
and where $|\x| \equiv \left(\sum_i x_i^2\right)^{1/2}$ is the length of the vector $\x$, then the number of iterations required to converge (ignoring the accumulation of round-off error) is bounded by \citep{ConjGrad}:
\beq
n \le \frac{1}{2}\sqrt{\kappa}\ln\left(\frac{2}{\varepsilon}\right)
\eeq
where $\kappa$ is the condition number of the matrix (not to be confused with $\kappa$ used elsewhere as a spectral index), defined as the ratio of its largest eigenvalue to its smallest:
\beq
\kappa(\C) \equiv \frac{\lambda_{\max}(\C)}{\lambda_{\min}(\C)}
\eeq
Because $n$ only depends logarithmically on $\varepsilon$, the convergence of the conjugate gradient method is exponential.  In order to make the algorithm converge in only a few iterations, it is necessary to ensure that $\kappa$ is not too large. This turns out to be a major hurdle that we must overcome, because we will routinely need to deal with covariance matrices with $\kappa(\C)\approx 10^{8}$ or worse.  This dynamic range problem is unavoidable; it comes directly from the ratio of the brightest foregrounds, typically hundreds of kelvin, to the noise and signal, typically tens of millikelvin.  That factor, about $10^4$, enters squared into the covariance matrices, yielding condition numbers of roughly $10^{8}$.  In Section \ref{FastPrecon} we will explain the efforts we undertake to mitigate this problem.

\subsection{Foreground and Noise Covariance Matrices} \label{FastCov}

Before we can go about ensuring that the conjugate gradient method converges quickly, we must understand the detailed structure of the constituent matrices of $\C$.  In particular, we will show that these matrices can all be multiplied by vectors in $\BigO(N\log N)$ time.  We will first examine the new kind of foreground we want to include, resolved point sources, which will also provide a useful example for how the foreground covariances can be quickly multiplied by vectors. 

\subsubsection{Resolved Point Sources} \label{RFast}

Unlike LT, we do not assume that bright point sources have already been cleaned out of our map.  Rather we wish to unify the framework for accounting for both resolved and unresolved foregrounds by inverse covariance weighting. This will allow us to directly calculate how our uncertainties about the fluxes and spectral indices of these point sources affect our ability to measure the 21\,cm power spectrum.

In contrast to the unresolved point sources modeled by $\U$, we model $N_R$ bright resolved point sources as having known positions\footnote{If the data cube is not overresolved, this assumption should be pretty good.  If a point source appears to fall in two or more neighboring pixels, it could be modeled as two independent point sources in this framework.  An even better choice would be to include the correlations between the two pixels, which would be quite strong.  Modeling those correlations could only improve the results, since it would represent including additional information about the foregrounds, though it might slow down the method slightly.  Not accounting for position uncertainty will cause the method to underestimate the ``wedge'' feature \citep{Dattapowerspec,juddjackiemiguel1,VedanthamWedge,MoralesPSShapes,CathWedge}.} with different fluxes $S_n$ (at reference frequency $\nu_*$) and spectral indices $\kappa_n$, neither of which is known perfectly.  We assume that resolved point source contributions to $\x$ are uncorrelated with each other, so we can define an individual covariance matrix $\mathbf{R}_n$ for each point source.  This means that our complete model for $\R$ is:
\beq
\mathbf{R} \equiv \sum_n \mathbf{R}_n.
\eeq
 
Following LT, we can express the expected brightness temperature in a given voxel along the line of sight of the $n^\text{th}$ point source by a probability distribution for flux, $p_{S_n}(S')$, and spectral index, $p_{\kappa_n}(\kappa')$, that are both Gaussians with means $S_n$ and $\kappa_n$ and standard deviations $\sigma_{S_n}$ and $\sigma_{\kappa_n}$, respectively.  Following the derivation in LT, this yields:
\begin{eqnarray} \label{spectralIndexAveraging}
\left<x_i\right>_n & = & \delta_{ii_n} (1.4\times 10^{-3}\mbox{K}) \left(\frac{\Omega_{pix}}{1 \mbox{ sr}}\right)^{-1} \eta_i^{-2} \nonumber \int_{-\infty}^{\infty} \left(\frac{S'}{1 \mbox{ Jy}}\right) p_{S_n}(S') dS' \int_{-\infty}^{\infty} \eta_i^{-\kappa'} p_{\kappa_n}(\kappa') d\kappa'
\\ & = & \delta_{ii_n}(1.4\times 10^{-3}\mbox{K}) \left(\frac{S_n}{1 \mbox{ Jy}}\right) \left(\frac{\Omega_{pix}}{1 \mbox{ sr}}\right)^{-1} \times \eta_i^{-2-\kappa_n} \mbox{exp}\left[\frac{\sigma_{\kappa_n}^2}{2}(\mbox{ln}\eta_i)^2\right],
\end{eqnarray}
where again $\eta_i \equiv \nu_i / \nu_*$.  Here $\delta_{ii_n}$ is a Kronecker delta that forces $\left<x_i\right>$ to be zero anywhere other than the line of sight corresponding to the $n^\text{th}$ resolved point source.  Likewise, we can write down the second moment:
\begin{align}
\left< x_i x_j \right>_n =&\text{ } \delta_{ii_n}\delta_{jj_n}(1.4\times 10^{-3}\text{K})^2 (\eta_i\eta_j)^{-2} \times \nonumber \\
& \left(\int_{-\infty}^{\infty} \left(\frac{S'}{1 \mbox{ Jy}}\right)^2 p_{S_n}(S') dS' \right)\left(\frac{\Omega_{pix}}{1 \mbox{ sr}}\right)^{-2} \left(\int_{-\infty}^{\infty} (\eta_i\eta_j)^{-\kappa'} p_{\kappa_n}(\kappa') d\kappa' \right) \nonumber \\
=& \text{ } \delta_{ii_n}\delta_{jj_n}(1.4\times 10^{-3}\text{K})^2 (\eta_i\eta_j)^{-2-\kappa_n} \times \nonumber \\
& \left(\frac{S_n^2 + \sigma^2_{S_n}}{(1\text{ Jy})^2}\right) \left(\frac{\Omega_{pix}}{1 \mbox{ sr}}\right)^{-2} \exp\left[ \frac{\sigma^2_{\kappa_n}}{2}(\text{ln} \eta_i \eta_j )^2 \right] 
\end{align}
where we assume $\sigma_{S_n} \approx 5\%$ of $S_n$ and $\sigma_{\kappa_n} \approx 0.2$. 

We know that $\left<x_i\right>_n\left<x_j\right>_n$ can be quickly multiplied by a vector because it is a rank 1 matrix.  Therefore, in order to show that all of $\R$ can be quickly multiplied, we recast $\left< x_i x_j \right>_n$ as the product of matrices that can be multiplied by a vector in $\BigO(N\log N)$ or faster.  If we then ignore the constants and just look at the parts of this matrix that depend on coordinates, we have that:
\begin{align}
\left< x_i x_j \right>_n \propto & \text{ }\delta_{ii_n}\delta_{jj_n} (\eta_i \eta_j)^{-2-\kappa_n} \exp\left[ \frac{\sigma^2_{\kappa_n}}{2}(\text{ln} \eta_i \eta_j )^2 \right] 
 \nonumber \\
= & \text{ } \delta_{ii_n} (\eta_i )^{-2-\kappa_n} \exp\left[ \sigma^2_{\kappa_n}(\text{ln}\eta_i)^2 \right] \times \nonumber \\
& \exp\left[-\frac{\sigma^2_{\kappa_n}}{2}\left(\text{ln}\frac{\eta_i}{\eta_j}\right)^2\right]  \times \nonumber \\
&\delta_{jj_n} (\eta_j )^{-2-\kappa_n}  \exp\left[ \sigma^2_{\kappa_n}(\text{ln}\eta_j)^2 \right] \label{2ndmoment}.
\end{align}
This matrix can be separated into the product of three matrices: one diagonal matrix that only depends on $\eta_i$, an inner matrix that includes the logarithm of a quotient of $\eta_i$ and $\eta_j$ in the exponent, and another diagonal matrix that only depends on $\eta_j$.  The diagonal matrices can be multiplied by a vector in $\BigO(n_z)$.  Moreover, because our cubes have redshift ranges $\Delta z < 0.5$ the frequencies at $i$ and $j$ are never very far apart, we can make the approximation that:
\begin{align}
\text{ln}\left(\frac{\eta_i}{\eta_j}\right) = \text{ln}\left(\frac{\nu_i}{\nu_j}\right) \approx \frac{\nu_0 + \Delta \nu_i}{\nu_0 + \Delta \nu_j} -1 = \frac{1}{\nu_0}(\Delta \nu_i - \Delta \nu_j) \label{freqApprox}
\end{align}
where $\Delta\nu_i \equiv \nu_i - \nu_0$ and $\nu_0$ is a constant reference frequency close to both $\nu_i$ and $\nu_j$.  We choose the center frequency of the data cube to be $\nu_0$. We can see now by combining Equations \ref{2ndmoment} and \ref{freqApprox} that the inner matrix in our decomposition of the second moment depends only on the magnitude of the difference between $\nu_i$ and $\nu_j$.  In the approximation that the physical size of the data cube is small enough that frequencies map linearly to distances, this shows that $\R_n$ respects translational invariance along the line of sight.
 
Because the entries in this inner part of $\R_n$ only depend on differences in frequencies, the inner matrix is a diagonal-constant or ``Toeplitz'' matrix.  Toeplitz matrices have the fortuitous property that they can be multiplied by vectors in $\mathcal{O}(N\mbox{log}N)$, as we explain in Appendix \ref{ToeplitzAppendix}.  Therefore, we can multiply $\R_n$ by a vector in $\BigO(n_z \log n_z)$ and we can multiply $\R$ by a vector faster than $\BigO(N \log N)$.

We can understand this result intuitively as a consequence of the fact that the inner part of $\R_n$ is translationally invariant along the line of sight.  Matrices that are translationally invariant in real space are diagonal in Fourier space.  That we need to utilize this trick involving circulant and Toeplitz matrices is a consequence of the fact that our data cube is neither infinite nor periodic.

\subsubsection{Unresolved Point Sources and the Galactic Synchrotron Radiation}\label{Ufast}
Let us now take what we learned in Section \ref{RFast} (and Appendix \ref{ToeplitzAppendix}) to see if $\U$ can also be quickly multiplied by a vector.  Looking back at Equation \ref{Uperp}, we can see that our job is already half finished; $(U_\perp)_{ij}$ only depends on the absolute differences between $(r_\perp)_i$ and $(r_\perp)_j$.  Likewise, we can perform  the exact same trick we employed in Equations \ref{2ndmoment} and \ref{freqApprox} to write down the relevant parts of $(U_\|)_{ij}$ from Equation \ref{Uexact} with the approximation that $\Delta \nu_i$ is always small relative to $\nu_0$:
\beq
(U_\|)_{ij} \propto  \exp\left[-\frac{\sigma^{2}_{\kappa_n}}{2\nu_0^{2}}(\Delta\nu_i - \Delta\nu_j)^{2}\right].
\eeq

In fact, we can decompose $\U$ as a tensor product of three matrices sandwiched between two diagonal matrices:
\beq
\U = \D_\U [\U_x \otimes \U_y \otimes \U_z] \D_\U. \label{formOfU}
\eeq
where all three inner matrices are Toeplitz matrices.  When we wish to multiply $\U$ by a vector, we simply pick out one dimension at a time and multiply every segment of the data by the appropriate Toeplitz matrix (e.g. every line of sight for $\U_z$). All together, the three sets of multiplications can be done in $\BigO(n_x n_y n_f \log n_f) + \BigO(n_x n_f n_y \log n_y) + \BigO(n_y n_f n_x \log n_x) = \BigO(N \log N)$ time.

Moreover, since $\G$ has exactly the same form as $\U$, albeit with different parameters, $\G$ too can be multiplied by a vector in $\BigO(N \log N)$ time by making the same approximation that we made in Equation \ref{freqApprox}.

\subsubsection{Instrumental Noise}\label{Nfast}
Lastly, we return now to the form of $\N$ we introduced in Section \ref{AdrianN}.  To derive a form we combine Equations \ref{NPowerSpectrum} and \ref{Ndef}.  The details are presented in Appendix \ref{NoiseAppendix}, so here we simply state the result:
\beq
\N = \F_\perp^\dagger \widetilde{\N} \F_\perp,
\eeq
where $\F_\perp$ and $\F_\perp^\dagger$ are the unitary discrete 2D Fourier and inverse Fourier transforms and where:
\beq
\widetilde{N}_{lm} =  \frac{\lambda^4 T_\text{sys}^2 j_0^2(k_{x,l} \Delta x /2) j_0^2(k_{y,l} \Delta y /2) }{A^2_\text{ant} (\Omega_\text{pix})^2 n_x n_y\Delta\nu}   \frac{\delta_{lm}}{t_l}.  \label{Nfourier}
\eeq
Here, $A_\text{ant}$ is the effective area of a single antenna, $\Delta\nu$ is the frequency channel width, $l$ and $m$ are indices that index over both $uv$-cells and frequencies, and $t_l$ is the total observation time in a particular $uv$-cell at a particular frequency. 

Because this matrix is diagonal, we have therefore shown that $\N$, along with $\R$, $\U$, and $\G$, can be multiplied by a vector in $\BigO(N \log N)$.  We have summarized the results for all four matrices in Table \ref{fastCovBasis}.
\begin{sidewaystable}
	\begin{center}   
	\begin{tabular}{| l | l | l |}
	\hline
   \textbf{Covariance Matrix} & \textbf{Parallel to the Line of Sight } & \textbf{Perpendicular to the Line of Sight} \\ \hline
	$\R$: Resolved Point Sources & Toeplitz symmetry & Diagonal in real space  \\ \hline
	$\U$: Unresolved Point Sources & Toeplitz symmetry & Toeplitz symmetry \\ \hline
	$\G$: Galactic Synchrotron Radiation & Toeplitz symmetry & Toeplitz symmetry \\ \hline
   $\N$: Instrumental Noise & Diagonal in real space & Diagonal in Fourier space \\ \hline
   \end{tabular}
	\caption[Summary of covariance models and their properties.]{Due to the symmetries or approximate symmetries our models for the foreground and noise covariance matrices, they can all be multiplied by a vector in $\BigO(N\log N)$ time or faster. Summarized above are the reasons why each matrix can be quickly multiplied by.  Either the matrices respect translation invariance and thus Toeplitz symmetry, their components are uncorrelated between lines of sight or  frequencies, making them diagonal in real space, or they are uncorrelated and thus diagonal in Fourier space.  In each case, the symmetries rely on the separabiltiy of the modeled covariance matrices into the tensor product of parts parallel or perpendicular to the line of sight. \label{fastCovBasis}}
  	\end{center}
\end{sidewaystable}

\subsubsection{Eliminating Unobserved Modes with the Psuedo-Inverse}\label{psuedoinverse}
In our expression for the noise covariance in Equation \ref{Nfourier}, we are faced with the possibility that $t_l$ could be zero for some values of $l$, leading to infinite values of $N_{ij}$.  Fourier modes with $t_l = 0$ correspond to parts of the $uv$-plane that are not observed by the instrument, i.e. to modes containing no cosmological information. We can completely remove these modes by means of the ``psuedo-inverse'' \citep{Maxgalaxysurvey1}, which replaces $\C^{-1}$ in the expression $\C^{-1}(\x - \langle \x \rangle)$ and optimally weights all observed modes (this removal can itself be thought of as an optimal weighting---the optimal weight being zero).  The psuedo-inverse involves $\Proj$, a projection matrix ($\Proj^\dagger = \Proj$ and $\Proj^2 = \Proj$) whose eigenvalues are 0 for modes that we want to eliminate and 1 for all other modes. It can be shown \citep{Maxgalaxysurvey1} that the quantity we want to calculate for inverse variance weighting is not $\C^{-1}(\x - \langle \x \rangle)$ but rather the quantity where:
\beq
\C^{-1} \longrightarrow \Proj \left[\Proj \C \Proj + \gamma (\I - \Proj) \right]^{-1} \Proj.
\eeq
In this equation, $\gamma$ can actually be any number other than 0.  The term in brackets in the above equation replaces the eigenvalues of the contaminated modes of $\C$ with $\gamma$.  The outer $\Proj$ matrices then project those modes out after inversion.  In this paper, we take $\gamma = 1$ as the convenient choice for the preconditioner we will develop in Section \ref{FastPrecon}.

The ability to remove unobserved modes is also essential for analyzing real data cubes produced by an interferometer.  Interferometers usually produce so-called ``dirty maps,'' which are corrected for the effects of the primary beam but have been convolved by the synthesized beam, represented by the matrix $\mathbf{B}$:
\beq
\x_{\text{dirty map}} = \mathbf{B} \x.
\eeq
To compute $\x$ for our quadratic estimator, we need to invert $\mathbf{B}$.  Since the synthesized beam matrix is diagonal in Fourier space, this would be trivial were it not for unobserved baselines that make $\mathbf{B}$ uninvertable.  This can be accomplished with the psuedoinverse as well, since the modes that would have been divided by 0 when inverting $\mathbf{B}$ are precisely the modes that we will project out via the psuedoinverse.  We can therefore comfortably take 
\beq
\x = \F^\dagger_\perp \Proj [\Proj \widetilde{\mathbf{B}} \Proj + \gamma(\I - \Proj)]^{-1} \Proj \F_\perp \x_{\text{dirty map}},
\eeq
where $\mathbf{B} \equiv \F^\dagger_\perp \widetilde{\mathbf{B}} \F_\perp$ and $\widetilde{\mathbf{B}}$ is diagonal.

The psuedo-inverse formalism can be usefully extended to any kind of mode we want to eliminate.  One especially useful application would be to eliminate frequency channels contaminated by radio frequency interference or adversely affected by aliasing or other instrumental issues.

\subsection{Preconditioning for Fast Conjugate Gradient Convergence} \label{FastPrecon}
\begin{figure*} 
	\centering 
	\includegraphics[width=1\textwidth]{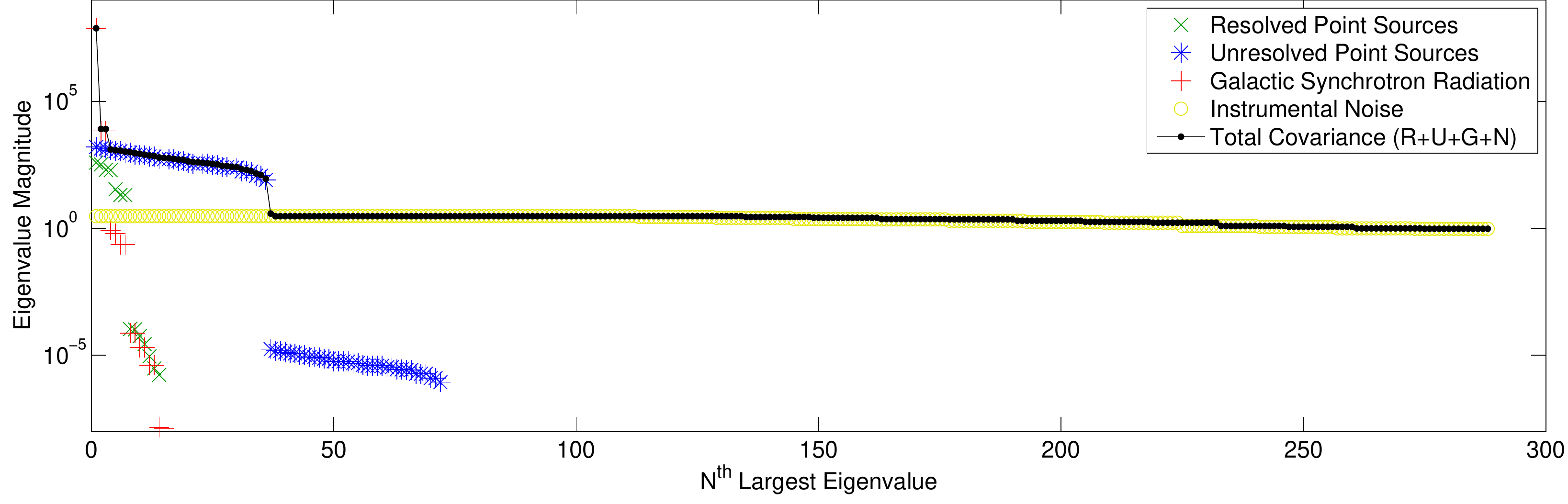}
	\caption[Covariance model components' eigenspectra.]{The distinct patterns in the eigenvalue spectrum of our covariance matrix provide an angle of attack for making the calculation of $\C^{-1} (\x - \langle \x \rangle)$ numerically feasible via preconditioning.  The plotted eigenvalue spectra of the covariance for a very small data cube exemplifies many of the important characteristics of the constituent matrices.  First, notice that the noise eigenvalue spectrum, while flatter than any of the others, is not perfectly flat.  The condition number of $N$ is related the ratio of the observing times in the most and least observed cell in the $uv$-plane.   Sometimes this factor can be $10^3$ or $10^4$.  Another important pattern to notice are the fundamental differences between the eigenvalue spectra of $\U$, $\G$, and $\R$.  First off, $\R$ has mostly zero eigenvalues, because $\R$ is a block diagonal matrix with most of its blocks equal to zero.  Second, despite the fact that $\U$ and $\G$ have nearly identical mathematical forms, $\U$ has stair-stepping eigenvalue spectrum while that of $\G$ is a much clearer exponential falloff.  This is due to the much stronger correlations perpendicular to the line of sight in $\G$. }
	\label{noPrecon}
\end{figure*}

We have asserted that the quantity $\mathbf{y} \equiv \C^{-1} (\x - \langle \x \rangle)$ can be estimated quickly using the conjugate gradient method as long as the condition number $\kappa(\C)$ is reasonably small.  Unfortunately, this is never the case for any realistic data cube we might analyze.  In Figure \ref{noPrecon} we plot the eigenvalues of $\C$ and its constituent matrices for a small data cube (only $6\times 6\times 8$ voxels) taken from a larger, more representative volume.  
In this example, $\kappa(\C) \approx 10^8$, which would cause the conjugate gradient method to require tens of thousands of iterations to converge.  This is typical; as we discussed in Section \ref{CGPSE}, values of around $10^{8}$ are to be expected.  We need to do better.

\subsubsection{The Form of the Preconditioner}

The core idea behind ``preconditioning'' is to avoid the large value of $\kappa(\C)$ by introducing a pair of preconditioning matrices $\Pre$ and $\Pre^\dagger$.  Instead of solving the linear system $\C\y=(\x - \langle \x \rangle)$, we solve the mathematically equivalent system: 
\beq
\C' \y' = \Pre (\x - \langle \x \rangle),
\eeq
where $\C' \equiv \Pre \C \Pre^\dagger$ and $\y' \equiv (\Pre^\dagger)^{-1}\y$.  If we can compute $\Pre(\x - \langle \x \rangle)$ and, using the conjugate gradient method on $\C'$, we can solve for $\y'$ and thus finally find $\y = \Pre^\dagger \y'$.  If $\Pre$ and $\Pre^\dagger$ are matrices that can be multiplied by quickly and if $\kappa(\C') \ll \kappa(\C)$, then we can greatly speed up our computation of $\y = \C^{-1}(\x - \langle \x \rangle)$.  Our goal is to build up preconditioning matrices specialized to the forms of the constituent matrices of $\C$.  We construct preconditioners for $\C = \N$, generalize them to $\C = \U + \N$, and then finally incorporate $\R$ and $\G$ to build the full preconditioner.  

The result is the following:
\beq
\C' = \F_\perp^\dagger \Pre_\U \Pre_\Gam \Pre_\N (\C) \Pre_\N^\dagger \Pre_\Gam^\dagger \Pre_\U^\dagger \F_\perp.
\eeq
Where $\Pre_\U$, $\Pre_\Gam$ and $\Pre_\N$ and preconditioners for $\U$, $\Gam \equiv \R + \G$, and $\N$ respectively.  A complete and pedagogical explanation of this preconditioner and the motivation for its construction and complex form can be found in Appendix \ref{PreconAppendix}.  The definitions of the matrices can be found in Equations \ref{PUmulti}, \ref{PreGammaMulti}, and \ref{PNdef} respectively.

\begin{figure} 
	\centering 
	\includegraphics[width=.6\textwidth]{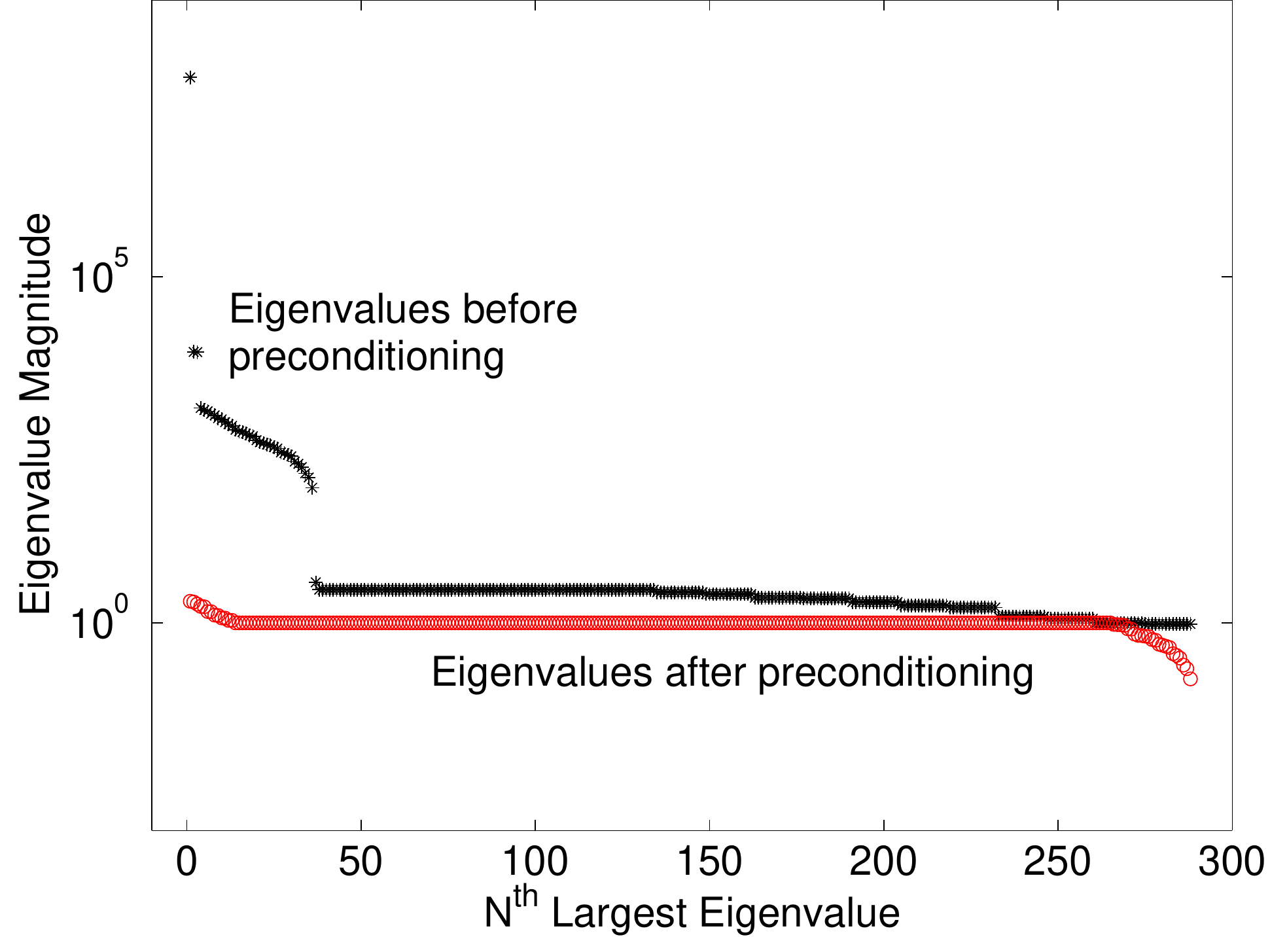}
	\caption[Results of the preconditioner.]{The preconditioner for the conjugate gradient method that we have devised significantly decreases the range of eigenvalues of $\C$.  Our preconditioner attempts to whiten the eigenvalue spectra of the constituent matrices of $\C$ sequentially, first $\N$, then $\R$ and $\G$ together, and finally $\U.$   By preconditioning, the condition number $\kappa(\C)$, the ratio of the largest to smallest eigenvalues, is reduced from over $10^8$ to about $10^1$.}
	\label{preconditionedEigs}
\end{figure}

Despite its complex form and construction, the procedure reduces $\kappa(\C)$ by many orders of magnitude.  In Figure \ref{preconditionedEigs}, in explicit contrast to Figure \ref{noPrecon}, we see a demonstration of that effect.

\subsubsection{Computational Complexity of the Preconditioner} \label{PreconComplexity}
In Appendix \ref{PreconAppendix}, we briefly discuss how the different steps in computing and applying this preconditioner scale with the problem size.  If any of them scale too rapidly with $N$, we can quickly lose the computational advantage of our method over that of LT.\footnote{This section may be difficult to follow without first reading Appendix \ref{PreconAppendix}.  However, the key results can be found in Table \ref{preconComplexity}.}  

First, let us enumerate the complexity of setting up the preconditioner for each matrix.  $\Pre_\N$ requires no setup since it only involves computing powers of the diagonal matrix $\widetilde{\N}$ (see Appendix \ref{PreconNSection}). $\Pre_\U$ requires the eigenvalue decomposition of $\U_z$, the component of $\U$ along the line of sight, which takes $\BigO (n_z^3)$ time (see Appendix \ref{PreconForU}).

We need the eigensystems of $\R$ and $\G$ to compute the eigensystem of $\Gam$ for $\Pre_\Gam$ (see Appendix \ref{PreconForGamma}). $\R$ requires performing one eigenvalue decomposition of an $n_z \times n_z$ matrix for every resolved point source; that takes $\BigO( N_R n_z^3)$ time. $\G$ simply requires three eigenvalue decompositions: one for each matrix like those that appear for $\U$ in Equation \ref{Uouterproduct} whose total outer product is $\G$.  Thus, the complexity is $\BigO(n_x^3) + \BigO(n_y^3) + \BigO(n_z^3)$.  

Next, we need to compute the eigenvalues of $\Gam_{\perp,k}$, the components of $\Gam$ perpendicular to the line of sight corresponding to each of the ``relevant'' (i.e. much bigger than the noise floor) eigenvalues of $\Gam$ along the line of sight (see Appendix \ref{PreconForGamma} for a more rigorous definition).  Using the notation we develop in Appendix \ref{PreconForU}, we denote the number of relevant eigenvalues of a matrix $\mathbf{M}$ as $m(\mathbf{M})$.  The number of times we need to decompose an $n_x n_y \times n_x n_y$ matrix is generally equal to the number of relevant eigenvalues of $\G_z$, since the number of relevant eigenvectors is almost always the same for $\G$ and $\R$.  So we have then a computational complexity of $\BigO(m(\G_z)(n_x n_y)^3)$.  Given the limited angular resolution of the experiment and the flat sky approximation, we generally expect $n_x$ and $n_y$ to be a good deal smaller than $n_f$, making this scaling more tolerable.  All these scalings are summarized in Table \ref{preconComplexity}.
\begin{table}
	\begin{center}   
	\begin{tabular}{| l | l |}
	\hline
   \textbf{Operation} & \textbf{Complexity} \\ \hline
	Compute $\U$ eigensystem & $\BigO(n_z^3)$ \\ \hline
	Compute $\G$ eigensystems & $\BigO(n_x^3) + \BigO(n_y^3) + \BigO(n_z^3)$ \\ \hline
	Compute $\R$ eigensystems & $\BigO(N_R n_z^3)$ \\ \hline
	Compute $\Gam$ eigensystems & $\BigO(m(\G_z)(n_x n_y)^3)$ \\ \hline \hline
	Apply $\Pre_\N$ & $\BigO(N\log N)$ \\ \hline
	Apply $\Pre_\U$ & $\BigO(N m(\U_z))$ \\ \hline
	Apply $\Pre_\Gam$ & $\BigO(N m(\G)) + \BigO(N N_R m(\R_n))$ \\ \hline
   \end{tabular}
	\caption[Preconditioner computational complexity summary.]{The computational complexity of setting up the preconditioner is, at worst, roughly $\BigO(N^2)$, though this operation only needs to be performed once. Even for large data cubes, this is not the rate-limiting step in power spectrum estimation.    The computational complexity of applying the preconditioner ranges from $\BigO(N \log N)$ to $\BigO(NN_R)$.  For large data cubes with hundreds of bright point sources, the preconditioning time is dominated by $\Pre_\Gam$, which is in turn dominated by preconditioning associated with individual point sources.  The computational complexity of the preconditioner therefore depends on the number of point sources considered ``resolved,'' which scales with both field of view and with the flux cut.  Here $N_R$ is the number of resolved point sources in our field of view, $n_d$ is the size of the box in voxels along the $d^{\text{th}}$ dimension, and $m$ is the number of relevant eigenvalues of a matrix above the noise floor that need preconditioning.\label{preconComplexity}}
  	\end{center}
\end{table}
     
\begin{figure} 
	\centering 
	\includegraphics[width=.6\textwidth]{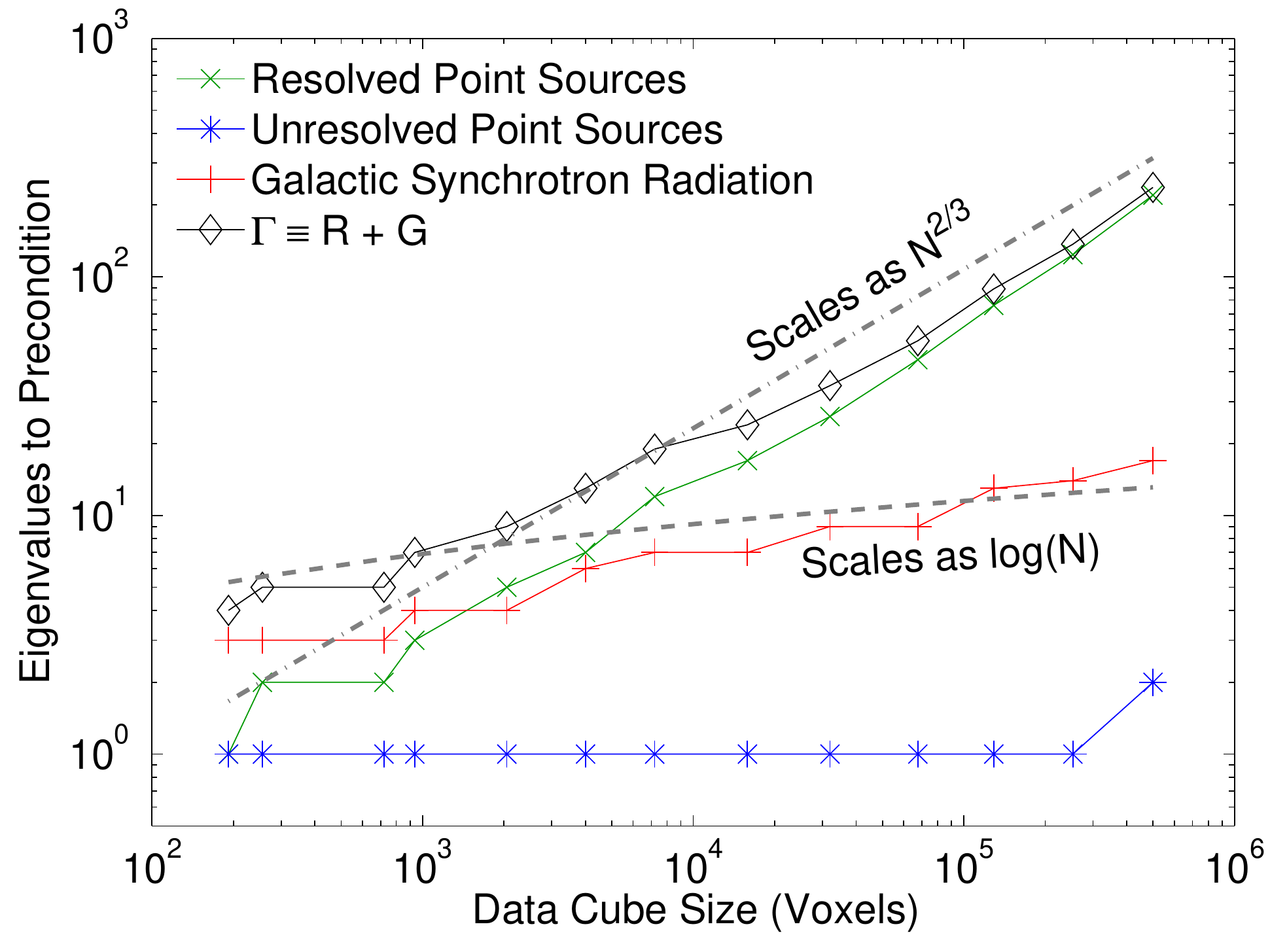}
	\caption[Preconditioner complexity scaling by component.]{For large data cubes and a fixed definition of what constitutes a ``bright'' point source, the complexity of preconditioning is dominated by the number of resolved point sources.  Specifically, the complexity of preconditioning for $\Gamma$ scales as $N^{2/3}$ because the number of resolved point sources is simply proportional to the solid angle of sky surveyed, which scales with the survey volume (and thus number of voxels, assuming fixed angular and frequency resolution) to the $\frac{2}{3}$ power. This also confirms our assertion that the number of important eigenvalues of $\G$ and $\U$ should scale logarithmically with data cube size (albeit with a different prefactor).  Each of the data cubes is taken from the same survey with the same ratio of width to depth.  The number of eigenvalues to precondition is computed assuming an eigenvalue threshold of $\theta = 1$.}
	\label{PreconScaling}
\end{figure}

Until now, all of our complexities have been $\BigO(N \log N)$ or smaller.  Because these small incursions into bigger complexity classes are only part of the set-up cost, they are not intolerably slow as long as $m(\G_z)$ is small. This turns out to be true because the eigenvalue spectra of $\R_n$ and $\G_z$ fall off exponentially, meaning that we expect the number of relevant eigenvalues to grow only logarithmically.  This is borne out in Figure \ref{PreconScaling} where we see exactly how the number of eigenvalues that need to be preconditioned scales with the problem size.

Let us now turn to a far more important scaling: that of multiplying the preconditioner by a vector.  The set-up needs to be done only once per Fisher matrix calculation; the preconditioning needs to happen for every iteration of the conjugate gradient method.  $\Pre_\N$ is the easiest; we only ever need to perform a Fourier transform or multiply by a diagonal matrix.  The complexity is merely $\BigO (N \log N)$.  $\Pre_\U$ only involves multiplying by vectors for each relevant eigenvalue of $\U_z$, so the total complexity is $\BigO(N m(\U_z))$.

Finally, we need to assess the complexity of applying $\Pre_\Gam$.  When performing the eigenvalue decomposition of $\Gam_{\perp,k}$, we expect roughly the same number of eigenvalues to be important that would have been important from $\R$ and $\G$ separately for that $k$ index.  Each of those eigenvectors takes $\BigO(N)$ time to multiply by a vector.  So we expect to deal with $m(\G)$ eigenvalues from $\G$ and one eigenvalue from each resolved point source for each relevant value of $k$, or about $N_R m(R_n)$.  Applying $\Pre_\Gam$ therefore is $\BigO(N m(\G)) + \BigO(N N_R m(\R_n))$.  If we keep the same minimum flux for the definition of a resolved point source and if we scale our cube uniformly in all three spatial directions, then $N_R \propto N^{2/3}$.  

This turns out to be the rate-limiting step in the entire algorithm.  If we decide instead to only consider the brightest $N_R$ to be resolved, regardless of box size, then applying $\Pre_\Gam$ reduces to $\BigO(N\log N)$.  Likewise, if we are only interested in expanding the frequency range of our data cube, the scaling also reduces to $\BigO(N\log N).$  We can comfortably say then that the inclusion of a model for resolved point sources introduces a complexity bounded by $\BigO(N\log N)$ and $\BigO(N^{5/3})$.  We can see the precise computational effect of the preconditioner when we return in Section \ref{CompScaling} to assess the overall scaling of the entire algorithm. These results are also summarized in Table \ref{preconComplexity}.

\subsubsection{Preconditioner Results} \label{PreconResults}
\begin{figure} 
	\centering 
	\includegraphics[width=.6\textwidth]{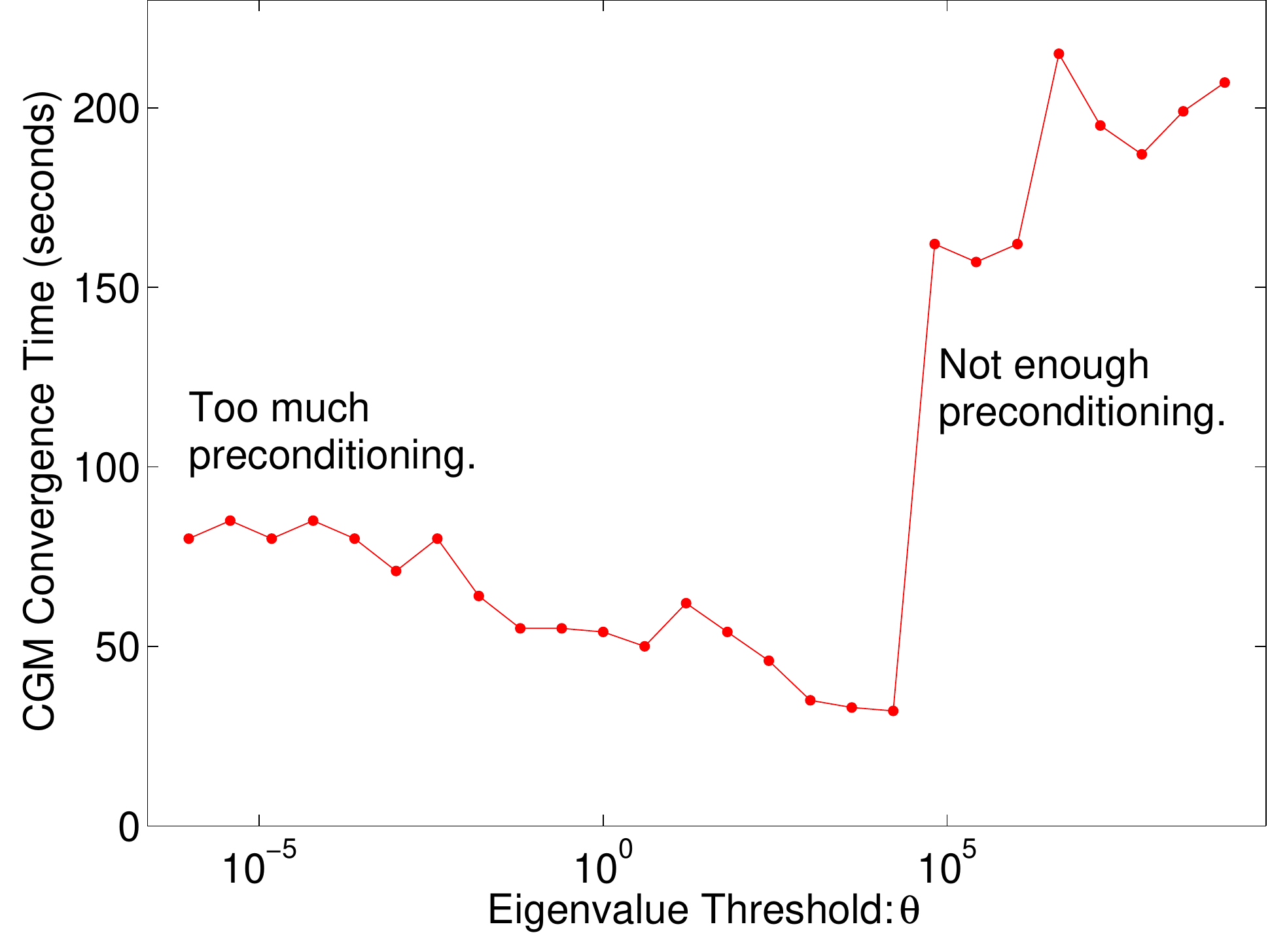}
	\caption[Conjugate gradient convergence time.]{This plot shows how computational time scales with $\theta$, the threshold for preconditioning, for the conjugate gradient method performed on an $N \approx 10^4$ voxel data cube.  It appears that for this particular covariance matrix, a minimum exists near $\theta = 10^4$.  At the minimum, the greater number of conjugate gradient iterations are balanced by quicker individual iterations (since each iteration involves less preconditioning).  We can see from this plot that there exists a critical value of $\theta$ around $5\times 10^4$ where the preconditioning of a small number of additional eigenvalues yields a large effect on the condition number of the resultant matrix.  Without preconditioning, sufficiently large values of $\kappa(\C)$ could also lead to the accumulation of roundoff error that prevents convergence of the conjugate gradient method.}
	\label{PreconPerformance}
\end{figure}

Choosing which eigenvalues are ``relevant'' in the constituent matrices of $\C$ and therefore need preconditioning depends on how these eigenvalues compare to the noise floor.  In Appendix \ref{PreconForU}, we define a threshold $\theta$ which distinguishes relevant from irrelevant eigenvalues by comparing them to $\theta$ times the noise floor.  Properly choosing a value for $\theta$, the threshold below which we do not precondition eigenvalues of $\Ubar$ and $\GammaBar$, presents a tradeoff.  We expect that that too low of a value of $\theta$ will precondition inconsequential eigenvalues, thus increasing the conjugate gradient convergence time.  We also expect that too large of a value of $\theta$ will leave some of the most important eigenvalues without any preconditioning, vastly increasing convergence time.  Both of these expectations are borne out by our numerical experiments, which we present in Figure \ref{PreconPerformance}.

In this work, we choose $\theta = 1$ (all foreground eigenvalues above the noise floor are preconditioned) for simplicity and to be sure that we are not skipping the preconditioning of any important foreground eigenvalues.  One might also worry that more iterations of the algorithm provides more opportunity for round-off error to accumulate and prevent convergence, as has sometimes proven the case in our numerical experiments.  For lengthy or repeated calculations of the Fisher matrix, it is wise to explore the performance of several levels of preconditioning, especially if it can garner us a another factor of 2 in speed.

\subsection{Fast Simulation of Foregrounds and Noise} \label{random}
We concluded Section \ref{fisherMC} with the fact that a Monte Carlo calculation of the Fisher matrix required the ability to compute $\widehat{\mathbf{q}}$ from many different realizations of the foregrounds and noise modeled by $\C$.  In Sections \ref{fastPSE} through \ref{FastPrecon}, we have shown how to quickly calculate $\widehat{\mathbf{q}}$ from a data vector $\x$ using Equation \ref{estimator}.

But where does $\x$ come from?  When we want to estimate the 21\,cm temperature power spectrum of our universe, $\x$ will come from data cubes generated from real observations.  But in order to calculate $\F$, which is essential both to measuring $\widehat{\mathbf{p}}$ and estimating the error on that measurement, we must first be able to create many realizations of $\x$ drawn from our models for noise and foregrounds that we presented in Section \ref{FastCov}.

A mathematically simple way to draw $\x$ from the right covariance matrix is to create a vector $\mathbf{n}$ of independent and identically distributed random numbers drawn from a normal distribution with mean 0 and standard deviation 1.  Then, it is easy to see that 
\beq
\x \equiv \C^{1/2} \mathbf{n} \label{CovHalfPower}
\eeq
is a random vector with mean $\mathbf{0}$ and covariance $\C$.  Unfortunately, computing $\C^{1/2}$ is just as computationally difficult as computing $\C^{-1}$.

In this last section of our presentation of our fast method for power spectrum estimation and statistics, we will explain how a vector can be created randomly with covariance $\C$.  We do so by creating vectors randomly from each constituent matrix of $\C$, since each contribution to the measured signal is uncorrelated.  In Section \ref{randomResults}, we will demonstrate numerically that these simulations can be performed quickly while still being accurately described by the underlying statistics.

\subsubsection{Resolved Point Sources} \label{Rrandom}
The simplest model to reproduce in a simulation is the one for resolved point sources, because the covariance was created from a supposed probability distribution over their true fluxes and spectral indices.  We start with a list of point sources with positions and with a specified but uncertain fluxes and spectral indices.  These fluxes can either come from a simulation, in which case we draw them from our source count distribution (Equation \ref{sourceCounts}) and spectral indices from a Gaussian distribution, or from a real catalog of sources with its attendant error bars.  The list of sources does not change over the course of calculating the Fisher matrix. 

In either case, calculating a random $\x_\R$ requires only picking two numbers, a flux and a spectral index, for each point source and then calculating a temperature in each voxel along that particular line of sight.  The latter is easy, since we assume it is drawn from a Gaussian.  The former can be quickly accomplished by numerically calculating the cumulative probability distribution from Equation \ref{sourceCounts} and inverting it.  Each random $\x_\R$ is therefore calculable in $\BigO(N_R n_z) < \BigO(N)$ time.

\subsubsection{Unresolved Point Sources} \label{Urandom}
We next focus on $\U$, which is more difficult.  Our goal is to quickly produce a vector with specified mean and covariance.  LT has already established what value we want for $\left<\x_\U\right>$ and $\left<\x_\G\right>$ with a calculation very similar to Equation \ref{spectralIndexAveraging}.  We need to figure out how to produce a vector with zero mean and the correct covariance.

One way around the problem of calculating $\C^{1/2}$ is to take advantage of the eigenvalue decomposition of the covariance matrix.  That is because if $\C = \mathbf{Q\Lambda Q}^\trans$, where $\mathbf{Q}$ is the matrix that transforms into the eigenbasis and $\mathbf{\Lambda}$ is a diagonal matrix made up of the eigenvalues, then $\C^{1/2} = \mathbf{Q\Lambda}^{1/2}\mathbf{Q}^\trans$.    We already found the few important eigenvalues of $\U$ for our preconditioner (see Section \ref{PreconForU}), so does this technique solve our problem?

Yes and no.  In the direction parallel to the line of sight, this technique works exceedingly well because only a small number of eigenvectors correspond to non-negligible eigenvalues.  We can, to very good approximation, ignore all but the largest eigenvalues (which correspond to the first few ``steps'' in Figure \ref{noPrecon}.)  We can therefore generate random unresolved point source lines of sight in $\BigO(n_z m(\U_z))$ with the right covariance.

A problem arises, however, when we want to generate $\x_\U$ with the proper correlations perpendicular to the line of sight.  Unlike the extremely strong correlations parallel to the line of sight, these correlations are quite weak.  Weak correlations entail many comparable eigenvalues; in the limit that point sources were uncorrelated, $\U_x \otimes \U_y \rightarrow \I_{\perp}$ and all the eigenvalues would be 1 (though the eigenvectors would of course be much simpler too).  Utilizing the same technique as above would require a total complexity of $\BigO(N n_x n_y)$ time, which is slower than we would like.
 
However, the fact that both $\U_x$ and $\U_y$ are Toeplitz matrices allows us to use the same sort of trick we employed to multiply our Toeplitz matrices by vectors in Section \ref{RFast} to draw random vectors from $\U_x \otimes \U_y$ \citep{ToeplitzSimulation}.  It turns out that the circulant matrix in which we embed our covariance matrix must be positive-semidefinite for this technique to work.  Although there exists such an embedding for any Gaussian covariance matrix, only Gaussians with coherence lengths small compared to the box size can be embedded in a reasonably small circulant matrix---exactly the situation we find ourselves in with $\U_\perp$.  As such, we can generate random $\x_\U$ vectors in $\BigO(N m(\U_z) \log(n_x n_y)) \approx \BigO(N \log N)$. 

\subsubsection{Galactic Synchrotron Radiation} \label{Grandom}
The matrix $\G$ differs from $\U$ primarily in the coherence length perpendicular to the line of sight.  Unlike $\U$, $\G$ has only a small handful of important eigenvalues, which means that random $\x_\G$ vectors can be generated in the same way we create line of sight components for $\x_\U$ vectors, which we described above.  Since $m(\G)$ is so small (see Figure \ref{noPrecon}) and grows so slowly with data cube size (see Figure \ref{PreconScaling}), we can create random $\x_\G$ vectors in approximately $\BigO(N)$.

\subsubsection{Instrumental Noise} \label{Nrandom}
Finally, we turn to $\N$, which is also mathematically simple to simulate.  First off, $\left< \x_\N \right> = 0$.  Next, because $\N$ is diagonal in the Fourier basis, we can simply use Equation \ref{CovHalfPower}.  Because $\N = \F_\perp^\dagger \widetilde{\N} \F_\perp$,
\beq
\N^{1/2} = \F_\perp^\dagger \widetilde{\N}^{1/2} \F_\perp,
\eeq
which is computationally easy to multiply by $\mathbf{n}$ because $\widetilde{\N}$ is a diagonal matrix.  The most computationally intensive step in creating random $\x_\N$-vectors is the fast Fourier transform, which of course scales as $\BigO(N\log N)$.

\subsubsection{Data Simulation Speed and Accuracy} \label{randomResults}
\begin{figure} 
	\centering 
	\includegraphics[width=.6\textwidth]{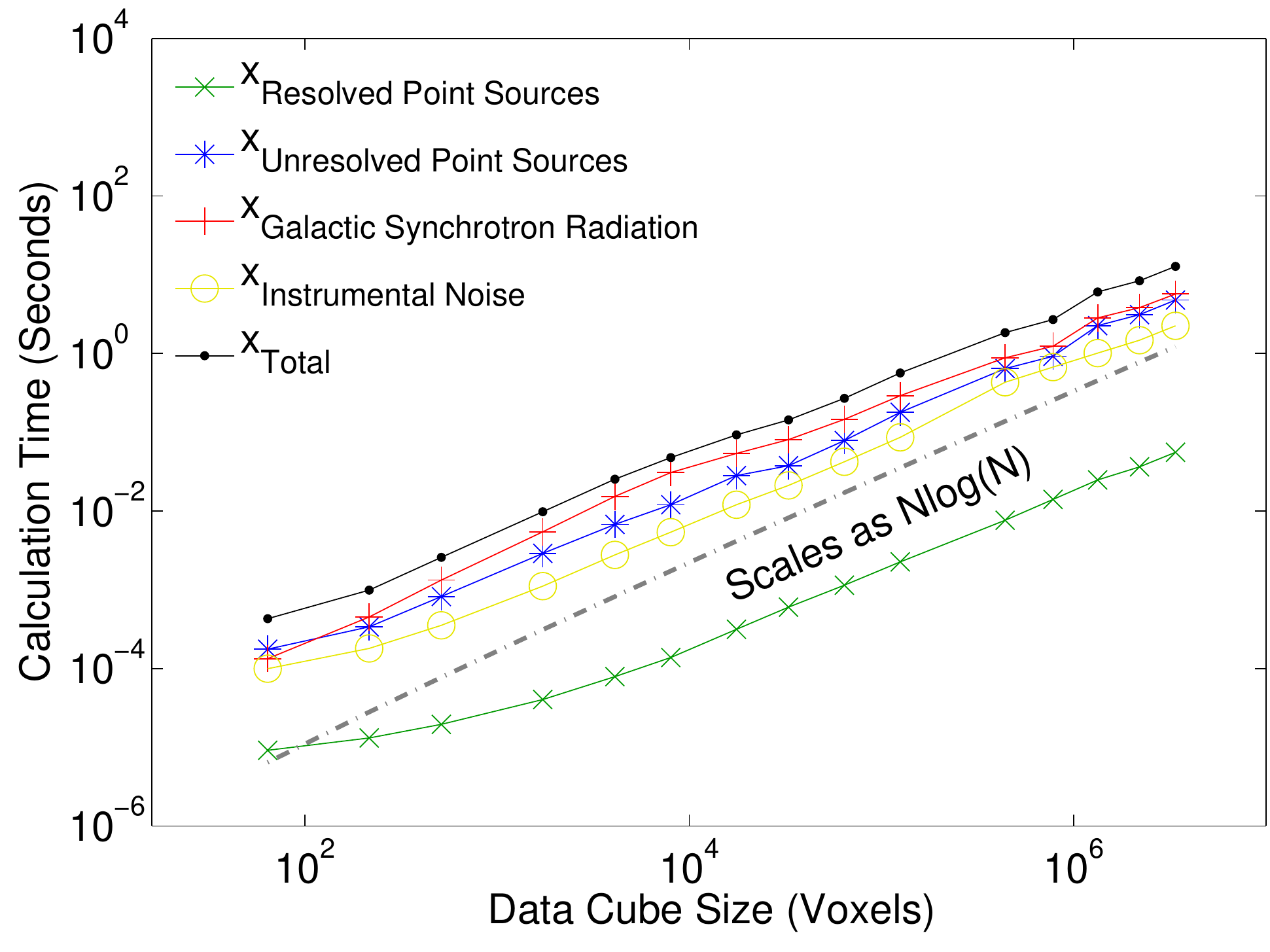}
	\caption[Simulation speed scaling from covariance model components.]{In order to estimate the Fisher matrix via a Monte Carlo, we need to draw random data cubes from our modeled covariance.  Here we show that we can do so in $\BigO(N \log N)$ by plotting computational time as a function of problem size for generating a random $\x$ for each of the constituent sources of $\x$.  In practice, generating random $\x$ vectors is never the rate-limiting step in calculating $\F$.}
	\label{RandomFieldScaling}
\end{figure}

Before we conclude this section and move on to the results of our method as a whole, we verify what we have claimed in the above sections: namely that we can quickly generate data cubes with the correct covariance properties.  Figure \ref{RandomFieldScaling} verifies the speed, showing that the algorithm is both fast and well-behaved for large data cubes.

\begin{figure} 
	\centering 
	\includegraphics[width=.6\textwidth]{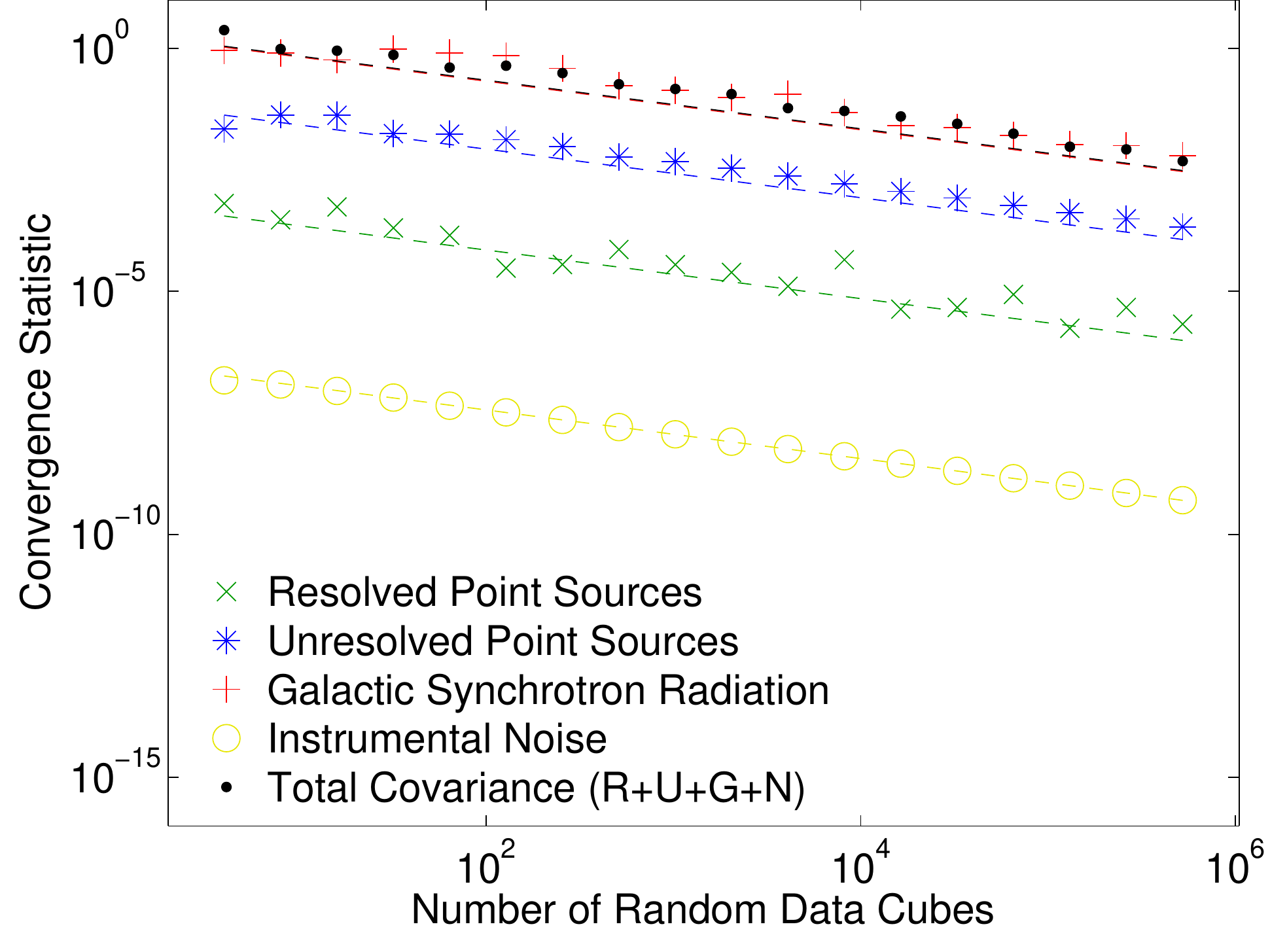}
	\caption[Covariance simulation convergence verification.]{We verify that our technique for quickly generating random data cubes actually reproduces the correct statistics by generating a large number of such cubes and calculating their sample covariances.  Plotted here is the error statistic detailed in Equation \ref{errorStatisic}.  The color-matched dotted lines are the expected convergences for correlated Gaussians from Equation \ref{expectedConvergence}.}
	\label{RandomFieldConvergence}
\end{figure}

In order to show that the sample covariance of a large number of random $\x$ vectors converges to the appropriate covariance matrix, we must first define a convergence statistic, $\varepsilon$.  We are interested in how well the matrix converges relative to the total covariance matrix $\C$.  For example, for $\R$ we choose:
\beq
\varepsilon(\widehat{\R}) \equiv \sqrt{\frac{\sum_{ij}\left|\widehat{R}_{ij} - R_{ij}\right|^2}{\sum_{ij}\left|C_{ij}\right|^2}} \label{errorStatisic}
\eeq
where $\widehat{\R}$ is the sample covariance of $n$ random $\x_\R$ vectors drawn from $\R$.  If each $\x$ is a Gaussian random vector then the expected RMS value of $\varepsilon$ is:
\beq
\sqrt{\left< \varepsilon(\widehat{\R})^2 \right>} = \frac{1}{\sqrt{n}}\left[\frac{\sum_{i,j}R_{ij}^2 + (\text{tr} \R)^2}{\sum_{ij}C_{ij}^2}  \right]. \label{expectedConvergence}
\eeq
In Figure \ref{RandomFieldConvergence}, we see that all four constituent matrices of $\C$ converge like $n^{-1/2}$, as expected, with very nearly the prefactor predicted by Equation \ref{expectedConvergence}.  We can be confident, therefore, in both the speed and accuracy of our technique for generating random vectors.

\subsection{Method Accuracy and Convergence} \label{fisherConverge}
Before we move on to discuss some of the results of our method, it is worthwhile to check that no unwarranted approximations prevent it from  converging to the exact form of the Fisher information matrix in Equation \ref{fisherTrace}.  Since calculating $\F$ exactly can only be done in $\BigO(N^3)$ time, we perform this test in two parts.
 
\begin{figure} 
	\centering 
	\includegraphics[width=.6\textwidth]{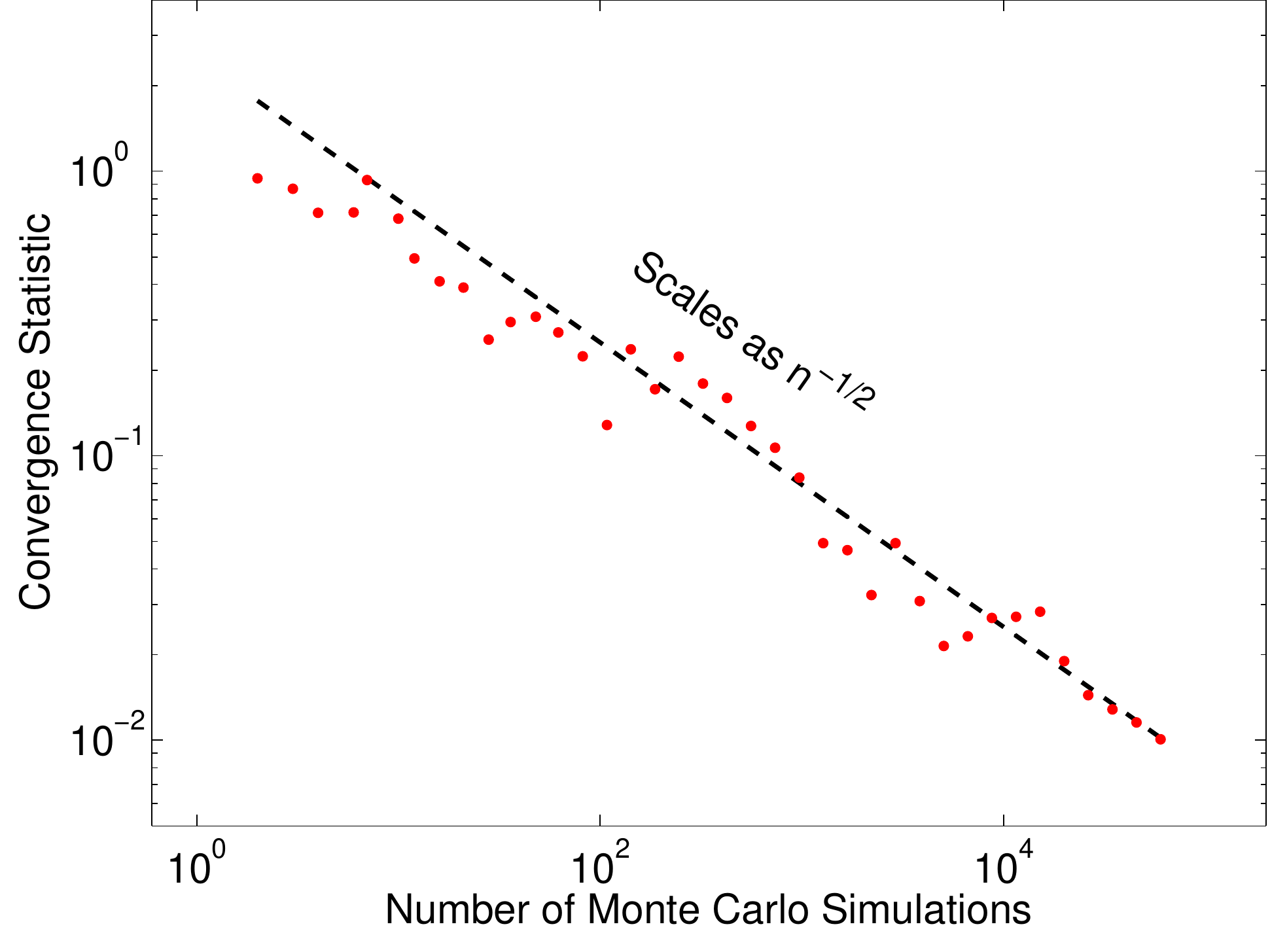}
	\caption[Monte Carlo convergence of the Fisher matrix.]{Our Monte Carlo converges to the correct Fisher matrix as $n^{-1/2}$, as expected.  In this plot, we compare the sample covariance of many $\widehat{\mathbf{q}}$ vectors generated from small data cubes to an exact calculation of $\F$ by calculating the relative error of their diagonals.}
	\label{MCFisherConvergence}
\end{figure}

First, we measure convergence to the exact Fisher matrix for a very small data cube with only $6\times 6 \times 8$ voxels.  Taking advantage of Equation \ref{covq}, we generate an estimate of $\F$, which we call $\widehat{\F}$, from the sample covariance of many independent $\widehat{\mathbf{q}}$ vectors.  We compare these $\widehat{\F}$, which we calculate periodically along the course of the Monte Carlo, with the $\F$ that we calculated directly using Equation \ref{fisherTrace}.  As we show in Figure \ref{MCFisherConvergence}, the sample covariance of our $\widehat{\mathbf{q}}$ vectors clearly follows the expected $n^{-1/2}$ convergence to the correct result.

\begin{figure} 
	\centering 
	\includegraphics[width=.6\textwidth]{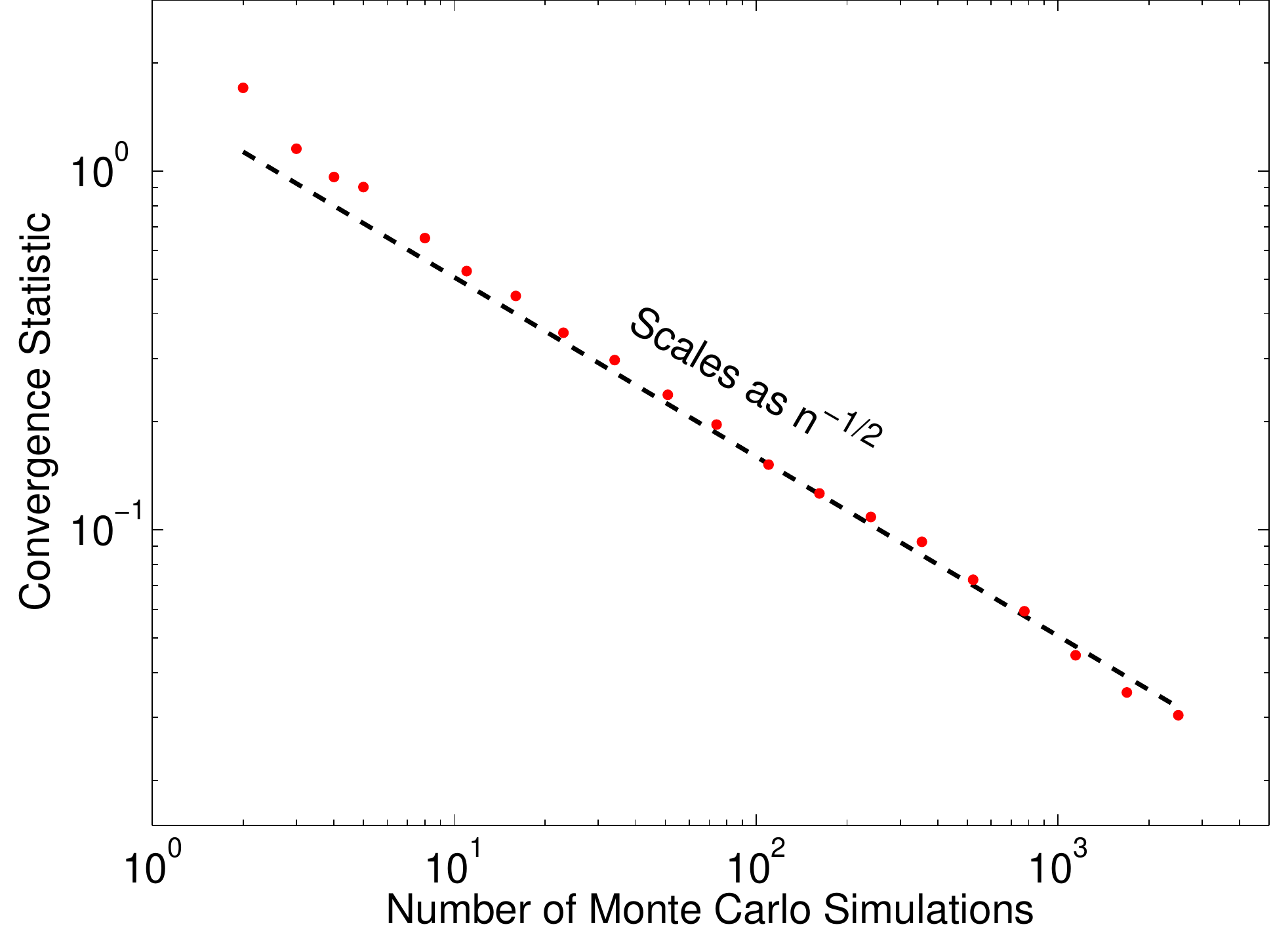}
	\caption[Monte Carlo convergence of the quadratic estimator.]{For large data cubes, the convergence of the sample covariance of our $\widehat{\mathbf{q}}$ vectors to $\F$ also nicely follows the expected $n^{-1/2}$ scaling.  We perform this analysis on a data cube with $1.5\times10^5$ voxels analogously to that which we performed in Figure \ref{MCFisherConvergence}, except that we use the sample covariance of double the number of Monte Carlo iterations as our ``true'' Fisher matrix.  This explains the artificially fast convergence we see in the last few points of the above plot.}
	\label{MCFisherConvergence2}
\end{figure}

However, we are more concerned with the accuracy of the method for large data cubes which cannot be tackled by the LT method.  Unfortunately, for such large data cubes, we cannot directly verify our result except in the case where $\C = \I$.  In concert with other tests for agreement  with LT, we also check that the method does indeed converge as $n^{-1/2}$ by comparing the convergence of subsets of the $\widehat{q}^\alpha$ vectors up to $n/2$ Monte Carlo iterations to the reference Fisher matrix, which we take to be the sample covariance of all $n$ iterations.  As we show in Figure \ref{MCFisherConvergence2}, our expectation is borne out numerically.

\subsection{Method Summary}

We have constructed a technique that accelerates the LT technique to $\BigO(N\log N)$ and extends it to include bright point sources in, at worst, $\BigO(N^{5/3})$.  We do so by generating random data vectors with the modeled foreground and noise covariances and calculating the Fisher information matrix via Monte Carlo.  We are able to calculate individual inverse variance weighted power spectrum estimates quickly using the conjugate gradient method with a specially adapted preconditioner.  

Our method makes a number of assumptions, most of which are not shared with the LT method.  Our method can analyze larger data sets but at a slight loss of generality. Although we have mentioned these assumptions throughout this work, it is useful to summarize them in one place:

\begin{itemize}
\item Our method relies on a small enough data cube perpendicular to the line of sight that it can be approximated as rectilinear (see Figure \ref{flatsky}).  

\item We approximate the natural log of the quotient of frequencies in the exponent of our point source covariance matrix by a leading-order Taylor expansion (Equation \ref{freqApprox}).  This assumption makes the foreground covariances translationally invariant along the line of sight and thus amenable to fast multiplication using Toeplitz matrix techniques.  This is a justified assumption as long as the coherence length of the foregrounds is much longer than the size of the box along the line of sight. 

\item Our ability to precondition our covariance matrix for the conjugate gradient method depends on the approximation that the correlation length of $\U$ perpendicular to the line of sight, due to weak spatial clustering of point sources, is not much bigger than the pixel size. For the purposes of preconditioning, we approximate $\U_\perp$ to be the identity (see Section \ref{PreconForU}).  The longer the correlation length of $\U_\perp$, the longer the conjugate gradient algorithm will take to converge.

\item Likewise, the speed of the preconditioned conjugate gradient algorithm depends on the similarities of the eigenmodes of the covariances for  $\R$, $\U$, and $\G$ along the line of sight.  The more similar the eigenmodes are (though their accompanying eigenvalues can be quite different) the more the preconditioning algorithm can reduce the condition number of $\C$.  We believe that this similarity is a fairly general property of models for foregrounds, though the introduction of a radically different foreground model might require a different preconditioning scheme.

\item We assume that the number of Monte Carlo iterations needed to estimate the Fisher information matrix is not so large that that it precludes analyzing large data cubes.  Because the process of generating more artificial $\widehat{\mathbf{q}}$ vectors is trivially parallelizable, we do not expect getting down to the requisite precision on the window functions to be an insurmountable barrier. 
\end{itemize}

One common theme among these assumptions, especially the last three, is that the approximations we made to speed up the algorithm can be relaxed as long as we are willing to accept longer runtimes.  This reflects the flexibility of the method, which can trade off speed for accuracy and vice versa.


\section{Results}\label{results}
Now that we are confident that our method can accurately estimate the Fisher information matrix and can therefore calculate both power spectrum estimates from data and the attendant error bars and window functions, we turn to the first end-to-end results of the algorithm.  In this Section, we demonstrate the power our method and the improvements that it offers over that of LT.  First, in Section \ref{CompToLT11} we show that our technique reproduces the results of LT in the regions of Fourier space where they overlap.  Then in Section \ref{CovMatBuildup}, we highlight the improvements stemming from novel aspects of our method, especially the inclusion of the pixelization factor $|\Phi(\kv)|^2$ in $\Q^\alpha$ and $\N$ and the separation of point sources into resolved point sources ($\R$) and unresolved point sources ($\U$), by showing how different parts of our algorithm affect $\F$.  In Section \ref{CompScaling} we examine just how much faster our algorithm is than that of LT, and lastly, in Section \ref{MWA} we forecast the cosmological constraining power of the 128-tile deployment of the MWA.

\subsection{Comparison to Liu \& Tegmark} \label{CompToLT11}

\begin{figure*}
\centering 
	\includegraphics[width=1\textwidth]{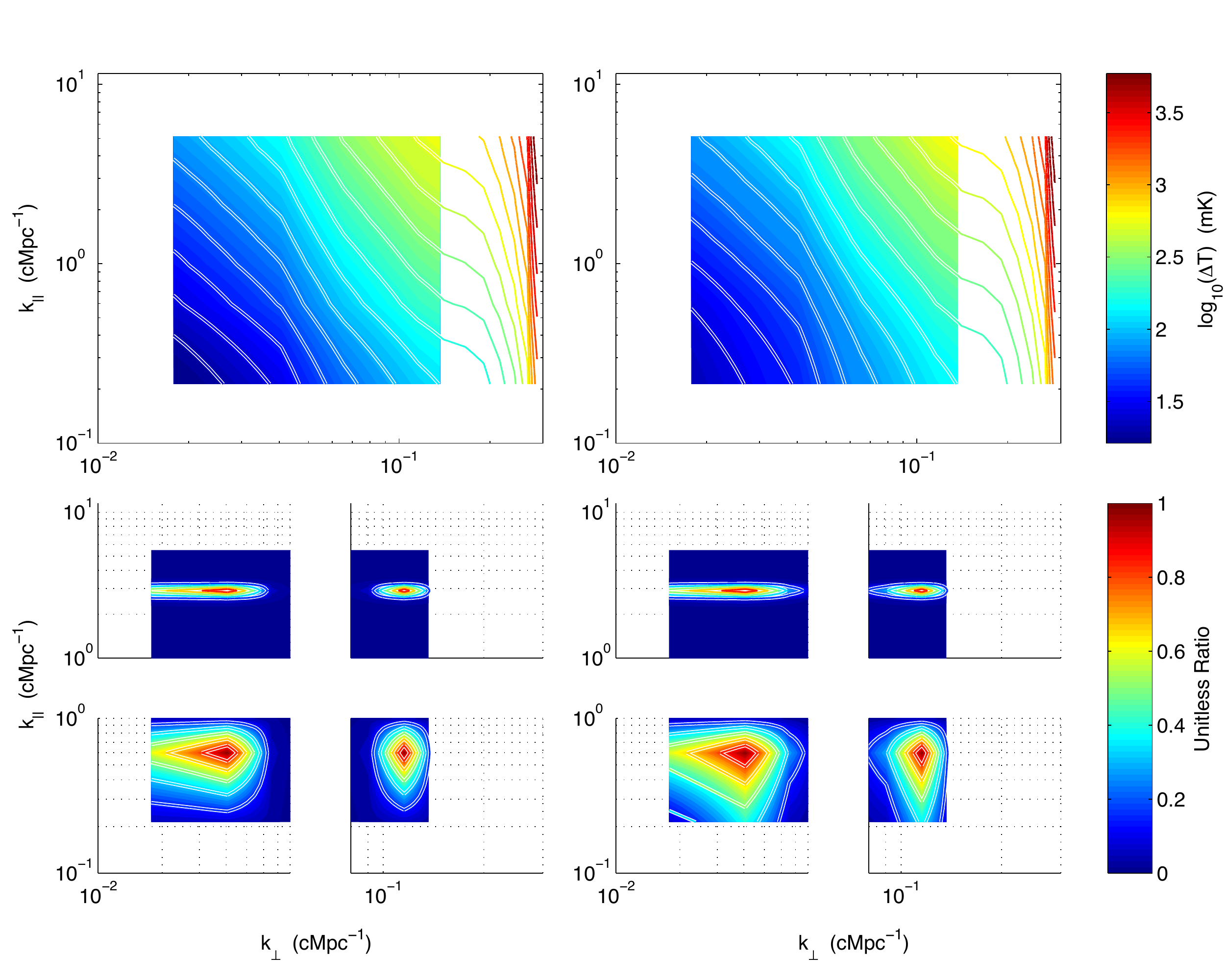}
	\caption[Comparison to exact method of Liu \& Tegmark.]{Our method faithfully reproduces the results of LT in the regions of Fourier space accessible to both methods.  Here we recreate both the vertical error bar contours from LT's Figure 8 (top two panels) and a few selected window functions from LT's Figure 2 (bottom two groups of panels).  The shaded regions represent the LT results; the white-outlined, colored contours are overplotted to show our results. Both are on the same color scale.  Following LT, we have plotted both the case without foregrounds ($\C = \N$, left two panels) and the case with foregrounds ($\C = \N + \U + \G$, right two panels), which allows us to get a sense for the effects of foregrounds on our power spectrum estimates, error bars, and window functions. }  
	\label{ComparisonToAdrian}
\end{figure*}

First we want to verify that our method reproduces that of LT in the regions of Fourier space accessible to both methods.  Figure \ref{ComparisonToAdrian} provides an explicit comparison to LT's Figures 2 and 8.  These plots show the shaded regions representing their method and over-plotted, white-outlined contours representing ours.  Both are on the same color scale.  These plots show error bars in temperature units in $k_\perp$-$k_\|$ space and a selection of window functions in both the case where $\C=\N$ and $\C = \U + \G + \N$.  They are generated from the same survey geometry with identical foreground and noise parameters.  In the regions where the methods overlap, we see very good agreement between the two methods.

In addition to the modes shown in the shaded regions in Figure \ref{ComparisonToAdrian}, the LT method can access Fourier modes longer than the box size, which we cannot.  This is no great loss---these modes are poorly constrained by a single data cube.  Moreover, they are generally those most contaminated by foregrounds; the low-$k_\perp$ modes will see heavy galactic synchrotron contamination while the low-$k_\|$ modes will be contaminated by types of the foregrounds.  We imagine that very low-$k_\perp$ Fourier modes, those that depend on correlations between data cubes that cannot be joined without violating the flat-sky approximation, will still be analyzed by the LT method.  Because our method can handle many more voxels, it excels in measuring both medium and high-$k$ modes that require high spectral and spatial resolution.

\subsection{Novel Effects on the Fisher Matrix} \label{CovMatBuildup}

\begin{figure*}	\centering 
	\includegraphics[width=.95\textwidth]{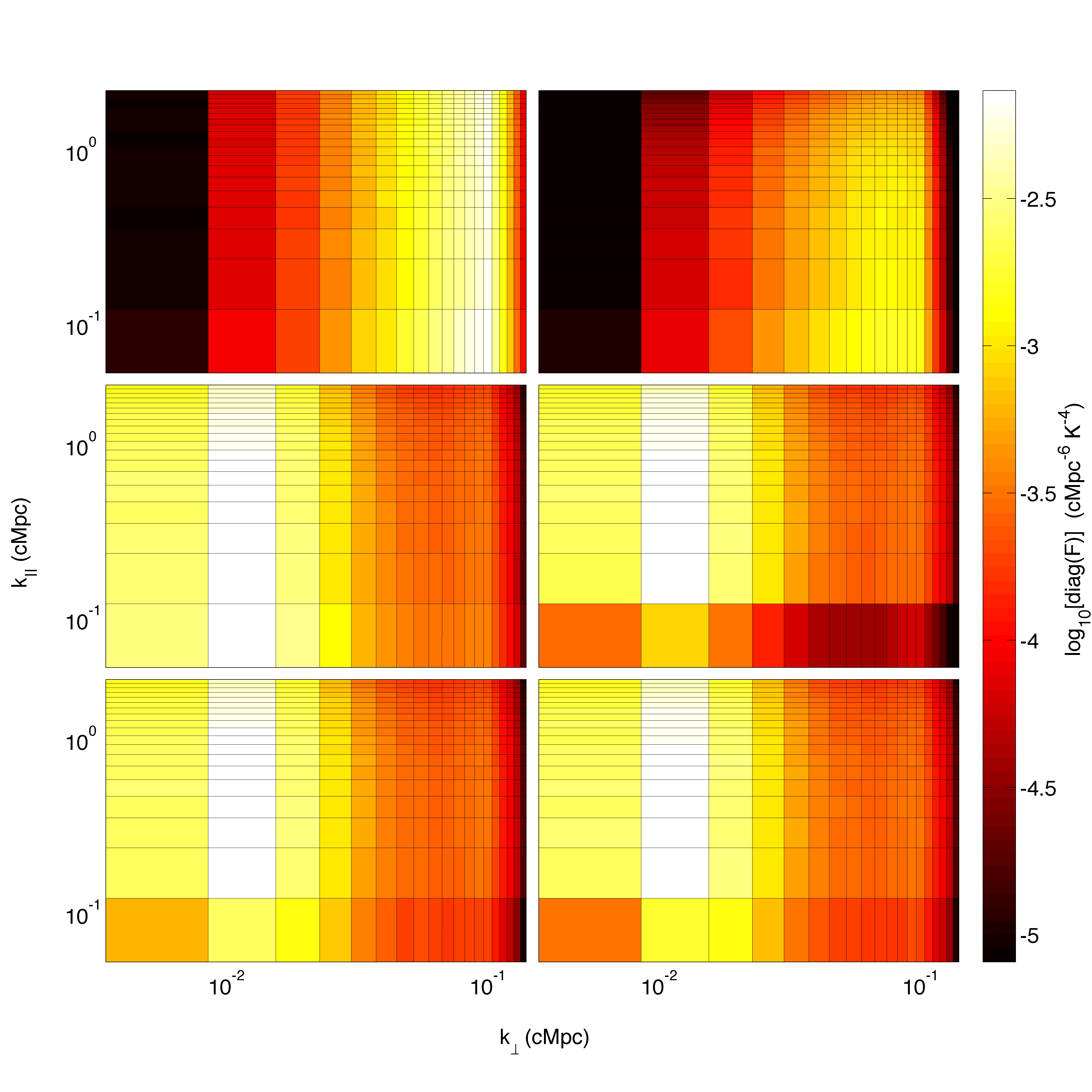}
	\caption[Fisher diagonal for progressively complex covariance models.]{The Fisher information matrix provides a useful window into understanding the challenges presented to measuring the 21\,cm power spectrum by the various contaminants.  In this figure, we add these effects one by one, reading from left to right and top to bottom, to see how the diagonal of the Fisher matrix (expanded along the $k_\perp$ and $k_\|$ directions), is affected.  Brighter regions represent, roughly speaking, more information and thus smaller error bars (in power spectrum units).  We comment in more detail on each panel individually in Section \ref{CovMatBuildup}, including upon the advantages of the novel aspects of our technique. \emph{Top left:} the covariance matrix is taken to be the identity and pixelization effects on $\Q^\alpha$ are ignored.  \emph{Top right:} the pixelization factor $|\Phi(\kv)|^2$ is included and not set to $1$.  \emph{Middle left:} the noise expected from 1000 hours of observation with the MWA 128-tile configuration is included. \emph{Middle right:} all point sources (up to 200 Jy) are modeled as unresolved; all information about their positions is ignored.  \emph{Bottom left:} resolved point sources are included in the model, with all point sources dimmer than 100 mJy considered unresolved. \emph{Bottom right:} in addition to bright point sources, galactic synchrotron is also included. }\label{fisher_buildup} 

\end{figure*}

A simple way to understand the different effects that our forms of $\C$ and $\Q^\alpha$ have on the Fisher information matrix, especially the novel inclusions of $\R$ and $|\Phi(\kv)|^2$, is to build up the Fisher matrix component by component.  In Figure \ref{fisher_buildup} we do precisely that by plotting the diagonal elements of $\F$.  These diagonal elements are related to the vertical error bars on our band powers.  Large values of $F^{\alpha\alpha}$ correspond to band powers about which we have more information.

In the top two panels of Figure \ref{fisher_buildup}, we show the first novel effect that our method takes into account.  In them, we can see how modeling the finite size of our voxels affects the information available in the case where $\C = \I$ (the color scale for these two panels only is arbitrary).  In the top left panel, we have set $|\Phi(\kv)|^2 = 1$, which corresponds to the delta function pixelization of LT.  We see that the amount of information depends only on $k_\perp$.  This is purely a binning effect: our bins in $k_\perp$ are concentric circles with constant increments in radius; higher values of $k_\perp$ incorporate more volume in Fourier space, except at the high-$k_\perp$ edge where the circles are large enough to only include the corners of the data cube. In the top right panel, we see that including $|\Phi(\kv)|^2\neq 1$ affects our ability to measure high-$k$ modes, which depends increasingly on our real space resolution and is limited by the finite size of our voxels.  

In the middle left panel, we now set $\C = \N$. In comparison to $\C = \I$, the new covariance matrix (and thus new vector $\x$ for the Monte Carlo calculation of $\F$), shifts the region of highest information to a much lower value of $k_\perp$.  Though there are fewer Fourier modes that sample this region, there are far more baselines in the array configuration at the corresponding baseline length. Our noise covariance is calculated according to our derivation in Section \ref{Nfast} for 1000 hours of observation with the 128-tile deployment of the MWA \citep{TingaySummary}.

We next expand to $\C = \U + \N$ for the middle right panel, where we have classified all point sources as ``unresolved.''  In other words, we take $S_\text{cut}$ in Equation \ref{I2fluxintegral} to be large (we choose 200 Jy, which is representative of some of the brightest sources at our frequencies).  As we expect, smooth spectrum contamination reduces our ability to measure power spectrum modes with low values of $k_\|$.  This is because of the exponentially decaying eigenvalue spectrum of $\U_\|$, most of which is smaller than the eigenvalues of $\N$. The effect is seen across $k_\perp$ because the characteristic clustering scale of unresolved point sources is smaller than the pixel size; localized structure in real space corresponds to unlocalized power in Fourier space.

\begin{figure*}	\centering 
	\includegraphics[width=.95\textwidth]{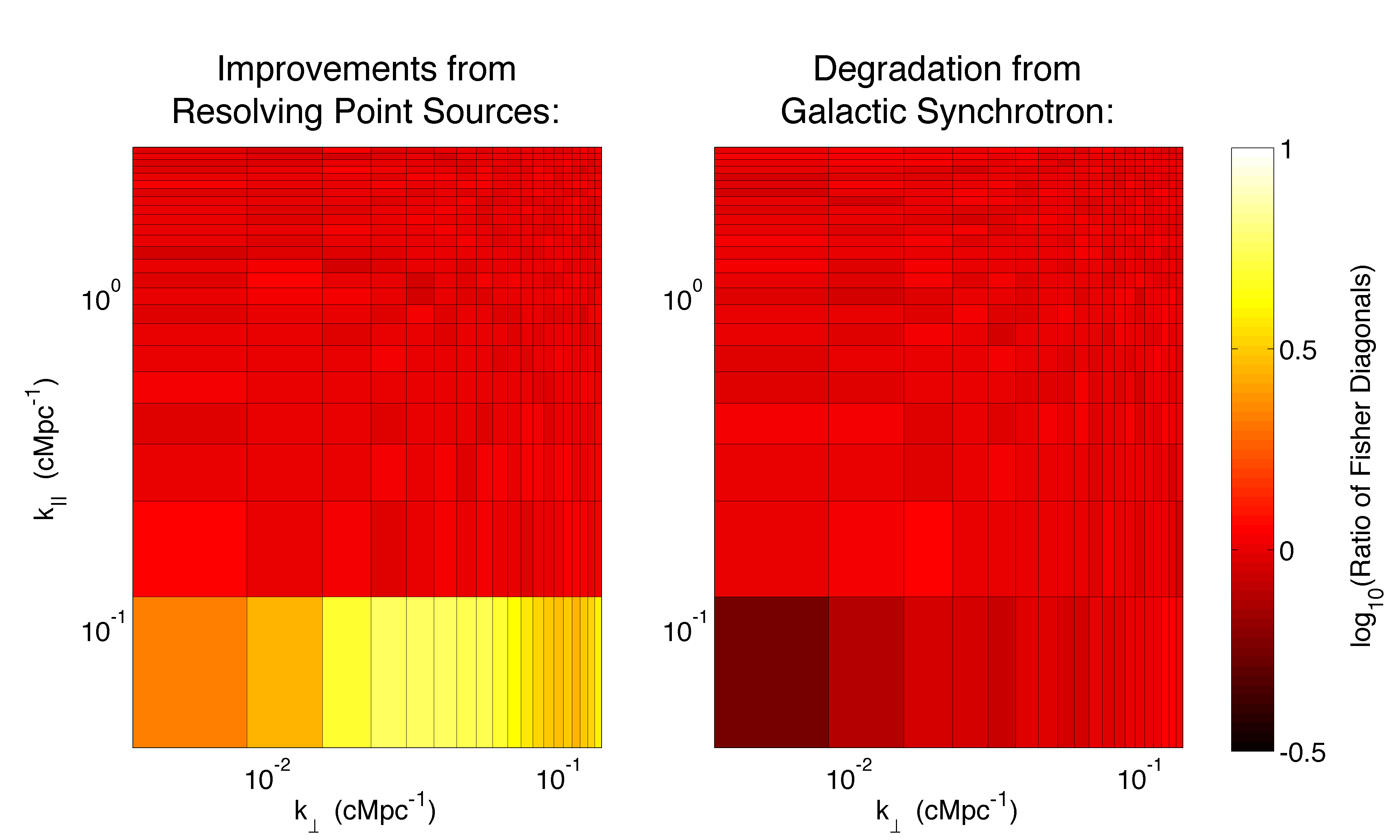} 
	\caption[Fisher matrix comparison of different covariance models.]{By comparing the Fisher matrices arising from the various covariance models we explore in Section \ref{CovMatBuildup} and Figure \ref{fisher_buildup}, the precise improvements are brought into sharper focus.  In the left panel, we can see the information we gain by explicity including resolved point sources.  Shown here is the ratio of the bottom left panel of Figure \ref{fisher_buildup} to the middle right panel.  By taking into account precise position, flux, and spectral information and uncertainties, we improve our ability to measure the power spectrum at the longest scales parallel to the line of sight, effectively ameliorating the effects of foregrounds.  In the right panel, we see the remarkably small effect that the galactic synchrotron radiation has on our abilty  the measure the 21\,cm power spectrum.  Shown here is the ratio of the bottom right panel of Figure \ref{fisher_buildup} to the botton left.  Because we take spatial information into account, the strong spatial and spectral coherence of the signal from our Galaxy is confined to the bottom left corner of the $k_\|$-$k_\perp$ plane.} \label{fisherRatios} 
\end{figure*}

In the bottom left panel, we have included information about the positions of roughly 200 resolved point sources above 100 mJy, with random fluxes drawn from our source count distribution (Equation \ref{sourceCounts}) and random spectral indices drawn from a Gaussian centered on $\kappa_n = 0.5$ with a width of $0.15$.  By doing this, we reduce $S_\text{cut}$ in our model for $\U$ down to 100 mJy.  Including all this extra information---positions, fluxes, flux uncertainties, spectral indices, and spectral index uncertainties---provides us with significantly more Fisher information at low-$k_\|$ where foregrounds dominate and thus smaller errors on those modes.  Additionally, by incorporating resolved point sources as part of our inverse covariance weighting, we no longer have to worry about forward propagating errors from any external point-source subtraction scheme. In the left panel of Figure \ref{fisherRatios} we see the ratio of this panel to the middle right panel.  

Finally, in the bottom right panel of Figure \ref{fisher_buildup} we show the effect of including Galactic synchrotron radiation.  Adding $\G$ has the expected effect; we already know that $\G$ has only a few important eigenmodes which correspond roughly to the lowest Fourier modes both parallel and perpendicular to the line of sight.  As a result, we only see a noticeable effect in the bottom left corner of the $k_\perp$-$k_\|$ plane; we include the ratio of the two figures in the right panel of Figure \ref{fisherRatios} for clarity.  Otherwise, our Galaxy has very little effect on the regions of interest.  In fact, the similarity between the this panel and the middle left panel tells us something very striking: in the regions of Fourier space that our data most readily probes, foregrounds (once properly downweighted) should not prove an insurmountable obstacle to power spectrum estimation. 

The set of plots in Figure \ref{fisher_buildup} is useful for developing a heuristic understanding of how noise and foregrounds affect the regions in which we can most accurately estimate the 21\,cm power spectrum.  With it, we can more easily identify the precise regions of $k$-space that we expect to be minimally contaminated by foregrounds and can thus tailor our instruments and our observations to the task of measuring the 21\,cm power spectrum.  

\subsection{Computational Scaling of the Method} \label{CompScaling}
\begin{figure}  
	\centering 
	\includegraphics[width=.5\textwidth]{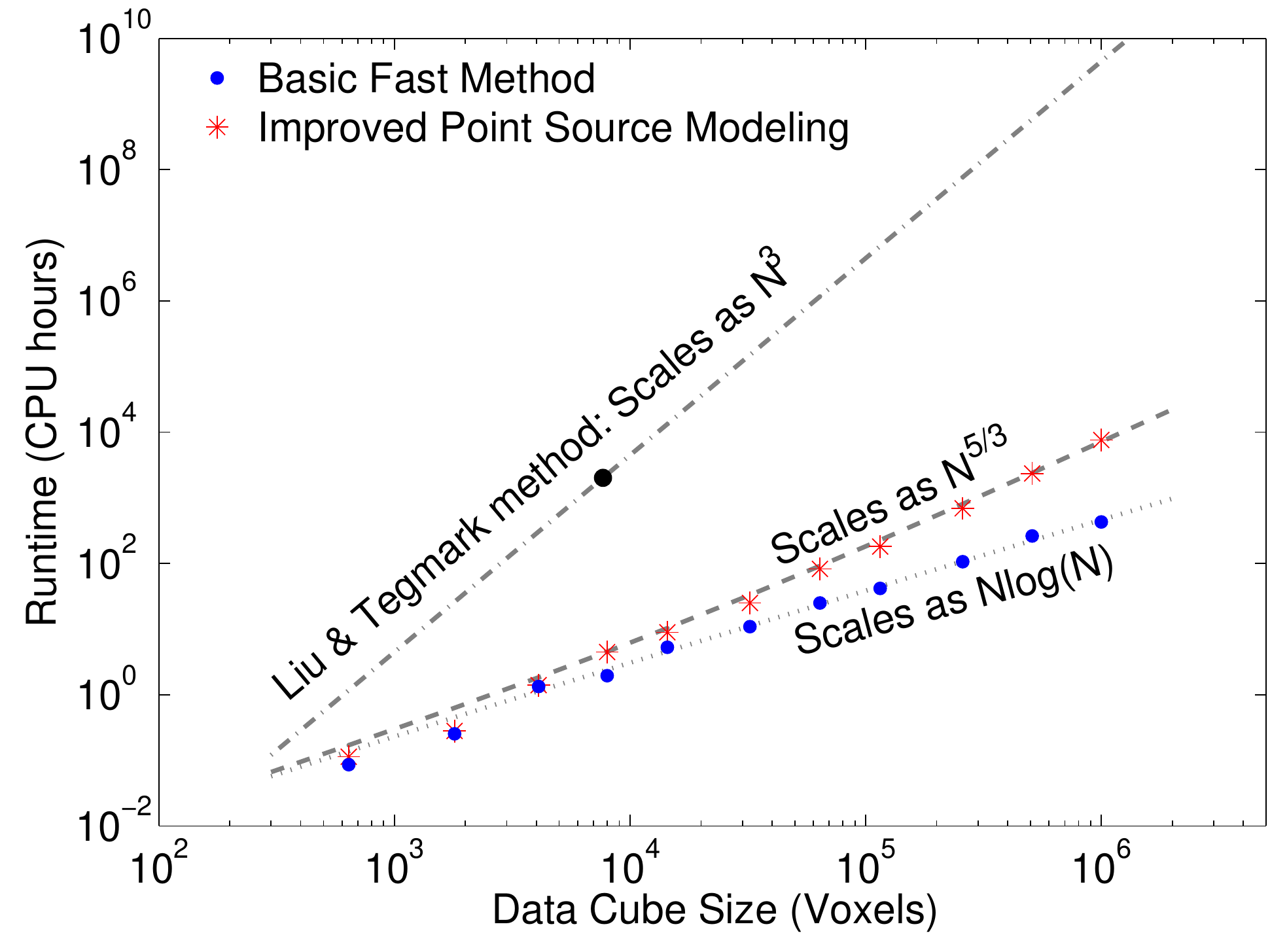}
	\caption[Total runtime results.]{Our algorithm scales with the number of voxels, $N$, as $\BigO(N \log N)$ in the best case and as $\BigO(N^{5/3})$ in the worst, depending on the treatment of bright point sources.  If we choose to ignore all information about the location, brightness, and spectra of bright point sources, we can estimate the power spectrum in $\BigO(N \log N)$.  If we choose to take into account this extra information, the algorithmic complexity increases to $\BigO(N N_R)$, where $N_R$ is the number of bright, resolved point sources.  For a fixed minimum flux for ``bright'' sources, this leads to $\BigO(N^{5/3})$ complexity for uniform scaling in all three dimensions. Both scenarios represent a major improvement over the LT method, which scales as $\BigO(N^3)$.}  
	\label{ComputationalTimeResults} 
\end{figure}

Now that we understand how our technique works, we want to also see that it works as quickly as promised by and achieves the desired computational speed up over the LT method.  Specifically, we want to show that we can achieve the theoretical $\BigO(N \log N)$ performance in reproducing the results of LT.  We also want to better understand the computational cost of including resolved point sources so as to compare that cost to benefits outlined in Section \ref{CovMatBuildup}. We have therefore tested the algorithm's speed for a wide range of data cube sizes; we present the results of that study in Figure \ref{ComputationalTimeResults}.

In this figure, we show the combined setup and runtime for power spectrum estimates including 1000 Monte Carlo simulations of $\widehat{\mathbf{q}}$ for estimating the Fisher matrix on a single modern CPU.  For each successive trial, we scale the box by the same ratio in all three dimensions.  Because we maintain a fixed flux cut, increasing the linear size of the box by a factor of two increases the number of resolved point sources in the box by a factor of 4 and the number of voxels by a factor of 8.  With any more than a few point sources, the computational cost becomes dominated by point sources, leading to an overall complexity of $\BigO(N N_R)$.  In this case, the largest data cubes include about 400 point sources over a field of about 50 square degrees, accounting for about 15\% of the lines of sight in the data cube.  

For any given analysis project, these exists a trade-off between including additional astrophysical information into the analysis and the computational complexity of that analysis; at some point the marginal cost of a slower algorithm exceeds marginal benefit of including more bright point sources.  It is beyond the scope of this paper to prescribe a precise rubric for where to draw the line between resolved and unresolved point sources.  However, we can confidently say that the algorithm runs no slower than $\BigO(N^{5/3})$ and can often run at or near $\BigO(N \log N)$ if only the brightest few point sources are treated individually. 

\subsection{Implications for Future Surveys}\label{MWA}
Though the primary purpose of this paper is to describe an efficient method for 21\,cm power spectrum analysis, our technique enables us immediately to make predictions about the potential performance of upcoming instruments.  In this section we put all our new machinery to work in order to see just how well the upcoming 128-tile deployment of the MWA can perform.

\begin{figure} 
	\centering 
	\includegraphics[width=.48\textwidth]{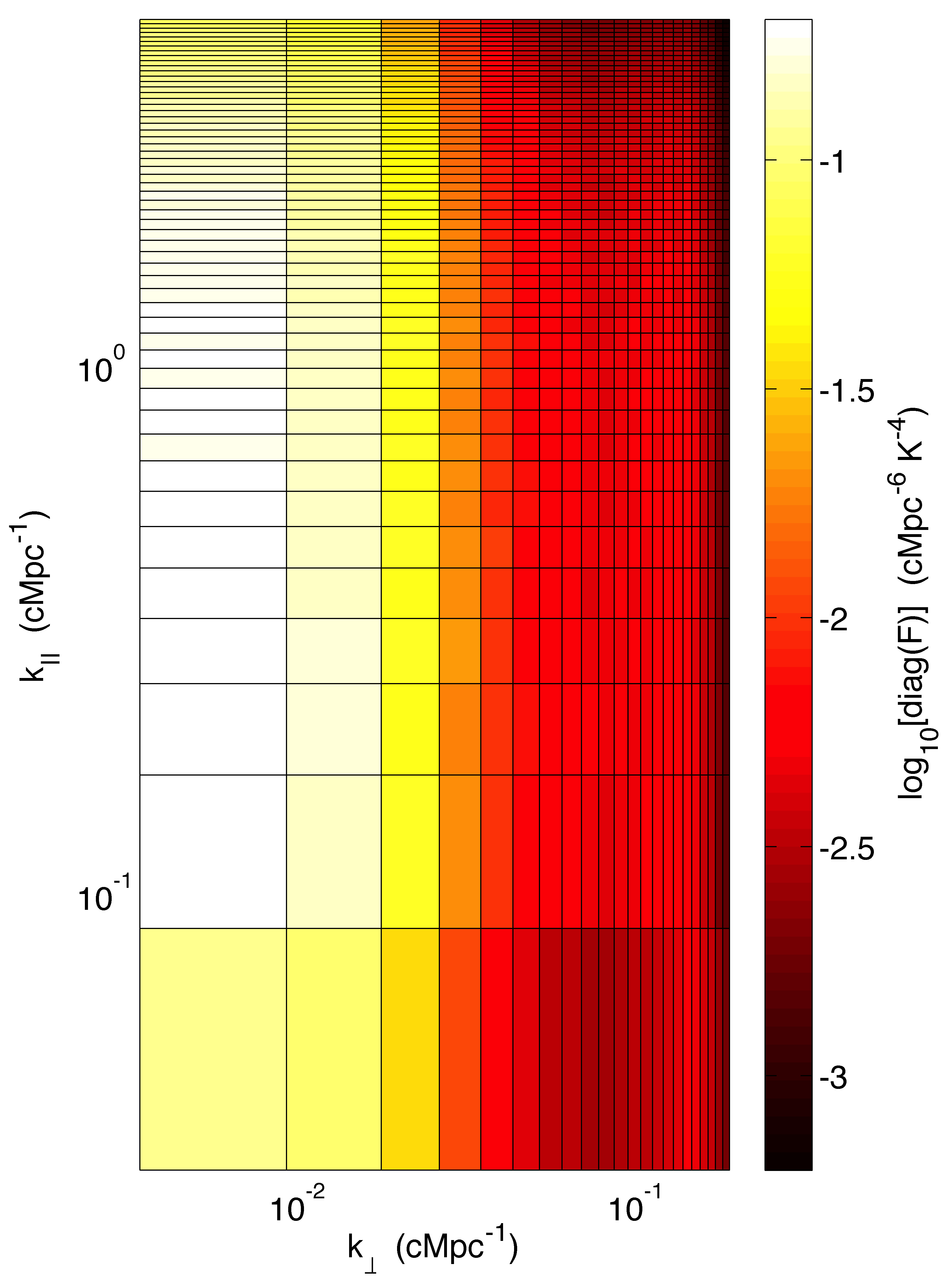}
	\caption[Fisher matrix results for 1000 hours of MWA observing.]{The diagonal of the Fisher matrix predicted for 1000 hours of observation with the MWA with 128 tiles shows the region of power spectrum space least contaminated by noise and foregrounds.  Noise, and thus array layout, dominates the shape of the region of maximum information, creating a large, vertical region at a value of $k_\perp$ corresponding to the typical separation between antennas in the compact core of the array.  The contaminating effects of the foregrounds are clearly visible at low-$k_\|$.}
	\label{MWAFisher}
\end{figure}

\begin{figure} 
	\centering 
	\includegraphics[width=.48\textwidth]{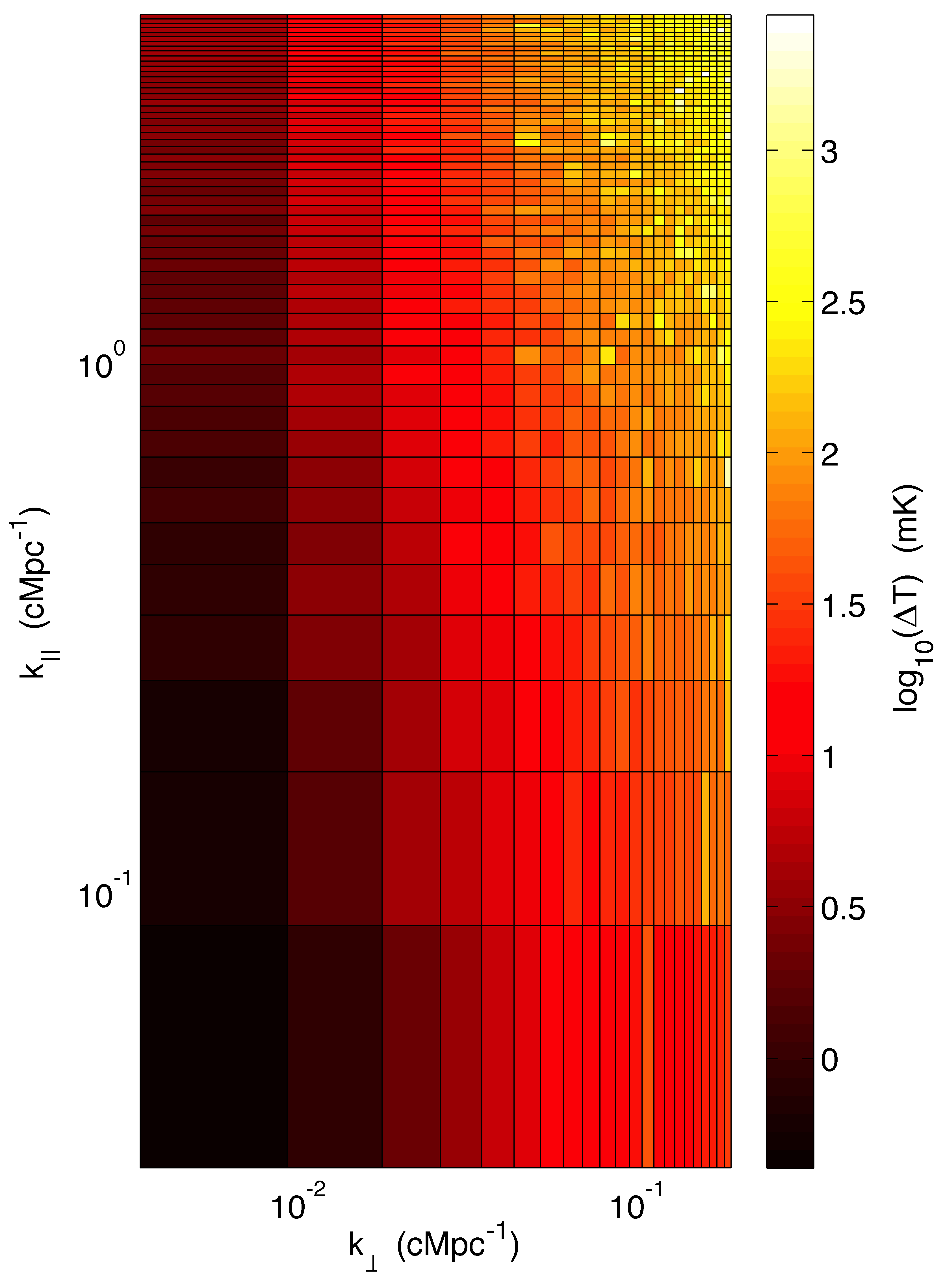}
	\caption[Expected errors for 1000 hours of MWA observing.]{The expected error bars in temperature units on decorrelated estimates of the power spectrum highlight a sizable region of $k$-space where we expect to be able to use the MWA with 128 tiles to detect a fiducial 10 mK signal with a signal to noise ratio greater than 1.  Perhaps surprisingly, the smallest error bars are still on the smallest $k$ modes acessible by our method, though some of them are contaminated by large foregrounds.  This is because our conversion to temperature units includes a factor of $(k_\perp^2 k_\|)^{1/2}$, which accounts for the difference between this Figure and Figure \ref{MWAFisher}.  From the shape of the region of smallest error, we can better appreciate the extent to which noise and our array layout determines where in $k$-space we might expect to be able to detect the EoR.  The noisiness at high-$k$ is due to Monte Carlo noise and can be improved with more CPU hours.} 
	\label{MWAerror}
\end{figure}

\begin{figure*} 
	\centering 
	\includegraphics[width=.7\textwidth]{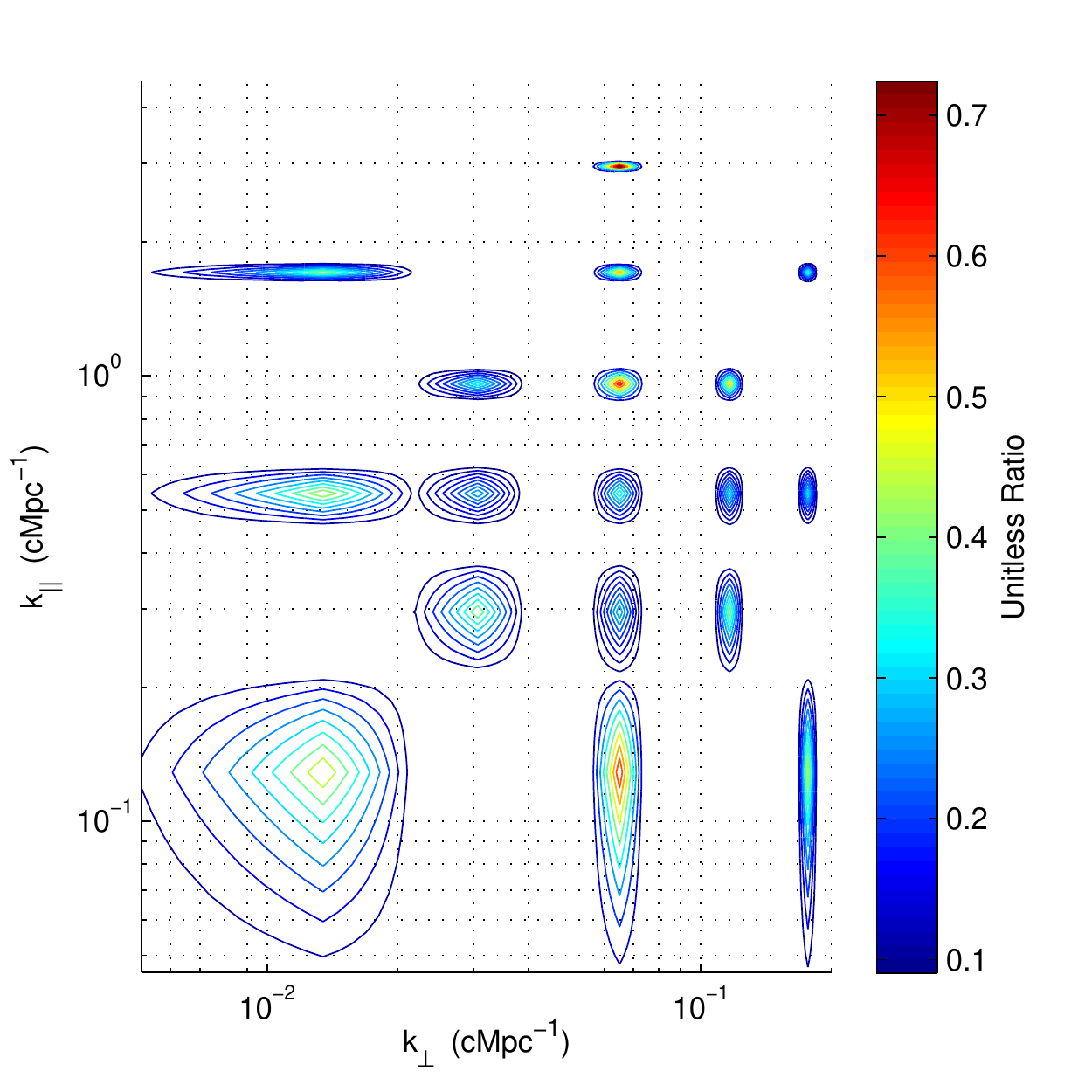}
	\caption[MWA expected window functions.]{We can see from a sampling of window functions that our band power spectrum estimates $\widehat{p}^\alpha$ represent the weighted averages of $p^\alpha$ over a narrow range of scales, especially at higher values of $k_\perp$ and $k_\|$.  The widest window functions can be attributed to binning (with linearly binned data, low-$k$ bins look larger on logarithmic axes) and to foregrounds.  This is good news, because it will enable us to accurately make many independent measurements of the power spectrum and therefore better constrain astrophysical and cosmological parameters.}
	\label{MWAwindow}
\end{figure*}

We envision 1000 hours of integration on a field that is $9^\circ$ on each side, centered on $z=8$ with $\Delta z = 0.5$.  With a frequency resolution of 40 kHz and an angular resolution of 8 arcminutes, our data cube contains over $10^6$ voxels.  We completed over 1000 Monte Carlo iterations on our 12 core server in about one week.  We use the foreground parameters outlined above in Sections \ref{AdrianModels} and \ref{FastCov}.  In Figures \ref{MWAFisher} through \ref{MWAwindow}, we show the diagonal elements of the Fisher matrix we have calculated, the temperature power spectrum error bars, and a sampling of window functions. 

In Figure \ref{MWAFisher} we plot the diagonal elements of the Fisher matrix, which are related directly to the power spectrum errors.  Drawing from the discussion in Section \ref{CovMatBuildup}, we can see clearly the effects of the array layout (and thus noise), of foregrounds (included resolved point sources), and of pixelization.  Interestingly, until pixelization effects set in at the highest values of $k_\|$, the least contaminated region spans a large range of values of $k_\|$.  One way of probing more cosmological modes is to increase the frequency resolution of the instrument.  The number of modes accessible to the observation scales with $(\Delta \nu)^{-1}$, though the amplitude of the noise scales scales with $(\Delta \nu)^{-1/2}$. As long as the noise level is manageable and the cosmological signal is not dropping off too quickly with $k$, increasing the frequency resolution seems like a good deal.

In Figure \ref{MWAerror} we show the vertical error bars that we expect on power spectrum estimates in temperature units.  The most important fact about this plot is that there is a large region where we expect that vertical error bars will be sufficiently small that we should be able to detect a 10 mK signal with signal to noise greater than 1.  This is especially the case at fairly small values of $k$, which is surprising since these $k$ modes were supposed to be the most contaminated by foregrounds.  There are two reasons why this happens.  

First, the conversion to temperature units (Equation \ref{TemperatureUnits}) introduces a factor of $(k_\perp^2 k_\|)^{1/2}$ that raises the error bars for larger values of $k$.  Second, the strongest foreground modes overlap significantly with one of the $k=0$ modes of the discrete Fourier transform, which we exclude for our power spectrum estimate (this is just the average value of the cube in all three directions, which is irrelevant to an interferometer that is insensitive to the average).  

Another way to think about it is this: because the coherence length of the foregrounds along the line of sight is much longer than the size of any box small enough to comfortably ignore cosmological evolution, we expect that the most contaminated Fourier mode will be precisely the one we ignore.  Unlike the LT method, our method cannot easily measure modes much longer than the size of the data cube.  Along the line of sight, these modes have very wide window functions and are the most contaminated by foregrounds.  Perpendicular to the line of sight, these modes are better measured by considering much larger maps where the flat sky approximation no longer holds.  For the purposes of measuring these low-$k$ modes, the LT method can provide a useful complement to ours.  Large-scale modes from down-sampled maps can be measured by LT; smaller-scale modes from full-resolution maps can be measured by our method.  Then both can be combined to estimate the power spectrum across a large range of scales. 

And finally, in Figure \ref{MWAwindow} we show many different window functions for a selection of values of $k_\perp$ and $k_\|$ that spans the whole space.  In general, these window functions are quite narrow, meaning that each band power measurement probes only a narrow range of scales.  The widest windows we see look wide for two reasons.  First, linearly separated bins appear wider at low-$k$ when plotted logarithmically.  Second, foregrounds cause contaminated bins to leak into nearby bins, especially at low-$k_\|$ and moderate $k_\perp$.  We saw hints of this effect in Figure \ref{ComparisonToAdrian} when comparing noise-only simulations to simulations with both noise and foregrounds. 

In the vast majority of the $k_\perp$-$k_\|$ plane, the window functions seem to be dominated by the central bin and neighbors.  Except for edge cases, no window function has contributions exceeding 10\% from bins outside the central bin and its nearest neighbors.  This means that we should be happy with our choice of Fourier space binning, which was designed to have bin widths equal to those of our data cube before zero padding.  We also know that significantly finer binning would be inappropriate, so we do not have to worry about the tradeoff between fine binning of the power spectrum and the inversion of the Fisher matrix.  Therefore, with the 128-tile deployment of the MWA, we can be confident that our estimates of the power spectrum correspond to distinct modes of the true underlying power spectrum.


\section{Conclusions}

With this paper, we have presented an optimal algorithm for 21\,cm power spectrum estimation which is dramatically faster than the Liu \& Tegmark (LT) method \cite{LT11}, scaling as $\BigO(N \log N)$ instead of $\BigO(N^3)$, where $N$ is the number of voxels in the 3D sky map.  By using the inverse variance weighted quadratic estimator formalism adapted to 21\,cm tomography by the LT method, we preserve all accessible cosmological information in our measurement to produce the smallest possible error bars and narrow window functions.  Moreover, our method can incorporate additional information about the brightest point sources and thus further reduce our error bars at the cost of some---but by no means all---of that computational advantage. Our method is highly parallelizable and has only modest memory requirements; it never needs to store an entire $N\times N$ matrix.  

Our method achieves this computational speed-up for measuring power spectra, error bars, and window functions by eliminating the time-consuming matrix operations of the LT method.  We accomplish this using a combination of Fourier, spectral, and Monte Carlo techniques which exploit symmetries and other physical properties of our models for the noise and foregrounds.

We have demonstrated the successful simulation of error bars and window functions for the sort of massive data set we expect from the upcoming 128-tile deployment of the MWA---a data set that cannot be fully utilized using only the LT method.  Our forecast predicts that 1000 hours of MWA observation should be enough to detect the fiducial 10 mK signal across much of the $k_\|$-$k_\perp$ plane accessible to the instrument.  Moreover, we predict that the horizontal error bars on each band power estimate will be narrow, allowing each estimate to probe only a small range of scales.

Our results suggest several avenues for further research.  Of course, the most immediate application is to begin analyzing the data already being produced by interferometers like LOFAR, GMRT, MWA, and PAPER as they start accumulating the sensitivity necessary to zero in on a detection of the EoR.  The large volume of data these instruments promise to produce might make it useful to explore ways of further speeding up the Monte Carlo estimation of the Fisher matrix.  There is significant redundancy in our calculated Fisher matrix because the window function shapes vary only relatively slowly with $k$-scale.  We believe that one can reduce the number of Monte Carlo simulations needed to attain the same accuracy by adding a postprocessing step that fits the Fisher matrix to a parametrized form. This should work best in the regions of the $k_\|$-$k_\perp$ plane that are fairly uncontaminated by foregrounds, where Fisher matrix elements are expected to vary most smoothly.  It may also be possible to speed up the Monte Carlo estimation of the Fisher matrix using the trace evaluation technology of \citep{UeLiFast}.

The forecasting power of our method to see whether a particular observing campaign might reveal a particular aspect the power spectrum need not be limited to measurements of the EoR.  Our method provides an opportunity to precisely predict what kind of measurement, and what kind of instrument, might be necessary for observing 21\,cm brightness temperature fluctuations during the cosmic dark ages.  Our method should prove useful for weighing a number of important design considerations: What is the optimal array configuration? What is the optimal survey volume?  What about angular resolution?  Spectral resolution?  And in what sense are these choices optimal for doing astrophysics and cosmology?  
 
To help answer such questions, our technique could be used to compare the myriad of ideas for and possible implementations of future projects like HERA and the SKA and even to help find an optimal proposal.  For example, one plan for achieving large collecting area is building a hierarchically regular array (a so-called ``Omniscope'') that takes advantage of FFT correlation \citep{FFTT2} and redundant baseline calibration \citep{redundant}.  There exist many array configurations that fit into this category and it is not obvious what the optimal Omniscope might look like.

The quest to detect a statistical signal from the Epoch of Reionization is as daunting as it is exciting.  It is no easy task to find that needle in a haystack of noise and foregrounds.  However, now that we are for the first time armed with a method that can extract all the cosmological information from a massive data set without a prohibitive computational cost, we can feel confident that a sufficiently sensitive experiment can make that first detection---not just in theory, but also in practice.





\begin{subappendices}

\section{Appendix: Toeplitz Matrices} \label{ToeplitzAppendix}

In this appendix, we briefly review how to rapidly multiply by Toeplitz matrices. We need to employ the advantages of Toeplitz matrices because the assumption that our covariance matrices are diagonal in real space or in Fourier space, as was the case in \citep{UeLiFast}, break down for covariance matrices with coherence lengths much larger than the box size.
 
A ``Toeplitz'' matrix is any matrix with the same number for every entry along its main diagonal and with every other diagonal similarly constant \citep{Toeplitz}.  In general, a Toeplitz matrix is uniquely defined by the entries in its first row and its first column: if $i \ge j$ then $T_{ij} = T_{1+i-j,1}$ and if $i \le j$ then $T_{ij} = T_{1,1-i+j}$.   If the first row of a matrix is repeated with a cyclic permutation by one to the right in each successive row, then it is a special kind of Toeplitz matrix called a ``circulant'' matrix.  Circulant matrices are diagonalized by the discrete Fourier transform \citep{Toeplitz}.  Given a circulant matrix $\mathbf{C}$ with first column $\mathbf{c}$, the product of $\mathbf{C}$ and some arbitrary vector $\mathbf{v}$ can be computed in $\mathcal{O}(N\log N)$ time because
\begin{equation} \label{circulant}
\mathbf{Cv} = \mathbf{F}^\dagger \mbox{ diag}(\mathbf{Fc}) \mbox{ } \mathbf{Fv},
\end{equation}
where $\mathbf{F}$ is the unitary, discrete Fourier transform matrix \citep{Toeplitz}.  Reading Equation \ref{circulant} from right to left, we see that every matrix operation necessary for this multiplication can be performed in $\mathcal{O}(N\mbox{log}N)$ time or better.  

Conveniently, any symmetric Toeplitz matrix can be embedded in a circulant matrix twice its size.  Given a symmetric Toeplitz matrix $\mathbf{T}$, we can define another symmetric Toeplitz matrix $\mathbf{S}$ with an arbitrary constant along its main diagonal.  If we specify that the rest of the first row (besides the first entry) is the reverse of the rest of the first row of $\T$ (again ignoring the first entry), the fact that the matrix is Toeplitz and symmetric completely determines the other entries.  For example,
\begin{equation}
\text{if } \mathbf{T} = \left( \begin{array}{ccc} 5 & 3 & 2 \\ 3 & 5 & 3 \\ 2 & 3 & 5 \end{array} \right), \text{ then } \mathbf{S} = \left( \begin{array}{ccc} 0 & 2 & 3 \\ 2 & 0 & 2 \\ 3 & 2 & 0 \end{array} \right).
\end{equation}
It is straightforward to verify that the matrix $\mathbf{C}$, defined as
\begin{equation}
\mathbf{C} \equiv \left( \begin{array}{cc} \mathbf{T} & \mathbf{S} \\ \mathbf{S} & \mathbf{T} \end{array} \right),
\end{equation}
is a circulant matrix. We can now can multiply $\mathbf{C}$ by a zero-padded vector so as to yield the product of the Toeplitz matrix and the original vector, $\mathbf{Tx}$, that we are looking for:
\begin{equation}
\mathbf{C} \left( \begin{array}{c} \mathbf{v} \\ \mathbf{0} \end{array} \right) = \left( \begin{array}{c} \mathbf{Tv} \\ \mathbf{Sv} \end{array} \right).
\end{equation}
Therefore, we can multiply any Toeplitz matrix by a vector in $\BigO(N \log N)$.

\section{Appendix: Noise Covariance Matrix Derivation} \label{NoiseAppendix}
In this appendix, we derive the the form of $\N$, the noise covariance matrix, in Equation \ref{Nfourier} by combining the form of $P^N(\kv,\lambda)$, the noise power spectrum, in Equation \ref{NPowerSpectrum} with Equation \ref{Ndef}, which relates $\N$ to $P^N(\kv,\lambda)$.  To accomplish this, we simplify $P^N(\kv,\lambda)$ into a form that is more directly connected to our data cube.   We then approximate the integrals in Equation \ref{Ndef} by assuming that the $uv$-coverage is piecewise constant in cells corresponding to our Fourier space grid.\footnote{We also assume that measurements in nearby $uv$-cells are uncorrelated, which may not be true if the baselines are not coplanar; instead $\N$ would have to be modeled as sparse rather than diagonal in angular Fourier space.}

To simplify $P^N(\kv,\lambda)$, we first note that  because the term $B(\kv,\lambda)$ in Equation \ref{NPowerSpectrum} represents the synthesized beam and is normalized to peak at unity, we can reinterpret the factor of $(f^\text{cover})^{-1} B^{-2}(\kv,\lambda)$ as an inverted and normalized $uv$-coverage.  When $f^\text{cover} = 1$, the array has uniform coverage.  We want to replace the factor $(f^\text{cover})^{-1} B^{-2}(\kv,\lambda)$ with a quantity directly tied our choice of pixelization of the $uv$-plane and written in terms of the simplest observational specification: the total time that baselines spend observing a particular $uv$-cell, $t(\kv,\lambda)$.  We already know that the noise power is inversely proportional to that time because more time yields more independent samples.

To relate $t^{-1}(\kv,\lambda)$ to $(f^\text{cover})^{-1} B^{-2}(\kv,\lambda)$, we want to make sure that the formula yields the same answer for peak density in the case of a complete coverage.  In other words, we want to find the constant $t_\text{max}$ such that
\beq
\frac{t_\text{max}}{t(\kv,\lambda)} = (f^\text{cover})^{-1} B^{-2}(\kv,\lambda). \label{timeConversion}
\eeq
The time spent in the most observed cell is related to the size of the cell in the $uv$-plane, the density of baselines in that cell, and the total integration time of the observation, $\tau$.  The cell size is determined by the pixelization of our data cube.  We have divided each slice of our data cube of size $L_x \times L_y$ into $n_x \times n_y$ pixels.  In Fourier space, this corresponds a pixel size of
\beq
\Delta k_x = \frac{2\pi}{L_x} = \frac{2\pi}{d_M \Delta\theta_x n_x},
\eeq
where $\Delta\theta_x$ and is the angular pixelization in the $x$ direction.  An equivalent relation is true for the $y$ direction.  Since $\Delta u = \Delta k_x d_M / (2\pi)$, we have that the area in $uv$-space of each of our grid points is
\beq
\Delta v \Delta u = \frac{1}{\Delta\theta_x n_x}\frac{1}{\Delta\theta_y n_y} = \frac{1}{\Omega_\text{pix} n_x n_y}.
\eeq

The maximum density of baselines is the density of the autocorrelations,\footnote{If the use of autocorrelations (which most observations throw out, due to their unfavorable noise properties) is troubling, then it is helpful to recall that for a large and fully-filled array, the $uv$-density of the shortest baselines is approximately the same as the $uv$-density of the autocorrelations.} which is
\beq
n_\text{max} = N_\text{ant} \left( \frac{A_\text{ant}}{\lambda^2}\right)^{-1},
\eeq
where the quantity $(A_\text{ant}/\lambda^2)$ is the area in the $uv$-plane associated with a single baseline \citep{Matt3}.  We thus have that
\beq
t_\text{max} \equiv  n_\text{max} \Delta u \Delta v  \tau  = \frac{N_\text{ant} \lambda^2 \tau}{A_\text{ant} \Omega_\text{pix} n_x n_y}.
\eeq
Now we can substitute Equation \ref{timeConversion} into Equation \ref{NPowerSpectrum} to get a more useful form of $P^N(\kv,\lambda)$:
\begin{align}
P^N(\kv,\lambda) &= \frac{\lambda^4 T_\text{sys}^2 y d^2_M}{A^2_\text{ant} \Omega_\text{pix} n_x n_y}\frac{1}{t(\kv_\perp,\lambda)}.
\end{align}

In general, $t(\kv_\perp,\lambda)$ depends in a nontrivial way on the array layout.  As such, the integral expression for $\N$ in Equation \ref{Ndef} with this form of $P^N(\kv,\lambda)$ is only analytically tractable along the line of sight.  Integrating $k_z$, we get that
\begin{align}
N_{ij} =& \frac{\delta_{z_i z_j}}{\Delta z} \frac{\lambda^4 T_\text{sys}^2 y d^2_M}{A^2_\text{ant} \Omega_\text{pix} n_x n_y} \int  j_0^2(k_x \Delta x /2) j_0^2(k_y \Delta y /2) \times  \nonumber \\ &e^{i k_x (x_i-x_j) + i k_y(y_i-y_j)} \frac{1}{t(\kv_\perp,\lambda_i)} \frac{dk_x dk_y}{(2\pi)^2}. 
\end{align}
We note that $\N$ is uncorrelated between frequency channels, as we would expect.

Along the other two dimensions, we will approach the problem by approximating the integrand as piecewise constant in Fourier cells, turning the integral into a sum and the $dk$ into a $\Delta k$.  We will use the index $l$ to run over all Fourier modes perpendicular to the line of sight.  Using the fact that the line of sight voxel length $\Delta z = y \Delta \nu$ and that $L_x L_y =  \Omega_\text{pix} d_M^2 n_x n_y$, we have that
\begin{align}
N_{ij} = & \frac{\lambda^4 T_\text{sys}^2 \delta_{z_i z_j}}{A^2_\text{ant} (\Omega_\text{pix} n_x n_y)^2\Delta\nu} \sum_{l}^{\text{all x \& y}} \bigg[  j_0^2(k_{x,l} \Delta x /2)  \times  \nonumber \\ &j_0^2(k_{y,l}  \Delta y /2) e^{i k_{x,l}(x_i-x_j)} e^{i k_{y,l}(y_i-y_j)} \frac{1}{t_l(\lambda_i)}  \bigg].
\end{align}
Next, we can turn this form into one that is more clearly computationally easy to multiply by a vector by introducing another Kronecker delta:
\begin{align}
N_{ij} = & \frac{\lambda^4 T_\text{sys}^2 \delta_{z_i z_j}}{A^2_\text{ant} (\Omega_\text{pix} n_x n_y)^2\Delta\nu} \sum_{l}^{\text{all x \& y}} e^{i k_{x,l}x_i} e^{i k_{y,l}y_i} \nonumber \\ & \sum_{m}^{\text{all x \& y}} \bigg[  j_0^2(k_{x,l} \Delta x /2) j_0^2(k_{y,l} \Delta y /2) e^{-i k_{x,m}x_j} e^{-i k_{y,m}y_j} \frac{\delta_{lm}}{t_l(\lambda_i)}  \bigg].
\end{align}
Finally, if we extend $l$ and $m$ to index over all frequency channels and all Fourier modes perpendicular to the line of sight, we can write down the noise covariance matrix as $\N = \F_\perp^\dagger \widetilde{\N} \F_\perp$ where $\F_\perp$ and $\F_\perp^\dagger$ are the discrete, unitary 2D Fourier and inverse Fourier transforms and where $\widetilde{\N}$ can be written as
\beq
\widetilde{N}_{lm} =  \frac{\lambda^4 T_\text{sys}^2 j_0^2(k_{x,l} \Delta x /2) j_0^2(k_{y,l} \Delta y /2) }{A^2_\text{ant} (\Omega_\text{pix})^2 n_x n_y\Delta\nu}   \frac{\delta_{lm}}{t_l}.
\eeq
The result, therefore, is a matrix that can be multiplied by a vector in $\BigO(N\log N)$.

\section{Appendix: Construction of the Preconditioner} \label{PreconAppendix}

In this final appendix, we show how to construct the preconditioner that we use to speed up the conjugate gradient method for multiplying $\C^{-1}$ by a vector. We devise our preconditioner by looking at $\C$ piece by piece, building up pairs of matrices that make our covariances look more like the identity matrix.  We start with $\C = \N$, generalize to $\C = \U + \N$, and then finally incorporate $\R$ and $\G$ to build the full preconditioner.

\subsection{Constructing a Preconditioner for $\N$} \label{PreconNSection}
Our first task is to find a pair of preconditioning matrices that turn $\N$ into the identity:
\beq
\Pre_\N \N \Pre_\N^\dagger = \I.
\eeq
Because $\N = \F_\perp^\dagger \widetilde{\N} \F_\perp$, and because $\widetilde{\N}$ is a diagonal matrix, we define $\Pre_\N$ and $\Pre_\N^\dagger$ as follows:
\begin{align}
\Pre_\N &= \widetilde{\N}^{-1/2} \F_\perp, \nonumber \\
\Pre_\N^\dagger &= \F_\perp^\dagger \widetilde{\N}^{-1/2}. \label{PNdef}
\end{align}
Since applying $\Pre_\N$ only requires multiplying by the inverse square root of a diagonal matrix and Fourier transforming in two dimensions, the complexity of applying $\Pre_\N$ to a vector is less than $\mathcal{O}(N\log N)$.

\subsection{Constructing a Preconditioner for $\U$}\label{PreconForU}
The matrix $\U$ (Equation \ref{formOfU}) can be written as the tensor product of three Toeplitz matrices, one for each dimension, bookended by two diagonal matrices, $\D_\U$.  Furthermore, since $\D_\U$ depends only on frequency (as we saw in Section \ref{Ufast}),  its effect can be folded into $\U_z$ such that
\beq \label{Uouterproduct}
\D_\U [\U_x \otimes \U_y \otimes \U_z] \D_\U \equiv \U_x \otimes \U_y \otimes \U_z'.
\eeq 
It is generally the case that $\U_x$ and $\U_y$ are both well approximated by the identity matrix.  This reflects the fact that the spatial clustering of unresolved point sources is comparable with the angular resolution of the instrument.  This assumption turns out to be quite good for fairly compact arrays, since for an array with 1 km as its longest baseline---the sort of compact array thought to be optimal for 21 cm cosmology---we expect an angular resolution on the order of 10 arcminutes, which is comparable to the fiducial value of 7 arcminutes that LT took to describe the clustering length scale for unresolved point sources.  That value appears to be fairly reasonable given the results of \citep{BernardiForegrounds,GhoshForegrounds}.   For the purposes of devising a preconditioner only, we can therefore adopt the simplification that
\beq
\U \approx \I_x \otimes \I_y \otimes \U_z,
\eeq
where we have dropped the prime for notational simplicity.  Looking back at Figure \ref{noPrecon}, this form of $\U$ neatly explains the stair-stepping behavior of the eigenvalues: for every eigenvalue of $\U_z$, $\U$ has $n_x\times n_y$ similar eigenvalues.
 
Since only a few eigenvalues of $\U_z$ are large, it is pedagogically useful to first address a simplified version of the preconditioning problem where $\U_z$ is approximated as a rank 1 matrix by cutting off its spectral decomposition after the first eigenvalue.  We will later return to include the other relevant eigenvalues.  We therefore write $\U$ as follows:
\beq
\U \approx \I_x \otimes \I_y \otimes \lambda \ev_z \ev_z^\dagger.
\eeq
where $\ev_z$ is the normalized eigenvector of $\U$.

Let us now take a look at the action of $\Pre_\N$ and $\Pre_\N^\dagger$ on $\U + \N$:
\begin{align}
\Pre_\N& (\U + \N) \Pre_\N^\dagger \nonumber \\
&= \I + \widetilde{\N}^{-1/2} \F_\perp  (\I_x \otimes \I_y \otimes \lambda \ev_z \ev_z^\dagger)\F_\perp^\dagger \widetilde{\N}^{-1/2} \nonumber\\
 &= \I + \widetilde{\N}^{-1/2} (\I_x \otimes \I_y \otimes \lambda \ev_z \ev_z^\dagger)\widetilde{\N}^{-1/2} \nonumber \\ &\equiv \I + \Ubar.
\end{align}
Our next goal, therefore, is to come up with a new matrix $\Pre_\U$ that, when applied to $\I + \Ubar$ gives us something close to $\I$.

We now take a closer look at $\Ubar$.  Since it is a good approximation to say that $\widetilde{\N}$ only changes perpendicular to the line of sight,\footnote{Were it not for the breathing of the synthesized beam with frequency, $\widetilde{\N}$, would only change perpendicularly to the line of sight.  Since it is a small effect when considered over a modest redshift range, we can ignore it in the construction of our preconditioner.  After all, we only need to make $\Pre\C\Pre^\dagger$ close to $\I$.} we can rewrite $\Ubar$:
\begin{align}
\Ubar &\approx (\widetilde{\N}_{\perp}^{-1/2}\otimes \I_z) (\I_x \otimes \I_y \otimes \lambda {\ev}_z {\ev}_z^\dagger)(\widetilde{\N}_{\perp}^{-1/2}\otimes \I_z)  \nonumber \\ &= (\widetilde{\N}_{\perp}^{-1}) \otimes (\lambda \ev_z {\ev}_z^\dagger), 
\end{align}
where $\widetilde{\N}_{\perp}$ is still a diagonal matrix, though only in two dimensions, generated from a baseline distribution averaged over frequency slices.  We now form a pair of preconditioning matrices, $\Pre_\U$ and $\Pre_\U^\dagger$ of the form $\Pre_\U \equiv \I - \beta \Proj$ where $\Proj$ has the property that $\Proj\Ubar = \Ubar$ and that $\Ubar\Proj^{\dagger} = \Ubar$.  The matrix that fits this description is:
\begin{align}
\Proj &= \widetilde{\N}^{-1/2} (\I_x \otimes \I_y \otimes {\ev}_z {\ev}_z^\dagger)\widetilde{\N}^{1/2} \nonumber \\&\approx (\I_x \otimes \I_y \otimes {\ev}_z {\ev}_z^\dagger) = \frac{1}{\lambda}\U,
\end{align}
since $\widetilde{\N}$ only affects the $x$ and $y$ components and thus passes through the inner matrix.  This also means that $\Proj = \Proj^\dagger$ and that $\Proj = \Proj^2$.  The result for $\Pre_\U (\I + \Ubar)\Pre_\U^\dagger$ is
\begin{align}
(\I - \beta \Proj)(\I + \Ubar)(\I - \beta \Proj^\dagger) = \I + \Ubar - 2\beta\Ubar - 2\beta\Proj + \beta^2\Ubar + \beta^2\Proj. \label{ubarquadratic}
\end{align}

The trick now is that for each $uv$-cell, $\Ubar$ has only one eigenvalue, which we call $\lbar_l$ (again using $l$ as an index over both directions perpendicular to the line of sight):
\beq 
\lbar_{l} = \frac{\lambda}{(\widetilde{N}_\perp)_{ll}}.
\eeq
Likewise, $\Proj$ only has one eigenvalue: 1. By design, these eigenvalues correspond to the same eigenvector.  Since our goal is to have the matrix in equation \eqref{ubarquadratic} be the identity, we need only pick $\beta$ such that:
\beq
1 = 1 + \lbar_{l} - 2\beta\lbar_{l} - 2\beta + \beta^2\lbar_{l} + \beta^2. \label{BetaEquation}
\eeq
Solving the quadratic equation, we get 
\begin{align}
\Pre_\U \equiv  \I - \bigg[ \sum_{l} \left(1-\sqrt{\frac{1}{1+\lbar_{l}}}  \right) {\ev}_z {\ev}_z^\dagger \text{ } \otimes \stackrel{\leftrightarrow}{\mathbf{\delta}}_{x,x_l} \otimes \stackrel{\leftrightarrow}{\mathbf{\delta}}_{y,y_l} \bigg], \label{PU}
\end{align}
where the pair of $\stackrel{\leftrightarrow}{\mathbf{\delta}}$ matrices pick out a particular $uv$-cell.  If we want to generalize to more eigenvectors of $\U_z$, we simply need to keep subtracting off sums of matrices on the right hand side of Equation \eqref{PU}:
\begin{align}
\Pre_\U \equiv \I - \sum_k \bigg[ \sum_{l} \left(1-\sqrt{\frac{1}{1+\lbar_{l,k}}}  \right) {\ev}_{z_k} {\ev}_{z_k}^\dagger \otimes \stackrel{\leftrightarrow}{\mathbf{\delta}}_{x,x_l} \otimes \stackrel{\leftrightarrow}{\mathbf{\delta}}_{y,y_l} \bigg], \label{PUmulti}
\end{align}
This works because every set of vectors corresponding to a value of $k$ is orthogonal to every other set.  Each term in the above sum acts on a different subspace of $\C$ independent of all the other terms in the sum.

If the relevant vectors ${\ev}_{z_k}$ are precomputed, applying $\Pre_\U$ can be done in $\BigO(N m(\U_z))$ where $m(\U_z)$ is defined as the number of relevant eigenvalues of $\U_z$ that need preconditioning or, equivalently, the number of ``steps'' in the eigenvalues of $\U$ in Figure \ref{noPrecon} above the noise floor. We examine how $m(\U_z)$ scales with the size of the data cube in Section \ref{PreconComplexity}. Because the fall off of the eigenvalues is exponential \cite{AdrianForegrounds} we expect the scaling of $m$ to be logarithmic.

In general, we can pick some threshold $\theta \ge 1$ to compare to the largest value of $\lbar_{l,k}$ for a given $k$ and then do not precondition modes with eigenvalues smaller than $\theta$.  One might expect there to be diminishing marginal utility to preconditioning the eigenvalues nearest $1$.  We explore how to optimally cut off the spectral decomposition in Section \ref{PreconResults} by searching for a value of $\theta$ where the costs and benefits of preconditioning equalize.

\subsection{Constructing a Preconditioner for $\R$ and $\G$}\label{PreconForGamma}

We now turn our attention to the full matrix $\C$.  The fundamental challenge to preconditioning all of the matrices in $\C$ simultaneously is that the components of $\R$ and $\G$ perpendicular to the line of sight are diagonalized in completely different bases.  However, $\U$, $\G$, and $\R$ have very similar components parallel to the line of sight, due to the fact they all represent spectrally smooth radiation of astrophysical origin.

We can write down $\R$ as follows:
\begin{align}
\R = \sum_n\bigg[ \stackrel{\leftrightarrow}{\delta}_{x,x_n}\otimes\stackrel{\leftrightarrow}{\delta}_{y,y_n}\otimes  \left( \sum_k \lambda_{n,k} \ev_{z_{n,k}} \ev_{z_{n,k}}^\dagger \right) \bigg],
\end{align}
which can be interpreted as a set of matrices describing spectral coherence, each localized to one point source, and all of which are spatially uncorrelated.  And likewise, we can write down $\G$ as:
\beq
\G = \sum_{i,j,k} \left[ \lambda_{x_i}\lambda_{y_j}\lambda_{z_k} \ev_{x_i} \ev_{x_i}^\dagger \otimes \ev_{y_j} \ev_{y_j}^\dagger \otimes \ev_{z_k} \ev_{z_k}^\dagger \right].
\eeq
We now make two key approximations for the purposes of preconditioning.  First, we assume that all the $z_k$ eigenvectors are the same, so $\ev_{z_k} \approx \ev_{z_{n,k}}$ for all $n$, all of which are also taken to be the same as the eigenvectors that appear in the preconditioner for $\U$ in Equation \ref{PUmulti}.  Second, as in Section \ref{PreconForU}, we are only interested in acting upon the largest eigenvalues of $\R$ and $\G$.  To this end, we will ultimately only consider the largest values of $\lambda_{n,k}$ and $\lambda_{i,j,k} \equiv \lambda_{x_i}\lambda_{y_j}\lambda_{z_k}$, which will vastly reduce the computational complexity of the preconditioner.

Our strategy for overcoming the difficulty of the different bases is to simply add the two perpendicular parts of the matrices and then decompose the sum into its eigenvalues and eigenvectors.  We therefore define 
\beq
\Gam \equiv \R + \G
\eeq
(choosing the symbol $\Gam$ because it looks like $\R$ and sounds like $\G$).  Given the above approximations, we can reexpress $\Gam$ as follows:
\begin{align} \label{GammaCalc}
\Gam \approx \sum_k \left(\Gam_{\perp,k} \otimes \ev_{z_k} \ev_{z_k}^\dagger\right),
\end{align}
where we have defined each $\Gam_{\perp,k}$ as:
\begin{align}
\Gam_{\perp,k} \equiv \left( \sum_n  \lambda_{n,k} \stackrel{\leftrightarrow}{\delta}_{x,x_n}\otimes\stackrel{\leftrightarrow}{\delta}_{y,y_n} \right) \left( \sum_{i,j} \lambda_{i,j,k} \ev_{x_i} \ev_{x_i}^\dagger \otimes \ev_{y_j} \ev_{y_j}^\dagger \right).
\end{align}
Due to the high spectral coherence of the foregrounds, only a few values of $k$ need to be included to precondition for $\Gam$. Considering the limit on angular box size imposed by the flat sky approximation and the limit on angular resolution imposed by the array size, this should require at most a few eigenvalue determinations of matrices no bigger than about $10^4$ entries on a side.  Moreover, those eigenvalue decompositions need only be computed once and then only partially stored for future use.  In practice, this is not a rate-limiting step, as we see in Section \ref{PreconComplexity}.

We now write down the eigenvalue decomposition of $\Gam$:
\beq
\Gam = \sum_k \left( \sum_l \lambda_{l,k} \ev_{\perp,l}\ev_{\perp,l}^\dagger \right) \ev_{z_k} \ev_{z_k}^\dagger.
\eeq
Before we attack the general case, we assume that only one value of $\lambda_{l,k}$ is worth preconditioning---we generalize to the full $\Pre_\Gam$ later.  We now know that if we have a matrix that looks like $\I + \Ubar$ we can make it look like $\I$.  So can we take $\I + \Ubar + \GammaBar$, where $\GammaBar \equiv \Pre_\N \Gam \Pre_\N^\dagger$, and turn it into $\I + \Ubar$?  Looking at $\GammaBar$,
\begin{align}
\GammaBar  =& \widetilde{\N}^{-1/2} \F_\perp \lambda_\Gam \ev \ev^\dagger \F_\perp^\dagger \widetilde{\N}^{-1/2} \nonumber \\ =& \lambda_\Gam (\widetilde{\N}_\perp^{-1/2}  \widetilde{\ev}_\perp \widetilde{\ev}_\perp^\dagger \widetilde{\N}_\perp^{-1/2}) \otimes \ev_z \ev_z^\dagger,
\end{align}
where $\lambda_\Gam$ is the sole eigenvalue we are considering and where $\widetilde{\ev}_\perp \equiv \F_\perp \ev_\perp.$

Again, we will look at a preconditioner of the $\Pre_\Gam = \I - \beta\Proj$ where:
\beq
\Proj \equiv \left(\widetilde{\N}_\perp^{-1/2} \widetilde{\ev}_\perp \widetilde{\ev}_\perp^\dagger \widetilde{\N}_\perp^{1/2}\right) \otimes \ev_z \ev_z^\dagger.
\eeq
This time, the $\widetilde{\N}^{\pm 1/2}_\perp$ matrices do not pass through the eigenvectors to cancel one another out.  We now exploit the spectral similarity of foregrounds and the fact that $\widetilde{\ev}_\perp^\dagger \widetilde{\ev}_\perp = \ev^\dagger_z \ev_z  = 1$ to obtain 
\beq 
\Pre_\Gam \Ubar \Pre_\Gam^\dagger = \Ubar + \frac{\lambda_\U}{\lambda_\Gam}(\beta^2 - 2\beta)\GammaBar.
\eeq
This is very useful because it means that if we pick $\beta$ properly, we can get the second term to cancel the $\GammaBar$ terms we expect when we calculate the full effect of $\Pre_\Gam$ and $\Pre_\N$ on $\N + \U + \Gam$.  Noting that the sole eigenvalue of $\GammaBar$ is $\overline{\lambda}_\Gam \equiv \lambda_\Gam \widetilde{\ev}_\perp^\dagger \widetilde{\N}_\perp^{-1} \widetilde{\ev}_\perp$, we also define $\overline{\lambda}_\U \equiv \lambda_\U \widetilde{\ev}_\perp^\dagger \widetilde{\N}_\perp^{-1} \widetilde{\ev}_\perp$.  Multiplying our preconditioner by our matrices, we see that the the equality of the single eigenvalues yields another quadratic equation for $\beta$: 
\begin{align}
1 + \overline{\lambda}_\U = \text{ } 1 - 2\beta + \beta^2 + (\beta^2 - 2\beta + 1)\overline{\lambda}_\Gam + \overline{\lambda}_\Gam \frac{\lambda_\U}{\lambda_\Gam}(\beta^2 -2\beta).
\end{align}
Solving, we finally have our $\Pre_\Gam$ that acts on $\I+\Ubar+\GammaBar$ and yields $\I + \Ubar$:
\begin{align}
\Pre_\Gam = \I - \left(1 - \sqrt{\frac{1+\overline{\lambda}_\U}{1+\overline{\lambda}_\U+\overline{\lambda}_\Gam}} \right) \left[\left(\widetilde{\N}_\perp^{-1/2} \widetilde{\ev}_\perp \widetilde{\ev}_\perp^\dagger \widetilde{\N}_\perp^{1/2}\right) \otimes \ev_z \ev_z^\dagger \right]. \label{PreGamma}
\end{align}
Finally, generalizing to multiple eigenvalues and taking advantage of the orthonormality of the eigenvectors, we have
\begin{align}
\Pre_\Gam =\text{ }\I - \sum_{k,m} \bigg[\left(1 - \sqrt{\frac{1+\overline{\lambda}_{\U_k}}{1+\overline{\lambda}_{\U_k}+\overline{\lambda}_{\Gam_{k,m}}}} \right) \left(\left(\widetilde{\N}_\perp^{-1/2} \widetilde{\ev}_{\perp_m} \widetilde{\ev}_{\perp_m}^\dagger \widetilde{\N}_\perp^{1/2}\right) \otimes \ev_{z_k} \ev_{z_k}^\dagger \right) \bigg]. \label{PreGammaMulti}
\end{align}
The result of this somewhat complicated preconditioner is a reduction of the condition number of the matrix to be inverted by many orders of magnitude (see Figure \ref{preconditionedEigs}).

Lastly, we include Fourier transforms at the front and the back of the preconditioner, so that the result, when multiplied by a real vector, returns a real vector.  Therefore, the total preconditioner we use for $\C$ is:
\beq
\F_\perp^\dagger \Pre_\U \Pre_\Gam \Pre_\N (\R + \U + \N + \G) \Pre_\N^\dagger \Pre_\Gam^\dagger \Pre_\U^\dagger \F_\perp.
\eeq

\end{subappendices}

\chapter{Mapmaking for Precision 21\,cm Cosmology} \label{ch:Mapmaking}

\emph{The content of this chapter was submitted to \emph{Physical Review D} on October 8, 2014 and published \cite{Mapmaking} as \emph{Mapmaking for precision 21\,cm cosmology} on January 6, 2015.}

\section{Introduction} \label{sec:Intro}

The prospect of directly probing the intergalactic medium (IGM) during the cosmic dark ages, through the ``Cosmic Dawn'' and culminating with the Epoch of Reionization (EoR) has generated tremendous excitement in 21\,cm cosmology over the past few years. Not only could it provide the first direct constraints on the astrophysics of the first stars and galaxies, but it could make an enormous new cosmological volume accessible to tomographic mapping---enabling exquisitely precise new tests of $\Lambda$CDM \cite{Yi}. For recent reviews, see e.g. \cite{FurlanettoReview, miguelreview, PritchardLoebReview, aviBook}. 

More recently, that excitement has translated into marked progress toward a statistical detection of the 21\,cm signal in the power spectrum. The first generation of experiments, including the Low Frequency Array (LOFAR \cite{LOFARinstrument}), the Donald C. Backer Precision Array for Probing the Epoch of Reionization (PAPER \cite{PAPER}), the Giant Metrewave Radio Telescope (GMRT \cite{GMRT}), and the Murchison Widefield Array (MWA \cite{TingaySummary,BowmanMWAScience}) have already begun their observing campaigns. Both PAPER \cite{DannyMultiRedshift} and the MWA \cite{X13} have released upper limits on the 21\,cm power spectrum across multiple redshifts. PAPER has already begun to use their results to constrain some models of the thermal history of the IGM \cite{PAPER32Limits}.

Still, the observational and analytical challenges that lie ahead for the field are considerable. The sensitivity requirements for a detection of the 21\,cm power spectrum necessitate large collecting areas and thousands of hours of observation across multiple redshifts \citep{MiguelNoise,Judd06,LidzRiseFall,LOFAR2,AaronSensitivity}. Of no less concern is the fact that the cosmological signal is expected to be dwarfed by foreground contaminants---synchrotron radiation from our Galaxy and other radio galaxies---by four or more orders of magnitude in brightness temperature at the frequencies of interest \citep{Angelica,LOFAR,BernardiForegrounds,PoberWedge,InitialLOFAR1,InitialLOFAR2}. 
 
The problem of power spectrum estimation in the presence of foregrounds has been the focus on considerable theoretical effort over the past few years \cite{paper1, paper2, JelicRealistic,CathWedge,LT11,DillonFast}. \citet{LT11} adapted inverse-covariance-weighted quadratic estimator techniques developed for Cosmic Microwave Background \cite{Maxpowerspeclossless} and galaxy survey \cite{Maxgalaxysurvey1} power spectrum analysis to 21\,cm cosmology. \citet{DillonFast} showed how those methods, which nominally take $\BigO(N^3)$ steps, where $N$ is the number of voxels in a 3D map or ``data cube'', could be accelerated to as fast as $\BigO(N\log N)$. 
 
However, both of those works took as their starting point data cubes containing signal, foregrounds, and noise. Neither considered the important impact that an interferometer has, not just on the noise in our maps, but on the maps themselves. An instrument-convolved map or ``dirty map'' has fundamentally different statistical properties than the underlying sky and the effects of the instrument cannot in general be fully undone. \citet{X13} discussed this problem approximately by assuming that point spread functions (PSFs) or ``synthesized beams'' depended only on frequency. Generally speaking, that is not true; PSFs are direction-dependent and typically not invertible. In this work, we relax the assumption that went into \citet{LT11} and \citet{DillonFast} while retaining the goals they strove for: minimal information loss, rigorously understood statistics, and well-controlled approximations that make the analysis computationally feasible.
  
For any near-future 21\,cm measurement, interferometric maps are essentially an intermediate data compression step. The ultimate goal is to turn time-ordered data coming from the instrument---namely, visibilities---into statistical measurements that constrain our models of astrophysics and cosmology. So why even bother making a map if we are only going to take Fourier transforms of it and look at power spectra?  The answer to that question depends on which strategy we pursue for separating the cosmological signal from foregrounds. There are two major approaches, which we will review presently.

Over the last few years, it has been realized that a region of cylindrical Fourier space\footnote{Points in cylindrical or ``2D'' Fourier space are denoted by $k_\|$, modes along the light of sight, and $k_\perp$, modes perpendicular to the line of sight. Cylindrical Fourier space takes advantage of isotropy perpendicular to the line of sight while keeping modes along the line of sight separate, since they are measured in a fundamentally different way.} should be essentially free of foreground contamination \cite{Dattapowerspec,AaronDelay,VedanthamWedge,MoralesPSShapes,Hazelton2013,CathWedge,ThyagarajanWedge,EoRWindow1,EoRWindow2}. We call this region the ``EoR window'' (see Figure \ref{fig:EoRWindow}). Observations of the EoR window thus far have found it noise dominated \cite{PoberWedge,X13}. For slowly varying spectral modes (i.e. low $k_\|$), the edge of the window is set by a combination of the intrinsic spectral structure of foreground residuals and the spectral structure introduced by the instrument. Fundamentally, an interferometer is a chromatic instrument and the fact that the shape of its point spread functions depends on frequency creates complex spectral structure in 3D maps of intrinsically smooth foregrounds \cite{EoRWindow1,EoRWindow2}. 
\begin{figure}[] 
	\centering 
	\includegraphics[width=.6\textwidth]{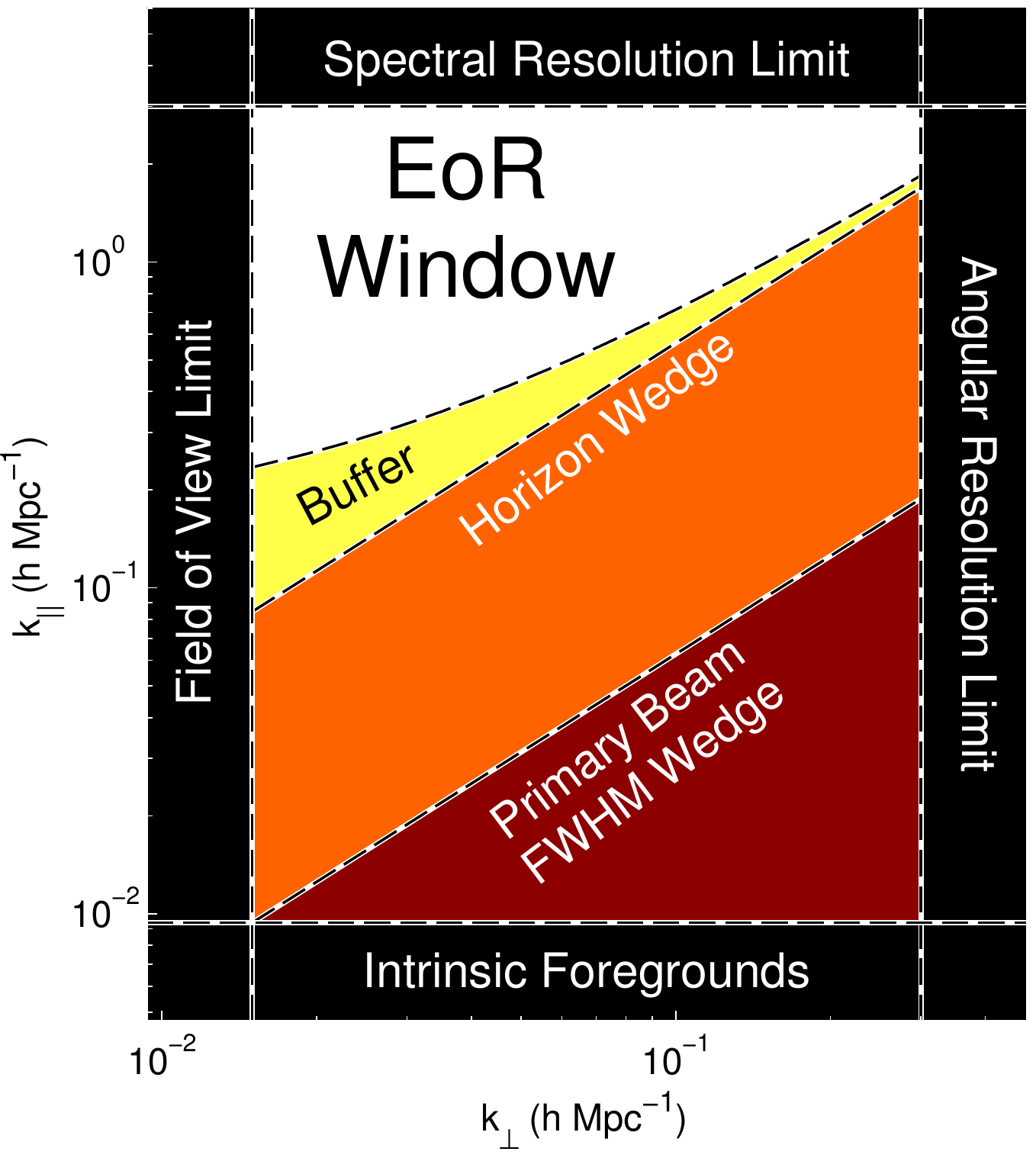}
	\caption[Simplified representation of the EoR window.]{The ``EoR window'' is a region of Fourier space believed to be essentially foreground free and thus represents a major opportunity for detecting the 21\,cm signal. Along the horizontal axis, the window is limited by the field of view, which sets the largest accessible modes, and the angular resolution of the instrument, which sets the smallest.  Along the vertical axis, the window is limited by the spectral resolution of the instrument and by the intrinsic spectral structure of galactic and extragalactic foregrounds, which dominate the spectrally smooth modes. The EoR window is further limited by ``the wedge,'' which results from the modulation of spectrally smooth foregrounds by the instrument's frequency-dependent and spatially varying point spread function. Much of the power in the wedge should fall below the wedge line associated with the primary beam while the horizon line serves as a hard cutoff for flat-spectrum foregrounds \cite{AaronDelay}. Limited ``suprahorizon'' emission has been observed and can be attributed to intrinsic spectral structure of the foregrounds \cite{PoberWedge}, so it is possible we need a small buffer beyond the horizon to be certain that the window is foreground free. Without foreground subtraction, foregrounds are expected to dominate over the cosmological signal throughout the wedge.}
	\label{fig:EoRWindow}
\end{figure} 

Fortunately, there is a theoretical limit to the region of Fourier space where instrumentally induced spectral structure can contaminate the power spectrum. It is set by the delay associated with a source at the horizon (which is the maximum possible delay) for any given baseline \cite{AaronDelay}. This region of cylindrical Fourier space is known colloquially as ``the wedge.'' Furthermore, we expect that most of the foreground emission should appear in the main lobe of the primary beam, setting a soft limit on foreground emission at lower $k_\|$ (see Figure \ref{fig:EoRWindow}). 

The simplest approach to power spectrum estimation in the presence of foregrounds, and likely the most robust, is to simply excise the entire section of Fourier space that could potentially be foreground-dominated. This conservative approach takes the perspective that we have no knowledge about the detailed spatial or spectral structure of the foregrounds and therefore that the entire region under the wedge is hopelessly contaminated.  If that were the case, the optimal strategy would simply be to project out those modes. This ``foreground avoidance'' strategy has been used to good effect by both PAPER \cite{PAPER32Limits,DannyMultiRedshift} and the MWA \cite{X13}, though neither made sensitive enough measurements to be sure that foregrounds are sufficiently suppressed inside the EoR window to make a detection without subtracting them. Considerable work has already been done with methods of estimating the power spectrum that minimize foreground contamination from the wedge into the window \cite{X13,EoRWindow2}. 

Foreground avoidance, however, comes at a significant cost to sensitivity. The more aggressive alternative is ``foreground subtraction'', a strategy that tries to remove power associated with foregrounds and expand the EoR window. The idea behind foreground subtraction is twofold. First, we remove our best guess as to which part of the data is due to foreground contamination. Second, we treat residual foregrounds as a form of correlated ``noise,'' downweighting appropriately in the power spectrum estimator and taking into account biases introduced. In the limiting case where we know very little about the foregrounds, foreground subtraction becomes foreground avoidance. 
 
For the upcoming Hydrogen Epoch of Reionization Array (HERA), \citet{PoberNextGen} compared the effects of foreground avoidance to foreground subtraction. If the window can be expanded from delay modes associated with the horizon to delay modes associated with the full width at half maximum of the primary beam, the sensitivity to the EoR signal improves dramatically. Over one observing season with a 547-element HERA, the detection significance of a fiducial EoR signal improves from 38$\sigma$ to 122$\sigma$. For smaller telescopes, this might mean the difference between an upper limit and a solid detection. More importantly, the errors on the measurements of parameters that describe reionization from the power spectrum improve from about 5\% to less than 1\% when employing extensive foreground subtraction. That would be the most sensitive measurement ever made of the direct effect of the first stars and galaxies on the IGM. Simply put, there is much that might be gained by an aggressive foreground subtraction approach.

That said, it will not be easy. In order to expand the EoR window and reduce the effect of foregrounds, one must model them very carefully. Likely we will want to use outside information like high-resolution surveys to try to measure source fluxes to be much better than a percent. Even more importantly, one must take our own uncertainty about these models into account. If we do not, we risk mistakenly claiming a detection. We must propagate both our best estimates for the foregrounds and our uncertainty in our models through the instrument, which is the source of the wedge itself. 

Both galactic and extragalactic foregrounds have complex spatial structure. Any precise model for their emission is direction dependent. More importantly, our model for the statistics of our uncertainty about their emission, is also direction dependent. The covariance of residual foregrounds, especially of bright sources, is most simply and compactly expressed in real space \cite{DillonFast}.  

\begin{figure}[] 
	\centering 
	\includegraphics[width=.35\textwidth]{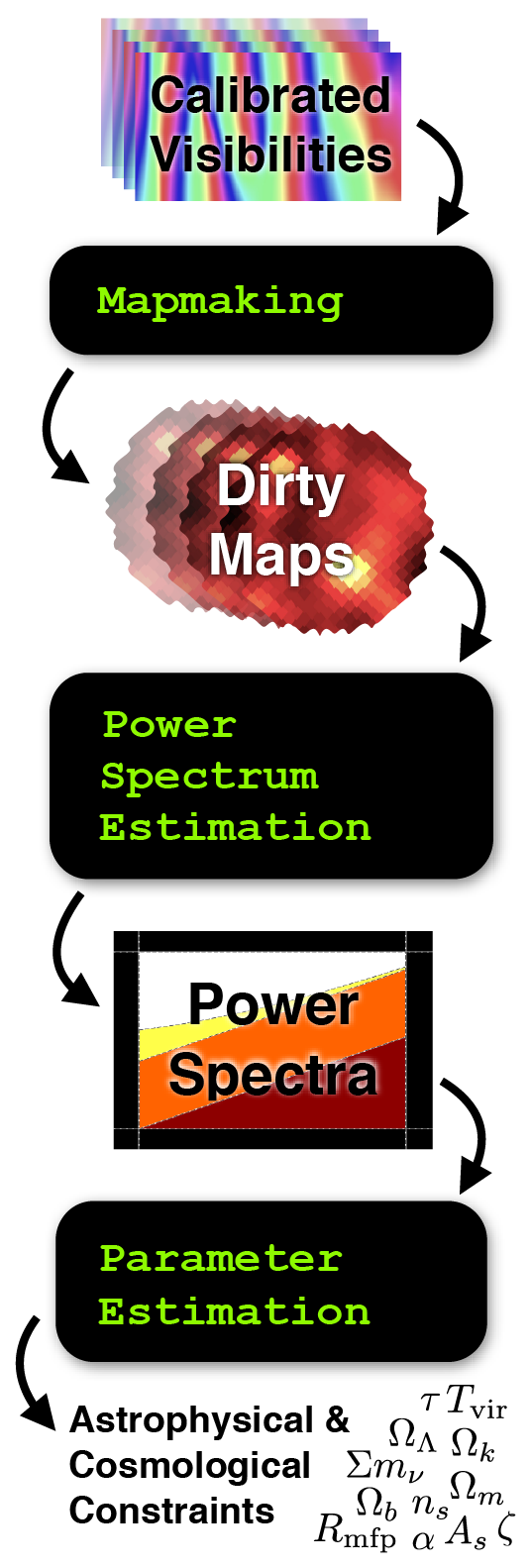}
	\caption[Schematic representation of an optimal analysis pipeline.]{Mapmaking is the first in a series of steps that reduce the volume of data while trying not to lose any astrophysical or cosmological information. The goal of this work is to address that first data-compressional step---turning calibrated visibilities into a stack of dirty maps or a data cube---with any eye toward the next step---power spectum estimation in the presence of dominant astrophysical foregrounds.  This data compression is achieved by combining together different observations a single, relatively small set of maps. Power spectra represent the cosmological signal even more compactly by taking advantage of homogeneity and isotropy and serve as the natural data product to connect to simulations and theory and thus constrain cosmological and astrophysical parameters.}
	\label{fig:pipeline}
\end{figure}

We can now finally answer the question of why we should make maps if we are ultimately interested in power spectra. We need maps as an intermediate data product because they allow us to prepare our data in a highly compressed form that puts us in a natural position to carefully pick apart the signal from the foregrounds and the noise. Forming power spectra directly with visibilities, by comparison, requires treating each local sidereal time separately and vastly increases the data volume. In Figure \ref{fig:pipeline} we put mapmaking into the larger context of data reduction all the way from calibrated visibilities to cosmological and astrophysical constraints. The goal of each step is to reduce the volume of data while keeping as much cosmological information as possible, allowing for quantification of errors, and making the next step easier.

The science requirements of our maps are very different from those that motivated most interferometric mapmaking in radio astronomy to date. Usually, radio astronomers are interested in the astrophysics of what we call ``foregrounds'' and focus on detailed images and spectra. For us it is especially important to understand how our maps are related statistically to the true sky, whose underlying statistics we would like to characterize using the power spectrum. Because interferometers do not uniformly or completely sample the Fourier plane, the relationship between our maps and the true sky is complicated. The PSFs of our maps depend both on frequency and on position on the sky. In order to estimate power spectra from maps accurately, we need to know precisely both the relationship of our dirty maps to the true sky and the covariance of our dirty maps that relates every pixel at every frequency to every other.\footnote{It is worth mentioning that the techniques developed here do not apply only to 21\,cm tomography. Any power spectrum made with maps produced from interferometric data needs to take into account the effects of the frequency-dependent and spatially varying PSF on both the signal and the contaminants. This includes intensity mapping of CO and CII and interferometric measurements of the CMB. Higher-order statistics, like the bispectrum and trispectrum, also need precise knowledge of the relationship between the true sky and the dirty maps.} Current imaging techniques do not compute these quantities. It is the main point of this paper to show why and how that must be done. 

Both \cite{EoRWindow1} and \cite{EoRWindow2} focused on a similar point about the important effect of the instrument on the power spectrum. There, the authors derived a framework for rigorously quantifying the errors and error correlations associated with instrument-convolved data and showed how the wedge feature arose even in a rigorous and optimal framework. However, because they formed power spectra directly from visibilities without using maps as an intermediate data-compression step, their tools are impractical for use with large data sets.

In this work, we have two main goals. First, we would like to mathematically understand how the instrument gives rise to a complicated PSF and how that PSF can be self-consistently incorporated into the inverse-covariance-weighted power spectrum estimation techniques (e.g. \cite{LT11} and \cite{DillonFast}). In Section \ref{sec:Mapmaking}, we discuss the theory of mapmaking as an intermediate step between observation and power spectrum estimation. Then, in Section \ref{sec:method}, we investigate how to put that theory into practice. We use HERA as a case study in carrying out the calculation of dirty maps and their statistics. Although the computational cost of performing those calculations is naively quite large, we develop and analyze three main ways reducing it dramatically:
\begin{itemize}
\item We explore how restricting our maps to independent facets on the sky lets us reduce the number of elements in our PSF matrices and the difficulty of calculating them (Section \ref{sec:faceting}).
\item We show how individual timesteps can be combined and analyzed simultaneously, approximately accounting for the rotation of the sky over the instrument (Section \ref{sec:snapshots}). 
\item We show how the point spread functions, while not translationally invariant,  vary smoothly enough spatially that the associated matrix operations can take advantage of certain symmetries for a computational speedup (Section \ref{sec:PSFfitting}).
\end{itemize}
We will show how each of these approximations works and analyze them to understand the trade-off between speed and accuracy in each case.
 
\section[Precision Mapmaking And Map Statistics in Theory]{Precision Mapmaking And Map Statistics\\ in Theory} \label{sec:Mapmaking}
 
Making maps from interferometric data has a long history and a great number of techniques have been developed with different science goals in mind \cite{ThompsonMoranSwenson}. Most focus on deconvolution, the removal of point source side lobes (or the side lobes of extended sources represented as multiple components) after their convolution with the synthesized beam. This is the basic idea behind the CLEAN algorithm \cite{CLEAN} and its many descendants, including \cite{Schwab1984,Cotton2004,Cotton2005,Bhatnagar2008,Carozzi2009,Mitchell2008,bernardi,Bart,Smirnov2011,Li2011,FHD,WSCLEAN}. Some of these, notably that of \citet{FHD}, take inspiration from \cite{TegmarkCMBmapsWOLosingInfo}, in that they use the framework of ``optimal mapmaking'' for forming dirty maps without losing any cosmological information contained in the visibilities. Additionally \cite{Richard} and \cite{ShawCoaxing}, which use the optimal mapmaking formalism in the $m$-mode basis to exploit the observational symmetries of a drift scanning interferometer, are also closely related to the work presented here.

A notable exception is \cite{SutterBayesianImaging}, which develops a method of Bayesian deconvolution via Gibbs sampling in the relatively simplified case of a gridded $uv$-plane, which can then be used for power spectrum estimation \cite{GibbsPSE}. This method  not only calculates a map but also gives error estimates on each pixel in that map. This is an especially promising technique for finding sources and quantifying the errors on our measurements of their fluxes and spectral indices. We take a different tack and do not focus on deconvolution at all.

In this work, we are interested not just in a dirty map but also in the statistical properties of that map. As in previous work, we want to know how sources are convolved with the instrument. But we also want to know how that instrumental convolution affects our covariance models for everything in the map, including signal, noise, and foregrounds. A complete understanding of the relationship between the true sky and our dirty maps will allow us to comprehensively model these important statistical quantities. Current imaging methods simply do not compute that relationship and the resulting noise covariance matrix. However, these are required for methods of power spectrum estimation in order to properly weight data in the presence of correlated noise and foregrounds and to account for missing modes. The importance of this was realized by \cite{FFTT2}, though we will use a different computational approach to speed up the calculations. 

We begin this section by summarizing the relevant physics behind interferometry in Section \ref{sec:interferometry}. We then review the optimal mapmaking formalism in Section \ref{sec:OMM}. Finally, in Section \ref{sec:powerspectra} we work out the consequences of proper map statistics for the inverse-covariance-weighted quadratic power spectrum estimation formalism, including how they affect the models of the covariance of cosmological signal, noise, and foreground residuals.

\subsection{Interferometric Measurements} \label{sec:interferometry}

When we make maps from interferometric data, we are interested in computing a map estimator or ``dirty map,'' which we call $\widehat{\x}$, and understanding its relationship to $\x$, the true, discretized sky.\footnote{We write these quantities as vectors as a compact way of combining indices over both angular dimensions on the sky and over frequency.} We do not have access to $\x$ directly; we can only make inferences about it by making a set of complex ``visibility'' measurements which we call $\y$. Each measurement made with our instrument is a linear combination of the true sky added to instrumental noise. Therefore, we can represent all our measurements with
\beq
\y = \A \x + \n, \label{eq:measurement}
\eeq
where $\A$ represents the interferometric response of our instrument over all times, frequencies, and baselines and where each $n_i$ is the instrumental noise on the $i$th visibility. The matrix $\A$ has the dimensions of the number of measured visibilities (for every baseline, frequency, and integration) by the number of voxels in the 3D sky (all pixels at all frequencies).

The statistics of $\n$ are fairly simple. It has zero mean and the noise on each visibility is generally treated as independent of that on every other visibility. Therefore,
\begin{align}
\langle n_i \rangle &= 0 \\
N_{ij} \equiv \langle n_i n_j^* \rangle &= \sigma^2_i \delta_{ij}.
\end{align}
The form of $\A$ is considerably more complicated, it can be written in the form of Equation \eqref{eq:measurement} because a visibility is a weighted integral over the whole sky which can be approximated to any desired precision by a finite matrix operation.

The visibility measured by a noise-free instrument with arbitrarily fine frequency resolution at frequency $\nu$ and baseline $\base_m$ in response to a sky specific intensity $I(\rhat,\nu)$ defined continuously over all points on the sky $\rhat$ is
\begin{align}
V(\base_m,\nu) = \int B_m\left(\rhat,\nu \right)
I(\rhat,\nu) \exp\left[-2 \pi i  \frac{\nu}{c} \base_m \cdot \rhat \right] d\Omega . \label{eq:AnalyticVisibility}
\end{align}
Here $B_m(\rhat,\nu)$ is the product of the complex primary beams of the two antenna elements that form the $m$th baseline. In this equation and in the rest of this section, we will ignore the polarization of the sky and the fact that there are different beams for each polarization, assuming homogenous antenna elements. We do this for simplicity; the results are straightforwardly generalizable to a complete treatment of polarization, which we will explore in  Appendix \ref{app:PolAndBeams}. In that appendix, we will also look at how heterogenous arrays straightforwardly incorporated into our framework as well.

Given a finite number of measurements, we are interested in the relationship between visibilities and a discretized true sky, $\x$. In frequency, that discretization comes from the spectral response of our instrument---we can only measure a limited number of frequency channels. Spatially, we need to choose our pixelization of the sky. Let us define a 3D pixelization function $\psi_i(\rhat,\nu)$ that incorporates both these kinds of pixelization. It is defined so that,
\beq
x_i = \int  \psi_i(\rhat,\nu) \frac{c^2}{2k_B\nu^2} I(\rhat,\nu) d\Omega d\nu ,\label{eq:PixelizationDefinition}
\eeq
where the extra factor of $c^2/2k_B\nu^2$ converts from units of specific intensity to brightness temperature. For simplicity, we define $\psi_i(\rhat,\nu)$ to be the unitless top-hat function, normalized such that  
\beq
\int \psi_i(\rhat,\nu) \frac{d\Omega}{\Delta \Omega} \frac{d\nu}{\Delta \nu} = 1 \label{eq:pixelization_defintiion}
\eeq
where $\Delta \nu$ is the frequency resolution of the instrument and $\Delta \Omega$ is the angular size of the pixels. Other choices of $\psi_i(\rhat,\nu)$ are perfectly acceptable, in which case $\Delta \nu$ and $\Delta \Omega$ become characteristic spectral and spatial sizes of pixels.

Therefore we can rewrite Equation \eqref{eq:AnalyticVisibility} as a sum:
\begin{align}
V(\base_m,\nu_n) \approx \sum_k \Delta\Omega \frac{2 k_B \nu_n^2}{c^2} x_k(\nu_n)
B_m(\rhat_k,\nu_n) \exp\left[-2 \pi i  \frac{\nu_n}{c} \base_m \cdot \rhat_k \right]. \label{eq:VisibilitySum}
\end{align}
Here we have chosen to break apart the index $i$ into a spatial subindex, $k$, and a spectral subindex, $n$. The sum is over all spatial pixels. This approximation relies on choosing a frequency and angular resolution small enough that $B(\rhat,\nu)$ and $ \exp\left[-2 \pi i  (\nu / c) \base_m \cdot \rhat \right]$ can be approximated as constants inside of a single spatial pixel and frequency channel. Since $V(\base_m,\nu_n)$ is an entry in $\y$, Equation \eqref{eq:VisibilitySum} gives us the elements of $\A$ by relating $\y$ to $\x$ for a single observation and a single baseline. Of course, the full matrix $\A$ that goes into Equation \eqref{eq:measurement} gives us a relationship between the true sky and every visibility at every frequency and at every local sidereal time. The basic physics, however, is captured by Equation \eqref{eq:VisibilitySum}.

\subsection{The Optimal Mapmaking Formalism}\label{sec:OMM}

Given a set of visibilities (or any time-ordered data) of the form in Equation \eqref{eq:measurement}, there is a well known technique for forming estimators of the true sky without losing any information about the discretized sky contained in the time-ordered data \cite{TegmarkCMBmapsWOLosingInfo}. Those estimators, known as ``optimal mapmaking'' estimators, take the general form
\beq
\widehat{\x} = \D \A^\dagger \N^{-1} \y \label{eq:OMM}
\eeq
where $\D$ can be any invertible normalization matrix. Especially for long observations, $\y$ is a much larger vector than  $\widehat{\x}$. Mapmaking represents a major data compression step. 

\begin{figure*}[] 
	\centering 
	\includegraphics[width=1\textwidth]{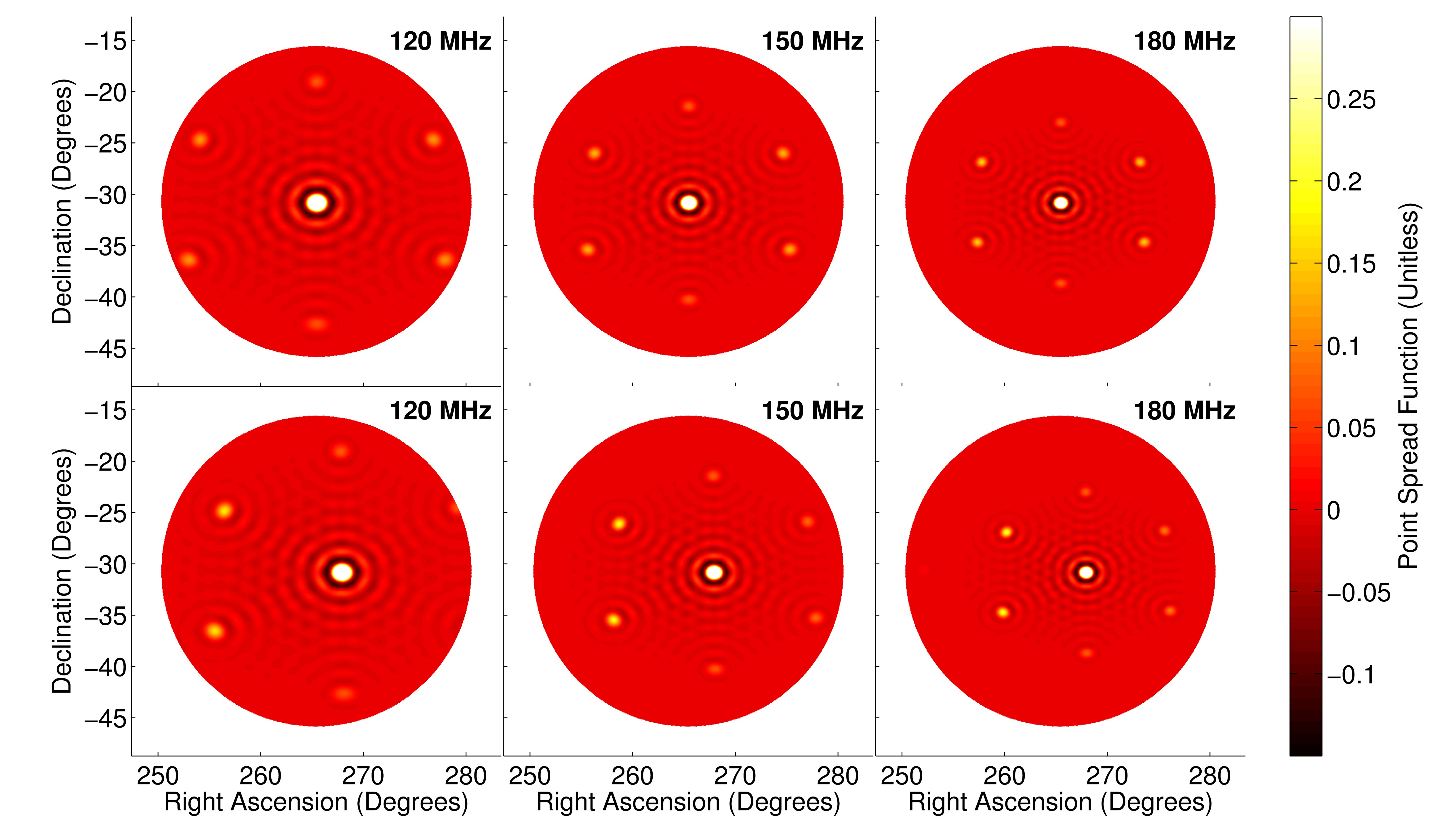}
	\caption[Example PSFs as a function of frequency and position.]{The point spread function (or equivalently, the synthesized beam) of a dirty map varies both as a function of position on the sky and as a function of frequency. In the top row, we show the point spread functions at three frequencies corresponding to the center of the primary beam calculated for HERA. They exhibit clear diffraction rings and fairly strong side lobes due to tje fact that the minimum separation between antennas is significantly longer than the wavelength. The hexagonal pattern is due to the geometry of the array. In the bottom row, we look at off-center point spread functions. These also have side lobes, though they are asymmetric due to the primary beam and the projected layout of the array and thus a clear example of the translational variation of the PSF. All six can be thought of as single rows of different frequency blocks of the full matrix of point spread functions, $\PSF$. Each PSF peaks at 1, but we have saturated the color scale to show detail. In Section \ref{sec:method}, we will explain in detail how these PSFs are calculated.}
	\label{fig:PSFs}
\end{figure*}

The expected value of the estimator is
\begin{align}
\langle \widehat{\x} \rangle &= \langle \D \A^\dagger \N^{-1} (\A \x + \n) \rangle \nonumber \\ 
&= \D \A^\dagger \N^{-1} (\A \x + \langle \n \rangle) \nonumber \\ 
&= \D \A^\dagger \N^{-1} \A \x. \label{eq:<xhat>}
\end{align}
In general, the expected value of $\widehat{x}$ is not the same as the true sky but is rather some complicated linear combination of pixels on the true sky. We define 
\beq
\PSF \equiv \D \A^\dagger \N^{-1} \A \label{eq:PSFdef}
\eeq
to be the matrix of point spread functions. Each column of this matrix tells us how each pixel on the true sky gets mapped to all the pixels of the dirty map. If we want to normalize the PSF to always have a central value of 1, we can achieve that by a judicious choice of $\D$. In this work, we make that choice of PSF normalization. Recall that $\D$ can be any invertible matrix. Since we are not trying to make images that look as much as possible like the true sky but rather just to keep track of exactly how our dirty maps are related to the true sky, making a very simple choice for $\D$ is sensible.\footnote{The choice of $\D=\left[\A^\dagger \N^{-1} \A\right]^{-1}$  was used by WMAP \cite{WMAPconjugategrad} because it makes  $\PSF = \Eye$, but that matrix is generally not invertible in radio interferometry. Whenever one cannot make that choice of $\D$, $\PSF$ is not the identity and one must keep track of its effects.} Therefore, we use our freedom in choosing $\D$ to make it a diagonal matrix---effectively a per-pixel normalization. In Figure \ref{fig:PSFs} we plot an example of the central portions of two different rows of $\PSF$ at three different frequencies. 

\subsection{Connecting Maps to Power Spectra}\label{sec:powerspectra}

As we discussed earlier, we are interested in mapmaking in order to reduce the volume of our data without losing any sky information or the ability to remove foregrounds. From the map, the next step is to further compress the data by calculating a power spectrum, which can be directly compared with theoretical predictions. To connect the mapmaking formalism to 21\,cm power spectrum estimation, we will review the statistical estimator formalism for calculating power spectra while not losing any cosmological information. In the process, we will enumerate the quantities that we need to calculate in order to estimate a power spectrum from $\widehat{\x}$. Then we will show the form that those quantities take in terms of $\widehat{\x}$, $\PSF$, and $\D$.

\subsubsection{Power Spectrum Estimation Reivew}

Fundamentally, a power spectrum estimate is a quadratic combination of the data. To calculate a power spectrum, roughly speaking, one simply Fourier transforms real-space data, squares, and then averages in discrete bins to form ``band powers.'' In a real-world measurement with noise and foreground contamination, we need a more sophisticated technique. 

Because we have a finite amount of data, we must discretize the power spectrum we estimate by approximating $P(\mathbf{k})$ as a piecewise constant function described by a set of band powers $\mathbf{p}$ using 
\beq
P(\mathbf{k}) \approx \sum_\alpha p_\alpha \chi_\alpha(\mathbf{k}), \label{eq:bandpowers}.
\eeq
Here $\chi_\alpha(\mathbf{k})$ is a characteristic function which equals 1 inside the region described by the band power $p_\alpha$ and vanishes elsewhere.

Since the power spectrum is a quadratic quantity in the data, an estimator $\widehat{\p}$ of the band power spectrum $\p$ (which is discretized by approximating the power spectrum as piecewise-constant) takes the form
\beq
\widehat{p}_\alpha = (\widehat{\x} - \mean)^\trans \mathbf{E}_\alpha (\widehat{\x} - \mean) - b_\alpha.
\eeq
Here $\E_\alpha$ very generally represents the operations we want to perform on the data and $\mean \equiv \langle \widehat{\x} \rangle$ is the ensemble average over many realizations of the same exact observation, each with different noise, and $\mathbf{b}$ removes additive bias from noise and residual foregrounds in the power spectrum.

Just as estimators of the form in Equation \eqref{eq:OMM} do not lose any information about the true sky contained in the visibilities, there exists an optimal quadratic estimator for power spectra that does not lose cosmological information \cite{Maxpowerspeclossless}.\footnote{This entails certain assumptions, most notably that the noise, residual foregrounds, and signal are all completely described by their means and covariances---in other words that they are Gaussian. We know that this is not exactly true in the case of residual foregrounds and signal, though it is generally assumed to be a pretty good approximation for the purposes of the first generation of 21\,cm measurements \cite{LT11}.} Those estimators take the form
\beq
\widehat{p}_\alpha = \frac{1}{2}M^{\alpha\beta} (\widehat{\x} - \mean)^\trans \C^{-1} \mathbf{C},_\beta \C^{-1} (\widehat{\x} -\mean) - b_\alpha. \label{eq:QE}
\eeq
In this equation, $\M$ is an invertible normalization matrix, analogous to $\D$ and $\C$ is the covariance of $\widehat{\x}$ (not of the true sky $\x$) and is defined as 
\beq
\C \equiv \langle \widehat{\x} \widehat{\x}^\trans \rangle - \langle \widehat{\x} \rangle \langle \widehat{\x} \rangle ^\trans.
\eeq
Each $\mathbf{C},_\beta$ matrix, which encodes the Fourier transforming and binning steps of the power spectrum, is defined such that
\beq
\C = \C^\text{contaminants} + \sum_\beta p_\beta \mathbf{C},_\beta.
\eeq
Here $\C^\text{contaminants}$ represents the covariance of anything that appears in $\widehat{\x}$ that is not the 21\,cm cosmological signal. In other words, the set of $\mathbf{C},_\beta$ matrices tells us how the covariance of $\widehat{\x}$ responds to changes in the underlying band powers, $\p$. We will explain the precise form of $\C,_\beta$ shortly.

\subsubsection{The Statistics of the Mapmaking Estimator} 

All of the quantities we are interested in calculating when estimating the power spectrum, including the bias term, the errors on our band powers, the error covariance between band powers, and the ``window functions'' that encode the relationship between $\widehat{\p}$ and $\p$, are derived from our models of $\mean$ and $\C$ (see e.g. \cite{Maxpowerspeclossless, LT11, DillonFast, X13} for the exact forms of these quantities). In this section, we will see how those quantities depend on the mapmaking algorithm and are inextricably linked to the response of the interferometer. 

We have already shown that $\langle \widehat{\x} \rangle = \PSF \x$ in Equations \eqref{eq:<xhat>} and \eqref{eq:PSFdef}. When we are making a map, this is sufficient---there is a ``true'' sky and we are trying to estimate a quantity related to it from noisy data in a well-understood way. In the context of power spectrum estimation, simply averaging down instrumental noise is not enough. Because we are interested in the statistical properties of the Universe as a whole, we are trying to use multiple independent spatial modes to learn about at the underlying statistics of $\x$, taking advantage of homogeneity and isotropy. Though there is only one true sky, we treat it as a random field with Gaussian statistics. Therefore,
\begin{align}
\mean = \langle \widehat{\x} \rangle  = \PSF \langle \x \rangle = \PSF\left[ \langle \x^S \rangle + \langle \x^N \rangle + \langle \x^{FG} \rangle  \right] = \PSF  \langle \x^{FG} \rangle. \label{eq:mean}
\end{align}
Here we have explicitly separated our model for the sky into three statistically independent parts: the 21\,cm signal, the noise, and the foregrounds. Only the foregrounds have nonzero mean.\footnote{The mean of the cosmological signal is zero only because it is usually defined as the fluctuations from the mean brightness temperature of the global 21\,cm signal. For our purposes, the global signal is a contaminant and can be treated as part of the diffuse foregrounds without loss of generality.} Because they are statistically independent, the covariance can be separated into the sum of three matrices.\footnote{It should be noted that each of these covariance matrices is the covariance of the instrument-convolved sky and not the true sky, in contrast to the notation in \cite{DillonFast} which, by treating an idealized scenario, ignored the distinction.} Hence,
\begin{align}
\C = \C^S + \C^N + \C^{FG}.
\end{align}
We will now show how all of these are calculated in the context of optimal mapmaking.

\subsubsection{The Signal Covariance}

First, let us turn to the signal covariance, $\C^S$. To understand what this really means, we need to first explain what we mean by $\x^S$. Imagine a continuous 21\,cm temperature field as a function of position in comoving coordinates, $x^S(\mathbf{r})$. Each element of the vector $\x^S$ is given by
\beq
x^S_i \equiv \int \psi_i(\mathbf{r})x^S(\mathbf{r}) \frac{d^3r}{\Delta V}, 
\eeq
where $\psi_i(\mathbf{r})$ encloses exactly the same volume as $\psi_i(\rhat,\nu)$ and $\Delta V \equiv \int \psi_i(\mathbf{r}) d^3r$ is the comoving volume of a voxel. The continuous 21\,cm power spectrum, $P(\mathbf{k})$ is defined by
\beq
\left< \left[\widetilde{x}^S(\mathbf{k})\right]^* \widetilde{x}^S(\mathbf{k}') \right> \equiv (2\pi)^3 \delta(\mathbf{k} - \mathbf{k}') P(\mathbf{k}), 
\eeq
where $\widetilde{x}^S(\mathbf{k})$ is the Fourier transform of $x^S(\mathbf{r})$. It follows then that
\beq
\langle x^S_i x^S_j \rangle - \langle x^S_i \rangle \langle x^S_j \rangle = \int \widetilde{\psi}_{i}(\mathbf{k})\widetilde{\psi}_{j}^{*}(\mathbf{k})P(\mathbf{k})\frac{d^3k}{(2\pi)^{3}}. \label{eq:signalcov}
\eeq

By combining Equations \eqref{eq:signalcov} and \eqref{eq:bandpowers}, we can write down the covariance of $\mathbf{x}^S$:
\beq
\langle x^S_i x^S_j \rangle - \langle x^S_i \rangle \langle x^S_j \rangle \approx \sum_\alpha p_\alpha Q^\alpha_{ij}, \label{eq:covx}
\eeq
where
\beq
Q^\alpha_{ij} \equiv \int \widetilde{\psi}_{i}(\mathbf{k})\widetilde{\psi}_{j}^{*}(\mathbf{k})\chi_\alpha(\mathbf{k})\frac{d^3k}{(2\pi)^{3}}.
\eeq
Finally, using the fact that $\langle \widehat{\x} \rangle = \PSF \x$ determines also the relationship between the cosmological components of $\x$ and $\widehat{\x}$, we find that
\beq
\C^S \approx \PSF \left[ \sum_\alpha p_\alpha \Q_\alpha \right]\PSF^\trans \label{eq:CS}
\eeq
and therefore that 
\beq
\C,_\alpha \approx \PSF \Q_\alpha \PSF^\trans. \label{eq:CcommaAlpha}
\eeq

\subsubsection{The Noise Covariance}
While $\langle \widehat{\x}^N \rangle = \langle \x^N \rangle = 0$, the instrumental noise still contributes to the covariance. Our mapmaking formalism makes it straightforward to track how the noise on individual visibilities, $\sigma^2_i$, translates into correlated noise between pixels in a dirty map, which is described by $\C^N$. Let us imagine that $\x = 0$ and our instrument measured just noise for each visibility. If we compute the covariance of $\widehat{\x}$ in this case we will have $\C^N$, since $\C^{S}$ and $\C^{FG}$ represent our knowledge about the sky. This is true because there are no cross terms that correlate noise with foregrounds or signal.

Therefore, since our usual inverse-covariance-weighted map estimator now gives us
\beq
\widehat{\x}^N = \D \A^\dagger \N^{-1} \n,
\eeq
it follows that
\begin{align}
\C^N &= \left< \widehat{\x}^N \left(\widehat{\x}^N \right)^\trans \right> =  \left< \D \A^\dagger \N^{-1} \n \n^\dagger \N^{-1} \A \D^\trans \right> \nonumber \\
&= \D \A^\dagger \N^{-1} \left< \n \n^\dagger \right> \N^{-1} \A \D^\trans \nonumber \\ 
&= \D \A^\dagger\N^{-1} \A \D^\trans = \PSF \D^\trans. \label{eq:CN}
\end{align}
This is a gratifyingly simple result; calculating $\PSF$ yields $\C^N$ virtually for free. It also allows us to avoid the common assumption (made for example by \cite{X13}, \cite{LT11} and, \cite{DillonFast}) that instrumental noise is uncorrelated between pixels in a gridded $uv$-plane. Correlations between $uv$ pixels introduced by the primary beam are fully taken into account in our framework because, like in \cite{EoRWindow1}, $\C^N$ contains all the relevant information about the instrument and the mapmaking process.

\subsubsection{The Foreground Covariance}
Finally, we come to the statistics of the foregrounds. The reason that we treat $\x^{FG}$ as a random field even though there is really only one set of true foregrounds is that we want to represent both our best guess at the foregrounds and our uncertainty about that guess. When we write $\langle \x^{FG} \rangle$ in Equation \eqref{eq:mean}, we really mean our best guess as to the true foregrounds---the average of our incomplete knowledge about their positions, fluxes, spectral indices, and angular extents. Therefore we need to calculate
\beq
\mean = \langle \widehat{\x}^{FG} \rangle = \PSF \langle \x^{FG} \rangle = \PSF \x^{FG}_\text{model}
\eeq
to use in our quadratic estimator in Equation \eqref{eq:QE}.
 
Previous work (e.g.  \cite{LT11,DillonFast}) built explicit models of the foreground uncertainty by looking at the first and second moments of $\x^{FG}$ and not at $\widehat{\x}^{FG}$. We can take that work and generalize it straightforwardly. If $\C^{FG}_\text{model}$ is a model of foregrounds that takes into account our uncertainties about fluxes, spectral indices, and angular correlations, like the one developed in \cite{LT11} and \cite{DillonFast}, then the foreground covariance of the estimator is
\beq
\C^{FG} = \PSF \C^{FG}_\text{model} \PSF^\trans. \label{eq:CFG}
\eeq

\begin{figure}[] 
	\centering 
	\includegraphics[width=.6\textwidth]{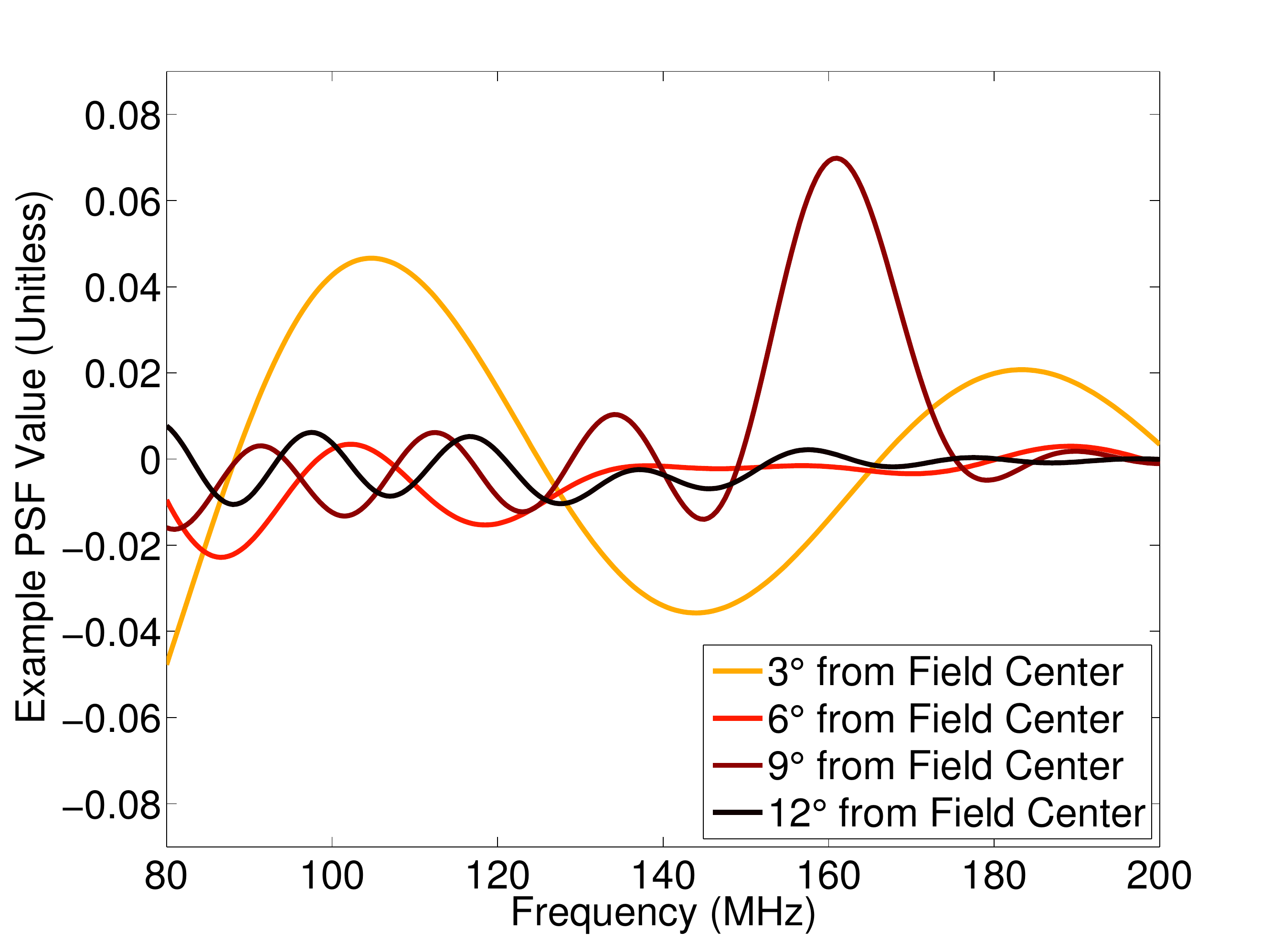}
	\caption[Example of PSF chromaticity.]{The position and frequency dependence of the synthesized beam is the origin of the ``wedge'' feature and plays a key role in determining which Fourier modes are foreground dominated in any power spectrum estimate. Here we show four different example lines of sight through a single frequency-dependent PSF, namely the one we showed for HERA in the top row of Figure \ref{fig:PSFs}. The structure we see means that intrinsically flat spectrum sources will appear far more complicated in a dirty map. We can also see that emission further from the zenith has more complicated spectral structure---an observation that helps explain the wedge. Any attempt at foreground subtraction will require detailed knowledge of this spectral behavior, both for our models for foregrounds and for our models of our uncertainty about foreground fluxes and spectral indices.}
	\label{fig:chromaticity} 
\end{figure}

This equation compactly illustrates a key difference between the analysis methods developed by \citet{LT11} and \citet{DillonFast} and any future work that takes into account the inherent frequency dependence of foregrounds in dirty maps---the focus of this work. Intrinsic foregrounds are believed to be dominated by only a few Fourier modes \cite{AdrianForegrounds}. That means that the expression of our uncertainty about the level of foreground contamination and thus our ability to subtract foreground, $C^{FG}_\text{model}$, should also be dominated by a few Fourier modes. However the PSF's spectral and spatial structure moves power from those low $k_\|$ modes up into the wedge. In Figure \ref{fig:chromaticity}, we plot a few representative lines of sight of a field-centered PSF of a zenith-pointed instrument at different distances from field center. 
Even a flat-spectrum source would see considerable structure introduced on many spatial scales along the line of sight, especially far from the zenith. This is the origin of the wedge \cite{MoralesPSShapes} and, as \cite{EoRWindow1} pointed out, it can be fully understood as a consequence of the fact that frequency appears in the exponent of Equation \eqref{eq:AnalyticVisibility}. An interferometer is an inherently chromatic instrument. 

To summarize, in order to optimally estimate a 21\,cm power spectrum from the results of an optimal mapmaking routine, we must properly take into account the relationship between the dirty map and the true sky. To do this, we will need:
\begin{enumerate}
\item Our estimated dirty map, $\widehat{\x}$.
\item The normalization matrix for that map, $\D$, and the matrix of point spread functions, $\PSF$. Those require knowledge of the instrument, the observing strategy, and the noise in our measurements.
\item A model for the cosmological signal, which will allow us to properly account for sample variance.
\item A ``best guess'' for the foregrounds and a model for our uncertainty about that best guess.
\end{enumerate}
With all these components, we can go from visibilities, through the data-compressing mapping step, and all the way to band powers in a self-consistent way while minimizing the loss of cosmological information and maintaining a full understanding of the error properties of our measurements.

\section[Precision Mapmaking in Practice: Methods, Trade-Offs, and Results]{Precision Mapmaking in Practice: Methods, \\Trade-Offs, and Results} \label{sec:method}

The theoretically optimal mapmaking method outlined in Section \ref{sec:Mapmaking} poses immense computational challenges. To make it useful for real-world application, we need to find and assess ways of simplifying it while maintaining its precision and statistical rigor.

Because this work serves in large part to generalize the work of \cite{DillonFast}, it is essential to continue to assess that the proposed algorithms are computationally feasible, despite the large size of these data sets and the potentially cost-prohibitive matrix operations involved. That work showed that as long as $\C$ could be decently preconditioned and then multiplied by a vector quickly, we could estimate the power spectrum in a way that scaled favorably with the data volume---between $\BigO(N\log N)$ and $\BigO(N^{5/3})$, where $N$ is the number of voxels in a data volume. This was accomplished using various numerical tricks, taking advantage of translational invariance, the fast Fourier transform, various symmetries, and the flat-sky approximation.

Without any approximations, the vectors and matrices we introduced in Section \ref{sec:Mapmaking} are very big. $\PSF$, for example relates the whole true sky to the whole dirty map---for every frequency, it has as many entries as the number of pixels squared. The time-ordered data vector is very big too---it has entries for every baseline, at every frequency, for every integration. That means that $\A$ is enormous, since it maps from $\x$ to $\y$. We quantify exactly the exact scale of the problem of data volume and computational difficulty in Section \ref{sec:challenges}, but it is clear that calculating every vector and matrix quantity we have enumerated in Section \ref{sec:Mapmaking} is not feasible.

When making maps, there are at least six ways to make $\widehat{\x}$ and $\PSF$ smaller or easier to calculate or use. Three have to do with the geometry of $\widehat{\x}$; three have to do with approximate methods of calculating $\widehat{\x}$ or $\PSF$:
\begin{enumerate}
\item We can make faceted maps of only very small parts of the sky at a time.
\item We can pixelize the sky more coarsely.
\item We can average together neighboring frequencies, lowering the frequency resolution.
\item We can average together neighboring timesteps before computing $\PSF$.
\item We can make $\PSF$ smaller by taking advantage of the finite sizes of the primary and the synthesized beams.
\item We can make $\PSF$ sparser by approximately fitting it in some basis.
\end{enumerate}
Roughly speaking, the first three approaches affect the kind of maps we want to make and the information content in them. The last three affect the quality of the maps we make or the fidelity with which an approximate version of $\PSF$ represents the relationship between $\widehat{\x}$ and $\x$. The exact properties of the desired maps depends upon the power spectrum estimation technique used. For example, if we want to measure high $k_\perp$ modes, we need high angular resolution and therefore a lot of pixels.

In this work, we take a specific case of the first three---choices motivated by the particular array we assess and the desire not to lose much cosmological information. We then evaluate quantitatively the trade-offs inherent in approaches that affect the quality of $\widehat{\x}$ and any approximation to $\PSF$. We begin by specifying both the array (Section \ref{sec:HERA}) and the sky model (Section \ref{sec:skymodel}) that we use for the case study we present. In that context, we can quantify the computational challenges involved in mapmaking in Section \ref{sec:challenges}. 

From there, we examine the three ways of making the mapmaking problem easier for a given kind of map. In Section \ref{sec:faceting} we look at truncating $\PSF$ and how that affects our understanding of the relationship between the dirty map and the true sky. In Section \ref{sec:snapshots} we look at the optimal way to perform time averaging and the trade-offs involved. Then we look at finding a sparse approximation to $\PSF$ in Section \ref{sec:PSFfitting}, which is important because multiplication by all three parts of $\C$ also requires multiplication by $\PSF$. We discuss a way of accomplishing that in the spirit of \cite{DillonFast}.\footnote{The question of preconditioning for rapid conjugate gradient convergence, which was addressed in \cite{DillonFast} in the context of estimators based on $\x$ rather than $\widehat{\x}$, is left for future work. That question cannot be answered until the exact form of the $\widehat{\x}$ is chosen. We may choose estimators with a tapering function, such as those suggested by \cite{EoRWindow1} and \cite{EoRWindow2}. We may also choose to project out certain modes from the dirty map, as we discuss in Appendix \ref{app:projection}.} All of these speed-ups require small approximations and we  assess the effect of those approximations quantitatively. Finally, in Section \ref{sec:computationalSummary} we summarize those results and what we can confidently say so far about the accuracy requirements for approximating $\widehat{\x}$ and $\PSF$ for the purposes of 21\,cm power spectrum estimation.
 
\subsection{HERA: A Mapmaking Case Study} \label{sec:HERA}

\begin{figure}[] 
	\centering 
	\includegraphics[width=.6\textwidth]{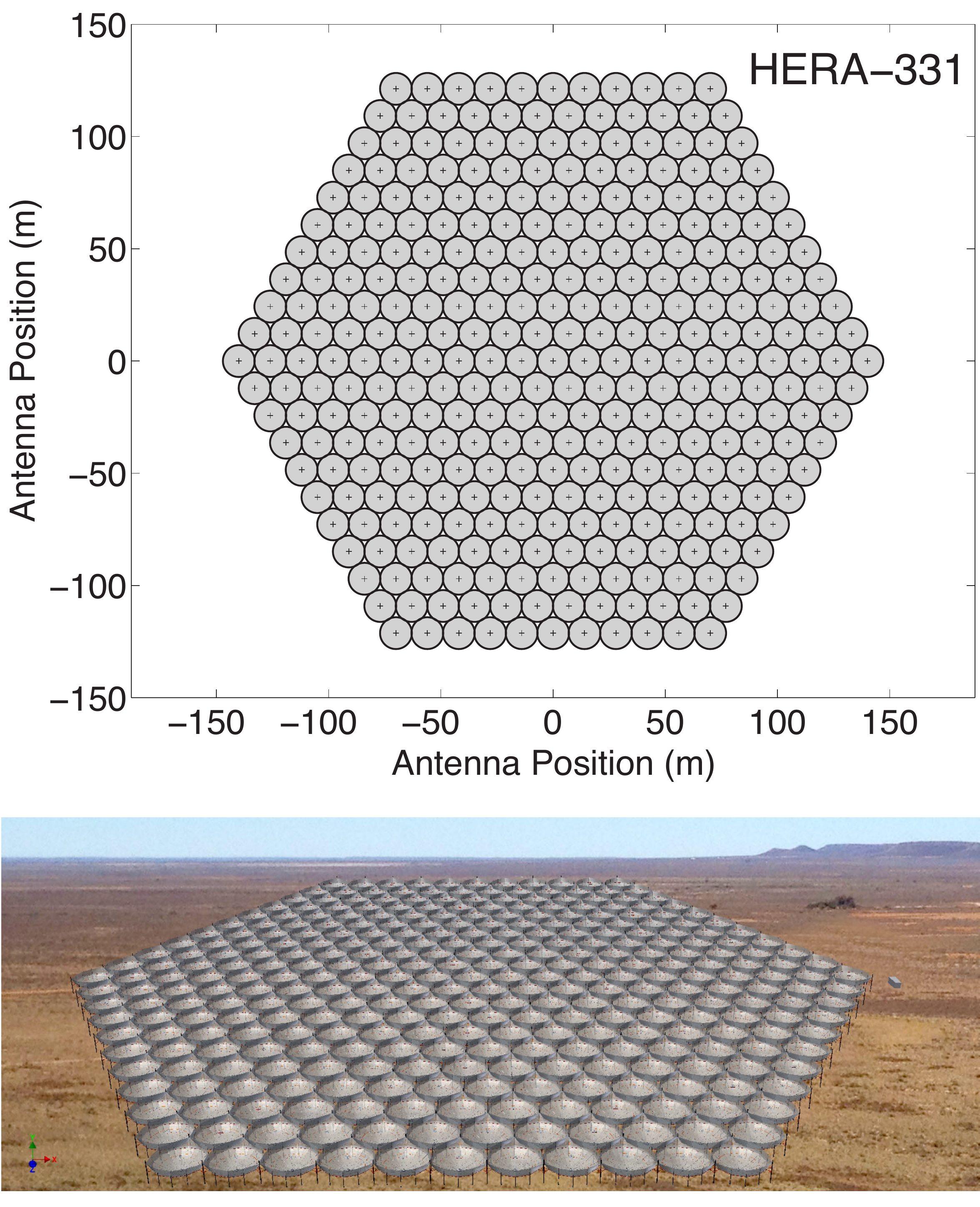}
	\caption[Configuration of the HERA array.]{We test our method on simulated visibilities from the planned Hydrogen Epoch of Reionization Array (HERA). The array, seeen schematically in the top panel, consists of 331 14\,m parabolic dishes, arranged in a close-packed hexagonal configuration. In the bottom panel, we show a rendering of the final array, which will feature more than $0.05\text{\,km}^2$ of collecting area (a standard shipping container, on the right side of the image, is shown for comparison.)}
	\label{fig:HERA-331}
\end{figure}

To test our mapmaking method and our techniques for speeding it up, we need to simulate the visibilities that a real instrument would see. We choose the planned design of the recently commenced Hydrogen Epoch of Reionization Array (HERA) as a particularly timely and relevant case study. HERA will have 331 parabolic dishes, each 14\,m in diameter. They will be fixed to point at the zenith with crossed dipole antennas suspended at prime focus. They will be arranged into a maximally dense hexagonal packing (see Figure \ref{fig:HERA-331}), both to maximize sensitivity to cosmological modes \cite{AaronSensitivity,PoberNextGen} and for ease and precision of calibration \cite{redundant,MITEoR_IEEE,MITEoR}.\footnote{Plans for HERA also include outrigger antennas at much greater distances from the hexagonal core to enable low signal-to-noise, high angular resolution imaging. Though they will be useful for making high-resolution maps and modeling astrophysical foregrounds, they do not add significantly to the cosmological sensitivity of the instrument. Since we are focused on maps as a data-compression step between visibilities and power spectra, we ignore them in this analysis.} In this work, our calculations assume perfect calibration of the instrument and (unless otherwise stated) perfect antenna placement.

HERA also has two advantages that make our algorithms easier to carry out on a relatively small number of computers. First, although it has 331 elements, it only has 630 unique baselines. That is because a highly-redundant array with $N$ baselines has $\BigO(N)$ unique baselines, as opposed to minimally redundant arrays, which have $\BigO(N^2)$ baselines. That is why the MWA has an order of magnitude more baselines than HERA, even though it has only 128 elements. Second, it has a relatively small primary beam, in contrast to both MWA and PAPER. In this work, we model it fairly accurately as a Gaussian beam with a full width at half maximum of $10^\circ$ at 150\,MHz. It should be noted that the method described in this work is independent of the interferometric design. HERA happens to be both a particularly convenient and relevant example.

\subsection{Testing Mapmaking with a Specific Sky Model} \label{sec:skymodel}

As we find ways to compute mapmaking statistics quickly and accurately, we need to answer a key question: do we understand the relationship between our dirty map $\widehat{\x}$ and the input sky model from which we simulated visibilities? It is not important how much our dirty maps look like the sky itself. We just want to make sure that we keep track of everything the instrument and our mapmaking algorithm has done to the data so we can take it into account properly when start estimating power spectra.

We therefore need an input sky model for two reasons. First, we need to be able to use Equation \eqref{eq:AnalyticVisibility} to compute visibilities and thus $\widehat{\x}$. Next, we also want to compute the matrix of point spread functions $\PSF$ corresponding to the same set of observations and multiply it by our true sky model $\x$. The error metric we use therefore is
\beq
\varepsilon = \frac{\left| \widehat{\x}_\text{exact} - \widehat{\x}_\text{approx} \right|}{\left|\widehat{\x}_\text{exact} \right|}. \label{eq:errorMetric}
\eeq 
To be clear, this does not measure the difference between our dirty map and the true sky. It is merely a measure of the discrepancy between what the instrument and our mapmaking routine did to the sky in order to form the dirty map ($\widehat{\x}_\text{exact}$) and what we think we know about those effects ($\widehat{\x}_\text{approx}$) when we write down $\mean$ and $\C$.
  
One advantage to this metric is that it is often relatively easy to calculate $\widehat{\x}_\text{exact}$, at least up to $\D$ which we can factor out of the numerator of Equation \eqref{eq:errorMetric}, compared to calculating $\PSF$. That is because calculating $\A^\dagger \N^{-1} \y$ is as computationally difficult as calculating a single row of $\PSF$. In the following sections, we will be examining ways of computing $\PSF$ faster. Sometimes (e.g. in Sections \ref{sec:faceting} and \ref{sec:PSFfitting}) that means an approximate $\PSF$ but an exact $\widehat{\x}$, in which case $\widehat{\x}_\text{approx} = \PSF_\text{approx} \x$. Other times (e.g. in Section \ref{sec:snapshots}) that means a method for computing $\widehat{\x}$ that also makes $\PSF$ easier to compute. In that case, Equation \eqref{eq:errorMetric} compares the approximate method for computing $\widehat{\x}$ with the exact one. 

We have chosen a sky model with two components: 1) bright point sources and 2) diffuse emission from our Galaxy and other dim, confusion-limited galaxies. Since each frequency is measured and analyzed independently (meaning that $\A$ is sparse and can be written compactly in blocks), we will perform all the simulations at a representative  frequency of 150\,MHz. While the simulations properly weight visibilities based on how many times each unique baseline was measured, we do not include any noise in our calculation of the quantities in Equation \eqref{eq:errorMetric}. We also assume that all baselines at a given frequency have the same noise properties, though that assumption can be straightforwardly relaxed.

\subsubsection{Point Sources}   

\begin{figure}[] 
	\centering 
	\includegraphics[width=.6\textwidth]{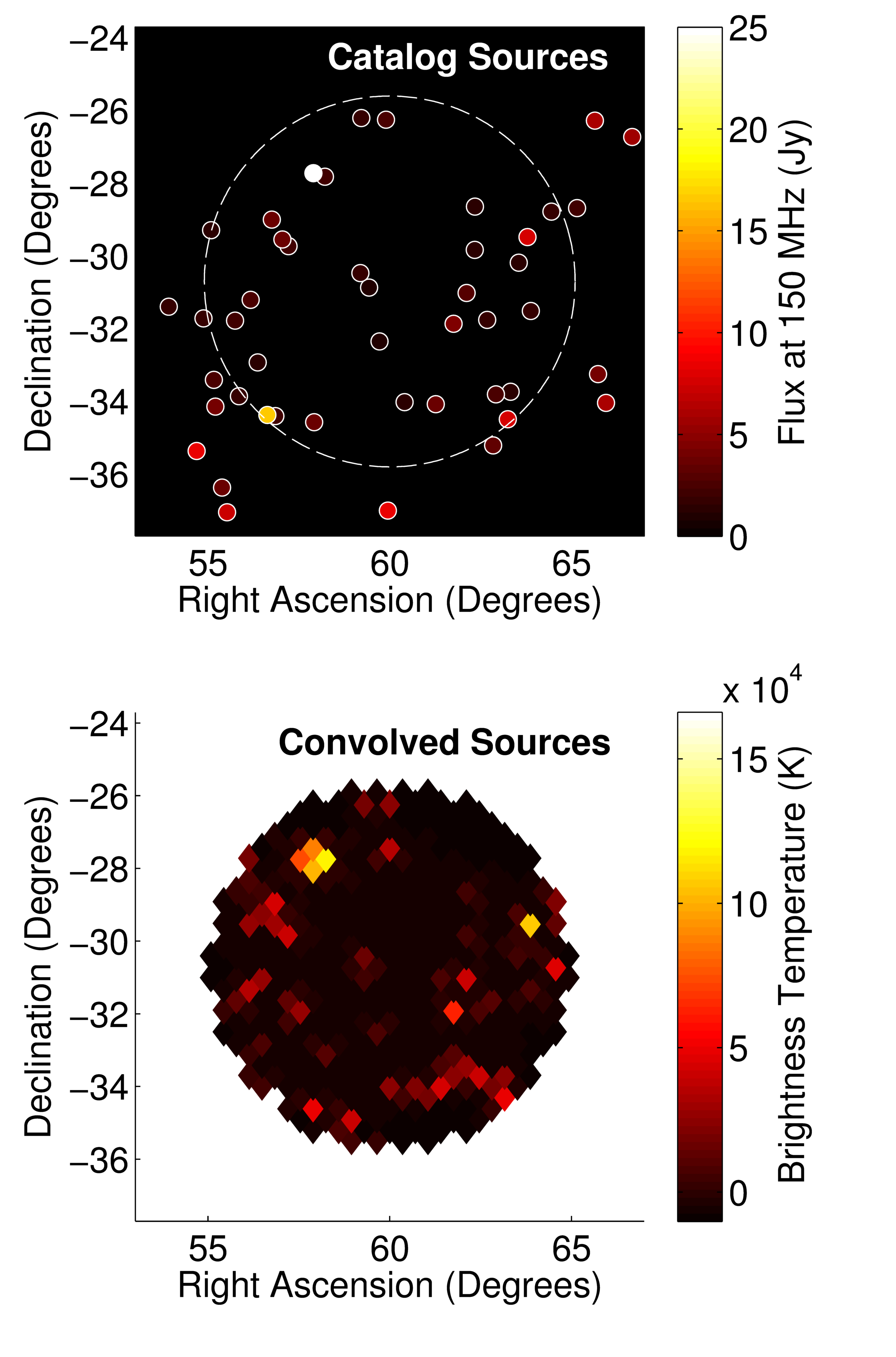}
	\caption[Point source component of the sky model.]{To test our mapmaking method and our approximate techniques for making it much faster, we need a fiducial sky model. One component of that model is bright point sources, which are taken from the MWA Commisioning Survey Catalog \cite{MWACS}. In the top panel, we show the spatial distribution and intrinsic flux of all point sources whose primary-beam-weighted fluxes are above 1\,Jy. In the bottom panel, we show $\widehat{\x}=\PSF\x$, the PSF-convolved and discretized dirty map with HEALPix $N_\text{side}=128$. Since the point spread functions are computed at the locations corresponding to each point source, the bottom panel is exact.}
	\label{fig:PointSources}
\end{figure}

Our sky model includes bright point sources above 1\,Jy with specified positions, fluxes, and spectral indices. These are taken from the MWA Commissioning Survey Catalog \cite{MWACS}, which is complete to below 1\,Jy for a large fraction of the sky. The included spectral indices are used to extrapolate their fluxes at 150\,MHz down from the survey frequency of 180\,MHz. For the calculation of visibilities using Equation \eqref{eq:AnalyticVisibility}, they are treated as true point sources with Dirac delta function spatial extent. In Figure \ref{fig:PointSources}, we show a representative sample of those point sources and what they look like in the dirty map, $\widehat{\x}$.

The sky model for point sources is completely independent of our pixelization. Since we know the location of all the point sources, we can think of $\x$ as having a discretized component covering the whole sky in pixels---which we will use for analyzing diffuse emission---and a set of Dirac delta function fluxes at the positions of the point sources. The sky model for point sources is completely independent of our pixelization. This is completely compatible with the definition of our pixelization in Section \ref{sec:interferometry}, it is just that some pixels have finite area and some have infinitesimal area. It is the pixels with finite volume that we care about for 21\,cm power spectrum estimation, but the infinitesimal ``pixels'' matter for foreground subtraction. Likewise, $\PSF$ has two blocks: one that maps pixels on the true sky to pixels on the dirty map and one that maps points on the true sky to pixels on the dirty map.

\subsubsection{Diffuse Emission}

In the case of point sources, we might hope to use precise locations on the sky to refine our models of $\mean$ and $\C$ and do a better job of separating foregrounds from the 21\,cm signal. That is simply not possible with diffuse synchrotron emission from our Galaxy or with the confusion-limited emission from  relatively dim radio galaxies. Fundamentally, our best guess at that emission and its statistics will have to be discretized and pixelized. Uncertainty about how many confusion-limited point sources appear in a single pixel introduces shot noise, which can be modeled \cite{LT11,DillonFast}. 

In this work, we are interested in errors caused by assumptions and approximations in our mapmaking routine whose effects are not taken into account when estimating power spectra. In order to write down a vector $\x$ that we can use to compute $\widehat{\x}$ and thus $\varepsilon$ with Equation \eqref{eq:errorMetric}, we can either treat the emission as constant in the pixel or we can treat the emission as a ``point source'' at the center of each pixel. For computational simplicity, we choose the latter. With relatively small pixels, there is no practical difference between the two. Since we are concerned about translating our models for foreground residuals in the true sky into models in the dirty map, the pixelization here is not an approximation so much as a consequence of the discretized models for foreground residuals we need for power spectrum estimation. It is possible to construct $\PSF$ to have different angular resolutions of $\x$ and $\widehat{\x}$, if one would like to incorporate a high-resolution diffuse foreground covariance model. The more information we can incorporate about the foregrounds, the smaller our uncertainties get and the better foreground subtraction works.

\begin{figure}[] 
	\centering 
	\includegraphics[width=.6\textwidth]{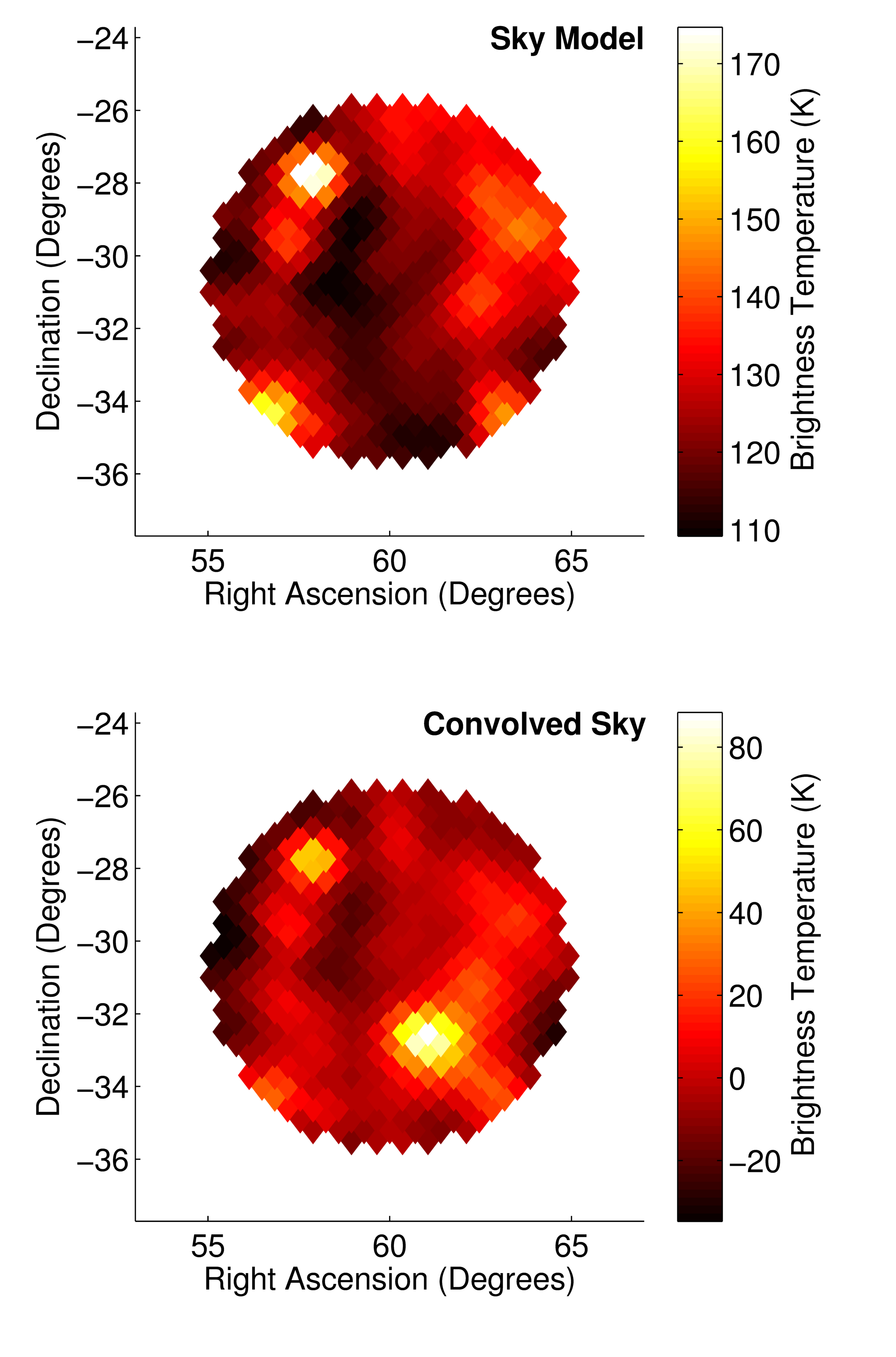}
	\caption[Diffuse component of the sky model.]{The sky model we use to evaluate our mapmaking algorithm and the accuracy of the approximations we make also includes diffuse emission from our Galaxy and faint radio galaxies. For our model of diffuse emission, we use the Global Sky Model of \cite{GSM}. In the top panel, we show a small part of our model for the true diffuse emission. Since we are not trying to model fine spatial information or the precise locations of point sources with our diffuse models, we pixelize the emission identically to the pixelization of our dirty map. In the bottom panel, we show that dirty map. It looks fairly different from the true sky, largely because of the appearance of a side lobe from a bright object outside the field. This occurs because the $\PSF$ maps a very large region of the sky to a small one shown here. The effects of faceting and side lobes will be explored further in Section \ref{sec:faceting}.}
	\label{fig:GSM}
\end{figure}

We use the popular HEALPix software package \cite{HEALPIX} for discretizing the celestial sphere into regularly spaced, equal-area pixels. As a model for the emission itself, we use the Global Sky Model of \citet{GSM} (see Figure \ref{fig:GSM}). The precise model we choose for this work matters only insofar as it is relatively realistic and representative of the true sky. That said, building good foreground models is an important ongoing endeavor relevant to power spectrum estimation and foreground subtraction \cite{ChrisMWA,PoberWedge,JacobsFluxScale,InitialLOFAR1,InitialLOFAR2}.
 
\subsection{Computational Challenges of Mapmaking} \label{sec:challenges}

We already alluded to the fact that we need to investigate various simplifications and approximations to make the calculation of $\widehat{\x}$ and $\PSF$ tractable. Let us take the time to see exactly where the problem lies.

Consider the matrix $\A$ where $\y = \A \x + \n$. $\A$ maps a discretized sky into time-ordered data. If we want to slightly over-resolve the sky with HERA, we might choose a HEALPix map with $N_\text{side} = 256$, which gives an angular resolution of about $0.2^\circ$. That is almost $10^6$ pixels at each of about 1000 different frequencies (assuming 100\,kHz resolution and 100\,MHz of simultaneous bandwidth). If we measure all our visibilities every two seconds for 1000 total hours at all 1000 frequencies, that is $10^{14}$ visibilities, so naively, $\A$ is a $10^{14} \times 10^9$ matrix. That is a problem.

Of course, there are many standard simplifications. Each frequency is treated completely independently during mapmaking, so we can treat $\A$ as either block diagonal or as a family of 1000 much smaller matrices, $\A(f)$. Redundant baselines measure the same sky, so their visibilities can be combined together, reducing both instrumental noise and the number of visibilities by a factor of almost 100 in the case of HERA. Getting 1000 hours of nighttime observation takes about 100 days, so we can LST-bin, reducing both noise variance and data volume by another two orders of magnitude. Since each time-step is independent of all others, we can further break $\A$ into about 10,000 pieces for each integration.  

We still have $10^7$ different $\A$ matrices, each $10^3 \times 10^6$. This size is challenging but acceptable for either simulating visibilities or calculating $\A^\dagger \N^{-1} \y$. However, it is simply too big for the calculation of $\PSF$, which would require the computationally infeasible task of multiplying together two matrices of this size $10^7$ times, each multiplication taking roughly $10^{15}$ operations. In the following sections, we will look at ways of reducing the number of $\A(f)$ matrices and making each $\A(f)$ smaller, especially during the calculation of $\PSF$.

\subsection{Faceting and First Mapmaking Results} \label{sec:faceting}

The matrix of point spread functions $\PSF$ is defined by the relation $\langle\widehat{\x} \rangle = \PSF \x$. It can be thought of as a transformation from one pixelized real space---that of the true sky---to another---that of the dirty map. For even a modest angular resolution, that is an enormous matrix. Do we really need to know the relationship between every pixel in the sky and every pixel in the dirty map? 

\subsubsection{Why We Facet} 

\begin{figure}[]  
	\centering 
	\includegraphics[width=.6\textwidth]{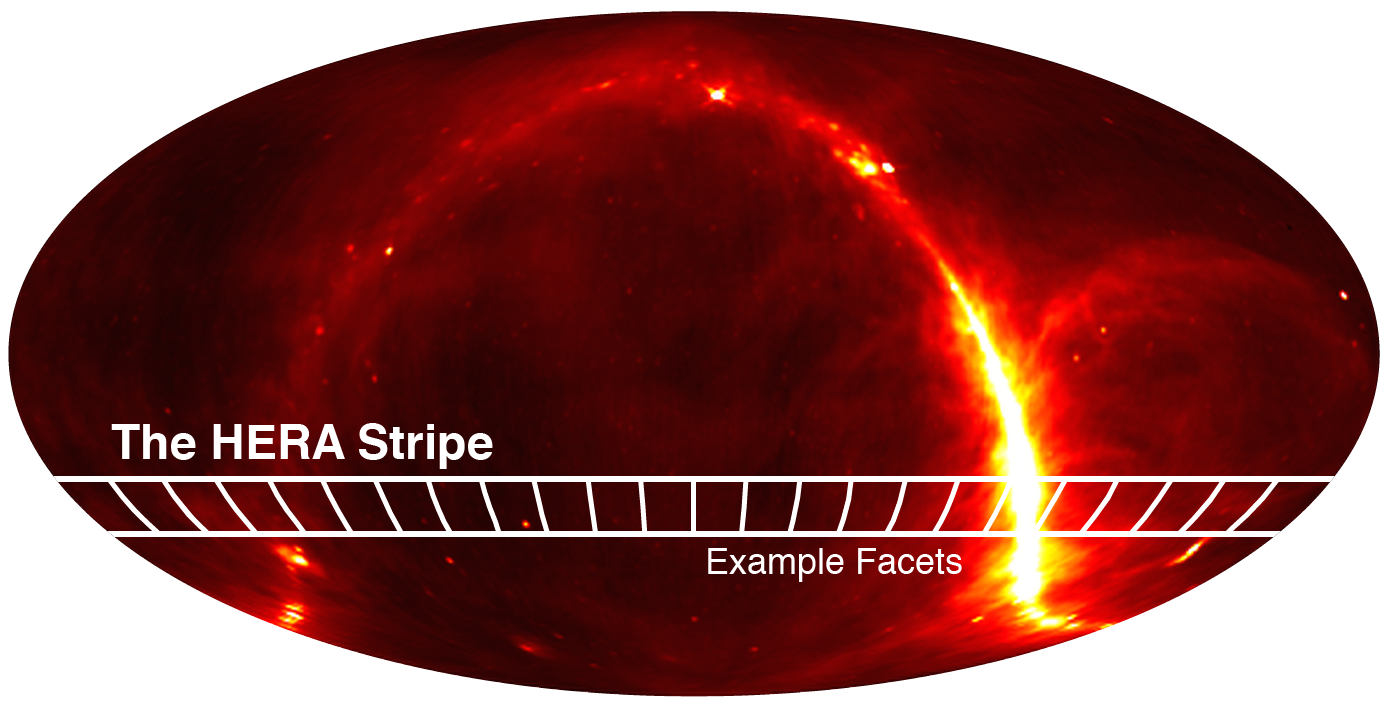}
	\caption[The division of the HERA stripe into facets.]{The faceted approach we use to speed up optimal mapmaking and power spectrum estimation will be especially useful for HERA because it is limited to only observe an approximately $10^\circ$ stripe of constant declination, centered on the array's latitude of approximately $-30.7^\circ$. It is fairly natural to split up the observation into roughly $10^\circ\times10^\circ$ facets, each analyzed separately. This makes $\PSF$ much easier to compute and lets us use the flat-sky approxmation, a requirement for implementing the power spectrum methods of \citet{DillonFast}. Very little cosmological information is lost in this process; only the longest spatial modes are thrown out and they should be dominated by galactic emission.}
	\label{fig:HERA_Stripe}
\end{figure}

Breaking up the field of view into a number of smaller facets is a standard technique in radio astronomy, especially when one wants to minimize the effects of noncoplanar baselines \cite{CornwellWProj}. For purposes of 21\,cm cosmology, there are two good reasons to consider relatively small regions of the sky one at a time. The first is HERA's observing strategy. Because it statically points at the zenith, HERA scans a fixed stripe in declination about $10^\circ$ degrees wide.  It seems reasonable that we can analyze parts of the stripe independently, making maps and computing power spectra for each small facet. In Figure \ref{fig:HERA_Stripe}, we show an example of what that faceting might look like.

The only significant disadvantage to faceting is that we lose the ability to measure modes in the power spectrum with wavelengths perpendicular to the line of sight that are larger than the facet. Doing so properly and with precisely quantified error properties would require calculating covariance between facets, which is effectively the same as not faceting at all. This is not such a great hardship. Due to the survey geometry, only the long modes oriented along the HERA stripe could have been measured at all. They are longer than the shortest baseline, meaning that they can only be sampled after considerable sky rotation.  The same $|\mathbf{k}|$ modes can be also be accessed along the line of sight, except those at very low spectral wave-numbers, which are bound to be foreground dominated.

The other major upside to faceting is that, if we want to use the fast power spectrum techniques developed in \cite{DillonFast}, we need to take our maps and chop them up into facets anyway. That is because any fast algorithm that takes advantage of the fast Fourier transform (e.g. that in \cite{DillonFast}) and translational invariance relies on rectilinear data cubes, which is only an accurate approximation for small fields where the flat-sky approximation holds. Happily, that rough size is also about $10^\circ$. For other instruments, the choice of facet size is less obvious and depends on the computational demands of both mapmaking and power spectrum estimation. Bigger facets preserve more information, but they can be more computationally expensive than they are cosmologically useful. The exact right choice for other interferometers is a matter for future work.

\subsubsection{Faceted Mapmaking Method And Results}

So, instead of using $\D\A^\dagger \N^{-1} \y$ to calculate $\widehat{\x}$, we instead redefine $\widehat{\x}$ using
\beq
\widehat{\x} = \D \K_\text{facet}\A^\dagger \N^{-1} \y,
\eeq
where $\K_\text{facet}$ maps the full sky to a small portion of the sky, thus making $\PSF$ asymmetric. Doing this for every facet basically amounts to only mapping the parts of the sky that are ever near the center of the primary beam. This provides a computational simplification by a factor of $4\pi / (\Omega_\text{facet} N_\text{facets})$, which for HERA is about an or order of magnitude.  An instrument that can see the whole sky would see no computational benefit just from breaking the sky in facets.

The real computationally limiting step is the calculation of $\PSF$. Since we are only interested in the dirty map of a facet, we care only about source flux that could have contributed to that dirty map. That means that we can truncate each point spread function some distance from the facet center. Flux outside that truncation radius is assumed not to contribute significantly. In other words, 
\beq
\PSF = \D \K_\text{facet} \A^\dagger \N^{-1} \A \K_\text{PSF}^\trans
\eeq
where $\K_\text{PSF}$ is the same as $\K_\text{facet}$ except that it cuts off at some larger radius than the facet size. We get to choose exactly what radius we want to assume that no outside flux contributes to the facet. This is a completely tunable approximation and it becomes exact in the limit that that radius encompasses the whole sky. 

Therefore, instead of mapping the whole sky to the whole sky, the matrix of point spread functions now maps some moderate portion of the sky to a somewhat smaller part of the sky. Since $\N$ is diagonal, both the time it takes to calculate $\PSF$ and the memory it takes to store it are reduced by very large factor. If the truncation region is 4 times the $10^\circ$ facet size, for example, then that savings is a factor of about $10^4$.

This new definition of $\widehat{\x}$ means that $\D$ is now a much, much smaller matrix---it has only as many elements as there are pixels in the facet. And since we are only interested in the correlation between pixels in the map, the noise covariance is now
\beq
\C^N = \PSF \K_\text{facet}^\trans \D^\trans,
\eeq
which is much smaller and still quite simple.

\begin{figure*}[] 
	\centering 
	\includegraphics[width=1\textwidth]{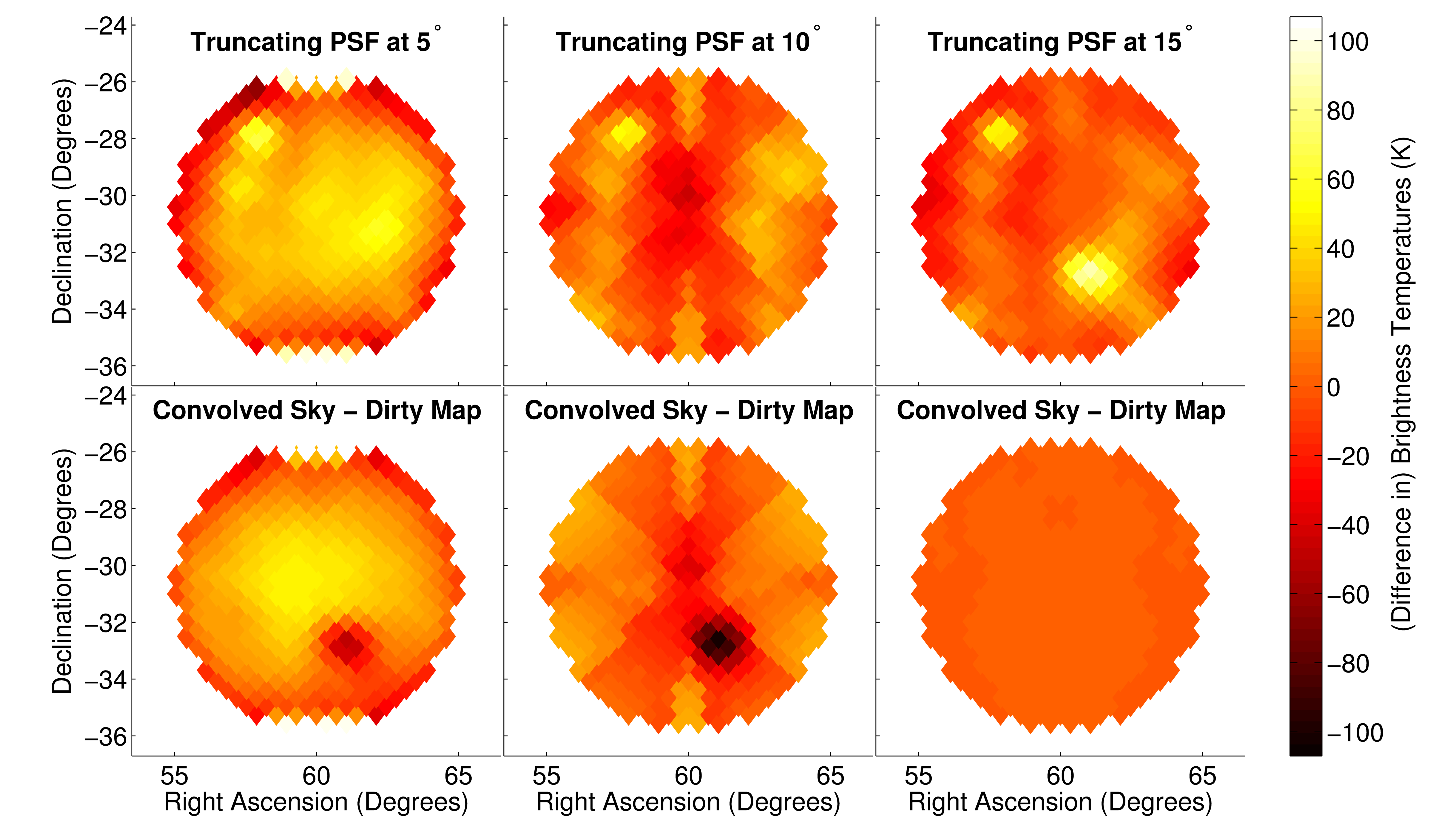}
	\caption[Illustration of the PSF truncation radius approximation.]{In order to accurately reproduce dirty maps, we must include in our $\PSF$ matrix the effect flux from outside the facet that appears in the side lobes of off-facet sources. Here we demonstrate that effect by looking at how the approximate PSF-convolved sky, $\PSF_\text{approx}\x$, evolves as we expand the distance from the center of the facet at which the point spread function is approximated to not contribute. In the top row, we plot $\PSF_\text{approx}\x$ while on the bottom row we plot $\PSF_\text{approx}\x - \widehat{\x}_\text{exact}$. ($\PSF_\text{exact}\x = \widehat{\x}_\text{exact}$ is shown in the bottom panel of Figure \ref{fig:GSM}.) Since the visibilites that go into computing $\widehat{\x}$ derive from a full-sky calculation, side lobes are automatically included. The bright spot we see on the top right panel, which appears as a dark spot on the bottom left and bottom middle panels, is a prominent side lobe from a very bright source outside the facet, but within $15^\circ$ of the facet center. This explains what we saw in Figure \ref{fig:GSM} and the dramatic improvement in the error we see in the right-hand panels.}
	\label{fig:PSF_Progress}
\end{figure*}

We illustrate the effect of the PSF truncation radius in Figure \ref{fig:PSF_Progress}, showing the large impact that increasing the truncation radius has on our calculations of $\widehat{\x}_\text{approx} = \PSF_\text{approx} \x$ and therefore of $\varepsilon$. We find that once the PSF includes both the central peak of the synthesized beam and the first major side lobes, the convergence of $\widehat{\x}_\text{approx}$ to $\widehat{\x}_\text{exact}$ is very quick.

\begin{figure}[] 
	\centering 
	\includegraphics[width=.6\textwidth]{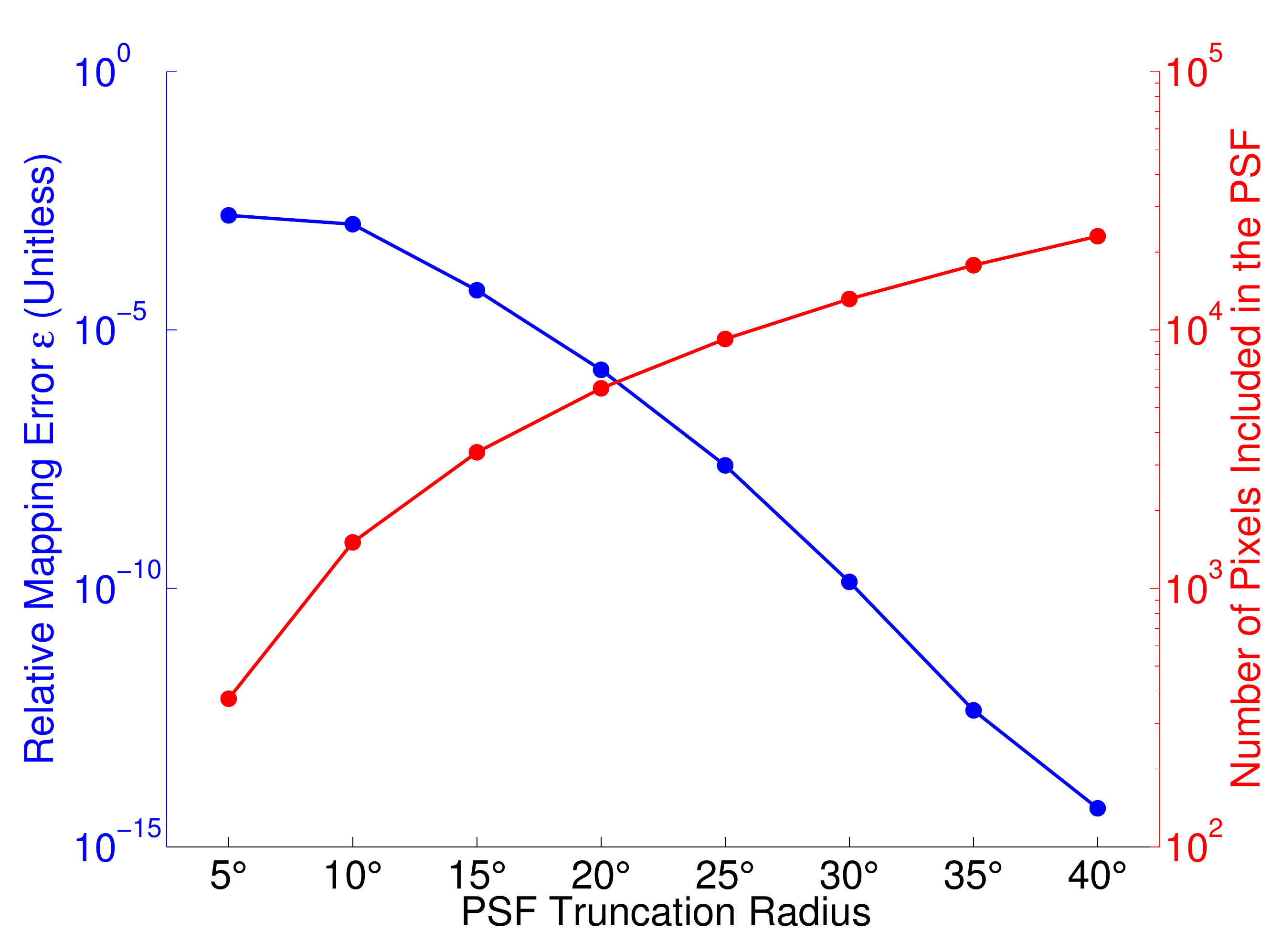}
	\caption[Error vs. complexity trade-off for PSF truncation.]{The error introduced by the approximation that the PSF can be truncated past a certain distance from the facet center gets very small very quickly. Here we show how both that error, which we define in Equation \eqref{eq:errorMetric}, and the number of pixels in each point spread function depend on the truncation radius. The number of pixels, and thus the computational difficulty of computing the matrix of point spread functions, $\PSF$, scales as the truncation radius squared---there are simply more pixel values to calculate. In general, the approximation works because the point spread functions are relatively compact. HERA's design is especially helpful here with its dense grid of baselines and its relatively small primary beam. Other arrays may need larger truncation radii to acheive the same accuracy.}
	\label{fig:PSF_Size_Test}
\end{figure}

We further tested the expected convergence of the algorithm for a fixed facet size and variable $\K_\text{PSF}$ using the sky model from Section \ref{sec:skymodel}. Our results, which we show in Figure \ref{fig:PSF_Size_Test}, again demonstrate that the PSF truncation radius does not need to be much larger than the facet, if the facet is comparable in size to the primary beam. The exact level of error introduced by faceting will, in general, depend upon the compactness of both the primary and synthesized beams. The approximation that the point spread function is Gaussian might make the plotted relative error a bit optimistic, though the side lobes in the real HERA primary beam are quite small.

In summary, faceting allows us to decrease the time it takes to calculate the $\PSF$ and the memory required to store it by a factor of $(4\pi)^2 / (\Omega_\text{facet} \Omega_\text{PSF})$, where $\Omega_\text{PSF}$ is the angular size of the region left by $\K_\text{PSF}$. In the case of HERA, that works out to about 10,000 times faster and smaller.

\subsubsection{Mitigating Nonredundancy} \label{sec:nonredundant}

Making maps in facets also has one extra advantage useful in addressing a common complication presented by real-world arrays. If we assume in our analysis that every baseline of a given designed separation actually has that separation, we will be ignoring errors that can be a decent fraction of a wavelength. And though HERA is a zenith-pointed array for which noncoplanar effects are small, they are not zero and can be quite large for other instruments like the MWA. Noncoplanarity creates nonredundancy. 

However, as long as we know precise positions of all of our antennas (which is far easier than making the array perfectly redundant) we can use the fact that we are only mapping a single facet at a time to reduce those phase errors near the center of our map. We can think of each baseline corresponding to some unique baseline $\base$ as
\beq
\base_m = \base + \Delta \base_m,
\eeq
where the residuals are caused by inexact antenna placement. That means that Equation \eqref{eq:VisibilitySum} becomes 
\begin{align}
V(\base_m,\nu_n) \approx \sum_k &\Delta\Omega \frac{2 k_B \nu_n^2}{c^2} x_k(\nu_n) B(\rhat_k,\nu_n) \times \nonumber \\
& \exp\left[-2 \pi i  \frac{\nu_n}{c} \left(\base + \Delta \base_m\right) \cdot \rhat_k \right]. \label{eq:nonredundant}
\end{align}
We need the right-hant side of this equation to be the same for all $\base_m$ corresponding to the unique baseline $\base$, otherwise we lose the redundancy bonus we discussed in Section \ref{sec:challenges}. 

We can achieve this approximately for small $\Delta \base_m$ because our facets are relatively small. Let us define $\Delta \rhat_k \equiv \rhat_k - \rhat_0$ where $\rhat_0$ points to the center of the facet and $\Delta\rhat_k$ is generally not a unit vector. We can expand the exponent of Equation \eqref{eq:nonredundant} as
\begin{align}
&\left(\base + \Delta \base_m\right)\cdot \left( \rhat_0 + \Delta\rhat_k \right)  \nonumber \\
= & \mbox{ } \base\cdot \rhat_0 + \base\cdot\Delta\rhat_k + \Delta\base_m\cdot\rhat_0 + \Delta\base_m \cdot \Delta\rhat_k.
\end{align}
The first two terms in the expansion are $\base\cdot\rhat_k$ and normally appear in $\A$. The last term, which second order in this expansion, is approximated to be zero. Even if $\base \cdot \rhat_0$ is small, the last term is in general much smaller than the second term. We can, however, correct for the middle term by multiplying both sides of Equation \eqref{eq:nonredundant} by a constant phase factor, since
\beq
V(\base, \nu_n) \approx \exp\left[2 \pi i  \frac{\nu_n}{c} \Delta \base_m \cdot \rhat_0 \right] V(\base_m, \nu_n).
\eeq
As was our goal, the $\PSF$ matrix that results from taking the above equation to be exactly true is the same as if we had not had any antenna placement errors or noncoplanarity. Rephasing lets us mitigate the effect of known errors without having to calculate a vastly more complicated $\PSF$, which treats all baselines completely independently, even if they are supposed to be redundant.

Effectively, our approximate correction cancels out the phase error at the exact center of the facet and thus minimizes its effect throughout the facet. For example, for $10^\circ$ facets at 150\,MHz, a 4\,cm antenna placement error (roughly the level seen in \cite{MITEoR}) leaves only a $0.63^\circ$ phase error in the visibility after rephasing. The error might be a bit worse when calculating the parts of $\PSF$ near the truncation radius. For very large fields, as \cite{CornwellWProj} addressed, this becomes a bigger problem and we may need to break each set of baselines that was supposed to be redundant into a few groups, each closer to exactly redundant, and treat each group separately. The exact effect on the accuracy of the dirty maps from this small correction is left to future work when the exact antenna placement of HERA or a similar array is known. 

\subsection{Grouping Visibilities into Snapshots} \label{sec:snapshots}

Standard interferometric mapmaking techniques accumulate visibilities in the $uv$-plane via sky rotation and thereby combine minutes or even hours of visibilities together \cite{Schwab1984,Cotton2004,Cotton2005,Bhatnagar2008,Carozzi2009,Mitchell2008,bernardi,Bart,Smirnov2011,Li2011,FHD,WSCLEAN,TrottObservingModes}. We would like to find a way of reducing the number of rows in $\A$ for the purpose of calculating $\PSF$ by grouping integrations into ``snapshots'' that are each analyzed as a single timestep when we calculate $\PSF$. How can we average together multiple visibilities over a range of times while approximating the $\PSF$ as having been calculated at only the middle timestep of each snapshot? 

Once again, we can use our freedom to rephase both the visibilities and the $\A$ matrix as we did in Section \ref{sec:nonredundant}. The idea is to try to remove, as much as possible, the effect of sky rotation from the visibilities. Consider again Equation \eqref{eq:AnalyticVisibility}, now with explicit time dependence:
\begin{align}
V(\base,\nu,t) = \int & B\left(\rhat,\nu \right)
I(\rhat,\nu,t) \times \nonumber \\
&\exp \left[-2 \pi i  \frac{\nu}{c} \base \cdot \rhat \right] d\Omega.
\end{align}
While the sky rotates, the primary beam is fixed relative to the ground. 

By contrast, let us consider a new reference frame with angle vector $\rhat'$, which rotates with the sky:
\begin{align}
V(\base,\nu,t) = \int&  B\left(\rhat',\nu,t \right)
I(\rhat',\nu) \times \nonumber \\
&\exp\left[-2 \pi i  \frac{\nu}{c} \base(t) \cdot \rhat' \right] d\Omega'.
\end{align}
Now the beam and the baseline vector have picked up an explicit time dependence while the sky has lost its time dependence. Let us assume that the primary beam is varying very slowly spatially---generally a good assumption since the primary beam is much larger than the spatial scales probed by most baselines. 

Let us think of $V(\base,\nu,t)$ as the visibility measured for the middle integration of a snapshot. A visibility measured a bit later during that snapshot would look like
\begin{align}
V(\base,\nu,t+\Delta t) \approx& \int d\Omega' B\left(\rhat',\nu,t \right)
I(\rhat',\nu) \times \nonumber \\
&\exp\left[-2 \pi i  \frac{\nu}{c} (\base(t)+\Delta \base) \cdot \rhat' \right],
\end{align}
where $\Delta \base$ is the difference between $\base(t+\Delta t)$ and $\base(t)$ in the primed coordinate system. The dot product is basis independent, so 
\beq
(\base(t)+\Delta \base) \cdot \rhat' = \base\cdot \left(\rhat + \Delta \rhat(\rhat) \right),
\eeq
where the right-hand side is back in the frame that is stationary relative to the Earth. $\Delta \rhat(\rhat)$, which is not a unit vector, is the amount of sky rotation between times $t$ and $t+\Delta t$. It is approximately constant across the facet for fairly short snapshots and moderately sized facets, meaning that we can pull it out of the integral. We can therefore undo much of the effect of sky rotation using the approximation that
\beq
V(\base,\nu,t+\Delta t) \approx e^{i \Delta\phi} V(\base,\nu,t)
\eeq
where
\beq
\Delta\phi \equiv -2 \pi  \frac{\nu}{c}  \base \cdot (\rhat_0(t+\Delta t) - \rhat_0(t))  
\eeq
and where again, $\rhat_0(t)$ points to the facet center. 

\begin{figure*}[] 
	\centering 
	\includegraphics[width=1\textwidth]{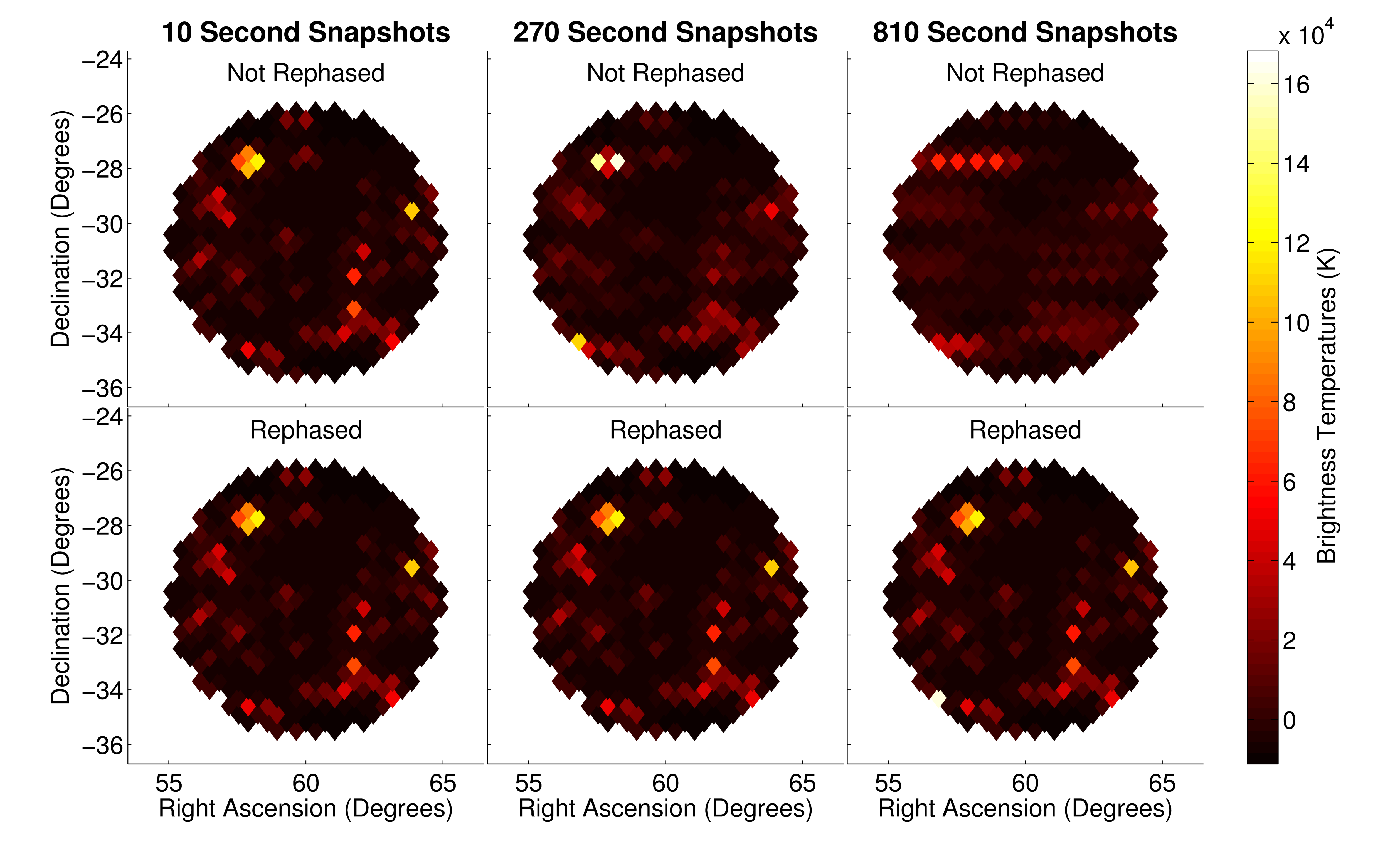}
	\caption[Illustration of the snapshotting approximation.]{One way to make the calculation of the matrix of point spread functions, $\PSF$, faster is to combine many consecutive integrations together into snapshots. When we compute $\PSF$, we effectively assume that all the associated visibilities we have grouped into one snapshot were taken exactly at the snapshot's middle time. Usually, this is a poor approximation. As we can see from the top row, where we have simply added together 10 second integrations to snapshots of increasing length, we are effectively spreading out sources in right ascension as the sky rotates overhead. However, if we use our freedom to rephase visibilities individually, we can dramatically reduce the error associated with forming snapshots. For example, the bottom right panel only exhibits error on the order of a few percent compared to the exact single-integration dirty maps in the left-hand panels.  The result is related to the traditional radio astronomy technique of ``fringe stopping.''}
	\label{fig:Snapshot_Progress}
\end{figure*}

We can therefore add together many visibilities taken at different times and approximately treat them as if there were all taken at the middle integration in the snapshot by rephasing them. This is very similar to the ``fringe-stopping'' technique from traditional radio astronomy, which seeks to counteract the effect of the rotation of the earth at the location of a source \cite{ThompsonMoranSwenson}. As we saw in Section \ref{sec:nonredundant}, the effect of rephasing visibilities cancels out in $\PSF$, since the extra term in $\A$ gets canceled out in $\A^\dagger.$ That is why we only have to perform the calculation of $\PSF$ once per snapshot rather than once per integration. We show in Figure \ref{fig:Snapshot_Progress} a marked improvement, especially in the case of long snapshots, between naively adding together visibilities as if the sky were not rotating overhead and adding together rephased visibilities.

\begin{figure}[] 
	\centering 
	\includegraphics[width=.6\textwidth]{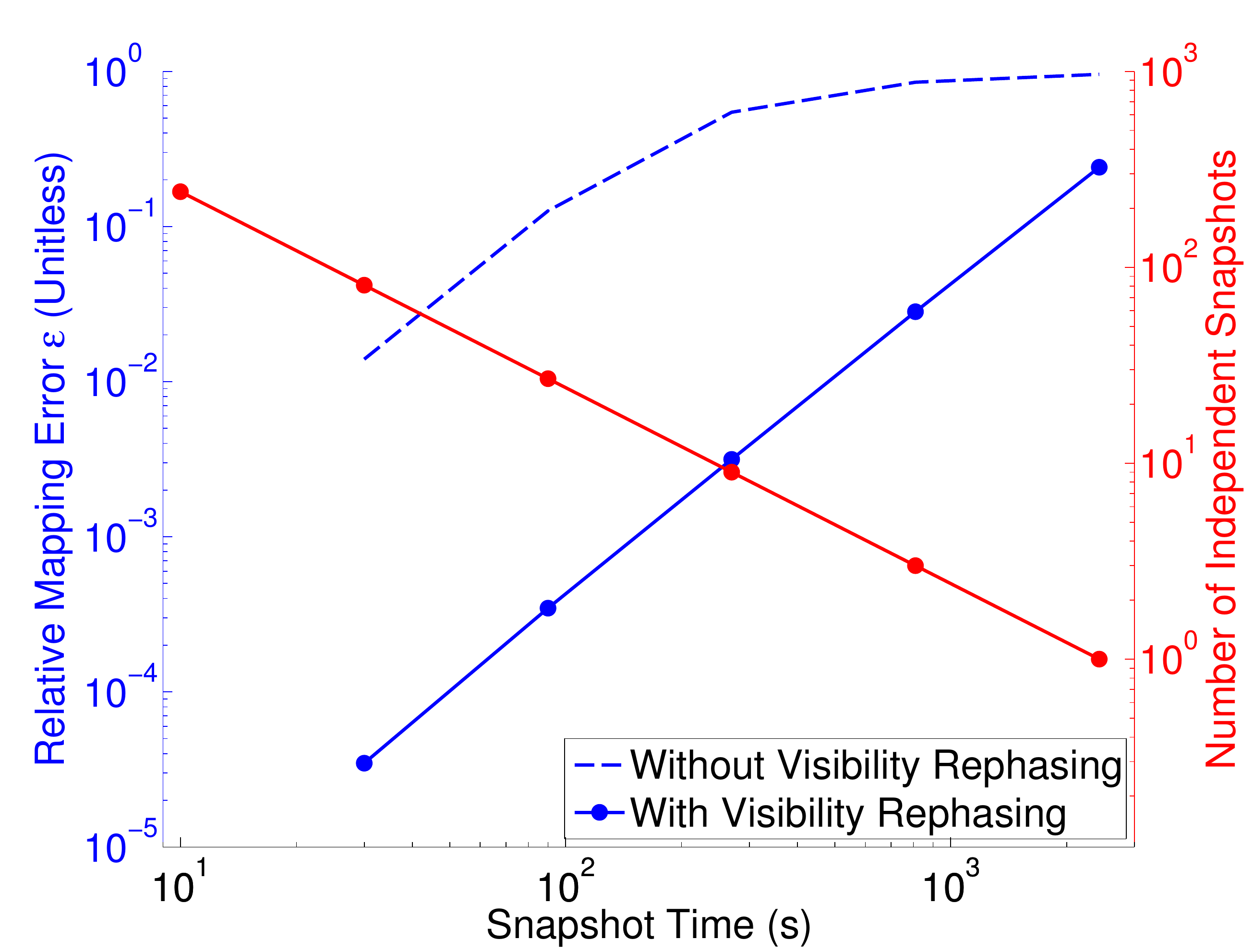}
	\caption[Error vs. complexity trade-off for the snapshotting approximation.]{The error introduced by approximating the observation as having taken place at at only a few discrete times, many seconds or minutes apart, can be mitigated by appropriately rephasing visibilities before combining them. Here we show quantitatively how the length of snapshots---all multiples of the 10 second integration time used in our simulation---introduces small errors. We calculate the relative error $\varepsilon$ between dirty maps calculated with a given integration time and those calculated exactly using only one integration per snapshot. We also show how the computational difficulty of calculating $\PSF$ is affected, since it scales linearly with the number of independent snapshots considered.}
	\label{fig:Snapshot_Time_Test}
\end{figure}

In Figure \ref{fig:Snapshot_Time_Test} we show quantitatively how the error increases as snapshots get longer. Here we care how these approximate dirty maps compare to the exact dirty maps made when each 10\,s integration is treated completely separately. We also found it important to rephase the visibilities to the exact middle of the snapshot, which creates a first-order cancellation that removes some of the error associated with this approximation.

Based on the results we show in Figure \ref{fig:Snapshot_Time_Test}, it is likely that we can cut another one to two orders of magnitude off the total number of operations we need to perform to calculate $\PSF$, making that calculation considerably easier. For a given accuracy goal, it is also possible to make the calculation of $\PSF$ even simpler by forming snapshots with different durations for baselines of different lengths, keeping $\Delta \phi$ small. 

\subsection{PSF Fitting} \label{sec:PSFfitting}

Now that we have found accurate and well-understood approximations that make computing $\PSF$ computationally feasible, we need to worry about multiplying a vector by $\PSF$. This is a necessary step in any power spectrum estimation scheme adapted from \cite{DillonFast}, since $\PSF$ appears in Equations \eqref{eq:CS}, \eqref{eq:CcommaAlpha}, \eqref{eq:CN}, and \eqref{eq:CFG}. In general, the number of operations in this calculation scales with the number of pixels in the facet, the number of pixels in the PSF, and the number of frequency channels, i.e. as $\BigO(N_\text{facet} N_\text{PSF} N_f)$. This is slower than we would like, so we will endeavor to show how it can be sped up.

If the point spread function were constant across the field---if it looked the same in the top and bottom rows of Figure \ref{fig:PSFs}---then the solution would be simple. We could calculate only one PSF and then use it to fill out all of $\PSF$. Then, if we approximate HEALPix pixelization as a regular grid---which is true in the flat-sky approximation---we can write $\PSF$ using Toeplitz matrices. A Toeplitz or ``constant-diagonal'' matrix represents a translationally invariant relationship.\footnote{Toeplitz matrices have a number of nice properties, including the fact that an $N\times N$ Toeplitz matrix can be multiplied by a vector in $\BigO(N\log N)$ operations. This is because the translational invariance lets us use the fast Fourier transform. See \cite{Toeplitz} for a review of these matrices and their properties or \cite{DillonFast} for a previous application to 21\,cm cosmology of the same relevant properties.} A Toeplitz matrix $\T$ has the property that each element only depends on its distance from the diagonal of the matrix, or in other words that
\beq
T_{ii'} = t_{i-i'}. \label{eq:Toeplitz}
\eeq
We can imagine that, if any part of the PSF can be fully represented by its displacement from the facet center, then we can write $\PSF$ for each frequency and facet as a tensor product of two matrices, each describing translational invariance along one of the two principal axes of the HEALPix grid.\footnote{We define the axes by taking the center pixel and computing the linearly independent vector directions towards the nearest two pixels. It is not a problem that these two directions are not orthogonal---the FFT can be performed along nonorthogonal directions, as pointed out by \cite{FFTT2}.} If we index along those axes with $i$ and $j$ in the dirty map and $i'$ and $j'$ in the true sky, then for a single frequency the matrix of point spread functions can be written as
\beq
P_{ii'jj'} = t_{i-i'}s_{j-j'} 
\eeq
or as
\beq
\PSF = \mathbf{T} \otimes \mathbf{S}
\eeq
where $\mathbf{T}$ and $\mathbf{S}$ are Toeplitz matrices.

And yet we can easily see from Figure \ref{fig:PSFs} that point spread functions do not respect translational invariance. In the bottom row where the PSFs are displaced from the center of the facet, the side lobes nearer the edge of the primary beam are downweighted relative to those nearer the center. This is a consequence of optimal mapmaking, which downweights the contribution from regions of the sky that the telescope is less sensitive to. However, we expect that the physical effects that lead to a translationally varying PSF, like the primary beam and the projected array geometry, should change smoothly over the field. So while the PSF is translationally varying, perhaps its translational variation can be modeled with a small number of parameters.

If we calculate $\PSF$, the matrix of point spread functions that maps every pixel in some extended facet to every pixel on the facet of interest, we can model this translational variation by reorganizing $\PSF$. We have chosen our normalization $\D$ so that the specific point spread function mapping the sky onto a given pixel has a value of 1 at the center pixel of its main lobe.  But what about all the pixels displaced exactly pixel northeast from the center of the main lobe in all the PSFs? Or ten pixels? 

\begin{figure*}[] 
	\centering 
	\includegraphics[width=1\textwidth]{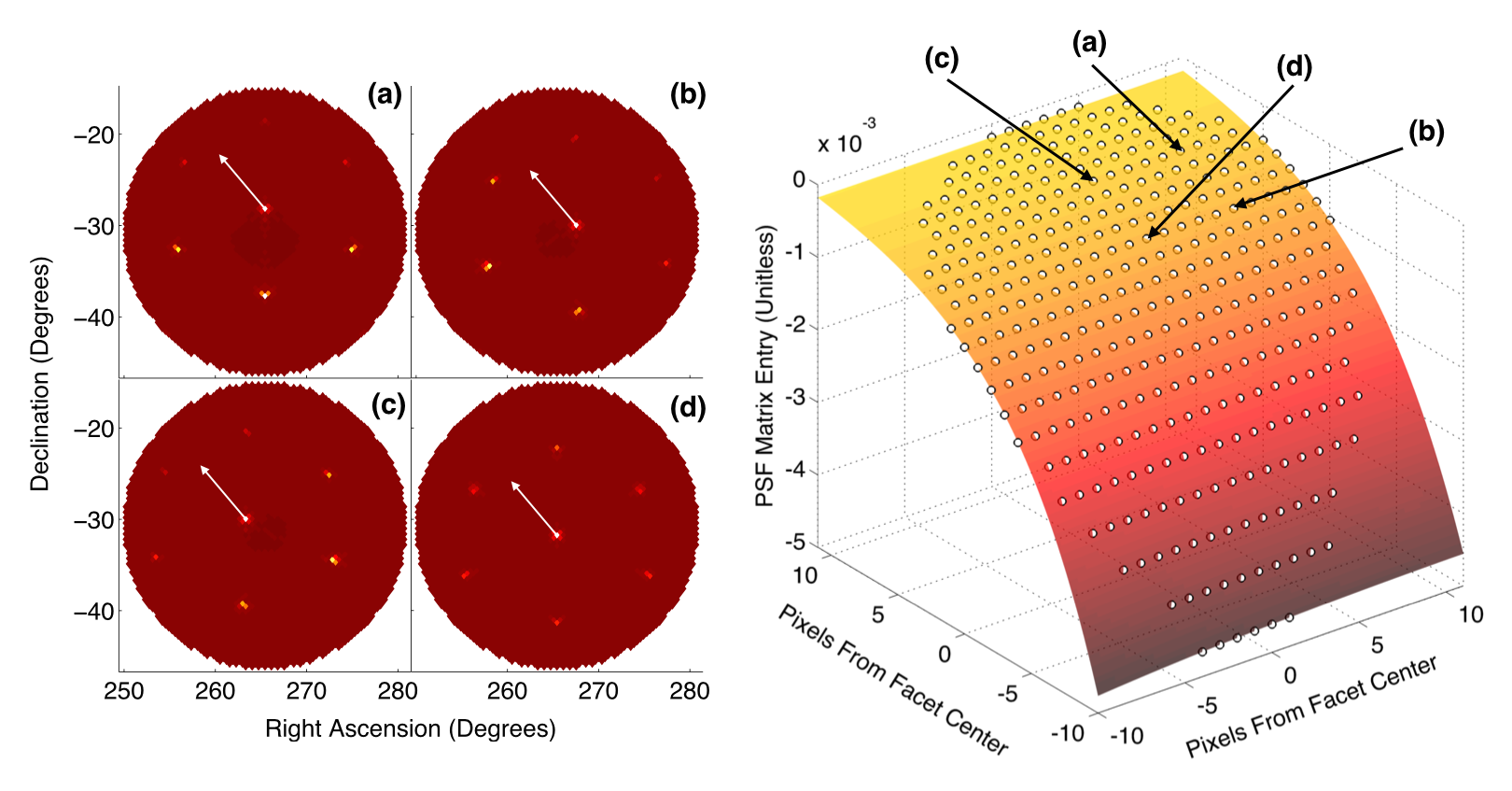}
	\caption[Example of the PSF's smoothly varying position dependence.]{Though our point spread functions are not translationally invariant---a fact we saw clearly in Figure \ref{fig:PSFs}---their translational variation is fairly smooth and can be captured by a relatively low order polynomial. In this figure, we examine a typical example consisting of all the entries in $\PSF$ displaced exactly 15 pixels along one of the two principal axes of the pixelization from the center of the main lobe of the synthesized beam. This displacement is represented by the four identical white arrows on top of the point spread functions in the left-hand panel. All such entries in $\PSF$ (white circles in the right-hand planel) are plotted as a function of the displacement of the corresponding main lobe from the facet center. The points indicated by the white arrows in panels (a) through (d) are the same as the white circles indicated on the right hand plot. We then fit those points as a low-order 2D polynomial (in this case, as a quartic), which we plot as a colored plane cutting through them. The fit on the right hand side is merely one in a family of fits to each possible displacement vector from the main lobe of the PSF. Fitting the translational variation of the PSF in this way is potentially very useful, since a sparse representation of $\PSF$, the matrix of point spread functions, would allow us to quickly multiply it by a vector. Though this is not important for mapmaking, it is important for estimating power spectra from the dirty maps and mapping statistics produced by our method. }
	\label{fig:PSF_Fitting_Demo}
\end{figure*}

We expect these all to be similar, but also to vary slowly over the facet---though exactly how is not obvious \emph{a priori}. In the right-hand panel of Figure \ref{fig:PSF_Fitting_Demo} we plot the points on the PSFs displaced exactly 15 pixels along one of the two principal axes from the centers of their main lobes (illustrated by the left-hand panel). The $x$ and $y$ axes of the plot tell us which pixel a given PSF is centered on. As we expected, the variation over the facet is very smooth and is well approximated by a low-order polynomial. If we had instead plotted a displacement of 0, the right-hand panel would have been a perfectly flat plan of all ones because of the definition of $\D$.

How can we take advantage of the sparsity of information needed to describe $\PSF$ to write it as the sum of matrices that can be quickly multiplied by a vector? Let us first consider the simpler, 1D case. Instead of the translational invariance that leads to matrices of the form in Equation \eqref{eq:Toeplitz} where the main diagonal and all parallel off diagonals are constant, instead we model them all as polynomials:
\beq
P^\text{1D}_{ii'} = \sum_n t_{n,i-i'}(i+i')^n. \label{eq:polyToeplitz}
\eeq
This is a polynomial expansion in $(i+i')$, the distance along a diagonal, with coefficients $t_{n,i-i'}$ that make up a Toeplitz matrix. Again, primed indices tell us where on the true sky and unprimed indices tell us where in the faceted dirty map. The polynomial fit coefficients are a function of specific displacement of the main lobe of the PSF, hence the index $i-i'$. However, to fit all PSF values for the same displacement, we need to multiply those coefficients by the displacement from the center of the facet to the correct polynomial power. Our hope is that we can approximate $\PSF$ with a relatively low-order polynomial.

Expanding this out and cutting off the series after the second order in $n$, we get that
\beq
\PSF^\text{1D} \approx \T_0 + \J\T_1 + \T_1\J + \J^2\T_2 + 2\J\T_2\J + \T_2 \J^2
\eeq
where each $\T_n$ is a Toeplitz matrix and $\J$ is a diagonal matrix with integer indices centered on zero as its entries:
\beq 
\J \equiv  \text{diag}\left(...,-4, -3, -2, -1, 0, 1, 2, 3, 4, ... \right).
\eeq
Terms in the expansion that involve $(i')^n$ look like $\J^n$ to the right of $\T_n$, since they index into a vector multiplied by $\PSF^\text{1D}$ on the right, like the true pixelized sky. Likewise, terms that involve $i^n$ have a $\J^n$ matrix on the left.

In 2D, the situation is a bit more complicated. For clarity, let us treat $\PSF$ as a 4-indexed object, mapping two spatial dimensions to two other spatial dimensions. We approximate $\PSF$ as a polynomial sum of the form
\beq
P_{ii'jj'} = \sum_{n,m} t_{n,m,i-i',j-j'}(i+i')^n(j+j')^m.
\eeq
Now $\T_{n,m}$ is a ``block Toeplitz'' matrix, essentially a Toeplitz matrix of Toeplitz matrices. Thankfully, multiplying by the matrix by a vector of size $N_\text{PSF}$ still only scales as $\BigO(N_\text{PSF} \log N_\text{PSF})$ \cite{BlockToeplitz}. Expanding this to second order yields quite a few more terms:
\begin{align}
&\PSF \approx  \overbrace{\T_{0,0}}^{\text{$0^\text{th}$ Order}} + \nonumber \\
&\overbrace{\T_{1,0}(\J\otimes\Eye) + (\J\otimes\Eye) \T_{1,0} }^{\text{$1^\text{st}$ Order}} \nonumber +\\ 
&\T_{0,1} (\Eye\otimes\J) + (\Eye\otimes\J)\T_{0,1} +\nonumber \\
&\overbrace{\T_{2,0}(\J^2\otimes\Eye) + 2(\J\otimes\Eye)\T_{2,0}(\J\otimes\Eye) + (\J^2\otimes\Eye)\T_{2,0}}^{\text{$2^\text{nd}$ Order}} + \nonumber \\
&\T_{1,1}(\J\otimes\J) + (\J\otimes\Eye)\T_{1,1}(\Eye\otimes\J) + \nonumber \\
&(\Eye\otimes\J)\T_{1,1}(\J\otimes\Eye) + (\J\otimes\J)\T_{1,1} + \nonumber \\
&\T_{0,2}(\Eye\otimes\J^2) + 2(\Eye\otimes\J)\T_{0,2}(\Eye\otimes\J) + (\J^2\otimes\Eye)\T_{0,2}.
\end{align}
Here, we adopt the convention that all tensor products have the matrices in the $i$ or $i'$ dimension on the left-hand side of the $\otimes$ symbol and $j$ or $j'$ matrices on the right-hand side. In fact, it turns out that the exact number of polynomial terms is 
\beq
N_\text{poly} = \frac{1}{24}\left(24 + 50\omega + 35\omega^2 + 10\omega^3 + \omega^4\right), \label{eq:NPolyTerms}
\eeq
where $\omega \equiv \max(n+m)$ is the highest order polynomial considered. 

\begin{figure}[] 
	\centering 
	\includegraphics[width=.6\textwidth]{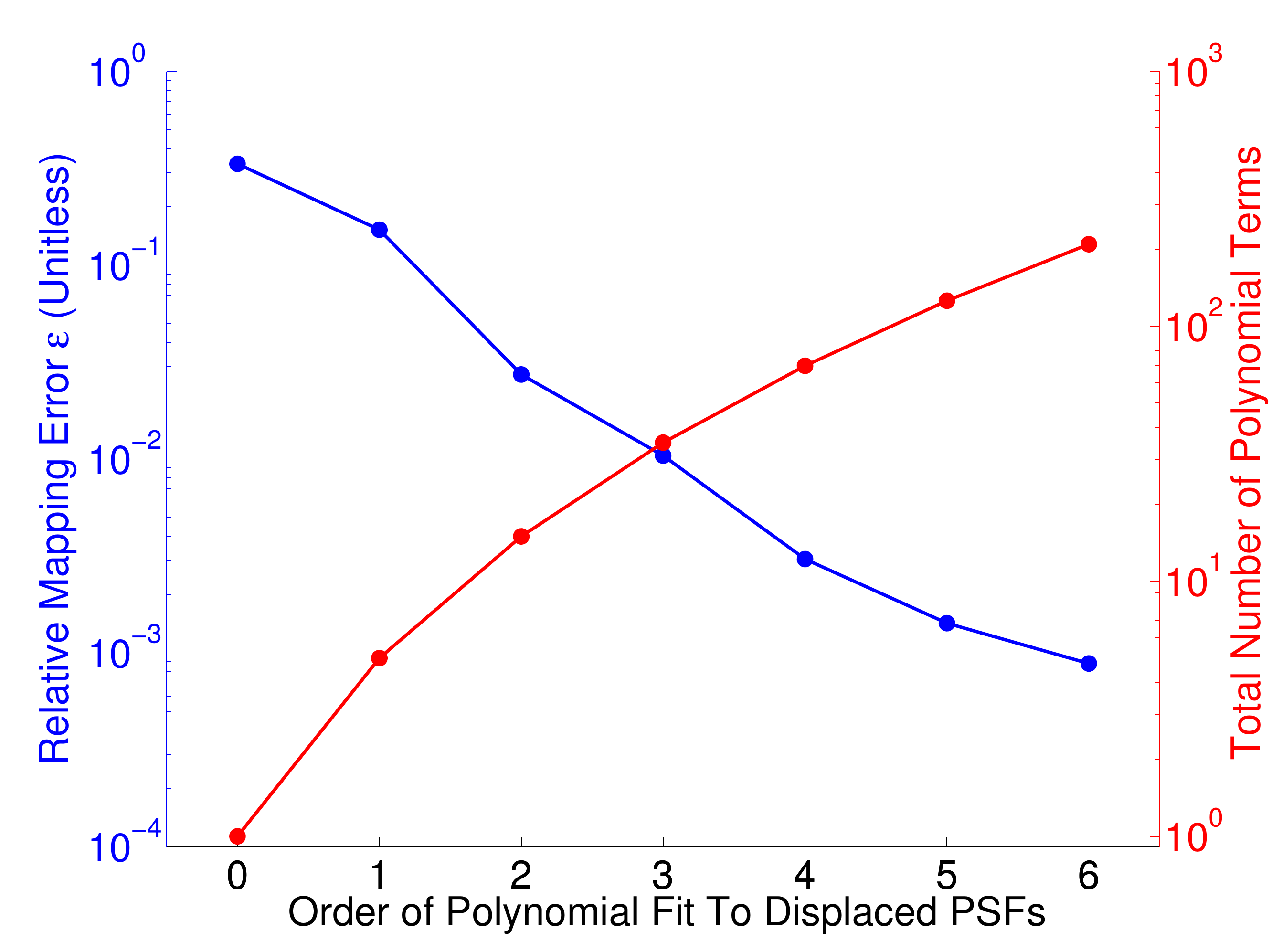}
	\caption[Error vs. complexity trade-off for PSF polynomial fitting.]{Approximating the translational variation of the point spread function with a low order polynomial can produce fairly small errors at a relatively low accuracy cost. Here we show the accuracy of multiplying a polynomially approximated $\PSF$ with the true sky compares to a direct calculation (using a large PSF truncation radius and no snapshotting). The errors are not negligible and the use of this approximation requires a carful examination of the accuracy requirements of the dirty maps. This technique saves time when the total number of terms in a polynomial/Toeplitz expansion of $\PSF$ is considerably smaller than the number of pixels in a facet. Unfortunately, that number of terms grows quartically with the polynomial order, meaning that very high orders and thus very high accuracy are not computationally useful.}
	\label{fig:PSF_Fitting_Order}
\end{figure}

The good news is that this fitting works pretty well at relatively low order, such as cubic or quartic. In Figure \ref{fig:PSF_Fitting_Order} we calculate the relative error between a dirty map computed by convolving the pixelized ``true'' sky with a very accurate $\PSF$ (one computed with a large truncation radius and no snapshotting) and one computed with a polynomial fit to the translationally varying component of $\PSF$.  We find that the method outlined above can faithfully reproduce the dirty map to high precision.

Increasing accuracy, however, comes at a steep cost. While multiplication of $\PSF$ by a vector for a single frequency can be performed in $\BigO(N_\text{facet}N_\text{PSF})$, multiplication of a polynomially-approximated $\PSF$ takes $\BigO(N_\text{poly}N_\text{PSF}\log N_\text{PSF})$. Since $N_\text{poly}$ scales with the fourth power of the maximum order, it gets expensive very quickly. Thus the method outlined above is especially useful when $\sim 1\%$ to $0.1\%$ errors are acceptable or when facets are exceptionally big or of exceptionally high resolution. 

It is possible to reduce that cost by attacking the problem with a hybrid approach. We find that the biggest fitting errors come far from the facet center, especially in the brightest side lobes. This makes sense, since it is where the notion of a fixed ``displacement'' from the main lobe of the PSF runs up against the limits of the flat-sky approximation. One could use this technique to incorporate the effects of most of $\PSF$, zeroing out the contributions from side lobe displacements. Then we could take the remainder of the $\PSF$ into account by simple matrix multiplication, achieving the same error with many fewer polynomial terms.

With big facets or at high resolution, PSF fitting serves another function. If the computational cost of mapmaking and  power spectrum estimation is dominated by the matrix multiplication $\A^\dagger \N^{-1} \A$ in the calculation of $\PSF$, we can choose to calculate only a representative sample of the entries in $\PSF$ (i.e. only some of the points on the right-hand side of Figure \ref{fig:PSF_Fitting_Demo}). Then we would rely on the fact that the polynomial fit is overdetermined to back out the missing entries.\footnote{It is worth noting that although a large number of terms might be needed to multiply $\PSF$ by a vector, there are not nearly so many free parameters in the fits. The number of free parameters needed to find a best-fit surface like that in Figure \ref{fig:PSF_Fitting_Demo} only scales like the square of the highest polynomial order.}

Whether or not to use the polynomial approximation to the $\PSF$ will depend on the exact telescope configuration and the nature of the mapmaking and power spectrum estimation problems at hand. If we want to try to precisely subtract foregrounds and work deep within the wedge, the polynomial approximation might not be good enough. However, if instead our power spectrum estimation strategy is to focus on isolating the EoR window and projecting out foreground-dominated modes entirely, it is less important that we very precisely understand the effect of the instrument. In that case, it is more likely that the polynomial PSF fitting approach outlined above will be useful. We explore these two approaches in the context of the mapmaking formalism in Appendix \ref{app:projection}.

\subsection{Computational Methods Summary} \label{sec:computationalSummary}

In the previous three sections, we explored three different ways of speeding up either the calculation of $\PSF$ or the multiplication of $\PSF$ by a vector. In Table \ref{tab:summary} we summarize those results. In general, we find that PSF truncation and snapshotting have the most utility for HERA. PSF fitting, in the fiducial scenario we considered, is the least helpful. However, for a telescope with much higher angular resolution than HERA, PSF fitting is likely to be more useful, since multiplication of a vector by $\PSF$ scales quadratically with the number of pixels in the facet.
\begin{sidewaystable}
  \centering
  \begin{tabular}{| c |  c | c | c |} 
\hline
\textbf{Approximation Parameter} & \textbf{PSF Truncation Radius} & \textbf{Snapshot Time} & \textbf{PSF Fitting Order} \\ \hline 
 \textbf{Improves} & Size of $\PSF$ & Steps in computing $\PSF$ & Multiplying by $\PSF$ \\ \hline
 \textbf{More exact when...} & ...larger & ...smaller & ...larger \\ \hline
 \textbf{Cost scaling} & Quadratic & Inverse Linear & Quartic \\ \hline
 \textbf{Acheives 1\% Error\footnote{Assumes HEALPix $N_\text{side}=256$, 2\,s integrations, and $10^\circ$ diameter facets.} for HERA at} & $5^\circ$ or\footnote{This depends on whether point sources are included, since they are mostly inside the facet in our simulations, depressing the error at small truncation radius.} $15^\circ$ & 10 minutes & $3^\text{rd}$ order \\ \hline
 \textbf{Speed-up at 1\% error} & $\sim$60 or $\sim$500 & $\sim$300 & $\sim$5 \\ 
\hline
  \end{tabular}
  \caption[Summary of PSF approximation techniques.]{Summary of the techniques we use to approximate the calculation of multiplication by the matrix of point spread functions, $\PSF$, in order to dramatically improve the speed of those operations. Faceting alone makes calculating $\PSF$ faster by a factor of 500 in our fiducial scenario. Combining PSF truncation and snapshotting brings the calculation of $\PSF$ well within the realm of feasibility. The benefit to fitting the PSF with polynomials and Toeplitz matrices is relatively small for our scenario, but it gets much better for higher resolution instruments.}
  \label{tab:summary}
\end{sidewaystable}

While these results are specific to HERA, we can draw a few general conclusions. For HERA at 150\,MHz, the first side lobes are about $13^\circ$ from the main lobe of the synthesized beam. At $13^\circ$ from the zenith, the primary beam is down by 20\,dB. In general, it is likely we will only be able to truncate the PSF in regions where the primary beam is small, meaning that a telescope with a broader primary beam will benefit less from cropping in a way that scales quadratically with the PSF truncation radius and therefore also the PSF's full width at half maximum. By contrast, larger primary beams are more slowly varying spatially, meaning that longer snapshots are likely to achieve the same error. If the primary beam is relatively smooth, that benefit scales inverse-linearly with the size of the primary beam.

Though we used 1\% as a somewhat arbitrary point of comparison in Table \ref{tab:summary}, it remains an open question how good our models of the $\PSF$ have to be. The only comprehensive way to answer this question is through a full end-to-end simulation of the signal, noise, and foregrounds all passed through a simulated instrument, a mapmaking code, and then power spectrum estimation. That sort of quantitative answer is outside the scope of this paper. However, it is worthwhile to enumerate the ways in which we need to use $\PSF$ to make maps and estimate power spectra and to examine the accuracy requirements for those tasks. By our count, $\PSF$ appears in six key places in the power spectrum estimation process:
\begin{enumerate}
\item When we calculate $\widehat{\x}$, we need $\PSF$ to define $\D$. However, looking closely at Equation \eqref{eq:QE} shows that $\D$ actually cancels out---the factor of $\D$ in each $\widehat{\x}$ and the two in $\C,_\beta$ are canceled by the two in each $\C^{-1}$. Therefore, it does not matter whether we get $\D$ right or not, as long as we are consistent about what we use for it. This makes sense, $\D$ was supposed to be an arbitrary choice, so as long as it is invertible, there is no way to get it ``wrong'' per se.
\item $\PSF$ also appears in our models for the parts of $\mean$ and $\C^{FG}$ corresponding to bright point sources in $\C^{FG}$. Accounting properly for bright point sources has the highest bang for the buck, in the sense that it is relatively straightforward to model both their means and covariances in the dirty map. In Section \ref{sec:skymodel}, we discussed how we could account for bright point sources with well-characterized positions, fluxes, and spectral indices by calculating a column in $\PSF$ that maps the point source to the entire facet in the dirty map. For that calculation, the PSF truncation radius is irrelevant because we account for the brightest sources in a separate part of the PSF independent of the HEALPix grid. Since we calculate only a moderate number of columns of $\PSF$, we do not even have to combine integrations into the snapshot. For bright point sources, it is not much extra effort to get $\PSF$ almost exactly right.
\item By contrast, diffuse emission from confusion-limited and galactic synchrotron emission in $\mean$ and $\C^{FG}$ depends, as we have argued, on knowing how $\PSF$ maps a large part of the true sky onto the facet. It is in this context that approximate versions of $\PSF$ are the most useful, but also where they are potentially the most worrisome. Galactic and confusion-limited foregrounds are still orders of magnitude stronger than the cosmological signal and understanding them precisely is very important. Forming $\mean$ from these foregrounds should be comparatively easy---all we need to do is take our sky model, compute visibilities, and then pass it through our mapmaking routine. We do not even need to calculate the full $\PSF$ matrix. Writing down $\C^{FG}$ is substantially more difficult, since $\C^{FG} = \PSF \C^{FG}_\text{model} \PSF^\trans$. Exactly how well we need to know $\PSF$ in order for $\C^{FG}$ to accurately reflect the foreground uncertainty depends on the specific instrument, the foreground model, and our uncertainty about that model. A quantitative answer requires detailed covariance modeling outside the scope of this work and is therefore left for future investigation.
\item Modeling noise properly is extremely important since inside the EoR window only noise and signal should matter. A slight mismodeling of noise due to an error in the calculation of $\C^N$ could lead to an erroneous detection. If however we perform mapmaking twice from a cross power spectrum of interleaved timesteps, we can eliminate noise bias \cite{X13,PAPER32Limits}. If we do that, it is acceptable (albeit not optimal) to be very conservative in our model of the instrumental noise, effectively increasing the error bars due to noise without biasing our measurement. If we adopt this conservative stance, then we can confidently use an approximate form of $\PSF$ when calculating $\C^N$.
\item Modeling $\C^S$ is mostly important for the calculation of sample variance. In any foreseeable experiment, this is a small contribution to the error. Getting $\C^S$ slightly wrong is unlikely to be the dominant error associated with approximating $\PSF$.
\item The $\C,_\beta$ family of matrices is necessary for telling us how to translate properly weighted dirty maps into power spectra. We  need $\PSF$ to be as accurate as the precision with which we would like to measure the power spectrum. 
\end{enumerate}
In general, the question of exactly how accurately we need to know $\PSF$---and by extension, exactly how well we need to understand our instruments---is an open question for future investigation.

\section{Summary and Future Directions} \label{sec:Summary}

In this work, we showed how to make precise maps with well-understood statistics specifically for 21\,cm power spectrum estimation. We investigated how to connect the framework of optimal mapmaking to that of inverse-variance weighted quadratic power spectrum estimation in order to understand what sort of maps and map statistics we need for power spectrum estimation. We showed that in addition to the dirty map estimator $\widehat{\x}$, we need the matrix of point spread functions, $\PSF$, and the noise covariance matrix which takes a gratifyingly simple form: $\C^N = \PSF \D^\trans$ where $\D$ is an invertible normalization matrix that we can choose to be diagonal. 

This analysis technology will allow us to consistently integrate our best understanding of an instrument with our best models for noise, foregrounds, and the cosmological signal. Not only does this approach help prevent the loss of cosmological information, but it will allow for a precise measurement of the 21\,cm power spectrum and for the confident and robust description of the errors in our estimates. 

In the main part of this work, we focused on the matrix of point spread functions, $\PSF$, which relates the true sky to our dirty maps. We calculated simulated dirty maps and PSFs for HERA, the upcoming Hydrogen Epoch of Reionization Array. While calculating $\PSF$ exactly is computationally prohibitive, we explored three methods for approximating $\PSF$. First, we explored how making maps in facets with truncated PSFs can dramatically reduce the computational cost of calculation $\PSF$ for only a small hit to accuracy. Next we showed how to combine consecutive integrations while controlling for the errors introduced by the process. It turns out that observations many minutes apart can be combined with minimal error. Lastly, we showed how the multiplication of $\PSF$ by a vector---a necessary step for power spectrum estimation---might be sped up by approximating its translational variance as slowly varying. Though the cost scaling of this approximation is steep, we find this technique especially promising when moderate errors are tolerable or for instruments with high angular resolution. 

Just as importantly, all these methods have tunable knobs---they can be made more accurate at the cost of speed or memory. Though our specific, quantitative results are only applicable to HERA, the accuracy trade-offs and the computational scalings we find should be quite general. In that sense, we hope that this work serves as a versatile guide to mapmaking in the context of 21\,cm cosmology.
 
Much work remains to be done to develop a clear and computationally tractable pathway from visibilities all the way to power spectra with rigorous errors and error correlations. Even after connecting this work to an appropriately updated version of the \citet{DillonFast} algorithm, one still needs to assess the effect of our approximations, as well as a number of important data analysis choices, on power spectrum estimates and ultimately on cosmological parameter constraints. Though the errors incurred by each can be made arbitrarily small, it is difficult to say yet what level of approximation is tolerable. This is an open question for future work.

We would like to see a full end-to-end simulation, starting with the 21\,cm signal, passing through the instrument, and ending with power spectra and their statistics. Such a full-scale test could prove the effectiveness of these techniques and clarify exactly what the approximations utilized both in this work and in \citet{DillonFast} do to our measured power spectra. A power spectrum estimation technique that passes such a test with realistic foregrounds and noise will be the one to produce trustworthy cosmological measurements.
 


\begin{subappendices}

\section[Appendix: Polarization and Heterogenous Primary Beams]{Appendix: Polarization and Heterogenous \\ Primary Beams} \label{app:PolAndBeams}

In Section \ref{sec:interferometry}, we worked out the relationship between visibilities and the true sky in terms of the matrix $\A$. For the sake of simplicity, we made two assumptions that, in this appendix, we would like to relax.

First, we ignored the effect of polarization. Though the 21\,cm signal is unpolarized, astrophysical foregrounds are generally polarized. And because the primary beams of the two orthogonal polarizations measured by a single element are different, the polarization of sources is important. This is especially important for sources with high rotation measures \cite{MoorePolarization}. Second, we ignored the possibility that not every element has the same primary beam. It is possible that an array is intentionally constructed with multiple kinds of elements. It is also generally true that different elements will behave slightly differently, just due to the variations in their construction. If we are able to measure that variation---which is no small task---we would like to take it into account.

Let us begin with polarization. There are a number of different conventions for expressing polarization \cite{FFTT}, but one relatively straightforward one is to replace $I(\rhat,\nu)$ with a four-element vector $\Eye(\rhat,\nu)$ containing Stokes I, Q, U, and V parameters. Instead of one visibility per baseline and frequency, we now measure four, one for each of the pairs of polarizations of antennas, $xx$, $xy$, $yx$, and $yy$. In this case, $B(\rhat,\nu)$ becomes $\mathbf{B}(\rhat,\nu)$, a $4\times 4$ matrix that describes the response of each type of visibility to each polarization and direction.

Otherwise, not much changes. The sky vector we are estimating gets four times bigger and the number of visibilities also gets four times bigger (though there are simplifications in practice, since $xy$ and $yx$ visibilities are just complex conjugates of one another and can be averaged together to reduce noise). The $\A$ matrix is not fundamentally different. Though it may seem like this makes the problem of computing $\PSF$ 64 times harder, that is fortunately not the case.

Fundamentally, we want to estimate the cosmological signal from our best guess at the Stokes I map. Foregrounds can have I, Q, and U components---astrophysical sources are not circularly polarized. So what we really want is a $\PSF$ matrix that maps I, Q, and U on the true sky, through $xx$ and $yy$ visibilities, to a dirty map of Stokes I. That is only six times more difficult than the calculations outlined above. If we do not want to model our foreground residual as polarized, then $\PSF$ is only twice as complicated as before---we just need to calculate $\mean$ through a more complicated mapmaking procedure involving the expanded definition of $\A$.

The issue with polarization is in many ways similar to the problem of heterogeneous primary beams. After all, the two polarization's dipoles generally have two different primary beams. Since the calculation of $\A^\dagger \N^{-1} \A$ is the computationally limiting step in our method, it is not significantly more difficult to treat multiple kinds of primary beam products $B(\rhat,\nu)$ when calculating $\A$, each row having a potentially different $B(\rhat,\nu)$. This gives us a straightforward way to account for arrays that include multiple types of antenna elements.

Of greater concern is the fact that every element in a real array has a slightly different beam---even if it was designed to be homogenous. For a minimally redundant array, this does not matter. If we know the correct primary beam for every antenna, we can write down $\A$ exactly. For a highly redundant array like HERA, antenna heterogeneity breaks the redundancy of baselines. If we want to include all measured visibilities in our maps, we may need to treat visibilities involving the most discrepant antennas separately. If we had to go further and treat every visibility separately, that would make $\PSF$ two orders of magnitude more difficult to calculate for HERA. If we can measure primary beams for all of our antennas, it would be worthwhile to simulate the error associated with the approximation that they are all the same. This is left to future work. Fortunately, it is theoretically possible to take into account slight variations between elements in the framework we have outlined.

\section[Appendix: A Foreground Avoidance Approach to Power Spectrum \\ Estimation]{Appendix: A Foreground Avoidance Approach \\ to Power Spectrum Estimation} \label{app:projection}

The power spectrum estimation method we outlined in Section \ref{sec:powerspectra} is a promising way to enlarge the EoR window and gain the additional sensitivity forecasted by \cite{PoberNextGen}. However, it is not the simplest approach. Instead of directly modeling foregrounds, we could choose to simply throw out all the modes that we believe to be foreground contaminated. The foreground avoidance approach was pioneered by \cite{AaronDelay} and used to produce the best current limits on the 21\,cm power spectrum by \cite{PAPER32Limits} and \cite{DannyMultiRedshift}. This choice should be more robust to foreground mismodeling than subtraction, since we are merely trying to isolate foreground free regions of Fourier space from the effects of regions we have given up on. Where exactly we draw the line between wedge and window is a question that deserves further investigation with both simulations and real data. 

One might ask why foreground avoidance estimators are interesting when the whole point of making maps like ours was to compress the data in a space where foregrounds were most naturally subtracted. There are a few reasons. First, foreground avoidance is simpler than foreground subtraction. If we are going to try to subtract foregrounds, it is worthwhile to first perform the simpler, more robust procedure so we have a baseline for comparison. Second, even if we are only interested in mitigating the effect of foregrounds by avoiding them, this method gives a proper accounting for $\C^N$, $\C^S$, and $\C,_\alpha$, without making any of the approximations previously relied upon about there being no correlations between $uv$ cells or that uniform weighted maps have no PSF. Third, the technique is fairly directly comparable to that of \cite{AaronDelay} without the additional assumption that delay modes for a given visibility map neatly to band powers or the computational challenges of \cite{EoRWindow1,EoRWindow2}. And finally, we may also want to implement a hybrid approach, similar in spirit to \cite{PAPER32Limits}, where we project out modes deep into the wedge but try to subtract foregrounds nearer the edge of the wedge.\footnote{This is similar in spirit to what WMAP did \cite{WMAP1PowerSpectrum}. They first masked out the galaxy and the brightest point sources, then they performed foreground subtraction in the map and foreground residual bias subtraction in the angular power spectrum. For us, the major difference is that the we do both } 

Therefore, it is worthwhile to write down the general framework for foreground avoidance in the context of optimal mapmaking. The idea is relatively simple. Let's define a new dirty map estimator, $\widehat{\x}'$, defined as
\beq
\widehat{\x}' \equiv \Proj \widehat{\x}
\eeq
where $\Proj$ is a projection matrix that has eigenvalues of 0 or 1 only.  As with all projection matrices, $\Proj = \Proj^\trans = \Proj^2$. The matrix $\Proj$ Fourier transforms the data cube, sets all modes outside the EoR window to zero, and Fourier transforms back. It also means that we need to replace $\C$ with $\C'$ where
\beq
\C' = \Proj \C \Proj.
\eeq 
 
By construction, the projection eliminates the foregrounds in $\mu$, meaning that
\begin{equation}
\Proj \langle \widehat{\x} \rangle = \Proj \mu \approx 0.
\end{equation}
Likewise, the part of the covariance associated with the foregrounds should also go to zero. Hence,
\beq
\Proj \C^{FG} \Proj \approx 0, 
\eeq
which means that 
\beq
\C' = \Proj \left[ \C^S + \C^N \right] \Proj.
\eeq
This also changes $\C_{,\alpha}$ which now takes the form
\beq
\C_{,\alpha}' = \Proj \C_{,\alpha} \Proj = \Proj \PSF \Q_\alpha \PSF^\trans \Proj.
\eeq

Of course, the new covariance has many zero eigenvalues, which means that it is not invertible.  That is not a problem since we can replace $(\C')^{-1}$ by its ``pseudoinverse'' \cite{Maxgalaxysurvey1}, defined as
\beq
(\C')^{-1}_\text{psuedo} = \Proj \left[ \Proj \C' \Proj + \gamma(\Eye - \Proj)\right]^{-1}\Proj
\eeq
where $\gamma$ can be any (numerically reasonable) nonzero number without changing the result.  The pseudoinverse reflects the idea that we want to completely throw out any power in possibly foreground-contaminated modes but also that we want to express infinite uncertainty in the modes---in other words, to give them no weight. This will accurately account for the fact that we have no information about these modes.

Putting all that together, our new quadratic estimator $\widehat{\p}$ is
\begin{align}
\widehat{p}_\alpha =& \frac{1}{2} M_{\alpha \beta} \widehat{\x}^\trans    (\C')^{-1}_\text{psuedo}  \PSF \Q_\beta \PSF^\trans  
(\C')^{-1}_\text{psuedo}  \widehat{\x} - b_\alpha,
\end{align}
where we have used the fact that $\Proj^2 = \Proj$. The estimator is not lossless, but it can still be unbiased in the region of Fourier space not projected out and have rigorously defined and calculable error properties.

\end{subappendices}



\part{Early Results from New Telescopes}
\chapter[{Overcoming Real-World Obstacles in 21\,cm Power Spectrum \\ Estimation: A Method Demonstration and Results from Early \\Murchison Widefield Array Data}]{Overcoming Real-World Obstacles in 21\,cm Power Spectrum Estimation: \\A Method Demonstration and Results from Early Murchison Widefield Array Data}\label{ch:MWAX13}

\emph{The content of this chapter was submitted to \emph{Physical Review D} on April 25, 2013 and published \cite{X13} as \emph{Overcoming real-world obstacles in 21\,cm power spectrum estimation: A method demonstration and results from early Murchison Widefield Array data} on January 15, 2014.}

\section{Introduction}
\label{sec:Intro}

In recent years, $21\,\textrm{cm}$ tomography has emerged as a promising probe of the Epoch of Reionization (EoR).  As a direct measurement of the three-dimensional distribution of neutral hydrogen at high redshift, the technique will allow detailed study of the complex astrophysical interplay between the intergalactic medium and the first luminous structures of our Universe.  This will eventually pave the way towards the use of $21\,\textrm{cm}$ tomography to constrain cosmological parameters to exquisite precision, thanks to the enormity of the physical space within its reach (please see, e.g., \citet{FurlanettoReview,miguelreview,PritchardLoebReview,aviBook} for recent reviews).

To date, observational efforts have focused on measurements of the $21\,\textrm{cm}$ power spectrum.  Such a measurement is exceedingly difficult.  Sensitivity requirements are extreme, requiring thousands of hours of integration and large collecting areas \citep{MiguelNoise,Judd06,LidzRiseFall,LOFAR2,AaronSensitivity}.  Adding to this challenge is the fact that raw sensitivity is insufficient---what counts is sensitivity to the cosmological signal above expected contaminants like galactic synchrotron radiation, which are three to four orders of magnitude brighter at the relevant frequencies \citep{Angelica,LOFAR,BernardiForegrounds,PoberWedge}. 
 
To deal with these challenges, numerous techniques have been proposed and implemented for foreground mitigation and power spectrum estimation.  These include foreground removal via parametric fits \citep{xiaomin,Judd08,paper2,paper1}, non-parametric methods \citep{Harker,Chapman1,Chapman2}, principal component analyses \citep{GMRT,AdrianForegrounds,GBT,newGMRT}, filtering \citep{nusserforegrounds,PetrovicOh,AaronDelay}, frequency stacking \citep{ChoForegrounds}, and quadratic methods \citep{LT11,DillonFast,Richard}.  In almost all of these proposals, foregrounds are separated from the cosmological signal by taking advantage of the differences in their spectra.  Foregrounds are dominated by continuum processes and thus have smooth spectra.  On the other hand, because the cosmological line-of-sight distance maps to the observed frequency of the redshifted $21\,\textrm{cm}$ line, the rapid fluctuations in the brightness temperature distribution that are expected from theory will map to a measured cosmological signal with jagged, rapidly fluctuating spectra.  When these spectral differences are considered in conjunction with instrumental characteristics, one can identify an ``EoR window'': a region in Fourier space where power spectrum measurements are expected to be relatively free from foregrounds \citep{Dattapowerspec,AaronDelay,VedanthamWedge,MoralesPSShapes,CathWedge,ThyagarajanWedge}.  This is shown schematically in Figure \ref{fig:EoRWindow}, where we have used early Murchison Widefield Array (MWA) data to estimate the power spectrum as a function of $k_\perp$ (Fourier mode perpendicular to the line-of-sight) and $k_\parallel$ (Fourier mode parallel to the line-of-sight).  
\begin{figure}[!ht] 
	\centering 
	\includegraphics[width=.6\textwidth]{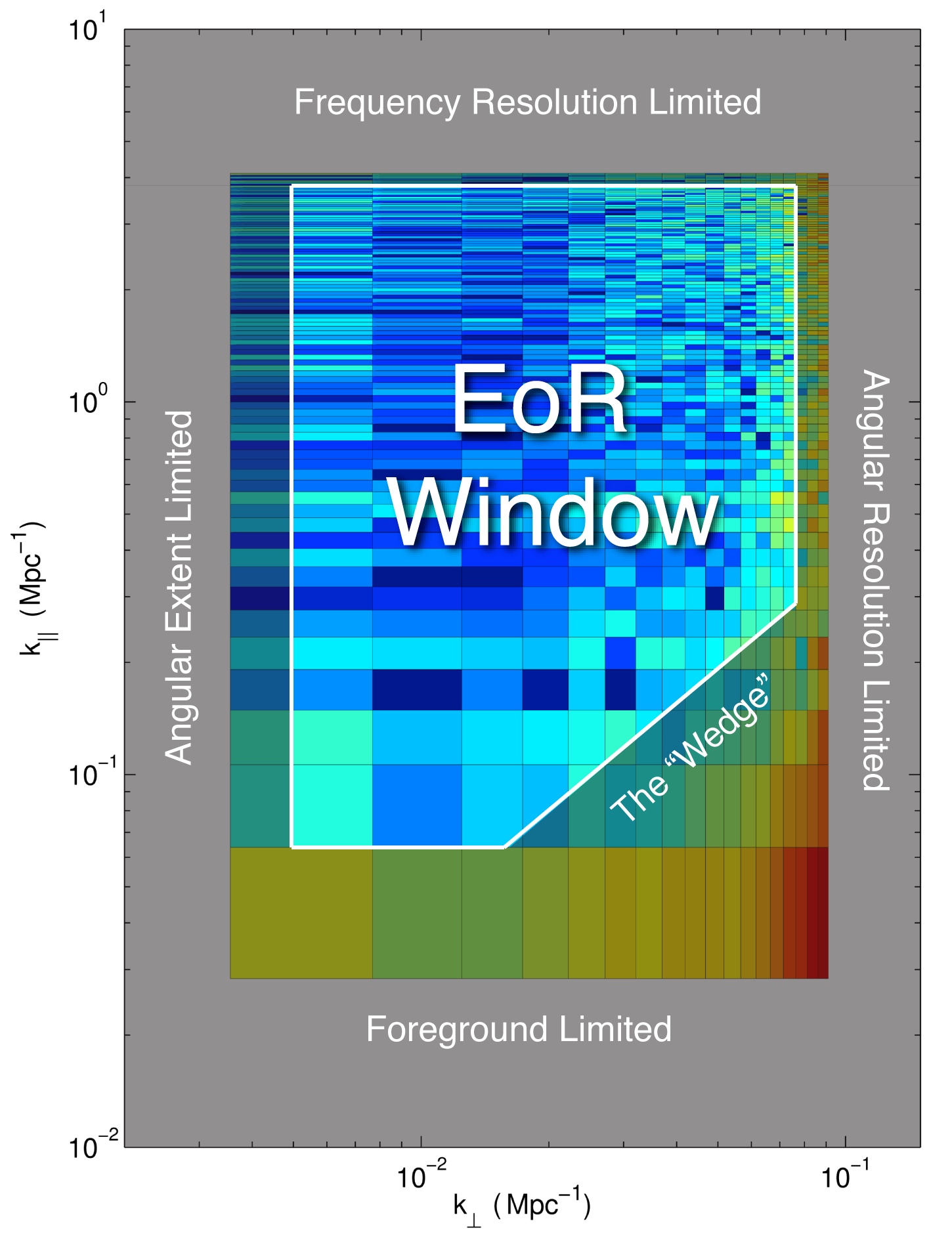}
	\caption[Illustration of the EoR Window.].{The ``EoR window,'' a region of Fourier space with relatively low noise and foregrounds, is thought to present the best opportunity for measuring the cosmological $21\,\textrm{cm}$ power spectrum during the Epoch of Reionization.  Here we show an example power spectrum from early MWA data, as a function of $k_\perp$ (Fourier components perpendicular to the line of sight) and $k_\parallel$ (Fourier components parallel to the line of sight).  More details on how we have calculated and plotted $P(k_\perp,k_\parallel)$ are found in Section \ref{sec:WorkedExample}.  We schematically highlight the instrumental and foreground effects that that delimit the EoR window---the coldest part of this power spectrum.  At low and high $k_\perp$, measurements are limited by an instrument's ability to probe the largest and smallest angular scales, respectively.  Limited spectral resolution causes similar effects at the highest $k_\parallel$.  As spectrally smooth sources, foregrounds inhabit primarily the low $k_\parallel$ regions.  Thanks to chromatic instrumental effects, however, there is a slight encroachment of foregrounds towards higher $k_\parallel$ at higher $k_\perp$, in what has been colloquially termed the ``wedge'' feature.}
	\label{fig:EoRWindow}
\end{figure} 
More details regarding this figuew are provided in Section \ref{sec:WorkedExample}; for now we simply wish to draw attention to the existence of a relatively contaminant-free region in the middle of the $k_\perp$-$k_\parallel$ plane.  This clean region is what we denote the EoR window.

The EoR window is generally considered the sweet spot for an initial detection of the cosmological $21\,\textrm{cm}$ power spectrum, and constraints are likely to degrade away from the window.  At high $k_\perp$ (i.e., the finest angular features on the sky), errors increase due to the angular resolution limitations of one's instrument.  For an interferometer, this resolution is roughly set by the length of the longest baseline.  Conversely, the shortest baselines define the largest modes that are observable by the instrument.  Errors therefore also increase at the lowest $k_\perp$ where again there are few baselines.

A similar limitation defines the boundary of the EoR window at high $k_\parallel$.  Since the spectral nature of $21\,\textrm{cm}$ measurements mean that different observed frequencies map to different redshifts, the highest $k_\parallel$ modes are inaccessible due to the limited spectral resolution of one's instrument.  At low $k_\parallel$, one probes spectrally smooth modes---precisely those that are expected to be foreground contaminated.  Thus there is another boundary to the EoR window at low $k_\parallel$.

A final delineation of the EoR window is provided by the region labeled as the ``wedge'' in Figure \ref{fig:EoRWindow}.  The wedge feature is a result of an interplay between angular and spectral effects.  Simulations have shown that the wedge is the effect of chromaticity in one's synthesized beam (which is inevitable when an interferometer is used to survey the sky).  This chromaticity imprints unsmooth spectral features on measured foregrounds, resulting in foreground contamination beyond the lowest $k_\parallel$ modes even if the foregrounds themselves are spectrally smooth.  Luckily, this sort of additional contamination follows a reasonably predictable pattern in the $k_\perp$-$k_\parallel$ plane, and in the limit of intrinsically smooth foregrounds, the wedge can be shown to extend no farther than the line
\begin{equation}
\label{eq:wedge}
k_\parallel = \left[ \sin \theta_{\textrm{field}} \frac{D_M (z) E(z)}{D_H (1+z)} \right] k_\perp,
\end{equation}
where $D_H \equiv c/H_0$, $E(z) \equiv \sqrt{\Omega_m (1+z)^3 + \Omega_\Lambda}$, $D_M(z) \equiv \int_0^z dz^\prime / E(z^\prime)$, $\theta_{\textrm{field}}$ is angular radius of the the field-of-view, and $c$, $H_0$, $\Omega_m$, and $\Omega_\Lambda$ have their usual meanings \citep{Dattapowerspec,VedanthamWedge,MoralesPSShapes,CathWedge}. Intuitively, the foreground-contaminated wedge extends to higher $k_\parallel$ at higher $k_\perp$ because the high $k_\perp$ modes are probed by the longer baselines of an interferometer array, which have higher fringe rates that more effectively imprint spectral structure in the measured signals.  For an alternate but equivalent explanation in terms of delay modes, please see the illuminating discussion in \citet{AaronDelay}.

The concept of an EoR window is important in that it provides relatively strict boundaries that separate fairly foreground-free regions of Fourier space from heavily foreground-contaminated ones.  It therefore provides one with the option of practicing foreground avoidance rather than foreground subtraction.  If it turns out that foregrounds cannot be modeled well enough to be directly subtracted with the level of precision required to detect the cosmological signal, foreground avoidance becomes an important alternative, in that the only way to robustly suppress foregrounds is to preferentially make measurements within the EoR window.  Likely, some combination of the two strategies---foreground subtraction and foreground avoidance---will prove useful for the detection of the 21 cm power spectrum. Of course, measurements within the EoR window are still contaminated by instrumental noise, but fortunately the noise integrates down with further observation time (as long as calibration errors and other instrumental systematics can be sufficiently minimized).  Observationally, it is encouraging that the EoR window has now been shown to be free of foregrounds to better than one part in a hundred in power \citep{PoberWedge}.  

As experimental sensitivities increase, however, one must take care to preserve the cleanliness of the EoR window to an even higher dynamic range.  There are several ways in which our notion of the EoR window may be compromised.  First, as experiments integrate in time and acquire greater sensitivity, we may discover that our approximation of spectrally smooth foregrounds is insufficiently good for a detection of the (faint) cosmological signal.  In other words, foreground sources may have small but non-negligible high $k_\parallel$ components in their spectra that have thus far gone undetected.  This would translate into a smaller-than-expected EoR window.  In addition, even intrinsically smooth foregrounds may appear jagged in a real measurement because of instrumental effects such as imperfect calibration.  The precise interferometer layout may also result in unsmooth artifacts that arise from combining data from non-redundant baselines \citep{Hazelton2013}.  Finally, suppose that the aforementioned effects are negligible and that the assumption of spectrally smooth foreground emission continues to hold.  The EoR window still cannot be taken for granted because non-optimal data analysis techniques may result in unwanted foreground artifacts in the region.   For the EoR window to exist at all, it is essential that power spectra are estimated in a rigorous fashion, with well-understood statistics.  

The goal of this paper is to minimize unwanted data analysis artifacts by establishing methods for power spectrum estimation that are both robust and as optimal as possible.  Previous efforts have rarely met both criteria: either the methods are robustly applicable to data with real-world artifacts but fail to achieve optimized (or even rigorously computable) error properties, or provide an optimal framework but ignore real-world complications.  In this paper we extend the rigorous framework described in \citet{LT11} and \citet{DillonFast} to deal with real-world effects.  The result is a computationally feasible approach to analyzing real data that not only preserves the cleanliness of the EoR window, but also rigorously keeps track of all relevant error statistics.

To demonstrate the applicability of our approach, we apply our techniques to early data from the Murchison Widefield Array (MWA).  These data were derived from $\sim 22\,\textrm{hours}$ of tracked observations using an early, 32-element prototype array.  The results are therefore not designed to be cosmologically competitive, but instead illustrate the rigor that will be required for an eventual detection of the EoR while also providing new measurements on the ``wedge'' feature that delineates the EoR window.

This paper is organized as follows.  In Section \ref{sec:Methods} we discuss various real-world obstacles that must be dealt with when analyzing real data, and how one can overcome them while maintaining statistical rigor.  We then apply our methods to MWA data in Section \ref{sec:WorkedExample} as a ``worked example'', highlighting the importance of various subtleties of power spectrum estimation.  In Section \ref{sec:earlyResults} we present some results from the data, emphasizing the agreement between theoretical expectations and our observations of the foreground wedge (particularly regarding the frequency dependence of the wedge).  We also present upper limits on the cosmological $21\,\textrm{cm}$ power spectrum over the broad redshift range of $z=6.2$ to $z=11.1$.  Finally, we summarize our conclusions in Section \ref{sec:Conc}.
 
\section[Systematic Methods for Dealing with Real-World Obstacles]{Systematic Methods for Dealing with\\ Real-World Obstacles}
\label{sec:Methods}

To understand the gap between an analysis framework for idealized observations and any real-world data set, we enumerate and address six different obstacles that rather universally affect real data.  Our goal in this section is to meet the challenges presented by these obstacles while maintaining as many of the advantages of the optimal framework as possible, which we reiterate in Section \ref{sec:IdealObs}, especially the ability to minimize and precisely quantify the uncertainties in the measurements.  In the following sections, we address the problems presented by large data volumes (Section \ref{sec:DataVolume}), uncertainties in the properties of contaminants such as foregrounds (Section \ref{sec:crossPower}), incomplete $uv$ coverage (Section \ref{sec:IncompleteUV}), radio frequency interference (RFI) flagging (Section \ref{sec:RFI}), foreground leakage into the EoR window (Section \ref{sec:decorr}), and binning to spherically averaged power spectra (Section \ref{sec:cylindToSph}).

\subsection[A Systematic Framework for Analyzing Idealized Observations]{A Systematic Framework for Analyzing \\Idealized Observations}
\label{sec:IdealObs}
In this section, we briefly review the formalism of \citet{LT11} for optimal power spectrum estimation, which was adapted for $21\,\textrm{cm}$ tomography from similar techniques used in galaxy survey and cosmic microwave background analysis \citep{Maxpowerspeclossless,BJK,Maxgalaxysurvey1,Maxgalaxysurvey2}.  For now, we do not include real-world effects such as missing data from RFI flagging, and the purpose of later sections is to extend the formalism to take into account these complications.

In $21\,\textrm{cm}$ tomography, one typically wishes to measure both the spherically-binned power spectrum $P_{\textrm{sph}}(k)$, defined by
\begin{equation}
\langle \widetilde{T}^* (\mathbf{k}) \widetilde{T} (\mathbf{k}^\prime) \rangle \equiv (2 \pi)^3 P_{\textrm{sph}}(k) \delta (\mathbf{k} - \mathbf{k}^\prime),
\end{equation}
and the cylindrically-binned power spectrum $P_{\textrm{cyl}} (k_\perp, k_\parallel)$, defined by
\begin{equation}
\langle \widetilde{T}^* (\mathbf{k}) \widetilde{T} (\mathbf{k}^\prime) \rangle \equiv (2 \pi)^3 P_{\textrm{cyl}}(k_\perp, k_\parallel) \delta (\mathbf{k} - \mathbf{k}^\prime),
\end{equation}
with $\widetilde{T} (\mathbf{k})$ signifying the spatial Fourier transform of the $21\,\textrm{cm}$ brightness temperature field $T(\mathbf{r})$, $\mathbf{k}$ denoting the spatial wavevector with magnitude $k$, and components $k_\perp$ and $k_\parallel$ as the components perpendicular and parallel to the line-of-sight, respectively.  The angled brackets $\langle \cdots \rangle$ represent an ensemble average.  The spherical power spectrum is useful for comparing to theoretical models, since it is obtained by angularly averaging over spherical shells in Fourier space, and thus makes the cosmologically relevant assumption of isotropy.  The cylindrical power spectrum is useful for identifying instrumental and foreground effects, which possess a cylindrical symmetry rather than a spherical one.  Typically, the cylindrical power spectrum is produced first as a tool for foreground isolation (i.e., to identify the EoR window), and then subsequently binned into a spherical power spectrum.  This section concerns the estimation of the cylindrical power spectrum.  Optimal binning techniques to go from the cylindrical spectrum to the spherical spectrum are discussed in Section \ref{sec:cylindToSph}.

In estimating a power spectrum from data, one must necessarily discretize the problem.  We make the approximation that the power spectra are piecewise constant functions, such that we can describe them in terms of a vector of bandpowers with components $p^\alpha$, where
\begin{equation}
p_\alpha \equiv P_{\textrm{cyl}} (k^\alpha_\perp, k^\alpha_\parallel).
\end{equation}
It is the bandpowers and their error properties that one wishes to estimate from the data, which come in the form of a data vector $\mathbf{x}$.  Intuitively, one can think of the data vector as a list of the $21\,\textrm{cm}$ brightness temperatures measured at various locations in a three-dimensional ``data cube''.  Rigorously, we define each element of the data vector (i.e., each voxel of the data cube) as
\begin{equation}
\label{eq:pixDef}
\mathbf{x}_i \equiv \int T(\mathbf{r}) \psi_i (\mathbf{r}) d^3 \mathbf{r},
\end{equation}
with $\psi_i (\mathbf{r})$ being the pixelization kernel and $T(\mathbf{r})$ as the (continuous) three-dimensional $21\,\textrm{cm}$ brightness temperature field\footnote{Of course, instrumental noise and foregrounds do not properly reside in a cosmological three-dimensional volume: noise is introduced in the electronics of the system, whereas foregrounds are ``nearby'' and only appear in the same location in the data cube as our cosmological signal by virtue of their frequency dependence.  However, there is a gain in convenience and no loss of generality in assigning a noise and foreground contribution to each voxel, pretending that those contaminants also live in the observed cosmological volume.}.  In this paper we take the $i^{th}$ pixelization kernel $\psi_i (\mathbf{r})$ to be a boxcar function centered on the $i^{th}$ voxel of the data.\footnote{This choice, following \cite{DillonFast}, is motivated by the fact that the covariance between each pixel in this basis for both noise and foregrounds can be written in an algorithmically convenient way.}

To estimate the $\alpha^{th}$ bandpower from the data vector, we first form a quadratic estimator of the form
\begin{align}
\label{eq:unnormBands}
q_\alpha \equiv \frac{1}{2}( \mathbf{x} - \mathbf{m} )^t \mathbf{C}^{-1} \mathbf{C}_{,\alpha} \mathbf{C}^{-1}( \mathbf{x} -\mathbf{m} ) - \frac{1}{2} \textrm{tr} [ \mathbf{C}_{\textrm{junk}} \mathbf{C}^{-1} \mathbf{C}_{,\alpha} \mathbf{C}^{-1} ],
\end{align}
where $\mathbf{m} \equiv \langle \mathbf{x} \rangle$ is the mean of the data, $\mathbf{C} \equiv \langle \mathbf{x} \mathbf{x}^t \rangle - \langle \mathbf{x} \rangle \langle \mathbf{x} \rangle^t$ is its covariance, $\mathbf{C}_{\textrm{junk}}$ is the component of the covariance ``junk''/contaminants (to be defined in the following section), and $\mathbf{C}_{,\alpha}$ is the derivative of the covariance with respect to the $\alpha^{th}$ bandpower.  Since we are approximating the power spectrum as piecewise constant, we have
\begin{equation}
\mathbf{C} =\mathbf{C}_{\textrm{junk}} +  \sum_{\alpha} p_\alpha \mathbf{C}_{,\alpha}.
\end{equation}
Combined with Equation \eqref{eq:pixDef}, this expression can be used to derive explicit forms for $\mathbf{C}_{,\alpha}$, which reveals that the matrix essentially Fourier transforms and bins the data \citep{LT11,DillonFast}.  Intuitively, $\mathbf{C}_{,\alpha}$ can be thought of as the response in the data covariance $\mathbf{C}$ to the bandpower $p_\alpha$.  Thus, as long as one selects  an appropriate form for $\mathbf{C}_{,\alpha}$, the formalism of this section can also be used to directly measure the spherical power spectrum.  However, as we discussed above, in this paper we choose to first estimate the cylindrical power spectrum as an intermediate diagnostic step, to quantify and mitigate foregrounds better.
 
Once the $q_\alpha$s have been formed, they need to be normalized using a suitable invertible matrix $\mathbf{M}$ to form the final bandpower estimates:
\begin{equation}
\label{eq:normingBands}
\mathbf{\widehat{p}} = \mathbf{M} \mathbf{q},
\end{equation}
where we have grouped the bandpower estimates $\widehat{p}_\alpha$ into a vector $\widehat{\mathbf{p}}$ (and similarly grouped the coefficients $q_\alpha$ and $\mathbf{q}$), 
with the hat ( $\widehat{ }$ ) signifying the fact that we have formed an \emph{estimator} of the true bandpowers\footnote{Note that $\mathbf{q}$, $\mathbf{\widehat{p}}$, and $\mathbf{M}$ live in a different vector space than $\mathbf{x}$, $\mathbf{C}$, and $\mathbf{C}_{,\alpha}$.  The former are in a vector space where each component refers to a different bandpower, whereas the latter are in one where different components refer to different voxels.}.  We shall discuss different choices of $\mathbf{M}$ in Section \ref{sec:decorr}.

To understand the uncertainty in our estimates, we compute several error properties.  The first is the covariance matrix of the final measured bandpowers:
\begin{equation}
\label{eq:CovP}
\boldsymbol \Sigma \equiv \langle \mathbf{\widehat{p}} \mathbf{\widehat{p}}^t \rangle - \langle \mathbf{\widehat{p}} \rangle \langle \mathbf{\widehat{p}} \rangle^t = \mathbf{M} \mathbf{F} \mathbf{M}^t,
\end{equation}
where we have introduced the Fisher matrix $\mathbf{F}$, which has components
\begin{equation}
F_{\alpha \beta} = \frac{1}{2} \textrm{tr}[ \mathbf{C}^{-1} \mathbf{C}_{, \alpha} \mathbf{C}^{-1} \mathbf{C}_{,\beta} ]. \label{eq:fisherDef}
\end{equation}
The Fisher matrix also allows us to relate our estimated bandpowers $\mathbf{\widehat{p}}$ to the true bandpowers $\mathbf{p}$ via the window function matrix $\mathbf{W}$:
\begin{equation}
\label{eq:phatWp}
\langle \mathbf{\widehat{p}} \rangle = \mathbf{W} \mathbf{p},
\end{equation}
where $\mathbf{W}$ can be shown to take the form
\begin{equation}
\mathbf{W} = \mathbf{M} \mathbf{F}.
\end{equation}
If we choose $\mathbf{M}$ such that the rows of $\mathbf{W}$ each sum to unity, Equation \eqref{eq:phatWp} shows that each bandpower estimate can be thought of as a weighted average of the truth, with weights given by each row (each window function).  Even with this normalization requirement, there are still many choices for $\mathbf{M}$.  We discuss the various options and tradeoffs in Section \ref{sec:decorr}.

Whatever the choice of $\mathbf{M}$, our estimator has optimal error properties in the sense that if $\mathbf{\widehat{p}}$ in Equation \eqref{eq:phatWp} is used to constrain parameters in some theoretical model, those measured parameters will have the smallest possible error bars given the observed data \citep{Maxpowerspeclossless}.  Our goal in the following sections will be to ensure that both these small error bars and our ability to rigorously compute them are preserved in the face of real-world difficulties.

\subsection{A Real-World Obstacle: Data Volume}
\label{sec:DataVolume} 
Perhaps the most glaring difficulty presented by the ideal technique outlined above is its computational cost.  Much of that cost arises from the inversion of the data covariance matrix $\mathbf{C}$ in Equations \eqref{eq:unnormBands} and \eqref{eq:fisherDef}, in addition to the multiplication of $\C$ and matrices of the same size.  Both of these operations scale like $\BigO(N^3)$, where $N$ is the number of voxels in each data vector.  The computational cost makes taking full advantage of current generational interferometric data prohibitive, not to mention upcoming observational efforts that expect to produce $10^6$ or more voxels of data. 

One would like to retain the information theoretic advantages of the quadratic estimator method and its ability to precisely model errors and window functions, without $\BigO(N^3)$ complexity.  The solution to this problem, developed and demonstrated in \cite{DillonFast}, comes from taking advantage of a number of symmetries and approximate symmetries of the survey geometry and the covariance matrix, $\C$, and can accelerate the technique to $\BigO(N\log N)$.

The fast method relies on assembling the data into a data cube with rectilinear voxels amenable to manipulation with the Fast Fourier Transform.  This is equivalent to the assertion that each voxel represents an equal volume of comoving space, an approximation that relies on two restrictions on the data cube geometry.  First, the range of frequencies considered must be small enough that $D_c(z)$ (the line-of-sight comoving distance, equal to $D_M(z)$ above in a spatially flat universe) is linear with $\nu$.  Generally, one should limit oneself to analyzing the power spectrum of redshift ranges short enough that the evolution of the power spectrum during reionization can be neglected.  This range, suggested by \cite{Yi} to be $\Delta z \lesssim 0.5$, makes the approximation of a linear relationship between $\nu$ and $D_c(z)$ better than one part in $10^3$ at the redshifts of interest to 21 cm cosmology. 

Second, the assumption of equal volume voxels relies on the flat sky approximation.  To achieve this the area surveyed can be broken into a number of subfields, each a few degrees on a side, for which the curvature of the sky can be neglected.  As long as the angular extent of the data cube is smaller than $\sim 10^\circ$, the flat sky approximation is correct to a few parts in $10^3$.

By analyzing a rectilinear volume of the universe, all steps in calculating the band powers $q_\alpha$ can be performed quickly by exploiting various symmetries and taking advantage of the Fast Fourier Transform.  The model for $\C$ can be broken up into a number of independent matrices representing signal, noise, and foregrounds.  Each of these models, developed by \cite{LT11}, is well approximated by a sparse matrix in a convenient combination of real and Fourier spaces \citep{DillonFast}.  As a result, multiplication of a vector by $\C$ can be performed in $\BigO(N\log N)$.  \citet{DillonFast} showed how that speed-up can be parlayed into a method for quickly calculating $q_\alpha$ using the Conjugate Gradient Method.  The rapid convergence of the iterative method for calculating $\C^{-1}\x$ can be ensured by the application of a preconditioner which relies on the spectral smoothness of foregrounds and the fact that they are well described by only a few eigenmodes \citep{AdrianForegrounds}. Then, by randomly simulating many data vectors from the covariance $\C$ and calculating $q_\alpha$ from each, the Fisher matrix can be estimated from the fact that 
\beq
\F = \langle \mathbf{q} \mathbf{q}^t \rangle - \langle \mathbf{q} \rangle \langle \mathbf{q}^t \rangle,
\eeq 
which follows from Equation \eqref{eq:CovP}.  All of this together allows for fast, optimal power spectrum estimation---including error bars and window functions---despite the challenge presented by an enormous volume of data.

\subsection{A Real-World Obstacle: Uncertain Contaminant Properties}
\label{sec:crossPower}
If one had perfect knowledge of the foreground contamination in the data cube, the problem of foreground contamination would be trivial; one would simply perform a direct subtraction of the foregrounds from the data vector $\mathbf{x}$.  Unfortunately, our knowledge of foregrounds is far from perfect, particularly at the level of precision required for a direct detection of the cosmological $21\,\textrm{cm}$ signal.  Because of this, the estimator shown in Equation \eqref{eq:unnormBands} in fact combines several different foreground subtraction steps in an attempt to achieve the lowest possible level of foreground contamination:
\begin{enumerate}
\item A direct subtraction of a foreground model from the data vector.  This is given by $\mathbf{x} - \mathbf{m}$.  To see this, note that the data vector can be thought of as being comprised of the cosmological $21\,\textrm{cm}$ signal $\mathbf{x}_{21}$, the foregrounds $\mathbf{x}_{\textrm{fg}}$, and the instrumental noise $\mathbf{n}$.  On the other hand, the mean data vector
\begin{equation}
\mathbf{m} \equiv \langle \mathbf{x} \rangle = \langle \mathbf{x}_{21} \rangle + \langle \mathbf{x}_{\textrm{fg}} \rangle + \langle \mathbf{n} \rangle = \langle \mathbf{x}_{\textrm{fg}} \rangle.
\end{equation}
contains only the foreground contribution, because we are interested in the \emph{fluctuations} of the $21\,\textrm{cm}$ signal, so the cosmological signal has zero mean, as does the instrumental noise (in the absence of instrumental systematics).  Note that because the mean here is the mean in the ensemble average sense (as opposed to just the spatial mean), $\mathbf{m}$ represents a full spatial and spectral model of the foregrounds.
\item Since the foregrounds also appear in the covariance matrix, the action of $\mathbf{C}^{-1}$  is to downweight foreground-contaminated modes, exploiting foreground properties such as smooth frequency dependence.
\item Subtracting the term $ \frac{1}{2} \textrm{tr} [ \mathbf{C}_{\textrm{junk}} \mathbf{C}^{-1} \mathbf{C}_{,\alpha} \mathbf{C}^{-1} ]$ eliminates the bias from contaminants.
\item Finally, the binning of the cylindrical power spectrum to the spherical power spectrum provides yet more foreground suppression.  Foregrounds are distributed in select regions on the $k_\perp$-$k_\parallel$ plane (i.e., outside the EoR window) in patterns that do not lie along contours of constant $k = \sqrt{k_\perp^2 + k_\parallel^2}$.  Thus, when binning along such contours to produce a spherical power spectrum, one can selectively downweight parts of the contour with greater foreground contamination, which constitutes a form of foreground cleaning.  Roughly speaking, this corresponds to taking advantage of the fact that foregrounds have a cylindrical symmetry in Fourier space, whereas the signal is spherically isotropic \citep{MiguelJackie1}.  We do note, however, that the formalism we introduce in Section \ref{sec:cylindToSph} is general enough to use any geometric differences between foregrounds and signal.
\end{enumerate}
Of these foreground mitigation strategies, the first and third are direct subtractions (in amplitude and power, respectively), whereas the second and the fourth act through weightings. The former group represent operations that are particularly vulnerable to incorrectly modeled foregrounds.  To see this, recall that the foregrounds are expected to be larger than the cosmological signal by three or four orders of magnitude \citep{Angelica,LOFAR,BernardiForegrounds,PoberWedge}.  Thus, when performing direct subtractions, low-level, unaccounted-for inaccuracies in the foreground model can translate into extremely large biases in the final results.  In addition, significant numerical errors may arise from the subtraction of two large numbers (the data and the foregrounds) to obtain a small number (the measured cosmological signal).

Our goal for the rest of the section is to immunize ourselves against biases from direct subtractions.  Of the direct subtraction steps list above, the Step 1 is likely to be relatively harmless for two reasons.  First, it is immediately followed by the $\mathbf{C}^{-1}$ downweighting.  The downweighting mitigates the effects of inaccuracies in modeling, for the $\mathbf{C}^{-1}$ tends to gives less weight to precisely the modes that have the largest foreground amplitudes, and therefore would be the most susceptible to modeling errors in the first place.  In addition, the uncertainty in foreground properties in those regions of the $k_\perp$-$k_\parallel$ plane result in large error bars there, providing a convenient marker of the untrustworthy parts of the plane, effectively demarcating the boundaries of the EoR window.  For these two reasons, Step 1 is unlikely to be an issue, at least not inside the EoR window.

More worrisome is Step 3, where the power spectrum bias of contaminants is subtracted off.  If we define ``contaminants'' to be ``everything but the cosmological $21\,\textrm{cm}$ signal'', there are two potential sources of bias: foregrounds and noise.  The subtraction of these biases is not followed by a downweighting analogous to the application $\mathbf{C}^{-1}$ in Step 1.  Moreover, whereas one could argue that the foreground bias is likely to be large only outside the EoR window, the noise bias will spread throughout the $k_\perp$-$k_\parallel$ plane.  This noise bias will also be quite large, as current experiments are firmly in the regime where the signal-to-noise is below unity.  It would therefore be advantageous to avoid bias subtractions altogether if possible.

To avoid having to subtract foreground bias, we simply redefine what we mean by contaminants/junk.  If we modify our mission to be one where we are measuring the power spectrum of total sky emission instead of the power spectrum of the cosmological $21\,\textrm{cm}$ signal, the foreground contribution to the bias term no longer exists, as foregrounds now count as part of the signal we wish to measure.  Of course, nothing has really changed, for we have simply ignored the subtraction of the foreground bias by redefining what we mean by ``contaminants''.  The method is still optimal for measuring the power spectrum of the sky emission---though now it will not provide the absolute best possible limits on the EoR power spectrum.  Within the EoR window, this should result in little degradation of our final constraints, for in this region foreground contamination is expected to be negligible, and the power spectrum of the cosmological signal should be essentially identical to the power spectrum of total sky emission.  In any case, this is an assumption that can be checked in the final results, and represents a conservative assumption throughout Fourier space since foreground power is necessarily positive.  As detailed low-frequency foreground observations are conducted, it may be possible to achieve more sensitivity in foreground contaminated regions by taking advantage of more detailed maps and developing more faithful models. This task is left to future power spectrum estimation studies.

In contrast, escaping to the safe confines of the EoR window alone is not sufficient to eliminate the instrumental noise portion of the bias term, for the instrumental noise bias pervades the entire $k_\perp$-$k_\parallel$ plane.  To eliminate the noise bias, one can choose to compute not the auto-power spectrum of a single data cube with itself, but instead to compute the cross-power spectrum of two data cubes that are formed from data from interleaved (i.e., odd and even) time samples.  Since the instrumental noise is uncorrelated in time, this has the effect of automatically removing the instrumental noise bias\footnote{The reader may object to this by (correctly) pointing out that there exist errors that are correlated in time, with calibration errors being a prime example.  The result would be a cross-power spectrum that still retained a bias.  However, this does not invalidate the cross-power spectrum approach, in the following sense.  While biases will make our estimates of the power spectrum imperfect, these estimate will not be incorrect---the final (biased) power spectra will still represent perfectly rigorous upper limits on the cosmological power, provided we are conservative about how we estimate our error bars.  We will discuss how to make such conservative error estimates later on in this section and in Section \ref{sec:CovarModeling}.}.

More explicitly, we can form a bandpower estimate of the cross-power spectrum by simply computing
\begin{equation}
\label{crossEstimator}
\widehat{p}^ \textrm{cross}_\alpha = \mathbf{x}^t_1 \mathbf{E}^\alpha \mathbf{x}_2,
\end{equation}
where $\mathbf{x}_1$ and $\mathbf{x}_2$ are the data vectors for the two time inter-leaved data cubes, and for notational brevity we have defined $\mathbf{E}^\alpha \equiv \frac{1}{2} \sum_\beta M_{\alpha \beta} \mathbf{C}^{-1} \mathbf{C}_{,\beta} \mathbf{C}^{-1}$.  For notational cleanliness we will omit the $-\mathbf{m}$ term in our power spectrum estimator for this section only, with the understanding that $\x$ signifies the data vector \emph{after} the best-guess foreground model has already been subtracted.  In a similar fashion, $\x_\textrm{fg}$ refers to the foreground residuals, post-subtraction.

To see that the cross-power spectrum has no noise bias, let us decompose the data vectors $\mathbf{x}_i$ into the sum of $ \mathbf{s}$ and $\mathbf{n}_i$, the signal and noise components respectively, where the signal component has no index because it does not vary in time (note also that following the discussion above, any true sky emission counts as signal, so that $\mathbf{s} \equiv \mathbf{x}_{21} + \mathbf{x}_{\textrm{fg}}$).  Inserting this decomposition into the preceding equation and taking the expectation value of the result gives
\begin{align}
\langle \widehat{p}^ \textrm{cross}_\alpha \rangle =& \langle (\mathbf{s} + \mathbf{n}_1)^t \mathbf{E}^\alpha (\mathbf{s} + \mathbf{n}_2)\rangle \nonumber \\
=& \langle \mathbf{s}^t \mathbf{E}^\alpha \mathbf{s} \rangle + \langle \mathbf{n}_1 \rangle^t \mathbf{E}^\alpha \mathbf{s} \mathbf{s}^t \mathbf{E}^\alpha \langle \mathbf{n}_2 \rangle  + \langle \mathbf{n}_1 \mathbf{E}^\alpha \mathbf{n}_2 \rangle \nonumber \\
=& \langle \mathbf{s}^t \mathbf{E}^\alpha \mathbf{s} \rangle,
\end{align}
where the last equality holds because the instrumental noise has zero mean, i.e. $\langle \mathbf{n}_i \rangle =0$, and no cross-correlation between different times, i.e. $\langle \mathbf{n}_1 \mathbf{n}_2 \rangle = 0$.  The resulting estimator depends only on the power spectrum of the signal, and there is no additive bias.

Importantly, however, we emphasize that while we have eliminated noise \emph{bias} by computing a cross-power spectrum, we have not eliminated noise \emph{variance}.  In other words, the instrumental nosie will still contribute to the error bars.  To see this, consider the variance in our estimator, which is given by
\begin{align}
\boldsymbol \Sigma_{\alpha \beta}^{\textrm{cross}} =& \langle \widehat{p}^ \textrm{cross}_\alpha  \widehat{p}^ \textrm{cross}_\beta \rangle - \langle \widehat{p}^ \textrm{cross}_\alpha \rangle \langle \widehat{p}^ \textrm{cross}_\beta \rangle \nonumber \\ 
=& \langle  \mathbf{x}^t_1 \mathbf{E}^\alpha \mathbf{x}_2  \mathbf{x}^t_1 \mathbf{E}^\beta \mathbf{x}_2 \rangle - \langle  \mathbf{x}^t_1 \mathbf{E}^\alpha \mathbf{x}_2\rangle \langle  \mathbf{x}^t_1 \mathbf{E}^\beta \mathbf{x}_2 \rangle
\end{align}
The second term simplifies to
\begin{equation}
 \langle \widehat{p}^ \textrm{cross}_\alpha \rangle \langle \widehat{p}^ \textrm{cross}_\beta \rangle = \sum_{ijkl} \langle \mathbf{x}^i_1 \mathbf{x}^j_2 \rangle \langle \mathbf{x}^k_1 \mathbf{x}^l_2 \rangle \mathbf{E}_{ij}^\alpha \mathbf{E}_{kl}^\beta.
\end{equation}
Similarly, the first term is equal to
\begin{align}
 \langle \widehat{p}^ \textrm{cross}_\alpha  \widehat{p}^ \textrm{cross}_\beta \rangle = \sum_{ijkl} &\langle  \mathbf{x}^i_1 \mathbf{x}^j_2 \mathbf{x}^k_1 \mathbf{x}^l_2 \rangle \mathbf{E}_{ij}^\alpha \mathbf{E}_{kl}^\beta \nonumber \\
 = \sum_{ijkl}& \bigg(\langle  \mathbf{x}^i_1 \mathbf{x}^j_2 \rangle \langle \mathbf{x}^k_1 \mathbf{x}^l_2 \rangle + \langle  \mathbf{x}^i_1 \mathbf{x}^k_1 \rangle \langle \mathbf{x}^j_2 \mathbf{x}^l_2 \rangle + \langle  \mathbf{x}^i_1 \mathbf{x}^l_2 \rangle \langle \mathbf{x}^j_1 \mathbf{x}^k_2 \rangle\bigg) \mathbf{E}_{ij}^\alpha \mathbf{E}_{kl}^\beta,
\end{align}
where in the last equality we assumed Gaussian distributed data to simplify the four-point correlation.\footnote{In principle, $\x$ may exhibit departures from Gaussianity, since foregrounds are typically not Gaussian-distributed.  However, there are several reasons to expect deviations from non-Gaussianity to be unimportant.  First, the most flagrantly non-Gaussian foregrounds are typically those that are bright.  When we analyze real data in Section \ref{sec:WorkedExample}, we alleviate this problem by analyzing only a relatively clean part of the sky.  In addition, recall that in this section, $\x$ represents the data \emph{after} a best-guess model of foregrounds has been subtracted from the original measurements.  Thus, the crucial probability distribution to consider is not the foregrounds themselves, but rather the \emph{deviations} from the foregrounds, which are likely to be better-approximated by a Gaussian distribution.}  Our bandpower covariance is now
\begin{align}
\boldsymbol \Sigma_{\alpha \beta}^{\textrm{cross}} =  \sum_{ijkl}  \bigg( \langle  \mathbf{x}^i_1 \mathbf{x}^k_1 \rangle \langle \mathbf{x}^j_2 \mathbf{x}^l_2 \rangle + \langle  \mathbf{x}^i_1 \mathbf{x}^l_2 \rangle \langle \mathbf{x}^j_1 \mathbf{x}^k_2 \rangle\bigg) \mathbf{E}_{ij}^\alpha \mathbf{E}_{kl}^\beta.
\end{align}
The first term in this expression consists only of auto-correlations, which contain both noise and signal:
\begin{equation}
\langle \mathbf{x}_1 \mathbf{x}_{1}^t \rangle = \langle (\mathbf{s}+\mathbf{n}_1) (\mathbf{s}^t + \mathbf{n}_1^t) \rangle -\langle \mathbf{s}\rangle \langle \mathbf{s} \rangle^t= \mathbf{S} + \mathbf{N} = \mathbf{C},
\end{equation}
where we have defined $\mathbf{C}$ to be the total data covariance (as defined in Section \ref{sec:IdealObs}), $\mathbf{S} \equiv \langle \mathbf{s} \mathbf{s}^t \rangle-\langle \mathbf{s} \rangle \langle \mathbf{s} \rangle^t$ is the sky signal covariance (as per the discussion earlier in this section), and $\mathbf{N} \equiv \langle \mathbf{n}_1 \mathbf{n}_1^t \rangle = \langle \mathbf{n}_2 \mathbf{n}_2^t \rangle$ is the instrumental noise covariance.  We have assumed that there is no correlation\footnote{Note that this assumption has nothing to do with whether or not the instrument is sky-noise dominated.  A sky-noise dominated instrument will have instrumental noise whose \emph{amplitude} depends on the sky temperature, but the actual noise fluctuations will still be uncorrelated with the sky signal.} between the sky emission and the instrumental noise, so that $  \langle \mathbf{s} \mathbf{n}^t_1 \rangle = \langle \mathbf{s} \mathbf{n}^t_2 \rangle=0$.

The second term in our bandpower covariance consists only of cross-correlations, and thus contains no noise covariance:
\begin{equation}
\langle \mathbf{x}_1 \mathbf{x}_2^t \rangle = \langle (\mathbf{s}+\mathbf{n}_1) (\mathbf{s}^t + \mathbf{n}_2^t) \rangle = \mathbf{S}.
\end{equation}
Putting everything together, we obtain
\begin{equation}
\label{properCrossCovar}
\boldsymbol \Sigma_{\alpha \beta}^{\textrm{cross}}  = \textrm{tr}\! \left[ \mathbf{C} \mathbf{E}^\alpha \mathbf{C} \mathbf{E}^\beta \right]+ \textrm{tr}\! \left[ \mathbf{S} \mathbf{E}^\alpha \mathbf{S} \mathbf{E}^\beta \right].
\end{equation}
This, then, is the error covariance of our cross power spectrum estimator.  It gives less variance than the expression for the auto power spectrum, which in the notation of this section takes the form
\begin{equation}
\label{autoCovar}
\boldsymbol \Sigma_{\alpha \beta}^{\textrm{auto}}  = 2 \textrm{tr}\! \left[ \mathbf{C} \mathbf{E}^\alpha \mathbf{C} \mathbf{E}^\beta \right].
\end{equation}
Despite this difference between equations \ref{properCrossCovar} and \ref{autoCovar}, one may conservatively opt to use the above covariance matrix for the auto-power spectrum to estimate error bars even when using Equation \eqref{crossEstimator} to estimate the power spectrum itself.  In fact, it may be prudent to make this choice because there exists the possibility that the noise between interleaved time samples may not be truly uncorrelated, making the true errors closer to those described by $\boldsymbol \Sigma^{\textrm{auto}}$.  In our worked example with MWA data in Section \ref{sec:WorkedExample}, we will conservatively use Equation \eqref{autoCovar} to estimate the errors of our cross-power spectrum.  The task of characterizing the noise properties of the instrument thoroughly enough to eliminate this assumption is left to future work on a larger data set.

In summary, uncertainties in noise and foreground properties make it desirable to avoid trying to extract weak signals by performing subtractions between two large numbers (the contamination-dominated data and the possibly inaccurate contaminant models).  Mathematically, the greatest concern comes with the subtraction of the noise and foreground biases from power spectra estimates.  To deal with the residual noise bias, one may evaluate cross-power spectra between interleaved time samples rather than auto-power spectra.  To deal with the foreground bias, one can conservatively elect to simply leave it in when placing upper limits on the cosmological signal, and rely on the robustness of the EoR window to separate out the foregrounds from the cosmological $21\,\textrm{cm}$ signal.  In effect, one can practice foreground avoidance rather than foreground subtraction, since the former (if it is sufficient for a detection of the cosmological signal) will be more robust than the latter in the face of foreground uncertainties.  Finally, as a brute-force safeguard, to quantify such uncertainties, one can always vary the foreground model used in power spectrum estimation, as we do in Section \ref{sec:CovarModeling} when we apply our methods to the worked example of MWA data.

\subsection{A Real-World Obstacle: Incomplete $uv$-Coverage}
\label{sec:IncompleteUV}

While the methods of the previous section allow one to alleviate the effects of foreground modeling uncertainty, it is impossible to avoid the fact that real interferometers are imperfect imaging instruments.  This is because a real interferometer will inevitably have $uv$-coverage that is non-ideal in two ways.  First, the coverage is non-uniform, resulting in images that have been convolved with non-trivial synthesized beam kernels.  Second, the $uv$-coverage is incomplete, in that certain parts of the $uv$-plane are not sampled at all.  The idealized methods of Section \ref{sec:IdealObs} deals with neither problem, and in this section with augment the formalism to rectify this.

Assume for a moment that $uv$ coverage is complete (so that there are no ``holes'' in the $uv$-plane), but not necessarily uniform.  In such a scenario, one has measured an unevenly weighted sample of the Fourier modes of the sky.  The effect of this non-trivial weighting needs to be accounted for when measuring the power spectrum, since $uv$ coordinates roughly map to $k_\perp$.  A failure to do so would therefore result in the final power spectrum estimate being multiplied by some function of $k_\perp$ corresponding to the $uv$ distribution.

Put another way, the $uv$ distribution of an interferometer defines its synthesized beam, the kernel with which the true sky has been convolved in the production of our image data cube.  The equations of Section \ref{sec:IdealObs} assume that this convolution has already been undone.  Thus, we must first perform this step, which in our notation may be written as
\begin{equation}
\label{eq:Deconv}
\mathbf{x} = \mathbf{B}^{-1} \mathbf{x}^\prime,
\end{equation}
where $\mathbf{x}^\prime$ represents the convolved data vector, $\mathbf{B}$ is the convolution matrix encoding the effects of the synthesized beam, and $\mathbf{x}$ is the processed data vector that is fed into Equation \eqref{eq:unnormBands}.  Note that this application of $\mathbf{B}^{-1}$ is meant to undo only the effects of the synthesized beam, not the primary beam.

The above method assumes that the matrix $\mathbf{B}$ is invertible.  In practice, this will likely not be the case as parts of the $uv$ plane will be missed by the interferometer, resulting in a singular $\mathbf{B}$ matrix.  In what follows, we will present two different ways to deal with this.  The first is to modify the equations of Section \ref{sec:IdealObs} so that they accept the convolved images (the ``dirty maps'') as input.  Since all the statistical information relevant to the power spectrum are encoded in the covariance matrix, we simply have to make the replacement
\begin{equation}
\mathbf{C} \equiv \langle \mathbf{x} \mathbf{x}^t \rangle - \langle \mathbf{x} \rangle \langle \mathbf{x} \rangle^t  \longrightarrow \langle \mathbf{x}^\prime \mathbf{x}^{\prime \,t} \rangle - \langle \mathbf{x}^\prime \rangle \langle \mathbf{x}^\prime \rangle^t.
\end{equation}
This amounts to
\begin{equation}
\label{eq:BCB}
\mathbf{C} \longrightarrow \mathbf{B} \left( \langle \mathbf{x} \mathbf{x}^t \rangle - \langle \mathbf{x} \rangle \langle \mathbf{x} \rangle^t \right) \mathbf{B}^t = \mathbf{B} \mathbf{C} \mathbf{B}^t.
 \end{equation}
Of course, changing the covariance matrix also changes $\mathbf{C}_{,\alpha}$, and we must propagate this change.  Differentiating the preceding equation with respect to the bandpower $p_\alpha$ gives the substitution
\begin{equation}
\label{eq:BCBderiv}
\mathbf{C}_{,\alpha} \longrightarrow \mathbf{B} \mathbf{C}_{,\alpha} \mathbf{B}^t.
\end{equation}
Since $\mathbf{C}_{,\alpha}$ is the response of the data covariance $\mathbf{C}$ to the bandpower $p_\alpha$, this is simply a statement of the fact that if our data consists of dirty maps, the revised $\mathbf{C}_{,\alpha}$ matrix should encode the response of a dirty map's data covariance to the bandpower.  With the substitutions given by Equations \eqref{eq:BCB} and \eqref{eq:BCBderiv}, the rest of the equations of Section \ref{sec:IdealObs} can be used unchanged.  In the limit of an invertible $\mathbf{B}$ matrix, it is straightforward to show that this is equivalent to using Equation \eqref{eq:Deconv}.

The second method for dealing with a singular $\mathbf{B}$, which was proposed in Ref. \cite{DillonFast}, is to replace the ill-defined inverse matrix $\mathbf{B}^{-1}$ with a pseudoinverse given by
\begin{equation}
\boldsymbol \Pi \left( \mathbf{B} + \gamma \mathbf{U} \mathbf{U}^\dagger \right)^{-1} \boldsymbol \Pi,
\end{equation}
where $\gamma$ is a non-zero but otherwise arbitrary real number, and $\boldsymbol \Pi$ is a projection matrix given by
\begin{equation}
\boldsymbol \Pi \equiv \mathbf{I} - \mathbf{U} (\mathbf{U}^\dagger \mathbf{U})^{-1} \mathbf{U}^\dagger.
\end{equation}
The matrix $\mathbf{U}$ specifies which modes on the sky are missing in the data as a result of unobserved pixels on the $uv$-plane.  It is constructed by computing the responses (on the sky) of each unobserved $uv$ pixel individually and storing each response as a column of $\mathbf{U}$.  As an example, in the flat-sky approximation the $\mathbf{U}$ matrix would have a sinusoid in each column, corresponding to the fringes that would have been observed by the interferometer had data not been missing in a particular $uv$ pixel.  If these modes were present in the covariance model (which might be the case, for example, if the covariance were constructed by modeling data from a different interferometer with different $uv$ coverage), then the inverse covariance $\mathbf{C}^{-1}$ in our estimator needs to be similarly replaced with the pseudoinverse:
\begin{equation}
\boldsymbol \Pi \left( \mathbf{C} + \gamma \mathbf{U} \mathbf{U}^\dagger \right)^{-1} \boldsymbol \Pi.
\end{equation}
Importantly, the pseudoinverse can be quickly multiplied by a vector using the previously discussed conjugate gradient method. Its usage therefore does not sacrifice any of the speedups that were identified in Section \ref{sec:DataVolume} for dealing with large data volumes.

\subsection{A Real-World Obstacle: Missing Data from RFI}
\label{sec:RFI}
In any practical observation, the presence of narrowband RFI will mean that certain RFI-contaminated frequency channels will need to be flagged as outliers and omitted from a final power spectrum analysis.  The result, once again, is the presence of gaps in the data, only this time the missing modes are complete frequency channels.  However, the pseudoinverse formalism of the previous section is quite flexible in that modes of any form can be projected out of the analysis.  Thus, to correctly account for RFI-flagged data, one simply uses the pseudoinverse in exactly the same way as one does to account for missing $uv$ data.

\subsection[A Real-World Obstacle: Foreground Leakage into the \\EoR Window]{A Real-World Obstacle: Foreground Leakage into the \\EoR Window}
\label{sec:decorr}
As Equation \eqref{eq:phatWp} showed, estimates of the power spectrum are not truly local, in the sense that every bandpower estimate $\widehat{p}_\alpha$ corresponds to a weighted average of the true power spectrum, with weights specified by the window functions.  \citet{LT11} showed that these window functions can be quite broad, particularly in regions with high foreground contamination.  There is thus the danger that foreground power could leak into the EoR window.  Because the foregrounds are so much brighter than the cosmological signal, even a small amount of leakage could compromise the cleanliness of the EoR window.

Fortunately, one can exert some control over the shape of the window functions\footnote{The term ``window function'' should not be confused with the term ``EoR window''.  The former refers to the weights that specify the linear combination of the true bandpowers that each bandpower estimate represents, as per Equation \eqref{eq:phatWp}.  The latter refers to the region on the $k_\perp$-$k_\parallel$ plane that naturally has very low levels of foreground contamination, as illustrated in Figure \ref{fig:EoRWindow}.} by making wise choices regarding the form of $\mathbf{M}$ in Equation \eqref{eq:normingBands}, which in turn gives the window functions via $\mathbf{W} = \mathbf{M} \mathbf{F}$.  As discussed above, $\mathbf{M}$ must be chosen such that the rows of $\mathbf{W}$ sum to unity.  Beyond that requirement, however, an infinite number of choices are admissible.  One choice would be $\mathbf{M} = \mathbf{F}^{-1}$, which gives $\mathbf{W} = \mathbf{I}$ (i.e., delta function windows).  This would certainly minimize the amount of leakage into the EoR window, but it comes at a high price: the resulting error bars on the power spectrum measurement---the diagonal elements of $\boldsymbol \Sigma$ from Equation \eqref{eq:CovP}---tend to be large, reflecting the data's inability to make highly localized measurements in Fourier space when the survey volume is finite.

On the other extreme, the error bars predicted by $\boldsymbol \Sigma$ can be shown to be their smallest possible if $\mathbf{M}$ is taken to be diagonal \citep{THX2df}.  However, this gives broader window functions, for it is via the smoothing/binning effect of these broad window functions that the small errors can be achieved.  One can also argue that the level of smoothing dictated by this approach is excessive, since the resulting bandpowers have positively correlated errors.  (To see this, note that up to a row-dependent normalization, the error covariance matrix takes the form $\boldsymbol \Sigma \sim \mathbf{F}$.  Since all elements of a Fisher matrix must necessarily be non-negative, this implies that all cross-covariances of the estimated bandpowers have positively correlated errors unless $\mathbf{F}$ is diagonal, which is rarely the case).

As a compromise option, we advise $\mathbf{M} \sim \mathbf{F}^{-1/2}$ (again after a normalization of each row so that the window functions sum to unity).  This choice gives window functions that are narrower than those for a diagonal $\mathbf{M}$ while maintaining reasonably small error bars.  In addition, an inspection of Equation \eqref{eq:CovP} reveals that this method gives a diagonal $\boldsymbol \Sigma$, which means that errors between different bandpowers are uncorrelated.

In Section \ref{sec:MethodDemo}, we use MWA data to demonstrate the crucial role that the $\mathbf{M} \sim \mathbf{F}^{-1/2}$ choice plays in preserving the cleanliness of the EoR window.\footnote{Of course, there exist other choices that are more elaborate than the three considered in this paper.  For example, with exquisite foreground and instrumental modeling, one could imagine first decorrelating to delta-function windows by setting $\M = \F^{-1}$ in an attempt to ``perfectly'' contain the foregrounds to regions outside the EoR window, and then to re-smooth the bandpowers within the window to reduce the variance.  This is a promising avenue for future investigation, but for this paper our goal is simply to apply the $\F^{-1/2}$ decorrelator to real data (see Section \ref{sec:MethodDemo}) to demonstrate the feasibility of containing foregrounds using decorrelation techniques.}

\subsection[A Real-World Obstacle: Ensuring that Binning doesn't Destroy Error Properties]{A Real-World Obstacle: Ensuring that Binning \\doesn't Destroy Error Properties}\label{sec:cylindToSph}
 
In previous sections, we have discussed how one can preserve all the desirable properties of the power spectrum estimator of Section \ref{sec:IdealObs} in the face of all the real-world complications presented in Sections \ref{sec:DataVolume} through \ref{sec:decorr}.  The result is a rigorous yet practical estimator for the cylindrical power spectrum $P_\textrm{cyl} (k_\perp, k_\parallel)$.  We now turn to the problem of binning the cylindrical power spectrum into the cosmologically relevant spherical power spectrum $P_\textrm{sph} (k)$, with a special emphasis on the preservation of the information content of our estimator.

Just as with the cylindrical power spectrum, we parameterize the spherical power spectrum as piecewise constant, so that all the information is encoded in a vector of bandpowers $\mathbf{p}^{\textrm{sph}}$, so that:
\begin{equation}
p_\alpha^\textrm{sph} \equiv P_{\textrm{sph}} (k^\alpha).
\end{equation}
The spherical bandpowers are related to estimates of the cylindrical bandpowers $\widehat{\mathbf{p}}^\textrm{cyl}$ by the equation
\begin{equation}
\widehat{\mathbf{p}}^\textrm{cyl} = \mathbf{A} \mathbf{p}^\textrm{sph} + \boldsymbol \varepsilon,
\end{equation}
where $\mathbf{A}$ is a matrix of size $N_\textrm{cyl} \times N_\textrm{sph}$ of 1s and 0s that relates $k_\perp$-$k_\parallel$ pairs to $k$ bins, with $N_\textrm{cyl}$ and $N_\textrm{sph}$ equal to the number of cells in the $k_\perp$-$k_\parallel$ plane and the number of spherical $k$ bins respectively.  The vector $\boldsymbol \varepsilon$ is a random vector of errors on $\widehat{\mathbf{p}}^\textrm{cyl}$.  It has zero mean (assuming that one has taken the care to avoid additive bias in our estimator of the cylindrical bandpowers, as discussed above), but non-zero covariance equal to $\boldsymbol \Sigma^\textrm{cyl} \equiv \langle \boldsymbol \varepsilon \boldsymbol \varepsilon^t \rangle$, where $\boldsymbol \Sigma^\textrm{cyl}$ is given by either Equation \eqref{properCrossCovar} or \eqref{autoCovar}, depending on whether the cylindrical bandpowers were computed using cross or auto-power spectra.  (The methods presented in this section are applicable either way).
 
Our goal is to construct an optimal, unbiased estimator of $\mathbf{p}^\textrm{sph}$ from $\widehat{\mathbf{p}}^\textrm{cyl}$.  This is a solved problem \citep{TegmarkCMBmapsWOLosingInfo}, and the best estimator $\widehat{\mathbf{p}}^\textrm{sph}$ is given by
\begin{equation}
\label{eq:Cylind2SphEst}
\widehat{\mathbf{p}}^\textrm{sph} = [ \mathbf{A}^t \boldsymbol \Sigma_\textrm{cyl}^{-1} \mathbf{A} ]^{-1} \mathbf{A}^t \boldsymbol \Sigma_\textrm{cyl}^{-1} \widehat{\mathbf{p}}^\textrm{cyl},
\end{equation}
with the final error covariance on the spherical bandpowers given by
\begin{equation}
\label{eq:Cylind2SphCovar}
\boldsymbol \Sigma^\textrm{sph}_{\alpha \beta} \equiv \langle \widehat{\mathbf{p}}^\textrm{sph}_\alpha \widehat{\mathbf{p}}^\textrm{sph}_\beta \rangle -  \langle \widehat{\mathbf{p}}^\textrm{sph}_\alpha \rangle \langle  \widehat{\mathbf{p}}^\textrm{sph}_\beta \rangle =  [ \mathbf{A}^t \boldsymbol \Sigma_\textrm{cyl}^{-1} \mathbf{A} ]^{-1} .
\end{equation}
Since the $\mathbf{A}$ matrix has (by construction) a single 1 per row and zeros everywhere else, an inspection of Equation \eqref{eq:Cylind2SphCovar} reveals that a diagonal $\boldsymbol \Sigma^{\textrm{cyl}}$ implies a diagonal $\boldsymbol \Sigma^\textrm{sph}$.  In other words, the estimator given by Equation \eqref{eq:Cylind2SphEst} preserves the decorrelated nature of the $\M \sim \F^{-1/2}$ version of the cylindrical power spectrum estimator defined in Section \ref{sec:decorr}.  This will not be the case for an arbitrary estimator (such as one that is formed from taking uniformly weighted Fast Fourier Transforms, then squaring and binning).  We also emphasize that if one does not choose to use decorrelated cylindrical bandpower vectors, Equations \eqref{eq:Cylind2SphEst} and \eqref{eq:Cylind2SphCovar} require that one keep full track of the off-diagonal terms of $\boldsymbol \Sigma_\textrm{cyl}^{-1}$.  Without it, a consistent propagation of errors to the spherical power spectrum is not possible, and may even lead to a mistakenly claimed detection of the cosmological signal, as we discuss in Section \ref{sec:MethodDemo} and in Appendix \ref{sec:ErrorCovAppendix}.

Just as with the cylindrical power spectra, we would like to compute the window functions.  The definition of the spherical window functions are exactly analogous to that provided in Equation \eqref{eq:phatWp} for the cylindrical power spectrum, so that
\begin{equation}
\langle \mathbf{\widehat{p}}^\textrm{sph} \rangle = \mathbf{W}^\textrm{sph} \mathbf{p}^\textrm{sph}.
\end{equation}
Taking the expectation value of Equation \eqref{eq:Cylind2SphEst}, we have
\begin{align}
\langle \mathbf{\widehat{p}}^\textrm{sph} \rangle &= [ \mathbf{A}^t \boldsymbol \Sigma_\textrm{cyl}^{-1} \mathbf{A} ]^{-1} \mathbf{A}^t \boldsymbol \Sigma_\textrm{cyl}^{-1} \langle \widehat{\mathbf{p}}^\textrm{cyl} \rangle \nonumber \\
 &= [ \mathbf{A}^t \boldsymbol \Sigma_\textrm{cyl}^{-1} \mathbf{A} ]^{-1} \mathbf{A}^t \boldsymbol \Sigma_\textrm{cyl}^{-1}  \mathbf{W}^\textrm{cyl} \mathbf{A} \mathbf{p}^{\textrm{sph}},
\end{align}
where we have used the definition of the cylindrical window functions to say that $\langle \widehat{\mathbf{p}}^\textrm{cyl} \rangle = \mathbf{W} \mathbf{p}^\textrm{cyl}$, as well as the fact that $\mathbf{p}^\textrm{cyl} = \mathbf{A} \mathbf{p}^\textrm{sph}$ (with no error term because we are relating the true cylindrical bandpowers to the true spherical bandpowers).  Inspecting this equation, we see that
\begin{equation}
\mathbf{W}^\textrm{sph} = [ \mathbf{A}^t \boldsymbol \Sigma_\textrm{cyl}^{-1} \mathbf{A} ]^{-1} \mathbf{A}^t \boldsymbol \Sigma_\textrm{cyl}^{-1}  \mathbf{W}^\textrm{cyl} \mathbf{A}.
\end{equation}
Therefore, by measuring the width of the spherical window functions (rows of $\mathbf{W}^\textrm{sph}$), one can place rigorous horizontal error bars on the final spherical power spectrum estimate.

\subsection{Summary of the issues}
In the last few sections, we have provided techniques for dealing with a number of real-world obstacles.  These include:
\begin{enumerate}
\item Taking advantage of the flat-sky approximation and the rectilinearity of data cubes, as well as the conjugate gradient algorithm for matrix inversion to allow large data sets to be analyzed quickly.
\item Using cross-power spectra rather than auto-power spectra in order to eliminate noise bias.
\item Replacing inverses with pseudoinverses to deal with data that has missing spatial modes (due to incomplete $uv$ coverage) and missing frequency channels (due to RFI).
\item Performing power spectrum decorrelation to avoid the leakage of foreground power into the EoR window.
\item Binning of cylindrical power spectra into spherical power spectra in a way that preserves desirable error properties.
\end{enumerate}
Crucial to this is the fact that these techniques all operate under a self-consistent framework.  This allows faithful error propagation that accurately captures how various real-world effects act together.  For example, it was shown in \cite{DillonFast} that properly accounting for pixelization effects in Equation \eqref{eq:pixDef} results in low Fisher information at high $k_\parallel$, providing a marker for parts of the $k_\perp$-$k_\parallel$ plane that cannot be well-constrained because of finite spectral resolution.  The identification of such a region would be trivial if one had spectrally contiguous data, for then one would simply say that the largest measurable $k_\parallel$ was roughly $1/\Delta L_\parallel$, where $\Delta L_\parallel$ is the width of a single frequency channel mapped into a cosmological line-of-sight distance.  However, such a straightforward analysis no longer applies when there are RFI gaps in the data at arbitrary locations.  In contrast, the unified framework presented in this paper allows all such complications to be folded in correctly.

\section{A Worked Example: Early MWA Data}
\label{sec:WorkedExample}
Now that we have bridged the gap between theoretical techniques for analyzing ideal data and the numerous challenges presented by real data, we are ready to bring together our methods, specify a covariance model, and estimate power spectra from MWA 32-tile prototype (MWA-32T) data.  The data were taken between the 21st and 29th of March 2010, the first observing campaign during which data were taken that were scientifically useful.  The observations are described in more detail by \cite{ChrisMWA}.  Real data affords us two opportunities.  In this section, we look at the data to examine and quantify the differences between power spectrum estimators and the pitfalls associated with choice of estimator.  In Section \ref{sec:earlyResults}, we take advantage of everything we have developed to arrive at interesting new foreground results and a limit on the 21 cm brightness temperature power spectrum.

\subsection{Description of Observations}

All of the data used for this paper were taken on the MWA-32T system.  This system has since been upgraded to a 128-tile instrument (MWA-128T; \citet{TingaySummary, BowmanMWAScience}), but in this paper we focus exclusively on MWA-32T data, reserving the MWA-128T data for future work.

The MWA-32T instrument consisted of 32 phased-array ``antenna tiles'' which served as the primary collecting elements.  Each tile contained 16 dual linear-polarization wideband dipole antennas which were combined to form a steerable beam with a full width at half maximum (FWHM) size of $\sim25^\circ$ at 150~MHz.  The array had an approximately circular layout with a maximum baseline length of $\sim340$~m, and a minimum baseline length of 6.6~m, although the shortest operating baseline during this observational campaign was 16~m.  After digitization, filtering, and correlation, the final visibilities had a 1~second time resolution and 40~kHz spectral resolution over a 30.72~MHz bandwidth.    The instrumental capabilities are summarized in Table~\ref{tab:mwa32t}.

\begin{table*}
\begin{center}
\begin{tabular}{| l|l|}
\hline
Field of View (Primary Beam Width) & $\sim25^\circ$ at 150~MHz \\ \hline
Angular Resolution & $\sim 20'$ at 150~MHz \\ \hline
Collecting Area & $\sim 690~{\rm m}^2$ towards zenith at 150~MHz  \\ \hline
Polarization & Linear X-Y \\ \hline
Frequency Range & 80~MHz to 300~MHz \\ \hline
Instantaneous Bandwidth & 30.72~MHz \\ \hline
Spectral Resolution & 40~kHz \\
\hline
\end{tabular}
\caption{MWA-32 Instrument Parameters\label{tab:mwa32t}}
\end{center}
\end{table*}

For our worked example, we concentrate on March 2010 observations of the MWA ``EoR2'' field.  It is centered located at ${\rm R.A.(J2000)} = 10^{\rm h}\ 20^{\rm m}\ 0^{\rm s}$,
${\rm decl.(J2000)} = -10^\circ\ 0'\ 0''$, and is one of two fields at high Galactic latitude that have been identified by the MWA collaboration as candidates for deep integrations, owing to their low brightness temperature in low frequency measurements of Galactic emission \citep{Haslam,Angelica}.  For further details about the observational campaign or the EoR2 field, please see \citet{ChrisMWA}, which was based on the same set of observations as the ones used in this paper.  

Observations covered three 30.72~MHz wide bands, centered at $123.52~{\rm MHz}$, $154.24~{\rm MHz}$ and $184.96~{\rm MHz}$, corresponding to a redshift range of $6.1 < z < 12.1$ (the redshift range of the results presented in this work is slightly smaller because of data flagging)  for the 21~cm signal.  The $123.52~{\rm MHz}$ and $154.24~{\rm MHz}$ bands were observed for approximately 5 hours each, and the $184.96~{\rm MHz}$ band was observed for approximately 12 hours.

These early data from the prototype have provided us with a set of test data that enabled development of extensive analysis methods and software on which the results of this paper are based.  The early prototype had shortcomings (e.g., mismatched cables, receiver firmware errors, correlator timing errors) that compromised the calibration to some extent, raising the apparent noise level.  Additionally, the instrument was only operating with $\lesssim 29$ tiles, and with a 50\% duty cycle throughout the course of these observations.  We account for this in Section \ref{sec:CovarModeling} by determining the magnitude of the noise empirically, in order to be able to place rigorously conservative upper limits on the cosmological power spectrum.  We expect that data from later prototype campaigns and from the full array will produce result closer to theoretical expectations.

\subsection{Mapmaking}
\label{sec:Mapmaking}
Before the data can be used as a worked example for our power spectrum estimator, however, we must convert the measured visibilities into a data cube of sky images at every frequency in our band. In other words, we must form the data vector $\mathbf{x}$, defined by Equation \eqref{eq:pixDef}, which serves as the input for our power spectrum pipeline.

To form the data vector, we performed the following steps.  First, we performed a reduction procedure similar to that described in \citet{ChrisMWA} for the initial flagging and calibration of the data.  Hydra~A was identified as the dominant bright source in the field, and used for calibration assuming a point source model.  The Hydra~A source model was then subtracted from the {\em uv} data.  As this same source model was also used for gain and phase calibration, this can be thought of as a ``peeling'' source removal procedure \citep{Noordam2004,vanderTol2007,Mitchell2008,Intema2009} on a single source.  Alternatively, in the absence of gridding artifacts, this is equivalent to imaging the point-source model and subtracting it from the data as part of the direct foreground subtraction step discussed in the first step of Section \ref{sec:crossPower} \citep{TegmarkCMBmapsWOLosingInfo}.

The subtracted data were imaged using the CASA task {\tt clean} without deconvolution to produce ``dirty'' images.  No multi-frequency synthesis was performed, so that the full 40~kHz spectral resolution of the data would be available.  The visibilities were gridded using w-projection kernels \citep{CornwellWProj} with natural (inverse-variance) weighting to produce maps at each frequency with a cell size of $3'$ over a $25.6^\circ$ field of view.  The resulting cubes contained $\sim200$ million voxels, with 512 elements along each spatial dimension and 768 elements in the frequency domain.  It is important to note that the pre-flagging performed on the data resulted in the flagging of entire frequency bands (which means that there are gaps in the final data cube).  Cubes were generated  for each 5~minute snapshot image.

The individual snapshot data cubes were combined using the primary beam inverse-variance weighting method described in \citet{ChrisMWA}.  The weighting and primary beams were simulated separately for each 40~kHz frequency channel in each 5~minute snapshot.  The combined maps and weights were saved, along with the effective point spread function at the center of the field.  Two additional data cubes were created by averaging alternating 5~minute snapshots (i.e. even numbered snapshots were averaged into one cube, and odd numbered snapshots were averaged into the other) so that they were generated from independent data, but with essentially the same sky and {\em uv} coverage properties.

A further flux scale calibration of the integrated cubes was performed using three bright point sources: MRC~1002-215, PG~1048-090, and PKS~1028-09 to set the flux scale on a channel-by-channel basis.  A two dimensional Gaussian fitting procedure was used to fit the peak flux of each of these sources in each 40~kHz channel of the data cube. Predictions for each source were derived by fitting a power law to source measurements from the 4.85~GHz Parkes-MIT-NRAO survey \citep{Griffith1995}, the 408~MHz Molonglo Reference Catalog \citep{Large1981}, the 365~MHz Texas Survey \citep{Douglas1996}, the 160~MHz and 80~MHz Culgoora Source List \citep{Slee1995} and the 74~MHz VLA Low-frequency Sky Survey \citep{Cohen2007}.  A weighted least-squares fit was then performed to calculate and apply a frequency-dependent flux scaling for the cube to minimize the square deviations of the source measurements from the power law models. 

An additional flagging of spectral channels was performed based on the root-mean-square (RMS) noise in each spectral channel of the cube.  A smooth noise model was determined by median filtering the RMS channel noise as a function of frequency (bins of 16 channels were used in the filtering).  Any channel with $5\sigma$ or larger deviations from the smoothed noise model was flagged.  Upon inspection, these additional flagged channels were observed to be primarily located at the edges of the coarse digital filterbank channels, which were corrupted due to an error in the receiver firmware.  After this procedure, approximately one third of the spectral channels were found to have been flagged.

Each individual map covered $25.6^\circ \times 25.6^\circ$ at a resolution of $3'$ with 768 frequency channels (40~kHz frequency resolution).  To decrease the computational burden of the covariance estimation, each map was subdivided into 9 subfields, and the pixels were averaged to a size of $15'$.  The data cubes were mapped to comoving cosmological coordinates using WMAP-7 derived cosmological parameters, with $\Omega_{\rm M}=0.266$, $\Omega_\Lambda=0.734$, $H_0=71~{\rm km}~{\rm s}^{-1}~{\rm Mpc}^{-1}$, and $\Omega_k \equiv 0$ \citep{WMAP7Cosmology}.  

At this point, the data cubes were ready to be used as input data to our power spectrum estimator, i.e., we had arrived at the final form of the data vector $\mathbf{x}$.  However, estimating power spectra and error statistics using the formalism of Section \ref{sec:Methods} also requires a covariance model, which we construct in the next section.

\subsection{Covariance Model}
\label{sec:CovarModeling} 
We follow \cite{LT11} and \cite{DillonFast} in modeling the covariance matrix $\C$ as the sum of independent parts attributable to noise and foregrounds.  We leave off the signal covariance because it only contributes to the final error bars by accounting for cosmic variance---a completely negligible effect in comparison to foreground and noise-induced errors.  We adopt a conservative model of the extragalactic foregrounds by treating them as a Poisson random field of sources with fluxes less than 100 Jy, after the manual removal of Hydra A.  By treating all extragalactic foregrounds as ``unresolved,'' we effectively throw out information about which lines of sight are most contaminated by bright foregrounds.  As \cite{DillonFast} showed, future analyses can improve on our limits by including more information about the foregrounds. We begin with the parameterized covariance model of \cite{LT11}, 
\begin{align}
C_{ij}^\text{unresolved} =& \left(1.4 \times 10^{-3} \text{ }\frac{\text{K}}{\text{Jy}} \right)^2 \left( \int_0^{S_\text{cut}} S^2 \frac{dn}{dS}dS \right) \left(\frac{\nu_i \nu_j}{\nu_*^2} \right)^{-2-\bar{\kappa}} \times \nonumber \\ 
&    \exp\left[ \frac{\sigma_{\bar{\kappa}}^2}{2} \left(\ln \left[\frac{\nu_i \nu_j}{\nu_*^2} \right] \right)^2 \right] \times   \exp \left[ \frac{(\mathbf{r}_{\perp i} - \mathbf{r}_{\perp j})^2}{2 \sigma_\perp^2}\right] \bigg(\Omega_\text{pix}\bigg)^{-1}
\label{eq:Umodel}
\end{align}
where $\nu_* = 150 \,\textrm{MHz}$ is a reference frequency, $\nu_i$ is the frequency of the $i$th voxel, which has an angular distance of $\mathbf{r}_{\perp i}$ from the field center.  The spectral index is $\bar{\kappa} = 0.5$, the uncertainty in the spectral index is $\sigma_\kappa = 0.5$, the clustering correlation length is $\sigma_\perp = 7'$, $\Omega_\text{pix}$ is the angular size of each pixel, the flux cut $S_\text{cut} = 100$ Jy, and $dn/dS$ is the differential source count from \cite{dimatteo1},
\begin{align}
\frac{dn}{dS} = (4000 \text{ Jy}^{-1} \text{sr}^{-1})\nonumber \times \begin{cases} \left(\frac{S}{0.880 \text{ Jy}}\right)^{-2.51} & \text{for S $>$ 0.880 Jy} \\ \left(\frac{S}{0.880 \text{ Jy}}\right)^{-1.75} & \text{for S $\leq$ 0.880 Jy.} \end{cases}
\end{align}
We adapt this model for the fast power spectrum estimation method outlined in Section \ref{sec:DataVolume} by calculating the translationally invariant approximation to this model in the manner described in \cite{DillonFast}.

For the Galactic synchrotron, we also follow \cite{LT11} and \cite{DillonFast} for the parameterization of the synchrotron emission covariance.  Namely, we adopt $\bar{\kappa} = 0.8$, $\sigma_\kappa = 0.1$, $\sigma_\perp = 30^\circ$, and replace the first three terms of the covariance in Equation \ref{eq:Umodel} with $T^2_\text{synch} = (335.4 \text{ K})^2$.

Our model for the instrumental noise is adopted from \cite{DillonFast}, with one key difference: the overall normalization.  For each subband, we let the noise covariance matrix scale by a free multiplicative constant.  This is equivalent to treating the combination $T_\text{sys}^2 / ( A_\text{ant}^2 t_\text{obs})$ as a free parameter.  We then fit for that parameter by requiring the RMS difference between the two time slices---which should be free of sky signal---for the densely sampled inner region of $uv$ space and rescaling our noise covariance matrix to match.  The spatial structure of the covariance was left unchanged.  Even though the data is somewhat nosier than suggested by a first principles calculation assuming fiducial values for system temperature and antenna effective area, this empirical renormalization allows for an honest account of the errors introduced by instrumental effects.

To verify that our parameterization of the foregrounds was reasonable, we varied these parameters over an order of magnitude and found that they had little effect on our final power spectrum estimates, except at the lowest values of $k$.  There are two reasons for this: first, since we are only measuring the power spectrum of the sky, we need not worry about precisely subtracting foregrounds.  Second, because the noise in our instrument is still more than two orders of magnitude from the cosmological signal, in the EoR window our band power measurements will be noise dominated and agnostic to our foreground model.  Future analyses might include a more thorough treatment of the foregrounds, especially by utilizing the full power of the \citet{DillonFast} method to include information about the positions, fluxes, and spectral indices of individual point sources.

\subsection{Evaluating Power Spectrum Estimator Choices}
\label{sec:MethodDemo}

With both a data vector $\mathbf{x}$ and a covariance matrix $\mathbf{C}$ in hand, we can now apply the methods of Section \ref{sec:Methods} to estimate power spectra.  In doing so, we deal with real-world obstacles using all of the techniques that we have developed.  In this section, we show why all this is necessary.

In Section \ref{sec:decorr} we touted the choice of power spectrum estimator $\widehat{\mathbf{p}} = \M \mathbf{q}$ with $\M \sim \F^{-1/2}$ as a compromise solution between the choice with the smallest error bars, $\M \sim \Eye$, and the choice with the narrowest window functions, $\M \sim \F^{-1}$.  In the race to detect the power spectrum from the EoR, one might be tempted to aggressively seek out the smallest possible errors.  This could prove a deleterious choice, as we will now show using MWA-32T data.
 
\begin{figure}[!ht] 
	\centering 
	\includegraphics[width=1\textwidth]{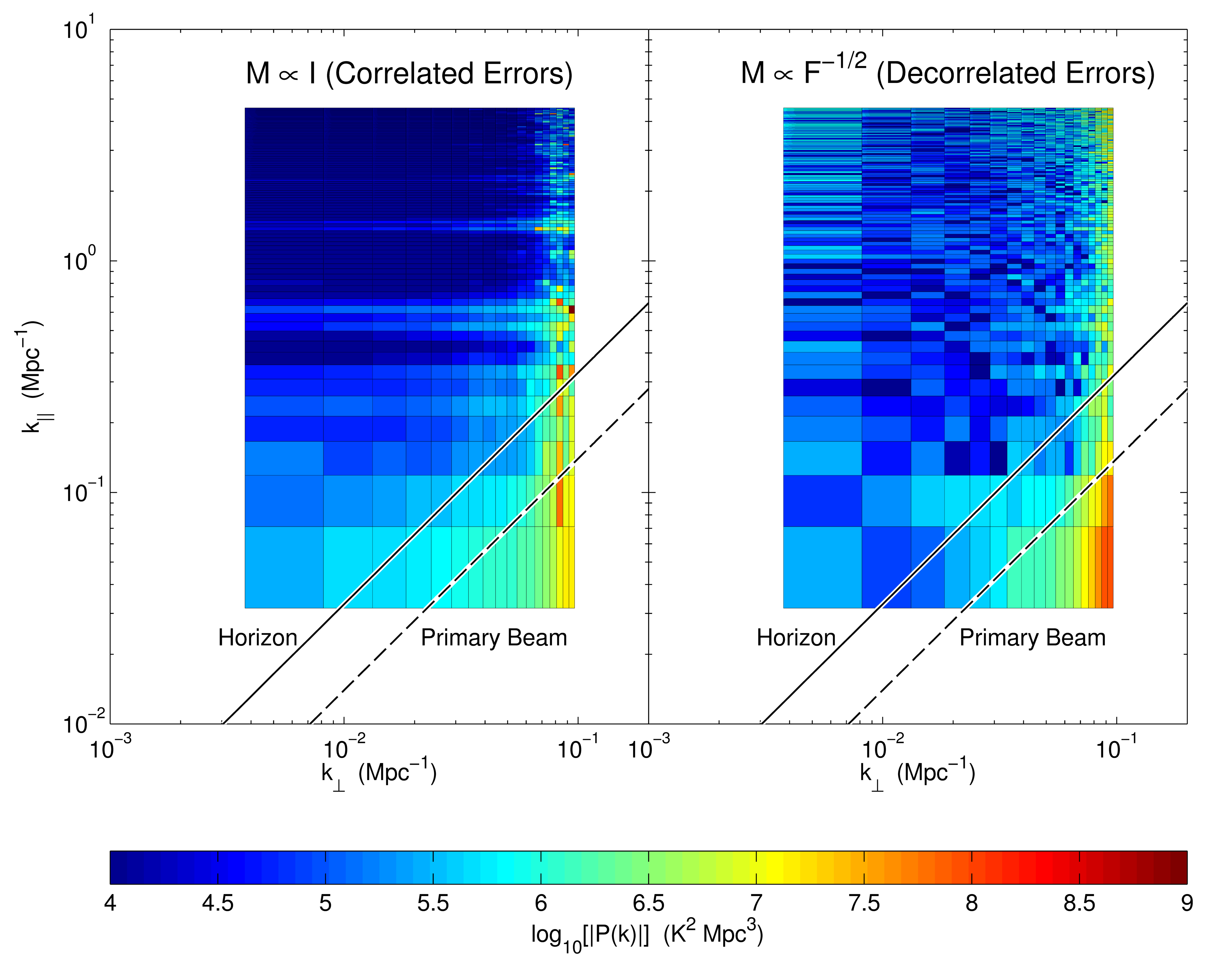}
	\caption[Effect of minimum-variance vs. decorrelated estimators on $|P(\kvec)|$.] {Unless one chooses a power spectrum estimator with decorrelated errors, foregrounds and other instrumental effects can leak significantly into the EoR window. Here we show the \textit{absolute value} of the cylindrical power spectrum estimate from the subband centered on 158 MHz ($z = 8.0$)  and averaged over all 9 fields.  On the left, we have set $\M \sim \Eye$.  On the right, $\M \sim \F^{-1/2}$.  We expect contamination from smooth spectrum foregrounds interacting with the chromatic synthesized beam to occupy the ``wedge'' portion of Fourier space, defined in Equation \eqref{eq:wedge}.  Optimistically, the wedge is delimited by the extent of the main lobe of the primary beam; conservatively, we should not see bright foreground contamination beyond the horizon.  In the regions where the power spectrum is noise dominated, we expect little structure in the $k_\|$ direction in the EoR window above some moderate value of $k_\|$.  In the left panel, we see considerably more $k_\|$ structure in the form of horizontal bands, attributable to foreground contamination and instrumental effects, that has leaked into the putative EoR window.}
	\label{fig:2DPkComparison}
\end{figure}

First, in Figure \ref{fig:2DPkComparison} we compare cylindrical power spectra, $\widehat{\mathbf{p}}$, generated using two different estimators of the power spectrum that we presented in Section \ref{sec:decorr}.\footnote{In our comparison of choices for $\M$, we drop the $\M \sim \F^{-1}$, $\delta$-function windows choice.  In addition to proving the noisiest estimator, it suffers from strong anti-correlated errors.  We adopt the perspective that the important comparison is between the ``obvious'' choice, the minimum variance $\M \sim \Eye$, and our preferred choice with decorrelated errors, $\M \sim \F^{-1/2}$.}
On the left, we have used $\M \sim \Eye$, the estimator with the smallest error bars, and on the right we have used $\M \sim \F^{-1/2}$, the estimator with decorrelated errors.  In both cases, we have plotted the absolute value of the power spectrum estimates (which can be negative because they are cross-power spectra).  Because the two estimates are related to one another by an invertible matrix, they contain the same cosmological information. In a sense, the $\M \sim \F^{-1/2}$ method is the most honest estimator of the power spectrum because the band powers form a mutually exclusive and collectively exhaustive set of measurements.
In other words, they represent all the all the power spectrum information from the data, divided into independent pieces. 

Moreover, just because two sets of estimators have the same information content does not mean that they are equally useful for distinguishing the cosmological power spectrum from foreground contamination.   In Figure \ref{fig:2DPkComparison}, the minimum variance estimator for the power spectrum introduces considerable foreground contamination into the EoR window, demarcated by the expected angular extent of the wedge feature (which we introduced in Section \ref{sec:Intro} and will discuss in greater detail in Section \ref{sec:WedgeResults}).  Even highly suspect features at high $k_\perp$ where $uv$ coverage is spottiest seem to get smeared across $k_\perp$ and into the EoR window.  We cannot simply cut out the wedge from our cylindrical-to-spherical binning and expect a clean measurement of the power spectrum in the EoR window.
 
\begin{figure}[!h]  
	\centering 
	\includegraphics[width=1\textwidth]{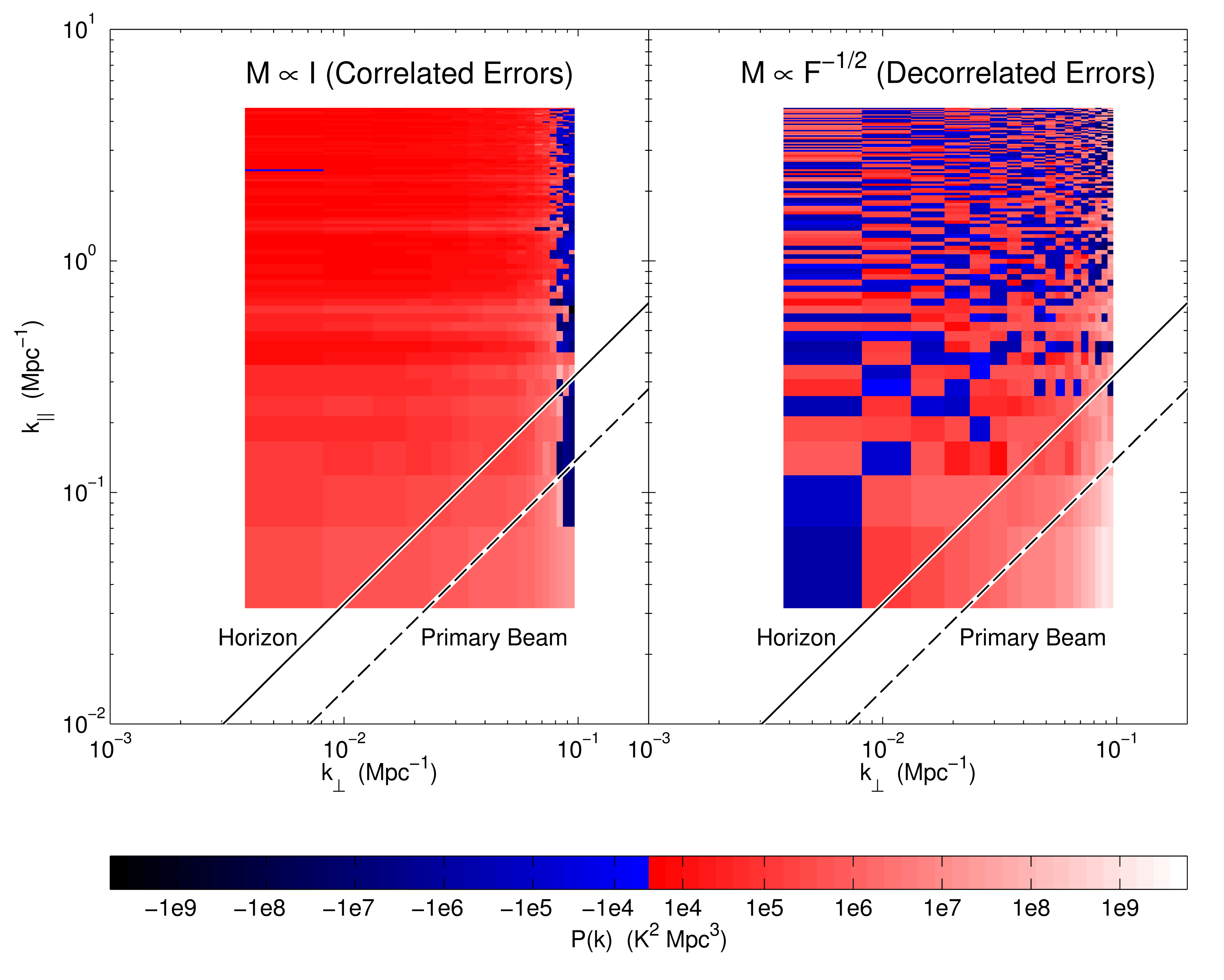}
	\caption[Effect of minimum-variance vs. decorrelated estimators on sign($P(\kvec)$).]{One advantage of calculating the cross power spectrum of interleaved time-slices of data is that we can easily tell which regions of Fourier space are noise dominated. Here we reproduce the power spectra from Figure \ref{fig:2DPkComparison} without taking the absolute value of $P(k)$.  By plotting with a discontinuous, sinh$^{-1}$ color scale, it is easy to see that the EoR window for our decorrelated power spectrum estimate (right panel) has roughly an equal number of positive and negative band power estimates---exactly what we would expect from a noise dominated region.  By contrast, our power spectrum estimate with correlated errors (left panel) shows positive power over almost all of Fourier space, indicating ubiquitous leakage of contaminants into the EoR window.}
	\label{fig:2DArcsinhPkComparison}
\end{figure}

Looking closely at Figure \ref{fig:2DPkComparison}, one might notice that some regions of the EoR window on the lefthand panel still seem very clean---cleaner perhaps that the same regions in the righthand panel.  To examine that apparent fact, we plot $\widehat{p}_\alpha$ instead of $|\widehat{p}_\alpha|$ in Figure \ref{fig:2DArcsinhPkComparison}.  
To make the figure more intelligible, we have plotted colors based on an sinh$^{-1}$ color scale with a sharp color division at 0.  The sinh$^{-1}$ has the advantage of behaving linearly at small values of $\widehat{p}_\alpha$ and logarithmically at large positive or negative $\widehat{p}_\alpha$. 

What emerges is a striking difference between the two estimators.  For the reasons discussed in Section \ref{sec:crossPower}, we have chosen to estimate the cross power spectrum between two time-interleaved sets of observations.  As a result, we expect that instrumental noise should be equally likely to contribute positive power as it is to contribute negative power.  In noise dominated regions of the $k_\perp$-$k_\|$ plane, we expect about half of our measurements to be positive and about half to be negative.  That is exactly what we see in the EoR window of the $\M \sim \F^{-1/2}$ estimator.  However, the $\M \sim \Eye$ estimator in the lefthand panel clearly shows positive power throughout the entire supposed EoR window.  Though the magnitude of that power is not enormous---often it is well within the vertical error bars---the overall bias towards positive cross power means that sky signal is contaminating the EoR window.  This is precisely the problem we were worried about in Section \ref{sec:decorr} and the data have clearly manifested it.\footnote{Of course, as we noted in Section \ref{sec:decorr}, the choice of $\M \sim \F^{-1/2}$ is not unique in its ability to mitigate foreground leakage, and other choices certainly warrant future investigation.  Picking $\M \sim \F^{-1/2}$ is, however, a good choice for a first attempt at decorrelation, particularly given its various other desirable properties that we have described.  The important point here is that while $\M \sim \F^{-1/2}$ may not be necessarily optimal for containing foregrounds within the wedge, our results show that it is a reasonable one.  In contrast, the ``straightforward'' approach of normalizing the power spectrum with the diagonal choice $\M \sim \Eye$ is clearly ill-advised.}

This also explains why there appeared to be less power in the EoR window of the lefthand panel of Figure \ref{fig:2DPkComparison}; by taking the absolute value of the weighted average of positive and negative quantities, we expect to measure a smaller absolute value of the power.  However, as this figure clearly shows, that weighted average is biased by foreground leakage.  And, even though there still appears to be a region just inside the EoR window that retains positive band powers consistent with foregrounds, that small amount of leakage can be attributed to finite sized windows functions and to calibration uncertainties.  Regardless, it does not appear to be an insurmountable limitation to the cleanliness of the EoR window; rather, it suggests that we should be careful in how we demarcate the EoR window when calculating spherically-averaged power spectra.
 
\begin{figure}[!h]  
	\centering 
	\includegraphics[width=1\textwidth]{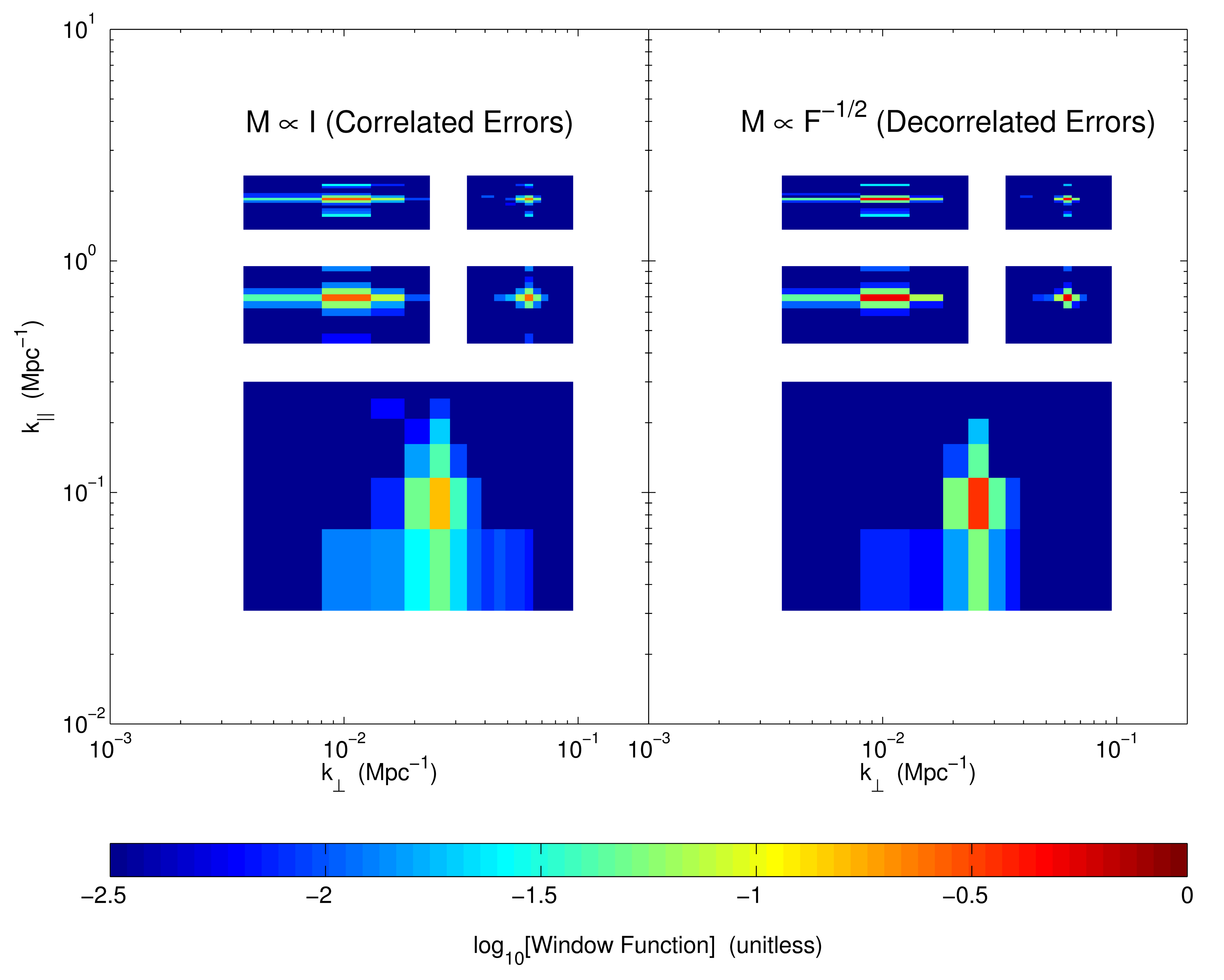}
	\caption[Effect of minimum-variance vs. decorrelated on 2D window functions.]{By using an estimator of the 21 cm power spectrum with  uncorrelated errors, we significantly narrow the window functions that relate the ensemble average of our estimator to the true, underlying power spectrum.  Here we show a sample of five cropped window functions for the power spectrum estimate in Figure \ref{fig:2DPkComparison}, each centered at their maxima, for both an estimator with correlated errors (left panel) and an estimator with uncorrelated errors (right panel).  Though the estimator with correlated errors produces smaller vertical error bars, it acheives this by ``over-smoothing'' many band powers together.  Narrow window functions let us independently measure many modes of the power spectrum. The band power measured with $\M \sim \F^{-1/2}$ is one of a set of mutually exclusive and collectively exhaustive pieces of information.} 
	\label{fig:2DWindowComparison}
\end{figure}

In addition to producing a cleaner EoR window, the decorrelated estimator of the power spectrum yields another advantage: narrower window functions.  Both the estimator with the minimum variance and estimator with decorrelated errors represent, in aggregate, the weighted average of the true, underlying band power spectrum, as we discussed in Section \ref{sec:IdealObs}. In Figure \ref{fig:2DWindowComparison}, we show the improvement that the decorrelated estimator offers over the minimum variance estimator by narrowing the window functions considerably.\footnote{While the choice of $\M \sim \F^{-1/2}$ ensures that the power spectrum estimator covariance is diagonal (recall, $\boldsymbol \Sigma = \M \F \M^t$ while $\mathbf{W} = \M \F$), it does not mean that the window functions are delta functions.  The off-diagonal terms of $\boldsymbol \Sigma$ might be zero even if the off-diagonal terms of $\mathbf{W}$ are not.}
We show five example window functions from the same subband that we plot in Figure \ref{fig:2DPkComparison}, cropped to fit together on one set of axes, each centered at their respective peaks.  Because the window functions are normalized to sum to 1, the breadth of each window function is reflected by the value of the central peak.  As we expected, the window functions are considerably narrower for our decorrelated power spectrum estimator.

\begin{figure}[!h] 
	\centering 
	\includegraphics[width=.6\textwidth]{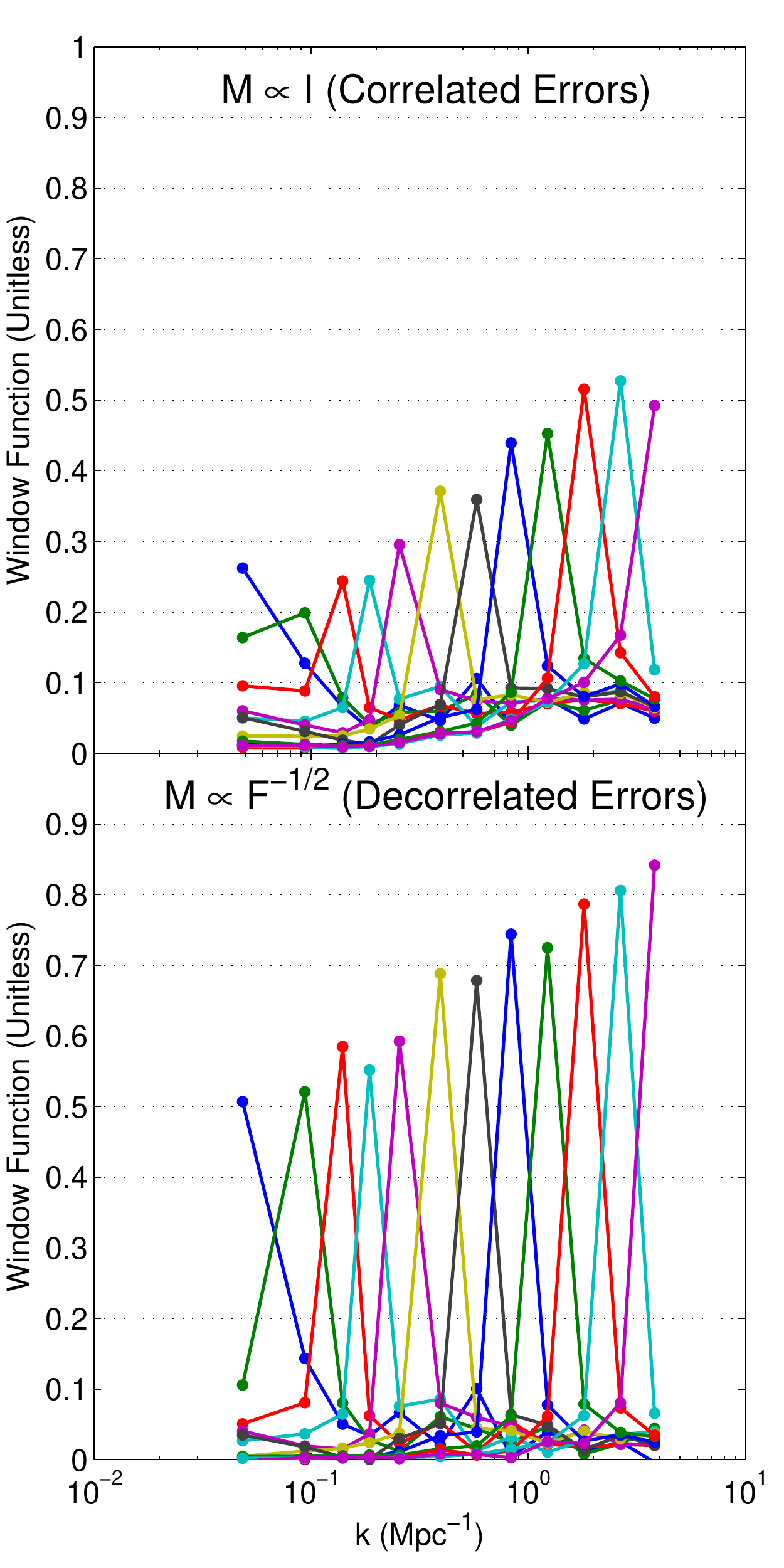}
	\caption[Effect of minimum-variance vs. decorrelated on 1D window functions.]{Even after optimally binning the cylindrical power spectra from Figure \ref{fig:2DPkComparison} to spherical power spectra, the choice of a power spectrum estimator with decorrelated errors produces much narrower window functions than the minimum variance technque.  In addition to maintaining a clean EoR window, the choice of $\M \sim \F^{-1/2}$ provides the additional benefit of allowing power spectrum modes to be measured more independently.}
	\label{fig:1DWindowComparison}
\end{figure}

Even after binning from cylindrical power spectra to spherical power spectra, the difference remains quite stark.  In Figure \ref{fig:1DWindowComparison} we see clearly that choosing a power spectrum estimator with decorrelated errors also considerably improves the window functions in one dimension as well as two.

Lastly, as we mentioned in Section \ref{sec:cylindToSph}, one of advantage of our method is that it keeps a full accounting of the error covariance, $\mathbf{\Sigma}$. When $\mathbf{M}$ is not chosen to make $\mathbf{\Sigma}$ diagonal, an improper accounting can lead to a suboptimal or simply incorrect propagation of errors.  In Appendix \ref{sec:ErrorCovAppendix} we work through an example of the consequences of assuming the independence of errors at various steps in the analysis.  This should serve as a warning of the importance of careful analysis; incorrectly assuming a diagonal $\mathbf{\Sigma}$ can lead to unnecessarily wide window functions, an overestimation of errors, or---worst of all---an underestimation of errors that could lead to an unjustified claim of a detection.

 
\section{Early Results}
\label{sec:earlyResults}
Having developed and demonstrated a technique that robustly preserves the EoR window while thoroughly and honestly keeping track of the errors on and correlations between our band power estimates, we can now confidently generate some interesting preliminary science results.  Because these data span the widest redshift range to date, we are able to investigate the behavior of the wedge feature over many frequencies.  Understanding the behavior of the EoR window over a large redshift range is important, since there is still considerable uncertainty about the timing and duration of the EoR.  Moreover, it is often argued that a tentative first detection of the cosmological signal will only be convincingly distinguishable from residual foregrounds if one can show that the $21\,\textrm{cm}$ brightness temperature fluctuations peak at some redshift, since theory predicts that the midpoint of reionization should be marked by such a peak \citep{LidzRiseFall,BittnerLoeb}.  It is therefore essential to characterize the EoR window (and by extension, residual foregrounds) over a broad frequency range. We also apply our methods from Section \ref{sec:Methods} to calculate spherically averaged power spectra over our entire redshift range, including error bars and window functions, thus setting a limit on the 21 cm brightness temperature power spectrum during the EoR.

\subsection{The Wedge} 
\label{sec:WedgeResults}

\begin{figure} [!h] 
	\centering 
	\includegraphics[width=1\textwidth]{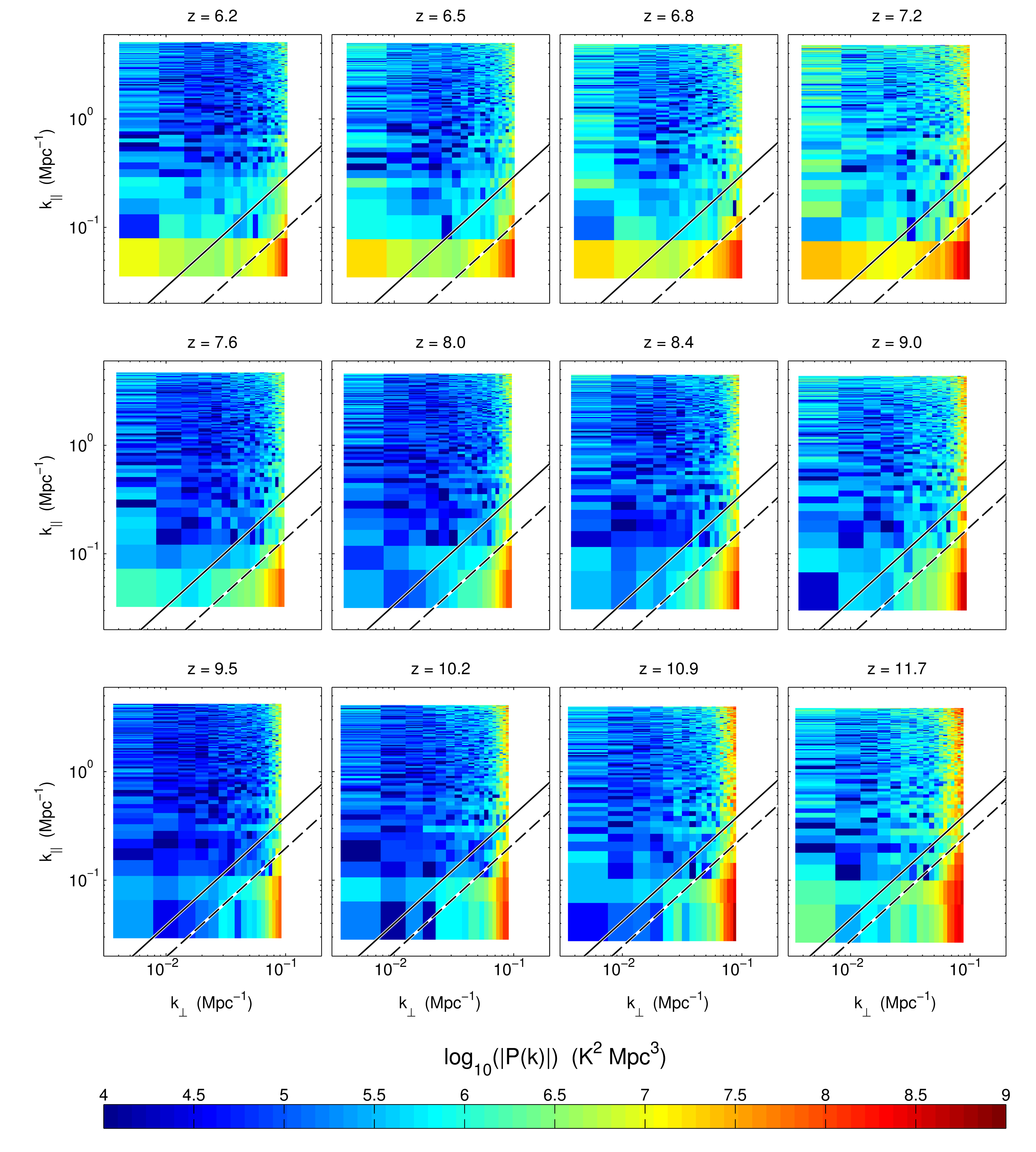}
	\caption[2D $P(\kvec)$ for all bands showing the evolution of the wedge.]{Examining cylindrically binned power spectra for each subband (each averaged over all nine subfields), reveals several important trends with frequency of the EoR window and the foregrounds.  Each row is a single simultaneously observed frequency band.  Since different bands were observed for different amounts of time, direct comparisons between rows is challenging.  However, several clear trends emerge.  For each band, moving to higher redshift (increasing wavelength) shows stronger foregrounds, a larger wedge (in part due to a wider primary beam), and a noisier EoR window (due to a higher system temperature).  In general the brightest foreground contamination is well demarcated by the wedge line in Equation \eqref{eq:wedge} for the primary beam (dotted line) and especially by the wedge line for the horizon (solid line).  In short, the wedge displays the theoretically expected frequency dependence.}
	\label{fig:allWedges}
\end{figure}

In Figure \ref{fig:allWedges}, we show all the cylindrical power spectra over the redshift range probed by our current observations.  
The spectra are sorted into three rows, each of which contain data coming from a single $30.72\,\textrm{MHz}$ wide frequency band.  All of the spectra were generated using the same techniques that were used to generate the example cylindrical power spectra in Section \ref{sec:MethodDemo} and thus contain all the desirable statistical properties discussed in Section \ref{sec:Methods}.  One sees that in every case the foregrounds are mostly confined to the wedge region in the bottom right corner of the $k_\perp$-$k_\parallel$ plane.  This builds upon the single frequency observations of \cite{PoberWedge}, demonstrating the existence of the EoR window across a wide range of frequencies relevant to EoR observations.

Having these measurements also allows us to examine the behavior of the EoR window as a function of frequency.  Consider first the high $k_\perp$ regions of the $k_\perp$-$k_\parallel$ plane.  The most striking feature here is the wedge.  Consistent with being dominated by foreground power, the wedge generally gets brighter with decreasing frequency within each wide frequency band, just as foreground emission is known to behave.  The extent of the wedge is also in line with theoretical expectations.  Recall from Equation \eqref{eq:wedge} that the wider the field-of-view, the farther up in $k_\parallel$ the wedge goes.  Since the field-of-view is defined by the primary beam, whose extent decreases with increasing frequency, one expects the wedge to have the largest area at the lowest frequencies.  This trend is clearly visible in the cylindrical power spectra of Figure \ref{fig:allWedges}, where the wedge extends to the highest $k_\parallel$ at the highest redshifts.  Importantly, the wedge is confined to its expected location across the entire range of the observations.  To see this, note that we have overlaid Equation \eqref{eq:wedge} on the plots, with the dashed line corresponding to $\theta_{\textrm{max}}$ equal to that of the first null of the primary beam, and the solid line with $\theta_{\textrm{max}} = \pi / 2$ (the horizon).  At all frequencies, the most serious contaminations lie within the first null, ensuring that the EoR window is foreground-free.

Foregrounds also enter indirectly into the instrumental noise-dominated regions because the MWA is sky-noise dominated.  Thus, as the brightest sources of emission in our observations, the foregrounds set the system temperature, and result in a higher instrumental noise at higher redshifts.  This trend can be seen within each wide frequency band (each row of Figure \ref{fig:allWedges}), although the slight interruption of this trend between bands suggests an additional source of noise.

At low $k_\perp$, theory suggests that foregrounds will contaminate a horizontally-oriented region at the bottom of the plot.  This is clearly seen in the highest frequency plots.  Interestingly, at lower frequencies the increasing instrumental noise plays more of a role, and the foreground contribution is less obvious in comparison (although it is still there).  While a naive reading of some of these low frequency plots (such as the one for $z=9.1$) might suggest that the EoR window extends to the lowest $k_\parallel$, such a conclusion would be misguided.  As we shall see in Section \ref{sec:Limits}, these modes are likely dominated by foregrounds (and therefore do not integrate down with further integration unlike instrumental noise dominated modes).  Moreover,  the error statistics (which self-consistently include foreground errors in our formalism) suggest that low $k_\parallel$ modes are less useful for constraining theoretical models, and that the true EoR window does in fact lie at higher $k_\parallel$, as suggested by theory.  Again, this highlights the importance of estimating power spectra in a framework that naturally contains a rigorous calculation of the errors involved.

\subsection{Spherical Power Spectrum Limits}
\label{sec:Limits}

\begin{figure} 
	\centering 
	\includegraphics[width=1\textwidth]{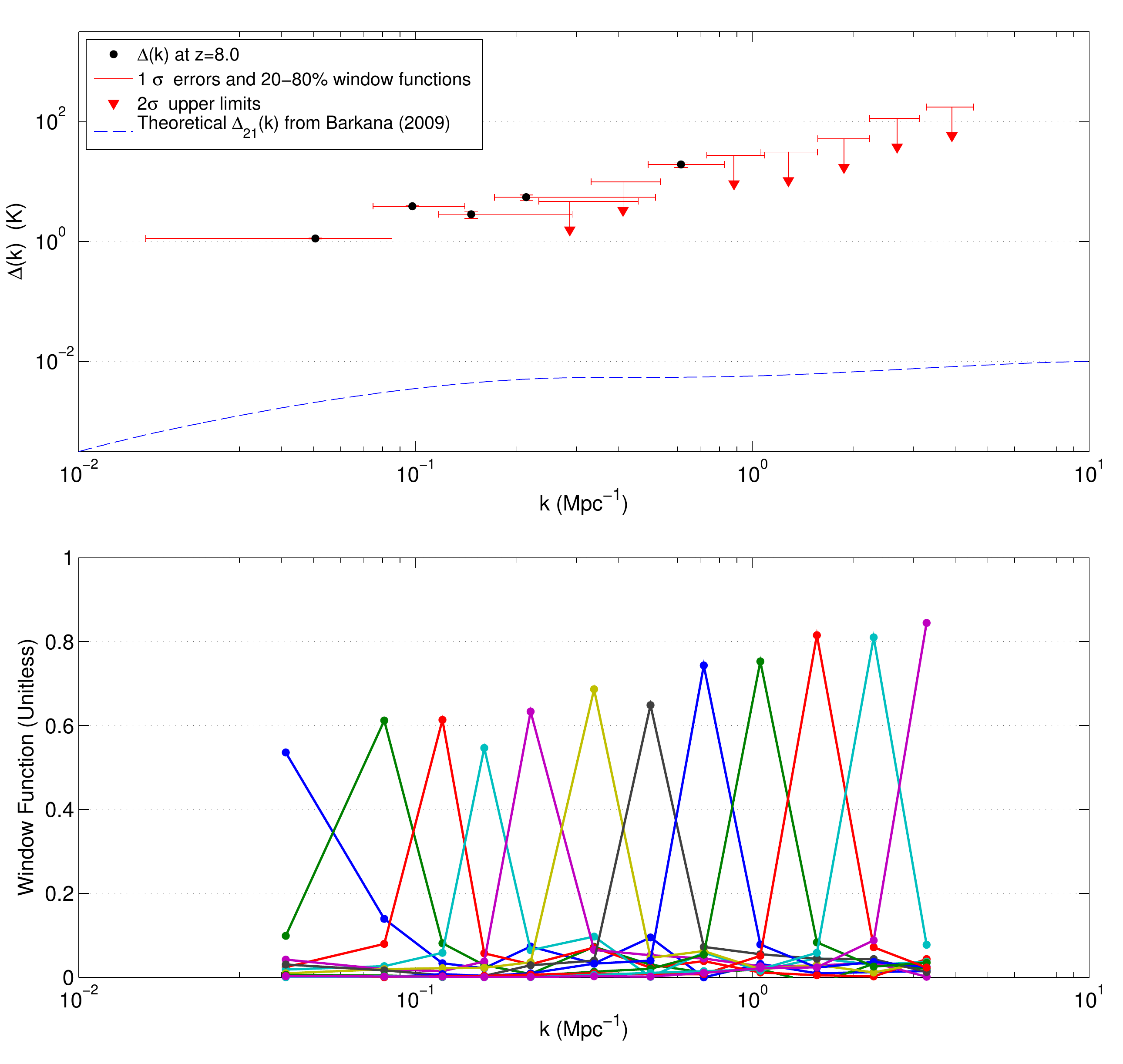}
	\caption[1D $P(k)$ for a single band with window functions.]{Our method allows for the estimation of the spherically binned power spectrum in temperature units, $\Delta(k)$, while keeping full acount of both vertical error bars and window functions (horizontal error bars) and making an optimal choice in the tradeoff between the two.  In the top panel, we have plotted our spherical power spectrum estimates of the subband centered on 158 MHz ($z = 8.0$), including $1\sigma$ errors on detections (which are often only barely visible), $2\sigma$ upper limits on non-detections, and horizontal error bars that span the middle three quintiles of the window functions (bottom panel).  At low $k$, the wide error bars are the expected consequence of foreground contamination \citep{LT11}.  Downward arrows represent measurements consistent with noise at the $2\sigma$ level. Even though the area under the primary beam wedge has been excised from the 2D-to-1D binning, the detection of foregrounds at low $k$, is expected due to the contribution of unresolved foregrounds over a wide range of $k_\perp$ \citep{DillonFast}. Our fiducial theoretical power spectrum is taken from \cite{BarkanaPS2009}.}
	\label{fig:singleDelta}
\end{figure}

Having confirmed that the EoR window behaves as expected, we will now proceed to place constraints on the spherical power spectrum.  In top panel of Figure \ref{fig:singleDelta} we show the result of binning the $z=10.3$ cylindrical power spectrum of Figure \ref{fig:allWedges}, using the optimal binning formulae presented in Section \ref{sec:cylindToSph}.
In addition, for ease of interpretation, we elect to plot
\begin{equation}
\Delta (k) \equiv \sqrt{ \frac{k^3}{2 \pi^2} P(k)}
\end{equation}
(which simply has units of temperature) rather than $P(k)$ itself.  

To quantify the errors in our spherical power spectrum estimate, we also bin the cylindrical power spectrum measurement covariances and window functions using the formulae of Section \ref{sec:cylindToSph}.  The resulting window functions are shown in the bottom panel, and give an estimate of the horizontal error bars.  Thinking of these window functions (which, recall, are normalized to integrate to unity) as probability distributions, the horizontal error bars shown in the top panel are demarcated by the 20th and 80th percentiles of the distribution.  (This corresponds to the full-width-half-maximum in the event that the window functions are Gaussians).  The vertical error bars were obtained by taking the square root of each diagonal element of the covariance matrix.  Since the methods of Section \ref{sec:cylindToSph} carefully preserved the diagonal nature of the bandpower covariance, each data point in Figure \ref{fig:singleDelta} represents a statistically independent measurement.  This would not have been the case had we not employed the decorrelation technique of Section \ref{sec:decorr}.

Immediately obvious from the plot is that there is a qualitative difference between the data points at low $k$ and those at high $k$.  In particular, the points at low $k$ are detections of the sky power spectrum, whereas the points at high $k$ are formal upper limits. This is not to say, of course, that the cosmological EoR signal has been detected at low $k$.  Rather, recall from Section \ref{sec:crossPower} that in an attempt to avoid having to make large bias subtractions, we elected to compute cross-power spectra of total sky emission rather than of the cosmological signal, with the expectation (largely confirmed in Section \ref{sec:WedgeResults}) that the intrinsic cleanliness of the EoR window would be sufficient to ensure a relatively foreground-free measurement at high $k_\parallel$.  Now, our survey volume is such that we are sensitive almost exclusively to regions in Fourier space where $k_\parallel \gg k_\perp$.  When binning along contours of constant $k$ in the cylindrical Fourier space, we have that $k \equiv \sqrt{k_\perp^2 + k_\parallel^2} \approx k_\parallel$, and therefore the low $k$ points of Figure \ref{fig:singleDelta} map to low $k_\parallel$.  The detections seen at low $k$ thus reside outside the EoR window and are almost certainly detections of the foreground power spectrum.  

Despite the fact that the low $k$ modes are foreground dominated, they still constitute a formal upper limit on the cosmological power spectrum, since the foreground power spectrum is necessarily positive.  In fact, our current, most competitive upper limit resides at the lowest $k$ values.  However, this is unlikely to continue to be the case as more data is taken with the MWA, for two reasons.  First, as foreground-limited measurements, the data points at low $k$ will not average down with further integration time.  In addition, the error statistics in the region are not particularly encouraging.  The window functions (and therefore the horizontal error bars) are seen to broaden towards lower $k$, reducing the ability of constraints at those $k$ to place limits on theoretical models.  (This is most easily seen by recalling that the window functions integrate to unity by construction, and thus the increase in their peak values towards higher $k$ implies a broadening of the window functions).  The broadening of the window functions is an expected consequence of foreground subtraction \citep{LT11} and thus will likely continue to limit the usefulness of the low $k$ regime unless future measurements can characterize foreground properties with exquisite precision.

In contrast, the points at high $k$ do reside in the EoR window.  The constraints here are limited by thermal noise, as we saw in Section \ref{sec:MethodDemo}.  Bolstering this view is the fact that the data here are consistent with zero, as one expects for a noise-dominated cross-power spectrum.  The limits here are given by the $2\sigma$ errors predicted by the Equation \eqref{eq:Cylind2SphCovar}.  As mentioned in Section \ref{sec:CovarModeling}, these errors are somewhat larger than what might be predicted by a theoretical sensitivity calculation.  However, they are consistent with rough estimates of the errors obtained from a calculation of root-mean-square values from the images produced in Section \ref{sec:Mapmaking}.  This suggests that the larger-than-expected errors are due to noisier-than-expected input maps, and not to any approximations made in the power spectrum estimation techniques presented in this paper.  The data on which these results are based are from the very first operation of the prototype array, and we expect better performance in later data.  Encouragingly, we note also that as noise-dominated constraints, the measurements at high $k$ will continue to improve with integration time. 
 
\begin{figure}  
	\centering 
	\includegraphics[width=1\textwidth]{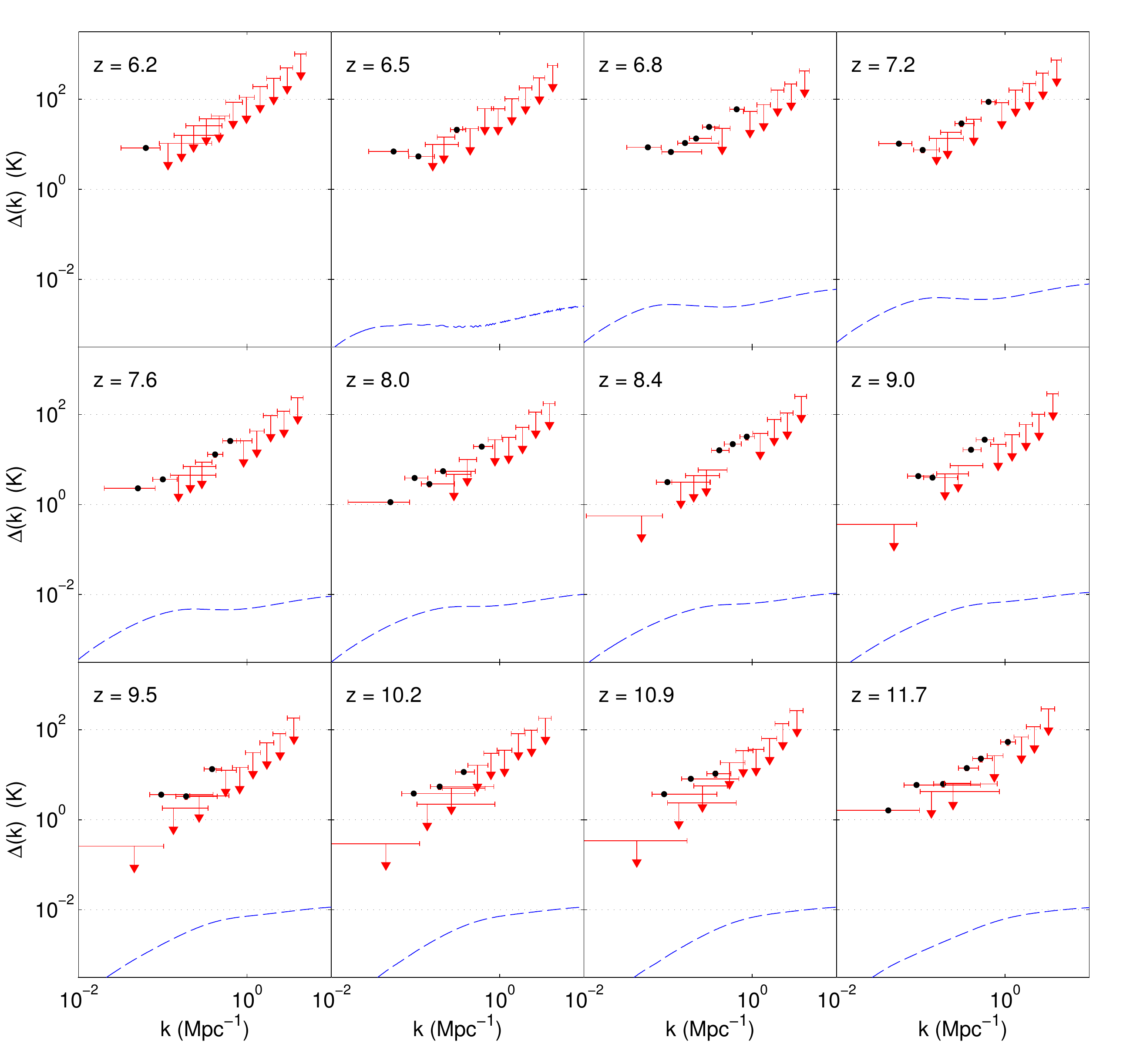}
	\caption[1D $P(k)$ for all bands showing best limits.]{Taking advantage of our fast yet thorough power spectrum estimation technique, we estimate $\Delta(k)$ for a wide range of $k$ and $z$, including both vertical and horizontal errors.  (For points that represent positive detections of foregrounds or systematic correlations, the vertical error bars are often barely visible).  Using the visual language of Figure \ref{fig:singleDelta}, we show here our spherical power spectrum limits as a function of both $k$ and $z$.  Each panel is a different subband.  The many detections can be attibuted to foregrounds (especially at low $k$), instrumental effects like those we saw in Figure \ref{fig:2DPkComparison} (especially at medium values of $k$), or both.	Our absolute lowest limit on the 21 cm brightness temperature power spectrum, $\Delta(k) < 0.3$ Kelvin at the 95\% confidence level, comes at $k=0.046$ cMpc$^{-1}$ and $z=9.5$ (or $\Delta (k) < 2 \,\textrm{K}$ at $z=9.5$ and $k= 0.134\,\textrm{cMpc}^{-1}$ if one discards the lowest $k$ bin to immunize oneself against foreground modeling uncertainties).}
	\label{fig:allDelta}
\end{figure}

In Figure \ref{fig:allDelta}, we show power spectrum limits across the entire frequency range of the MWA, along with some theoretical predictions generated using the models in \cite{BarkanaPS2009}.  At the lowest redshift, no theory curve is plotted because the model predicts that reionization is complete by then.  This yet again underscores the importance of making measurements over a broad frequency range---with access to a wide range of redshifts, future detections of the cosmological signal can be distinguished from residual foregrounds by measuring null signals at redshifts where reionization is complete.

Each redshift bin of Figure \ref{fig:allDelta} exhibits trends that are qualitatively similar to those discussed above for the $z=10.3$ case.  We see many apparent detections of correlations positive correlations between the two time-interleaved data cubes---more than can be attributed to foregrounds alone.  As we saw with the cylindrically binned power spectrum in Figure \ref{fig:2DPkComparison}, there is evidence of systematic and instrumental effects sending foreground power into the EoR window, leading to higher $k$ detections and large differences between neighboring $k$ bins.  With as new an instrument as the MWA was at time of this observation, this issues are understandable.  The exact physical origin of those systematics is beyond the scope of this paper, however they should serve as a reminder to stay vigilant for them in future datasets from a more battle-tested instrument.  However, because we see no evidence of strong anti-correlations between data cubes, we expect that the extra power introduced by systematics into the EoR window only the effect of worsening the limits we can set.  

Over all bands, our best limit is $\Delta (k) < 0.3 \,\textrm{K}$, occurring at $z=9.5$ and $k= 0.046\,\textrm{cMpc}^{-1}$.  However, as remarked in Section \ref{sec:CovarModeling}, the lowest $k$ bins can be rather sensitive to the covariance model, and if one excludes those bins, our best limit is $\Delta (k) < 2 \,\textrm{K}$, at $z=9.5$ and $k= 0.134\,\textrm{cMpc}^{-1}$.  While our limits may not be quite as low as other existing limits in the literature \citep{newGMRT,PAPER32Limits}, they are the only limits on the EoR power spectrum that span a broad redshift range from $z=6.2$ to $z=11.7$.  Moreover, these statistically rigorous limits will likely improve with newer and more sensitive data from the MWA.

\section{Conclusions}
\label{sec:Conc}
In this paper, we have accomplished three goals.  First, we adapted $21\,\textrm{cm}$ power spectrum estimation techniques from \citet{LT11} and \citet{DillonFast} with real-world obstacles in mind, so that they could be applied to real data.  With early MWA data, our generalized formalism was then used to demonstrate the importance of employing a statistically rigorous framework for power spectrum estimation, lest one corrupt the naturally foreground-free region of Fourier space known as the EoR window.  Finally, we used the MWA data to set limits on the EoR power spectrum.

In confronting real-world obstacles, our desire is to preserve the as much of the statistical rigor in previous matrix-based power spectrum estimation frameworks as possible.  To avoid having to perform direct subtractions of instrumental noise biases, we advocate computing cross-power spectra between statistically identical subsets of the data (in the case of the MWA worked example of this paper, these subsets were formed from odd and even time samples of the data).  This has the effect of eliminating noise bias in the power spectrum, although instrumental noise continues to contribute to the error bars.  To avoid direct subtractions of foreground biases, we simply look preferentially in the EoR window, where foregrounds are expected to be low.  Missing data, whether from incomplete $uv$ coverage or RFI flagging, can be dealt with using the pseudoinverse formalism.  Doing this allows the effects of missing data to be self-consistently propagated into error statistics such as the power spectrum covariance and the window functions.  In an effort to preserve the cleanliness of the EoR window, one should form decorrelated bandpower estimates, which have uncorrelated errors and reasonably narrow window functions.  Care must then be taken to preserve these nice properties via an optimal binning of cylindrical bandpowers into spherical bandpowers.

Using early MWA data to demonstrate these techniques, we have confirmed theoretical predictions for the existence of the EoR window and have extended previous observations done by other groups to a much wider frequency range.  This allowed us to check predicted trends of the EoR window as a function of frequency, all of which are consistent with theory.  Crucially, we found that without using the decorrelation technology of Section \ref{sec:decorr}, measurements in the EoR window are not in fact instrumental noise dominated, and contain a systematic bias that is indicative of foreground leakage from outside the EoR window.

The early MWA data has also allowed us to place limits on the cosmological EoR power spectrum.  Our best limit is $\Delta (k) < 0.3\,\textrm{K}$, at $z=9.5$ and $k=0.046\,\textrm{cMpc}$ (or $\Delta (k) < 2 \,\textrm{K}$ at $z=9.5$ and $k= 0.134\,\textrm{cMpc}^{-1}$ if one discards the lowest $k$ bin to immunize oneself against foreground modeling uncertainties).  This may not be competitive with other published observations, but generalizes them in an important way: instead of focusing on one particular frequency, our limits span a wide range of redshifts relevant to the EoR, going from $z=6.2$ to $z=11.7$.  In addition, these limits will almost certainly improve in the near future, using already-collected (but yet to be analyzed) data from the MWA-32T system, as well as soon-to-be-collected data from the MWA-128T system.  The rigorous statistical tools developed in this paper should be equally applicable to these newer data sets, ensuring that foreground contamination remains confined to outside the EoR window, safeguarding the potential of current generation experiments to make an exciting first detection of the EoR within the next few years.

\begin{subappendices}

\section[Appendix: On the Importance of Modeling the Full Error Covariance]{Appendix: On the Importance of Modeling the \\Full Error Covariance}
\label{sec:ErrorCovAppendix}

\begin{figure}
	\centering 
	\includegraphics[width=1\textwidth]{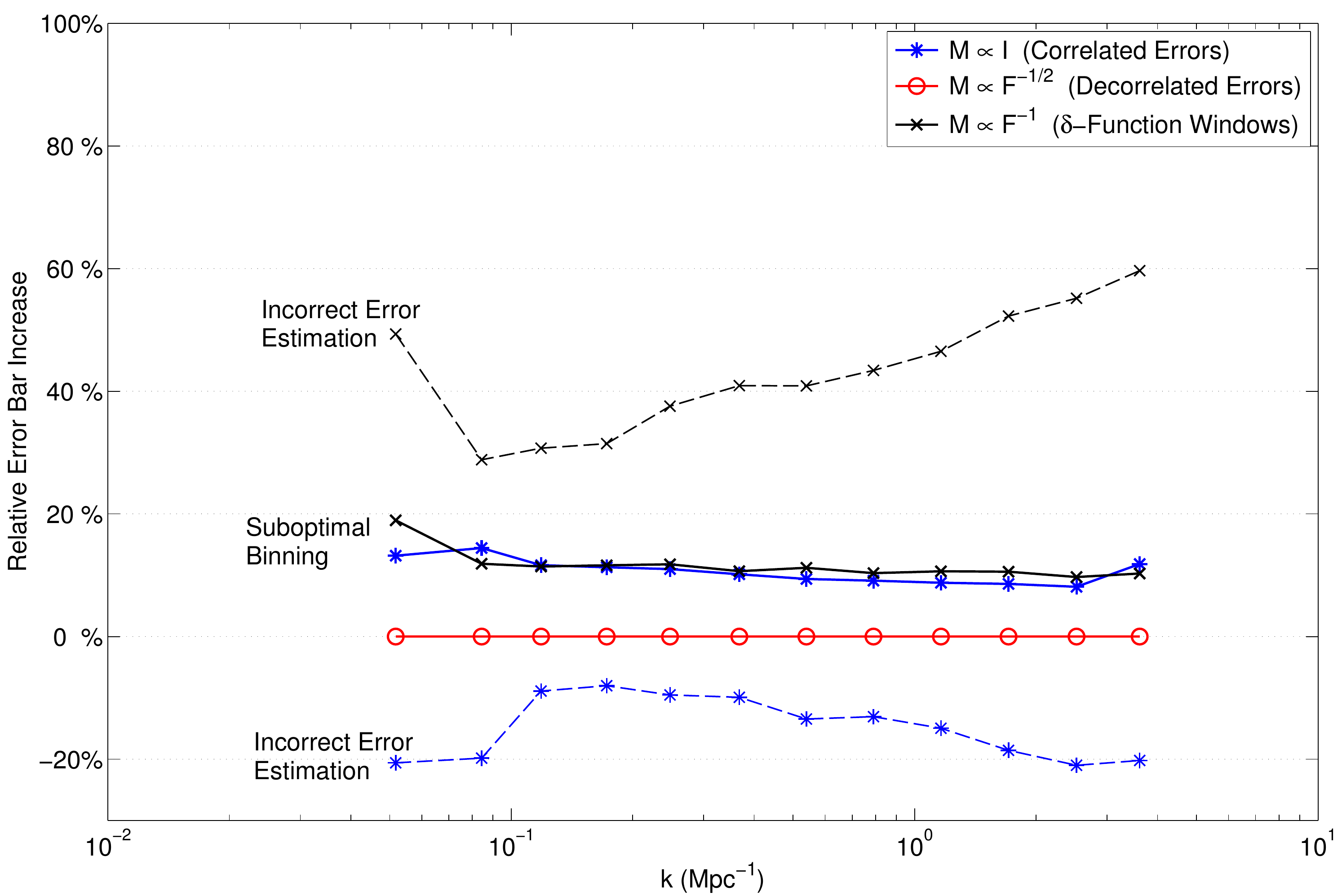}
	\caption[Effect of neglecting off diagonal covariance terms on errors.]{Neglecting the fact that the covariance of the power spectrum estimator is, in general, non-diagonal, can lead to two mistakes that can either unnecessarily enlarge our error bars or, even worse, unjustifiably shrink them.  In this figure, we first show an approximately 10\% increase in the vertical error bars on the power spectrum (solid lines) from a suboptimal inverse variance weighted binning scheme, rather than the inverse covariance weighted binning of Equation \eqref{eq:Cylind2SphEst}.  This problem is obviated by choosing an estimator with decorrelated errors and thus a diagonal covariance matrix.  If one simply assumes that the estimator covariance in Equation \eqref{eq:Cylind2SphCovar} is diagonal when it is not (dotted lines), one is led, depending on the choice of estimator, either to roughly 50\% larger error bars than necessary or, worse yet, artificially small error bars.  The last mistake, choosing an estimator with small error bars---despite its wide window functions---and then neglecting the off-diagonal terms in the estimator covariance, is potentially the most pernicious since it could lead to a claimed detection in the absence of signal.}
	\label{fig:VarVsCovarErrors}
\end{figure}

In Section \ref{sec:cylindToSph}, we argued that an inverse covariance weighted binning scheme for estimating spherical band powers produced optimal spherical power spectrum estimate.  In the case where $\M$ is chosen either for the smallest possible error bars or the narrowest possible window functions, the estimator covariance $\boldsymbol \Sigma^\text{cyl}$ is non-diagonal.  Assuming that the matrix is actually diagonal, at one or more steps in the binning and error propagation, can lead error bars that are overly conservative or---worse yet---error bars that are insufficiently conservative and might falsely lead to a claimed detection.  In Figure \ref{fig:VarVsCovarErrors}, we show the effects of making a suboptimal choice for binning. 

If one fully models the covariance matrix $\boldsymbol \Sigma^\text{cyl}$, including off-diagonal elements, but chooses to generate $\widehat{\mathbf{p}}^\text{sph}$ as an inverse variance (and not inverse covariance) weighted average of cylindrical band powers, neglecting off diagonal terms in the weighting, one's estimators will be noisier as a result (see the solid lines in Figure \ref{fig:VarVsCovarErrors}). These are the correct errors for the suboptimal choice of estimators. 

\begin{figure}
	\centering 
	\includegraphics[width=.6\textwidth]{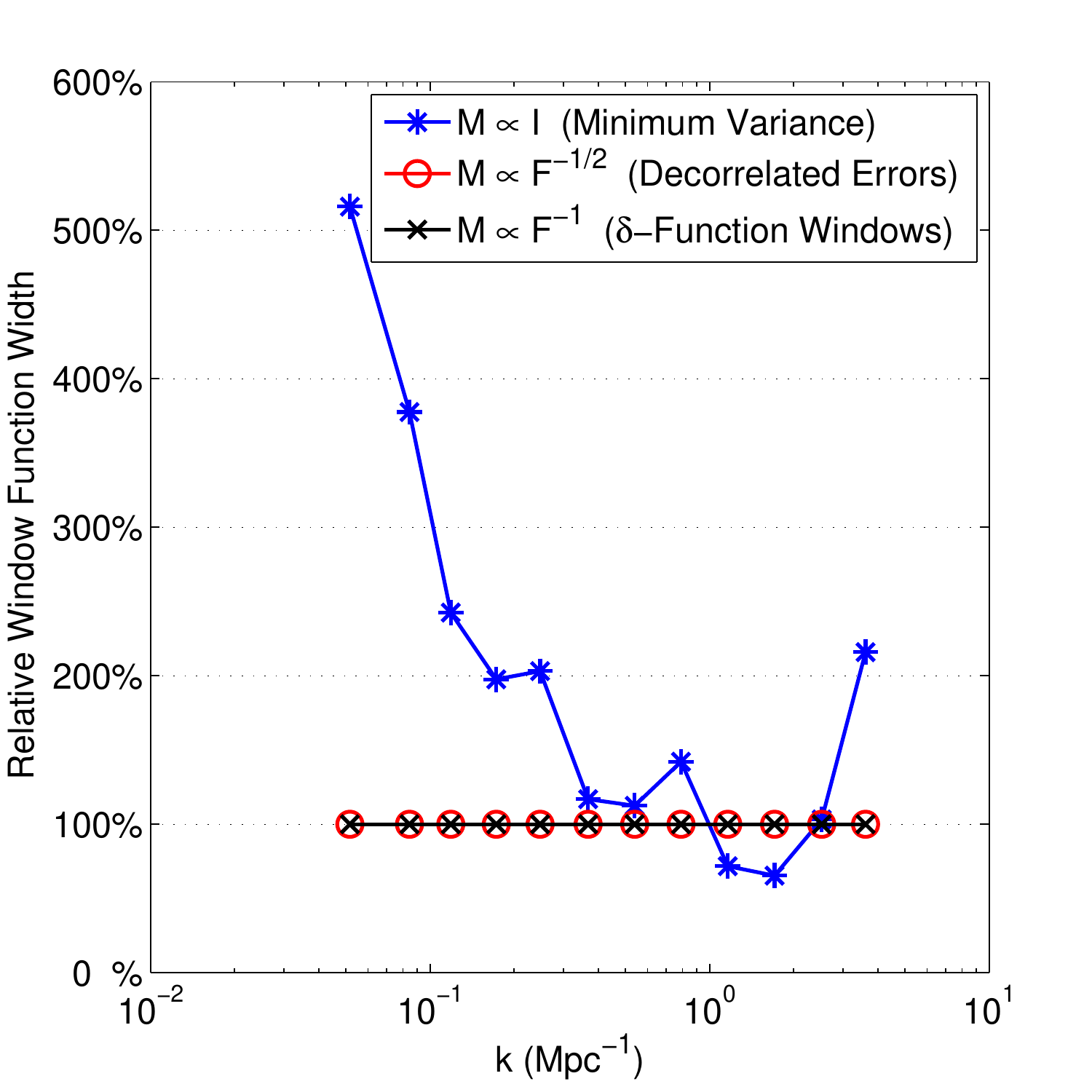}
	\caption[Effect of neglecting off diagonal covariance terms on window functions.]{Just as with the error bars in Figure \ref{fig:VarVsCovarErrors}, generating suboptimally binned spherical power spectrum estimates by neglecting off-diagonal terms in the estimator covariance can lead to wider window functions than necessary.  We illustrate the effect by comparing the width of the window functions between the 20th and 80th percentiles between the two binning schemes. This is important for the choice of power spectrum estimator with the smallest error bars and widest window functions ($\M \sim \Eye$).  In the case where our power spectrum estimator has uncorrelated errors, there are no off-diagonal terms in the estimator covariance and both binning schemes are identical.  In the case of the estimator with $\delta$-function window functions, suboptimal binning does not affect the window functions---though it still affects the vertical errors (see Figure \ref{fig:VarVsCovarErrors}).}
	\label{fig:VarVsCovarWindows}
\end{figure}

Even worse, if one assumes that $\boldsymbol \Sigma^\text{cyl}$ is diagonal when it is not, one is led either to overestimate the error bars, in the case of $\M \sim \F^{-1}$, or underestimate them, as would be the case when $\M \sim \Eye$.  This is because the former case general exhibits anti-correlated errors while the latter suffers from correlated errors.  The last scenario is the most troubling: by aggressively choosing the estimator with the smallest vertical error bars ($\M \sim \Eye$) and then neglecting the correlations between errors, one will underestimate the error bars and might be lead to falsely claiming a detection.  In this case, the estimator is suboptimal and the errors are incorrect.  Additionally, as we show in Figure \ref{fig:VarVsCovarWindows}, if one were to calculate the the window functions under the assumption that $\boldsymbol \Sigma^\text{cyl}$ is diagonal, one would find window functions several times boarder than they would otherwise be.  
  
Thankfully, choosing the cylindrical power spectrum estimator with decorrelated errors avoids the subtle difference between inverse variance and inverse covariance weighted binning.  The $\M \sim \F^{-1/2}$ decorrelated estimator preserves the EoR window and allows for easy, optimal binning of uncontaminated regions into spherical band power spectrum estimates.

\end{subappendices}


\chapter[Empirical Covariance Modeling for 21\,cm Power Spectrum \\Estimation: A Method Demonstration and New Limits from Early Murchison Widefield Array 128-Tile Data]{Empirical Covariance Modeling for 21\,cm Power Spectrum Estimation: \\A Method Demonstration and New Limits from Early Murchison Widefield Array 128-Tile Data} \label{ch:EmpiricalCovariance}

\emph{The content of this chapter was submitted to \emph{Physical Review D} on March 10, 2015.}


\newcommand{\modelResidualCorr}{0.116}
\newcommand{\limitDeltaSq}{3.7\times 10^4}
\newcommand{\limitk}{0.18}
\newcommand{\limitz}{6.8}
\newcommand{\wedgeBuffer}{0.02}

\section{Introduction} \label{sec:intro}

Tomographic mapping of neutral hydrogen using its 21\,cm hyperfine transition has the potential to directly probe the density, temperature, and ionization of the intergalactic medium (IGM), from redshift 50 (and possibly earlier) through the end of reionization at $z\sim 6$. This unprecedented view of the so-called ``cosmic dawn'' can tightly constrain models of the first stars and galaxies \cite{FurlanettoReview, miguelreview, PritchardLoebReview, aviBook} and eventually yield an order of magnitude more precise test of the standard cosmological model ($\Lambda$CDM) than current probes \cite{Yi}. 

Over the past few years, first generation instruments have made considerable progress toward the detection of the power spectrum of the 21\,cm emission during the epoch of reionization (EoR). Telescopes such as the Low Frequency Array (LOFAR \cite{LOFARinstrument}), the Donald C. Backer Precision Array for Probing the Epoch of Reionization (PAPER \cite{PAPER}), the Giant Metrewave Radio Telescope (GMRT \cite{newGMRT}), and the Murchison Widefield Array (MWA \cite{MWAdesign,TingaySummary,BowmanMWAScience}) are now operating, and have begun to set limits on the power spectrum. GMRT set some of the earliest limits \cite{newGMRT} and both PAPER \cite{DannyMultiRedshift} and the MWA \cite{X13} have presented upper limits across multiple redshifts using small prototype arrays. PAPER has translated its results into a constraint on the heating of the IGM by the first generation of x-ray binaries and miniquasars \cite{PAPER32Limits} and has placed the tightest constraints so far on the power spectrum \cite{PAPER64Limits} and the thermal history of the IGM \cite{PoberPAPER64Heating}.

Despite recent advances, considerable analysis challenges remain. Extracting the subtle cosmological signal from the noise is expected to require thousand hour observations across a range of redshifts \citep{MiguelNoise,Judd06,LidzRiseFall,LOFAR2,AaronSensitivity,ThyagarajanWedge}. Even more daunting is the fact that the 21\,cm signal is probably at least 4 orders of magnitude dimmer than the astrophysical foregrounds---due to synchrotron radiation from both our Galaxy and from other galaxies \citep{Angelica,LOFAR,BernardiForegrounds,PoberWedge,InitialLOFAR1,InitialLOFAR2}. 

Recently, simulations and analytical calculations have established the existence of a region in cylindrical Fourier space---in which three-dimensional (3D) Fourier modes $\vec{k}$ are binned into $k_\|$ modes along the line of sight and  $k_\perp$ modes perpendicular to it---called the ``EoR window'' that should be fairly free of foreground contamination \cite{Dattapowerspec,AaronDelay,VedanthamWedge,MoralesPSShapes,Hazelton2013,CathWedge,ThyagarajanWedge,EoRWindow1,EoRWindow2}. Observations of the EoR window confirm that it is largely foreground-free \cite{PoberWedge,X13} up to the sensitivity limits of current experiments. The boundary of the EoR window is determined by the volume and resolution of the observation, the intrinsic spectral structure of the foregrounds, and the so-called ``wedge.'' 

Physically, the wedge arises from the frequency dependence of the point spread function (PSF) of any interferometer, which can create spectral structure from spectrally smooth foregrounds in our 3D maps (see \cite{EoRWindow1} for a rigorous derivation). Fortunately, in $k_\|$-$k_\perp$ space, instrumental chromaticity from flat-spectrum sources is restricted to the region below 
\beq  
k_\| = \theta_0 \frac{ D_M(z) E(z)}{D_H (1+z)} k_\perp, \label{eq:wedge}
\eeq
where $D_H \equiv c/H_0$, $E(z) \equiv \sqrt{\Omega_m(1+z)^3+\Omega_\Lambda}$, and $D_m(z) \equiv \int_0^z \mathrm{d}z'/E(z')$ with cosmological parameters from \cite{PlanckCosmoParams}. The size of the region is determined by $\theta_0$, the angle from zenith beyond which foregrounds do not significantly contribute. While most of the foreground emission we observe should appear inside the main lobe of the primary beam, foreground contamination from sources in the sidelobes are also significant compared to the signal \cite{PoberSidelobe,NithyaPitchfork,NithyaPitchfork}. A conservative choice of $\theta_0$ is therefore $\pi/2$, which reflects the fact that the maximum possible delay a baseline can measure corresponds to a source at the horizon \cite{AaronDelay}. Still, this foreground isolation is not foolproof and can be easily corrupted by miscalibration and imperfect data reduction. Further, slowly varying spectral modes just outside the wedge are also affected when the foreground residuals have spectral structure beyond that imprinted by the chromaticity of the interferometer.

To confidently detect the 21\,cm EoR power spectrum, we need rigorous statistical techniques that incorporate models of the cosmological signal, the foregrounds, the instrument, the instrumental noise, and the exact mapmaking procedure. With this information, one may use estimators that preserve as much cosmological information as possible and thoroughly propagate errors due to noise and foregrounds through the analysis pipeline.

The development of such statistical techniques has progressed rapidly over the past few years. The quadratic estimator formalism was adapted \cite{LT11} from previous work on the cosmic microwave background \cite{Maxpowerspeclossless} and galaxy surveys \cite{Maxgalaxysurvey1}. It was accelerated to meet the data volume challenges of 21\,cm tomography \cite{DillonFast} and refined to overcome some of the difficulties of working with real data \cite{X13}.  Further, recent work has shown how to rigorously incorporate the interferometric effects that create the wedge \cite{EoRWindow1,EoRWindow2,Mapmaking}, though they rely on precision instrument modeling, including exact per-frequency and per-antenna primary beams and complex gains. A similar technique designed for drift-scanning telescopes using spherical harmonic modes was developed in \cite{Richard,ShawCoaxing}, which also demonstrated the need for a precise understanding of one's instrument.

However, at this early stage in the development of 21\,cm cosmology, precision instrument characterization remains an active area of research \cite{sutinjo2014,AbrahamOrbcomm,NewburghCHIMECal,JonniePrimaryBeam}. We thus pursue a more cautious approach to foreground modeling that reflects our incomplete knowledge of the instrument by modeling the residual foreground covariance from the data itself. As we will show, this mitigates systematics such as calibration errors that would otherwise impart spectral structure onto the foregrounds, corrupting the EoR window. While not a fully Bayesian approach like those of \cite{SutterBayesianImaging} and \cite{GibbsPSE}, our technique discovers both the statistics of the foregrounds and the power spectrum from the data. Our foreground models are subject to certain prior assumptions but are allowed to be data motivated in a restricted space. However, by working in the context of the quadratic estimator formalism, we can benefit from the computational speedups of \cite{DillonFast}. This work is meant to build on those techniques and make them more easily applied to real and imperfect data.

This paper is organized into two main parts. In Section \ref{sec:methods} we discuss the problem of covariance modeling in the context of power spectrum estimation and present a method for the empirical estimation of that foreground model, using MWA data to illustrate the procedure. Then, in Section \ref{sec:results}, we explain how these data were taken and reduced into maps and present the results of our power spectrum estimation procedure on a few hours of MWA observation, including limits on the 21\,cm power spectrum.


\section{Empirical Covariance Modeling} \label{sec:methods}

Before presenting our method of empirically modeling the statistics of residual foregrounds in our maps, we need to review the importance of these covariances to power spectrum estimation. We begin in Section \ref{sec:review} with a brief review of the quadratic estimator formalism for optimal power spectrum estimation and rigorous error quantification. We then discuss in Section \ref{sec:CovTheory} the problem of covariance modeling in greater detail, highlighting exactly which unknowns we are trying to model with the data. Next we present in Section \ref{sec:empirical} our empirical method of estimating the covariance of foreground residuals, illustrated with an application to MWA data. Lastly, we review in Section \ref{sec:caveats} the assumptions and caveats that we make or inherit from previous power spectrum estimation work.


\subsection{Quadratic Power Spectrum Estimator Review}\label{sec:review}

The fundamental goal of power spectrum estimation is to reduce the volume of data by exploiting statistical symmetries while retaining as much information as possible about the cosmological power spectrum \cite{Maxpowerspeclossless}. We seek to estimate a set of band powers $\p$ using the approximation that
\beq
P(\kvec) \approx \sum_\alpha p_\alpha \chi_\alpha (\kvec),
\eeq
where $P(\kvec)$ is the power spectrum as a function of wave vector $\kvec$ and $\chi_\alpha$ is an indicator function that equals 1 wherever we are approximating $P(\kvec)$ by $p_\alpha$ and vanishes elsewhere.

Following \cite{LT11,DillonFast,X13}, we estimate power spectra from a ``data cube''---a set of sky maps of brightness temperature at many closely spaced frequencies---which we represent as a single vector $\xhat$ whose index iterates over both position and frequency. From $\xhat$, we estimate each band power as
\beq
\widehat{p}_\alpha = \frac{1}{2}M_{\alpha\beta} \left(\xhat_1 - \mean\right)^\trans \C^{-1} \mathbf{C},_\beta \C^{-1} \left(\xhat_2 -\mean\right) - b_\alpha. \label{eq:QuadEst}
\eeq
Here $\mean = \langle \xhat \rangle$, the ensemble average of our map over many different realizations of the observation, and $\C$ is the covariance of our map,
\beq
\C = \langle \xhat \xhat^\trans \rangle - \langle \xhat \rangle  \langle \xhat \rangle^\trans.
\eeq 
$\mathbf{C},_\beta$ is a matrix that encodes the response of the covariance to changes in the true, underlying band powers; roughly speaking, it performs the Fourier transforming, squaring, and binning steps one normally associates with computing power spectra.\footnote{For a derivation of an explicit form of $\C,_\beta$, see \cite{LT11} or \cite{DillonFast}.}  Additionally, $\M$ is an invertible normalization matrix and $b_\alpha$ is the power spectrum bias from nonsignal contaminants in $\xhat$. In this work, we follow \cite{X13} and choose a form of $\M$ such that $\boldsymbol\Sigma \equiv \text{Cov}(\widehat{\p})$ is diagonal, decorrelating errors in the power spectrum and thus reducing foreground leakage into the EoR window. In order to calculate $\M$ and $\boldsymbol\Sigma$ quickly, we use the fast method of \cite{DillonFast} which uses fast Fourier transforms and Monte Carlo simulations to approximate these matrices.

Finally, temporally interleaving the input data into two cubes $\xhat_1$ and $\xhat_2$ with the same sky signal but independent noise avoids a noise contribution to the bias $b_\alpha$ as in \cite{X13}. Again following \cite{X13}, we abstain from subtracting a foreground residual bias in order to avoid any signal loss (as discussed in \ref{sec:filter}).

\subsection{What Does Our Covariance Model Represent?} \label{sec:CovTheory}

Our brightness temperature data cubes are made up of contributions from three statistically independent sources: the cosmological signal, $\xhat^S$; the astrophysical foregrounds, $\xhat^{FG}$; and the instrumental noise $\xhat^N$. It follows that the covariance matrix is equal to the sum of their separate covariances:
\beq
\C = \C^S + \C^{FG} + \C^N.
\eeq

Hidden in the statistical description of the different contributions to our measurement is an important subtlety. Each of these components is taken to be a particular instantiation of a random process, described by a mean and covariance. In the case of the cosmological signal, it is the underlying statistics---the mean and covariance---which encode information about the cosmology and astrophysics. However, we can only learn about those statistics by assuming statistical isotropy and homogeneity and by assuming that spatial averages can stand in for ensemble averages in large volumes. In the case of the instrumental noise, we usually think of the particular instantiation of the noise that we see as the result of a random trial. 

The foregrounds are different. There is only one set of foregrounds, and they are not random. If we knew exactly how the foregrounds appear in our observations, we would subtract them from our maps and then ignore them in this analysis. We know that we do not know the foregrounds exactly, and so we choose to model them with our best guess, $\mean^{FG}$. If we define the cosmological signal to consist only of fluctuations from the brightness temperature of the global 21\,cm signal, then the signal and the noise both have $\mean^S = \mean^N = 0$. Therefore, we start our power spectrum estimation using Equation \eqref{eq:QuadEst} by subtracting off our best guess as to the foreground contamination in our map. But how wrong are we? 

The short answer is that we do not really know that either. But, if we want to take advantage of the quadratic estimator formalism to give the highest weight to the modes we are most confident in, then we must model the statistics of our foreground residuals. If we assume that our error is drawn from some correlated Gaussian distribution, then we should use that \emph{foreground uncertainty covariance} as the proper $\C^{FG}$ in Equation \eqref{eq:QuadEst}.

So what do we know about the residual foregrounds in our maps? In theory, our dirty maps are related to the true sky by a set of point spread functions that depend on both position and frequency \cite{Mapmaking}. This is the result of both the way our interferometer turns the sky into measured visibilities and the way we make maps to turn those visibilities into $\xhat$. In other words, there exists some matrix of PSFs, $\PSF$ such that
\beq
\langle \xhat \rangle = \PSF \x^\text{true}.
\eeq
The spectral structure in our maps that creates the wedge feature in the power spectrum is a result of $\PSF$.

We can describe our uncertainty about the true sky---about the positions, fluxes, and spectral indices of both diffuse foregrounds and points sources---with a covariance matrix $\C^{FG,\text{true}}$ \cite{LT11,DillonFast}, so that
\beq
\C^{FG} = \PSF \C^{FG,\text{true}} \PSF^\trans.
\eeq
This equation presents us with two ways of modeling the foregrounds. If we feel that we know the relationship between our dirty maps and the true sky precisely, then we can propagate our uncertainty about a relatively small number of foreground parameters, as discussed by \cite{LT11} and \cite{DillonFast}, through the $\PSF$ matrix to get $\C^{FG}$. This technique, suggested by \cite{Mapmaking}, relies on precise knowledge of $\PSF$. Of course, the relationship between the true sky and our visibility data depends both on the design of our instrument and on its calibration. If our calibration is very good---if we really understand our antenna gains and phases, our primary beams, and our bandpasses---then we can accurately model $\PSF$.

If we are worried about systematics (and at this early stage of 21\,cm tomography with low frequency radio interferometers, we certainly are), then we need a complementary approach to modeling $\C^{FG}$ directly, one that we can use both for power spectrum estimation and for comparison to the results of a more theoretically motivated technique. This is the main goal of this work.


\subsection{Empirical Covariance Modeling Technique} \label{sec:empirical}

The idea of using empirically motivated covariance matrices in the quadratic estimator formalism has some history in the field. Previous MWA power spectrum analysis \cite{X13} used the difference between time-interleaved data cubes to estimate the overall level of noise, empirically calibrating $T_\text{sys}$, the system temperature of the elements. PAPER's power spectrum analysis relies on using observed covariances to suppress systematic errors \cite{PAPER32Limits} and on boot-strapped error bars \cite{PAPER32Limits,DannyMultiRedshift}. A similar technique was developed contemporaneously with this work and was used by \cite{PAPER64Limits} to estimate covariances.

$\C^{FG}$ has far more elements than we have measured voxels---our cubes have about $2\times 10^5$ voxels, meaning that $\C^{FG}$ has up to $2\times 10^{10}$ unique elements. Therefore, any estimate of $\C^{FG}$ from the data needs to make some assumptions about the structure of the covariance. Since foregrounds have intrinsically smooth spectra, and since one generally attempts to model and subtract smooth spectrum foregrounds, it follows that foreground residuals will be highly correlated along the line of sight. After all, if we are undersubtracting foregrounds at one frequency, we are probably undersubtracting at nearby frequencies too. We therefore choose to focus on empirically constructing the part of $\C^{FG}$ that corresponds to the frequency-frequency covariance---the covariance along the line of sight.  If there are $n_f$ frequency channels, then that covariance matrix is only $n_f \times n_f$ elements and is likely dominated by a relatively small number of modes. 

In this section, we will present an approach to solving this problem in a way that faithfully reflects the complex spectral structure introduced by an (imperfectly calibrated) interferometer on the bright astrophysical foregrounds. As a worked example, we use data from a short observation with the MWA which we will describe in detail in Section \ref{sec:results}. We begin with a uniformly weighted map of the sky at each frequency, a model for both point sources and diffuse emission imaged from simulated visibilities, and a model for the noise in each $uv$ cell as a function of frequency.

The idea to model $\C^{FG}$ empirically was put forward by \citet{LiuThesis}. He attempted to model each line of sight as statistically independent and made no effort to separate $\C^{FG}$ from $\C^{N}$ or to reduce the residual noisiness of the frequency-frequency covariance.  

Our approach centers on the idea that the covariance matrix can be approximated as block diagonal in the $uv$ basis of Fourier modes perpendicular to the line of sight. In other words, we are looking to express $\C^{FG}$ as
\beq
C^{FG}_{uu'vv'ff'} \approx \delta_{uu'} \delta_{vv'} \widehat{C}_{ff'}(k_\perp), \label{eq:blockdiag}
\eeq
where $k_\perp$ is a function of $\sqrt{u^2+v^2}$. This is the tensor product of our best guess of the frequency-frequency covariance $\CHat$ and the identity in both Fourier coordinates perpendicular to the line of sight. In this way, we can model different frequency-frequency covariances as a function of $|u|$ or equivalently, $k_\perp$, reflecting that fact that the wedge results from greater leakage of power up from low $k_{\|}$ as one goes to higher $k_\perp$. This method also has the advantage that $\C$ becomes efficient to both write down and invert directly, removing the need for the preconditioned conjugate gradient algorithm employed by \cite{DillonFast}.

This approximation is equivalent to the assumption that the residuals in every line of sight are statistically independent of position. This is generally a pretty accurate assumption as long as the primary beam does not change very much over the map from which we estimate the power spectrum. However, because $\widehat{C}_{ff'}(k_\perp)$ depends on the angular scale, we are still modeling correlations that depend only on the distance between points in the map. 

While we might expect that the largest residual voxels correspond to errors in subtracting the brightest sources, the voxels in the residual data cube (the map minus the model) are only weakly correlated with the best-guess model of the foregrounds (we find a correlation coefficient $\rho =$ \modelResidualCorr, which suggests that sources are removed to roughly the 10\% level, assuming that undersubtraction dominates). As we improve our best guess of the model foregrounds through better deconvolution, we expect $\rho$ to go down, improving the assumption that foregrounds are block diagonal in the $uv$ basis. We will now present the technique we have devised in four steps, employing MWA data as a method demonstration.


\subsubsection{Compute sample covariances in $uv$ annuli.}

We begin our empirical covariance calculation by taking the residual data cubes, defined as
\beq
\xhat^\text{res} \equiv \xhat_1/2 + \xhat_2/2 - \mean,
\eeq
and performing a discrete Fourier transform\footnote{For simplicity, we used the unitary discrete Fourier transform for these calculations and ignore any factors of length or inverse length that might come into these calculations only to be canceled out at a later step.} at each frequency independently to get $\widetilde{\mathbf{x}}^\text{res}$. This yields $n_x\times n_y$ sample ``lines of sight'' ($uv$ cells for all frequencies), as many as we have pixels in the map. As a first step toward estimating $\CHat$, we use the unbiased sample covariance estimator from these residual lines of sight. However, instead of calculating a single frequency-frequency covariance, we want to calculate many different $\CHat^\text{res}$ matrices to reflect the evolution of spectral structure with $k_\perp$ along the wedge. We therefore break the $uv$ plane into concentric annuli of equal width and calculate $\CHat^\text{res}_{uv}$ for each $uv$ cell as the sample covariance of the $N^\text{LOS} - 2$ lines of sight in that annulus, excluding the cell considered and its complex conjugate. Since the covariance is assumed to be block diagonal, this eliminates a potential bias that comes from downweighting a uv cell using information about that cell. Thus,
\beq
\widehat{C}_{uv,ff'}^\text{res} = \hspace{-3mm}\sum_{\substack{\text{other } u',v' \\ \text{in annulus}}} \hspace{-3mm} \frac{\left(\widetilde{x}^\text{res}_{u'v'f} - \langle \widetilde{x}^\text{res}_f \rangle\right) \left(\widetilde{x}^\text{res}_{u'v'f'} - \langle \widetilde{x}^\text{res}_{f'} \rangle \right)^*}{N^\text{LOS}-2-1}, \label{eq:CovEstOtherAnnuli}
\eeq
where $\langle \widetilde{x}^\text{res}_f \rangle$ is an average over all $u'$ and $v'$ in the annulus. We expect this procedure to be particularly effective in our case because the $uv$ coverage of the MWA after rotation synthesis is relatively symmetric.

\begin{figure}[]  
	\centering 
	\includegraphics[width=.75\textwidth]{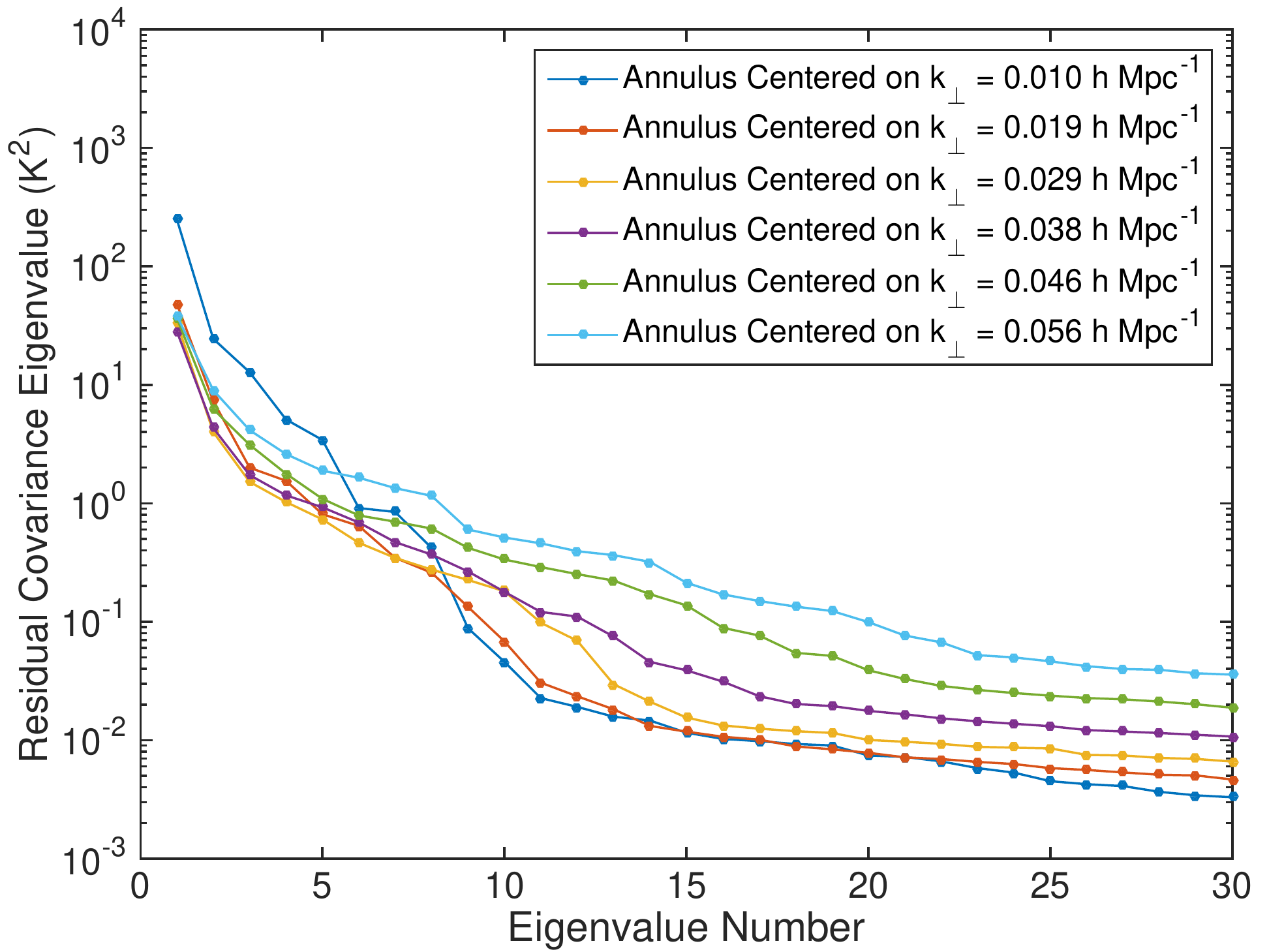}
	\caption[Residual covariance eigenvalues by $k_\perp$ annulus.]{The evolution of the wedge with $k_\perp$ motivates us to model foregrounds separately for discrete values of $k_\perp$. In this plot of the 30 largest eigenvalues of the observed residual covariance (which should include both noise and foregrounds) sampled in six concentric annuli, we see steeper declines toward a noise floor for the inner annuli than the outer annuli. This is consistent with the expected effect of the wedge---higher $k_\perp$ modes should be foreground contaminated at higher $k_\|$.}
	\label{fig:wedge_eigenvalues}
\end{figure}

As a sense check on these covariances, we plot their largest 30 eigenvalues in Figure \ref{fig:wedge_eigenvalues}. We see that as $|u|$ (and thus $k_\perp$) increases, the eigenspectra become shallower. At high $k_\perp$, the effect of the wedge is to leak power to a range of $k_\|$ values. The eigenspectrum of intrinsically smooth foregrounds should be declining exponentially \cite{AdrianForegrounds}. The wedge softens that decline. These trends are in line with our expectations and further motivate our strategy of forming covariance matrices for each annulus independently.  

Because we seek only to estimate the foreground portion of the covariance, the formal rank deficiency of $\CHat^\text{res}_{uv}$ is not a problem.\footnote{In fact, the rank of $\CHat^\text{res}_{uv}$ is $N_\text{LOS} - 3$ if $N^\text{LOS}-2 \le n_f$.} All we require is that the largest (and thus more foreground-dominated) modes be well measured. In this analysis, we used six concentric annuli to create six different frequency-frequency foreground covariances. Using more annuli allows for better modeling of the evolution of the wedge with $k_\perp$ at the expense of each estimate being more susceptible to noise and rank deficiency. 


\subsubsection{Subtract the properly projected noise covariance.}
 
The covariances computed from these $uv$ lines of sight include contributions from the 21\,cm signal and instrumental noise as well as foregrounds. We can safely ignore the signal covariance for now as we are far from the regime where sample variance is significant. We already have a theoretically motivated model for the noise (based on the $uv$ sampling) that has been empirically rescaled to the observed noise in the difference of time-interleaved data (the same basic procedure as in \cite{X13}). We would like an empirical estimate of the residual foreground covariance alone to use in $\C^{FG}$ and thus must subtract off the part of our measurement we think is due to noise variance.

To get to $\CHat^\text{FG}_{uv}$ from $\CHat^\text{res}_{uv}$, we subtract our best guess of the portion of $\CHat^\text{res}_{uv}$ that is due to noise, which we approximate by averaging the noise model variances in all the other $uv$ cells in the annulus at that given frequency, yielding
\beq
\widehat{C}_{uv,ff'}^\text{N} = \frac{1}{N^\text{LOS}} \hspace{-2mm}  \sum_{\substack{\text{other } u',v' \\ \text{in annulus}}}  \hspace{-2mm} \delta_{uu'} \delta_{vv'} \delta_{ff'} C^{N}_{uu'vv'ff'}.
\eeq
Note, however, that $\CHat_{uv}^\text{N}$ is full rank while $\CHat_{uv}^{res}$ is typically rank deficient. Thus a naive subtraction would oversubtract the noise variance in the part of the subspace of $\CHat_{uv}^\text{N}$ where $\CHat_{uv}^{res}$ is identically zero. Instead, the proper procedure is to find the projection matrices $\Proj_{uv}$ that discard all eigenmodes outside the subspace where $\CHat_{uv}^\text{res}$ is full rank. Each should have eigenvalues equal to zero or one only and have the property that
\beq
\Proj_{uv} \CHat_{uv}^\text{res} \Proj_{uv}^\trans = \CHat_{uv}^\text{res}.
\eeq
Only after projecting out the part of $\CHat^N_{uv}$ inside the unsampled subspace can we self-consistently subtract our best guess of the noise contribution to the subspace in which we seek to estimate foregrounds. In other words, we estimate $\CHat^\text{FG}_{uv}$ as
\beq
\CHat^\text{FG}_{uv} = \CHat_{uv}^\text{res} - \Proj_{uv} \CHat_{uv}^\text{N} \Proj_{uv}^\trans.
\eeq

\begin{figure*}[]  
	\centering 
	\includegraphics[width=1\textwidth]{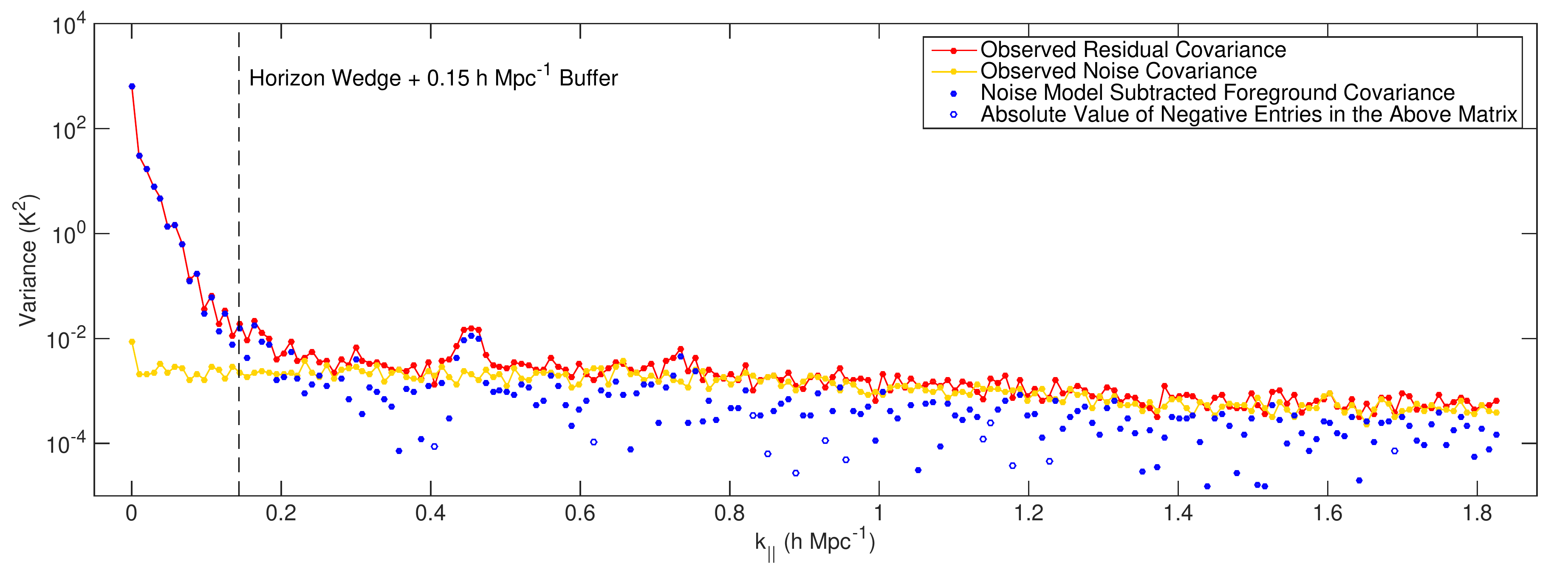}
	\caption[Residual, foreground, and noise variances in Fourier space.]{Examining the diagonal elements of the observed residual and inferred foreground covariance matrices in Fourier space reveals the effectiveness of subtracting model for the noise covariance. In red, we plot the observed residual covariance, which contains both foregrounds and noise. As a function of $k_\|$, the two separate relatively cleanly---there is a steeply declining foreground portion on the left followed by a relatively flat noise floor on the right. The theory that the right-hand portion is dominated by noise is borne out by the fact that it so closely matches the observed noise covariance, inferred lines of sight of $\x_1 - \x_2$, which should have only noise and no sky signal at all. The regions where they differ significantly, for example at $k_\| \sim 0.45$\,$h$\,Mpc$^{-1}$ , are attributable to systematic effects like the MWA's coarse band structure that have not been perfectly calibrated out. For the example covariances shown here (which correspond to a mode in the annulus at $k_\perp \approx 0.010$\,$h$\,Mpc$^{-1}$), we can see that subtracting a properly projected noise covariance removes most of the power from the noise-dominated region, leaving only residual noise that appears both as negative power (open blue circle) and as positive power (closed blue circles) at considerably lower magnitude.}
	\label{fig:fourier_diags}
\end{figure*}

We demonstrate the effectiveness of this technique in Figure \ref{fig:fourier_diags} by plotting the diagonal elements of the Fourier transform of $\CHat_{uv}^\text{res}$ and $\CHat_{uv}^\text{FG}$ along the line of sight. Subtracting of the noise covariance indeed eliminates the majority of the power in the noise dominated modes at high $k_\|$; thus we expect it also to fare well in the transition region near the edge of the wedge where foreground and noise contributions are comparable.

\subsubsection{Perform a $k_\|$ filter on the covariance.} \label{sec:filter}

\begin{figure}[] 
	\centering 
	\includegraphics[width=.6\textwidth]{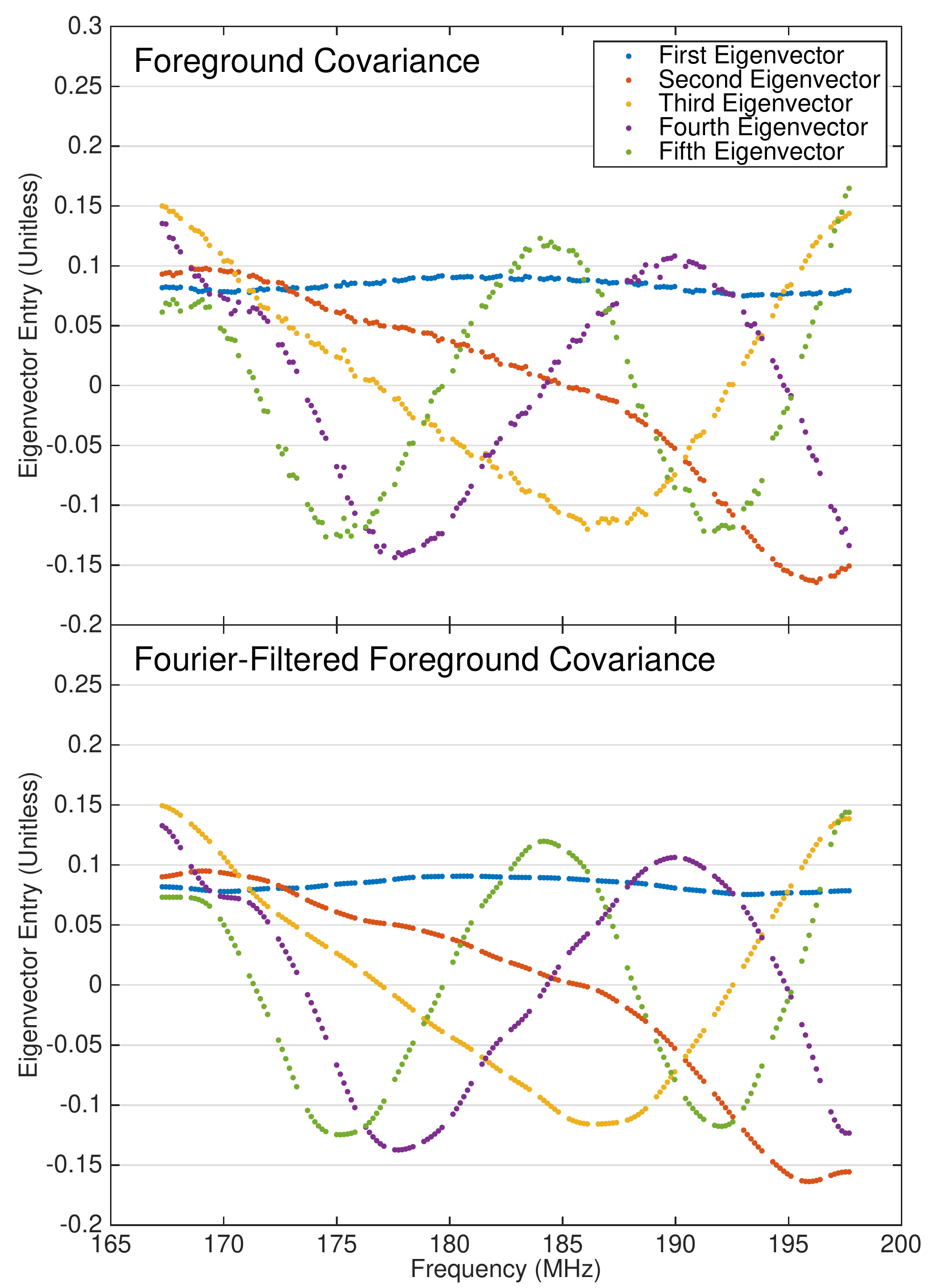} 
	\caption[Effect of Fourier filtering on foreground covariance eigenvectors.]{The foreground covariance we estimate from our limited data set is still very noisy, and we run the risk of overfitting the noise in our measurements if we take it at face value. In the top panel, we plot the eigenvectors corresponding to the five largest eigenvalues of $\CHat^\text{FG}$ for a mode in the annulus centered on $k_\perp \approx 0.010$\,$h$\,Mpc$^{-1}$.  In the bottom panel, we show dominant eigenvectors of the Fourier-filtered covariance. As expected, they resemble the first five Fourier modes. The missing data every 1.28 MHz are due to channels flagged at the edge of the coarse bandpass of the MWA's polyphase filter bank---the most difficult part of the band to calibrate.}
	\label{fig:eigenvectors}
\end{figure}

Despite the relatively clean separation of foreground and noise eigenvalues, inspection of some of the foreground-dominated modes in the top panel of Figure \ref{fig:eigenvectors} reveals residual noise. Using a foreground covariance constructed from these noisy foreground eigenmodes to downweight the data during power spectrum estimation would errantly downweight some high $k_\|$ modes in addition to the low $k_\|$ foreground-dominated modes. To avoid this double counting of the noise, we allow the foreground covariance to include only certain $k_\|$ modes by filtering $\CHat^\text{FG}_{uv}$ in Fourier space to get $\CHat^\text{FG,filtered}_{uv}$. Put another way, we are imposing a prior on which Fourier modes we think have foreground power in them. The resulting noise filtered eigenmodes are shown in the bottom panel of Figure \ref{fig:eigenvectors}.

In practice, implementing this filter is subtle. We interpolate $\CHat^\text{FG}$ over the flagged frequency channels using a cubic spline, then symmetrically pad the covariance matrix, forcing its boundary condition to be periodic. We then Fourier transform, filter, inverse Fourier transform, remove the padding, and then rezero the flagged channels. 

Selecting a filter to use is also a subtle choice. We first keep modes inside the horizon wedge with an added buffer. For each annulus, we calculate a mean value of $k_\perp$, and then use Equation \eqref{eq:wedge} to calculate the $k_\|$ value of the horizon wedge, using $\theta_0 = \pi/2$. Although the literature suggests a $0.1$ to $0.15$\,$h$\,Mpc$^{-1}$ buffer for ``suprahorizon emission'' due to some combination of intrinsic spectral structure of foregrounds, primary beam chromaticity, and finite bandwidth \cite{PoberWedge,PoberNextGen}, we pick a conservative $0.5$\,$h$\,Mpc$^{-1}$. Then we examine the diagonal of $\CHat^\text{FG}$ (Figure \ref{fig:fourier_diags}) to identify additional foreground modes, this time in the EoR window, due to imperfect bandpass calibration appearing as spikes. One example is the peak at $k_\| \sim 0.45$\,$h$\,Mpc$^{-1}$. Such modes contribute errant power to the EoR window at constant $k_\|$. Since these modes result from the convolution of the foregrounds with our instrument, they also should be modeled in $\C^{FG}$ in order to minimize their leakage into the rest of the EoR window. 

One might be concerned that cosmological signal and foregrounds theoretically both appear in the estimate of $\C^{FG}$ that we have constructed, especially with our conservative $0.5$\,$h$\,Mpc$^{-1}$ buffer that allows foregrounds to be discovered well into the EoR window. For the purposes of calculating $\C^{-1}(\widehat{\x}-\mean)$ in the quadratic estimator in Equation \eqref{eq:QuadEst}, that is fine since its effect is to partially relax the assumption that sample variance can be ignored. However, the calculation of the bias depends on being able to differentiate signal from contaminants \cite{Maxpowerspeclossless,LT11,DillonFast}. 

The noise contribution to the bias can be eliminated by cross-correlating maps made from interleaved time steps \cite{X13}. However, we cannot use our inferred $\C^{FG}$ to subtract a foreground bias without signal loss. That said, we can still set an upper limit on the 21\,cm signal. By following the data and allowing the foreground covariance to have power inside the EoR window, we are minimizing the leakage of foregrounds into uncontaminated regions and we are accurately marking those regions as having high variance. As calibration and the control of systematic effects improves, we should be able to isolate foregrounds to outside the EoR window, impose a more aggressive Fourier filter on $\C^{FG}$, and make a detection of the 21\,cm signal by employing foreground avoidance.


\subsubsection{Cut out modes attributable to noise.}

After suppressing the noisiest modes with our Fourier filter, we must select a cutoff beyond which the foreground modes are irrecoverably buried under noise. We do this by inspecting the eigenspectrum of $\CHat^\text{FG,filtered}_{uv}$. The true $\C^{FG}$, by definition, admits only positive eigenvalues (though some of them should be vanishingly small). 

By limiting the number of eigenvalues and eigenvectors we ultimately associate with foregrounds, we also limit the potential for signal loss by allowing a large portion of the free parameters to get absorbed into the contaminant model \cite{EricAdrianEmpiricalGlobalForegrounds,PAPER64Limits}. When measuring the power spectrum inside the EoR window, we can be confident that signal loss is minimal compared to foreground bias and other errors.

\begin{figure}[]  
	\centering 
	\includegraphics[width=.75\textwidth]{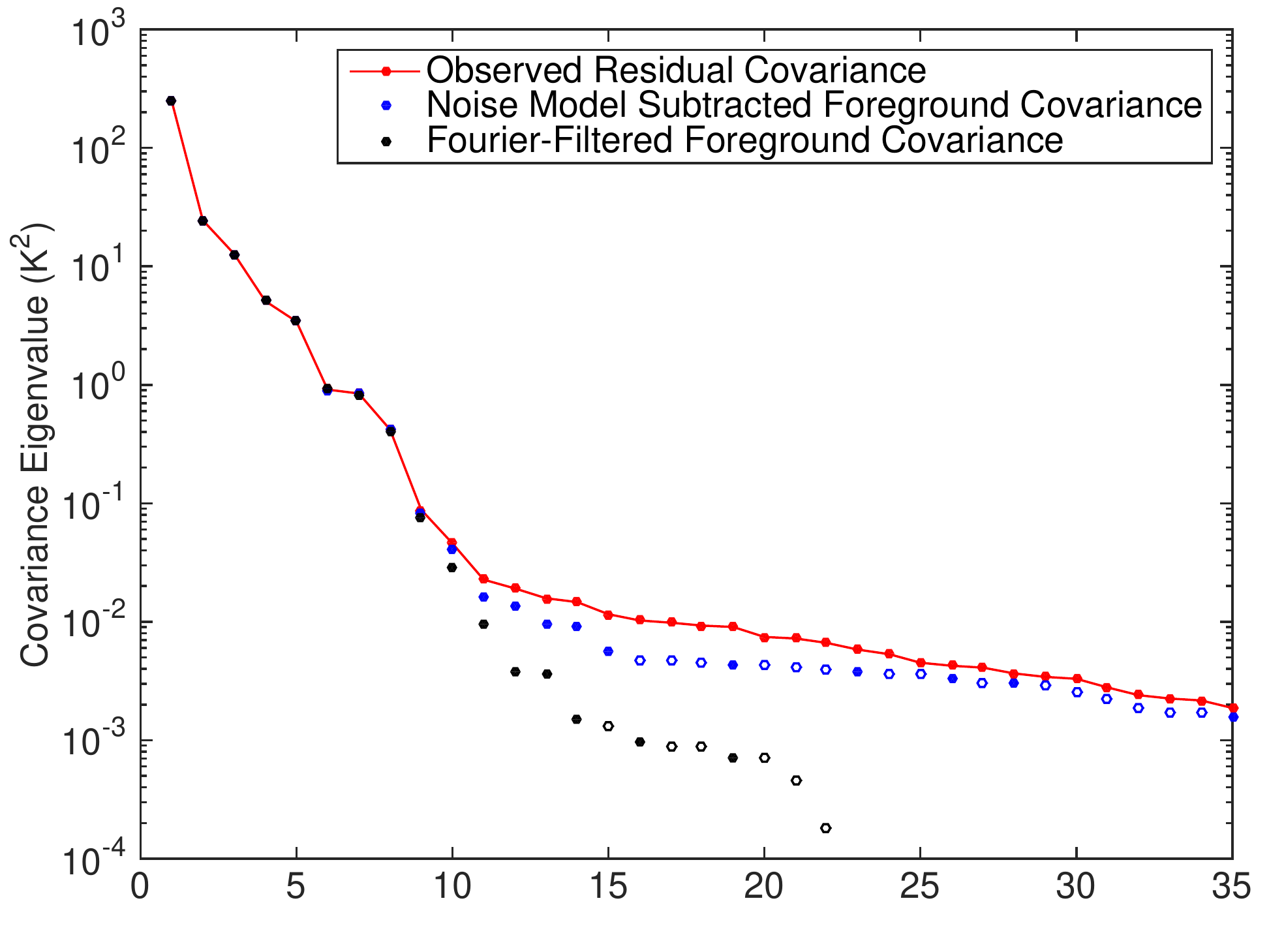}
	\caption[Eigenvalues of the empirical covariance at each processing step.]{The evolution of the eigenvalues of our estimated foreground covariance matrix for a mode in the annulus corresponding to $k_\perp \approx 0.010$\,$h$\,Mpc$^{-1}$ at each of the first three stages of covariance estimation. First we calculate a sample covariance matrix from the residual data cubes (shown in red). Next we subtract our best guess as to the part of the diagonal of that matrix that originates from instrumental noise, leaving the blue dots (open circles are absolute values of negative eigenvalues). Then we filter out modes in Fourier space along the line of sight that we think should be noise dominated, leaving the black dots. Finally, we project out the eigenvectors associated with eigenvalues whose magnitude is smaller than the largest negative eigenvalue, since those are likely due to residual noise. What remains is our best guess at the foreground covariance in an annulus and incorporates as well as possible our prior beliefs about its structure.}
	\label{fig:cov_steps}
\end{figure}

We plot in Figure \ref{fig:cov_steps} the eigenspectra of $\CHat_{uv}^\text{res}$, $\CHat^\text{FG}_{uv}$, and $\CHat^\text{FG,filtered}_{uv}$, sorted by absolute value. There are two distinct regions---the sharply declining foreground-dominated region and a flatter region with many negative eigenvalues. We excise eigenvectors whose eigenvalues are smaller in absolute value than the most negative eigenvalue. This incurs a slight risk of retaining a few noise dominated modes, albeit strongly suppressed by our noise variance subtraction and our Fourier filtering. Finally we are able to construct the full covariance $\CHat$ using Equation \eqref{eq:blockdiag}.

\subsection{Review of Assumptions and Caveats} \label{sec:caveats}

Before proceeding to demonstrate the effectiveness of our empirical covariance modeling method, it is useful to review and summarize the assumptions made about mapmaking and covariance modeling. Some are inherited from the previous application of quadratic power spectrum estimation to the MWA \cite{X13}, while others are necessitated by our new, more faithful foreground covariance. Relaxing these assumptions in a computationally efficient manner remains a challenge we leave for future work.

\begin{enumerate}[i.]
\item We adopt the flat sky approximation as in \cite{DillonFast,X13}, allowing us to use the fast Fourier transform to quickly compute power spectra. The error incurred from this approximation on the power spectrum is expected to be smaller than 1\% \cite{X13}.

\item We assume the expectation value of our uniformly weighted map is the true sky (i.e., $\langle \widehat{\x} \rangle = \x^\text{true}$) when calculating $\C,_\beta$ in Equation \ref{eq:QuadEst}, again following \cite{X13}. In general $\langle \widehat{\x} \rangle$ is related to $\x^\text{true}$ by $\PSF$, the matrix of point spread functions \cite{Mapmaking}. Here we effectively approximate the PSF as position independent. Relaxing this approximation necessitates the full mapmaking theory presented in \cite{Mapmaking} which has yet to be integrated into a power spectrum estimation pipeline. 

\item We approximate the foreground covariance as uncorrelated between different $uv$ cells (and thus block diagonal). At some level there likely are correlations in $uv$, though those along the line of sight are far stronger. It may be possible to attempt to calculate these correlations empirically, but it would be very difficult considering relative strength of line-of-sight correlations. It may also be possible to use a nonempirical model, though that has the potential to make the computational speedups of \cite{DillonFast} more difficult to attain.
 
\item We approximate the frequency-frequency foreground covariance as constant within each annulus, estimating our covariance for each $uv$ cell only from other cells in the same annulus. In principle, even if the foreground residuals were isotropic, there should be radial evolution within each annulus which we ignore for this analysis. 

\item The Fourier filter is a nontrivial data analysis choice balancing risk of noise double counting against that of insufficiently aggressive foreground downweighting.

\item In order to detect the 21\,cm signal, we assume that foregrounds can be avoided by working within the EoR window. Out of fear of losing signal, we make no effort to subtract a residual foreground bias from the window. This makes a detection inside the wedge impossible and it risks confusing foreground contamination in the window for a signal. Only analysis of the dependence of the measurement on $z$, $k$, $k_\|$, and $k_\perp$ can distinguish between systematics and the true signal. 

\end{enumerate}


\section{Results} \label{sec:results}

We can now demonstrate the statistical techniques we have motivated and developed in Section \ref{sec:methods} on the problem of estimating power spectra from a 3\,h observation with the 128-antenna MWA. We begin with a discussion of the instrument and the observations in Section \ref{sec:observation}. In Section \ref{sec:processing} we detail the data processing from raw visibilities to calibrated maps from which we estimate both the foreground residual covariance matrix and the power spectrum. Finally, in Section \ref{sec:results} we present our results and discuss lessons learned looking toward a detection of the 21\,cm signal.


\subsection{Observation Summary} \label{sec:observation}

The 128-antenna Murchison Widefield Array began deep EoR observations in mid-2013. We describe here the salient features of the array and refer to \cite{TingaySummary} for a more detailed description. The antennas are laid out over a region of radius 1.5\,km in a quasirandom, centrally concentrated distribution which achieves approximately complete $uv$ coverage at each frequency over several hours of rotation synthesis \cite{MWAsensitivity}. Each antenna element is a phased array of 16 wideband dipole antennas whose phased sum forms a discretely steerable 25$^\circ$ beams (full width at half maximum) at 150\,MHz with frequency-dependent, percent level sidelobes \cite{AbrahamOrbcomm}. We repoint the beam to our field center on a 30\,min cadence to correct for earth rotation, effectively acquiring a series of drift scans over this field. 

We observe the MWA ``EOR0'' deep integration field, centered at R.A.(J2000) $= 0^\text{h}\,0^\text{m}\,0^\text{s}$ and decl.(J2000) $= -30^\circ\,0'\,0''$. It features a near-zenith position, a high Galactic latitude, minimal Galactic emission \cite{GSM}, and an absence of bright extended sources. This last property greatly facilitates calibration in comparison to the ``EOR2'' field---a field dominated by the slightly resolved radio galaxy Hydra A at its center---which was used by \cite{ChrisMWA} and \cite{X13}. A nominal 3\,h set of EOR0 observations was selected during the first weeks of observing to use for refining and comparing data processing, imaging, and power spectra pipelines \cite{JacobsPipelines}. In this work, we use the ``high band,'' near-zenith subset of these observations with 30.72\,MHz of bandwidth and center frequency of 182\,MHz, recorded on Aug 23, 2013 between 16:47:28 and 19:56:32 UTC (22.712 and 1.872 hours LST). 


\subsection{Calibration and Mapmaking Summary} \label{sec:processing}

Preliminary processing, including radio frequency interference (RFI) flagging followed by time and frequency averaging, was performed with the \texttt{COTTER} package \cite{AndreMWARFI} on raw correlator data. These data were collected at 40\,kHz resolution with an integration time of 0.5\,s, and averaged to 80\,kHz resolution with a 2\,s integration time to reduce the data volume. Additionally, 80\,kHz at the upper and lower edges of each of 24 coarse channels (each of width 1.28\,MHz) of the polyphase filter bank is flagged due to known aliasing.

As in \cite{X13}, we undertake snapshot-based processing in which each minute-scale integration is calibrated and imaged independently. Snapshots are combined only at the last step in forming a Stokes $I$ image cube, allowing us to properly align and weight them despite different primary beams due to sky rotation and periodic repointing. 
While sources are forward modeled for calibration and foreground subtraction using the full position dependent PSF (i.e., the synthesized beam), we continue to approximate it as position independent (and equal to that of a point source at the field center) during application of uniform weighting and computation of the noise covariance. 

We use the calibration, foreground modeling, and first stage image products produced by the Fast Holographic Deconvolution\footnote{For a theoretical discussion of the algorithm see \cite{FHD}. The code is available at \url{https://github.com/miguelfmorales/FHD}.} (FHD) pipeline as described by \cite{JacobsPipelines}. The calibration implemented in the FHD package is an adaptation of the fast algorithm presented by \cite{Salvini2014} with a baseline cutoff of $b>50\lambda$. In this data reduction, the point source catalogs discussed below are taken as the sky model for calibration.  Solutions are first obtained per antenna and per frequency before being constrained to linear phase slopes and quadratic amplitude functions after correcting for a median antenna-independent amplitude bandpass.  The foreground model used for subtraction includes models both of diffuse radio emission \cite{AdamDiffuse} and point sources. In detail, the point source catalog is the union of a deep MWA point source survey within $20^\circ$ of the field center \cite{PattiCatalog1}, the shallower but wider MWA commissioning point source survey \cite{MWACS}, and the Culgoora catalog \cite{Slee1995}. Note that calibration and foreground subtraction of off-zenith observations are complicated by Galactic emission picked up by primary beam sidelobes, and are active topics of investigation \cite{PoberSidelobe,NithyaPitchfork,NithyaPitchforkConfirmation}. During these observations a single antenna was flagged due to known hardware problems, and 1--5 more were flagged for any given snapshot due to poor calibration solutions.

These calibration, foreground modeling, and imaging steps constitute notable improvements over \cite{X13}. In that work, the presence of the slightly resolved Hydra A in their EOR2 field likely limited calibration and subtraction fidelity as only a point source sky model was used. In contrast, the EOR0 field analyzed here lacks any such nearby radio sources. Our foreground model contains $\sim 2500$ point sources within the main lobe and several thousand more in the primary beam sidelobes in addition to the aforementioned diffuse map. A last improvement in the imaging is the more frequent interleaving of time steps for the cross power spectrum, which we performed at the integration scale (2\,s) as opposed to the snapshot scale (a few minutes). This ensures that both $\xhat_1$ and $\xhat_2$ have identical sky responses and thus allows us to accurately estimate the noise in the array from difference cubes. Assuming that the system temperature contains both an instrumental noise temperature and a frequency dependent sky noise temperature that scales as $\nu^{-2.55}$, the observed residual root-mean-square brightness temperature is consistent with $T_\text{sys}$ ranging from 450\,K at 167\,MHz to 310\,K at 198\,MHz, in line with expectations \cite{MWAsensitivity}.

As discussed in \cite{JacobsPipelines} and \cite{HazeltonEppsilon}, FHD produces naturally weighted sky, foreground model, ``weights,'' and ``variances'' cubes, as well as beam-squared cubes. All are saved in image space using the HEALPix format \cite{HEALPIX} with $N_\text{side}=1024$. Note that these image cubes are crops of full-sky image cubes to a $16^\circ\times16^\circ$ square field of view, as discussed below. The sky, foreground model, and weights cubes are image space representations of the measured visibilities, model visibilities, and sampling function, respectively, all originally gridded in $uv$ space using the primary beam as the gridding kernel. The variances cube is similar to the weights cube, except the gridding kernel is the {\textit square} of the $uv$ space primary beam. It represents the proper quadrature summation of independent noise in different visibilities when they contribute to the same $uv$ cell, and will ultimately become our diagonal noise covariance model. The FHD cubes from all ninety-four 112\,s snapshots are optimally combined in this ``holographic'' frame in which the true sky is weighted by two factors of the primary beam, as in \cite{X13}.

We perform a series of steps to convert the image cube output of FHD into uniformly weighted Stokes $I$ cubes accompanied by appropriate $uv$ coverage information for our noise model. We first map these data cubes onto a rectilinear grid, invoking the flat sky approximation. We do this by rotating the (RA,Dec) HEALPix coordinates of the EOR0 field to the north pole (0$^\circ$,90$^\circ$), and then projecting and gridding onto the $xy$ plane with $0.2^\circ\times0.2^\circ$ resolution over a $16^\circ\times16^\circ$ square field of view. To reduce the data volume while maintaining cosmological sensitivity, we coarse grid to approximately $0.5^\circ$ resolution by Fourier transforming and cropping these cubes in the $uv$ plane at each frequency. We form a uniformly weighted Stokes I cube $I_{\text{uni}}(\vec{\theta})$ by first summing the XX and YY data cubes, resulting in a naturally weighted, holographic stokes I cube $I_{\text{nat},h}(\vec{\theta})  = I_{XX,h}(\vec{\theta})+I_{YY,h}(\vec{\theta})$. Then we divide out the holographic weights cube $W_h({\vec{\theta}})$ in $uv$ space, which applies uniform weighting and removes one image space factor of the beam, and lastly divide out the second beam factor $B(\vec{\theta})$: $I_{\text{uni}}(\vec{\theta}) = \mathcal{F}^{-1}[\mathcal{F}I_{\text{nat},h}(\vec{\theta})/\mathcal{F}W_h(\vec{\theta})]/B(\vec{\theta})$, where $\mathcal{F}$ represents a Fourier transform and $B(\vec{\theta}) = [B_{XX}^2(\vec{\theta})+B_{YY}^2(\vec{\theta})]^{1/2}$. Consistent treatment of the variances cube requires $uv$ space division of {\it two} factors of the weights cube followed by image space division of {\it two} factors of the beam.

Lastly, we frequency average from 80 kHz to 160 kHz, flagging a single 160 kHz channel the edge of each 1.28\,MHz coarse channel due to polyphase filter bank attenuation and aliasing, which make these channels difficult to reliably calibrate. Following \cite{X13}, we also flag poorly observed $uv$ cells and $uv$ cells whose observation times vary widely between frequencies. In all cases, we formally set the variance in flagged channels and $uv$ cells in $\C^N$ to infinity and use the pseudoinverse to project out flagged modes \cite{X13}. 
 
  
\subsection{Power Spectrum Results}
 
We can now present the results of our method applied to 3\,h of MWA-128T data. We first study cylindrically averaged, two-dimensional (2D) power spectra and their statistics, since they are useful for seeing the effects of foregrounds and systematic errors on the power spectrum. We form these power spectra with the full 30.72\,MHz instrument bandwidth to achieve maximal $k_\|$ resolution.

\begin{figure}[] 
	\centering 
	\includegraphics[width=.75\textwidth]{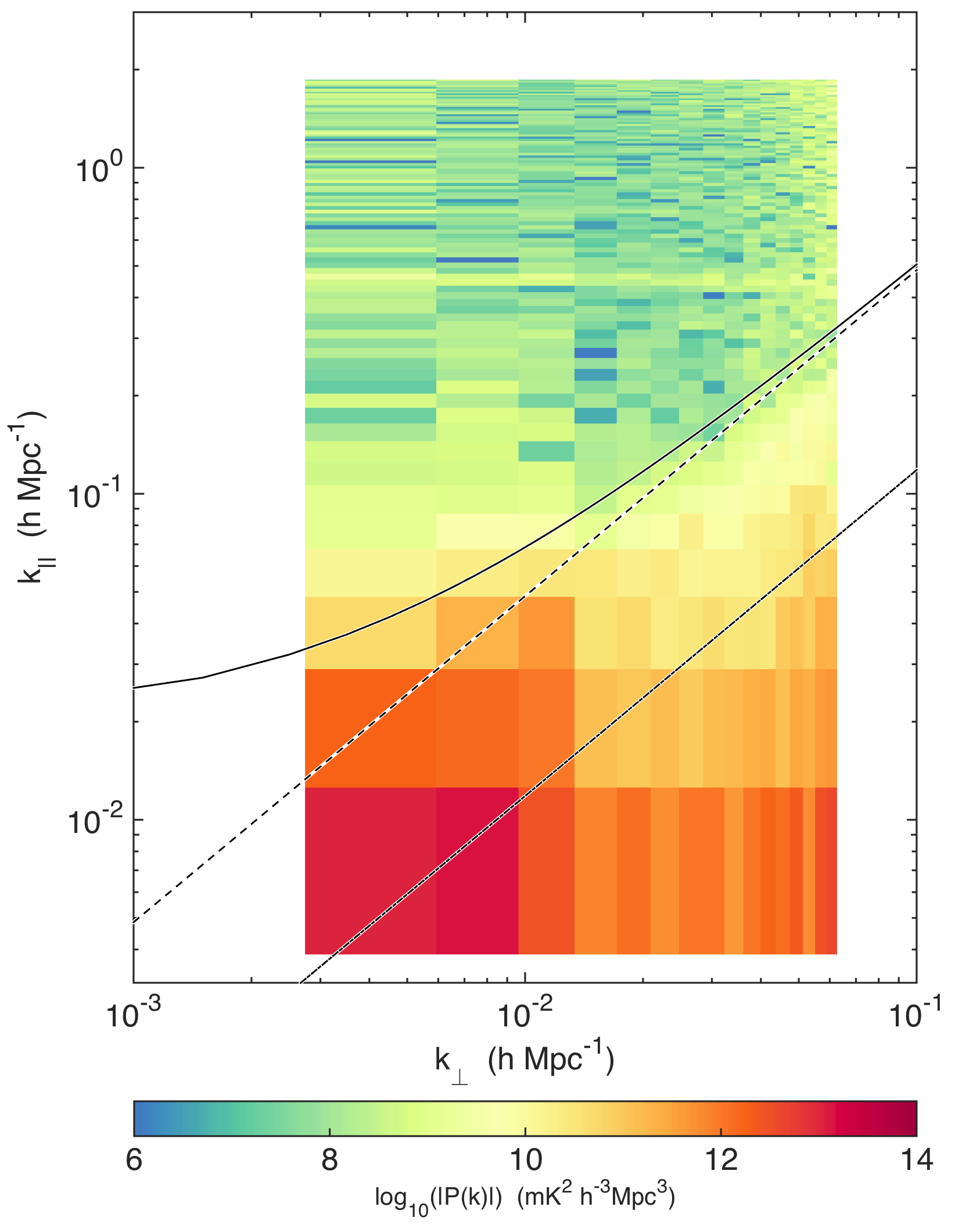}
	\caption[$|P(k_\perp,k_\|)|$ over the full observed bandwidth.]{Our power spectrum clearly exhibits the typical EoR window structure with orders-of-magnitude suppression of foregrounds in the EoR window. Here we plot our estimates for $|P(k_\perp,k_\|)|$ for the full instrumental bandwidth, equivalent to the range $z=6.2$ to $z=7.5$. Overplotted is the wedge from Equation \ref{eq:wedge} corresponding to the first null in the primary beam (dash-dotted line), the horizon (dashed line), and the horizon with a relatively aggressive \wedgeBuffer\,$h$\,Mpc$^-1$ buffer (solid curve). In addition to typical foreground structure, we also see the effect of noise at high and very low $k_\perp$ where baseline coverage is poor. We also clearly see a line of power at constant $k_\| \approx 0.45$\,$h$\,Mpc$^{-1}$, attributable to miscalibration of the instrument's bandpass and cable reflections \cite{HazeltonEppsilon}.}
	\label{fig:2dPk}
\end{figure}

We begin with the 2D power spectrum itself (Figure \ref{fig:2dPk}) in which several important features can be observed.
First, the wedge and EoR window are clearly distinguishable, with foregrounds suppressed by at least 5 orders of magnitude across most of the EoR window. At high $k_\perp$, the edge of the wedge is set by the horizon while at low $k_\perp$ the cutoff is less clear. There appears to be some level of suprahorizon emission, which was also observed with PAPER in \cite{PoberWedge} and further explained by \cite{EoRWindow1}. Consistent with Figure \ref{fig:wedge_eigenvalues} we see the strongest foreground residual power at low $k_\perp$, meaning that there still remains a very large contribution from diffuse emission from our Galaxy---potentially from sidelobes of the primary beam affecting the shortest baselines \cite{NithyaPitchfork,NithyaPitchforkConfirmation}.

We also see evidence for less-than-ideal behavior. Through we identified spectral structure appearing at $k_\| \sim 0.45$\,$h$\,Mpc$^{-1}$ in Figure \ref{fig:fourier_diags} and included it in our foreground residual covariance, that contamination still appears here as a horizontal line. By including it in the foreground residual model, we increase the variance we associate with those modes and we decrease the leakage out of those modes, isolating the effect to only a few $k_\|$ bins.

\begin{figure}[] 
	\centering 
	\includegraphics[width=.75\textwidth]{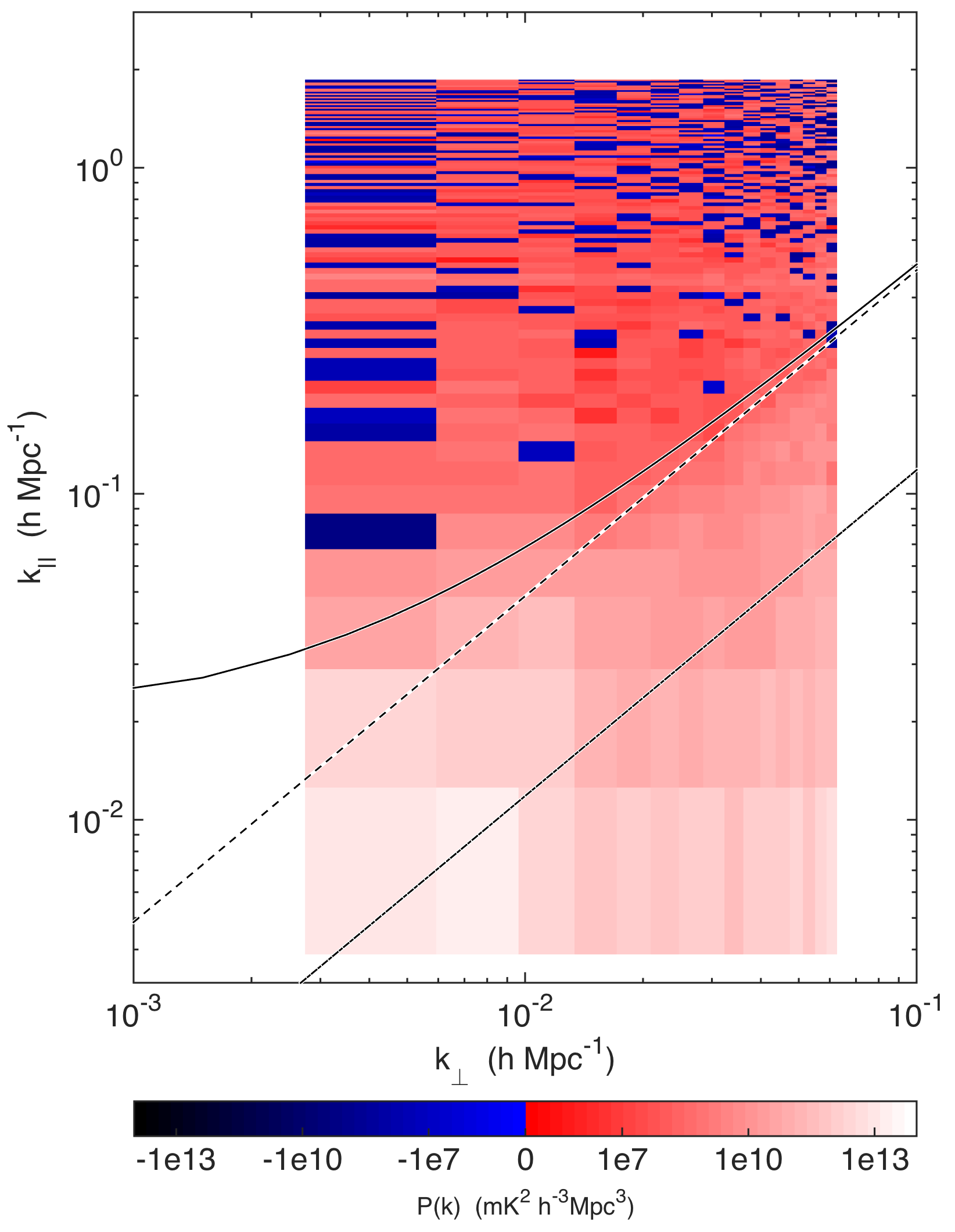}
	\caption[ArcSinh$(P(k_\perp,k_\|))$ over the full observed bandwidth.]{By using an estimator of the power spectrum with uncorrelated errors between bins, we can see that most of the EoR window is noise dominated in our power spectum measurement. Here we show the inverse hyperbolic sine of the power spectum, which behaves linearly near zero and logarithmically at large magnitudes. Because we are taking a cross power spectrum between two data cubes with uncorrelated noise, noise dominated regions are equally likely to have positive power as negative power. Since we do not attempt to subtract a foreground bias, foreground contaminated regions show up as strongly positive. That includes the wedge, the bandpass line at $k_\| \approx 0.45$\,$h$\,Mpc$^{-1}$  (see Figure \ref{fig:2dPk}), and some of the EoR window at low $k_\perp$ and relatively low $k_\|$, consistent with the suprahorizon emission seen in \cite{PoberWedge}.}
	\label{fig:2dPkArcSinh}
\end{figure}

While Figure \ref{fig:2dPk} shows the magnitude of the 2D power spectrum, Figure \ref{fig:2dPkArcSinh} shows its sign using a split color scale, providing another way to assess foreground contamination in the EoR window. 
Because we are taking the cross power spectrum between two cubes with identical sky signal but independent noise realizations, the noise dominated regions should be positive or negative with equal probability. This is made possible by our use of a power spectrum estimator normalized such that $\boldsymbol\Sigma \equiv \text{Cov}(\widehat{\p})$ is a diagonal matrix \cite{DillonFast}. This choice limits leakage of foreground residuals from the wedge into the EoR window \cite{X13}.

By this metric, the EoR window is observed to be noise dominated with only two notable exceptions. The first is the region just outside the wedge at low $k_\perp$ attributable to suprahorizon emission due to some combination of  intrinsic foreground spectral structure, beam chromaticity, and finite bandwidth. This suggests our aggressive 0.02\,$h$\,Mpc$^{-1}$ cut beyond the horizon will leave in some foreground contamination when we bin to form one-dimensional (1D) power spectra. As long as we are only claiming an upper limit on the power spectrum, this is fine. A detection of foregrounds is also an upper limit on the cosmological signal. More subtle is the line of positive power at $k_\| \sim 0.45$\,$h$\,Mpc$^{-1}$, confirming our hypothesis that the spike observed in Figure \ref{fig:2dPk} is indeed an instrumental systematic since it behaves the same way in both time-interleaved data cubes. There is also a hint of a similar effect at $k_\| \sim 0.75$\,$h$\,Mpc$^{-1}$, possibly visible in Figure \ref{fig:fourier_diags} as well. We attribute both to bandpass miscalibration due to cable reflections, complicated at these frequency scales by the imperfect channelization of the MWA's two-stage polyphase filter, as well as slight antenna dependence of the bandpass due to cable length variation \cite{HazeltonEppsilon}.

\begin{figure}[] 
	\centering 
	\includegraphics[width=.75\textwidth]{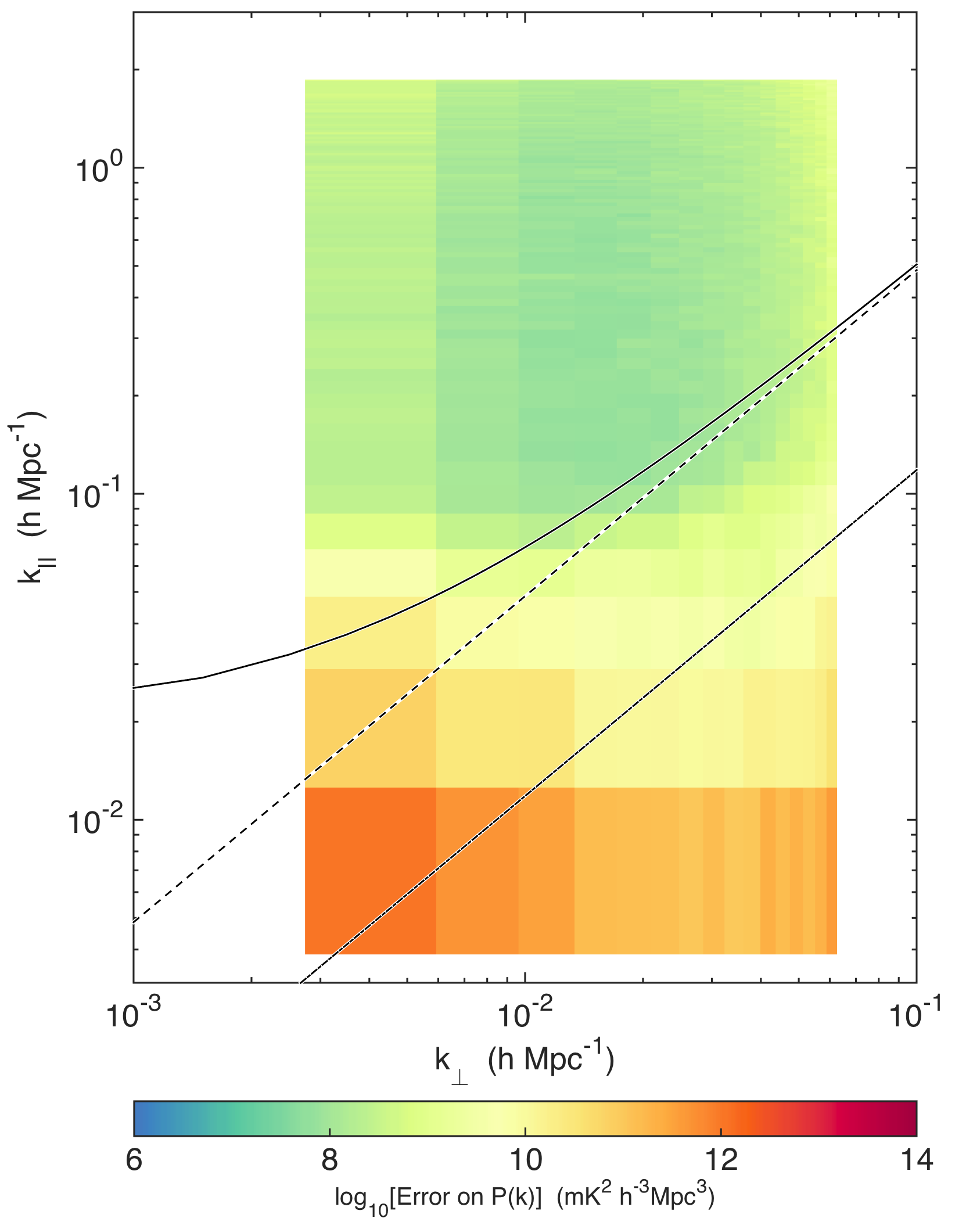}
	\caption[2D power spectrum variance due to noise and foregrounds.]{By including both residual foregrounds and noise in $\C$, our model for the covariance, we can calculate the expected variance on each band power in $\widehat{\p}$, which we show here. We see more variance at high (and also very low) $k_\perp$ where we have few baselines. We also see high variance at low $k_\|$ consistent with foregrounds. We see the strongest foregrounds at low $k_\perp$, which implies that the residual foregrounds have a very strong diffuse component that we have much to gain from better diffuse models to subtract. We also see that foreground-associated variance extends to higher $k_\|$ at high $k_\perp$, which is exactly the expected effect from the wedge. Both these observations are consistent with the structure of the eigenmodes we saw in Figure \ref{fig:wedge_eigenvalues}. Because we have chosen a normalization of $\widehat{\p}$ such that the $\text{Cov}(\widehat{\p})$ is diagonal, this is a complete description of our errors. Furthermore, it means that the band powers form a mutually exclusive and collectively exhaustive set of measurements.}
	\label{fig:2dError}
\end{figure}

Additionally, the quadratic estimator formalism relates our covariance models of residual foregrounds and noise to the expected variance in each band power \cite{LT11,DillonFast,X13}, which we plot in Figure \ref{fig:2dError}.
As we have chosen our power spectrum normalization $\M$ such that $\boldsymbol\Sigma \equiv \text{Cov}(\widehat{\p})$ is diagonal, it is sufficient to plot the diagonal of $\boldsymbol\Sigma^{1/2}$, the standard deviation of each band power. The EoR window is seen clearly here as well. There is high variance at low and high $k_\perp$ where the $uv$ coverage is poor, and also in the wedge due to foreground residuals. It is particularly pronounced in the bottom left corner, which is dominated by residual diffuse foregrounds.

\begin{figure}[] 
	\centering 
	\includegraphics[width=.75\textwidth]{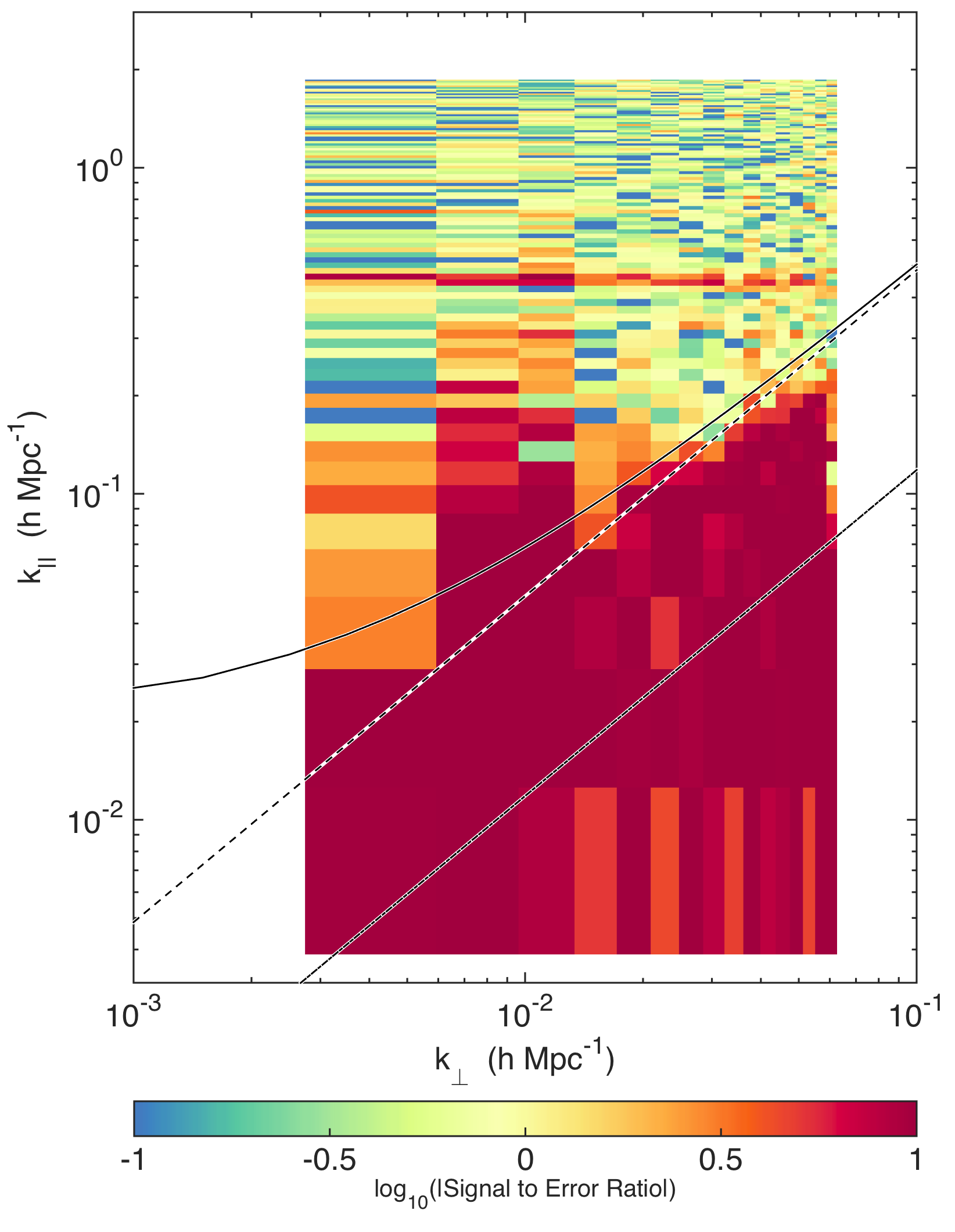}
	\caption[2D power spectrum signal-to-error ratio.]{The foregrounds' wedge structure is particularly clear when looking at the ratio of our measured power spectrum to the modeled variance, shown here. Though the variance in foreground residual dominated parts of the $k_\perp$-$k_\|$ plane are elevated (see Figure \ref{fig:2dError}), we still expect regions with signal to error ratios greater than one. This is largely due to the fact that we choose not to subtract a foreground bias for fear of signal loss. This figure shows us most clearly where the foregrounds are important and, as with Figure \ref{fig:2dPkArcSinh}, it shows where we can hope to do better with more integration time and where we need better calibration and foreground modeling to further integrate down.}
	\label{fig:2dSNR}
\end{figure}

As our error covariance represents the error due to both noise and foregrounds we expect to make in an estimate of the 21\,cm signal, it is interesting to examine the ``signal to error ratio'' in Figure \ref{fig:2dSNR}---the ratio of Figure \ref{fig:2dPk} to Figure \ref{fig:2dError}.
The ratio is of order unity in noise dominated regions---though it is slightly lower than what we might naively expect due to our conservative estimate of $\boldsymbol\Sigma$ \cite{X13}. That explains the number of modes with very small values in Figure \ref{fig:2dError}. In the wedge and just above it, however, the missubtracted foreground bias is clear, appearing as a high significance ``detection'' of the foreground wedge in the residual foregrounds. The bandpass miscalibration line at $k_\| \sim 0.45$\,$h$\,Mpc$^{-1}$ also appears clearly due to both foreground bias and possibly an underestimation of the errors. Hedging against this concern, we simply project out this line from our estimator that bins 2D power spectra into 1D power spectra by setting the variance of those bins to infinity.

Though useful for the careful evaluation of our techniques and of the instrument, the large bandwidth data cubes used to make Figures \ref{fig:2dPk} and \ref{fig:2dPkArcSinh} encompass long periods of cosmic time over which the 21\,cm power spectrum is expected to evolve. The cutoff is usually taken to be $\Delta z \lesssim 0.5$ \cite{Yi}. These large data cubes also violate the assumption in \cite{DillonFast} that channels of equal width in frequency correspond to equal comoving distances, justifying the use of the fast Fourier transform. Therefore, we break the full bandwidth into three 10.24\,MHz segments before forming spherically averaged power spectra, and estimate the foreground residual covariance and power spectrum independently from each. We bin our 2D power spectra into 1D power spectra using the optimal estimator formalism of \cite{X13}. In our case, since we have chosen $\M$ such that $\boldsymbol\Sigma$ is diagonal, this reduces to simple inverse variance weighting with the variance on modes outside the EoR window or in the $k_\| \sim 0.45$\,$h$\,Mpc$^{-1}$ line set to infinity.

\begin{figure*}[] 
	\centering 
	\includegraphics[width=1\textwidth]{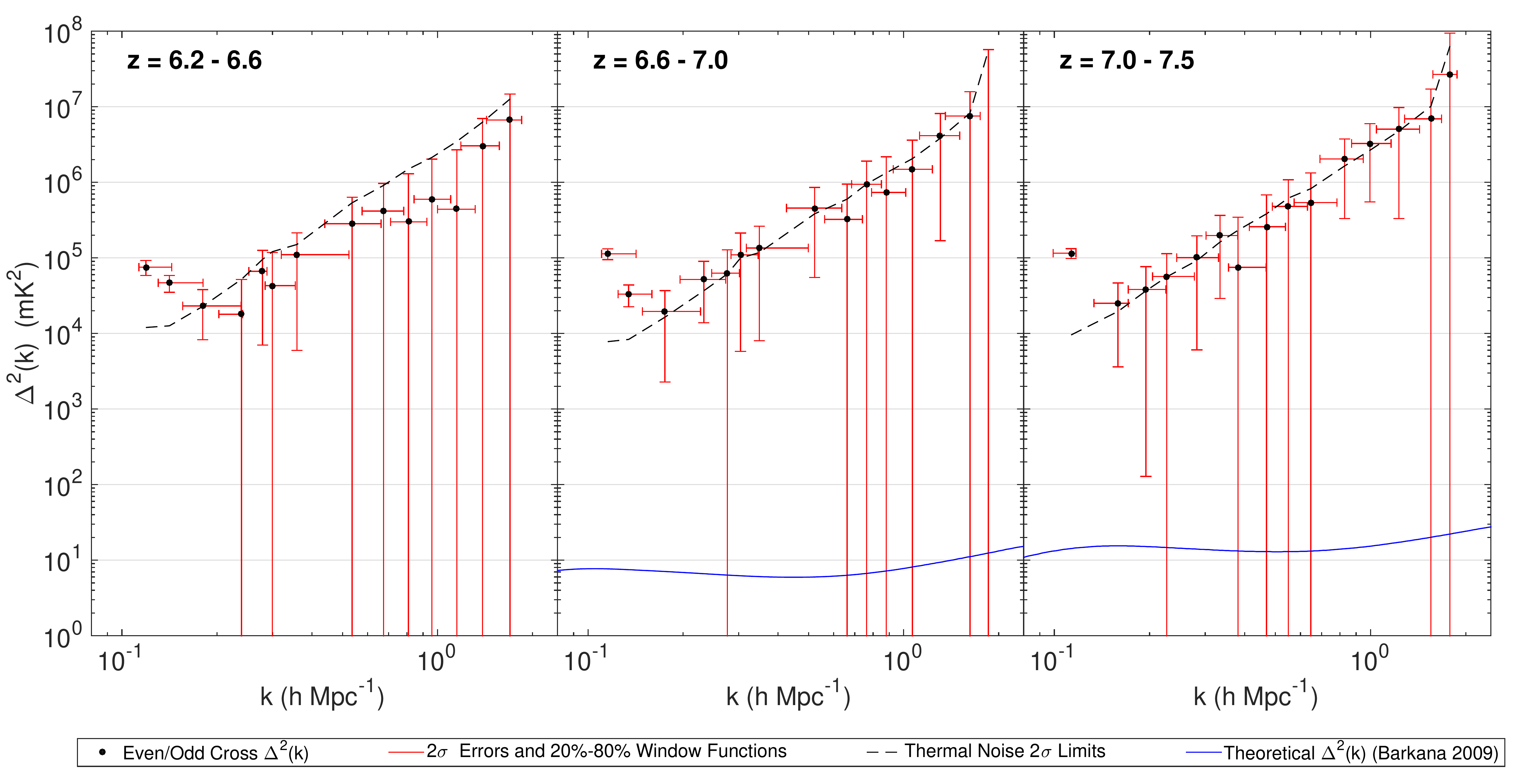}
	\caption[Final $P(k)$ uppter limits accross all measured redshifts.]{Finally, we can set confident limits on the 21\,cm power spectrum at three redshifts by splitting our simultaneous bandwidth into three 10.24\,MHz data cubes. The lowest $k$ bins show the strongest ``detections,'' though they are attributable to suprahorizon emission \cite{PoberWedge} that we expect to appear because we only cut out the wedge and a small buffer (0.02\,$h$\,Mpc$^{-1}$) past it. We also see marginal ``detections'' at higher $k$ which are likely due to subtle bandpass calibration effects like cable reflections. The largest such error, which occurs at bins around $k_\| \sim 0.45$\,$h$\,Mpc$^{-1}$ and can be seen most clearly in Figure \ref{fig:2dSNR}, has been flagged and removed from all three of these plots. Our absolute lowest limit requires $\Delta^2(k) < \limitDeltaSq$\,mK$^2$ at 95\% confidence at comoving scale $k = \limitk$\,$h$\,Mpc$^{-1}$ and $z = \limitz$, which is consistent with published limits \cite{newGMRT,X13,PAPER32Limits,DannyMultiRedshift,PAPER64Limits}. We also include a simplistic thermal noise calculation (dashed line), based on our observed system temperature. Though it is not directly comparable to our measurements, since it has different window functions, it does show that most of our measurements are consistent with thermal noise. For comparison, we also show the theoretical model of \cite{BarkanaPS2009} (which predicts that reionization ends before $z=6.4$) at the central redshift of each bin. While we are still orders of magnitude away from the fiducial model, recall that the noise in the power spectrum scales inversely with the integration time, not the square root.}
	\label{fig:1dDeltaSq}
\end{figure*}

In Figure \ref{fig:1dDeltaSq} we show the result of that calculation as a ``dimensionless'' power spectra $\Delta^2(k) \equiv k^3 P(k) /2\pi^2$.
We choose our binning such that the window functions (calculated as in \cite{X13} from our covariance model) were slightly overlapping. 

Our results are largely consistent with noise. Since noise is independent of $k_\|$ and $k\approx k_\|$ for most modes we measure, the noise in $\Delta^2(k)$ scales as $k^3$. We see deviations from that trend at low $k$ where modes are dominated by residual foreground emission beyond the horizon wedge and thus show elevated variance and bias in comparison to modes at higher $k$. Since we do not subtract a bias, even these ``detections'' are upper limits on the cosmological signal.

A number of barely significant ``detections'' are observed at higher $k$. Though we excise bins associated with the $k_\| \sim 0.45$\,$h$\,Mpc$^{-1}$ line, the slight detections may be due to leakage from that line. At higher $z$, the feature may due to reflections from cables of a different length, though some may be plausibly attributable to noise. Deeper integration is required to investigate further. 

Our best upper limit at 95\% confidence is $\Delta^2(k) < \limitDeltaSq$\,mK$^2$ at $k = \limitk$\,$h$\,Mpc$^{-1}$ around $z = \limitz$. Our absolute lowest limit is about 2 times lower than the best limit in \cite{X13}, though the latter was obtained at substantially higher redshift and lower $k$, making the two somewhat incomparable. Our best limit is roughly 3 orders of magnitude better than the best limit of \cite{X13} over the same redshift range, and the overall noise level (as measured by the part of the power spectrum that scales as $k^3$) is more than 2 orders of magnitude smaller. This cannot be explained by more antenna tiles alone; it is likely that the noise level was overestimated in \cite{X13} due to insufficiently rapid time interleaving of the data cubes used to infer the overall noise level.

Although one cannot directly compare limits at different values of $k$ and $z$, our limit is similar to the GMRT limit \cite{newGMRT}, $6.2\times 10^4$\,mK$^2$ at $k = 0.50$\,$h$\,Mpc$^{-1}$ and $z=8.6$ with 40\,h of observation, and remains higher than the best PAPER limit \cite{PAPER64Limits} of 502\,mK$^2$ between $k = 0.15$\,$h$\,Mpc$^{-1}$ and $k=0.50$\,$h$\,Mpc$^{-1}$ and $z=8.4$ with 4.5 months of observation.
 
In Figure \ref{fig:1dDeltaSq} we also plot a theoretical model from \cite{BarkanaPS2009} predicting that reionization has ended by the lowest redshift bin we measure. We remain more than 3 orders of magnitude (in mK$^2$) from being able to detect that particular reionization model, naively indicating that roughly 3000\,h of data are required for its detection. This appears much larger than what previous sensitivity estimates have predicted for the MWA (e.g.~\cite{MWAsensitivity}) in the case of idealized foreground subtraction. 
 
However, much of this variance is due to the residual foregrounds and systematics in the EoR window identified by our empirical covariance modeling method, not thermal noise (see Figure \ref{fig:2dError}). More integration will not improve those modes unless it allows for a better understanding of our instrument, better calibration, and better foreground models---especially of diffuse emission which might contaminate the highly sensitive bottom left corner of the EoR window. Eliminating this apparent ``suprahorizon'' emission, seen most clearly as detections in Figure \ref{fig:2dSNR} below $k \approx 0.2$\,$h$\,Mpc$^{-1}$, is essential to achieving the forecast sensitivity of the MWA \cite{MWAsensitivity}. If we can do so, we may still be able to detect the EoR with 1000\,h or fewer. This is especially true if we can improve the subtraction of foregrounds to the point where we can work within the wedge, which can vastly increase the sensitivity of the instrument \cite{MWAsensitivity,PoberNextGen}. On the other hand, more data may reveal more systematics lurking beneath the noise which could further diminish our sensitivity.


\section{Summary and Future Directions} \label{sec:summary}

In this work, we developed and demonstrated a method for empirically deriving the covariance of residual foreground contamination, $\C^{FG}$, in observations designed to measure the 21\,cm cosmological signal. Understanding the statistics of residual foregrounds allows us to use the quadratic estimator formalism to quantify the error associated with missubtracted foregrounds and their leakage into the rest of the EoR window. Because of the complicated interaction between the instrument and the foregrounds, we know that the residual foregrounds will have complicated spectral structure, especially if the instrument is not perfectly calibrated. By deriving our model for $\C^{FG}$ empirically, we could capture those effects faithfully and thus mitigate the effects of foregrounds in our measurement (subject to certain caveats which we recounted in Section \ref{sec:caveats}).

Our strategy originated from the assumption that the frequency-frequency covariance, modeled as a function of $|u|$, is the most important component of the foreground residual covariance. We therefore used sample covariances taken in annuli in Fourier space as the starting point of our covariance model. These models were adjusted to avoid double counting the noise variance and filtered in Fourier space to minimize the effect of noise in the empirically estimated covariances. Put another way, we combined our prior beliefs about the structure of the residual foregrounds with their observed statistics in order to build our models.

We demonstrated this strategy through the power spectrum analysis of a 3\,h preliminary MWA data set. We saw the expected wedge structure in both our power spectra and our variances. We saw that most of the EoR window was consistent with noise, and we understand why residual foregrounds and systematics affect the regions that they do. We were also able to set new MWA limits on the 21\,cm power spectrum from $z=6.2$ to $7.5$, with an absolute best 95\% confidence limit of $\Delta^2(k) < \limitDeltaSq$\,mK$^2$ at $k = \limitk$\,$h$\,Mpc$^{-1}$ and $z = \limitz$, consistent with published limits \cite{PAPER32Limits,DannyMultiRedshift}.

This work suggests a number of avenues for future research. Of course, improved calibration and mapmaking fidelity---especially better maps of diffuse Galactic structure---will improve power spectrum estimates and and allow deeper integrations without running up against foregrounds or systematics. Relaxing some of the mapmaking and power spectrum assumptions discussed in Section \ref{sec:caveats} may further mitigate these effects. A starting point is to integrate the mapmaking and statistical techniques of \cite{Mapmaking} with the fast algorithms of \cite{DillonFast}. The present work is based on the idea that it is simpler to estimate $\C^{FG}$ from the data than from models of the instrument and the foregrounds. But if we can eliminate systematics to the point where we really understand $\PSF$, the relationship between the true sky and our dirty maps, then perhaps we can refocus our residual foreground covariance modeling effort on the statistics of the true sky residuals using the fact that $\C^{FG} = \PSF \C^{FG,\text{true}} \PSF^\trans.$ Obtaining such a complete understanding of the instrument will be challenging, but it may be the most rigorous way to quantify the errors introduced by missubtracted foregrounds and thus to confidently detect the 21\,cm power spectrum from the epoch of reionization.

\chapter[MITEoR: A Scalable Interferometer for Precision 21\,cm \\Cosmology]{MITEoR: A Scalable Interferometer for Precision 21\,cm Cosmology}\label{ch:MITEOR}

\emph{The content of this chapter was submitted the \emph{Monthly Notices of the Royal Astronomical Society} on June 12, 2014 and published \cite{MITEoR} as \emph{MITEoR: a scalable interferometer for precision 21 cm cosmology} on October 8, 2014.}

\section{Introduction}

Mapping neutral hydrogen throughout our universe via its redshifted 21~cm line 
offers a unique opportunity to probe the cosmic ``dark ages,'' the formation of the first luminous objects, and the epoch of reionization (EoR). A suitably designed instrument with a tenth of a square kilometer of collecting area will allow tight constraints on the timing and duration of reionization and the astrophysical processes that drove it  \cite{PoberNextGen}. 
Moreover, because it can map a much larger comoving volume of our universe, it has the potential to overtake the 
Cosmic Microwave Background (CMB) as our most sensitive cosmological probe of inflation, dark matter, dark energy, and neutrino masses. For example  \cite{Yi}, a radio array with a square kilometer of collecting area, maximal sky coverage, and good foreground maps could improve the sensitivity to spatial curvature and neutrino masses by up
to two orders of magnitude, to $\Delta\Omega_k\approx 0.0002$ and $\Delta m_\nu\approx 0.007$ eV, and shed new light on the early universe by a $4\sigma$ detection of the spectral index running predicted by the simplest inflation models favored by the BICEP2 experiment  \cite{bicep2}.

Unfortunately, the cosmological 21~cm signal is so faint that none of the current experiments around the world (LOFAR  \cite{LOFAR}, MWA  \cite{TingaySummary}, PAPER  \cite{PAPER}, 21CMA  \cite{21cma}, GMRT  \cite{GMRT})
have detected it yet, although increasingly stringent upper limits have recently been placed  \cite{newGMRT,X13,PAPER32Limits}. A second challenge is that foreground contamination from our galaxy and extragalactic sources is perhaps four orders of magnitude larger than the cosmological hydrogen signal  \cite{GSM}. Any attempt to accurately clean it out from the data requires even greater sensitivity as well as more accurate calibration and beam modeling than the current state-of-the-art in radio astronomy (see  \citet{FurlanettoReview,miguelreview} for reviews).
 
Large sensitivity requires large collecting area. Since steerable single dish radio telescopes become prohibitively expensive beyond a certain size, the aforementioned experiments have all opted for interferometry, combining $N$ (generally a large number) independent antenna elements which are (except for GMRT) individually more affordable. The LOFAR, MWA, PAPER, 21CMA and GMRT experiments currently have comparable $N$. The problem with scaling interferometers to high $N$ is that  all of these experiments use standard hardware cross-correlators whose cost grows quadratically with $N$, since they need to correlate all $N(N-1)/2\sim N^2/2$ pairs of antenna elements. This cost is reasonable for the current scale $N\sim 10^2$, but will completely dominate the cost for $N\simgt 10^3$, making precision cosmology arrays with $N\sim 10^6$ as discussed in  \citet{Yi} infeasible in the near future, which has motivated novel correlator approaches such as  \citet{moff}.

For the particular application of 21~cm cosmology, however, designs with better cost scaling are possible, as described in  \citet{FFTT,FFTT2}: by arranging the antennas in a 
hierarchical rectangular or hexagonal grid and performing the correlations using Fast Fourier Transforms (FFTs), thereby cutting the cost scaling to $N\log N$. This is particularly attractive for science applications requiring exquisite sensitivity at vastly different angular scales, such as
21~cm cosmology (where short baselines are needed to probe the cosmological signal\footnote{It has been shown that the 21~cm signal-to-noise ratio ($S/N$) per resolution element in the $uv$-plane (Fourier plane) is $\ll 1$ for all current 21~cm cosmology experiments, and that their cosmological sensitivity therefore improves by moving their antennas closer together to focus on the center of the $uv$-plane and bringing its $S/N$ closer to unity
 \cite{MiguelNoise,Judd06,Matt3,Yi,Lidz2009}. Error bars on the cosmological power spectrum have contributions from both noise and sample variance, and it is well-known that the total error bars on a given physical scale (for a fixed experimental cost) are minimized when both contributions are comparable, which happens when the $S/N\sim 1$ on that scale. This is why more compact 21~cm experiments have been advocated.
This is also why early suborbital CMB experiments focused on small patches of sky to get $S/N\sim 1$ per pixel, and why galaxy redshift surveys target objects like luminous red galaxies that give $S/N\sim 1$ per 3D voxel.
}  
and long baselines are needed for point source removal).
Such hierarchical grids thus combine the angular resolution advantage of traditional array layouts with the cost advantage of a rectangular Fast Fourier Transform Telescope. If the antennas have a broad spectral response as well and their signals are digitized with high bandwidth, the cosmological neutral hydrogen gets simultaneously imaged in a vast 3D volume covering both much of the sky and also a vast range of distances (corresponding to different redshifts, \ie, different observed frequencies.)
Such low-cost arrays have been called {\it omniscopes} \cite{FFTT,FFTT2} for their wide field of view and broad spectral range.

Of course, producing such scientifically rich maps with any interferometer depends crucially on our ability to precisely calibrate the instrument, so that we can truly understand how our measurements relate to the sky. Traditional radio telescopes rely on a well-sampled Fourier plane to perform self-calibration using the positions and fluxes of a number of bright point sources. At first blush, one might think that any highly-redundant array would be at a disadvantage in its attempt to calibrate the gains and phases of individual antennas. However, we can use the fact that redundant baselines should measure the same Fourier component of the sky to improve the calibration of the array dramatically and quantifiably. In fact, we find that the ease and precision of redundant baseline calibration is a strong rationale for building a highly-redundant array, in addition to the improvements in sensitivity and correlator speed.

Redundant calibration is useful both for current generation redundant arrays like MITEoR and PAPER and for future large arrays that will need redundancy to cut down correlator cost. Omniscopes must be calibrated in real time, because they do not compute and store the visibilities measured by each pair of antennas,  but effectively gain their speed advantage by averaging redundant baselines in real time. Individual antennas therefore cannot be calibrated in post-processing. No calibration scheme used on existing low frequency radio interferometers has been demonstrated to meet the speed and precision requirements of omniscopes. Thus, the main goal of the MIT Epoch of Reionization experiment (MITEoR) and this paper is to demonstrate a successful redundant calibration pipeline that can overcome the calibration challenges faced by current and future generation instruments by performing automatic precision calibration in real time.

Building on past redundant baseline calibration methods by Wieringa  \cite{Wieringa} and others, some of us recently developed an algorithm which is both automatic and statistically unbiased, able to produce precision phase and gain calibration for all antennas in a hierarchical grid (up to a handful of degeneracies) without making any assumptions about the sky signal  \cite{redundant}. Once obtained, precision calibration solutions can in turn produce more accurate modeling of the synthesized and primary beams\footnote{For tile-based interferometers like the MWA and 21CMA, gain and phase errors in individual antennas (as opposed to tiles) do not typically get calibrated in the field, adding a fundamental uncertainty to the tile sky response.}  \cite{JonniePrimaryBeam}, which has been shown to improve the quality of the foreground modeling and removal which is so crucial to 21~cm cosmology. It is therefore timely to develop a pathfinder instrument that tests how well the latest calibration ideas works in practice. 

MITEoR is such a pathfinder instrument, designed to test redundant baseline calibration. We developed and successfully applied a real-time redundant calibration pipeline to data we took with our 64 dual-polarization antenna array during the summer of 2013 in The Forks, Maine. The goal of this paper is to describe the design of the MITEoR instrument, demonstrate the effectiveness of our redundant baseline calibration and absolute calibration pipelines, and use the calibration results to obtain an optimal scheme for estimating calibration parameters as a function of time and frequency. 

This paper is organized as follows. We first describe in Section \ref{secinstrument} the instrument, including the custom developed analog components, the 8 bit 128 antenna-polarization correlator, the deployment, and the observation history. 
In Section \ref{seccal}, we focus on precision calibration.
We explain and quantitatively evaluate relative redundant calibration, and address the question of how often calibration coefficients should be updated. We also examine the absolute calibration, including breaking the degeneracies in relative calibration, mapping the primary beam, and measuring the array orientation.
In Section \ref{secsummary}, we summarize this work and discuss implications for future redundant arrays such as HERA  \cite{PoberNextGen}.
\section{The MITEoR Experiment}\label{secinstrument}
\begin{figure}
\centerline{\includegraphics[width=.6\textwidth]{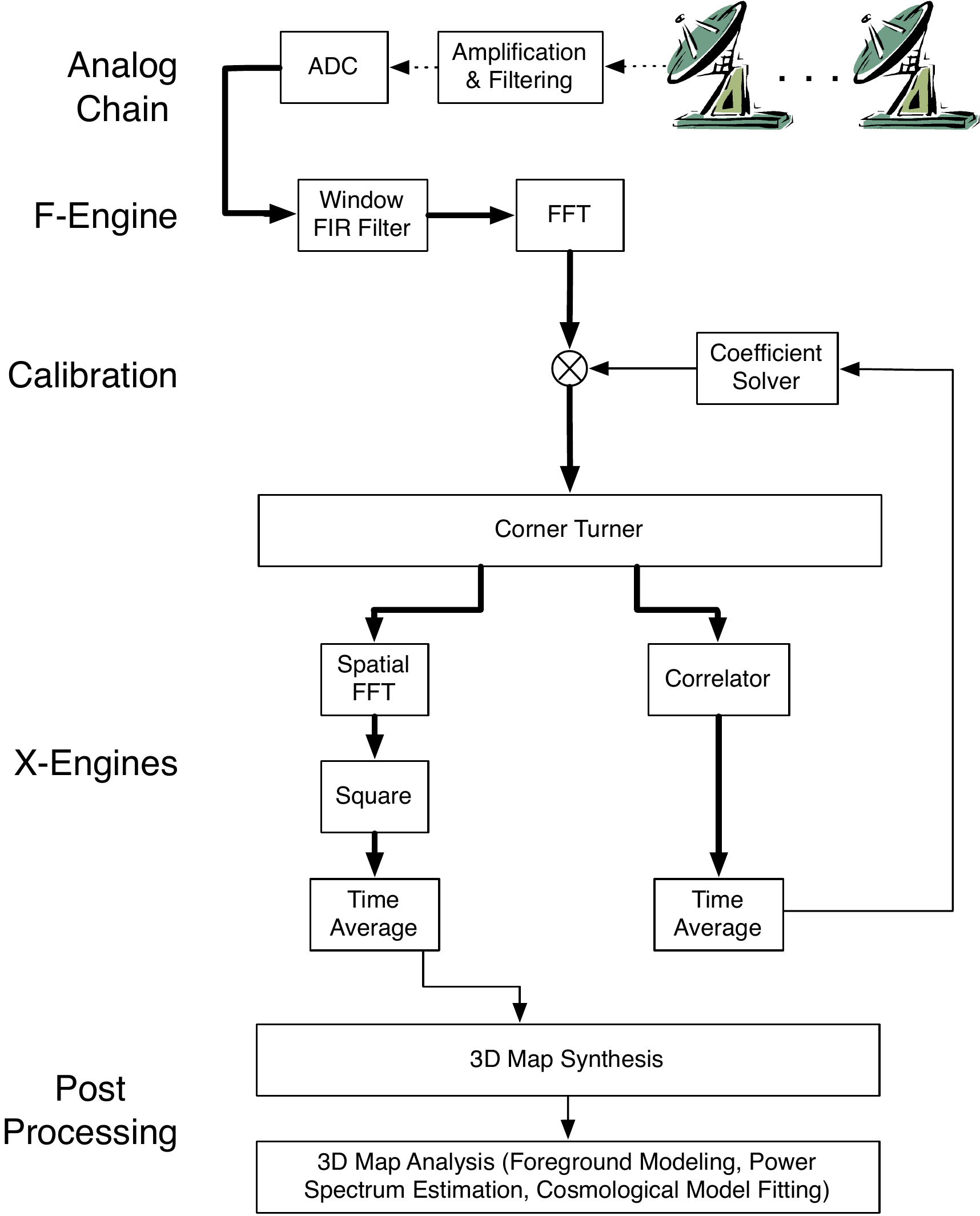}}
\caption[System schematic of an FFT telescope.]{
Data pipeline for a large omniscope that implements FFT correlator and redundant baseline calibration. First, a hierarchical grid of dual-polarization antennas converts the sky signal into volts, which get amplified and filtered by the analog chain, transported to a central location, and digitized every few nanoseconds.
These high-volume digital signals (thick lines) get processed by field-programmable gate arrays (FPGAs) which perform a temporal Fourier transform. The FPGAs (or GPUs) then multiply by complex-valued calibration coefficients that depend on antenna, polarization and frequency, then spatially Fourier transform, square and accumulate the results, recording integrated sky snapshots every few seconds and thus reducing the data rate by a factor $\sim 10^9$.  They also cross-correlate a small fraction of all antenna pairs, allowing the redundant baseline calibration software  \cite{redundant,LOFARcal} to update the calibration coefficients in real time and automatically monitor the quality of calibration solutions for instrumental malfunctions.
Finally, software running on regular computers combine all snapshots of sufficient quality into a 3D sky ball or ``data cube'' representing the sky brightness as a function of angle and frequency in Stokes (I,Q,U,V)  \cite{FFTT2}, and subsequent software accounts for foregrounds and measures power spectra and other cosmological observables.
\label{ArchitectureFig}
}
\end{figure}

In theory, a very large omniscope can be built following the generalized architecture in Figure \ref{ArchitectureFig}. On the other hand, it is crucial to demonstrate that automatic and precise calibration is possible in real-time using redundant baselines, since the calibration coefficients for each antenna must be updated frequently to allow the FFTs to combine the signals from the different antennas without introducing errors. In this section, we will present our partial implementation of this general design, including both the analog and the digital systems. Because the digital hardware is powerful enough to allow it, the MITEoR prototype correlates all 128 input channels with one another, rather than just a small sample as mentioned in the caption of Figure \ref{ArchitectureFig}. This provides additional cross-checks that greatly aid technological development, where instrumentation may be particularly prone to systematics. This also allows us to explore the question of exactly how often and how finely in frequency we must measure visibilities to solve for calibration coefficients, a question we return to in Section \ref{seccal}. Since we chose to implement a full correlator, an additional FFT correlator would bring no extra information (simply computing the same redundant-baseline-averaged visibilities faster), so we leave the digital implementation of an FFT correlator to future work.  In general, our mission is to empirically explore any challenges that are unique to a massively redundant interferometer array.  Once these are known, one can reconfigure the cross-correlation hardware to perform spatial FFTs, thereby obtaining an omniscope with $N\log N$ correlator scaling.

\subsection{The Analog System}\label{secanalog}

\begin{figure}
\centerline{\includegraphics[width=.6\textwidth]{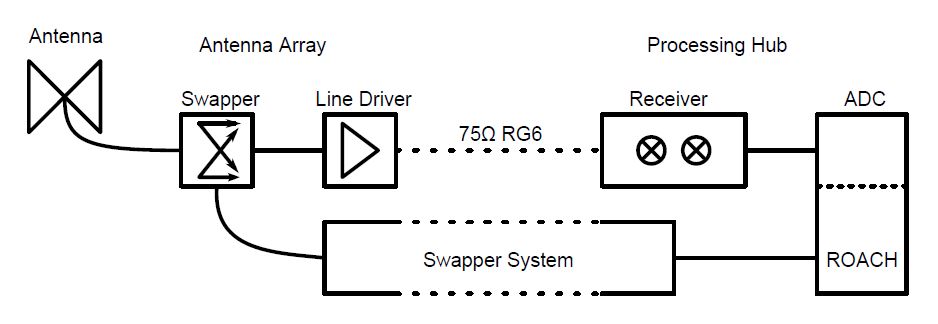}}
\caption[MITEoR analog system design.]{
System diagram of the analog system. The signal received with an MWA ``bow-tie'' antenna is first amplified by the built-in low noise amplifier, then Walsh-modulated in the swapper module controlled by the swapper system. The signal is amplified again in the line driver and sent to the processing rack through 50\,m long coaxial cables. In the processing rack, the signal first goes into the receiver, where it undergoes further amplification, frequency down-mixing and I/Q modulation from the 120-180\,MHz range to the 0-25\,MHz range. The analog chain ends with digitization on ADC connected to ROACH boards. \label{analog_schematic}
}
\end{figure}

MITEoR contains 64 dual-polarization antennas, giving 128 signal channels in total. The signal picked up by the antennas is first amplified by two orders of magnitude in power by the low noise amplifiers (LNAs) built-in to the antennas. It is then phase switched in the swapper system, which greatly reduces cross-talk downstream. The signal is then amplified again by about five orders of magnitude in the line-drivers before being sent over 50 meter RG6 cables to the receivers. The receivers perform IQ demodulation on a desired $50\,\textrm{MHz}$ band selected between $100\,\textrm{MHz}$ and $200\,\textrm{MHz}$, producing two channels with adjacent $25\,\textrm{MHz}$  bands, and sends the resulting signals into the digitization boards containing 256 analog-to-digital converters (ADCs) sampling at $50\,\textrm{MHz}$.  The swappers, line-drivers and receivers we designed are shown in Figure \ref{analog_schematic}.

When designing the components of this system, we chose to use commercially-available integrated circuits and filters whenever possible, to allow us to focus on
system design and construction. In some cases (such as with the amplifiers) the cost of the IC is less than the cost of enough discrete transistors to implement even a rough approximation of the same functionality. Less expensive filters
could be made from discrete components, but the characteristics of purchased modules tend to be  better due to custom inductors and shielding. When we needed to produce our own boards as described below, our approach was to design, populate and test them in our laboratory, then have them affordably mass-produced for us by Burns Industries\footnote{ \url{http://www.burnsindustriesinc.com}}.


\subsubsection{Antennas} 
The dual-polarization antennas used in MITEoR were originally developed for the Murchison Widefield Array  \cite{MWAdesign,TingaySummary}, and consist of two ``bow-tie''-shaped arms as can be seen in Figure \ref{ArrayFigure}. They are inexpensive, easy to assemble, and sensitive to the entire band of our interest. The MWA antennas were designed for the frequency range $80$-$300\,\textrm{MHz}$, and
have a built-in low noise amplifier with $20\,\textrm{dB}$ of gain. The noise figure of the amplifier is $0.2\,\textrm{dB}$, and the $20\,\textrm{dB}$
of gain means that subsequent gain stages do not contribute significantly to the noise figure\footnote{In a multi-stage amplifier, the contribution of each stage's noise figure is suppressed by a factor that is equal to the total gain of previous stages.}.

\subsubsection{Swappers (Phase Switches)}
As with many other interferometers, crosstalk within the receivers, ADCs, and cabling significantly affects signal quality. 
We observe the cross-talk to depend strongly on the physical proximity of channel pairs, reaching as high as about $-30$\,dB between nearest neighbor receiver channels. Our swapper system is designed to cancel out crosstalk during the correlator's time averaging by selectively inverting analog signals using Walsh modulation  \cite{nevadathesis}. The signal from each antenna-polarization is inverted 50\% of the time according to its own Walsh function, by an analog ZMAS-1 phase switch from Mini-Circuits located before the second amplification stage (line-driver), then appropriately re-inverted after digitization\footnote{Since the undesirable crosstalk signal is demodulated with a different Walsh function than it is modulated with, it will be averaged out due to orthogonality of Walsh functions.}. We perform the inversion once every millisecond, which is much longer than the ADC's 20\,ns sample time, and much shorter than the averaging time of a few seconds\footnote{The inversion cannot be too frequent, because we need to discard data during the analog inversion process which takes a few microseconds. At the same time, the inversion needs to be frequent enough to average out the cross-talk.}. This eliminates all crosstalk to first order  \cite{nevadathesis}.
\begin{figure}
\centerline{\includegraphics[width=.6\textwidth]{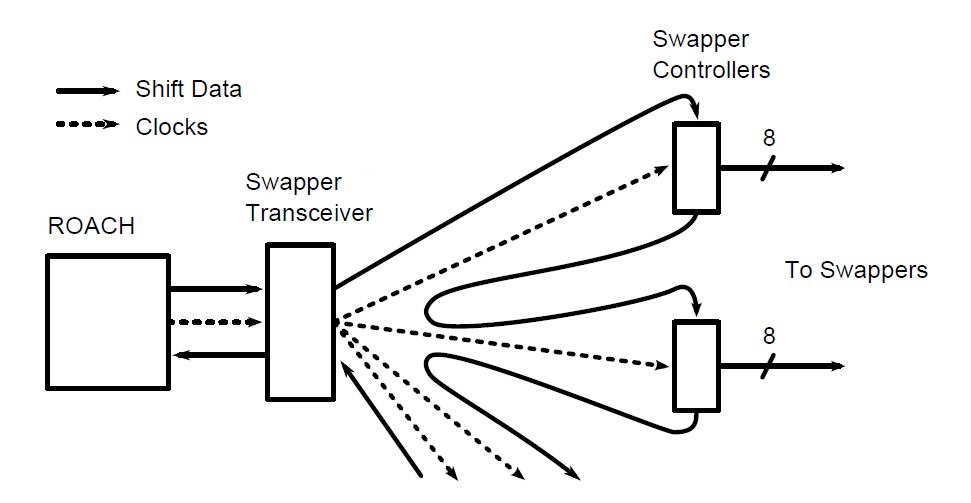}}
\vskip0.3cm
\centerline{\includegraphics[width=.6\textwidth]{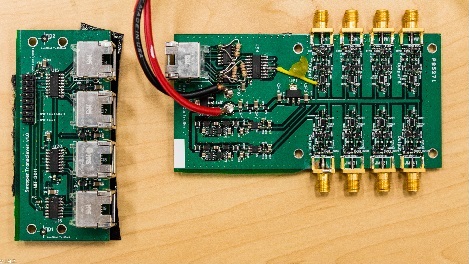}}
\caption[MITEoR swapper design.]{
System diagram of our swapper signal system and physical components of the swapper transceiver (lower left) and swapper controller (lower right). The swapper is designed to reduce crosstalk between neighboring channels.
\label{Swapper}
}
\end{figure}
If crosstalk reduction were the only concern, the ideal position for the swapper would be 
immediately after the antenna, in order to cancel as much crosstalk as possible. In practice, the swapper introduces a loss of about 3\,dB, so we perform the modulation after the LNA to avoid adding noise (raising the system temperature). To evaluate the effectiveness of the swapper modules, we sent a monotone signal into one single channel of the receivers while leaving other channels open, and measured the correlation between the signal channel and each empty channels with the swapper turned on and off. We then repeated this while varying the signal frequency over the full range of interest. As seen in Figure \ref{SwapperFig},  
the swapper system attenuates crosstalk in the receiver and ADC by as much as 50\,dB over the frequency band of interest, typically reducing it to being of order $-80$\,dB for strongly afflicted signal pairs.

\begin{figure*}
\centerline{\includegraphics[width=\textwidth]{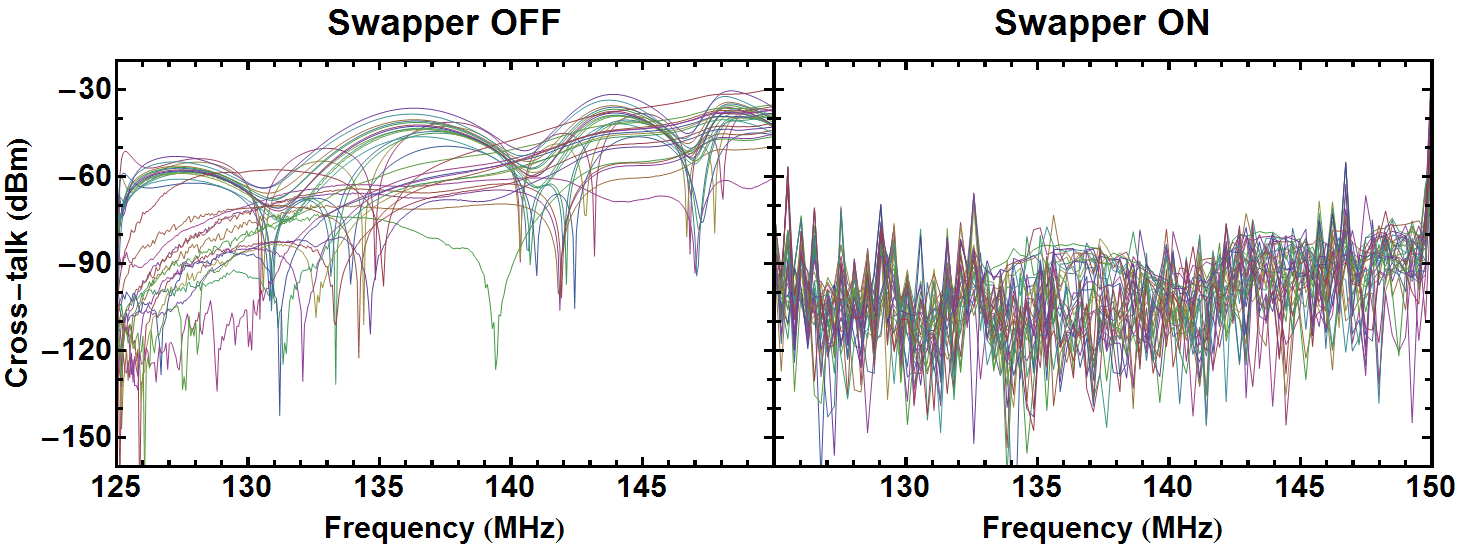}}
\vskip-2mm
\caption[Laboratory measurements of cross-talk with and without the swapper.]{
Plots of cross-talk power measured in the laboratory. The swapper suppresses crosstalk between channels by as much as 50\,dB. To measure these curves we fed a 0\,dBm sinusoidal signal into input channel 0 of the receivers and left the other 31 input channels open. We then measured the correlations between channel 0 and all 31 empty channels, due to crosstalk from channel 0. We repeated the procedure with input frequencies from 125-150\,MHz and obtained the results shown above.  
\label{SwapperFig}
}
\end{figure*}
\subsubsection{Line-Driver}

\begin{figure}
\vskip-2cm
\centerline{\includegraphics[width=.75\textwidth]{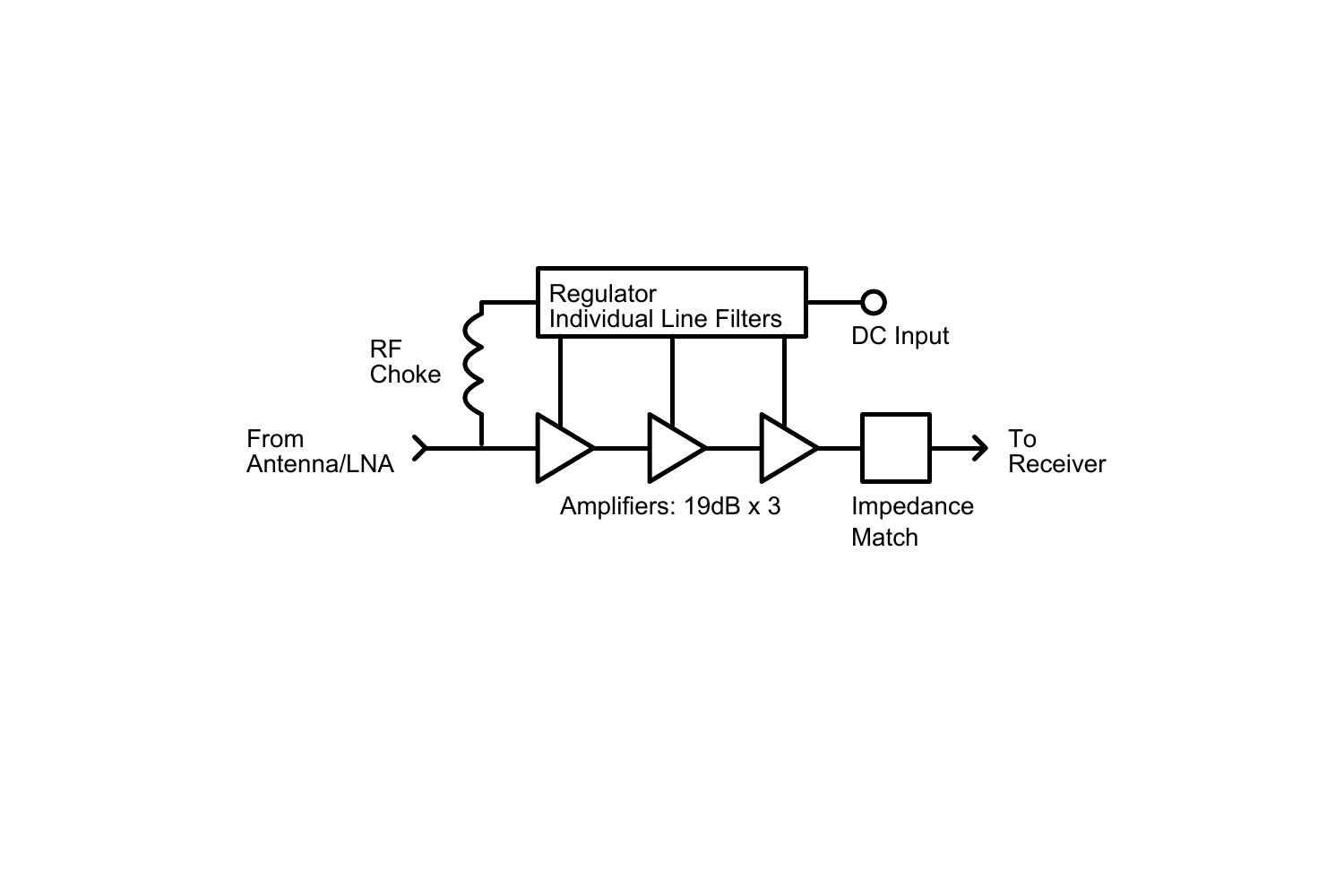}}
\vskip-2.7cm
\centerline{\includegraphics[width=.6\textwidth]{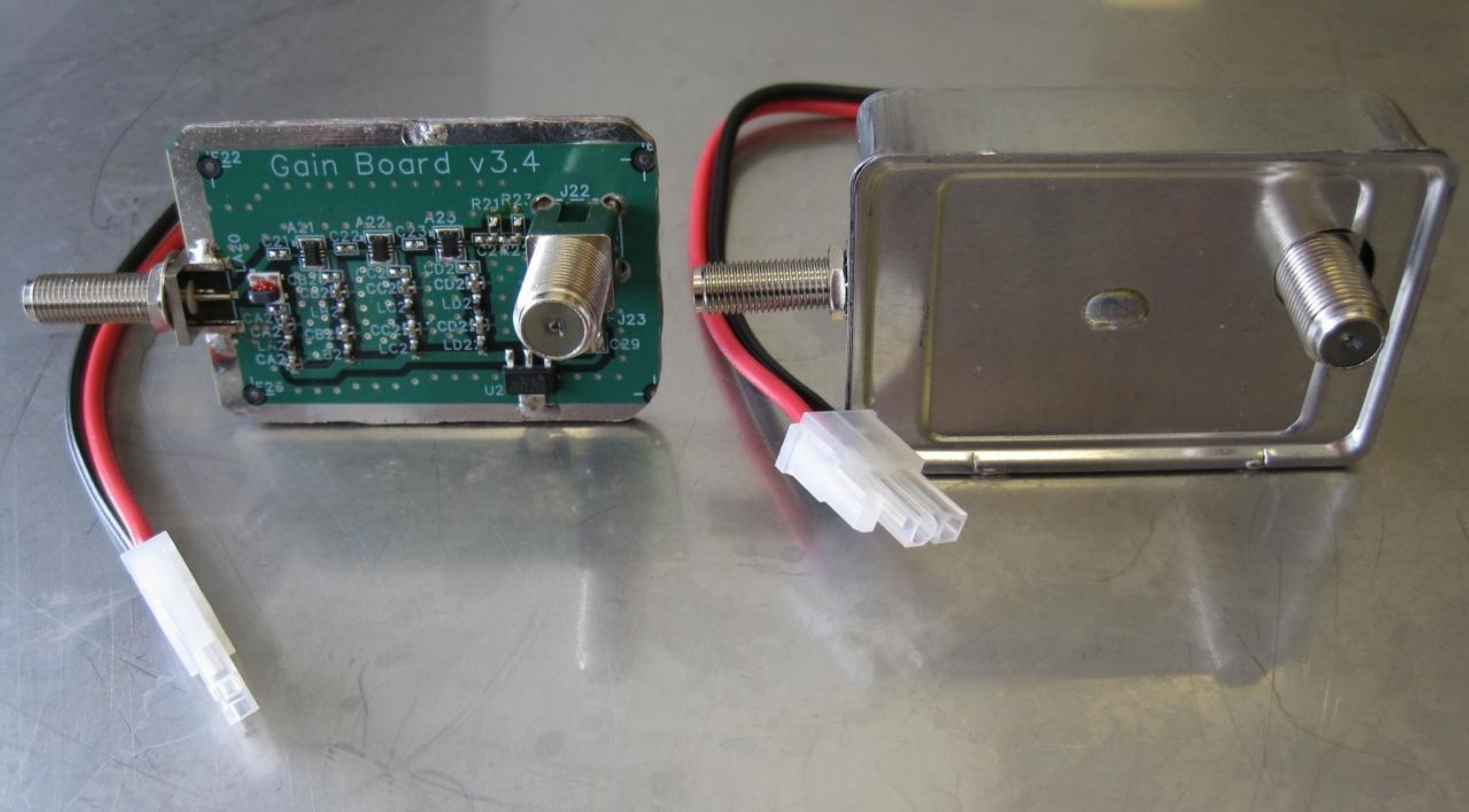}}
\caption[Schematic and photograph of the line drivers.]{
System diagram and physical components of the line drivers. The line driver we designed takes the signal in the 50$\Omega$ coaxial cable from the antenna LNA and amplifies it by 51\,dB, in order to overpower noise picked up in the subsequent 75$\Omega$ coaxial cable and further processing steps up to 50 meters away.  It operates on 5V DC and also provides DC bias power to the antenna's LNA through the 50 Ohm cable.
\label{LineDriverFig}
}
\end{figure}

A line-driver (Figure \ref{LineDriverFig}) amplifies a single antenna's signal from one of its two polarization channels while also powering its LNA. Line-drivers only handle a single channel to reduce potential crosstalk from sharing a printed circuit board. They are placed within a few meters of the antennas in order to reduce resistive
losses from powering the antenna at low voltage. Additional gain that they provide early in the analog chain helps the signal overpower
any noise picked up along the way to the processing hub, and maintains the low noise figure set up by the LNA. To further reduce potential radio-frequency interference (RFI), we chose to power the line-drivers with 58Ah 6V sealed lead acid rechargeable batteries during the final 64-antenna deployment, rather than 120\,VAC to 6\,VDC adapters (whose unwanted RF-emission may have caused occasional saturation problems during our earlier expeditions).

\subsubsection{Receiver}
\begin{figure}
\vskip-1.3cm
\centerline{\includegraphics[width=.65\textwidth]{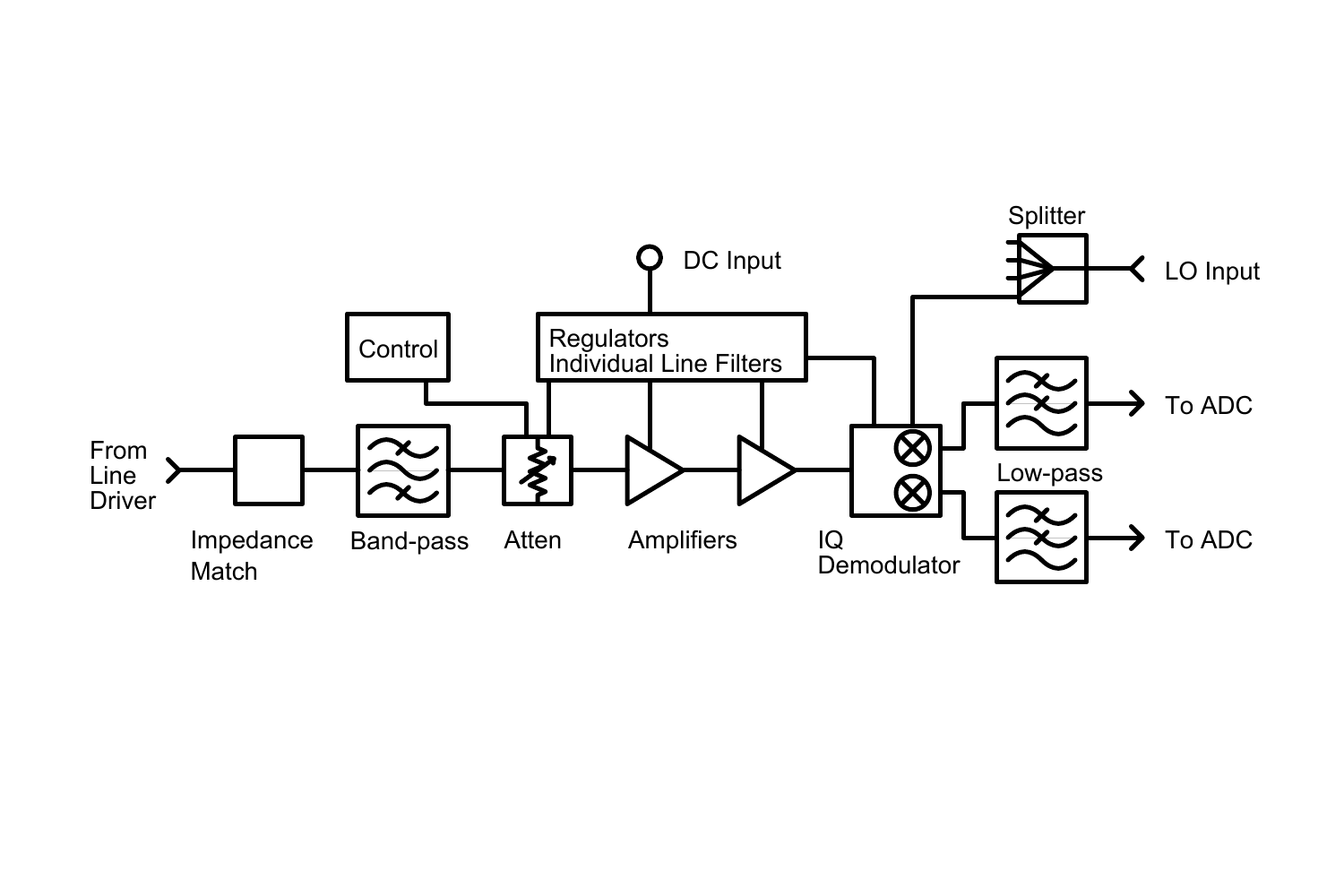}}
\vskip-1.9cm
\centerline{\includegraphics[width=.6\textwidth]{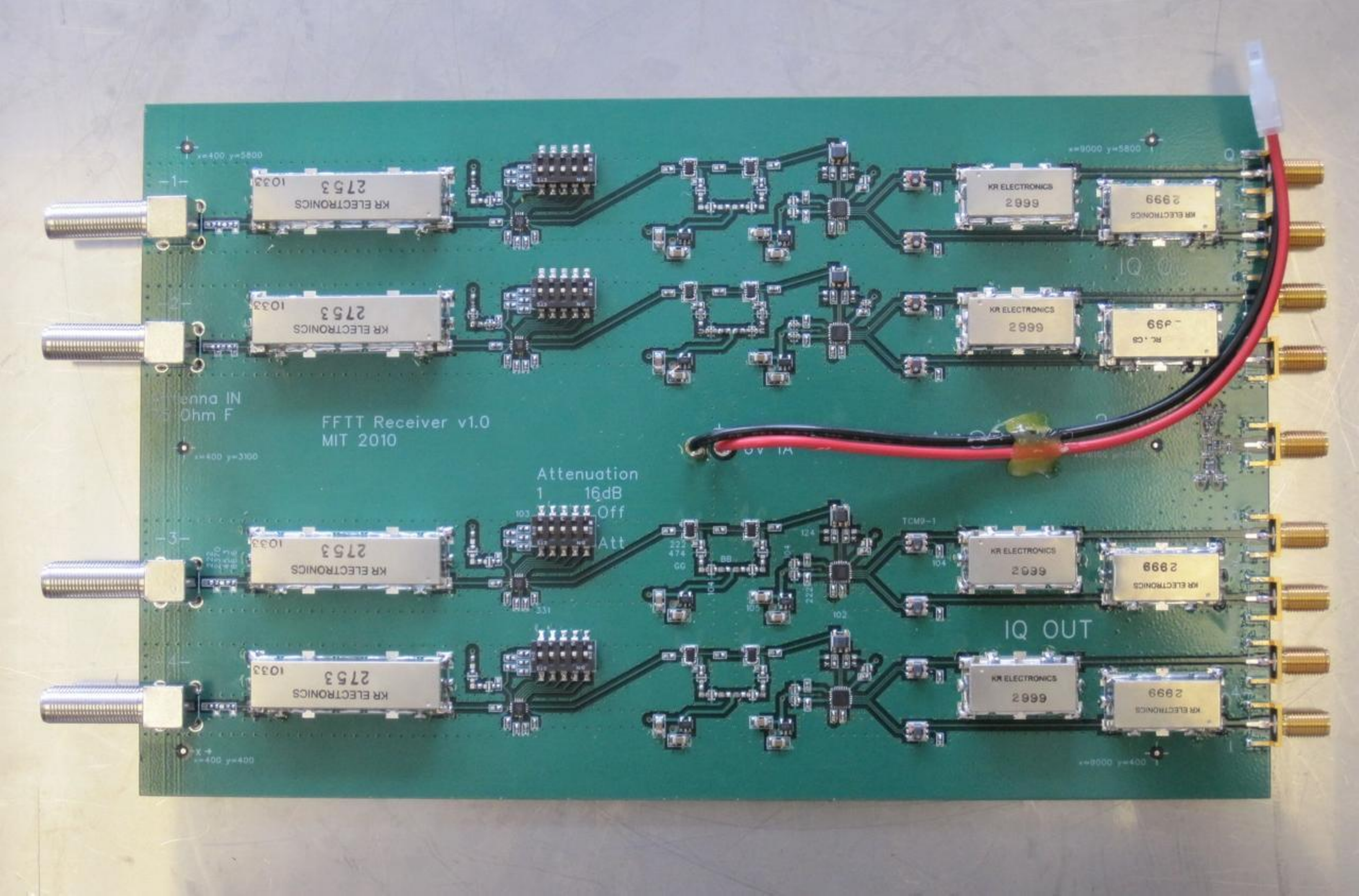}}
\caption[Schematic and photograph of the receiver boards.]{
System diagram and physical component of the receiver boards. The boards take the signals arriving from four line drivers, 
band-pass filter and amplify them, then use a local oscillator to frequency shift them from the band of interest to a DC-centered signal suitable for input to the ADC.
\label{ReceiverFig}
}
\end{figure}

Our receivers (Figure \ref{ReceiverFig}) take input from the line-drivers, bandpass filter the incoming signals, amplify their power level by 23\,dB, and IQ-demodulate them. The resulting signals go directly to an ADC for digitization. Receivers are placed near the ADCs to which they are connected to reduce cabling for local oscillator (LO) distribution and ADC connections. 
IQ demodulation is used, which doubles received bandwidth for a given ADC frequency at the cost of using two ADC channels, and has the advantage of requiring only a single  LO and low speed ADCs. The result is 40 MHz of usable bandwidth\footnote{Due to limitations in our FPGAs' computing power, only 12.5\,MHz of digitized data are correlated and stored at any instant.} anywhere in the range 110-190 MHz, with a 2-3 MHz gap centered around the LO frequency due to bandpass filters.
The receiver boards have five pins allowing their signals to be attenuated by any amount between 0\,dB and 31\,dB (in steps of 1\,dB) before the second amplification stage, to avoid saturation and non-linearities from RFI and to attain signal levels optimal for digitization.

\subsection{The Digital System}\label{secdigital}
\label{digi}
\begin{figure}
\centerline{\includegraphics[width=.6\textwidth]{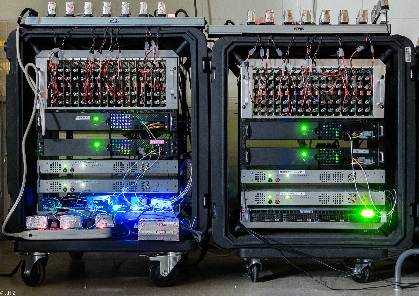}}
\caption[Photograph of the MITEoR computer rack.]{
The entirety of our 128 antenna-polarization digital correlator system, packaged in two portable shock mounted racks. The two black chassis and two silver chassis in the middle of each rack are F-engines (ROACH) and X-engines (ROACH2), respectively. Above the ROACHes are 32 receiver boards that input the signals from 128 line drivers via F-cables. The blue lit area below the ROACHes contains various clocking devices responsible for synchronization, whereas the chassis below the ROACHes on the right hand side is the 8TB data acquisition server.
\label{RackFig}
}
\end{figure}

\begin{figure}
\centerline{\includegraphics[width=.6\textwidth]{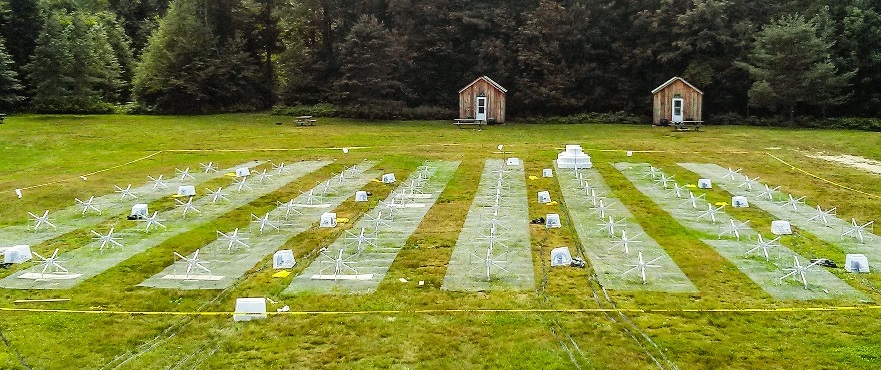}}
\caption[Photograph of the 2013 MITEoR deployment.]{Part of the MITEoR array during the most recent deployment in the summer of 2013. 64 dual-polarization antennas were laid on a 21~m by 21~m regular grid with 3~m separation. The digital system was housed in the back of a shielded U-Haul truck (not shown). 
\label{ArrayFigure}
}
\end{figure}

We designed MITEoR's digital system (Figure \ref{RackFig}) to be highly compact and portable. The entire system occupies 2 shock-mounted equipment racks on wheels, each measuring about 1\,m on all sides. It takes in data from 256 ADC channels (64 antennas with I and Q signals for polarizations), Fourier transforms each channel, reconstructs IQ demodulated channels back to 128 corresponding antenna channels, computes the cross-correlations of all pairs of the 128 antenna channels with 8 bit precision, and then time-averages these cross-correlations. Although standard 4 bit correlators suffice for most astronomical observation tasks, the better dynamic range of our 8 bit correlator allows us to observe faint astronomical signals at the same time as $10^3$ times brighter ORBCOMM satellites, whose enormous signal-to-noise has proved invaluable in characterizing various aspects of the system (see Sections \ref{secbeam}, \ref{secrotation}, and \ref{secsystematic}). The digital hardware is capable of processing an instantaneous bandwidth of $12.5\,\textrm{MHz}$ with $49\,\textrm{kHz}$ frequency bins. It averages those correlations and then writes them to disk every few seconds (usually either 2.6 or 5.3 seconds).
 
While one of the advantages of a massively redundant interferometer array is the ability to reduce costs by performing a spatial FFT rather than a full cross-correlation, we have not implemented FFT correlation in the current MITEoR prototype as the hardware is powerful enough to correlate all antenna pairs in real time (the feasibility of implementing FFT correlation on the ROACH platform has been demonstrated by  \citet{best2fftt}). Rather, the goal of MITEoR is to quantify the accuracy that automatic redundant baseline calibration can attain, thereby experimentally characterizing all of the unknowns in the system, such as unexpected analog chain systematics and other barriers to finding good calibration solutions.

We adopted the widely-used F-X scheme in MITEoR's digital system. We have 4 synchronized F-engines that take in data from 4 synchronized 64-channel ADC boards, which run at 12 bits and 50Ms/s. The F-engines perform the FFT and IQ reconstruction, and distribute the data onto 4 X-engines through 16 10GbE links. The 4 asynchronous X-engines each perform full correlation on 4 different frequency bands on all 128 antenna polarizations, and send the time averaged results to a computer for data storage.

To implement the computational steps of the MITEoR design, we used Field Programmable Gate Arrays (FPGAs). These devices can be programmed to function as dedicated pieces of computational hardware. Each F-engine and X-engine is implemented by one Xilinx FPGA (Virtex-5 for F-engines and Virtex-6 for X-engines). These FPGAs are seated on custom hardware boards developed by the CASPER collaboration\footnote{\url{https://casper.berkeley.edu/}}  \cite{4176933}. We also use the software tool flow developed by CASPER to design the digital system. The CASPER collaboration is dedicated to building open-source programmable hardware specifically for applications in astronomy. We currently use two of their newer devices, the ROACH\footnote{\url{https://casper.berkeley.edu/wiki/ROACH}} (Reconfigurable Open Architecture Computing Hardware) for the F-engines, and the ROACH 2\footnote{\url{https://casper.berkeley.edu/wiki/ROACH-2_Revision_2}} for the X-engines. The main benefit of using CASPER hardware is that it facilitates the time-consuming process of designing
and building custom radio interferometry hardware. The CASPER collaboration also offers a large open-source library of FPGA blocks for commonly used signal processing structures such as polyphase filter banks, FIR filters and fast Fourier transform blocks  \cite{4840623}. However, due to MITEoR's ambitious architecture, involving both extreme compactness, an 8-bit correlator, and tight inter-ROACH synchronization constraints, we custom-designed most of the digital FPGA blocks. The specifications of our latest correlator are listed in Table \ref{tabspecs}.

\subsection{MITEoR Deployment and Data Collection}\label{secdeploy}
\begin{table}
\begin{center}
\begin{tabular}{ | l | l | }
	\hline
	Antenna & MWA dual-pol bow-tie \\ \hline
	Antenna count & 64 $\times$ 2 polarizations \\ \hline
	Array configuration &  8 $\times$ 8 grid  \\ \hline
	ADC & 4 $\times$ 64-channel 50\,Msps \\ \hline
	F-engine & 4 ROACHes with Virtex-5\\ \hline
	X-engine & 4 ROACH2s with Virtex-6\\ \hline
	Correlator precision &  8 bits  \\ \hline
	Frequency range & 110-190\,MHz \\ \hline
	Instantaneous bandwidth & 12.5\,MHz  (50 MHz digitized)\\ \hline
	Frequency resolution & 49\,kHz \\ \hline
	Time resolution & $\geq$ 2.68\,s \\ \hline
  \end{tabular}
\end{center}
\caption[MITEoR specifications.]{
List of MITEoR specifications. We observed with two different 8 by 8  array configurations, one with 3\,m separation and one with 1.5\,m separation. We observed ORBCOMM band with 2.68\,s resolution, and we chose a resolution of 5.37\,s for other bands. 
\label{tabspecs}
}
\end{table}
We deployed MITEoR in The Forks, Maine, which our online research suggested might be one of the most radio quiet region in the United States at the frequencies of interest\footnote{The Forks has also been successfully used to test the EDGES experiment  \cite{EDGES}, and we found the RFI spectrum to be significantly cleaner than at the National Radio Astronomy Observatory in Green Bank, West Virginia at the very low (100-200 MHz) frequency range that is our focus: the entire spectrum at The Forks is below -100\,dBm except for one -89.5\,dBm spike at 150MHz.}\cite{rfi}. We deployed the first prototype in September 2010, and performed a successful suite of test observations with an 8-antenna interferometer. In May 2012, we completed and deployed a major upgrade of the digital system to fully correlate $N=16$ dual-polarization antennas. With the experience of this successful deployment, we further upgraded the digital system to accommodate $N=64$ dual-polarization antennas, which led to our latest deployment in July 2013 and the results we describe in this paper.

The MITEoR experiment was designed to be portable and easy to assemble. The entire experiment was loaded into a 17 foot U-Haul truck and driven to The Forks. It took a crew of 15 people less than 2 days to assemble the instrument and bring it to full capacity. A skeleton crew of 3 members stayed on site for monitoring and maintenance for the following two weeks, during which we collected more than 300 hours of data. Subsequently, a demolition crew of 5 members disassembled and packed up MITEoR in 6 hours and concluded the successful deployment.

During the deployment, we scanned through the frequency range $123.5$-$179.5\,\textrm{MHz}$, with at least 24 consecutive hours at each frequency. We used two different array layouts for most of the frequencies we covered. The observation began with the antennas arranged in a regular 8 by 8 grid, with 3 meter spacing\footnote{We aligned the antenna positions using a laser-ranging total station, and measured their positions with millimeter level precision. The median deviation from a perfect grid is 2\,mm in the N-S direction, 3\,mm E-W, and 28\,mm vertical, primarily caused by the fact that the deployment site had not been leveled.} between neighboring antennas, which we later reconfigured to an 8 by 8 regular grid with 1.5 meter spacing for a more compact layout (which provides better signal-to-noise ratio on the 21~cm signal). The total volume of binary data collected was 3.9TB, and in the rest of this paper, we demonstrate the results of our various calibration techniques using this data set.

\section{Calibration Results}\label{seccal}

As we have emphasized above, the precision calibration of an interferometer is essential to its ability to detect the faint cosmological imprint upon the 21~cm signal, and the key focus of MITEoR is to determine how well real-time redundant calibration can be made to work in practice. In this section we describe the calibration scheme that we have designed and implemented and quantify its performance. We first constrain the {\it relative} calibration between antennas, utilizing both per-baseline algorithms and redundant-baseline calibration algorithms  \cite{redundant}. We then build on these relative calibration results to constrain the {\it absolute} calibration of the instrument, including breaking the few degeneracies inherent to redundant calibration.

\subsection{Relative Calibration}\label{secrelcal}

\subsubsection{Overview}
The goal of relative calibration is to calibrate out differences among antenna elements caused by non-identical analog components, such as variations in amplifier gains and cable lengths, which may be functions of time and frequency. We parametrize the calibration solution as a time- and frequency-dependent multiplicative complex gain $g_i$ for each of the 128 antenna-polarizations. Calibrating the interferometer amounts to solving for the coefficients $g_i$ and undoing their effects on the data.
\begin{figure*}
\centering
\hskip-5mm\centerline{\includegraphics[width=\textwidth]{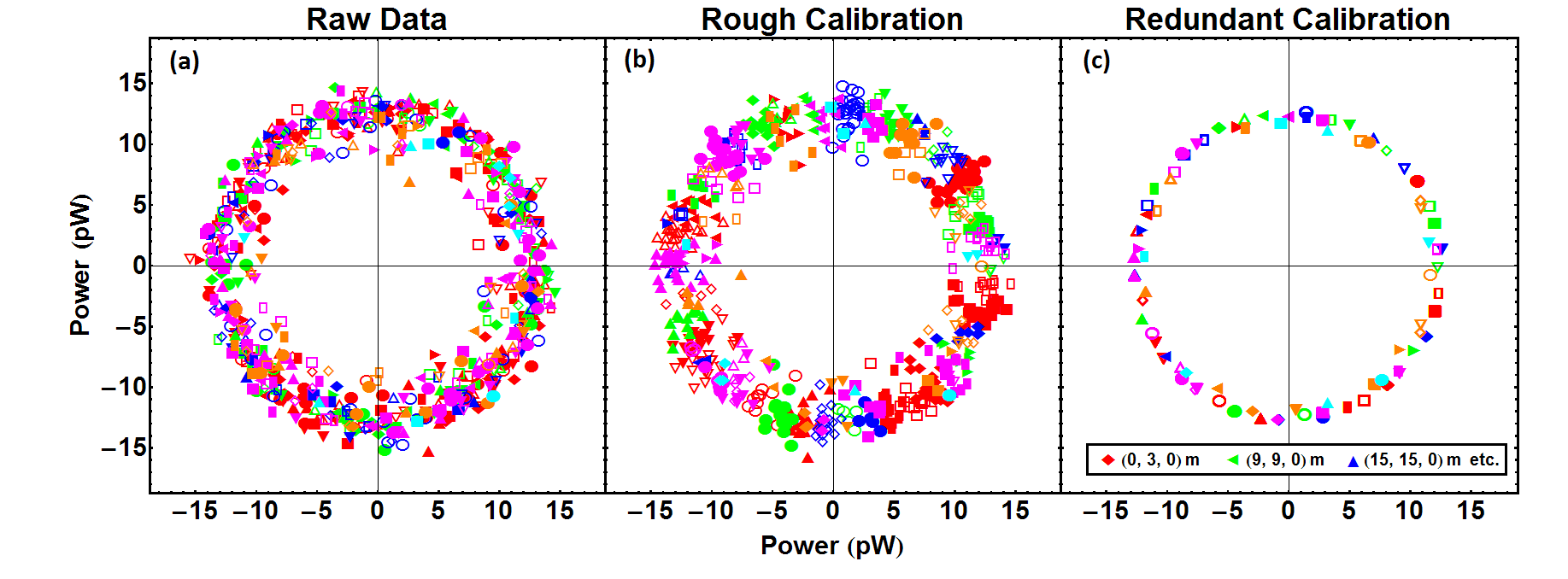}}
\caption[Redundant calibration during an ORBCOMM satellite pass.]{
Illustration of three stages in the redundant baseline calibration pipeline. Each panel is a complex plane, and each point is a complex visibility for a specific pair of antennas at 137.1\,MHz during the passage of an ORBCOMM satellite. Each unique combination of color and shape stands for one set of redundant baselines. In an ideal world, all identical symbols, such as all upright red triangles, should have the same value thus overlap exactly. Due to noise, they should cluster together around the same complex value. In panel (a) showing raw data, the redundant baselines have almost no clustering visible---for example, red filled circles can be found throughout the plot. After crude calibration in panel (b), we see most points falling into clustered segments---though the clustering is still far from exact. Finally in panel (c), after performing log calibration, we see that all points corresponding to each redundant baseline are almost exactly overlapping, with no visible deviation due to the high signal-to-noise. While the difference is not visible here, linear calibration can further improve log calibration results, as shown in Figure \ref{chisq}.
\label{omniviewer}
}
\end{figure*}
Our calibration scheme revolves around calibration methods that heavily utilize the redundancy of our array, whose efficacy we aim to demonstrate with MITEoR. The current redundant calibration pipeline involves three steps, as illustrated in Figure \ref{omniviewer}: 
\begin{enumerate}
\item Rough calibration computes approximate calibration phases using knowledge of the sky.
\item Logarithmic calibration (``logcal'') decomposes roughly calibrated data into amplitudes and phases and computes least square fits for amplitude and phase separately.
\item Linear calibration (``lincal'') takes the relatively precise but biased results from logcal and computes unbiased calibration parameters with even higher precision. 
\end{enumerate}
Although logcal and lincal have been previously proposed  \cite{Wieringa, redundant, lincal}, they both fail  in their original form if the phases of $g_i$ are not close to 0.\footnote{Logcal requires phase calibrations close to 0 to avoid phase wrapping issues, whereas lincal requires phase calibrations close to 0 to converge.} In practice, the phases of $g_i$ can be anywhere in the interval $[0, 2\pi)$. To overcome these practical challenges, we introduced various improvements to these algorithms. In the following sections, we describe our improvements to calibration algorithms in detail, and demonstrate the effectiveness of our calibration by obtaining $\chi^2/\mbox{DoF}\approx 1$ for the majority of our data. We then analyze these calibration parameters to construct a Wiener filter to optimally average them over time and frequency, which also tells us how frequently we need to calibrate in time and frequency.

\subsubsection{Rough Calibration}\label{secrough}
The goal of rough calibration is to obtain reliable initial phase estimates for the calibration parameters to enable the subsequent more sophisticated algorithms. This step does not have to involve redundancy, thus it can be done with any standard calibration techniques, for example self-calibration \cite{selfcalreview,selfcal}. The rough calibration algorithm that we describe below is computationally cheap and can robustly improve upon raw data even when a few antennas have failed.

At a given time and frequency, we have both the measured visibilities, $v_{ij}$, and $v^\text{model}_{ij}$, a rough model of the true sky signal\footnote{Since we are trying to obtain an initial estimate, the model does not have to be very accurate.}, where indices $i$ and $j$ represent antenna number. We first compute the phase difference between each measured visibility and its prediction. We then pick one reference antenna and subtract the phases of its visibilities from the phases of other visibilities to obtain a list of estimated phase calibration for each antenna. Finally, we take the median of these calibration phases to obtain a robust phase calibration estimator for each antenna. More concretely, we use the following procedure:
\begin{enumerate}
\item Construct a matrix $\mathbf{M}$ of phase differences where $M_{ij}=-M_{ji}= \arg(v_{ij}/v^\text{model}_{ij})$.
\item Define the first antenna as the reference by subtracting the first column of $\mathbf{M}$ from all columns to obtain $M'_{jk} = M_{jk} - M_{j0}$.
\item Obtain rough phase calibration parameters $\phi_k \equiv \arg(g_k)$ by computing the median angle of column $k$ in $M'$, defined as 
\begin{align}\label{eqmangle}
\phi_k
\equiv&\arg\,[\mbox{median}_j\{\exp(i M'_{jk})\}] \nonumber \\
=&\arg\,[\mbox{median}_j \{\cos(M'_{jk})\} \nonumber \\
&+i\,\mbox{median}_j \{\sin(M'_{jk})\}].
\end{align} 
\end{enumerate}
For stable instruments, the true calibration parameters have very small variation over days, so we can use one set of rough calibration parameters from a single snapshot in time for data from all other times. Thus we pick a snapshot at noon when each $v^\text{model}_{ij}$ can be easily computed from position of the Sun, and use the resulting raw calibration parameters as the starting point for logcal at all other times.

\subsubsection{Log Calibration and Linear Calibration}\label{secloglin}
To explain our redundant calibration procedure, we first need to briefly reintroduce the formalism developed in  \citet{redundant}. Suppose the $i^{\rm th}$ antenna measures a signal $s_i$ at a given instant. This signal can be expressed in terms of a complex gain factor $g_i$, the antenna's instrumental noise contribution $n_i$, and the true sky signal $x_i$ that would be measured in the limit of perfect gain and no noise:
\begin{equation}
s_i = g_i x_i + n_i.
\end{equation}
Under the standard assumption that the noise is uncorrelated with the signal, each baseline's measured visibility is the correlation between the two signals from the two antennas: 
\begin{align}
\label{eq:eq1}
v_{ij} &\equiv \langle s_i^* s_j \rangle \nonumber \\
&=g_i^* g_j\langle x_i^* x_j\rangle + g_i^*\langle x_i^* n_j\rangle + g_j\langle n_i^* x_j\rangle +\langle n_i^* n_j\rangle\nonumber \\ 
&= g_i^* g_jy_{i-j} + n_{ij}^{\text{res}},
\end{align}
where we have denoted the true correlation $\langle x_i^* x_j\rangle$ by $y_{i-j}$,\footnote{Following  \citet{redundant}, we use $y_{i-j}$ instead of $y_{ij}$ to emphasize that in a redundant array, the number of unique baseline visibilities can be much smaller than number of measured visibilities. The complete expression should be $y_{u(i,j)}$, where $u(i,j)$ means that baseline $ij$ corresponds to the $u$th unique baseline.} the noise from each antenna by $n_i$, the noise for each baseline by $n_{ij}^{\mbox{res}}$, and expectation values (effectively time averages) by angled brackets $\langle \dots \rangle$. In a maximally redundant array such as MITEoR, the number of unique baselines is much smaller than the total number of baselines. Therefore, we can treat all the $g_i$s and the $y_{i-j}$s as unknowns while keeping the system of equations \eqref{eq:eq1} overdetermined, enabling fits for both despite the presence of instrumental noise.

In  \citet{redundant}, some of us proposed logcal and lincal, and we have implemented both for calibrating MITEoR data. In log calibration, we take the logarithm of both sides of Equation \ref{eq:eq1} and obtain a linearized equation in logarithmic space. We then perform a least squares fit for the system of equations 
\begin{equation}
\log v_{ij} =  \log g_i^*+ \log g_j + \log y_{i-j},
\end{equation}
 where we solve for $\log g_i$ and $\log y_{i-j}$. Because the least squares fit takes place in log space whereas the noise is additive in linear space, the best fit results are biased. Linear calibration, on the other hand, is unbiased  \cite{redundant}. The lincal method performs a Taylor expansion of Equation \ref{eq:eq1} around initial estimates $g_{i}^0$ and $y_{i-j}^0$ and obtains a system of linearized equations
\begin{equation}
v_{ij} =  g_{i}^{0*} g_{j}^0y_{i-j}^0 +  g_i^{1*} g_j^0y_{i-j}^0+  g_i^{0*} g_j^1y_{i-j}^0+  g_i^{0*} g_j^0y_{i-j}^1,
\end{equation}
where we solve for $g_i^1$ and $y_{i-j}^1$.  For a detailed description of the logcal and lincal algorithms and their noise properties, we direct the reader to  \citet{redundant,lincal}. We now describe some essential improvements to these algorithms. 

Logcal was first thought to be unable to calibrate the phase component due to phase wrapping, since logcal has no way to recognize that $0^\circ$ and $360^\circ$ are the same quantity. Consider, for example a pair of redundant baselines that measure phases of $0.1^\circ$ and $359.9^\circ$ respectively. We can infer that they each only need a very small phase correction ($\pm0.1^\circ$) to agree perfectly. However, since logcal treats the difference between them as $359.8^\circ$ rather than $0.2^\circ$, it will calibrate by averaging $0.1^\circ$ and $359.9^\circ$ to $180^\circ$, which is completely wrong.

We made two improvements to the logcal method to guard against this. The first is to perform rough calibration beforehand, as described in Section \ref{secrough}. The second is to re-wrap the phases of $v_{ij}$. While rough calibration can make the phase errors relatively small\footnote{In our experience, they need to be less than about 20 degrees to ensure that the subsequent calibration steps converge reliably.}, that improvement alone is not sufficient, since $0^\circ$ and $360^\circ$ are still treated as different quantities. Thus we need to intelligently wrap the phases of the input vector before feeding it into logcal. This is done in two simple steps. For a snapshot of rough calibrated visibilities at given time and frequency, $v_{ij}$,  we first estimate the true phases of each group of redundant baselines, $\arg(y_{i-j})$, by computing median angles of measured phases using Eq. \ref{eqmangle}. Then for each measured phase, we add or subtract $2\pi$ until it is within $\pm\pi$ of $\arg(y_{i-j})$. This eliminates the phase wrapping problem. 

Unlike logcal, lincal is an unbiased algorithm, but it relies on a set of initial estimates for the correct calibration solutions to start with. The output of lincal can be fed back into the algorithm and it can iteratively improve upon its own solution.
However, the algorithm converges to the right answer only if the initial estimates are good. In practice, we find that three iterations of lincal typically produces excellent convergence, because the outputs of logcal are already decent estimates of the calibration solutions. Thus, by improving logcal, we also greatly improve lincal's effectiveness.

Our current calibration pipeline performs all steps of redundant calibration in less than 1 millisecond on a single processor core for a data slice at one time and one frequency channel, which is an order of magnitude faster than the rate data is saved onto disk. It is carried out by our open source \texttt{Omnical} package, coded in C++/Python.\footnote{The package supports the miriad file format and is easily adapted to work with other file formats. To obtain a copy, please contact jeff\_z@mit.edu.}
Thus there should be no computational challenge in performing the above described calibration procedure in real-time for any array with less than $10^3$ elements. For a future omniscope that has as many as $10^6$ elements, there are two ways to reduce the computational cost. The first is to calibrate less frequently in time and frequency, and we will discuss in detail the minimal sampling frequency in Section \ref{secwiener}. The other is to adapt a hierarchical redundant calibration scheme, where instead of calibrating all visibilities at the same time, one can calibrate the array in a hierarchical fashion whose computational cost scales only linearly with the number of elements. We discuss more details regarding hierarchical redundant calibration in Appendix \ref{apphierarchical}.
\begin{figure*}
\centering
\includegraphics[width = 180mm]{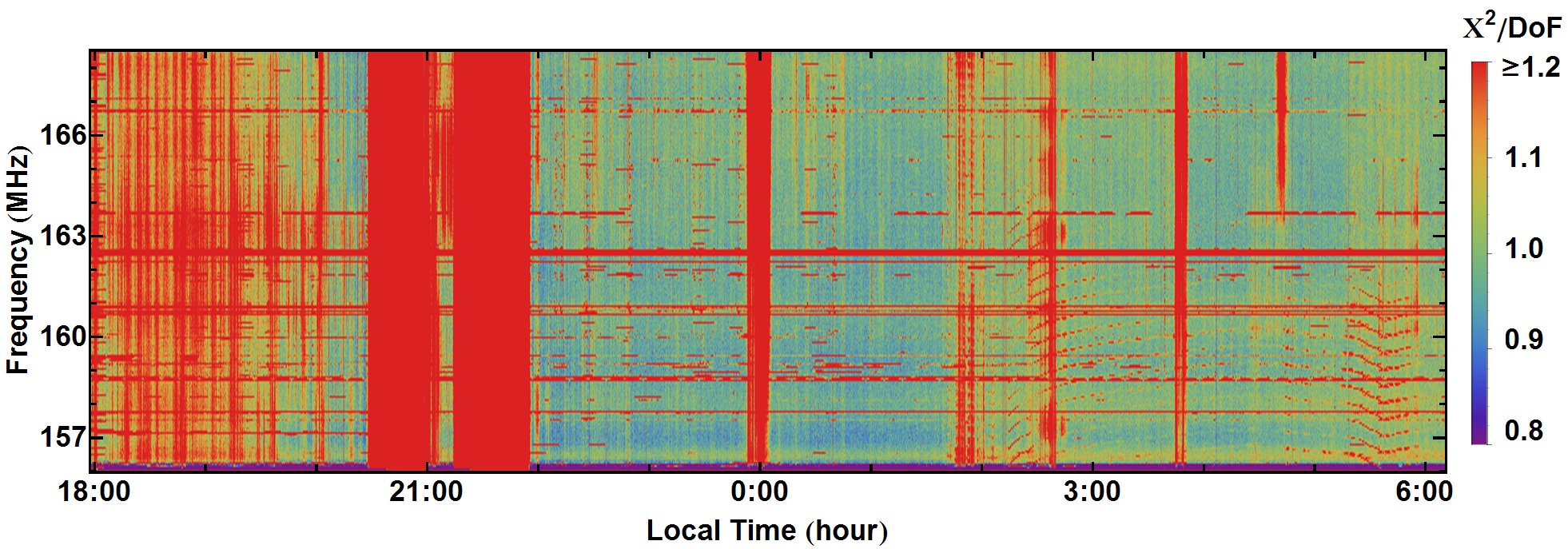}
\caption[$\chi^2$ as a function of time and frequency waterfall plot.]{
Waterfall plot of $\chi^2$/DoF for a day's data. This demonstrates the stability of our instrument as well as the effectiveness of using $\chi^2$/DoF as a indicator of data quality. We evaluate $\chi^2$/DoF every 5.3 seconds and every 49\,kHz. For the majority of the night time data, $\chi^2$/DoF is close to 1. We flag all data with $\chi^2$ larger than 1.2, which are marked red in this plot and account for 20\% of this data set. The amount of detailed structure in the flagged area (around 18:00 for example) shows the $\chi^2$ flagging technique's sensitivity to rapidly changing data quality.
\label{chisqwaterfall}
}
\end{figure*}

\subsubsection{$\chi^2$ and Quality of Calibration}\label{secchisq}

One of the many advantages of redundant calibration is it allows for the calculation of a $\chi^2$ for every snapshot to quantify how accurate the estimated visibilities are for each unique baseline, even without any knowledge of the sky. For a set of visibilities at a given time and frequency, $v_{ij}$, with calibration results $g_i$ and $y_{i-j}$, we define $\chi^2$ as
\begin{align}
\chi^2 & = \sum_{ij}{\frac{|v_{ij}-y_{i-j}g_i^*g_j|^2}{\sigma_{ij}^2}}\label{eqchisq},
\end{align}
where $\sigma_{ij}^2$ is the noise contribution to the variance of  the visibility $v_{ij}$. 
The effective number of degrees of freedom (DoF) is
\begin{align}
\mbox{DoF} & = N_{\text{measurements}} - N_{\text{parameters}}\nonumber\\
&= N_{\text{baselines}} - (N_{\text{antennas}} + N_{\text{unique baselines}}).
\end{align}
The numerator in Equation \ref{eqchisq} represents the deviation of measured data, $v_{ij}$, from the best fit redundant model, $y_{i-j}g_i^*g_j$. Thus $\chi^2$/DoF can be interpreted as the non-redundancy in measured data divided by the expected non-redundancy from pure noise. If the data agrees perfectly with the redundant model (with noise) and is free from systematics, then $\chi^2/\text{DoF}$ is drawn from a $\chi^2$ distribution with mean 1 and variance $2/\text{DoF}$ \cite{abramowitz1964handbook}.

\begin{figure}
\centering
\includegraphics[width = 83mm]{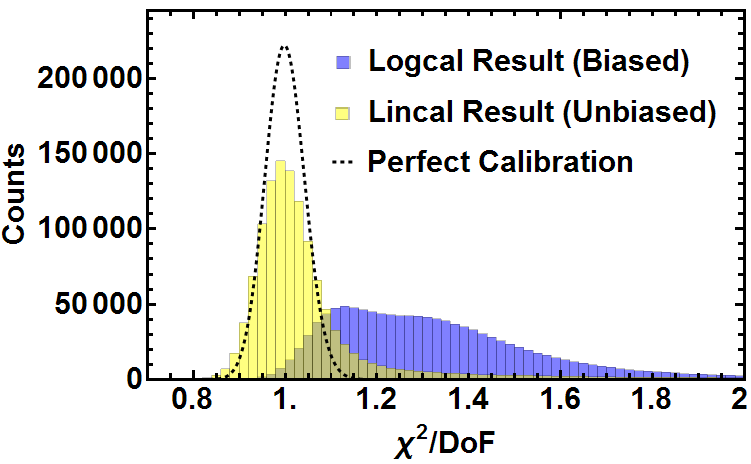}
\caption[$\chi^2$ histograph of redundantly calibrated data.]{
Histograms of the distributions of $\chi^2$/DoF of logcal results (mean 1.31) and lincal results (mean 1.05, median 1.01), together with the theoretical distribution of $\chi^2$/DoF (mean 1). They contain one night of data in a 12.5MHz frequency band (21:00-5:00 in Figure \ref{chisqwaterfall}). We evaluate $\chi^2$/DoF for every 5.3 seconds and every 49\,kHz. We set the flagging threshold to $\chi^2=1.2$, and 80\% of the lincal result is below the threshold (majority of the 20\% flagged data have $\chi^2$ much larger than 2, thus not shown in this figure). Among the data that is not flagged, 85\% is accounted for by the theoretical $\chi^2$ distribution. The right tail in lincal's distribution is due to the noise model sometimes underestimating the noise in order to minimize false negatives in the flagging process. The fact that $\chi^2$/DoF for lincal is so close to the theoretical distribution means that both the instrument and the calibration algorithms are working exactly as we expect.
\label{chisq}
}
\end{figure}

With a smooth model for $\sigma_{ij}$ which we describe below, we compute $\chi^2$/DoF for the results of rough calibration, log calibration, and linear calibration using all of our data. The $\chi^2$ distributions of our calibrations for one day's data are shown in Figures~\ref{chisqwaterfall} and~\ref{chisq}. Each calibration algorithm significantly reduces the $\chi^2$/DoF, and lincal's produces a distribution of $\chi^2$/DoF consistently centered around 1. We automatically flag any data with $\chi^2$/DoF larger than 1.2, which accounts for about 20\% of the data. Among the data that is not flagged, 85\% is accounted for by the theoretical $\chi^2$ distribution. The 15\% in the right tail is mostly attributable to a slightly optimistic noise model designed to avoid underestimating $\chi^2$. This close agreement between predicted and observed $\chi^2$-distributions for the lincal results suggests that except during periods that get automatically flagged, our instrument and analysis pipeline is free from significant systematic errors.
The fully automatic nature of our calibration pipeline and data quality assessment is encouraging for future instruments with data volume too large for direct human intervention.

Calculating $\chi^2$/DoF for flagging and data quality assessment requires an accurate model of noise in the measured visibilities. To compute the noise $\sigma_{ij}$, we approximate $\sigma_{ij}^2$ by $\langle\sigma^2\rangle$, where the average is over all baselines. This assumption that all antennas have the same noise properties drastically deceases the computational cost of calculating $\chi^2$/DoF. Because we have $10^3$ baselines, and the variation of $\sigma_{ij}$ between baselines is less than 20\% (due to slightly different amplifier gains), this approximation should cause only about a 1\% error in the final $\chi^2$/DoF.

To compute $\langle\sigma^2\rangle$, we perform linear regression on each visibility $v_{ij}$ over one minute to obtain its estimated variance $\sigma_{ij}^2$, and then average all $\sigma_{ij}^2$ to obtain $\sigma^2$. Thus we have $\langle\sigma^2\rangle$ at all frequencies every minute. Before we plug $\langle\sigma^2\rangle$ into Equation \ref{eqchisq}, we model it as a smooth and separable function: $\langle\sigma^2\rangle(f,t) = F(f)T(t)$, where $F(f)$ and $T(t)$ are polynomials. The smooth model has three advantages. The first is that it is physically motivated to model thermal fluctuation as a smooth and separable function. Secondly, a smooth noise model makes the $\chi^2$/DoF a much more sensitive flagging device. Theoretically, $\chi^2$/DoF should not rise above 1 when unwanted radio events such as radio frequency interference (RFI) occur,  because they are far field signals that do not violate any redundancy. However, since RFI events make both the signal and noise stronger, by demanding a smooth noise model, the $\langle\sigma^2\rangle$ we use will underestimate the noise during RFI events and give abnormally high $\chi^2$/DoF, which can then be successfully flagged with the $\chi^2/$DoF$\><1.2$ threshold. Thirdly, seasonal changes aside, the noise model is expected to largely repeat itself from day to day, so for future experiments that will operate for years, it suffices to use the same model repeatedly without recomputing $\sigma_{ij}$ {\it in situ} for all the data. Thus, by using a smooth noise model, one can drastically reduce the occurrence of false negatives (since it is better to flag good data than it is to fail to flag bad data) as well as the computational cost of calculating $\chi^2$/DoF.

\subsubsection{Optimal Filtering of Calibration Parameters }\label{secwiener}

While the above-mentioned estimates of the calibration parameters that we obtain from redundant baseline calibration vary over time and frequency, much of that variation is due to the noise in raw data. To minimize the effect of instrumental noise on the calibration parameters, we would like to optimally average information from nearby times and frequencies to estimate the calibration parameters for any particular measurement. 

As we will show below, the optimal method for performing this averaging is Wiener filtering. In the rest of this section, we first measure the  power spectrum of the calibration parameters over time and frequency, and make a determination of how to decompose this into  contributions from signal (true calibration changes) and noise. We  then weight the Fourier components in a way that is informed by their signal-to-noise ratio, and quantify how this Wiener filtering procedure improves upon more naive averaging over time and/or frequency. Finally, we discuss the implications for how regularly (in time and frequency) we should calibrate. It is important to note that while these methods are applied only to MITEoR below, they are applicable to any current or future experiment. 

We model the measured calibration parameter $g_i(f,t)$ for the $i^{\rm th}$ antenna 
as the sum of a true calibration parameter $s_i(f,t)$ (the ``signal'') and uncorrelated noise $n_i(f,t)$:
\begin{equation}
g_i(f,t) = s_i(f,t) + n_i(f,t).
\end{equation}
We choose our estimator $\hat{g_i}(f, t)$ of the true calibration parameter $s_i(f, t)$ to be a linear combination of the observed calibration parameters $g_i$ at different times and frequencies:
\begin{equation}
\hat{{g_i}}(f, t)\equiv\int\int W(f,t,f',t') g_i(f',t') df' dt'
\end{equation}
for some weight function $W$.
We optimize the estimator $\hat{{g_i}}$ by choosing the weight function $W$ that
minimizes the mean-squared estimation error $\langle|\hat{{g_i}}(f, t)-{s_i}(f, t)|^2\rangle$.
Assuming that the statistical properties of the signal and noise fluctuations are stationary over time\footnote{We perform this analysis on data over 12 MHz and two hours, where the signal and noise power are empirically found to be approximately time-independent.}, all correlation functions become diagonal in Fourier space:
\begin{eqnarray}
\langle\tilde{{s}_i}(\tau, \nu)^*\tilde{s}_i(\tau', \nu')\rangle &=& (2\pi)^2 \delta(\tau'-\tau)\delta(\nu'-\nu) S(\tau,\nu),\nonumber\\
\langle\tilde{n}_i(\tau, \nu)^*\tilde{n}_i(\tau', \nu')\rangle &=& (2\pi)^2 \delta(\tau'-\tau)\delta(\nu'-\nu) N(\tau,\nu),\nonumber\\
\langle\tilde{s}_i(\tau, \nu)^*\tilde{n}_i(\tau', \nu')\rangle &=& 0,
\end{eqnarray}
where tildes denote Fourier transforms and $S$ and $N$ are the power spectra of signal and noise, respectively.
This means that the optimal filter becomes a simple multiplication $\hat{\tilde{g}}=\tilde{W}\tilde{g}$
in Fourier space, corresponding to the 
weighting function $\tilde{W}(\tau, \nu)$ that minimizes the mean-squared error 
\begin{equation}
\langle |\tilde{W}(\tau, \nu)\tilde{g}_i(\tau, \nu) - \tilde{{s}_i}(\tau, \nu)|^2\rangle.
\end{equation}
Requiring the derivative of this with respect to $\tilde{W}$ to vanish gives the Wiener filter  \cite{Wiener1942}
\begin{equation}
\tilde{W}(\tau, \nu) = \frac{S(\tau, \nu)}{S(\tau, \nu)+N(\tau, \nu)}.
\end{equation}
\label{WienerEq}
Since $S$ and $N$ are to a reasonable approximation independent of the antenna number $i$, we have dropped all subscripts $i$ for simplicity.
Back in real space, the optimal estimator $\hat{g}_i$ for the $i^{\rm th}$ calibration parameter is thus $g_i$ convolved with 
the 2D inverse Fourier transform of $\tilde{W}$.

 \begin{figure}
\centering
\includegraphics[width=.6\textwidth]{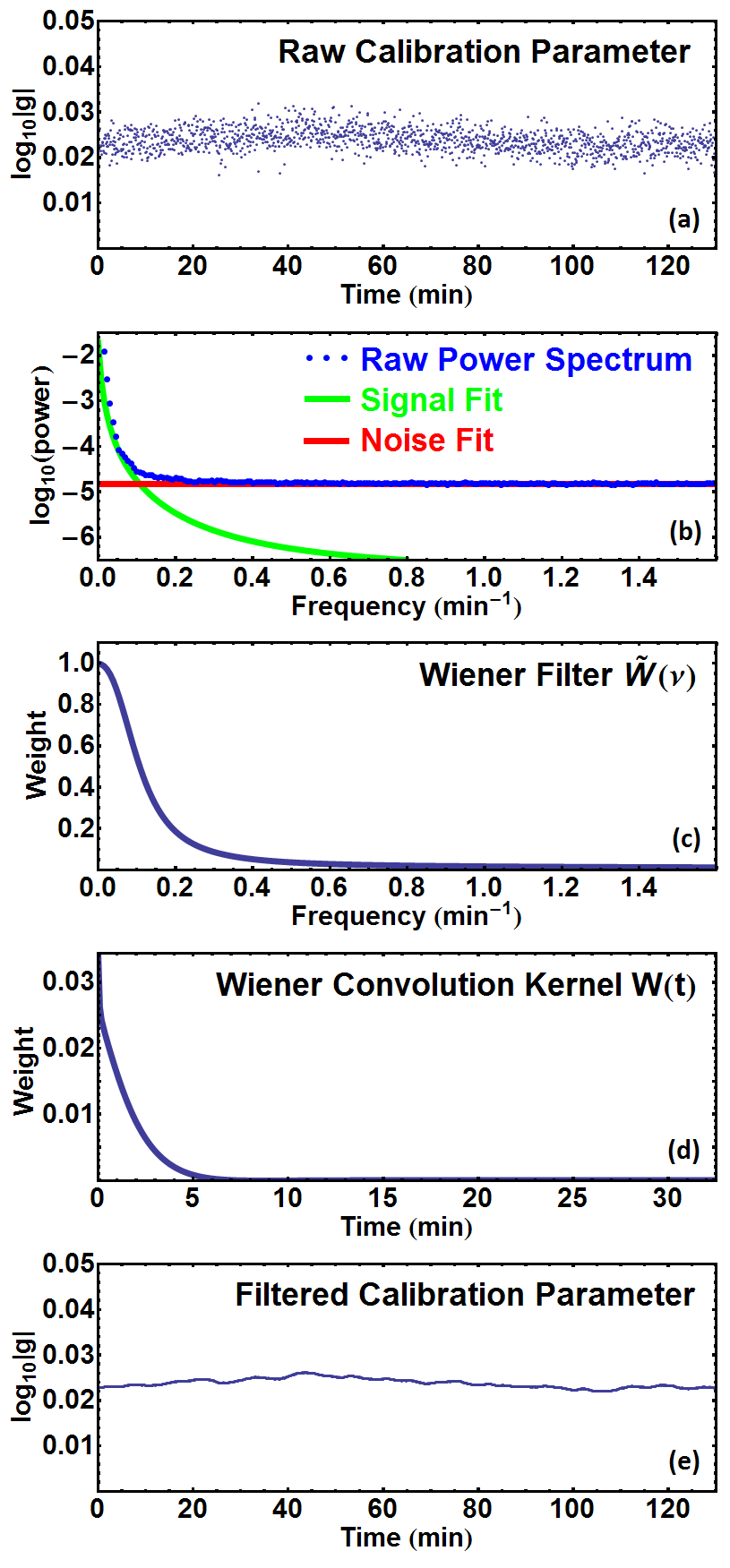}
\caption[Illustration of Weiner filtering calibration solutions.]{
Illustration of 1D Wiener filtering of calibration parameters at different times. Panel (a) shows the amplitude of calibration parameters measured for one antenna over two hours. Panel (b) shows that the average power spectrum across all antennas (blue dots) is well fit by a white noise floor (red horizontal line) plus a sum of two power laws (green curve). Panel (c) shows the Wiener filter in frequency domain computed using Eq. \ref{WienerEq} and the power spectra from panel (b). Panel (d) shows the Wiener convolution kernel in the time domain, the Fourier transform of the filter in Panel (c). Panel (e) shows the best estimates of the true calibration amplitude. The effectiveness of this filter is compared with that of other filters in Table \ref{WienerTab}.
\label{Wiener}
}
\end{figure}
To demonstrate this technique, we show the above process carried out in the time dimension in Figure \ref{Wiener}. In practice we perform the analysis on time and frequency dimensions simultaneously through a 2D FFT. 
The noise power spectrum $N(\nu)$ is seen to be constant to an excellent approximation, corresponding to white noise (uncorrelated noise in each sample). The signal power spectrum $S(\nu)$ is seen to be well fit by a combination of two power laws: 
$S(\nu)\approx (\nu/2.9\times10^{-5}\,\text{Hz})^{-2.7} +(\nu/4.8\times10^{-17}\,\text{Hz})^{-0.46}$.
The optimal convolution kernel $W$ is seen to perform a weighted average of the data on the timescale of roughly 200\,s and frequency scale of 0.15\,MHz, giving the greatest weight to nearby times and frequencies, resulting in an order-of-magnitude noise reduction.

To quantify the effectiveness of the obtained filter compared to  naive ``boxcar'' averages, we use the 2D power spectrum and noise floor of the calibration parameters obtained from real data to simulate many realizations of calibration parameters $g(f,t)=s(f,t)+n(f,t)$, apply various averaging/convolution schemes $W(f,t)$ on the simulated data, and compare their effectiveness by computing the RMS error $\langle |(W\star g)(f,t)-s(f,t)|^2\rangle$ normalized by $\langle{|n|^2}\rangle$. Due to our limited frequency bandwidth as well as frequent RFI contamination, power spectrum modeling in the frequency dimension is very challenging, so the frequency Wiener filter appears to be less effective than the time filter. In Table \ref{WienerTab} we list the normalized noise powers using frequency Wiener filter, time Wiener filter, 2D Wiener filter, as well as traditional boxcar averaging, and the 2D Wiener filter produces results three times less noisy than that of the traditional boxcar averaging.

We have described how to optimally average calibration parameters when we calibrate very regularly in time and frequency. For a future instrument such as an omniscope with $10^6$ antennas, calibration will pose a serious computational challenge, so it is important to know what the minimal frequency one needs to calibrate the instrument. The above analysis conveniently provides an answer to this question. As shown in the second panel of Figure \ref{Wiener}, the signal\footnote{We only show results for amplitude calibration parameters for brevity, as the phase calibration results have nearly identical power spectrum.} is band limited. By the Nyquist theorem, one needs to sample with at least double the frequency of signal bandwidth, so in our case we could measure the calibration parameters without aliasing problems as long as we calibrate once per minute. Calibrating more frequently than this simply helps average down the noise.
Although this one-minute timescale depends on the temporal stability characteristics of the amplifiers and other components used in our particular experiment, it provides a useful lower bound on what to expect from future experiments whose analog chains are even more stable. 

\begin{table}
\begin{center}
\begin{tabular}{ | l | l | }
	\hline
	Averaging method & Relative noise power \\ \hline
	No average & 1 \\ 
	Frequency Wiener filter & 0.33 \\ 
	Time Wiener filter & 0.12 \\ 
	Time and frequency Wiener filter & 0.09  \\ 
	Time and frequency boxcar average & 0.32 \\ \hline
  \end{tabular}
\end{center}
\caption[Effects of Wiener filtering calibration solutions.]{
Wiener filtering reduces the noise contribution to the calibration parameters by an order of magnitude. This table lists residual noise power (normalized by original noise power) after applying various filters to average the amplitude calibration parameters in time and/or frequency. The optimal two-dimensional Wiener filter indeed performed the best, lowering the noise power by an order of magnitude. In comparison, the naive boxcar average, using the characteristic scales of the optimal Wiener filter (200\,s and 0.15\,MHz), has more than 3 times residual noise power than the Wiener filtered result.
\label{WienerTab}
}
\end{table}
\subsection{Absolute Calibration}\label{secabscal}
The absolute calibration of the array involves two separate tasks. One is to find the overall gain and to break phase degeneracies  that redundant baseline calibration is unable to resolve, and the other to calibrate fixed properties of the instrument such as the orientation of the array and shape of the primary beam. The former is done by comparing the data to a sky model comprised of the global sky model (GSM)  \cite{GSM} and published astronomical catalogs (see  \citealt{JacobsFluxScale} for example).  The latter is done using bright point sources with known positions. While we can take advantage of the extremely high signal-to-noise data in the ORBCOMM channels (around 137~MHz), thanks to the dynamic range provided by our 8 bit correlator, it is important to note that all the algorithms described here are applicable to astronomical point sources as well. 

This section is divided into three parts. The first part describes how we use prior knowledge of the sky to break the degeneracies in redundant calibration results, a vital step to obtain usable data products. The second and third part each describe one aspect of absolute calibration using satellite data: primary beam measurement and array orientation.

\subsubsection{Breaking Degeneracies in Redundant Calibration}\label{secdegeneracy}
Redundant calibration alone cannot produce directly usable data products, due to the degeneracies intrinsic to the algorithms. There is one degeneracy in the amplitude of the calibration coefficients $g_i$, since scaling the amplitude of everything up by a common factor does not violate any redundancy (the degeneracies discussed here are per frequency and per time, as are the calibration solutions). There are three degeneracies in phase, corresponding to three degrees of freedom in a two dimensional linear field (see Appendix \ref{appdegeneracy} for a detailed discussion). In general, breaking these degeneracies requires prior knowledge of the sky. In this section, we briefly describe our algorithm that uses the global sky model (GSM) of  \citet{GSM} to remove these degeneracies. Doing so requires efficiently simulating the response of the instrument to the GSM; we summarize a fast algorithm for doing so in Appendix \ref{appfastgsm}. We defer detailed comparison of our data and the GSM to a future publication.

Our degeneracy removal procedure is an iterative loop that repeats two steps. The first step is to fit for the amplitude degeneracy factor. The knowledge of the GSM and bright point sources give us a set of model visibilities, $m_{ij}^a$, where index $a$ denotes different modeled components such as the GSM or Cygnus A. A linear combination of these models should be able to fit our measurements\footnote{We allow each model to have a separate weight to guard against potential calibration offsets between existing models.}. Thus we fit for the weights $w_a$ of the models by minimizing 
\begin{equation}
\left|v_{ij}-\sum_a w_a m_{ij}^a\right|^2.
\end{equation}
The second step is to break the degeneracy in redundant phase calibration by fitting for the degeneracy vector $\boldsymbol{\mathnormal{\Phi}}$ and the constant $\psi$ defined in Appendix \ref{appdegeneracy}. We assume that the error in the first step's fitting is mostly due to the phase degeneracies, so we take the $w_a$ from step one and fit for $\boldsymbol{\mathnormal{\Phi}}$ and $\psi$ by minimizing
\begin{equation}
\left|\arg(v_{ij})-\arg\left(\sum_a w_a m_{ij}^a\right)-\boldsymbol{d}_{i-j}\cdot\boldsymbol{\mathnormal{\Phi}}-\psi\right|^2,
\end{equation}
where $\boldsymbol{d}_{i-j}$ is the position vector for baseline $i-j$.

\begin{figure*}
\centering
\includegraphics[width=\textwidth]{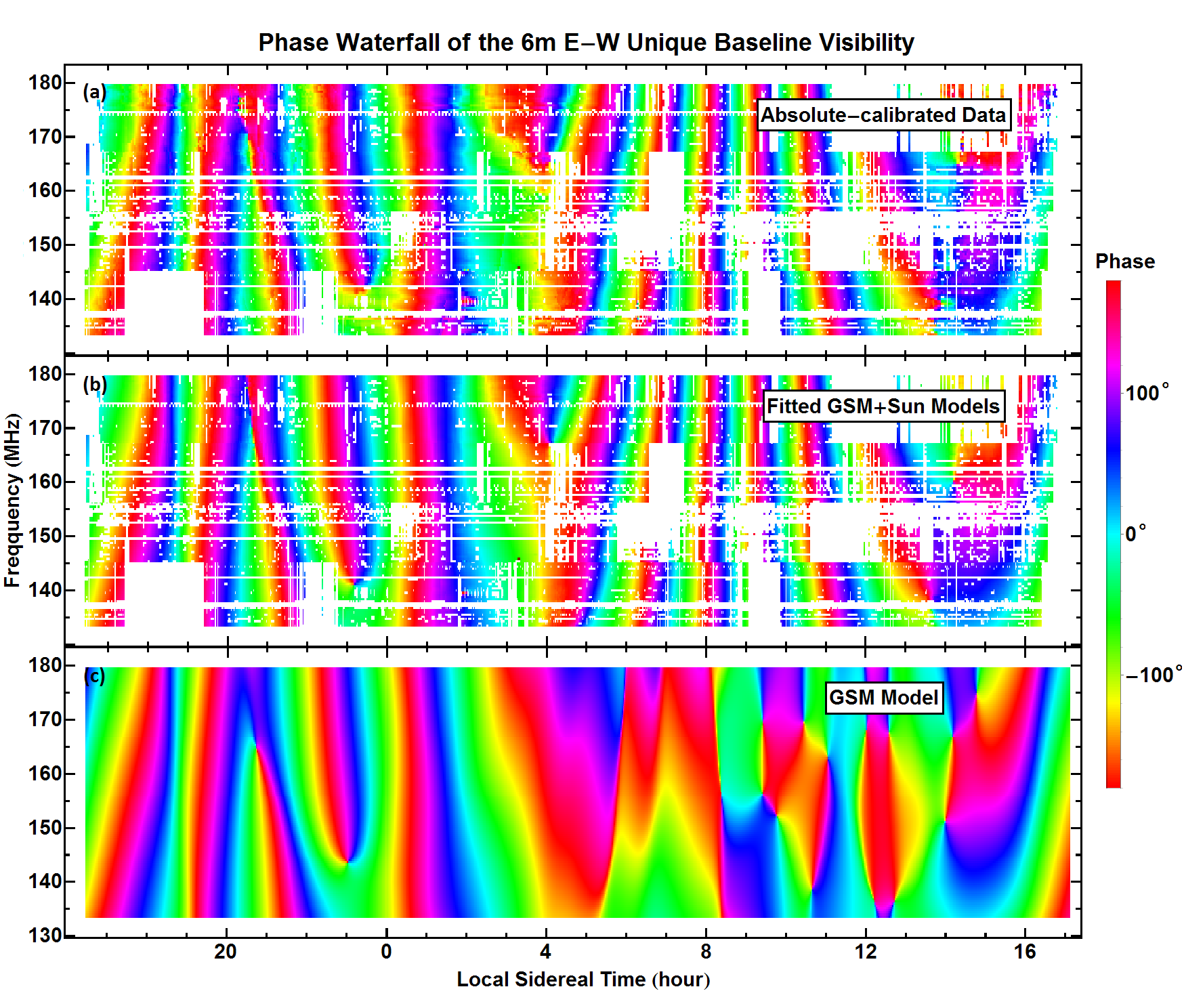}
\caption[Waterfall plot of a full day's data compared to a simulation.]{
Waterfall plots of phases on the 6~m E-W baseline. These show that our absolute calibration successfully matches the data (panel (a)) with a linear combination of global sky model and known point sources, including the Sun (panel (b)). Panel (c) shows the global sky model alone. The white areas are flagged out using $\chi^2$ criterion described in Figure \ref{chisqwaterfall}. Each plot is stitched together using four independently measured and calibrated frequency bands, aligning local sidereal time. Thus the discontinuities between hours 4 and 12 are due to the Sun rising at different local sidereal times on different days of our observing expedition.
\label{waterfall}
}
\end{figure*}

Note that the two fitting processes described above are not independent of one another, so we repeat these steps until convergence is reached. We find that in practice, the errors converge within two iterations. Our preliminary result is illustrated in Figure \ref{waterfall}, which shows that the data agree very well with current models.

\subsubsection{Beam Measurement Using ORBCOMM Satellites}\label{secbeam}

In general, \textit{in situ} measurements of antenna primary beams over large fields of view pose a challenge to 21\,cm cosmology, as primary beam uncertainties are intimately related to calibration, imaging, and catalog flux uncertainties  \cite{jacobsforeground}. Motivated by these difficulties,  \citet{JonniePrimaryBeam} present a solution that uses celestial point sources and assumes reflection symmetry of the beam, whereas Neben, Bradley, and Hewitt (in preparation) demonstrate high dynamic range beam measurement using the constellation of ORBCOMM satellites. Here, we present \textit{in situ} primary beam measurements of the MWA bow-tie antennas using the ORBCOMM constellation. We take advantage of both the high signal-to-noise ratio of ORBCOMM signals, and of our full cross-correlation measurements (rather than auto-correlations alone) to determine the beam.

In order to measure their primary beam profile $B_{\rm mwa}(\hat{\bf r})$, we compare measurements with MWA antennas to simultaneous measurements with simple center-fed dipoles, whose beam pattern $B_{\rm dipole}$ is known analytically. When there is a single extremely bright point source in the sky, such as an ORBCOMM satellite, we can compute the ratio of the visibilities of select baselines to obtain the ratio of the MWA antenna beam to the analytically known
dipole antenna beam --- thus determining the MWA antenna beam itself. To perform this analysis, two dipole antennas, one orientated along the $x$-polarization axis of the array and the other along the $y$-polarization axis, are added to the array and cross-correlated with all other MWA antennas.

The rationale behind this technique is as follows. At an angular frequency $\omega$, the electric field from a sky signal at the position of a receiving antenna can be encoded in the Jones vector $\boldsymbol{S}(\hat{\boldsymbol{k}})$, where $\hat{\boldsymbol{k}}$ is the position vector of the source \cite{collett2005field}. With a primary beam matrix $\mathbf{B}_j(\hat{\boldsymbol{k}})$, the signal measured by the $j^{\rm th}$ antenna at position $\boldsymbol{r}_j$ is
\begin{equation}
s_j =\int e^{-i [\boldsymbol{k}\cdot \boldsymbol{r}_j+\omega t]}\mathbf{B}_j({\hat{\boldsymbol{k}}})  \boldsymbol{S}(\hat{\boldsymbol{k}}) \; \mathrm{d} \Omega.
\end{equation}
When a single ORBCOMM satellite is above the horizon\footnote{There is typically more than one ORBCOMM satellite above the horizon at any one time, but they are coordinated so that they do not transmit in the same frequency band simultaneously.}, its signal strength is so dominant at its transmit frequency that $\boldsymbol{S}(\hat{\boldsymbol{k}})$ becomes well-approximated by a point source at the satellite's location. The measured signal can then be written as:
\begin{equation}
s_j \approx e^{-i [ \boldsymbol{k}_s \cdot \boldsymbol{r}_j+\omega t]} \mathbf{B}_j(\hat{\boldsymbol{k}}_s) \boldsymbol{S}_s,
\end{equation}
where $\boldsymbol{k}_s$ is the wave vector of the satellite signal, and $\boldsymbol{S}_s$ is the Jones vector encoding the satellite signal strength.

If we limit our attention to either $x$-polarization or $y$-polarization and approximate the off diagonal terms of $\mathbf{B}(\hat{\boldsymbol{k}})$ as zero, the visibility for two antennas can be written as
\begin{equation}
v_{jk} \approx S^2 B_j(\hat{\boldsymbol{k}}_s)^* B_k(\hat{\boldsymbol{k}}_s)e^{-i \boldsymbol{k}_s \cdot(\boldsymbol{r}_k-\boldsymbol{r}_j)}.
\end{equation}
If we take one visibility $v_{ij}$ formed by correlating a simple center-fed dipole with an MWA bow-tie antenna and another visibility $v_{kl}$ for the same baseline vector formed by correlating two MWA antennas, 
then their ratio is simply
\begin{equation}
 \frac{|v_{ij}|}{|v_{kl}|} \approx \frac{|B_{\rm mwa}(\hat{\boldsymbol{k}}_s)|}{|B_{\rm dipole}(\hat{\boldsymbol{k}}_s)|},
 \label{Eq.correlation ratio}
\end{equation}
because the satellite intensity $S$ and one MWA beam factor $B_{\rm mwa}$ all cancel out, and the phase factor $e^{-i \boldsymbol{k}_s \cdot(\boldsymbol{r}_j-\boldsymbol{r}_i)}$ is removed due to taking absolute values of the visibilities. This means that when a single point source dominates the sky, the ratio of visibility amplitudes is simply the ratio of the antenna beams at the direction of the point source. Since we already know the beam 
$B_{\rm dipole}$ of a center-fed dipole over a ground screen, we can directly infer the magnitude of MWA primary beam $|B_{\rm mwa}(\hat{\boldsymbol{k}}_s)|$.

\begin{figure}
\centerline{\includegraphics[width=.6\textwidth]{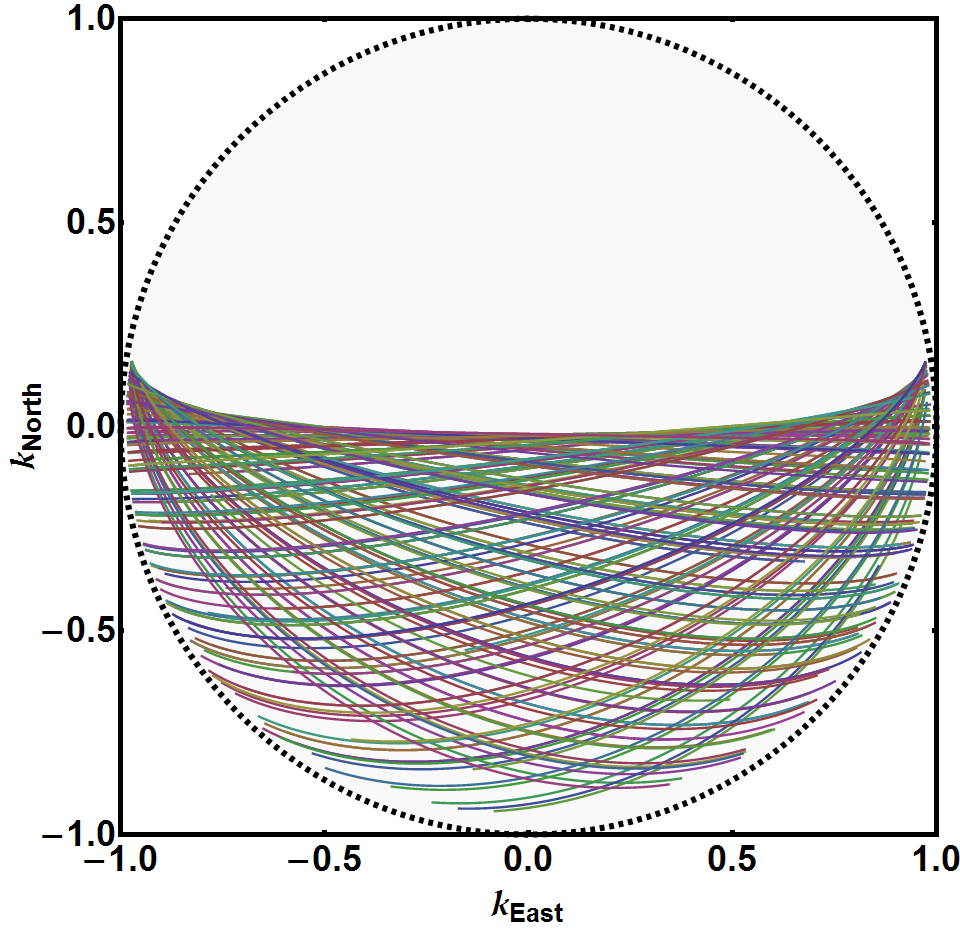}}
\caption[ORBCOMM trajectories through the primary beam.]{Projected trajectories of 248 passes of ORBCOMM satellites over 40 hours. With these passes we obtain sufficiently dense sampling of the MWA antennas primary beam that we can robustly map its response, especially at high elevations where the response is strongest. With a map of the southern half of the primary beam, we can use the reflection symmetry of the antennas to infer the entire beam at the ORBCOMM transmission frequencies. Each curve is a satellite pass projected onto the x-y plane, and the different colors specify sets of data taken at different times.}
\label{fig:sat passes}
\end{figure}

In order to fully map out the MWA primary beam, we need to take data during many satellite passes until we have direction vectors that densely cover the entire sky. Satellite signals from 27 ORBCOMM satellites at 5 frequencies in the range of 137.2-137.8 MHz were identified.  Their orbital elements are publicly available\footnote{We obtained the TLE files from CelesTrak, a company that archives TLEs of many civil satellites.}, so we can calculate $\hat{\boldsymbol{k}}_s(t)$ straightforwardly. With 40 hours of data taken at the frequencies of interest, we were able to obtain 248  satellite passes, shown in Figure \ref{fig:sat passes}.

\begin{figure}
\centering
\includegraphics[width=.6\textwidth]{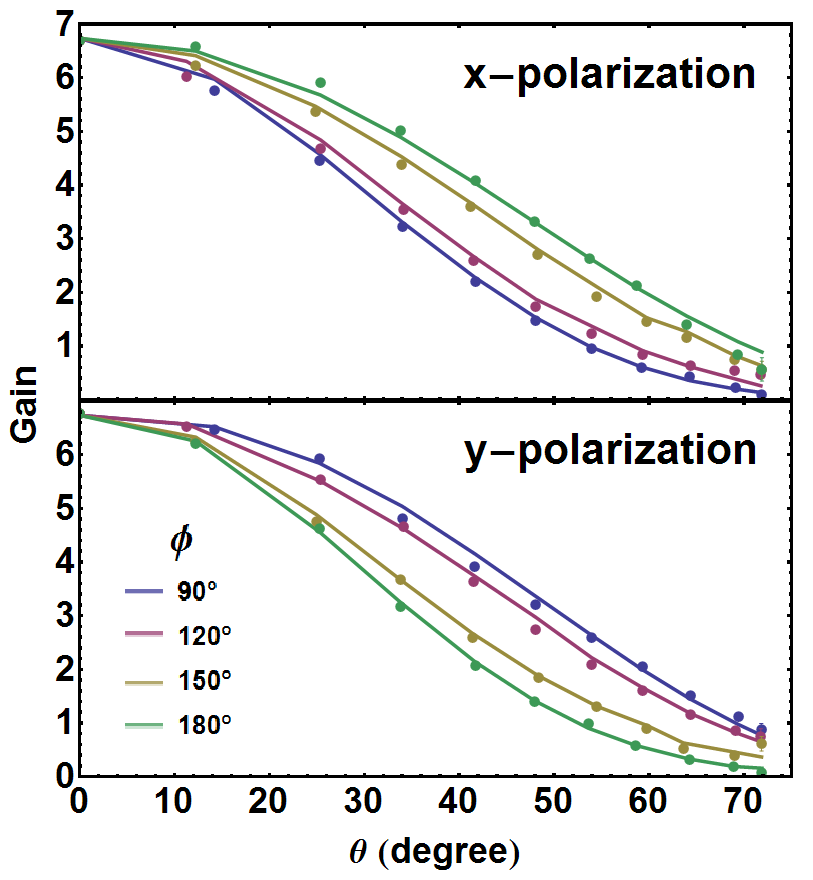}
\caption[Measured primary beam using ORBCOMM satellite passes.]{Measured MWA primary beam patterns compared to those obtained from numerical modeling. The two panels show the predictions (curves) and measurements (points) of the primary beam for the $x$-polarized and $y$-polarized MWA antennas. Each curve shows how the primary beam changes with the polar angle $\theta$ for a fixed
azimuth angle $\phi$. To reduce noise, the measurements have been averaged in 10 square degree bins.}
\label{fig:azimuth slice}
\end{figure}

We compared our measurements of the MWA primary beam using Equation \ref{Eq.correlation ratio} to numerical calculations using the FEKO electromagnetic modeling software package. Fixing an azimuth angle $\phi$, we can plot and compare the simulated and measured beam at different polar angles $\theta$ (the angle between the direction vector and zenith). Figure \ref{fig:azimuth slice} shows how the beam changes with $\theta$ for four different $\phi$-values, where $\phi=0$ correspond to North and increases clockwise.  Our measurements of the MWA beams are seen to agree well with the numerical predictions for both polarizations. The small differences between the predicted and measured beams are larger than the statistical noise, implying that the main limitation is not noise but one or more of the above-mentioned approximations, or approximations in the electromagnetic antenna modeling.

\subsubsection[Calibrating Array Orientation Using ORBCOMM \\Satellites and the Sun]{Calibrating Array Orientation Using ORBCOMM Satellites and the Sun}\label{secrotation}
\begin{figure}
\hskip-1.5mm\centerline{\includegraphics[width=.6\textwidth]{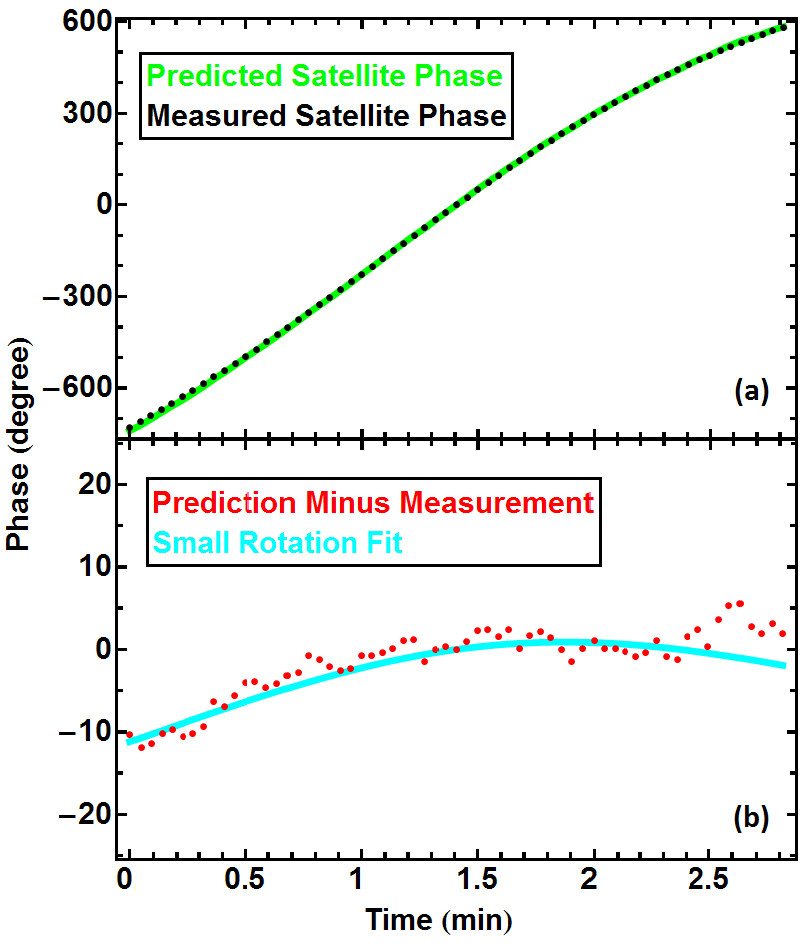}}
\caption[Calibration of array orientation using ORBCOMM.]{
Illustration of using the high signal-to-noise ORBCOMM data to calculate any small rotation in the array relative to the field-measured orientation. Panel (a) shows the rapidly wrapping phase of the raw data (black) from one baseline at the ORBCOMM frequency during the peak three minutes of a single satellite pass. In green, we see the predicted values computed with the field-measured array orientation and publicly available satellite positions. The residual between the model and the data is plotted in red points in panel (b). Finally, the cyan curve shows the best fit using small angle rotations of the array. In practice we use hundreds of satellite passes and all the baselines to obtain a single accurate fit for the true orientation of the array.
}\label{sat}
\end{figure}

The orientation of the array is very important, because the degeneracy removal process relies on the predicted measurement for each unique baseline, which in turn relies on precise knowledge of the baselines' orientations. Although we measured the relative position of each antenna to millimeter level precision with a laser-ranging total station, we did not measure the absolute orientation of the array to better than the $\sim 1^\circ$ accuracy obtainable with a handheld compass. To improve upon this crude measurement, we make use of both the known positions of both ORBCOMM satellites and the Sun. As we show in Figure \ref{sat}, the exceptional signal-to-noise in the ORBCOMM data allows us to fit for a small array rotation as a first order correction to a model based on our crude measurement.
Our method for finding the true orientation of the array is as follows. For a given baseline during the peak few minutes of an ORBCOMM satellite pass at frequency $\nu$, we measure a phase $\phi(t)$. We also know the satellite's position vector $\boldsymbol{k}(t)$.  However, we only have crude knowledge of baseline vector $\boldsymbol{d}_0$ in units of wavelength, where vectors are in horizontal coordinates with $x,y,z$ that correspond to south, east and up. We can therefore only compute a crude prediction of the phase measurement 
\begin{equation}
\phi_0(t) = 2\pi \boldsymbol{k}(t)\cdot\boldsymbol{d}_0.
\end{equation}
We assume that the difference between the measurement $\phi(t)$ and our crude prediction $\phi_0(t)$ is due to a small angle rotation of the baseline vector $\boldsymbol{d}_0$ around the axis $\pmb{\theta}=(\theta_x,\theta_y,\theta_z)$ by an angle $\theta = |\pmb{\theta}|$, ignoring a constant cable length delay.\footnote{Here it is important to use data before redundant baseline calibration to avoid phase degeneracy. We remove the phase delay from cables by allowing a constant offset that matches $\phi(t_M)$ with the crude prediction at time $t_M$ when the satellite has the strongest signal during the pass.} In the small $\theta$ regime, we have that
\begin{align}
\phi(t) - \phi_0(t) 
&= 2\pi \boldsymbol{k}(t)\cdot(\mathbf{R}(\pmb{\theta})\cdot\boldsymbol{d}_0 - \boldsymbol{d}_0) \nonumber \\
&\approx 2\pi \boldsymbol{k}(t)\cdot(\pmb{\theta}\times\boldsymbol{d}_0) \nonumber \\
&= 2\pi (\boldsymbol{d}_0\times\boldsymbol{k}(t))\cdot\pmb{\theta},
\end{align}
where $\mathbf{R}(\pmb{\theta})$ is the rotation matrix.
Since we have a set of equations each representing a different time, the problem of finding $\pmb{\theta}$ can be reduced to that of finding a least squares fit.
With 117 satellite passes, we obtained the following best fit for the array rotation around the vertical axis:
$$\theta_z^\text{sat}=0.66^{\circ}\pm0.0005^{\circ}_{\text{stat}}\pm0.07^{\circ}_{\text{sys}}.$$

While this method is very precise for solving the main problem we were worried about---the direction of North ($\theta_z$) which we approximated in the field with a handheld compass---it is less useful for measuring rotations of the array in the other two directions. Our instrument's absolute timing precision is only $\sim 0.5$ seconds, which makes it hard to distinguish rotations about the North-South axis from timing errors, as most ORBCOMM passes are East-West. This issue can of course be easily addressed in future experiments; for our experiment, we solve it using a more slowly moving bright point source: the Sun.

By using one day of solar data at 139.3\,MHz, we obtained
\begin{align*}
(\theta_x,\theta_y,\theta_z)^{\odot} =&(-0.08^{\circ}, -0.12^{\circ}, 0.672^{\circ})\\
\pm&(0.01^{\circ}, 0.03^{\circ}, 0.004^{\circ})_{\text{stat}}\\
\pm&(0.04^{\circ}, 0.003^{\circ}, 0.005^{\circ})_{\text{sys}}.
\end{align*}
Although solar data is noisier, in part because the Sun is not as bright as the ORBCOMM satellites in a given channel, timing errors are no longer important. These results agree with and complement the satellite-based results and allow us to confidently pin down the orientation of the array and thus improve the quality of the calibration of all of our data. The excellent agreement between the independent measurements 
$\theta_z^{\rm sat}\approx 0.66^\circ$ and
$\theta_z^{\odot}\approx 0.67^\circ$ 
provides encouraging validation of both the satellite and solar calibration techniques.

\subsection{Systematics}\label{secsystematic}
As we discussed in section \ref{secchisq}, most of our data are well-calibrated with $\chi^2/\mbox{DoF} <1.2$, which means that any systematic effects should lie well below the level of the thermal noise. In this section we aim to identify all the systematic effects present in the system, and describe our efforts to quantify and, whenever possible, remove them. The systematics can be categorized into two groups: 
\begin{enumerate}
\item Signal-dependent systematics that grow as the signal becomes stronger, such as cross-talk, antenna position errors and antenna orientation errors.
\item Signal-independent systematics, such as radio frequency interference (RFI) from outside or inside the instrument.
\end{enumerate}
Below we find a strict upper bound of $0.15\%$ for the signal-dependent component, as well as a signal-independent component which is easy to remove. 

\begin{figure}
\centerline{\includegraphics[width=.6\textwidth]{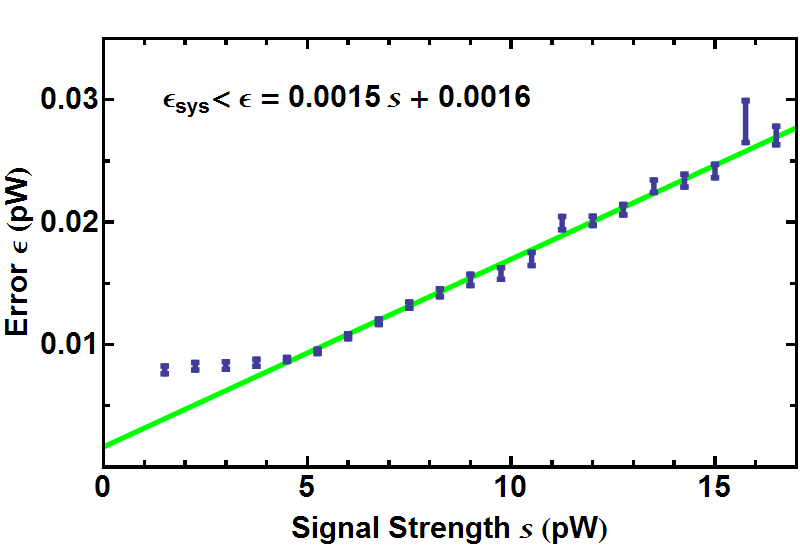}}
\caption[Investigation of signal-dependent systematic error.]{
Signal-dependent systematic error and its linear fit. By comparing the modeled and calibrated visibilities during ORBCOMM satellite passes, we conclude that signal-dependent systematic effects account for no more than $0.15\%$ of our measurement. We calculate the average fitting error per baseline $\epsilon = \langle|v_{ij}-y_{i-j}g_i^*g_j|\rangle$ and the average signal strength $s = \langle|y_{i-j}|\rangle$ binned over one day's data (blue points). The green line fits the data points above the noise floor. While many systematic errors, such as cross-talk, can contribute to the fitting error in addition to thermal noise, the best-fit slope of 0.0015 puts an upper bound on the sum of all signal-dependent errors. Since the ORBCOMM signal is so strong, any signal-independent systematic errors are negligible in this analysis. The high noise floor of $\sim 0.01$~pW is due to our digital tuning in the ORBCOMM frequency channels to maximize dynamic range.}\label{saterr}
\end{figure}

To quantify signal-dependent systematics, we again use ORBCOMM satellite data. Because the ORBCOMM signals are $10^3$ times brighter than astronomical signals, and we know that any signal-independent systematics must be weaker than the astronomical signals (otherwise they would have been blatantly apparent in the data), any signal-independent systematics must be negligible compared to the ORBCOMM signal. We therefore investigate how the discrepancies between calibrated visibilities and the models for each unique baseline depend on ORBCOMM signal strength.  We define the average fitting error per baseline at a given time and frequency to be
\begin{equation}
\epsilon = \langle|v_{ij}-y_{i-j}g_i^*g_j|\rangle,
\end{equation} 
which is a combination of antenna noise and systematic errors. If we compute $\epsilon$ at different times with different signal strength and compute its signal dependence, we can derive an upper bound on the signal dependence of systematic errors. To do this, we take all data at the ORBCOMM satellite frequency over a day and compute $\epsilon$ after performing redundant calibration. We then bin the $\epsilon$-values according to the average signal strength $s = \langle|y_{i-j}|\rangle$, and obtain the results shown in 
Figure \ref{saterr}\footnote{Another way of describing these data points is that, if we look at the third panel in Figure \ref{omniviewer}, we are plotting the average small spread in each unique baseline group versus the radius of the circle, and as the satellites pass over, both the circle size and the amount of average spread change over time, forming the data set in question.}.
The result is seen to be well fit by a constant noise floor plus a 
straight line $\epsilon\approx 0.0015 s$.
This slope implies that the combined effect of all signal-dependent systematic effects is 
at most 0.15\%.
This is merely an upper bound on the systematics, since it is possible that the increase in $\epsilon$ is mainly due not to systematics but to an increase in instrumental noise caused by an increase in the system temperature during the ORBCOMM passes.

There is one signal-dependent systematic that is not included in the above analysis: deviation from redundancy caused by imperfect positioning of antenna elements. This is because the data we used to derive the upper bound is always dominated by a single point source, the ORBCOMM satellite, and redundant calibration cannot detect any deviation of antenna position when the sky is dominated by a single point source\footnote{This is because for any arbitrary position deviation $\Delta \boldsymbol{r}_i$ for antenna $i$, one can add a phase equal to $\boldsymbol{k}\cdot\Delta \boldsymbol{r}_i$ to the calibration parameter $g_i$ to perfectly ``mask'' this deviation. Note that this ``mask phase'' depends on $\boldsymbol{k}$ and thus changes rapidly over time when the ORBCOMM satellite moves across the sky.}. We have two ways to quantify the error in our data due to antenna position errors. Firstly, the laser-ranging measurements of antenna positions in the field indicate an average of 0.037\,m deviation from perfect redundancy, which translates to about 2\% average error in phase on each visibility. Since the deviations are in random directions, the variance of phase error in the unique baseline fits should be brought down by a factor equal to the number of redundant baselines, resulting in phase errors much less than 1\% for most of the unique baselines. Secondly, although satellite calibration cannot detect position error in a given snapshot, over time the position errors would create very rapidly changing calibration parameters, which we do not observe in our data. Lastly, a formalism exists  \cite{redundant} to treat errors in antenna placement as small perturbations when redundantly  calibrating, although though we did not need to take advantage of this technique for the present paper.

We first identified a signal-independent systematic when we obtained consistent $\chi^2/\text{DoF}\sim 4$ for much of our data\footnote{This was before we obtained a consistent $\chi^2/\text{DoF}\sim 1$ in Section \ref{secchisq}, which occurred after we were able to remove the systematic described here.}, which means that the fitting error was on average twice as large as the thermal noise in each visibility. This implies a systematic (or a combination of systematics) at the level of $10^{-6}$\,pW/kHz, about 10\% of the total astronomical signal. Given the above analysis, we can exclude the possibility of any signal-dependent explanations such as cross-talk between channels or antennas. While we are unable to offer any conclusive explanation of this systematic, it appears consistent with persistent near-field RFI, perhaps originating from our electronics. Fortunately, we found this additive signal to vary only very slowly over time, typically remaining roughly constant over 5-minute periods, which made it easy to remove. After calibrating the data with logcal, we average the fitting errors $\epsilon_{ij} = \langle v_{ij}-y_{i-j}g_i^*g_j\rangle_t$ over time and subtract them from the data before we run logcal again. We perform the averaging over 5 minute segments, corresponding to 112 independent time samples, and iterate the calibration-subtraction process three times. This corresponds to less than a 1\% increase in the number of effective calibration parameters we fit for during logcal. Because many baselines probe the same unique baseline, the procedure described above exploits the redundancy of the array to robustly remove this slowly varying, signal-independent systematic, leaving us with $\chi^2/\text{DoF}\sim 1$.

\section{Summary and Outlook}\label{secsummary}

We have described the MITEoR experiment, a pathfinder ``omniscope'' radio interferometer with 64 dual-polarization antennas in a highly redundant configuration. We have demonstrated a real-time precision calibration pipeline with automatic data quality monitoring that uses $\chi^2$/DoF as a data quality metric to ensure that redundant baselines are truly seeing the same sky. We have also implemented various instrumental calibration techniques that utilize the ORBCOMM constellation of satellites to measure the primary beam and precise orientation of the array. Our success bodes well for future attempts to perform such calibration in real-time instead of in post-processing, and thus clears the way for FFT correlation that will make interferometers with $\gtrsim 10^3$ antennas cost-efficient by reducing the computational cost of correlating $N$ antennas from an $N^2$ scaling to an $N\log N$ scaling.  It also suggests that the extreme calibration precision required to reap the full potential of 21~cm cosmology is within reach. 

The various calibration techniques that MITEoR successfully demonstrates are now being incorporated into the much more ambitious HERA project\footnote{\url{http://reionization.org/}}  \cite{PoberNextGen}, a broad-based collaboration among US radio astronomers from the PAPER, MWA, and MITEoR experiments. Our results are also pertinent to the design of the SKA low-frequency aperture array\footnote{\url{http://skatelescope.org/}}. HERA plans to deploy around 331 14-meter dishes in a close-packed hexagonal array in South Africa, giving a collecting area of more than 0.05 km$^2$, virtually guaranteeing not only a solid detection of the elusive cosmological 21 signal but also interesting new clues about our cosmos. 


\begin{subappendices}

\section{Appendix: Phase Degeneracy in Redundant Calibration}\label{appdegeneracy}
Both of our redundant baseline calibration algorithms, logcal and lincal (see Section \ref{secloglin}), have the same set of phase degeneracies that require additional absolute calibration that must incorporate knowledge of the sky. When calibrating a given unique baseline, the quantity that logcal minimizes is
\begin{align}\label{eqlogdegen}
\sum_{jk} |(\theta_{j-k}-\phi_j+\phi_k)- \arg(v_{jk})|^2,
\end{align}
where we define $\theta_{j-k} \equiv \arg(y_{j-k}), \phi_j\equiv\arg(g_j)$. Similarly, lincal minimizes
\begin{align}\label{eqlindegen}
&\sum_{jk} |(y_{j-k}g_j^*g_k)-v_{jk}|^2\nonumber\\
=& \sum_{jk} \left||y_{j-k}g_j^*g_k|\exp\left[i(\theta_{j-k}-\phi_j+\phi_k)\right]-v_{jk}\right|^2.
\end{align}

Unfortunately, for all values of $\theta_{j-k}$ and $\phi_k$, one can add any linear field $\boldsymbol{\mathnormal{\Phi}}\cdot\boldsymbol{r}_j+\psi$ to the $\phi_j$ across the entire array while subtracting $\boldsymbol{\mathnormal{\Phi}}\cdot\boldsymbol{d}_j$ from the $\theta_{j-k}$ without changing the minimized quantities:
\begin{align}
 \theta'_{j-k}-\phi'_j+\phi'_k \equiv& ( \theta_{j-k} - \boldsymbol{\mathnormal{\Phi}}\cdot\boldsymbol{d}_{j-k}) -(\phi_j+\boldsymbol{\mathnormal{\Phi}}\cdot\boldsymbol{r}_j+\psi) \nonumber \\ 
&+(\phi_k+\boldsymbol{\mathnormal{\Phi}}\cdot\boldsymbol{r}_k+\psi) \nonumber \\ 
=& \theta_{j-k}-\phi_j+\phi_k.
\end{align}
Here $\boldsymbol{r}_j$ is the position vector of antenna $j$ and $\boldsymbol{d}_{j-k} \equiv \boldsymbol{r}_k-\boldsymbol{r}_j$ is the baseline vector for the unique baseline with best-fit visibility $y_{j-k}$. Thus, the quantities in expressions \ref{eqlogdegen} and \ref{eqlindegen} that the calibrations minimize are degenerate with changes to the linear phase field $\boldsymbol{\mathnormal{\Phi}}$ and the scalar $\psi$. This means that there are, in general, 4 degenerate phase parameters that need absolute calibration: one overall phase $\psi$ and three related to the three degrees of freedom of the linear function $\boldsymbol{\mathnormal{\Phi}}$ (which reduces to two for a planar array).

In an ideal instrument, the measured visibilities for a given unique baseline would be
\begin{align}
y_{i-j} = \int_{k_x,k_y}e^{i\boldsymbol{k}\cdot\boldsymbol{d}_{i-j}}S_B(k_x,k_y)dk_xdk_y,
\end{align}
where $\boldsymbol{k}=(k_x,k_y,k_z)$ is the wave vector of incoming radiation and $S_B(k_x,k_y)$ is the product of the incoming signal intensity and the primary beam in the direction $\hat{\boldsymbol{k}}$ normalized by $k k_z$ (which comes from the Jacobian of the coordinate transformation from spherical coordinates; see  \citealt{FFTT}). When the array is coplanar, we can take an inverse Fourier transform of $y_{i-j}$ and obtain an image of $S_B(k_x,k_y)$.  Above we saw that the best fit $y_{i-j}$ computed by logcal and lincal is multiplied by an unknown linearly varying phase $\boldsymbol{\mathnormal{\Phi}}\cdot\boldsymbol{d}_{i-j}$. 
Since multiplication in $uv$ space is a convolution in image space, this means that the image generated using these $y_{i-j}$ is the true image convolved with a Dirac delta function centered at $\boldsymbol{\mathnormal{\Phi}}$, which corresponds to a simple shift by the unknown vector 
$\boldsymbol{\mathnormal{\Phi}}$ in the $S_B(k_x,k_y)$ image space. 

\begin{figure}
\centering
\centerline{\includegraphics[width=.6\textwidth]{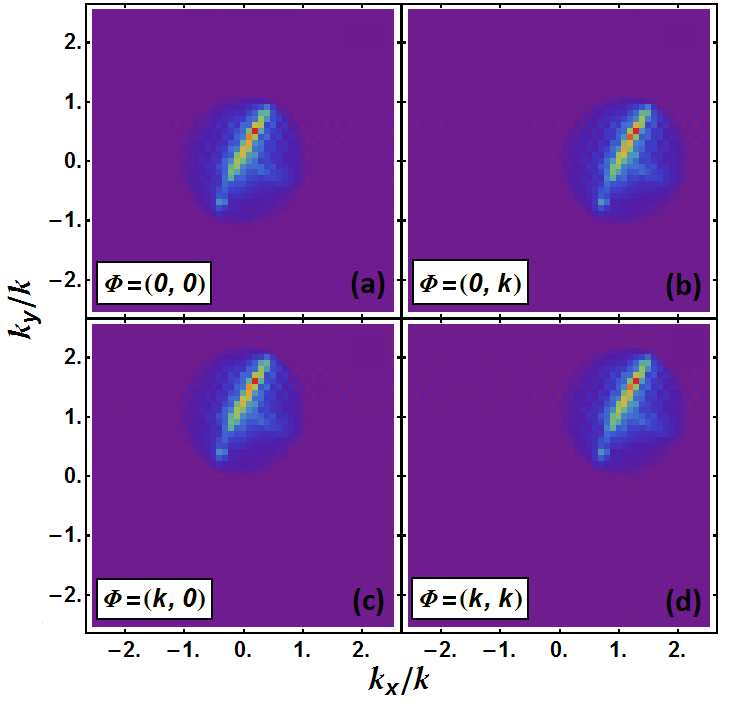}}
\caption[Illustration of phase degeneracies with short baselines.]{
Illustration of phase degeneracies shifting the sky image where the sky disk is demarcated. The linear phase degeneracy, which takes the form $\boldsymbol{\mathnormal{\Phi}}\cdot\boldsymbol{r}_i$ in each antenna for any $\boldsymbol{\mathnormal{\Phi}}$, corresponds to a shift of the reconstructed image.
These simulated images demonstrate shifts of fiducial sky image at 160~MHz caused by four different $\boldsymbol{\mathnormal{\Phi}}$, where the fiducial array's shortest baseline is 0.2~m.  Panel (a) shows the image obtained from visibilities with no $\boldsymbol{\mathnormal{\Phi}}$, and the sky image is centered at 0. In the other three panels, the sky image is shifted by amount $\boldsymbol{\mathnormal{\Phi}}$. Even if one has no knowledge of what the true sky is, it is still possible to determine 
$\boldsymbol{\mathnormal{\Phi}}$ from where the sky image is centered.
\label{shifts}
}
\end{figure}

To calibrate these last few overall phase factors, one can either make sure that bright radio sources line up properly in the image, or match phases between measured visibilities and predicted visibilities, as we described in Section \ref{secdegeneracy}. However, there may be another complementary way to remove this phase degeneracy without any reference to the sky. We know that physically the true image $S_B(k_x,k_y)$ is only non-zero within the disk
$|k_x^2+k_y^2|^{1/2}\le k$ centered around the origin, and a shift caused by $\boldsymbol{\mathnormal{\Phi}}$ would move this circle off center. This suggests that we should be able to reverse engineer $\boldsymbol{\mathnormal{\Phi}}$ by looking at how much the image circle has been shifted, without knowing what $S_B(k_x,k_y)$ is. Figure \ref{shifts} demonstrates how the image is shifted by $\boldsymbol{\mathnormal{\Phi}}$ using simulated data.

\begin{figure}
\centering
\includegraphics[width=.6\textwidth]{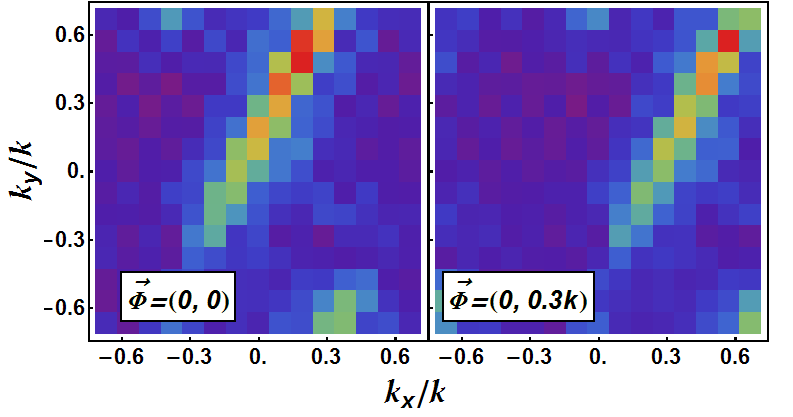} 
\caption[Illustration of phase degeneracies without short baselines.]{
Illustration of phase degeneracies shifting the sky image where the sky disk is not demarcated. With any practical array configuration, including that of MITEoR, distinguishing image shifts caused by the $\boldsymbol{\mathnormal{\Phi}}$-degeneracy becomes significantly more difficult. 
These images demonstrate shifts of fiducial sky image at 160~MHz just as in Figure \ref{shifts}, but with MITEoR's compact configuration where the shortest baseline is 1.5~m.  In the left panel, the image is obtained from visibilities with $\boldsymbol{\mathnormal{\Phi}}=(0,0)$, and in the right panel the sky image is shifted by and amount $\boldsymbol{\mathnormal{\Phi}}=(0,0.3k)$. Because the shortest baseline is too long ($0.8\lambda$), the Fourier transform of the visibilities only cover up to about 0.7 in $k_x$ and $k_y$, so in contrast with Figure \ref{shifts}, it is impossible to determine $\boldsymbol{\mathnormal{\Phi}}$ by merely looking at where the sky image is centered without prior knowledge of the sky.
\label{shiftsmiteor}
}
\end{figure}

Unfortunately, this simple approach to identifying and removing the effect of $\boldsymbol{\mathnormal{\Phi}}$ suffers from a few complications. By far the most important one is the requirement of very short baselines. In the example in Figure \ref{shifts}, the shortest separation between antennas is $0.21\lambda$, and it is easy to show that the sky disk is only clearly demarcated when the shortest separation is less than $0.5\lambda$\footnote{This is the 2D imaging counterpart of the well-known fact that, in signal processing, one must sample with a time interval shorter than $0.5\nu^{-1}$ to avoid aliasing in the spectrum of maximum frequency $\nu$.}. This sets a limit on the physical size of each element, which makes achieving a given collecting area proportionately more difficult. As Figure \ref{shiftsmiteor} shows, the deployed configuration of MITEoR cannot be used to reverse engineer the degeneracy vector $\boldsymbol{\mathnormal{\Phi}}$ without knowledge of the true sky.

\section{Appendix: A Hierarchical Redundant Calibration Scheme with $\mathcal{O}(N)$ Scaling}\label{apphierarchical}
One of the major advantages of an omniscope is its $N\log N$ cost scaling where $N$ is the number of antennas. However, existing calibration techniques, including the ones presented in this paper, require all of the visibilities to compute the calibration parameters. Since the cost for computing the visibilities alone scales as $N^2$, this is a lower bound to the computational cost of existing calibration schemes regardless of the actual algorithm. While current instruments with less than $10^3$ elements can afford full $N^2$ cross-correlation, such computation will be extremely demanding for a future omniscope with $10^4$ or more elements. Thus, to take advantage of the $N\log N$ scaling of an omniscope with large $N$, it is necessary to have a calibration method whose cost scaling is less than $N\log N$. In this section, we describe a such a method using a hierarchical approach, and show that its computational cost scales only linearly with the number of antennas. 

\begin{figure}
\centering
\centerline{\includegraphics[width=.6\textwidth]{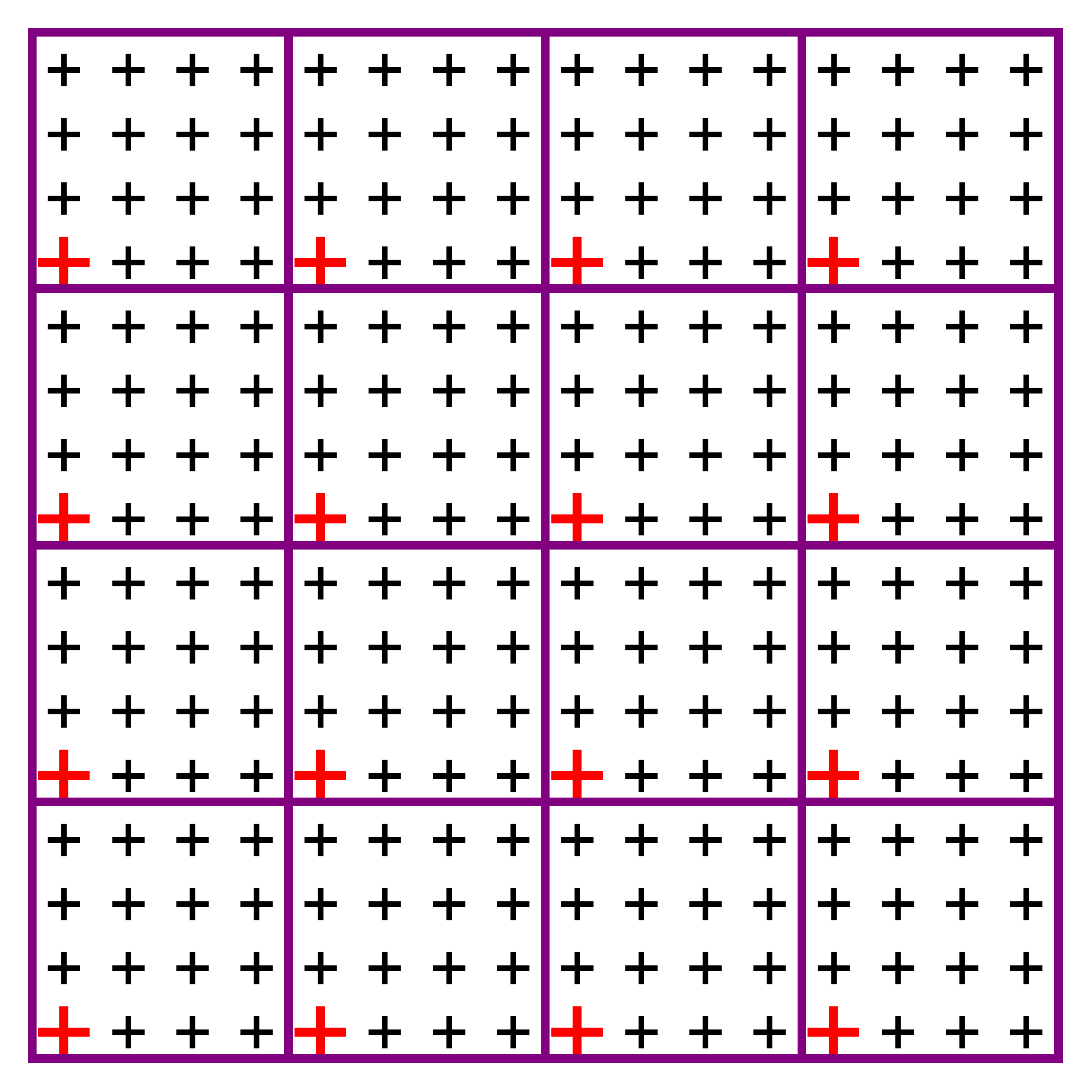}}
\caption[Schematic of heirarchical calibration method.]{
Example of the hierarchical calibration method for 256 antennas (marked by $+$-symbols) 
viewed  as a 2-level hierarchy of $4\times 4$ arrays $(m=16, n=2)$.
Our method first calibrates each sub-array independently with both relative and absolute calibrations. This produces calibration parameters for every antenna, up to one phase degeneracy $\psi$ per sub-array. We can remove these 16 phase degeneracies among sub-arrays by choosing one antenna out of each sub-array (marked red and thick) and performing calibration on these 16 antennas. Thus we have calibrated the whole 256 antenna array by performing 16-antenna calibration 16+1=17 times. This can be generalized to a hierarchy with more levels by placing 16 such 256-antenna arrays in a $4\times 4$ grid to get a 4096-antenna array, and then repeating to obtain arrays of exponentially increasing size. As shown in the text, the computational cost for this calibration method scales only linearly with the number of antennas.
\label{hierarray}
}
\end{figure}

Figure \ref{hierarray} illustrates the hierarchical calibration method for an example with a 
256 antennas in a $16\times 16$ regular grid, viewed as a 2-level hierarchy of $4\times 4$ grids.
More generally, consider an $n$-level hierarchy with $m$ sub-arrays at each level, containing a total
of $N=m^n$ antennas; the example in Figure \ref{hierarray} corresponds to $m=16$, $n=2$ and $N=256$.\footnote{It is easy to see that for a regular grid of $N$ antennas, $N$ need not an exact power of $m$ to obtain the scaling that we will derive.}
Let $B_m$ denote the computational cost of calibrating a basic $m$-antenna array\footnote{$B_m$ includes the cost to compute cross-correlations between the $m$ antennas, as well as both relative and absolute calibrations. The cost $B_m$ is unimportant for the scaling as long as it is independent of $n$.}
Let $C_n$ denote the computational cost of calibrating the entire $n$-level hierarchy containing all $N$ antennas.
We have $C_1=B_m$ by definition and 
\begin{equation}
C_{n+1}=m C_n + B_m
\end{equation}
since, as explained in the caption of Figure \ref{hierarray}, we can calibrate the $m$ sub-arrays at cost $C_n$ each and then calibrate the $m$ reference antennas (one from each sub-array) at cost $B_m$.
Solving this recursion relation gives 
\begin{eqnarray} 
C_n&=&B_m(1+m(1+m(1+m(1+\dotsi))))\nonumber\\
&=&B_m\sum_{k=0}^{n-1}{m^k}
=\frac{m^n - 1}{m-1}B_m\nonumber\\
&=&\frac{N-1}{m-1}B_m =\mathcal{O}(N).
\label{eq:hierarchy}
\end{eqnarray}

Eq.~\ref{eq:hierarchy} implies that for a fixed $m$, the computational cost for calibrating a $10^5$ antenna array will be 10 times that of a $10^4$ antenna array. Intuitively, the cost reduction comes from not computing cross-correlations among most pairs of antennas. In the simple case in Figure \ref{hierarray}, only one visibility is computed between each pair of sub-arrays, rather than 256 visibilities in a full correlation scheme. Because of the reduced number of cross-correlations computed, we expect the quality of calibration parameters to be worse than that in the full correlation case. Since both the precision of calibration parameters and the computational cost depend on $m$, one can tune the value of $m$ to achieve an optimal balance between precision and computational cost. 

\section{Appendix: Fast Algorithm to Simulate Visibilities Using Global Sky Model}\label{appfastgsm}

For both traditional self-calibration and the absolute calibration described in this paper, it is crucial to have accurate predictions for the visibilities. This requires simulation of both the contributions of bright point sources and diffuse emission, which can be added together due to the linearity of visibilities. While it is computationally easy to compute the contributions of point sources of known flux, it is much harder to compute visibilities from diffuse emission such as that modeled by the global sky model (GSM,  \citealt{GSM}). Dominated by Galactic synchrotron radiation, this diffuse emission is especially important for the low frequencies and angular resolutions typical of current 21~cm experiments. 

In general, we want to compute visibilities
\begin{align} 
y_{u}(f,t)&=\int s(\hat{\boldsymbol{k}}, f, t) B(\hat{\boldsymbol{k}},f)e^{i \boldsymbol{k}\cdot\boldsymbol{d}_{u}} d\Omega_k, \label{eq:gsmvisibility}
\end{align}
where $s(\hat{\boldsymbol{k}},f, t)$ is the magnitude squared of the global sky model at time $t$ in horizontal coordinates, and $B(\hat{\boldsymbol{k}},f)$ the magnitude squared of the primary beam at a given frequency. Performing the integral by summing over all $n_\text{pix}$ pixels in the GSM takes $\mathcal{O}(n_\text{pix} n_b n_f n_t)$ computational steps, where $n_b$ is the number of unique baselines one simulates, $n_f$ is the number of frequency bins, and $n_t$ is the number of visibilities one simulates for one sidereal day. 

The faster algorithm we describe here takes only $\mathcal{O}(n_\text{pix} n_b n_f)$ steps, by taking advantage of the smoothness of the primary beam as well as the diurnal periodicity in Earth's rotation. It applies only to drift-scanning instruments, so $B(\boldsymbol{k},f,t) = B(\boldsymbol{k},f)$ in horizontal coordinates, and is similar in spirit to the ideas proposed by  \citet{Richard}. 

The key idea is to decompose Equation \ref{eq:gsmvisibility} as follows:
\begin{equation} 
y_{u}(f,t)=
\sum_{\ell,m} a^f_{\ell m}\mathcal{B}_{\ell m}^{uf} e^{i m\phi(t)},
\end{equation}
where each $a_{\ell m}^f$ is a spherical harmonic component of the GSM at a given frequency, and each $\mathcal{B}_{\ell m}^{u f}$ is a spherical harmonic component of $B(\hat{\boldsymbol{k}},f)e^{i \boldsymbol{k}\cdot\boldsymbol{d}_{u}}$, both in equatorial coordinates. In this appendix, we describe precisely how to perform this decomposition and why it decreases the computational cost of calculating visibilities from the GSM.

\subsection{Spherical Harmonic Transform of the GSM}
The GSM of  \citet{GSM} is composed of three HEALPIX maps of size $n_\text{side}$ describing different frequency-independent sky principal components $s^c(\hat{\boldsymbol{k}})$ and the relative weights of each component $w^c(f)$ that encode the frequency dependence. We can decompose the spatial dependence into spherical harmonics,
\begin{align}
a_{\ell m}^c=\int Y^*_{\ell m}(\hat{\boldsymbol{k}})s^c(\hat{\boldsymbol{k}}) d\Omega_k
\end{align}
in $\mathcal{O}(n_\text{pix}^\frac{3}{2})$ steps, due to the advantage of HEALPIX format \citep{HEALPIX}. The frequency dependence of the spherical harmonic coefficients of the sky is given by 
\begin{align}
a^f_{\ell m}=\sum_c a_{\ell m}^c w^c(f),
\end{align}
and the total complexity of computing the coefficients $a^f_{\ell m}$  is $\mathcal{O}(n_\text{pix}^\frac{3}{2}) + \mathcal{O}(n_f)$.


\subsection{Spherical Harmonic Transform of the Beam and Phase Factors}

Next, we would like to compute the spherical harmonics components of $B(\hat{\boldsymbol{k}},f)e^{i \boldsymbol{k}\cdot\boldsymbol{d}_{u}}$:
\begin{align}
\mathcal{B}_{\ell m}^{u f}=\int Y^*_{\ell m}(\hat{\boldsymbol{k}})B(\hat{\boldsymbol{k}},f)e^{i \boldsymbol{k}\cdot\boldsymbol{d}_{u}}d\Omega_k.
\end{align}
Substituting the spherical harmonic decompositions of  $B(\hat{\boldsymbol{k}},f)$ and $e^{i \boldsymbol{k}\cdot\boldsymbol{d}_{u}}$ gives
\begin{align}
\mathcal{B}_{\ell m}^{u f}=&\int Y^*_{\ell m}(\hat{\boldsymbol{k}}) \sum_{\ell' m'}B_{\ell',m'}^{f}Y_{\ell' m'}(\hat{\boldsymbol{k}})\nonumber\\
&\times\sum_{\ell'' m''}4\pi i^{\ell''}j_{\ell''}\left(\dfrac{2\pi f}{c} d_u\right)Y^*_{\ell'' m''}(\hat{\boldsymbol{d}}_u) Y_{\ell'' m''}(\hat{\boldsymbol{k}}) d\Omega_k\nonumber\\
=& \sum_{\ell' m'}  \sum_{\ell'' m''}4\pi i^{\ell''}j_{\ell''}\left(\dfrac{2\pi f}{c} d_u\right)B_{\ell' m'}^{f}Y^*_{\ell'' m''}(\hat{\boldsymbol{d}}_u)\nonumber\\
&\times\int Y^*_{\ell m}(\hat{\boldsymbol{k}})Y_{\ell' m'}(\hat{\boldsymbol{k}}) Y_{\ell'' m''}(\hat{\boldsymbol{k}}) d\Omega_k\nonumber\\
=& \sum_{\ell' m'}  \sum_{\ell'' m''}4\pi i^{\ell''}j_{\ell''}\left(\dfrac{2\pi f}{c} d_u\right)B_{\ell' m'}^{f}Y^*_{\ell'' m''}(\hat{\boldsymbol{d}}_u)\nonumber\\
&\times\sqrt{\frac{\left(2\ell+1\right) \left(2\ell'+1\right) \left(2\ell''+1\right)}{4 \pi }} \nonumber\\
&\times(-1)^m
\left(
\begin{array}{ccc}
\ell & \ell' & \ell'' \\
 0 & 0 & 0 \\
\end{array}
\right) \left(
\begin{array}{ccc}
 \ell & \ell' &\ell'' \\
 -m & m' & m'' \\
\end{array}
\right),
\end{align}
where $j_\ell(x)$ is the spherical Bessel function, $\ell'm'$ represent quantum numbers when expanding the primary beam, $\ell''m''$ represent quantum numbers when expanding $e^{i \boldsymbol{k}\cdot\boldsymbol{d}_{u}}$, and the Wigner-3j symbols are results of integrating the product of three spherical harmonics. Because the Wigner-3$j$ symbols vanish unless $\ell-\ell'\leq\ell''\leq\ell+\ell'$ and $-m+m'+m''=0$, the above sum simplifies to
\begin{align}
\mathcal{B}_{\ell m}^{u f}&= \sum_{\ell' m'}  \sum_{\ell''=\ell-\ell'}^{\ell+\ell'}4\pi i^{\ell''}j_{\ell''}\left(\dfrac{2\pi f}{c} d_u\right)B_{\ell' m'}^{f}Y^*_{\ell'' m''}(\hat{\boldsymbol{d}}_u)\nonumber\\
&\times
\sqrt{\frac{\left(2\ell+1\right) \left(2\ell'+1\right) \left(2
 \ell''+1\right)}{4 \pi }} \nonumber\\
&\times(-1)^m
\left(
\begin{array}{ccc}
 \ell & \ell' & \ell'' \\
 0 & 0 & 0 \\
\end{array}
\right) \left(
\begin{array}{ccc}
 \ell & \ell' & \ell'' \\
 -m & m' & m''\\
\end{array}
\right),
\end{align}
where $m''=m-m'$.
Note that $\ell'$, $m'$ and $\ell''$ in this sum are all limited to the range of $\ell$-values where the spherical harmonics components for the primary beam are non-zero, so the complexity for this triple sum is $n_{B\text{pix}}^\frac{3}{2}$, where $n_{B\text{pix}}$ is the number of non-zero spherical harmonics components for the primary beam. Since the cost for each $\mathcal{B}_{\ell m}^{u f}$ is $n_{B\text{pix}}^\frac{3}{2}$, and there are $n_b n_f n_\text{pix}$ of them, the computational complexity of calculating all $\mathcal{B}_{\ell m}^{u f}$-coefficients scales like $\mathcal{O}(n_b n_f n_\text{pix} n_{B\text{pix}}^\frac{3}{2})$.


\subsection{Computing Visibilities}

By performing a coordinate transformation on Equation \ref{eq:gsmvisibility} from horizontal coordinates 
(corresponding to the local Horizon at the observing site)
to equatorial coordinates, the time dependence of $s(\hat{\boldsymbol{k}})$ is transferred to $B(\hat{\boldsymbol{k}})$ and $\boldsymbol{d}_u$. We can now calculate $y_u(f,t)$ by computing
\begin{align} 
y_u(f,t)&=\int s(\hat{\boldsymbol{k}}) B^{f t}(\hat{\boldsymbol{k}})e^{i \boldsymbol{k}\cdot\boldsymbol{d}_u(t)} d\Omega_k\nonumber\\
&=\sum_{\ell m} a^*_{\ell m}\mathcal{B}_{\ell m}^{u f t}.
\end{align}
Since the time dependence of $\mathcal{B}_{\ell m}^{u f t}$ is a constant rotation along the azimuthal direction, we can write the above as 
\begin{equation} 
y_u(f,t)=\sum_{\ell m} a^*_{\ell m}\mathcal{B}_{\ell m}^{u f} e^{i m\phi(t)}=\sum_{m}c_m^{uf} e^{i m\phi(t)}, \label{eq:fftvisibility}
\end{equation}
where we have defined 
\begin{equation} 
c_m^{u f} \equiv \sum_{\ell} a^*_{\ell m}\mathcal{B}_{\ell m}^{u f},
\end{equation}
which can be computed in $\mathcal{O}(n_b n_f n_\text{pix})$ steps. Given $c_m^{u f}$, it is clear that we can evaluate Equation \ref{eq:fftvisibility} using a fast Fourier Transform (FFT), whose cost is
\begin{align} 
\mathcal{O}(n_b n_f n_t \text{log}(n_t)).
\end{align}
 Note that this FFT in Equation \ref{eq:fftvisibility} has no $n_\text{pix}$ dependence, because we always need to zero-pad $c_m$ to length $n_t$ before the FFT.
%
In summary, the total complexity of all of the above steps is 
\begin{align} 
&\mathcal{O}\left(n_\text{pix}^\frac{3}{2}\right) + \mathcal{O}(n_f)+\mathcal{O}\left(n_b n_f n_\text{pix} n_{B\text{pix}}^\frac{3}{2} \right) \nonumber\\ 
&+\mathcal{O}\left(n_b n_f n_\text{pix}) +\mathcal{O}(n_b n_f n_t \text{lg}(n_t)\right)\nonumber\\
\approx& \mathcal{O}\left(n_b n_f n_\text{pix} n_{B\text{pix}}^\frac{3}{2} \right).
\end{align}
This does not scale with $n_t$, unlike the naive integration's $\mathcal{O}(n_b n_f n_\text{pix} n_t)$. Thus with a spatially smooth beam whose $ n_{B\text{pix}} \ll n_t^\frac{2}{3}$, the algorithm described here is  much faster than the naive numerical integration approach described at the beginning of this Appendix.

\end{subappendices}

\part{The Cosmic Dawn on the Horizon}
\chapter[Detecting the 21\,cm Forest in the 21\,cm Power Spectrum]{Detecting the 21\,cm Forest in the 21\,cm Power Spectrum} \label{ch:Forest}

\emph{The content of this chapter was submitted to the \emph{Monthly Notices of the Royal Astronomical Society} on October 30, 2013 and published \cite{AaronForest} as \emph{Detecting the 21\,cm forest in the 21\,cm power spectrum} on May 20, 2014.} 

\section{Introduction}

Observations of emission and absorption at 21 cm  from the neutral intergalactic medium (IGM) at high redshift will offer an unprecedented glimpse into the cosmic dark-ages up through the epoch of reionization (EoR), constraining both fundamental cosmological parameters and the properties of the first stars and galaxies \citep[for reviews]{FurlanettoReview,miguelreview,PritchardLoebReview}. Direct mapping of the 21 cm signal during the EoR is likely a decade or more away, requiring projected instruments such as the Square Kilometer Array (SKA). However, a first generation of experiments attempting to detect the power spectrum are already underway.  These include the Low Frequency ARray \citep[LOFAR]{InitialLOFAR1}, the Murchison Widefield Array \citep[MWA]{TingaySummary}, the Precision Array for Probing the Epoch of Reionization \citep[PAPER]{PAPER32Limits}, and the Giant Metre-wave Telescope \citep[GMRT]{newGMRT}. The MWA, PAPER, and LOFAR have the potential to achieve statistical detections of brightness temperature fluctuations within the next several years \citep{Judd06,Matt3,MWAsensitivity,Mesinger:2013c}.

Most theoretical investigations of observing neutral hydrogen in the
EoR have focused on IGM emission and absorption
against the Cosmic Microwave Background (CMB). It has
also been recognized by \citet{Carilli:2002tb,Furlanetto:2002,Xu:2009tc,Mack:2012tf,Ciardi:2013to} that the 21 cm forest, HI absorption in the spectra of background radio-loud (RL) active galactic nuclei (AGN), can be used to probe the IGM's thermal state.
  
 Studies of the forest have focused on its detection in the frequency spectra of a known RL source to glean information on the thermal properties of the absorbing IGM. The possibility for such a study depends on the existence of high redshift RL sources. As of 2013, the RL source distribution is only well constrained out to $z \sim 4$ (see \citet{DeZotti:2010} for review). Theoretical work suggests that at 100 $\MHz$ hundreds of $S \sim 1 \mJy$ sources with redshifts greater than 10 might exist within one of the $(30^{\circ})^2$ fields of view (FoV) offered by existing and upcoming wide field interferometers  \citep{HaimanQuasarCounts,WilmanExtragalacticSimulation} (hereafter H04 and W08 respectively). However the discovery of a suitable source at high redshift entails an extensive follow up program to measure  photometric redshifts of  radio selected candidates. 

Should sufficiently RL sources exist, a line of sight (LoS) detection of individual absorption features will require large amounts of integration time on a radio telescope with the collecting area comparable to the Square Kilometer Array (SKA). At reionization redshifts, \citep{Mack:2012tf} find that a $5 \sigma$ detection of an individual absorption feature with a $z \approx 9$ Cygnus A type source\footnote{ flux density at $151 \MHz$ of $S_{151} \approx 20 \mJy$ and spectral index of $\alpha \approx 1.05$ } would require years of integration on an SKA-like instrument. \citet{Ciardi:2013to} find that after 1000 hours of integration only 0.1\% of the LoS in an IGM simulation box contained regions of large enough optical depth to produce absorption features\footnote{ against a $S_{129} \approx 50 \mJy $ source at $z \sim 7$} observable by LOFAR. Hence a detection of the forest with a present day interferometer would require a very rare juxtaposition of an extremely loud RL source with an outlying optical depth feature. Even if this detection were achieved, it is unlikely that significant inferences on the thermal history could be made from only a handful of such observations.

 While detecting individual absorption features presents an enormous challenge, statistical methods have been demonstrated to reduce the necessary integration times. One target for a statistical detection is the increased variance in flux, along the LoS.  It is shown in  \citep{Mack:2012tf} that the integration time required for detecting this variance increase for a Cygnus A source, is only a few weeks with an SKA-like telescope, as apposed to the decades needed for detecting a single feature. \citet{Ciardi:2013to} find that LOFAR could detect the global suppression in the spectrum of a 50 mJy source at $z \sim 12$ with a 1000 hour integration, though they note that a detection by LOFAR is unlikely due to excessive RFI in the FM band ($80 \MHz \lesssim \nu \lesssim 108 \MHz$).

The possibility of a statistical detection of the forest using information from the wide FoV available to the current and upcoming generations of experiments has not yet been investigated. Observing the forest signature in the 21 cm power spectrum would integrate the signal from many high redshift sources within a FoV, reducing the sensitivity requirements of the instrument. Also, a power spectrum detection does not require a priori knowledge of high redshift sources. Hence the technique we describe can put constraints on both the properties of the IGM, such as the heating and reionization history, and the population of high redshift RL sources. It is likely that 21cm forest absorption features could be fruitfully explored using high-order statistical measures as well, but we do not consider those in this paper.

In this proof-of-concept, we begin to explore the characteristics and observability of the forest in the 21 cm power spectrum. We derive analytically the features that the global forest should introduce to the power spectrum and confirm their existence by combining semi-numerical simulations of the IGM, computed with 21cmFAST \citep{21CMFAST}, with the semi-empirical model of the high redshift population of RL sources from W08.
We find that in all heating scenarios studied, the contribution to the
21 cm fluctuations by the absorption of our RL sources is comparable to or dominates the contribution from the brightness temperature on small spatial scales $(k \gtrsim 0.50 \Mpci)$. To determine the detectability of the forest in the power spectrum, we perform sensitivity calculations for several radio arrays with designs similar to the MWA, including a future array with a collecting area of $\sim 0.1 \km^2$, similar to the planned Hydrogen Epoch of Reionization Array (HERA). In order to give the reader a sense of how the strength of this signal scales across a large range of of radio loud source populations, we extrapolate the expected S/N of the Forest using our analytic expression for the signal strength.

This paper is organized as follows.  In Section \ref{sec:theory} we provide the theoretical background and use a toy model to derive the morphology of the 21 cm forest power spectrum; relating its shape and amplitude to the optical depth power spectrum and the radio luminosity function. In Section \ref{sec:sim} we describe the semi-numerical simulations of the IGM along with the semi-empirical RL source distribution of W08 and how we combine them to simulate the wide field forest.  In Section \ref{sec:results} we discuss our results and identify the separate regions of $k$-space that may be used to independently constrain the thermal history of the IGM and the high redshift RL distribution. In Section \ref{sec:det} we explore the prospects for detecting the forest in  spherically averaged power spectrum measurements considering the sensitivity of current and future radio arrays. In Section \ref{sec:extr} we extrapolate our detectability results across a broad range of source populations and X-ray heating scenarios. 

Throughout this work we assume a flat universe with the cosmological parameters $h = 0.7$, $\Omega_\Lambda = 0.73$, $\Omega_M = 0.27$, $\Omega_b = 0.082$, $\sigma_8 = 0.82$, and $n=0.96$ as determined by the WMAP 7-year release \citep{WMAP7Cosmology}. All cosmological distances are in comoving units unless stated otherwise. 
\vskip 1.0 cm

\section{Theoretical Background} \label{sec:theory}

In this section we establish our notation and present a basic mathematical description of how forest absorption modifies the 21 cm brightness temperature signal. 

\subsection{Notation}
We adopt the Fourier transform convention
\begin{equation}
\widetilde{ f }({\bf k}) = \int d^3x   e^{-i {\bf k \cdot x}} f( {\bf x} ). 
\end{equation}
In addition, we often refer to cylindrical Fourier coordinates where $\kperp \equiv \sqrt{k_x^2 + k_y^2}$ and $\kpara \equiv |k_z|$. 
The power spectrum of a field $A$ over a comoving volume V is defined as
\begin{equation}\label{eq:ps}
P_A = \frac{1}{V} \langle | \widetilde{ \Delta A } | ^2 \rangle
\end{equation}
and the cross power spectrum between fields $A$ and $B$ over $V$ is given by
\begin{equation}
P_{A,B}= \frac{1}{V} \langle \widetilde{\Delta A} \widetilde{\Delta B}^* \rangle
\end{equation}
where 
\begin{equation}
\Delta A = A-\langle A \rangle
\end{equation}
and $\langle A \rangle $ is defined as the ensemble average of $A$ though in practice it is computed by averaging over some spatial or Fourier volume. In our discussion, we will also be referring to the one dimensional LoS power spectrum (not to be confused with the 1D spherical power spectrum) of a field A along a LoS column of comoving length L. 
\begin{equation}
P_A^{LoS}(k_z) = \frac{1}{L}  \int dz dz' \Delta A(z) \Delta A(z')  e^{i k_z(z-z')}
\end{equation}
Finally, we use $\Delta^2$ to denote the dimensionless power spectrum
\begin{equation}
\Delta^2 (k) \equiv \frac{k^3}{2 \pi^2} P(k)
\end{equation}
\subsection{The Forest's Modification of the Brightness Temperature}
 The forest absorption traces the optical depth of the IGM and will therefore introduce a signal on similar spatial scales as the 21 cm brightness temperature. We now discuss this signal in detail. The optical depth of a high redshift HI cloud is given by \citep{FurlanettoReview}
\begin{equation}\label{eq:tau}
\tau_{21} \approx .0092(1+\delta)(1+z)^{3/2} \frac{x_{HI}}{T_S} \left[ \frac{H(z)/(1+z)}{d v_{\parallel}/dr_{\parallel} } \right].
\end{equation}
$\delta$ is the fractional baryonic over-density, H(z) is the Hubble factor, $dv_\parallel / dr_\parallel$ is the velocity gradient along the LoS (including the Hubble expansion), and $x_{HI}$ is the neutral hydrogen fraction. The numerical factor in front of Equation (\ref{eq:tau}) is computed from fundamental constants and is independent of cosmology.
The spin temperature, $T_s$ is defined by the relative population densities of the two hyperfine energy levels, $n_1$ and $n_0$  \citep{Field:1958}
\begin{equation}
\frac{n_1}{n_0} = 3 \exp \left( {- \frac{h \nu_{21}}{k_B T_s}} \right).
\end{equation}
 Where, $h$ is Plank's constant, $k_B$ is the Boltzmann constant, and $\nu_{21} = 1420.41 \text{MHz}$ is the rest frame frequency of  the hyperfine transition radiation.

 Prior works on 21 cm tomography assume that the sky temperature at $\nu = \nu_{21}/(1+z)$ in the direction of an HI cloud is given by 
\begin{equation}\label{eq:sky_temp_noRL}
T_{sky} = \frac{T_s}{(1+z)} (1-e^{-\tau_{21}}) + \frac{T_{CMB}}{(1+z)}e^{-\tau_{21}} + T_{fg}.
\end{equation}
where $T_{CMB}$ is the comoving temperature of the cosmic microwave background radiation and $T_{fg}$ is the temperature of foreground emission including synchrotron radiation of the Galaxy, resolved point sources,  free-free emission, and radio emission from unresolved point sources below the confusion limit \citep{dimatteo1,LOFAR,GSM,xiaomin}. 

 The first term in Equation (\ref{eq:sky_temp_noRL}) includes both the 21 cm emission and self absorption of the HI cloud, hence it is multiplied by a factor of $(1-e^{-\tau_{21}})$. The second term describes the observed intensity of a background source shining through the cloud so its temperature is attenuated by $e^{-\tau_{21}}$. The third term describes radiation emitted by sources closer than the cloud so its intensity unaffected by $\tau_{21}$.

 21 cm experiments seek to measure the difference between the first two terms of Equation (\ref{eq:sky_temp_noRL}) and $T_{CMB}$. This difference is often referred to as the ``differential brightness temperature'' and is given by \citep{FurlanettoReview}
\begin{equation}\label{eq:dtb_no_RL}
\dtb = \frac{(T_s - T_{CMB})}{(1+z)}(1-e^{-\tau_{21}}) \approx \frac{T_s-T_{CMB}}{(1+z)}\tau_{21}.
\end{equation}
 We depart from previous work by considering the effect of radio loud sources behind the HI cloud whose combined observed\footnote{In accordance with much of the literature, we use the observed temperature for $T_{RL}$ and $T_{fg}$, rather than the comoving temperature as we have for $T_s$ and $T_{CMB}$. As a result, there are no factors of $(1+z)$ under $T_{RL}$ or $T_{fg}$.} brightness temperature we denote as $T_{RL}$. Including these background sources, Equation (\ref{eq:sky_temp_noRL}) becomes
\begin{equation}\label{eq:sky_temp}
T_{sky}' = \frac{T_{s}}{(1+z)} ( 1- e^{-\tau_{21}}) + \frac{T_{CMB}}{(1+z)}e^{-\tau_{21}} + T_{RL} e^{-\tau_{21}} +T_{fg} 
\end{equation}

$T_{fg}$ and $T_{RL}$  are expected to have predominantly smooth spectra which reside within a limited region of Fourier space known as the ``wedge'' \citep{Dattapowerspec,MoralesPSShapes,VedanthamWedge}. Smooth spectrum components may be removed by filtering\citep{AaronDelay} or subtraction \citep{Judd08,paper2,X13}, both employing the separation of the foregrounds and signal in the Fourier domain.

We will focus on the fluctuating signal, assuming that the smooth spectrum components of the foregrounds and background sources are properly avoided and/or subtracted. The effective differential brightness temperature now includes a contribution from the forest absorption features. 
\begin{equation}\label{eq:dtb_prime}
\dtb \to \dtb'  \approx \dtb   - T_{f}
\end{equation}
where $T_f = T_{RL} \tau_{21}$ is the ``forest temperature''. We can see how the power spectrum is transformed by the inclusion of $T_f$ by inserting Equation (\ref{eq:dtb_prime}) into Equation (\ref{eq:ps}) 
\begin{equation} \label{eq:ps_with_RL}
  P_b \to P_b' =  P_b  +  P_f- 2 \text{Re}(P_{f,b})
 \end{equation}  
Where $P_b \equiv P_{\dtb}$, $P_b' \equiv P_{\dtb'}$, $P_f \equiv P_{T_f}$ and $P_{f,b} \equiv P_{f,\dtb}$. Equation (\ref{eq:ps_with_RL}) sums up how the forest modifies the power spectrum that we expect to observe in upcoming 21 cm observations. Essentially, smooth spectrum power from $T_{RL}$ is leaked from the largest spatial modes to those occupied by  $\dtb$ via a convolution with the power spectrum of the optical depth field. The magnitude of this leakage will increase with the magnitude of the optical depth.
\subsection{The Morphology of the Forest Power Spectrum}
	The first thing one might ask concerning the forest contribution described in Equation (\ref{eq:ps_with_RL}) is how the magnitudes of the two contributions compare to each other and what their qualitative features are. While we will answer these questions with simulations it is useful to gain as much insight as we can through analytic methods. We start with $P_f$ which can be decomposed (see Appendix \ref{app:compare_pf} for a derivation) into a sum of auto power spectra $P_j$ originating from each individual RL source behind or within an imaged volume of IGM and their cross power spectra, $P_{j,k}$. 
	\begin{equation} \label{eq:sum_forest_sources}
 P_f  = \frac{1}{V} \left \langle \left| \widetilde{T_{RL} \tau_{21}}  \right|^2  \right \rangle = \displaystyle \sum_j      P_j +2\text{Re}\left(   \displaystyle \sum_{j < k }  P_{j,k} \right)
\end{equation}	
If all of the background sources are unresolved\footnote{a fair assumption given the large synthesized beams of interferometers and small angular extent of high redshift sources} then each $P_j$ is the absolute magnitude of the Fourier transform of a function that is a delta function in the perpendicular to LoS directions. As a result, each $P_j$ in Equation (\ref{eq:sum_forest_sources}) is constant in $k_\perp$. The cross multiplying $P_{j,k}$ terms are not so simple; however, we show in Appendix \ref{app:compare_pf} that in the absence of clustering, the cross sum only contributes to $P_f$ at the $10\%$ level for $\kpara \gtrsim 0.1 \Mpci$. At these scales, $P_f$ only has considerable structure along $\kpara$ 
\begin{equation}\label{eq:pf_simple}
P_f({\bf k}) \approx \sum_j P_j(\kpara) = \frac{D_M^2 \lambda^4}{4 k_B^2 \Omega_{cube}} P_{\tau_{21}}^{LoS}(k_\parallel)\langle \sum_j s^2_j \rangle
\end{equation}
where $\lambda=\lambda_{21}(1+z)$ is the observed wavelength of 21 cm light emitted from the center of the imaged volume, $D_M$ is the comoving distance to the data cube, and $\Omega_{cube}$ is the solid angle subtended by the cube. In the second step, we have expressed each $P_j$ in terms of the flux of each source, $s_j$, and the 1D power spectrum along the line-of-site to that source, $P_{\tau_{21}}^{LoS}$. In addition $P_f$ is positive so that it will always add to the power spectrum amplitude 

We can convert the sum in Equation (\ref{eq:pf_simple}) to an integral over the radio luminosity function
\begin{equation}\label{eq:integral}
 P_f \approx \frac{c D_M^2 \lambda^4}{4 k_B^2} P_{\tau_{21}}^{LoS}(\kpara)  \int_z^\infty \int_0^\infty s'^2 \rho(z,z',s') \frac{D_M^2(z')}{H(z)} dz'ds' 
\end{equation}
where $\rho(z,z',s') \equiv \frac{dN}{ds' dV_c}$ is the differential number of radio loud sources per comoving volume at redshift $z'$ per flux bin at observed frequency $\nu_{21}/(1+z)$ and $s'$ is the flux at $\nu = \nu_{21}/(1+z)$. 

Equation (\ref{eq:integral}) tells us that the amplitude of the forest power spectrum is set by the integral over the high redshift radio luminosity function multiplied by the average optical depth squared\footnote{By our definition, the power spectrum is the Fourier transform squared of $\Delta \tau_{21}$, not $\delta_{\tau_{21}} = \Delta \tau_{21}/\langle \tau_{21} \rangle$ which is often used in other work. Hence our power spectrum amplitude is set by $\langle \tau_{21} \rangle^2$} while the shape of the forest power spectrum is set by the 1D LoS power spectrum of optical depth fluctuations.

$P_{f,b}$ does not separate so conveniently but we can gain insight into whether it adds or subtracts to Equation (\ref{eq:ps_with_RL}) by considering the  physical phenomena that govern $T_f$ and $\dtb$. Expanding Equation (\ref{eq:dtb_no_RL}) and $T_f$ in terms of the IGM properties using Equation (\ref{eq:tau}) one can see that $P_{f,b}$ is the cross power spectrum between the two quantities:
	\begin{equation}\label{eq:dtb_full}
	\dtb \approx 9 x_{HI} (1+\delta)(1+z)^{1/2} \left[ 1 - \frac{T_{CMB}}{T_s} \right] \left[ \frac{H(z)}{dv_\parallel/dr_\parallel} \right] \mK
	\end{equation}
	and
	\begin{equation}\label{eq:tf_full}
	T_f  \approx 0.009 x_{HI} (1+\delta)(1+z)^{1/2}\frac{T_{RL}}{T_s}\left[ \frac{H(z)}{dv_\parallel/dr_\parallel} \right]
	\end{equation} 
	
	Before the reionization era, $x_{HI}$ is relatively homogenous so that fluctuations in $\dtb$ are governed primarily by those in $T_s$. Regions of the IGM with larger $T_s$ will have more positive $\dtb$ but smaller $T_f$. Because of this anti-correlation between $T_f$ and $\dtb$, $\text{Re}(P_{f,b})$ is negative during the pre-reionization era and the net effect will be for it to increase the power spectrum amplitude through its negative contribution in Equation (\ref{eq:ps_with_RL}). At lower redshifts, after X-rays have heated the IGM, $T_s \gg T_{CMB}$, and $\dtb$ becomes independent of $T_s$. As a result, $\dtb$ is always positively correlated with $x_{HI}$ as is $T_f$. $\text{Re}(P_{f,b})$ is positive with a net effect of subtracting from the power spectrum amplitude. We are unable to make any more progress analytically, but we will reexamine the cross power term in our simulation results below.
	
	We now move on to describe our simulations. We will return to our discussion of the power spectrum morphology in the context of our simulation results in Section \ref{sec:results}.


\section{Simulations}\label{sec:sim}

In this section we describe the semi-numerical simulations that we use to explore a range of IGM thermal histories along with the the semi-empirical RL source model that we employ to add the 21 cm forest signal. 

\subsection{Simulations of the Optical Depth of the IGM}

Our IGM simulations are run using a parallelized version of the public, semi-numerical 21cmFAST code\footnote{\url{http:/homepage.sns.it/mesinger/Sim}} described in \citet{21CMFAST}. Tests of the code can be found in \citet{Mesinger:2007hl,Zahn:2011,21CMFAST}. The simulation box is 750 Mpc on a side, with resolution of $500^3$. Different scenarios for $\tau_{21}$ can be obtained by exploring histories of the spin temperature, $T_s$ and/or the neutral fraction, $x_{HI}$.

 21cmFAST includes sources of both UV ionizing photons and X-rays.  The former dominate reionization (i.e. $x_{HI}$), except for extreme scenarios we do not consider in this work \citep{Furlanetto:2006, McQuinn:2012, Mesinger:2013vm}.
  Since a full parameter study is beyond the scope of this work, and since the bulk of the relevant signal is likely during the pre-reionization epoch,  we fix the ionizing emissivity of galaxies (and hence the reionization history), to agree with the Thompson scattering optical depth from WMAP \citep{WMAP7Cosmology}.  Instead we focus on the X-ray emissivity and its impact on $T_s$.

 $T_s$ is affected by a variety of processes. These include Ly-$\alpha$ photons which couple to the hyperfine transition through the Wouthuysen-Field effect \citep{Wouthuysen:1952,Field:1958}, particle collisions, and emission or absorption of CMB photons. The coupling of $T_s$ to these processes is described by (e.g. \citealt{FurlanettoReview}):
 \begin{equation}
 T_s^{-1} = \frac{T_{CMB}^{-1}+x_c T_k^{-1} + x_\alpha T_c^{-1}}{1+x_c + x_\alpha}
 \end{equation}
 where $T_k$ is the kinetic temperature of the HI gas, $T_c$ is the color temperature of Ly-$\alpha$ photons, and $x_c$ and $x_\alpha$ are the collisional and Ly-$\alpha$ coupling constants. Due to the high optical depth of the neutral IGM to Ly-$\alpha$ photons, the color temperature is very closely coupled to the kinetic temperature, $T_c \approx T_k$  \citep{Wouthuysen:1952,Hirata:2006} .

Although the self-annihilation of some dark matter candidates can contribute significantly \citep{Valdes:2013}, in fiducial models $T_k$ is predominantly determined by X-ray heating (e.g. \citealt{FurlanettoReview}). Hence, we explore a range of optical depth histories by running simulations for different galactic X-ray emissivities.

We use the fiducial model of X-ray heating described in \citet{Mesinger:2013vm}, adopting a spectral energy index of $\alpha=1.5$ and an obscuration threshold of 300 eV. We parameterize the X-ray luminosity by a dimensionless efficiency parameter, $f_X$.  Our fiducial model, $f_X=1$ corresponds to 0.2 photons per stellar baryon, or a total X-ray luminosity above $h\nu_0=0.3$ keV of $L_{\rm X, {\rm 0.3+keV}}\approx 10^{40}$ erg s$^{-1}$ ($\Msun$ yr$^{-1})^{-1}$. This choice is consistent with (a factor of $\sim$2 higher than) an extrapolation from the 0.5--8 keV measurement of \citet{Mineo:2013} that $L_{\rm X, {\rm 0.5-8keV}}\approx3\times10^{39}$ erg s$^{-1}$ ($\Msun$ yr$^{-1})^{-1}$. 

\begin{figure}
\centering
\includegraphics[width=.6\textwidth]{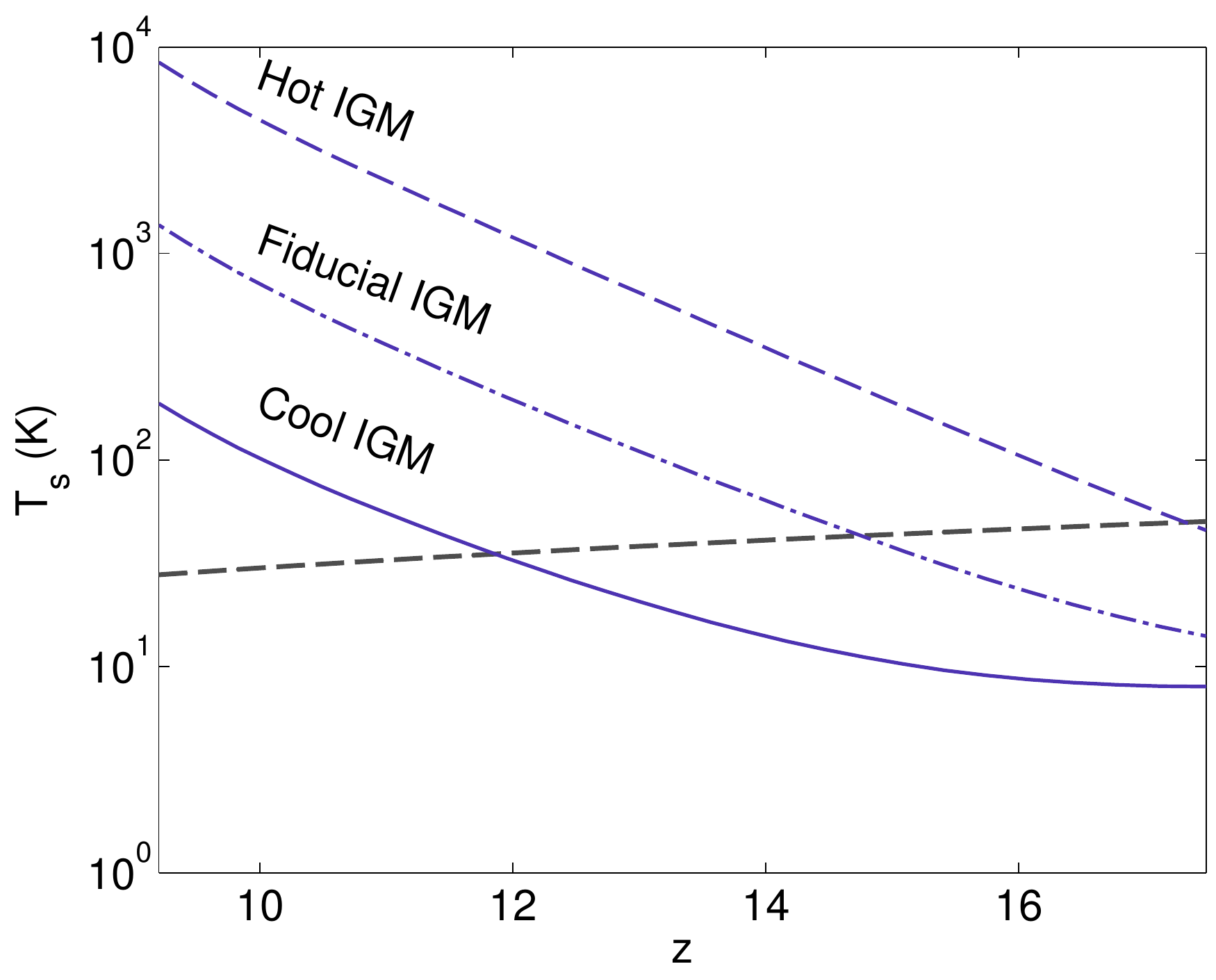}
\caption[Three scenarios for the thermal history of the IGM.]{The mean thermal evolution of our IGM simulations for our three models. ``cool IGM''- solid lines, ``fiducial IGM''- dashed-dotted lines, and ``hot IGM'' - dashed lines. $\langle T_s \rangle$ is plotted in lavendar. Varying $f_X$ effectively shifts $\langle T_s \rangle$ in redshift.} 
\label{fig:Thermal_Evolution}
\end{figure}

Summarized in Table \ref{tab:fxmodels} are our three values of $f_X$: a ``fiducial IGM'' model with $f_X=1$ corresponding to the fiducial value in \citet{Mesinger:2013vm}, a ``hot IGM'' model with $f_X = 5$, and a ``cool IGM'' model with $f_X = 0.2$. In Figure \ref{fig:Thermal_Evolution} We show the evolution of the mean spin and brightness temperatures from our simulations. Over the range of emissivities considered, the effect of varying $f_X$ is to shift the evolution of $\langle T_s \rangle$ in redshift.  Because $P_f$ varies as $\langle \tau_{21} \rangle^2 \sim \langle T_s \rangle^{-2}$ and $f_X$ simply shifts $\langle T_s \rangle$ in redshift, this relatively modest spread in $f_X$ is sufficient to understand a broader range of expected outcomes, as we shall see below.

\begin{table} 
\begin{center}
\begin{tabular}{|c|c|}   \hline
Number Name & $f_X$ \\ \hline
 Hot  IGM & $5.0$ \\ 
 Fiducial IGM & $1.0$ \\
 Cool IGM & $0.2$ \\ \hline
\end{tabular}
\caption{IGM Heating Parameters.}
\label{tab:fxmodels}
\end{center}
\end{table}

\subsection{The Model of the Radio Loud Source Distribution} 
We now review present constraints on the RL source distribution and describe the semi-empirical radio luminosity function that we use to simulate the global 21 cm forest. To gain perspective of how our choice of population model might compare to other theoretical work we determine which flux ranges are relevant to the sum in Equation (\ref{eq:pf_simple}) and compare the counts of sources in W08 to those in H04.  We also describe our method for combining the simulated radio sources with our simulations of the IGM. 

\subsubsection[Review of Constraints and Predictions of High\\ Redshift Radio Counts]{Review of Constraints and Predictions of High Redshift Radio Counts}

Constraints on the luminosity function of the most luminous radio loud sources are presently limited to $z \sim 4$ \citep{DeZotti:2010} Confirmed in these works, is that the comoving density of ultra steep spectrum sources peaks at $z \sim 2$ with little evidence for evolution out to $z \gtrsim 4.5$.

 To model the abundance of RL quasars with $6 \lesssim z \lesssim 20$ one must rely on theoretical extrapolations.  \citet{HaimanQuasarCounts} give estimates of source counts by assigning black hole masses to a halo mass function using the black hole mass-velocity dispersion relation of \citet{Wyithe:2003ev}. The RL fraction is derived assuming Eddington accretion, and the RL-i band luminosity correlation observed by \citet{Ivezic:2002}. 

 More sophisticated attempts at predicting the bolometric luminosities of high redshift quasars up to $z=11$ have been undertaken using hydrodynamic simulations with self consistent models for black hole growth and feedback \citep{DeGraf:2012ta}. Even with a more nuanced treatment of the luminosity distribution, the RL fraction at high redshift still remains a wide open question.  Indeed, the purpose of this work is to propose a technique for determining this population by showing that an empirically motivated RL population can have significant and observable features in the power spectrum for a range of thermal scenarios. 

\subsubsection{Our Choice of Population Model}

 We choose to work with the RL AGN population described in W08 in which sources are generated by sampling extrapolated radio luminosity functions biased to structure from a CDM simulation. Specifically, the radio luminosity function used is that ``Model C.'' from \citet{Willott:2001} which describes the high and low luminosity populations of AGN as Schechter functions. The redshift evolution of the low luminosity population is modeled as a power law in redshift while the high luminosity component as a gaussian with a mean of $z\approx 1.9$. Lists of source positions, fluxes, and morphologies from the Wilman simulation are downloadable through a web interface\footnote{\url{http://s-cubed.physics.ox.ac.uk/s3_sex}}. 

\begin{figure}
\centering
\includegraphics[width=.6 \textwidth]{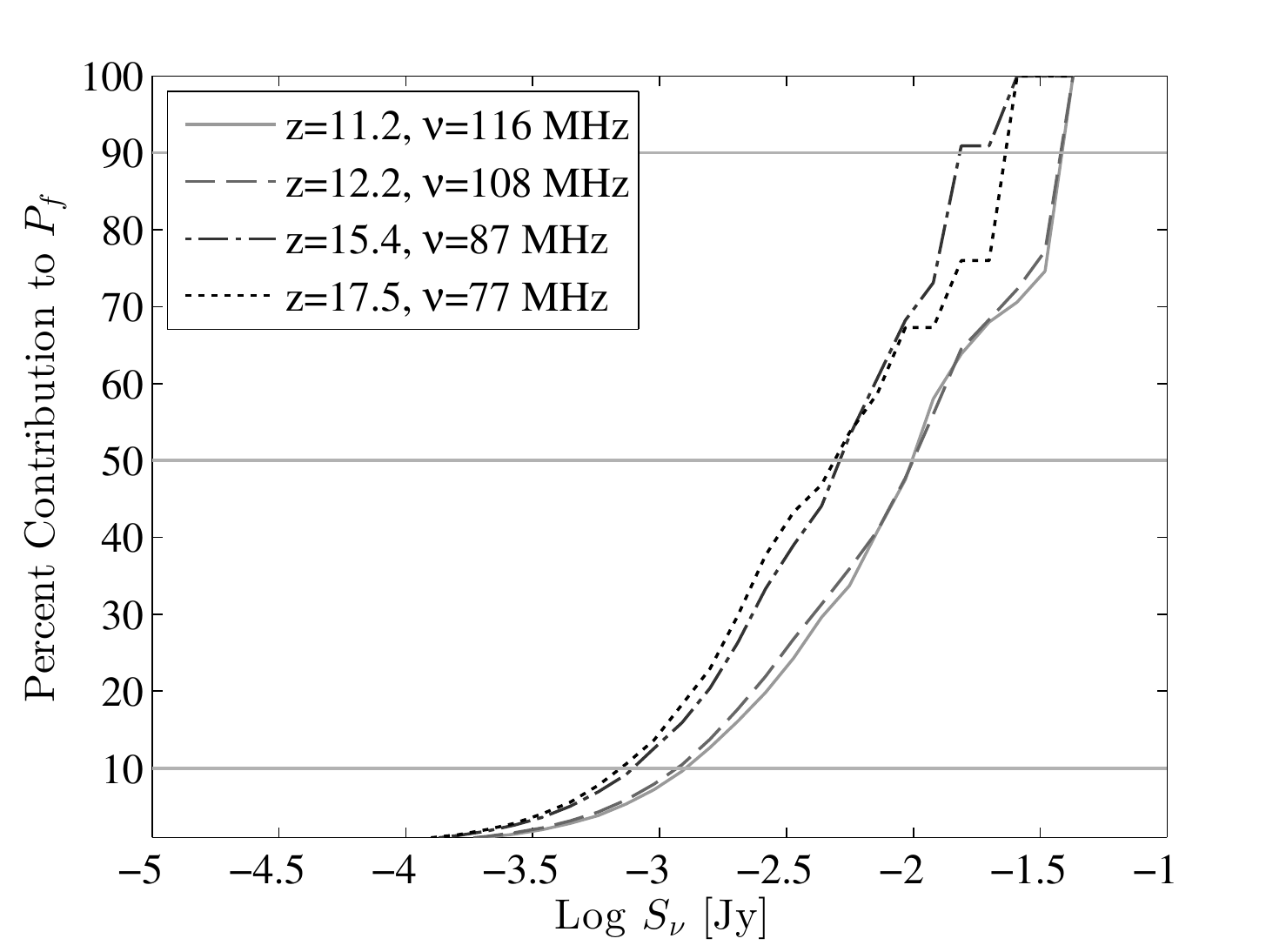}
\caption[Contribution of various fluxes of sources to the 21\,cm forest.]{The 21 cm forest is dominated by sources in the 1-10 mJy flux range. We plot the sum of fluxes squared, in Equation (\ref{eq:integral}), for $S < S_\nu$.  A detection of $P_f$ would constrain the high redshift source counts at these flux intervals.} 
\label{fig:flux_integral}
\end{figure}

Having chosen our population model, we can employ our formalism from Section \ref{sec:theory} to understand which sub population of the luminosity function contributes most to $P_f$.  In Figure \ref{fig:flux_integral}, we plot the percent contribution of sources below a threshold, $S_\nu$, to $P_f$ from the flux squared sum in Equation (\ref{eq:pf_simple}). One can see that roughly 75\% of the contribution to $P_f$ comes from sources with fluxes between $1-10\mJy$ at $80-115 \MHz$. At lower redshifts, the integral curves are increasingly dominated by higher fluxes as the sources with the greatest fluxes increase in number. The detection or lack of detection of the features we find using this simulation would either confirm or reject the W08 model for sources with $S_\nu$ between $1$ and $10 \mJy$.
While this paper is a study of observability for one model, in future work we will determine what range of  RL population this technique can constrain.

It is worth getting an order of magnitude idea of how our choice of the W08 semi-empirical model might compare to other theoretical predictions of the radio luminosity function. In Appendix \ref{app:compare} we compare the source counts in our semi-empirical prediction to the more physically motivated bottom up model in H04. The counts of W08 sources contributing to the bulk of $P_f$  tend to be more numerous than those in H04 by a factor of $\approx 10$ at $z \sim 12$ to $\approx 80$ for $z \sim 15-20$, underscoring the need for a full parameter space study. Even though such a study is beyond the scope of this paper, our extrapolated results in Figure \ref{fig:sn_params} show that the range of populations that the power spectrum can constrain depends heavily on the IGM's thermal history.

\subsection{Adding Sources to the Simulation}

  We simulate the theoretical power spectra accessible to upcoming observations by drawing 36 random sub-fields from the W08 simulations and combining them with 36 random 8MHz slices from our IGM simulations. The number of subfields is chosen to roughly correspond to the $\sim (30^{\circ})^2$ FoV of the MWA.

While our analytic approach in Section \ref{sec:theory} does not account for sources within the imaged volume, we incorporate them into our simulation by determining the location of DM halos down to masses of $5 \times 10^9 M_\odot$ through the excursion-set + perturbation theory approach outlined in \citet{Mesinger:2007hl}. We then populate these dark matter halos with RL sources, monotonically assigning the most luminous sources at 151 MHz\footnote{We order sources by their luminosity  at {\it observed} frequency of 151 MHz at regardless of their redshift which varies very little over the span of an $8 \MHz$ data cube so that we are approximately comparing their rest frame luminosities.} to the most massive halos. 
 Sources falling behind the cubes retain their original positions. All W08 sources are unresolved in our IGM simulation; hence, for each pixel the fluxes for all sources behind that pixel are summed together to give $S_{pix}$. This flux cube is converted to temperature using the Rayleigh-Jeans equation,
\begin{equation}  
T_{pix} = \frac{ \lambda^2 S_{pix} }{2 k_B \Omega_{pix}},
\end{equation}
where $\Omega_{pix}$ is the solid angle subtended be each simulation pixel\footnote{We show in Appendix \ref{app:compare_pf}, that the choice of pixel solid angle does not effect $P_f$}. Finally we introduce quasar absorption by multiplying this source cube by our $\tau_{21}$ cube $T_f \approx T_{pix} \tau_{21}$.

\section{Simulation Results} \label{sec:results}
In this section we present the results of our combined IGM-RL population model by computing the spherical and cylindrical power spectrum, $P(k)$, averaged over our 36 sub-cubes. We identify the regions of $k$-space in which the forest is dominant and might be used to constrain the high redshift radio luminosity function and discuss the morphology of the observed power spectra, verifying the essential results of Section \ref{sec:theory}.

\subsection{Computing Power Spectra}
Power spectra are computed using a direct Fast Fourier transform of each data cube multiplied by a kaiser window along the LoS with attenuation parameter $\beta = 3.5$. In averaging over bins of our spherical power spectra, we exclude the ``wedge'', the region of $k$-space heavily contaminated by foregrounds given by \citep{VedanthamWedge,MoralesPSShapes}
\begin{equation} \label{eq:wedge}
\kpara \le  \sin \frac{\Theta}{2} \left( \frac{D_M(z)}{D_H} \frac{E(z)}{(1+z)} \right) \kperp
\end{equation}
where $z$ is the redshift of a data cube's center frequency,  $D_M(z)$ is the comoving distance, $E(z)=H(z)/H_0$, and $\Theta$ is the FWHM of the primary beam which we calculate using a short dipole model of the MWA antenna element.  Table \ref{tab:mwa_params} gives the FWHM value of our primary beam model for several different frequencies. 

\begin{figure*}
\includegraphics[width=\textwidth]{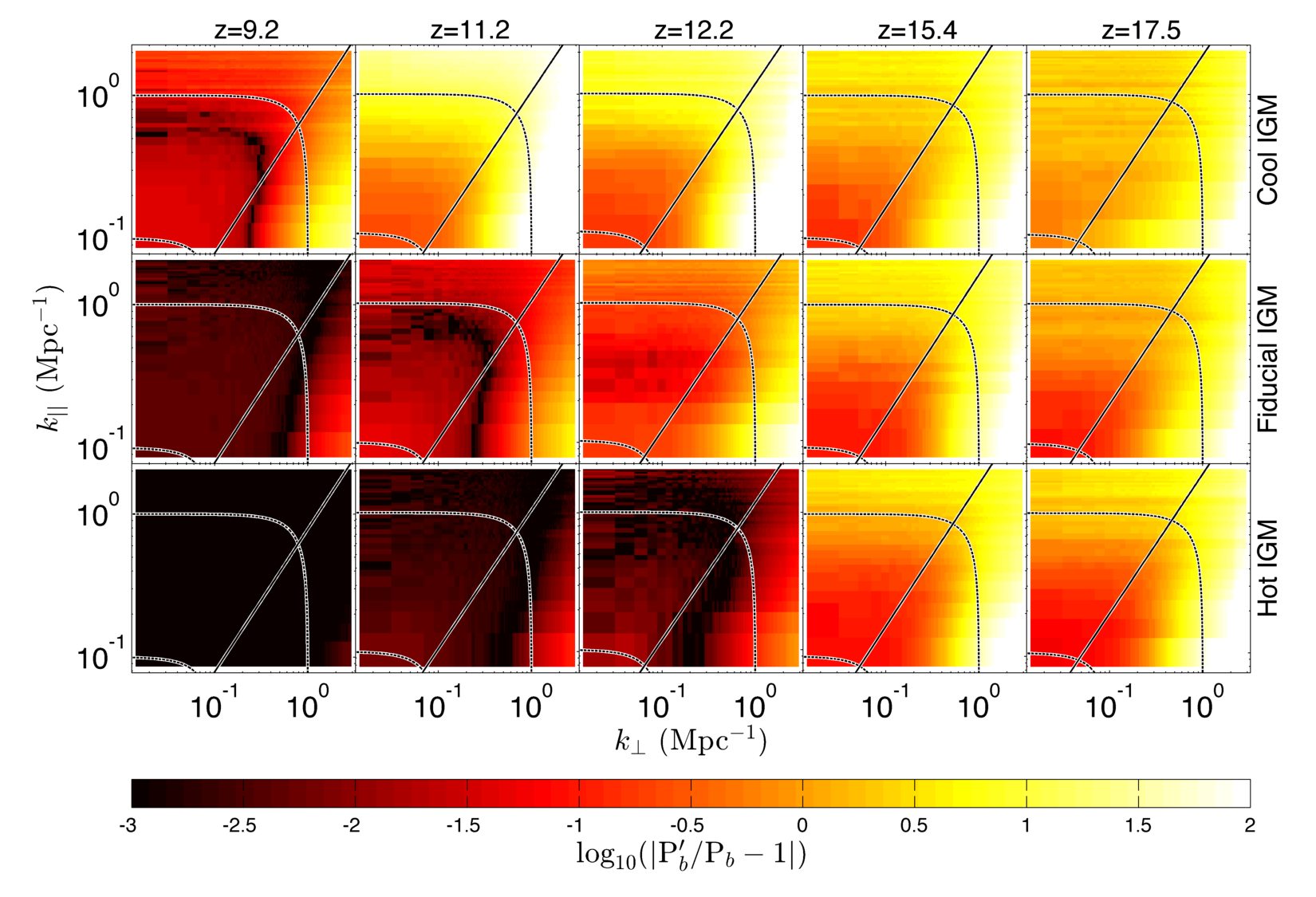}
\centering
\caption[Difference of power spectra with and without the 21\,cm Forest.]{For every heating scenario we study, there is some redshift and region within the EoR window for which 21 cm forest dominates the power spectrum. Here we show the fractional difference between the power spectrum with, ($P_b'$), and without ($P_b$) the forest for the redshifts (top to bottom) 9.2, 11.2, 12.2, 15.4, and 17.5. The diagonal lines denote the location of the ``wedge''.  By $z \gtrsim 12.2$ there is a substantial region  ($\kpara \gtrsim 0.5 \Mpci$) of  the Fourier volume that our simulations cover in which the forest dominates $P_b$ by a factor of a few.}\label{fig:forest_compare} 
\end{figure*}

\subsection{Simulation Output and the Location of the Forest in $k$-space}

We now discuss the power spectra output by our simulations and  the significant features produced by the forest. 

To isolate the the effect of the forest and to compare its significance to the brightness temperature power spectrum, $P_b$, we plot the fractional difference between $P_b'$, the power spectrum with the forest ,and $P_b$ in Figure \ref{fig:forest_compare}. We see that the forest introduces a significant feature, especially at the smallest scales. This feature is most prominent at high redshifts and less emissive heating models, when the IGM is cool. For our cool model, the forest feature dominates $P_b$ by over a factor of 100 for a wide range of redshifts. In the fiducial model, the dominant region is primarily at larger values of $\kpara$, though dominance by a factor of a few is visible at $z=12.2$ and $z=17.5$. In our hot model, a significant feature is visible only for $z \gtrsim 12.2$.

 For all heating scenarios, there are redshifts $z \gtrsim 12.2$ in which the same region of Fourier space contains a strong forest signal that dominates $P_b$ by a factor of at least a few.  Fortunately for those interested in the brightness temperature signal, the region $k \lesssim 10^{-1} \Mpci$ remains dominated by $P_b$. Hence at pre-reionization redshifts,  $k \lesssim 10^{-1} \Mpci$ can still be used to constrain cosmology and the thermal history of the IGM. With the thermal properties of the IGM determined, one may constrain  the high redshift RL population using the forest power spectrum signal at $k \gtrsim 0.5 \Mpci$.

\begin{figure*}
\centering
\includegraphics[width=\textwidth]{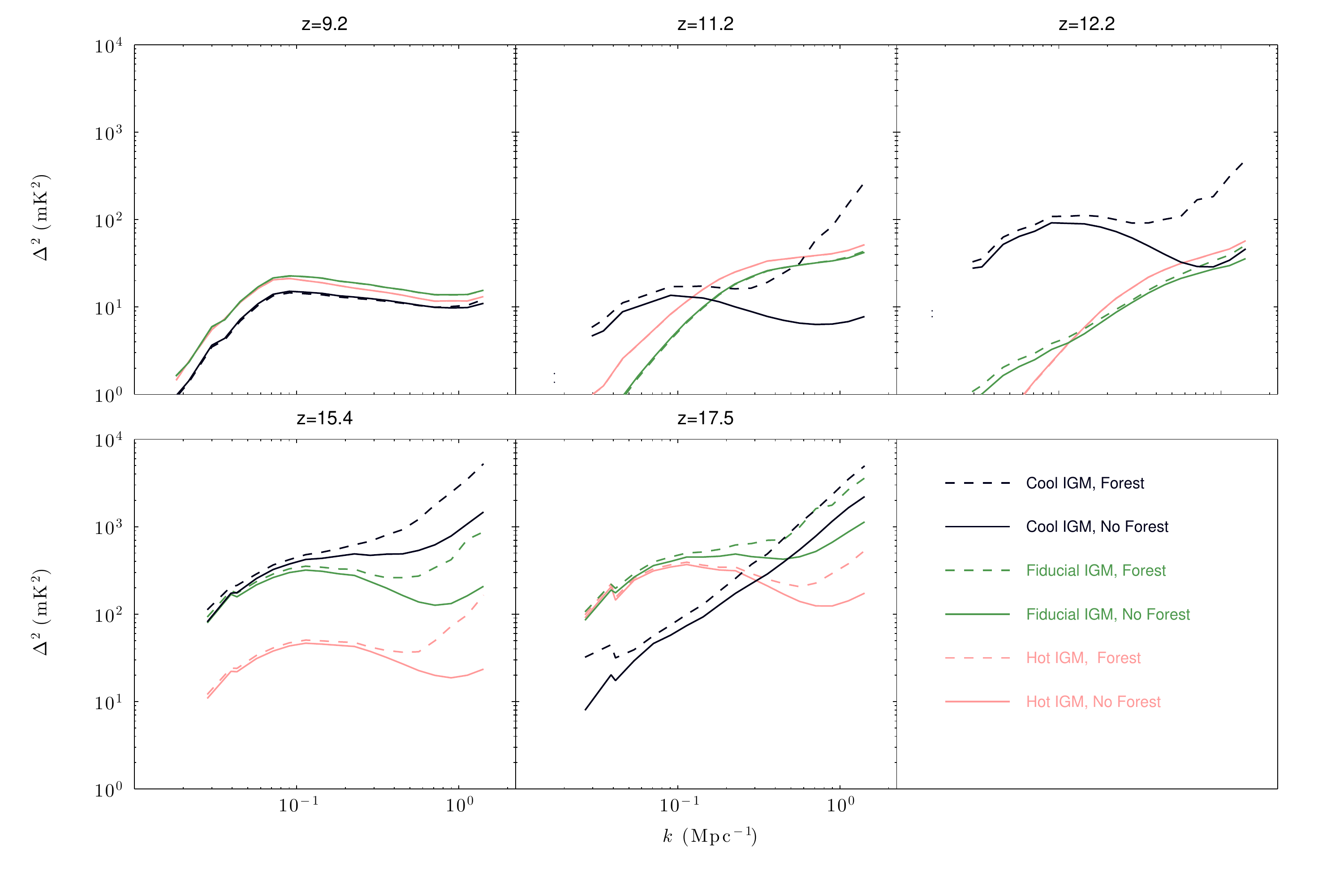}
\caption[1D power spectra with and without the 21\,cm forest.]{The 21 cm forest dominates the spherically averaged power spectrum for $k \gtrsim 0.5 \Mpci$. Plotted is the spherically averaged power spectrum with  (dashed lines) and without (solid lines) the presence of the 21 cm forest. In our cool model, the forest causes a significant power increase at $k \gtrsim 0.5 \Mpci$ at redshifts as low as $z = 11.2$.   At $z=15.4$ we see a significant feature in all thermal scenarios. Our cool IGM model experiences a reduction in the power spectrum amplitude at $z \gtrsim 17.5$ as it passes through the X-ray heating peak.  }
\label{fig:ps}
\end{figure*}

The first generation of interferometers will not be sensitive enough to measure the cylindrical power spectrum with high S/N but will rather measure the spherically averaged power spectrum. We compute spherically averaged power spectra from data cubes with and without the presence of forest absorption and excluding the wedge. We plot these power spectra in Figure \ref{fig:ps}. In all of the heating scenarios considered, the forest introduces significant power at $k \gtrsim 0.5\Mpci$ for $z \gtrsim 15.4$. Hence, it is in principle possible to constrain the distribution of RL AGN at high redshift for a range of heating scenarios.

We note that the high-$k$ region extends into our simulations' Nyquist frequency of $2.1 \Mpci$. We ensure that the  forest dominance is not an aliasing effect by running simulations on a 125 Mpc cube with six times higher resolution.  The results in the the overlapping $k$-space regions agree well with these larger volume, lower resolution simulations.

\begin{figure*}
\centering
\includegraphics[width=\textwidth]{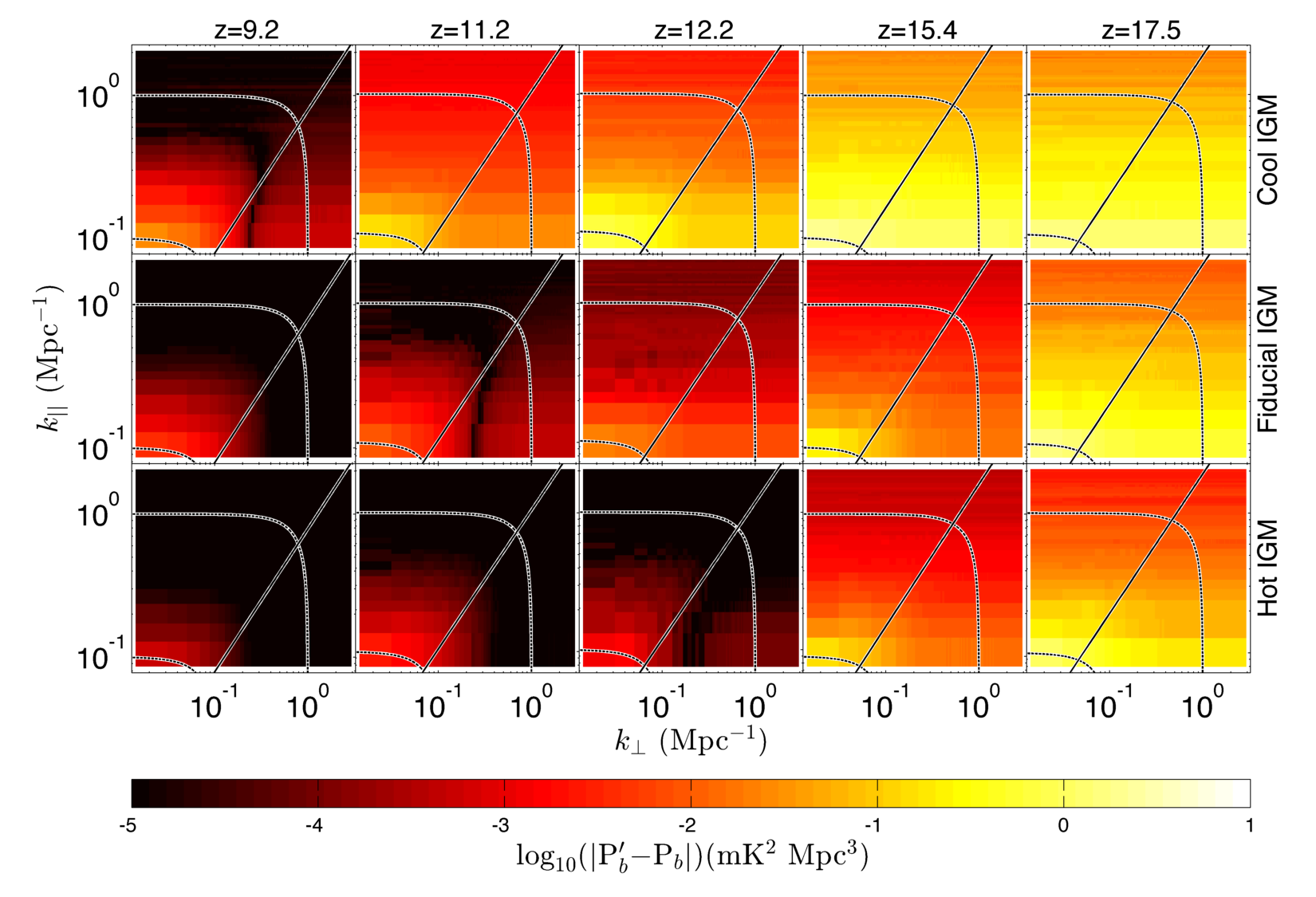}
\caption[Effect of cross-terms in 21\,cm forest power spectrum.]{We plot the magnitude of the difference between the 21 cm  power spectrum with and without the presence of the 21 cm forest including the auto-power and cross power terms of Equation (\ref{eq:ps_with_RL}). At high redshifts and low $f_X$, there is little $\kperp$ structure in $P_b'-P_b$, indicating that $P_f$ is the significant contributer. At lower redshifts and higher $f_X$, we see signficant $\kperp$ structure, indicating that in a heated IGM, $P_b' - P_b$ is dominated by $P_{f,b}$ which is somewhat spherically symmetric and negative at large $k$. The trough in the low redshift plots marks the region where $P_f-2\text{Re}(P_{f,b})$ transitions from negative (for small k) to positive (for large k). }
\label{fig:ps_modified}
\end{figure*}

\subsection{The Morphology of the Simulation results.}
We now explain the morphology of our simulation results and verify our analytic predictions in Section \ref{sec:theory}. 

We noted in Figure \ref{fig:forest_compare} that the 21 cm forest dominates the power spectrum both at large $\kperp$ and $\kpara$. The former observation is consistent with a forest power spectrum that is uniform in $\kperp$. In Figure \ref{fig:ps_modified} we show $|P_b'-P_b|$ and see that at high redshift and cool heating models, the forest power spectrum is mostly uniform in $\perp$ though at lower redshifts and hotter IGM, there is significant $\kperp$ structure. Since in section 
\ref{sec:theory} we showed that $P_f$ only varies along $\kpara$, this suggests that the cross power spectrum, $P_{f,b}$ is the prime contributor to $P_b' - P_b$ in a hot IGM, while $P_f$ is in a cool one. The trough at lower redshifts, at $k \sim 0.5 \Mpci$ is caused by the fact that $-2P_{f,b}$ is negative as we shall see below.

A potentially interesting consequence of the auto-terms invariance in $\kperp$ is a potential for contaminating the separation of powers analysis advocated in \citet{Barkana2}. We may Taylor expand $P_f$  

\begin{equation}
P_f(\kpara) = P_f(k \mu) = \sum_{n=1}^{\infty} \frac{1}{n!}\frac{\partial P_f}{\partial (\mu k)}_{\mu k = 0} ( \mu k)^n,
\end{equation}

so $P_f$ introduces signal over a wide range of powers of $\mu$ and has the potential to contaminate the cosmological $\mu^4$ and $\mu^6$ components of the brightness temperature power spectrum. On the other hand, the small $k$, where the perturbative expansion is most accurate, is dominated by the diffuse brightness temperature emission. In all but the coolest heating models, contamination will likely be small, since we can see in Figure \ref{fig:forest_compare} that $P_f \lesssim 0.1 P_b$ at $k \lesssim 0.1 \Mpci$. 

Decomposing the forest signal into powers of $\mu$ may be another way of distinguishing it from the brightness temperature. Even within the ``IGM dominated'' region. Detailed analysis on contamination of the cosmological signal and additional distinguishability offered by the angular dependence is beyond the scope of this paper will be the subject of future work. 

To be more quantitative, we turn our attention to right hand side of Equation (\ref{eq:pf_simple}) and verify our decomposition of the forest power spectrum into $P_{\tau_{21}}^{LoS}$ and the sum of background source fluxes. To do this, we find the summed squares of the fluxes (at the center frequency of the observation) of all sources falling in or behind our data cubes at several redshifts, multiply by the 1D LoS power spectrum of $\tau_{21}$ and compare with $\Delta^2_f$ computed from our simulation as outlined above. We find that Equation (\ref{eq:pf_simple}) consistently underpredicts the simulation amplitude by a factor of 2. However, when we remove the clustering of sources by randomly assigning source positions (rather than using the dark matter biased positions), Equation (\ref{eq:pf_simple}) agrees with simulation output within $5-20\%$ over the studied redshifts. Hence we rewrite Equation (\ref{eq:integral}) as 
\begin{equation}\label{eq:integral_fix}
 P_f \approx A_{cl} \frac{c D_M^2 \lambda^4}{4 k_B^2} P_{\tau_{21}}^{LoS}  \int_z^\infty \int_0^\infty s^2 \rho(z,z',s)  \frac{D^2_M(z')}{H(z')}  dz'ds 
\end{equation}
Where $A_{cl}$ is a constant of order unity that accounts for the boost in power due to clustering. We briefly explain this power boost in Appendix \ref{app:compare_pf}. In Figure $\ref{fig:mod_compare}$ we show the power spectrum, $\Delta^2_f$ computed from our simulation and the prediction from Equation (\ref{eq:pf_simple}) for several redshifts in our fiducial heating model. For $k \gtrsim 10^{-1} \Mpci$, Equation (\ref{eq:pf_simple}) agrees with our simulation at the 10\% level, indicating that we can ignore the cross terms in Equation (\ref{eq:sum_forest_sources}) and consider the forest power spectrum as the simple product of the 1D $\tau_{21}$ power spectrum and the integrated radio luminosity function. 

\begin{figure}
\centering
\includegraphics[width= \textwidth]{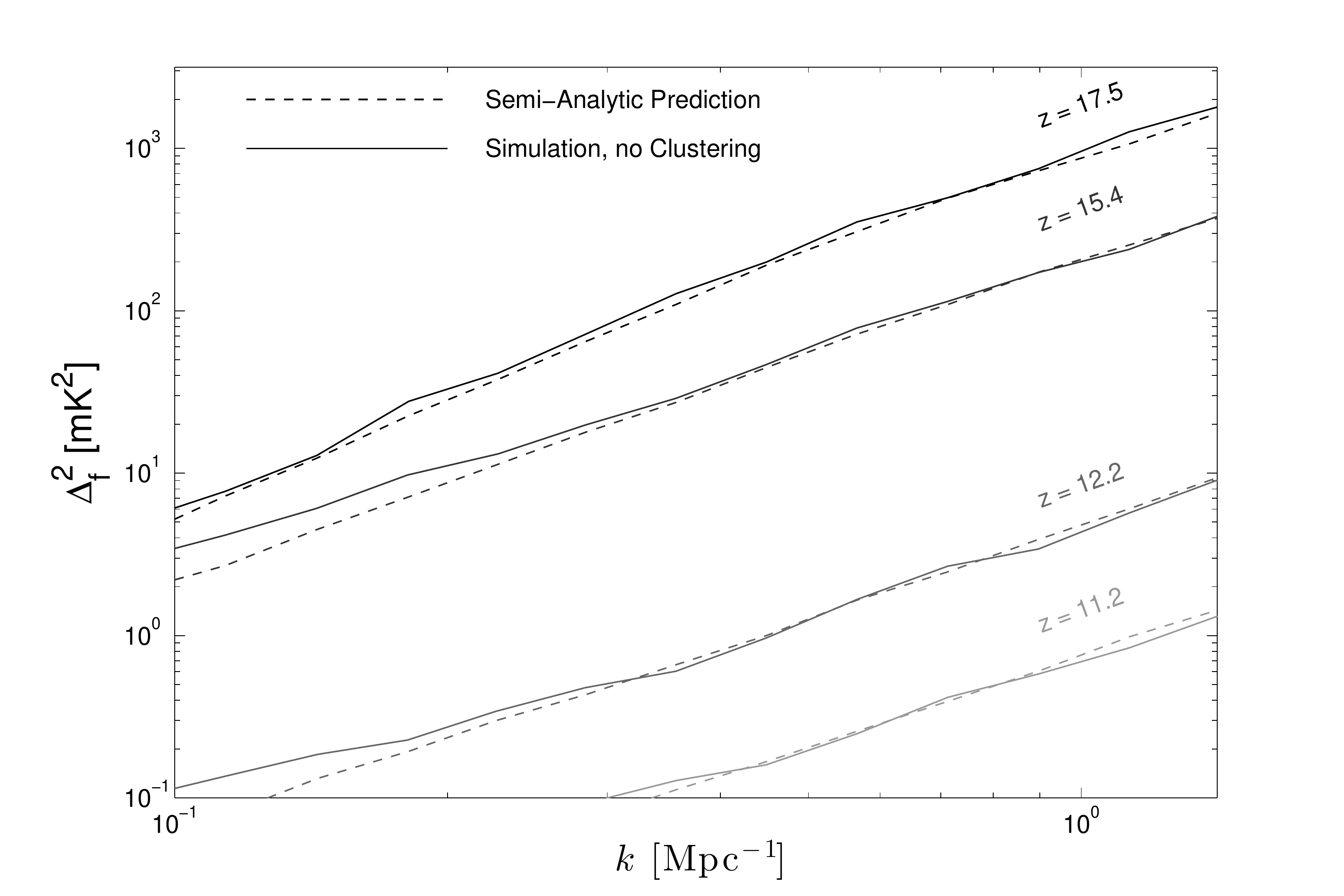}
\caption[Semi-analytic prediction results.]{Our semi-analytic prediction agrees well with unclustered simulation results. The semi-analytic prediction of Equation (\ref{eq:pf_simple}) is plotted with dashed lines and $\Delta^2_f(k)$ computed directly from our simulation without clustering in solid lines. This demonstrates that for $k\gtrsim 10^{-1} \Mpci$, the cross terms in Equation (\ref{eq:sum_forest_sources}) may be ignored and $P_f$ may be well approximated by the LoS power spectrum of $\tau_{21}$ multiplied by the summed squared fluxes for sources lying in and behind the data cube.}
\label{fig:mod_compare}
\end{figure}

\begin{figure}
\centering
\includegraphics[width=\textwidth]{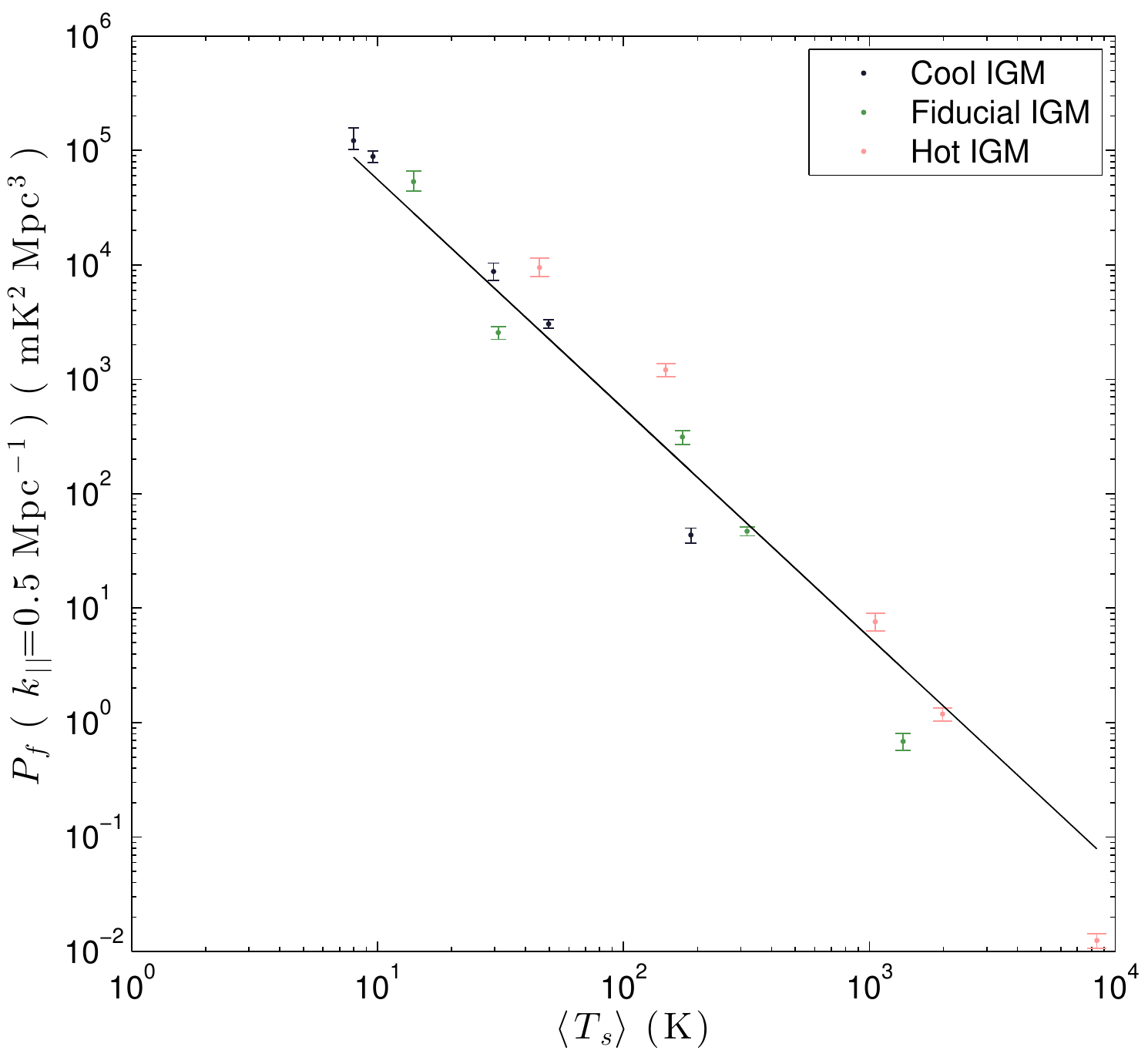}
\caption[Effect of spin temperature on the power spectrum.]{ We see that for a fixed quasar distribution, the magnitude of $P_f$ can be parameterized by $\left \langle T_s \right \rangle$ and that the amplitude is consistant with a simple power law. Here, we plot $P_f(k_\parallel)$ at $k_\parallel=0.5 \Mpci$ vs. $\langle T_s \rangle$ for all considered redshifts and $f_X$. The black line is the power law $\langle T_s \rangle^{-2}$ as one might expect for an amplitude set by  $\langle \tau_{21} \rangle^2$ (Equation (\ref{eq:integral})). Inasmuch of this simple trend, a modest spread in heating models gives us a decent understanding of the behavior of the amplitude for $P_f$. This relation holds for the quasar population considered here because the integral over the luminosity function does not change significantly over the redshifts we consider.}
\label{fig:ps_modifiedVsTs}
\end{figure}

A striking feature of Figure \ref{fig:ps_modified} is the apparent similarity of $P_f$ along diagonal sets of different redshifts and models. For example, the ``Cool IGM'' model at $z=12.2$ is very similar to the ``Fiducial IGM'' result at $z=15.4$ and the ``Hot IGM'' at $z=17.5$. It is suggestive that one can obtain the results of one particular thermal model by simply shifting another model in redshift, this translational invariance in redshift demonstrates that we may not need to simulate a broad range of heating models to understand the evolution of the forest power spectrum. Indeed, given our decomposition in Equation (\ref{eq:pf_simple}) where the amplitude of $P_f$ is proportional to $\langle \tau_{21} \rangle^2 \propto \langle T_s^{-1} \rangle^2$, we should expect $\langle T_s \rangle$ to be a more generally applicable parameterization than $f_X$ and redshift during the pre-reionization epoch.
 To show the importance of $\langle T_s \rangle$ as a parameter, we plot, in Figure \ref{fig:ps_modifiedVsTs}, the amplitude of $P_f$ at $k_\parallel = 0.5~\Mpci$ as a function of $\langle T_s \rangle$ for our three heating scenarios and redshifts. Across all thermal models and redshifts, the amplitude of $P_f$ is well described by a power law of $\langle T_s \rangle^{-2}$, consistent with the normalization predicted in Equation (\ref{eq:pf_simple}).

\begin{figure*}
\centering
\includegraphics[width=  \textwidth]{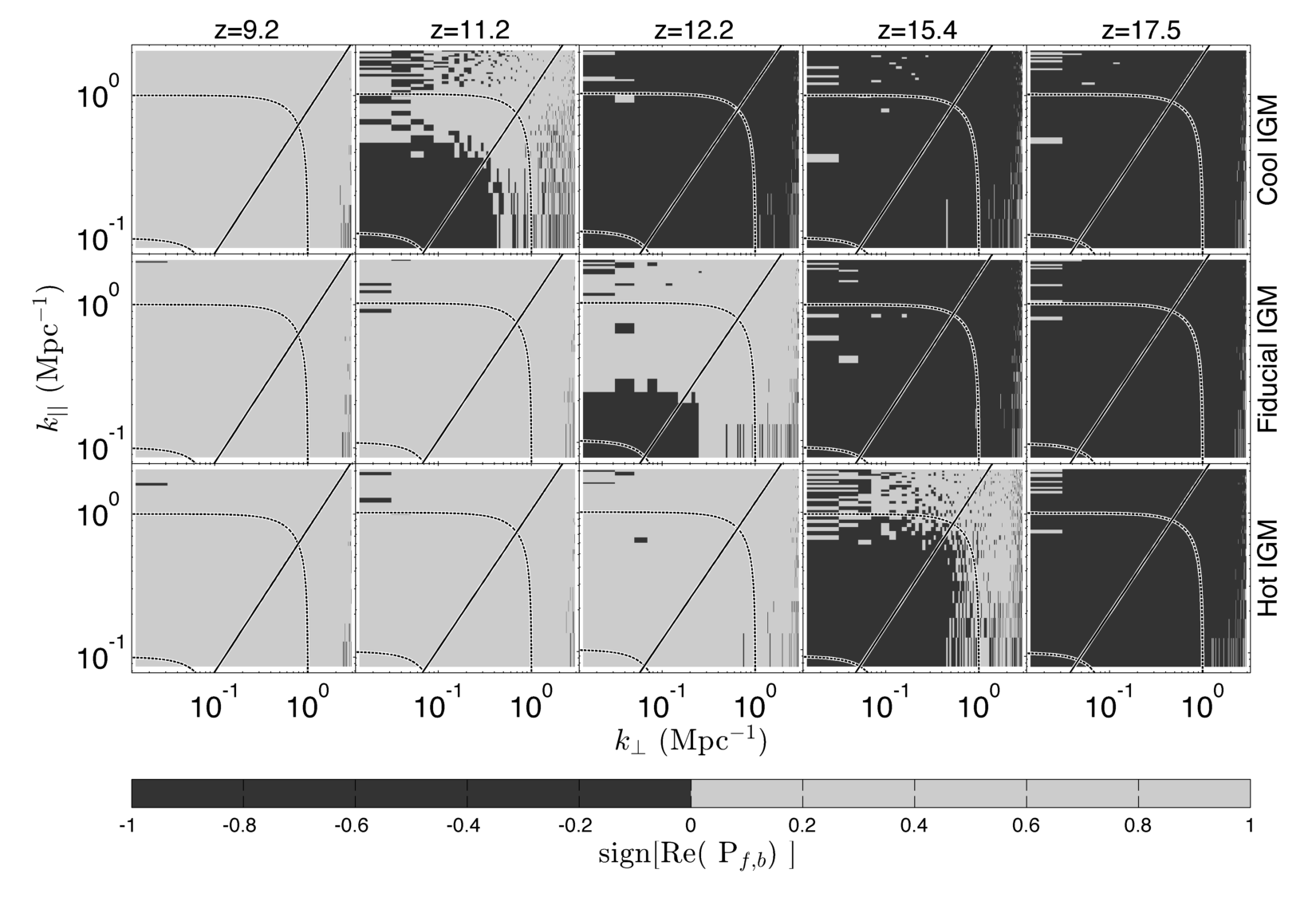}
\caption[Sign of the forest/brightness temperature cross power spectrum.]{The cross power spectrum, $\text{Re}(P_{f,b})$'s, sign is determined by the anti-correlation of $x_{HI}$ and $T_s$ during the pre-heating epoch and by $x_{HI}$ after heating has taken place. Here we show the sign of $\text{Re}(P_{f,b})$ for our three different heating models as a function of redshift. At pre-heating redshifts, $T_s$ is small and $x_{HI}$ is relatively uniform so that $\dtb$ and $T_f$ primarily depend on $T_s$ and anti-correlate so that $\text{Re}(P_{f,b})$ is negative. At low redshifts, $\dtb$ is independent of $T_s$ and fluctuations are primarily sourced by $x_{HI}$ so that $\dtb$ and $T_f$ are correlated and $\text{Re}(P_{f,b})$ is positive. Futhermore, heating proceeds in an ``inside-out'' manner so that the smallest scales become correlated first.}
\label{fig:ps_modified_sign}
\end{figure*}

Verifying our prediction on the sign of $P_{f,b}$ is our next task; we plot this quantity in Figure \ref{fig:ps_modified_sign} for all models and redshifts. At high redshift, $P_{f,b}$ is entirely negative due to the anti-correlation between $T_f$ and $\dtb$ and adds to the total amplitude of $P_b'$. As heating takes place, $T_s$ drops out of $\dtb$ and fluctuations in $\dtb$ are sourced predominantly by variations in $x_{HI}$ leading to positive correlation between $\dtb$ and $T_f$ for positive $P_{f,b}$.  As we see from the figures, this process is ``inside-out'', with large scales remaining anti-correlated longer than the small scales. Heating proceeds in an ``inside-out'' manner, and since there is an overlap between the completion of heating and onset of reionization, temperature fluctuations remain important on large scales \citep{PritchardFurlanettoXray,Mesinger:2013vm}.

\section{Prospects for Detection with an MWA-like Array} \label{sec:det}

We now turn to addressing the detectability of the power spectrum signature of the forest and its distinguishability from the power spectrum, $P_b$. Our strategy is to combine our simulations with random realizations of instrumental noise and galactic and extragalactic foregrounds.  With data cubes containing both our simulated signals and our random contaminants, we can then take advantage of the full quadratic estimator formalism developed by \citet{Maxpowerspeclossless}, adapted for 21 cm tomography by \citet{LT11}; hereafter LT11, and accelerated for large data sets by \citet{DillonFast}; hereafter D13.  In this section, we will explain those techniques and show what results when our simulations of the forest are added to realistic foregrounds and instrumental noise.

\subsection{Power Spectrum Estimation Methods}

To estimate the power spectrum of the forest, we apply the quadratic estimator formalism \citep{Maxpowerspeclossless}.  This formalism has the advantage that, in the approximation of foregrounds and noise that are completely described by their covariances, all cosmological information is preserved in going from three-dimensional data cubes to power spectra.  This formalism was adapted by LT11 for 21 cm power spectrum estimation and further refined and accelerated by D13.\footnote{For further details on this particular implementation of the quadratic estimator method, the reader is referred to D13.}  

In essence, the method relies an optimal and unbiased estimator of band powers in the $k_\perp$-$k_\|$ plane, $\widehat{\mathbf{p}}$, defined as
\be
\widehat{p}^\alpha = \sum_\beta M^{\alpha\beta} \left(\mathbf{x}^\trans\mathbf{C}^{-1}\mathbf{Q}^\beta\mathbf{C}^{-1}\mathbf{x} - b^\beta\right).
\ee
where $\mathbf{x}$ is a vector containing mean-subtracted data, $\mathbf{C}$ is the covariance of $\mathbf{x}$, including noise and contaminants, $\mathbf{Q}$ is a matrix that encodes the Fourier transforming, squaring, and binning necessary to calculate a band power, and $\mathbf{b}$ is the bias term.  The normalization matrix $\mathbf{M}$ is related to the Fisher information matrix $\mathbf{F}$.  Both $\mathbf{F}$ and $\mathbf{b}$ can be calculated via a Monte Carlo using the fact that
\be
b^\beta = \langle \mathbf{x}^\trans\mathbf{C}^{-1}\mathbf{Q}^\beta\mathbf{C}^{-1}\mathbf{x} \rangle  \equiv \langle \widehat{q}^\beta \rangle
\ee
and that
\be
\mathbf{F} = \text{Cov}(\widehat{\mathbf{q}}).
\ee

The ensemble average of each band power is related to the true band power $\mathbf{p}$ by a window function matrix, $\mathbf{W} = \mathbf{MF}$,
\be
\langle \widehat{\mathbf{p}} \rangle = \mathbf{Wp}.
\ee
The error on true band powers is also related to $\mathbf{M}$ and $\mathbf{F}$ through
\be \label{eq:cov}
\text{Cov}(\widehat{\mathbf{p}}) = \mathbf{MFM}^\trans.
\ee
Each quadratic estimator can thus be thought of as a weighted average of the true band powers with potentially correlated errors, both of which depend on one's choice of $\mathbf{M}$. Though any choice of $\mathbf{M}$ that makes $\mathbf{W}$ a properly normalized weighted average is reasonable, we adopt a form of $\mathbf{M}$ that makes the errors on $\widehat{\mathbf{p}}$ uncorrelated. \citet{X13}, argue that this choice of $\mathbf{M}$ dramatically reduces the contamination of the EoR window by residual foregrounds.  It also provides a set of band power estimates which can be considered both mutually exclusive and collectively exhaustive because they cover the whole $k_\perp$-$k_\|$ plane while not containing any overlapping information.

\subsection{Noise and Foreground Models} \label{sec:CovModels}
The method outlined above requires model means and covariances of the contaminants that contribute to $\mathbf{x}$, like noise and foregrounds.  Our model of the instrumental noise depends, first and foremost, on the design of the interferometer.  In this paper, we consider the MWA with 128 tiles whose locations are detailed in \citet{Beardsley:2012ws} as representative of the current generation of low frequency interferometers.  Additionally, we consider possible realizations of double and quadruple sized instruments (MWA-256T and MWA-512T, respectively), as representative of extensions to current generation interferometers or next generation, $A_{eff} \sim 0.1 \km^2$, arrays such as the Hydrogen Epoch of Reionization Array (HERA) \citep{Whitepaper5}.  As we will show, we generally do not need a square kilometer scale instrument to see the statistical effects of the forest.

\begin{figure}
\centering 
\includegraphics[width=\textwidth]{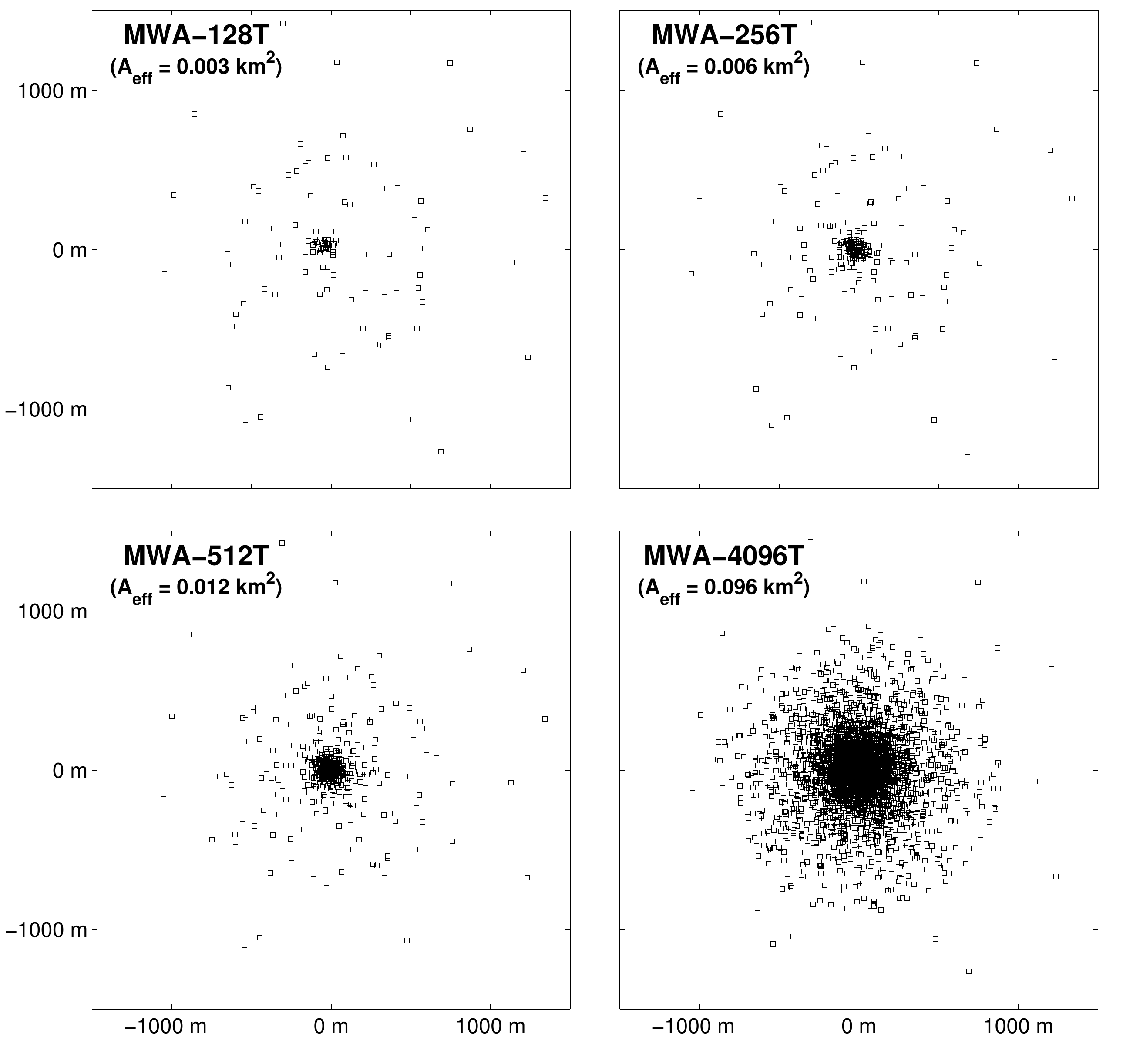}
\caption[Array layouts and sizes studied.]{Array layouts that we use to determine the detectibility and distinguishability of the 21 cm forest power spectrum signature. We chose to study two moderate extensions of the MWA-128T: MWA-256T and MWA-512T. In addition we study a 4096T array that is representative of a HERA scale instrument with $\sim 400$ times the collecting area of the MWA. Tile locations are drawn randomly from a distribution that is constant for the inner $50 \meter$ and drops as $r^{-2}$ for larger radii.}
\label{fig:ArrayLayouts}
\end{figure}

To generate our MWA-256T and MWA-512T designs with maximum sensitivity to 21 cm cosmology, we add antenna tiles to the current MWA-128T design within a dense core 900 m in radius.  These are drawn blindly from a probability distribution similar to that in \citet{juddjackiemiguel1}: uniform for $r<50$ m and decreasing as $r^{-2}$ above 50 m.  The tile locations of the arrays we use are shown in Figure \ref{fig:ArrayLayouts}.

Our model for the noise is adapted from D13 \footnote{The method of D13 is adapted with one correction: the form of the noise power spectrum adapted from \citep{FFTT} does not include the assumption that the field and beam sizes are the same.}.  In it, we incorporate observation times calculated in each $uv$-cell from 1000 hours of rotation synthesis at the latitude of the MWA.  The effective area of each tile is computed using a crossed dipole model while the system temperature is treated as the sum of receiver temperature, given by a power law fitted to two data points appearing in \citet{TingaySummary}, and sky temperature, measured in \citet{Rogers:2008vf}. In Table \ref{tab:mwa_params} we give our instrumental parameters at several different frequencies. 

\begin{table}
\centering
\begin{tabular}{|c|c|c|c|} \hline
f (MHz) &  FWHM (deg) & $A_{eff}$ (m$^2$) & $T_{sys}$ (K)\\ \hline
150 & 23 & 23& 290 \\ 
120 &30&24&  490\\ 
100	&34&24& 760\\ 
80  & 39&27&1300 \\ \hline
\end{tabular}
\caption{Instrumental Parameters}
\label{tab:mwa_params}
\end{table}

Similarly, our model of the foregrounds is the one application of the model developed by LT11 and D13.  For the sake of simplicity,\footnote{Breaking extragalactic foregrounds into a bright ``resolved'' population and a confusion-limited ``unresolved'' population only improves the error bars (D13), so our efficient choice is also a conservative one.} we model extragalactic foregrounds as a random field of point sources with fluxes up to 200 Jy.  They have an average spectral index of 0.5 and variance in their spectral indices of 0.5.  Their clustering has a correlation length scale of $7'$.  Likewise, we model Galactic synchrotron radiation as a random field with an amplitude of 335.4 K at 150 MHz, a coherence length scale of $30^\circ$, and a mean spectral index of 0.8 with an uncertainty in that index of 0.1.

As we have previously discussed, we conservatively cut out the region of $k_\perp$-$k_\|$ space that lies below the wedge.  Once the wedge has been excised, we optimally bin from 2D to 1D Fourier space with the inverse covariance weighted technique described by D13.

To create simulated observations, we divide our simulated volumes into 36 fields, each 750 Mpc on a side, which roughly fill the primary beam of our antenna tiles.  We add random noise and foregrounds to each field independently, taking advantage of the fast technique for foreground and noise simulations developed by D13.  Finally, we take the sample variance of the cosmic signal into account by  using our power spectrum results from Section \ref{sec:results} and by counting the number of independent modes probed by the instrument at each $k$ scale.

\subsection{Detectability Results}
We now present the results of our sensitivity calculation. We demonstrate that, given prior knowledge\footnote{Here, ``prior knowledge'' means that we know what the IGM power spectrum without the 21 cm forest  to within the error bars of our thermal senstivity.} of the X-ray heating history, a power spectrum measurement with a modest expansion of an MWA-like instrument is sufficient to distinguish between scenarios with or without the forest in our fiducial and cool heating models. Since the forest signal is detectable with smaller arrays only at smaller $k$, where $P_b$ dominates, its effect is likely degenerate with diffuse IGM emission. Observing this region for all considered models will require a HERA scale instrument with $A_{eff} \sim 0.1 \km^2$. 

\begin{figure*}
\centering 
\includegraphics[width=1\textwidth]{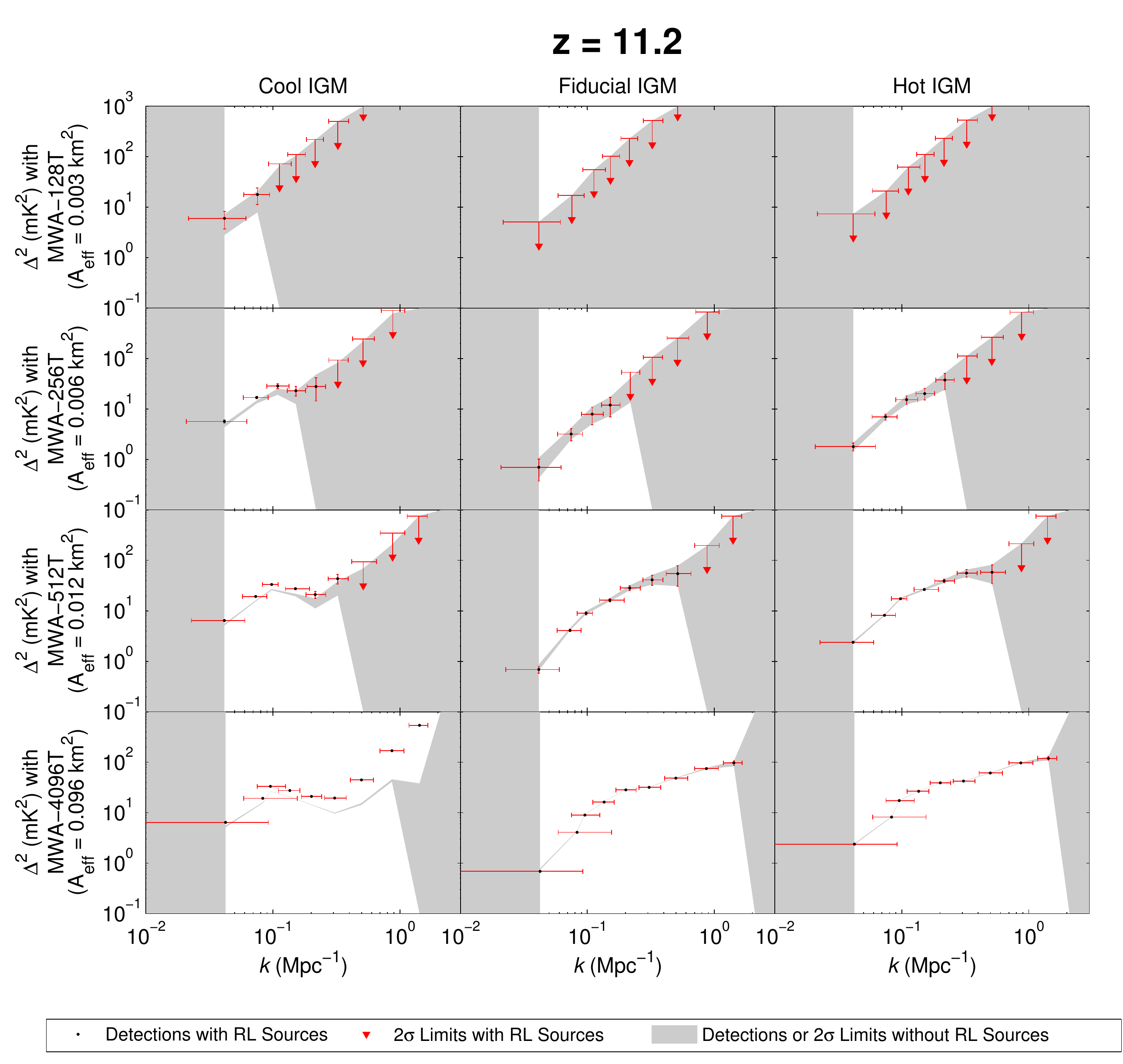}
\caption[Power spectrum sensitivity at $z=11.2$ for various arrays.]{Detections (black dots) and upper limits (red triangles) of the 21 cm Power Spectrum at z=11.2 for all of our arrays and heating models in the presence of 21 cm forest absorption from background RL sources. The grey fill denotes the 2 $\sigma$ region around the measured power spectrum with no RL sources present. To determine whether we can detect the forest imprint we ask, ``do the points and their error bars lie outside the gray shaded region?'' MWA-256T and MWA-512T would be capable of distinguishing power spectra with or without sources in our cool IGM model, however only 4096T is consistantly sensitive to the $k \gtrsim 0.5 \Mpci$ region where the forest dominates. Only for our cool IGM model, MWA-512T would sufficient to detect this upturn as well. Hence a moderate MWA extension would likely be able to constrain some RL populations given a cooler heating scenario while a HERA scale instrument will be able to constrain the W08 RL population using the Forest power spectrum even for more emissive heating scenarios. Note that the upturn in the gray region is not from increased power at high $k$ but larger error bars.}
\label{fig:PlotModelVsArray}
\end{figure*}

In order to determine the array size necessary to resolve the forest power spectrum, we first focus on $z=11.2$, the lowest redshift considered where there is significant signal for one of our thermal models and quasar counts are relatively high. In Figure \ref{fig:PlotModelVsArray} we shade the $2 \sigma$ region for a detection of $\Delta^2(k)$ with no 21 cm forest absorption present and mark detections of $\Delta^2(k)$ with 21 cm forest absorption with black dots. The $2 \sigma$ vertical error bars, given by the diagonal elements of Equation (\ref{eq:cov}) are marked in red. Also marked in red are the horizontal error bars which are given by the 20$^{th}$ and $80^{th}$ percentiles of the window functions. To determine whether we can detect the forest imprint, we ask ``are the points consistent with the gray shaded region?''

We see that MWA-256T and MWA-512T can distinguish cool models with and without the forest at greater than $2 \sigma$. However these detections are not within the region of Fourier space where the forest dominates $P_b'$.  As a result, though MWA expansions can resolve two models with or without the forest, it is unlikely that they will be able to distinguish a model with the forest from one with a slight variation in heating. If an independent measure of the global spin temperature can be obtained, the radio luminosity function might be constrained with a modest MWA extension. We note that MWA-4096T is only able to detect the forest in our cool model at $z=11.2$ since the optical depth in our more X-ray emissive models is far too small at this time. 

\begin{figure*}
\centering 
\includegraphics[width=\textwidth]{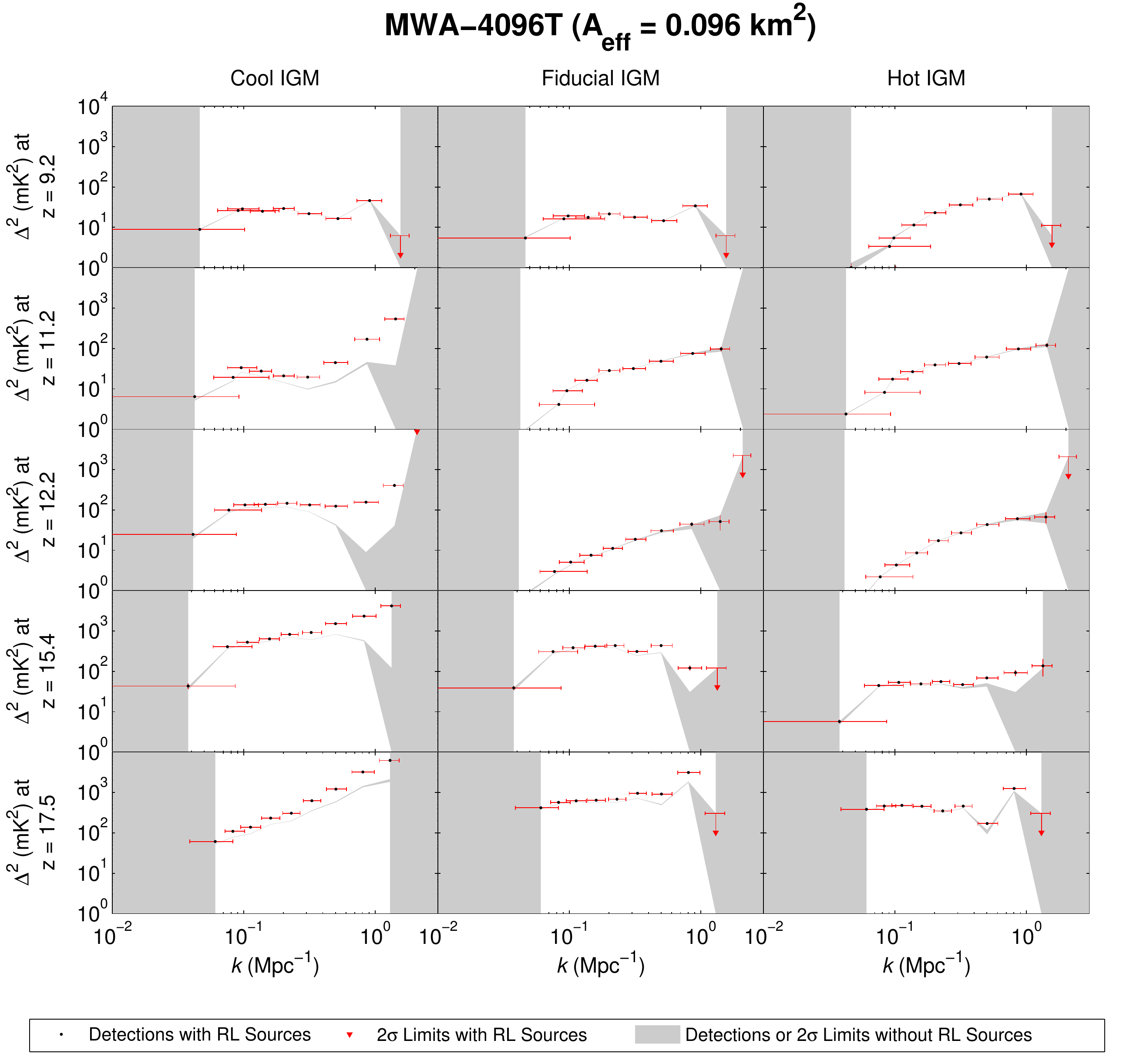}
\caption[Power spectrum sensitivity across redshifts for the largest array.]{These plots are identical to Figure \ref{fig:PlotModelVsArray} except the array is fixed to be MWA-4096T, representative of a HERA generation instrument, and redshift is varied. A HERA class instrument is able to resolve the upturn at $k \gtrsim 0.5 \Mpci$ that distinguishes the forest, and should be able to detect the 21 cm forest feature considered in this work for a variety of heating scenarios. The thermal noise error bars are to small to resolve by eye in most of these plots.}
\label{fig:PlotModelVsZ4096T}
\end{figure*}

To see more broadly what might be achieved by the next generation, we show in Figure \ref{fig:PlotModelVsZ4096T} the error bars and detections with and without the forest across all considered $f_X$ and $z$ for our HERA scale model. We find that z=15.4 is our ``sweet spot'' for the W08 distribution. 4096T is able to resolve the $k \gtrsim 0.5 \Mpci$ forest region for all of the IGM heating models that we investigate. For our cool and fiducial models, 4096T is also able to observe the forest region for a range of redshifts. These results show that a HERA scale array has the potential to constrain the IGM state by measuring $\Delta^2$ for $k \lesssim 10^{-1} \Mpci$, where the brightness temperature dominates, and the RL distribution in observing the region $k \gtrsim 0.5 \Mpci$ where the forest has a significant contribution.

Over the course of the IGM's evolution, there are times where the 21-cm power spectrum becomes particularly steep; for example, during the era immediately before the X-ray heating peak.  As a result, observing excess power at $k \gtrsim 0.5 \Mpci$ for a single redshift alone will likely not be sufficient to constrain the radio luminosity function. However, discerning the IGM thermal history with measurements of the power spectrum amplitude at $k \sim 0.1 \Mpci$ and observing an absence of flattening at high k, over the range of redshifts after the X-ray heating peak as shown in Figure \ref{fig:ps} should allow for constraints to be placed on the high-redshift radio luminosity function.

\subsection{Distinguishability Results}

In order to quantify how distinguishable our simulations with the forest are from our simulations without the forest for a given instrument, redshift, and heating model, we calculate the standard score of the $\chi^2$ sum of the power spectrum values across all k-bins,  
\begin{equation}\label{eq:zscore}
Z \equiv  \frac{\chi^2-N_k}{\sqrt{2N_k}},
\end{equation}
where  $N_k$ is the number of k bins, $\chi^2 \equiv \sum_k \left(\frac{P_b(k)'-P_b(k)}{\sigma_k} \right)^2$, and $\sigma_k$ are the diagonal elements of Equation (\ref{eq:cov}) for each model without the 21 cm forest present. Assuming statistical independence between $k$ bins, $Z$ is the number of standard deviations at which we can distinguish a model with the 21 cm forest from a model without it using the $\chi^2$ statistic. Unfortunately, this measure is somewhat naive since it does not account for potential degeneracies in the power spectrum amplitude from different thermal histories. However it enables us to quantitively compare outlooks across the numerous dimensions of redshift, array, and heating history. We consider a $Z \gtrsim 10$ to indicate significant distinguishability.  

\begin{figure*}
\centering 
\includegraphics[width=\textwidth]{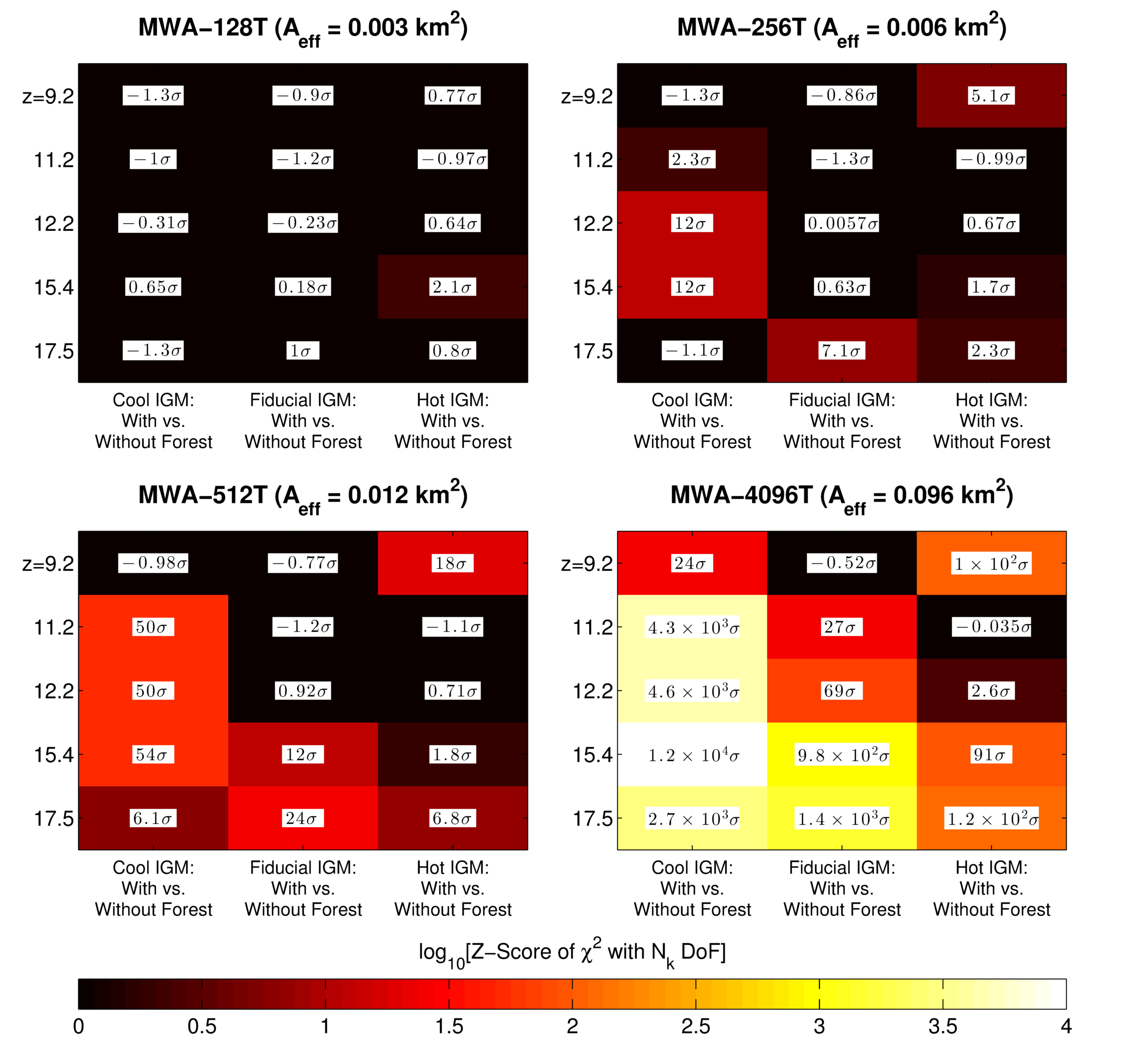}
\caption[Distinguishability of 21\,cm forest scenarios.]{The significance of distinguishability across all measured k bins (Equation (\ref{eq:zscore})) for all arrays, redshifts, and IGM heating models for a 1000 hour observation. An extension of MWA-128T is capable of distinguishing models with and without the 21 cm forest from the W08 RL population in our cool and fiducial heating scenarios. MWA-512T and HERA scale MWA-4096T are capable of distinguishing the forest in the power spectrum in all heating models considered in this work.}
\label{fig:PlotModelZScore}
\end{figure*}

In Figure \ref{fig:PlotModelZScore} we show  the value of Equation (\ref{eq:zscore}) for all models and arrays. Our first observation is that MWA-128T is not capable of distinguishing a model with the forest from a model without the forest for any of the considered $f_X$. MWA-256T would be capable of distinguishing the forest at all considered $z\gtrsim 9.2$ for our cool X-ray heating model at greater than $5 \sigma$ and in our fiducial heating model only at the highest considered redshift (which is near the X-ray heating peak). MWA-512T would be capable of resolving the forest at the two highest redshifts for our fiducial model and at all considered redshifts for our cool model. The hot model remains unobservable for all MWA expansion arrays but is accessible to a HERA scale instrument.

\begin{figure*}
\centering
\includegraphics[width=\textwidth]{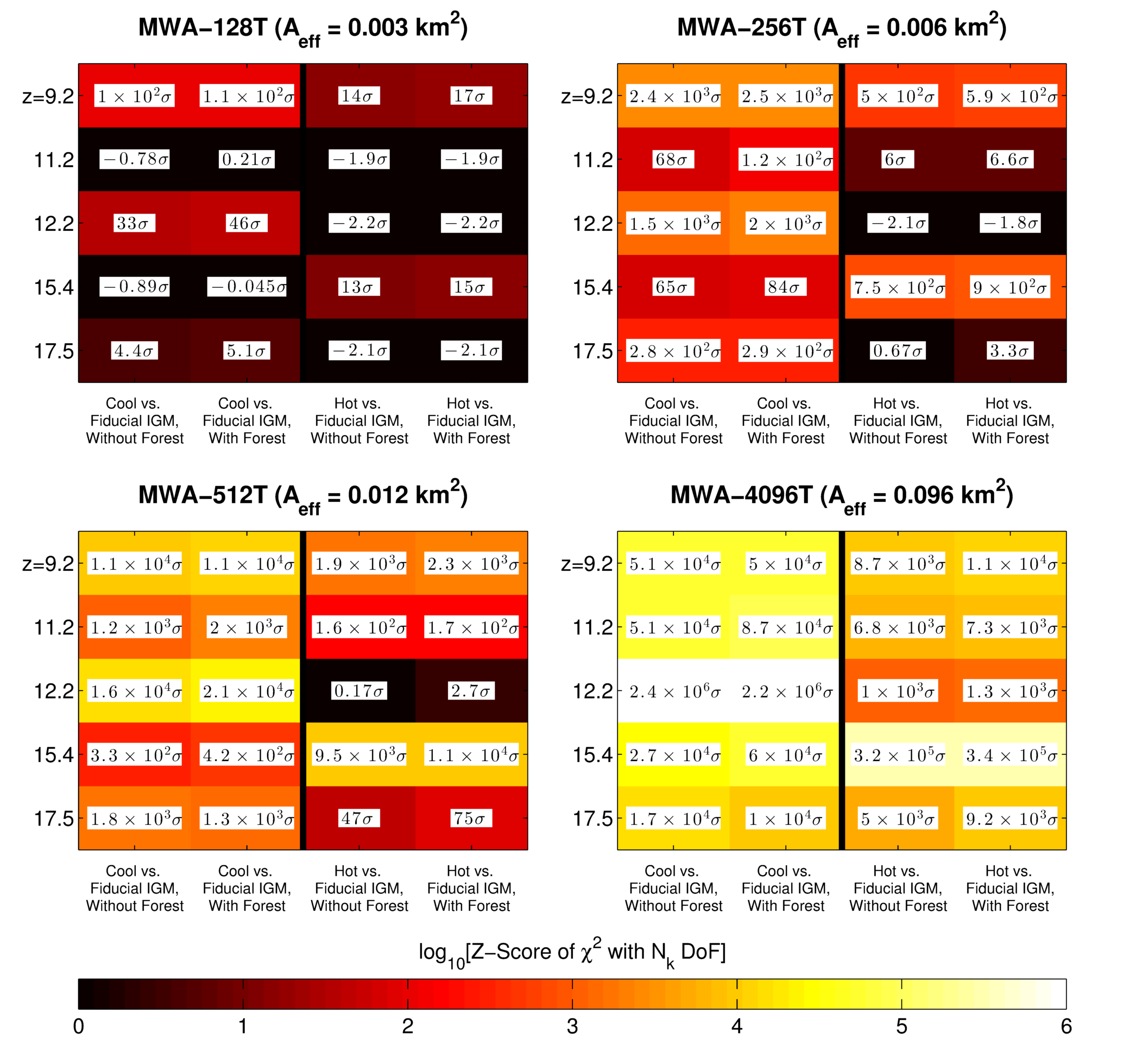}
\caption[Disciminating power between IGM thermal histories.]{ The 21 cm Forest moderately enchances the distinguishability between thermal scenarios and MWA scale interferometers can distinguish between the power spectra for reasonable X-ray heating histories. Here we show the cumulative z-score described in Equation (\ref{eq:zscore}), except now applied to the difference between different IGM heating models, for all arrays and redshifts. At low redshift, the forest decreases the distinguishability of different X-ray heating scenarios by subtracting from the higher amplitude model. When the positive auto-term dominates at high redshift, the forest increases the contrast between given heating models.}
\label{fig:PlotIGMZScore}
\end{figure*}

How the distinguishability between different heating models is affected by the presence of the 21 cm forest is explored in Figure \ref{fig:PlotIGMZScore}. In our 128T table, we see that a  detection of the IGM and constraints on low X-ray emissive histories are possible with the current generation of EoR experiments. There are several caveats worth noting however. First, the high S/N distinctions at $z=9.2$ are due to a detection of the reionization peak at redshifts in which reionization physics such as the uv-efficiency (which we have assumed fixed) become significant.  However, we note that this result contradicts the marginal detectability claimed in \citet{Mesinger:2013c} primarily due to the fact that we include bins with $k < 0.1 \Mpci$ in our standard score. Though these bins have large S/N they may be contaminated by more pessimistic foreground leakage than we consider here such as what is observed by \citet{PoberWedge}.  We also note that the increased sensitivity of combining k-bins allow for constraints on the fiducial X-ray model at $z \sim 15$. The peaks in detectability at $z \approx 9$ and $z \approx 17$ arise from the two peaked structure of the power spectrum in redshift with the low redshift peak corresponding the reionization, and the high redshift peak corresponding to x-ray heating \citep{PritchardFurlanettoXray}.  We see that the forest introduces a small enhancement to the distinguishability between hot and cool heating models. Since the forest adds positively to the power spectrum of a cool, optically thick IGM, its presence enhances the distinguishability between vigorous and cool heating. We find that a modest extension to the MWA can distinguish between hot and fiducial models over a wider range of redshifts and MWA-4096T is able to distinguish between all models over our entire considered redshift range. 

\section{The Detectability of the Forest over a Broad Parameter Space}\label{sec:extr}

\begin{figure*}
\includegraphics[width=\textwidth]{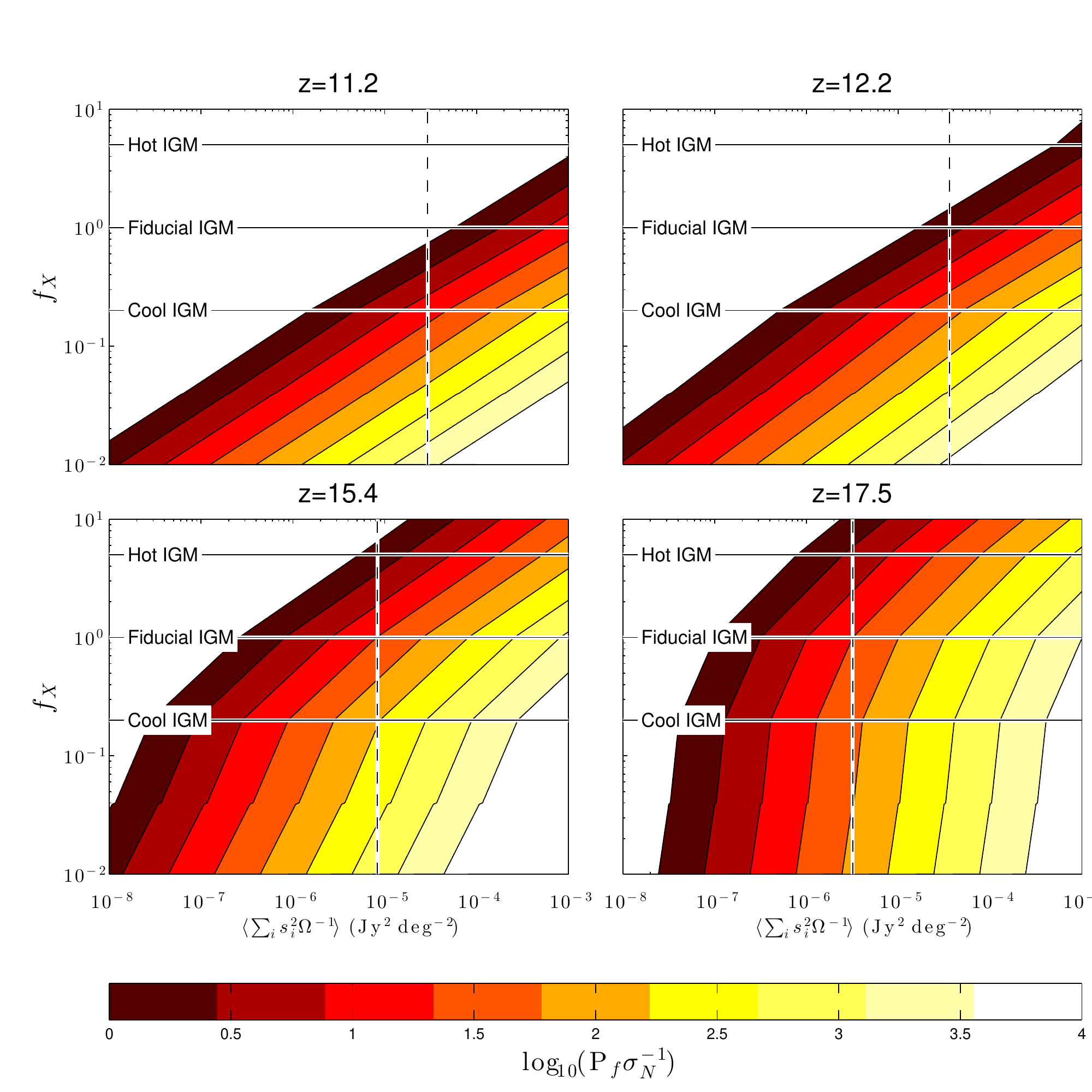}
\caption[Extrapolated observability the 21\,cm forest over many scenarios.]{The ratio of the 21 cm Forest power spectrum, $P_f(k=0.5 \Mpci)$ to thermal noise for 1000 hours of observation on a HERA scale interometer, extrapolated over a large range of X-ray efficiencies and flux squared densities. Vertical dashed black and white lines indicate the value of the simulation by \citep{WilmanExtragalacticSimulation} while the horizontal black and white lines indicate the $f_X$ efficiencies that we explicitly simulate in this paper. At the highest redshifts, $\langle T_s \rangle$ levels off and the detectability of the signal is independent of redshift. At late prereionizatoin redshifts, we see that the 21 cm Forest will only be detectable for heating efficiencies $\lesssim 1$.}
\label{fig:sn_params}
\end{figure*}

For the sake of simplicity, we focus on the detectability of the 21 cm Forest power spectrum from the single population model considered in \citet{WilmanExtragalacticSimulation}. In doing this, it is unclear over what range of radio loud populations the signal is observable. Fortunately, thanks to  Equation (\ref{eq:integral_fix}), we can give order of magnitude estimates of how the detectability of the Forest power spectrum scales with the radio loud source population and the heating history. According to Equation (\ref{eq:integral_fix}), the amplitude of the forest power spectrum, at prereioinization redshifts, scales as 
\begin{equation}\label{eq:scaling}
P_f \propto \frac{1}{\langle T_s \rangle^{2}} \frac{ \sum_i s^2_i(>z) }{\Omega } 
\end{equation}
 where $\sum_i s^2_i(>z) \Omega^{-1}$ is the average sum of source fluxes squared per solid angle. We will call this quantity the {\it flux squared density} of the source population. We take advantage of the simple scaling  in Equation (\ref{eq:scaling}) to extrapolate the amplitude of the Forest signal over a large range of heating models and redshifts. At each redshift, with our fiducial heating model and source population, we obtain a normalization factor for  $P_f$ at a single mode, $k = 0.5 \Mpci$. We then compute $\langle T_s \rangle $ for a large number of lower resolution, $(600 \Mpc)^3$ 21cmFAST simulations with $400^3$ pixels, varying the $f_X$ parameter by three orders of magnitude from $f_X = 10^{-2} - 10^1$. In Figure \ref{fig:sn_params}, we show the ratio of $P_f$ to the amplitude of thermal noise as a function of $f_X$ and the flux squared density of sources, marking the predicted flux squared density of \citet{WilmanExtragalacticSimulation} by a dashed black and white line. We find that the detectability of the forest power spectrum at $z \sim 10$ depends strongly on the thermal state of the IGM, with models significantly fainter than \citet{WilmanExtragalacticSimulation} undetectable except for cool heating histories with  $f_X \lesssim 10^{-1}$. On the other hand, for $z \gtrsim 15$, X-rays in all models have not had sufficient time to heat the IGM above the adiabatic cooling floor and the detectability of $P_f$ becomes significantly less dependent on $f_X$, allowing for a broader range of populations to be probed at higher $f_X$.

\section{Conclusions and Future Outlook}

Using semi-numerical simulations of the thermal history of the IGM, and a semi-empirical RL source distribution, we have shown that the 21 cm forest imprints a distinctive feature in the power spectrum that is, for the most part, invariant in $\kperp$ and, depending on the RL population and thermal history, potentially dominates over the cosmological 21 cm power spectrum at $k_\parallel \gtrsim 0.5 \Mpci$.  We have also derived a simple semi-analytic equation that directly relates the forest power spectrum of $\tau_{21}$ and the radio luminosity function. 
 
Using realistic simulations of power spectrum estimation and including the effects of foregrounds and noise, we have shown that a moderate extension of the MWA-128T instrument has the thermal sensitivity to detect the forest feature in the power spectrum for the W08 RL population with an X-ray efficiency of $f_x \lesssim 1$. For more vigorous heating scenarios, a HERA scale array will have the sensitivity to distinguish this feature. Our simulations also support the results of \citet{Christian:2013} and \citet{Mesinger:2013c}, that low emissivity heating scenarios can be constrained with existing arrays and an extensive examination of the heating history will be possible in the future with larger instruments.    

Signal-to-noise considerations alone do not tell us whether we will be able to distinguish the forest signal from the effects of IGM physics on the power spectrum, especially at small $k$ where a slight change in $f_X$ might shift the power spectrum amplitude up or down, mimicking the shift from the 21 cm forest. Fortunately, the region, $k \gtrsim 0.5 \Mpci$ is dominated by the forest power spectrum, $P_f$, for a range of redshifts in all of our heating models. Specifically, the 21 cm forest removes the $k \gtrsim 0.5 \Mpci$ flattening that occurs after the X-ray heating peak. Observations of the power spectrum over a range of redshifts, with a sensitivity similar to HERA or the SKA should be able to isolate the thermal history at $k \lesssim 0.1 \Mpci$ and constrain RL populations similar to that of W08 at $k \gtrsim 0.5 \Mpci$.

While this paper is a proof of concept, considering a single fiducial RL source distribution, it is possible that measurements with current generation instruments, or moderate extensions, can put constraints on more optimistic scenarios. On the other hand, there are many steep decline scenarios whose power spectrum signatures will be inaccessible even to future arrays. In section \ref{sec:extr} we illustrate the scaling of the detectability of the signal with source flux squared density and X-ray emissivity, finding that populations with order of magnitude smaller flux squared densities than W08 will require a relatively cool prereionization IGM  to be detectable.  In particular, we note that the H04 simulation is one to two orders of magnitude more pessimistic than the predictions of W08 at the highest considered redshifts and would not be detectable in the forest dominated region if $f_X \gtrsim 10^{-1}$. However, higher resolution simulations of the IGM indicate that $\Delta^2_{\tau_{21}}$ continues to climb to $k \sim 10 \Mpci$ while $P_b$ remains flat. Hence the result of a fainter radio luminosity function would be to shift the region of forest dominance to higher k rather than eliminating it, leaving the possibility of detection for a more powerful instrument such as the SKA. There also exists the possibility of separating $P_f$ using its LoS symmetry which might be exploited at $k \sim 0.1 \Mpci$ where EoR interferometers are most sensitive. Finally, we have not considered the absorption of mini halos which \citet{Mack:2012tf} show to substantially increase the variance along the line of sight towards sources (see their Figure 11). Since this variance is an integral of the power spectrum we are being conservative in neglecting them. The sensitivity of future instruments to the forest can be enhanced by increased frequency resolution, allowing them to probe the higher $\kpara$ modes where the forest is especially strong.  The parameter space of radio loud quasars is greatly unconstrained and the disparity between W08 and H04 simply underscores the need for future studies to explore this parameter space. The exploration of a range of RL populations for fixed arrays is left for future work.

In summary, we have shown that the 21 cm power spectrum not only contains information on the IGM in absorption and emission against the CMB but also includes detectible, and in many cases non-negligible signatures of the 21 cm forest. This absorption may be used to constrain the high redshift RL population and IGM thermal history with upcoming interferometers.

\begin{subappendices}

\section{Appendix: A Derivation of the Morphology of $P_f$}\label{app:compare_pf}
In Section \ref{sec:theory} we present a formula, Equation (\ref{eq:pf_simple}), for the the 21 cm forest power spectrum that is the sum of the auto power spectra along the line of sight to each background source. This equation is particularly convenient because it can easily be decomposed into an integral of the radio luminosity function and the optical depth power spectrum. In addition, its $k$-space morphology, which includes no structure in $\kperp$, is relatively simple. 
 In this appendix we derive Equation (\ref{eq:pf_simple}) by applying an analytic toy model to the auto and cross power spectrum contributions to $P_f$ described in Equation (\ref{eq:ps_with_RL}). For the sake of analytic tractability, we invoke a number of approximations. However our results describe $P_f$ very well for $k \gtrsim 10^{-1} \Mpci$. Our assumptions are
\begin{enumerate}
\item  The sources all have the same flux. The W08 simulation includes sources ranging from $1 \nJy$ to $\sim10 \mJy$ over the redshifts of interest. We see in Figure \ref{fig:flux_integral} that the integral of the source fluxes squared is dominated (at the 10\% level) by sources with $S_\nu$ between $1-10\mJy$ so modeling our population as having equal flux gives a decent order of magnitude approximation.
\item The sources are spatially uncorrelated. Clustering from the W08 dark matter bias is actually significant and boosts the results of our simulation, relative to Equation (\ref{eq:pf_simple}), by a factor of two without changing $P_f$'s predicted shape. We will thus absorb this clustering boost into a multiplicative factor of order unity.
\item The sources are unresolved. This will almost certainly be true in all interesting cases given the large synthesized beams of radio interferometers and the extreme distances to the sources. 
\item Source spectra are flat over the frequency interval of a data cube. This is true on the 10\% level over a $\sim 8 \MHz$ band for $S \sim \nu^{-.75}$ sources. Because this slow variation gives a very narrowly peaked convolution kernel in $k$-space, power spectra are not noticably effected by this assumption.
\item The source positions are completely uncorrelated with the cube optical depth field. In reality, the sources that fall within a data cube should be correlated with $\tau_{21}$. We find that correlating or not correlating in cube sources only changes the simulation output by approximately $10\%$.
\end{enumerate}

  We start by reiterating Equation (\ref{eq:sum_forest_sources}) where $P_f$ may be written as
\begin{align} \label{eq:sum_forest_app}
P_f  = \frac{1}{V} \left \langle \left| \widetilde{\Delta T_{RL} \tau_{21}  }\right|^2  \right \rangle  = \displaystyle \sum_{j}   P_j +2   \text{Re} \left( \displaystyle \sum_{j < k}  P_{j,k} \right)  \equiv \Sigma_{auto}+\Sigma_{cross},
\end{align}
where $P_j = \frac{1}{V} \langle | \widetilde{ \Delta T_j \tau_{21}}|^2 \rangle$ and $P_{j,k} = \frac{1}{V}\langle  \widetilde{ \Delta T_j \tau_{21}} \widetilde{ \Delta T_k \tau_{21} }^* \rangle$.
The first term in Equation (\ref{eq:sum_forest_app}) sums the power spectra of each of the absorbed background sources which is positive and the second term is the sum of their cross power spectra. 

  We will show that for the range of spatial scales perpendicular to the LoS, accessed by EoR interferometers, the auto power terms in Equation (\ref{eq:sum_forest_app}) dominate the cross power ones at $\kpara \gtrsim 10^{-1} \Mpci$. We show that the suppression of cross terms is due to two mechanisms: (1) the cross terms are proportional to the cross power spectra between widely separated lines of sight and (2) the cross terms are multiplied by randomly phased sinusoids which cancel out when summed.

\subsection[{The Suppression of the Cross Terms from LoS Cross Power \\Spectra}]{The Suppression of the Cross Terms from LoS Cross Power Spectra}

 To relate the sum in Equation (\ref{eq:sum_forest_app}) to the spectra and locations of the background sources, we assume that all sources are unresolved so that $T_j$ is a delta-function in the plane perpendicular to the LoS. Here, as in \citet{Matt3}, we will adopt observers coordinates $(\ell,m,\nu)$, rather than comoving coordinates $(x,y,z)$, to emphasize the fact that the the broad-spectrum source does not physically occupy a range of positions along the LoS. In such coordinates, the temperature field of each source can be written as $T_j( \ell, m, \nu)$ where $\ell$ and $m$ are the direction cosines from the north-south and east-west directions, and $\nu$ is the difference from the data cube's central frequency. $\tau_{21} T_j(\ell, m,\nu)$ is given by
\begin{equation}\label{eq:single_source}
\tau_{21} T_{j}(\ell,m,\nu)=   \Omega_{pix} \delta(\ell - \ell_j) \delta( m - m_j) \tau_{21}(\ell_j, m_j, \nu)T_j
\end{equation}
where $ \Omega_{pix}$ is the solid angle of a map pixel and $\delta( ... )$ is the Dirac delta function. For notational simplicity, we will use vector notation to denote direction cosines, ${\bf \ell}=(\ell,m)$ and their Fourier duals, ${\bf u} = (u,v)$. Taking the Fourier transform of  $\tau_{21} T_j({\bf \ell}, \nu)$ and summing over all sources gets
\begin{align} \label{eq:inter}
\widetilde{T}_f({\bf u},\eta)  = \sum_j \widetilde{\tau_{21}T_j}({\bf u},\eta) =   \Omega_{pix}  \sum_j T_j e^{2 \pi i ( {\bf \ell}_j \cdot {\bf u})}  \int \tau_{21}({\bf \ell}_j,\nu)  e^{-2 \pi i \eta \nu}  d \nu.
\end{align}
We take the modulus squared of Equation (\ref{eq:inter}) and multiply by the cosmology dependent variables, $D_M^2Y$ \citep{AaronSensitivity} that relate observers coordinates to the cosmological comoving coordinates that we've used to define our power spectrum in Equation (\ref{eq:ps}). We find that the sum of the auto terms in Equation (\ref{eq:sum_forest_app}) is
\begin{equation}
\Sigma_{auto}= \frac{ D_M^2 \Omega_{pix}^2}{\Omega_{cube}}  P_{\tau_{21}}^{LoS}(\kpara) \left \langle \sum_j T_j^2 \right \rangle.
\end{equation}
The sum of cross terms is
\begin{align}
 \Sigma_{cross} = 2 \frac{ D_M^2 \Omega_{pix}^2}{\Omega_{cube}}   \sum_{j<k}T_j T_k \Bigl[& \text{Re} \left( P_{\tau_{21;j,k}}^{LoS}(\kpara)\right)\langle \cos [ 2 \pi( {\bf u} \cdot  {\bf \Delta \ell_{j,k} })] \rangle \notag  \\
 &+ \text{Im} \left( P_{\tau_{21};j,k}^{LoS}(\kpara)\right) \langle \sin[ 2 \pi( {\bf u} \cdot {\bf \Delta \ell_{j,k} })] \rangle \Bigr] \notag \\
= 2  P_{\tau_{21}}^{LoS}(\kpara) \frac{ D_M^2 \Omega_{pix}^2 }{\Omega_{cube}} \sum_{j<k}T_j T_k \Bigl[& \frac{\text{Re}\left(P_{\tau_{21};j,k}^{LoS} (\kpara)\right)}{P_{\tau_{21}}^{LoS}(\kpara)}  \langle \cos [ 2 \pi  ( {\bf u} \cdot {\bf \Delta \ell_{j,k} })]  \rangle \notag  \\ 
  &+\frac{\text{Im}\left(P_{\tau_{21};j,k}^{LoS}(\kpara)\right)}{P_{\tau_{21}}^{LoS}(\kpara)} \langle \sin[ 2 \pi (  {\bf u} \cdot {\bf \Delta \ell_{j,k} } ) ] \rangle  \Bigr],\label{eq:cross_sum}
\end{align}

where ${\bf \Delta \ell_{j,k}}={\bf  \ell_j} - {\bf \ell_k}$. Here, we define the cross power spectrum between two lines of sight to be 
 
\begin{equation}
P_{\tau_{21};j,k}^{LoS}(k_z)=\frac{1}{L} \int d z d z'  e^{  i k_z (z-z')} \Delta \tau_{21} ( {\bf \ell}_j,z) \Delta \tau_{21}( {\bf \ell}_k  ,z').
\end{equation}

\begin{figure}
\centering
\includegraphics[width=.6\textwidth]{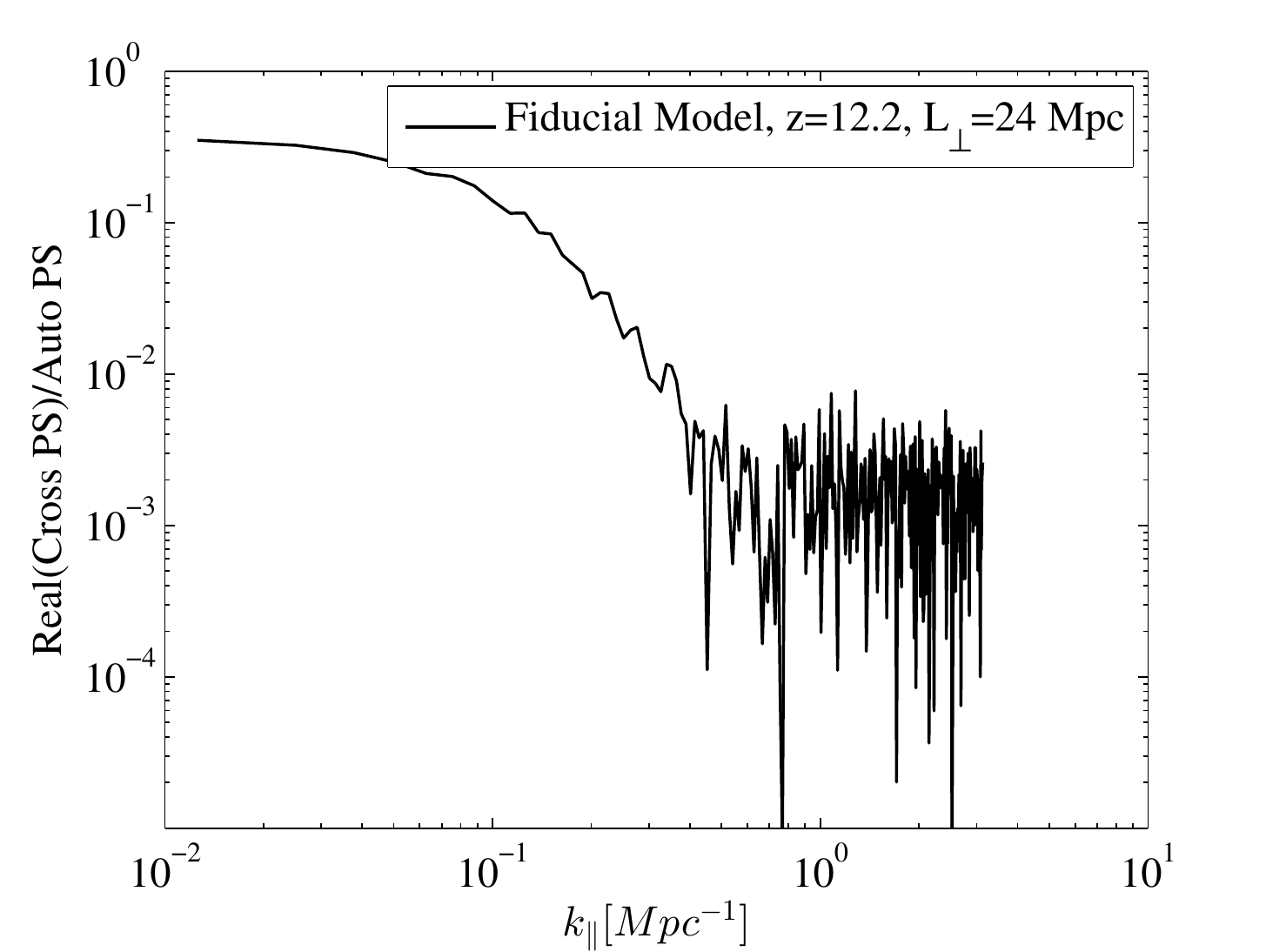}
\includegraphics[width=.6\textwidth]{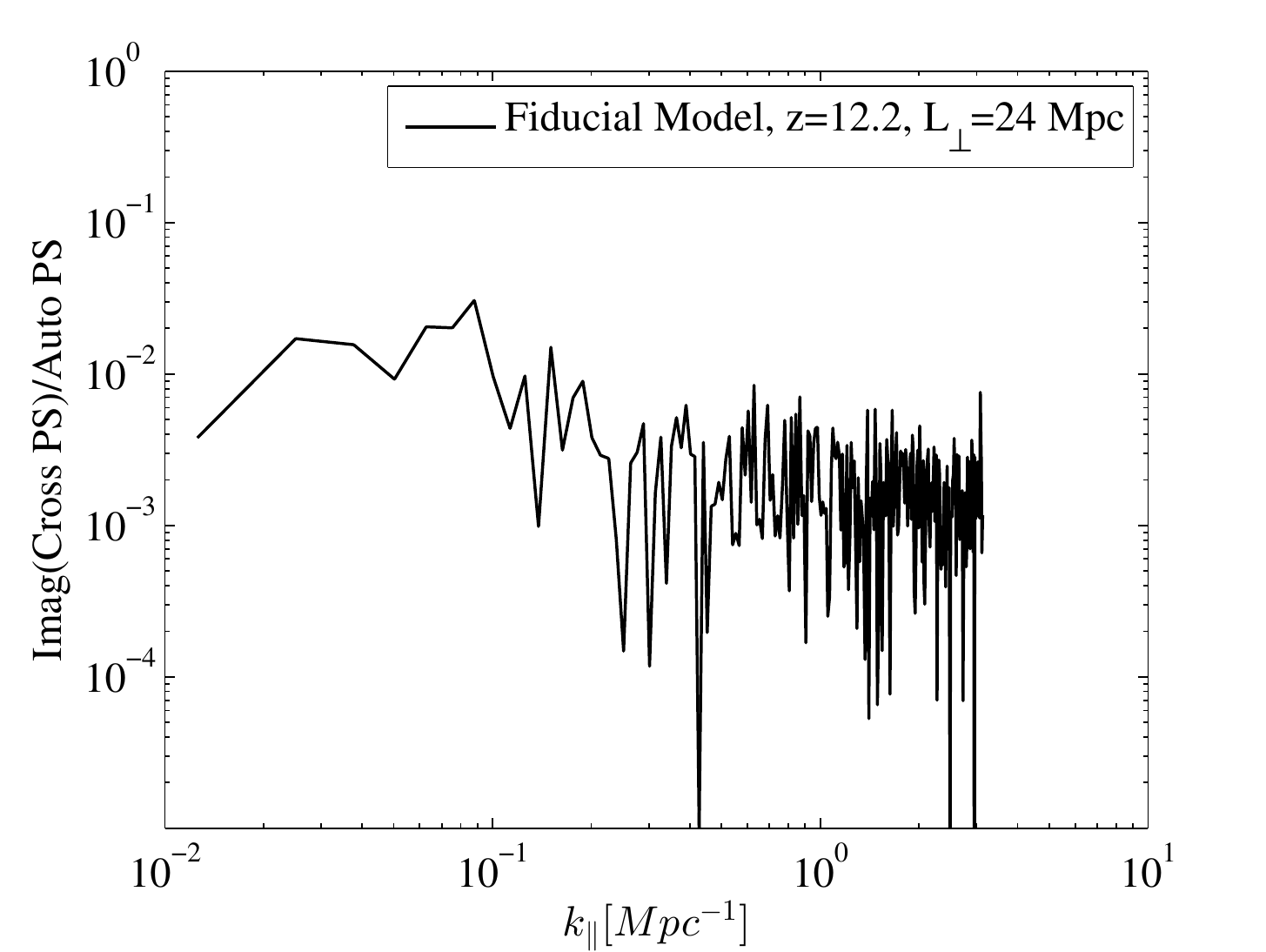}
\caption[Relative power of auto and cross power spectra.]{The LoS cross power spectra between spatially separated lines of site are on the order of $\sim 100-1000$ times smaller than auto power spectra. In the left figure, we plot the ratio of the real cross power spectra between lines of site separated by 24 Mpc to auto power spectra, and on the right we show the ratio of the imaginary cross power spectrum to the auto power spectrum. In both cases, for $\kpara \gtrsim 10^{-1} \Mpci$, the cross power spectra are on the order of 10-1000 times smaller. The real cross power spectrum becomes non negligible on scales comparable to the separation between the two lines of site.}
\label{fig:cl}
\end{figure}

 It is clear from Equation (\ref{eq:cross_sum}) that each summand in $\Sigma_{cross}$ is smaller than each term in $\Sigma_{auto}$ by a factor of the ratio between the LoS cross power spectra of spatially separated lines of sight and the LoS auto power spectrum. If lines of sight to each source are sufficiently separated, this ratio should be very small. In Figure \ref{fig:cl} we show the ratios of $\text{Re} \left( P_{\tau_{21};j,k}^{LoS} \right)/P_{\tau_{21}}^{LoS}$ and $\text{Im} \left( P_{\tau_{21};j,k}^{LoS} \right)/P_{\tau_{21}}^{LoS}$ from our fiducial model at $z=12.2$, separated by $L_\perp=24 \Mpc$ which is the mean distance in our data cube between 1000 background sources. Because two sufficiently separated lines of sight should be statistically independent except on the largest spatial scales, these ratios are on the order of $10^{-2}-10^{-3} $ for $\kpara \gtrsim 10^{-1} \Mpci$ .

\subsection{Supression of the Cross Terms from Summing the Random Source Phases}

 The factor of 100-1000 introduced by the ratio of the cross spectra to the auto spectra would be enough to suppress the cross terms if the number of sources were reasonably small. However the number of cross terms relative to the number of auto terms in Equation (\ref{eq:sum_forest_app}) goes as $(N-1)/2$ where $N$ is the number of contributing sources. Thus, even though the cross power spectrum between individual LoS pairs is small, naively summing 100-500 sources could still yield a significant contribution. We now show that summing over many randomly distributed source angles suppresses this. 
 
\begin{figure}
\centering
\includegraphics[width=1\textwidth]{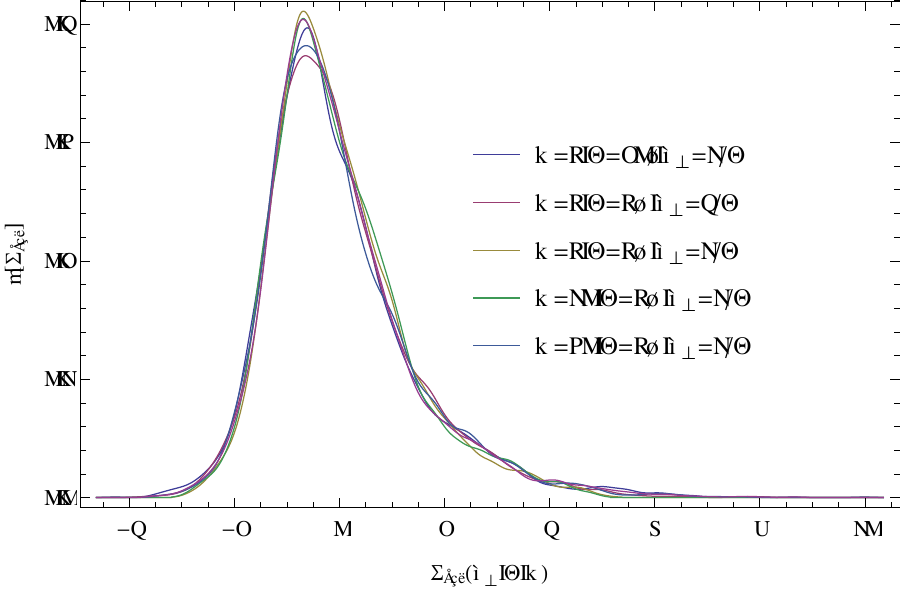}
\caption[Power spectrum dependence on observational and source parameters.]{Here we see that $P[\Sigma_{cross}(u_\perp,\Theta,N)]$ is invariant in N, $\Theta$, and $u_\perp$, and $N$ with a mean of approximately zero. The lines which indicate, $P[\Sigma_{cross}(u_\perp,\Theta,N)]$, are estimated from 10000 draws. Since $\langle \Sigma_{cross} \rangle \approx 0$ we expect the cross terms to contribute negligibly to $P_f$ in 3D Fourier Space.}
\label{fig:dist_upper}
\end{figure}

Since $\text{Im} \left( P_{\tau_{21};j,k}^{LoS} \right)/P_{\tau_{21}}^{LoS}$ is on the same order of, or smaller than  $\text{Re} \left( P_{\tau_{21};j,k}^{LoS} \right)/P_{\tau_{21}}^{LoS}$, we will use the real term on both the cosine and sine terms in Equation (\ref{eq:cross_sum}) to give an upper bound. Assuming that all sources have the same temperature, $T_j=T_k=T_0$, we may write
\begin{align}\label{eq:sum_cross_full}
\Sigma_{cross} \approx  2 \sum_{j<k}T_0^2 P_{\tau_{21};j,k}^{LoS} (\kpara) \Bigl[   \langle \cos [ 2 \pi (  {\bf u} \cdot {\bf \Delta \ell_{j,k} })] \rangle  +   \langle \sin [ 2 \pi(  {\bf u} \cdot {\bf \Delta \ell_{j,k} } )] \rangle \Bigr].
\end{align}
Similarly,
\begin{equation}
\Sigma_{auto} \approx N T_0^2 P_{\tau_{21}}^{LoS}
\end{equation}
Hence the ratio between $\Sigma_{cross}$ and $\Sigma_{auto}$ is given by
\begin{align}
\frac{\Sigma_{cross}}{\Sigma_{auto}} \approx  2 \frac{ \text{Re}\left( P_{\tau_{21};j,k}^{LoS}(\kpara) \right)}{N P_{\tau_{21}}^{LoS}(\kpara)}\sum_{j<k} \Bigl[ \langle \cos [ 2 \pi (  {\bf u} \cdot {\bf \Delta \ell_{j,k} })] \rangle +  \langle \sin [ 2 \pi (  {\bf u} \cdot {\bf \Delta \ell_{j,k} })]  \rangle \Bigr]
\end{align}
Because of the cylindrical symmetry, we need only concern ourselves with a uv cell at $v=0$ and simply write
\begin{align}
\Sigma_{cross} / \Sigma_{auto} &\approx 2  \frac{\text{Re}\left( P_{\tau_{21};j,k}^{LoS}(\kpara) \right)}{N P_{\tau_{21}}^{LoS}(\kpara)} \sum_{j<k}  \Bigl[  \langle \cos [ 2 \pi  u_\perp \Delta \ell_{j,k}]  \rangle + \langle \sin [ 2 \pi  u_\perp \Delta \ell_{j,k}] \rangle \Bigr] \notag \\
&=   \frac{\text{Re}\left( P_{\tau_{21};j,k}^{LoS}(\kpara) \right)}{ P_{\tau_{21}}^{LoS}(\kpara)} \left \langle \Sigma_{cos} \left( u_\perp,\Theta,N\right) \right \rangle,
\end{align}
where 
\begin{equation}\label{eq:cossum}
\Sigma_{cos} \left( u_\perp,\Theta,N \right) \equiv \frac{2}{N} \sum_{j<k}  \cos[ 2 \pi u_\perp \Delta \ell_{j,k}]  +  \sin[ 2 \pi u_\perp \Delta \ell_{j,k}]. 
\end{equation}
We can easily compute this ensemble average for any $u_\perp$ by drawing N different source positions  distributed randomly over the angular span of the field, $\Theta$, and summing over the sines and cosines of pair-wise angle differences. In Figure \ref{fig:dist_upper} we show $P[\Sigma_{cos}(u_\perp,\Theta,N)]$ for randomly distributed $\Delta \ell_{j,k}$ for a variety of $N$, $u_\perp$, and $\Theta$ where the minimal $u_\perp$ is set by the maximal scale accessible by an interferometers primary beam, $\sim 1/\Theta$. We calculate these distributions from 10000 random realizations. We see that the distribution of $\Sigma_{cos}$  is independent of $N,\Theta$, and $u_\perp$ and has a mean of $\approx 0$ (which is the quantity that sets the amplitude of $\Sigma_{cross}$.   As long as sources are randomly distributed, we can expect LoS cross power spectra to suppress the cross terms sum to below the $10\%$ level at $\kpara \gtrsim 10^{-1} \Mpci$, regardless of the number of terms.
 
We may finally write. 
\begin{align} \label{eq:sum_forest_sourcesapprox}
P_f ( {\bf k} ) \approx \displaystyle \sum_j   P_j(\kpara) = \frac{D_M^2 \Omega_{pix}^2}{\Omega_{cube}} \sum_j T_j^2 P_{\tau_{21}}^{LoS} =\frac{D_M^2 \lambda^4}{4 k_B^2 \Omega_{cube}} \sum_j s_j^2 P_{\tau_{21}}^{LoS}
\end{align}
where $\lambda=\lambda_{21}(1+z)$ is the wavelength at the center of the data cube, $P_j$ is the absorption power spectrum for the $j^{th}$ source, $s_j$ and $T_j$ are the flux and temperatures of the $j^{th}$ source, $\Omega_{cube}$ is the solid angle subtended by the observed volume, and $P_{\tau_{21}}^{LoS}$ is the 1D LoS power spectrum.

 We may therefor consider the absorption power spectrum resulting from the forest as simply the sum of the absorption power spectra of each individual source in the background of the source cube. Since all quantities in this sum are positive, we see that the amplitude of the power spectrum increases linearly with the number of sources present behind an observed volume.  Because the power spectra for unresolved sources are constant in $k_\perp$, $P_f$ will have a structure that is nearly constant in $k_\perp$. 
 
Hence, for $k \gtrsim 10^{-1} \Mpci$, Equation (\ref{eq:sum_forest_sources}) simplifies to a sum of the auto power spectra along the LoS to each source. We finish by briefly commenting on the of the effect of clustering which we have ignored but we find (after comparing Equation (\ref{eq:pf_simple}) to our simulations) is still significant. Clustering will cause a disproportionate number of sources to reside in close proximity on the sky. The effect of this is two fold. First, the clustered sources will tend to be behind correlated optical depth columns so that the cross terms between such sources will be better described by auto power spectra. Second, the phases between such sources will be small so that they will not sum to zero. In addition, they will not introduce significant $\kperp$ structure except at the smallest perpendicular scales. Hence the cross terms introduced by clustered sources will closely resemble the $\kpara$ invariant auto terms and simply increase the overall amplitude of $P_f$. We treat this increase by introducing a multiplicative constant of order unity, $A_{cl}$, in Equation (\ref{eq:integral_fix}).

\section{Appendix: A Comparison Between Two Source Models}\label{app:compare}
\begin{figure}
\centering
\includegraphics[width=.6\textwidth]{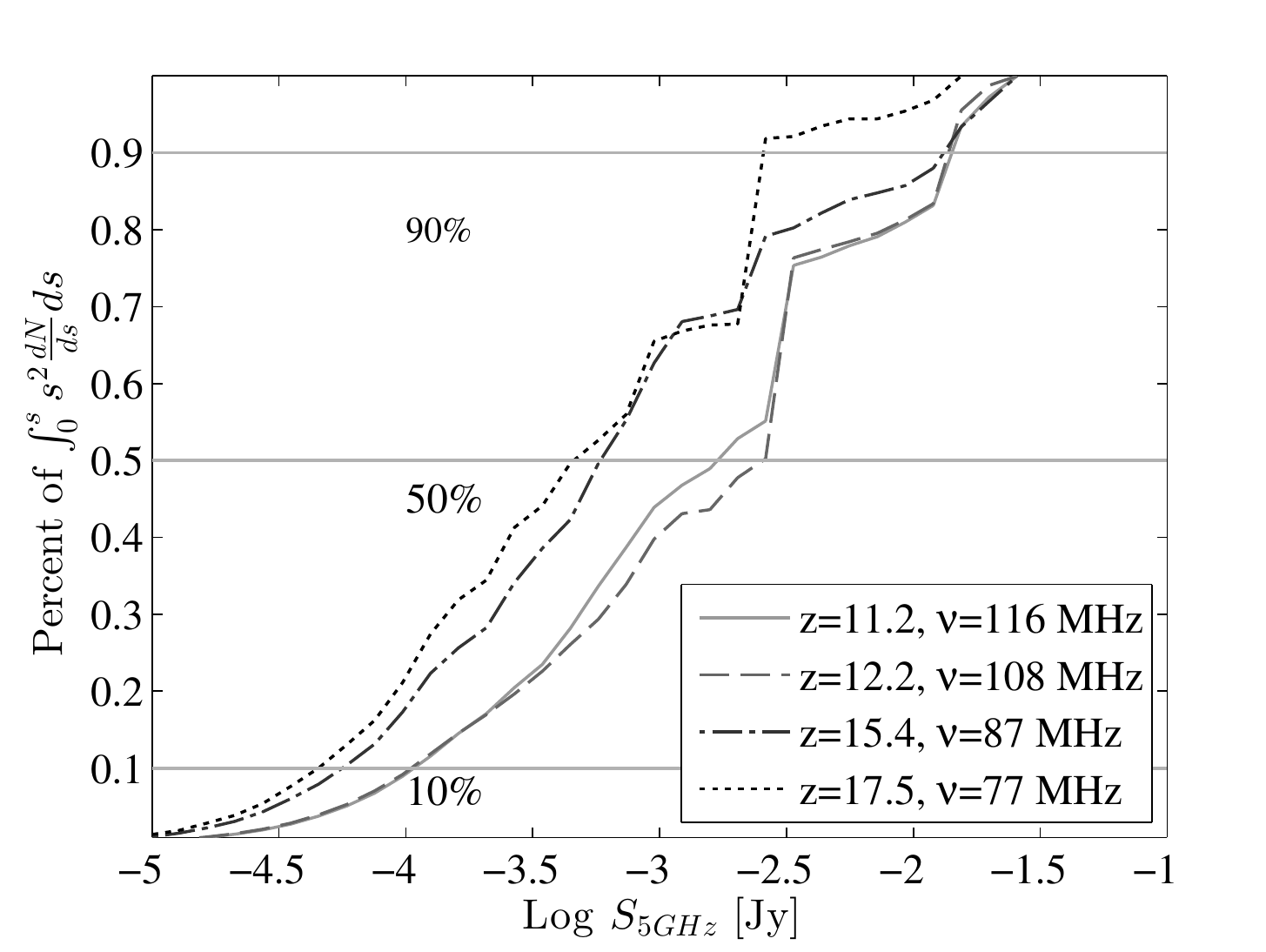}\\
\includegraphics[width=.6\textwidth]{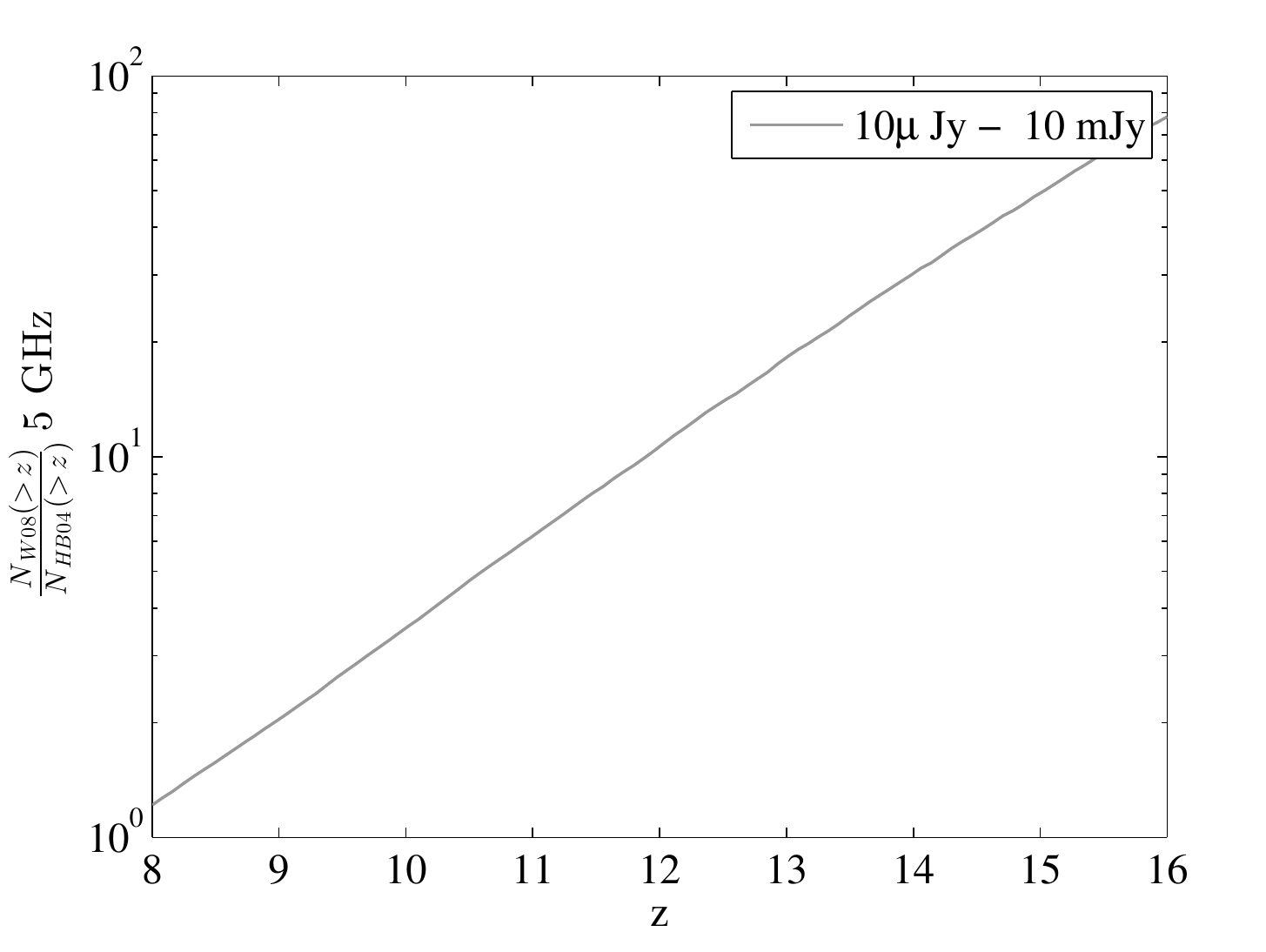}
\caption[Comparison of source flux population models.]{Left: The percentage of the integrated luminosity function in Equation (\ref{eq:integral}) as a function of the source fluxes at $5 \text{GHz}$ for comparison to the catalogue of H04. We see that most contributions to the forest power spectrum come in between $S_{5 \GHz} = 10 \muJy$ and $S_{5 \GHz} = 10 \mJy$. Right: The ratio of the number of sources with redshift greater than $z$ between $S_{5\GHz} = 10 \muJy$ and $10 \mJy$ as predicted by the W08 and H04. The W08 simulation over predicts the counts in H04 by a factor ten at $z \gtrsim 12$ and nearly $80$ at $z \gtrsim 16$, emphasizing the importance of exploring this widely unconstrained parameter space in future work.}
\label{fig:flux_counts}
\end{figure}

In this paper, we choose to work with the semi-empirical source population in the simulation by \citet{WilmanExtragalacticSimulation}. This choice was in part motivated by the lack of constraints at high redshift and the ease which which we could use data from the W08 simulation using its online interface. Another prediction in the literature for the high redshift radio luminosity function is made by \citet{HaimanQuasarCounts}. This model, like the one in W08, relies on a number of uncertain assumptions but is a more physically motivated bottom up approach which is derived from the cold dark matter power spectrum and assumptions about the black hole-halo mass relation and radio loud fraction. In this appendix, we attempt to understand how our choice of the Wilman source population compares to that in H04. To do this, we attempt to compare the source counts from W08 that contribute the most to $P_f$ to those of H04 who provide cumulative flux counts for $1-10 \GHz$ as a function of redshift. To compare the W08 sources, we compute the percentage of the radio luminosity function integral in Equation (\ref{eq:integral}) as a function of the extrapolated $S_{5 \GHz}$. On the left, in Figure \ref{fig:flux_counts}, a large fraction of $P_f$ is determined by W08 sources with $5 \GHz$ fluxes between $10 \muJy$ and $10 \mJy$. We show, in Figure \ref{fig:flux_counts}, the ratio of W08 and H04 source counts with $S_{5 \GHz}$ between $10 \muJy$ and $10 \mJy$. The H04 counts fall much faster with redshift than those of W08. At $z \sim 10-12$ the number of contributing sources is larger by a factor of $\approx 10$ and $\approx 80$ by $z \sim 16$.

This comparison is very approximate since different spectral indices are assumed in H04 and W08. However, we emphasize that the observability claims we make in this paper would not apply accurately to the H04 prediction. A more extensive exploration of parameter space will be necessary to determine what range of radio loud source populations may be constrained by the power spectrum technique. 

 Since $P_b$ is observed to be flat out to $k\approx 10 \Mpci$ while $P_f$ climbs, a more pessimistic source scenario has the effect of pushing the forest dominant region to higher $\kpara$ which does not preclude detection with a more powerful telescope such as the SKA.

\end{subappendices}


\chapter[What Next-Generation 21\,cm Power Spectrum Measurements \\Can Teach Us About the Epoch of Reionization]{What Next-Generation 21\,cm Power Spectrum Measurements Can Teach Us About the Epoch of Reionization}\label{ch:NextGen}

\emph{The content of this chapter was submitted to \emph{The Astrophysical Journal} on October 25, 2013 and published \cite{PoberNextGen} as \emph{What Next-Generation 21\,cm Power Spectrum Measurements Can Teach Us About the Epoch of Reionization}  on January 28, 2014.} 


\renewcommand{\topfraction}{0.85}
\renewcommand{\bottomfraction}{0.1}

\section{Introduction}

The Epoch of Reionization (EoR) represents a turning point in cosmic history,
signaling the moment when large scale structure has become significant
enough to impart a global change to the state of the baryonic universe.  In particular,
the EoR is the period when ultraviolet photons (likely from the first galaxies)
reionize the neutral hydrogen in the intergalactic medium (IGM).
As such, measurements
of the conditions during the EoR promise a wealth of information about
the evolution of structure in the universe. 
Observationally, the redshift of EoR is roughly constrained to be between
$z \sim 6 \mbox{--} 13$, with a likely extended duration; see 
\cite{FurlanettoReview}, \cite{PritchardLoebReview},
and \cite{aviBook} for 
reviews of the field.
Given the difficulties of optical/NIR observing at these
redshifts, the highly-redshifted 21\,cm line of neutral hydrogen has been
recognized as a unique probe of the conditions during the EoR 
(see \citet{miguelreview} and 
\citet{PritchardLoebReview} for recent reviews discussing this technique).  

In the last few years, the first generation of experiments targeting
a detection of this highly-redshifted 21\,cm signal from the EoR 
has come to fruition.  In particular, the LOw Frequency ARray 
(LOFAR; \citet{InitialLOFAR1,LOFAR2013})\footnote{http://www.lofar.org/}, 
the Murchison Widefield Array (MWA; \citet{TingaySummary,BowmanMWAScience})\footnote{http://www.mwatelescope.org/}, 
and the Donald C. Backer Precision Array for Probing the Epoch of Reionization
(PAPER; \citet{PAPER})\footnote{http://eor.berkeley.edu/}
have all begun long, dedicated campaigns with the goal of detecting the
21\,cm power spectrum.  Ultimately, the success or failure of these
campaigns will depend on the feasibility of controlling both instrumental
systematics and foreground emission.
But even if these challenges can be overcome, a positive detection of
the power spectrum will likely be marginal at best because of limited
collecting area.  Progressing from a detection to a characterization of the
power spectrum (and eventually, to the imaging of the EoR) will require
a next generation of larger 21\,cm experiments.

The goal of this paper is to explore the range of constraints that
could be achievable with larger 21\,cm experiments and, in particular, focus
on how those constraints translate into a physical understanding of the EoR.  
Many groups have
analyzed the observable effects of different reionization models on the 21\,cm
power spectrum; see e.g., 
\citet{ZFH}, \citet{furlanetto2},
\citet{Matt3},
\citet{Judd06}, \citet{juddjackiemiguel1}, 
\citet{trac_and_cen_2007}, \citet{LidzRiseFall}, and
\citet{iliev_et_al_2012}.  These studies did not include the
more sophisticated understanding of foreground emission that has arisen
in the last few years, i.e., the division of 2D cylindrical $k$-space into the
foreground-contaminated ``wedge'' and the relatively clean ``EoR window''
\citep{Dattapowerspec,VedanthamWedge,MoralesPSShapes,AaronDelay,CathWedge,ThyagarajanWedge}.
The principal undertaking of this present work is to reconcile
these two literatures, exploring the effects of both different 
EoR histories and foreground removal models on the recovery of astrophysical
information from the 21\,cm power spectrum.
Furthermore, in this work we present some of the first analysis focused 
on using realistic measurements to distinguish between different theoretical 
scenarios, rather than simply computing observable (but possibly degenerate) 
quantities from a given theory.
The end result is a set of generic
conclusions that both demonstrates the need for a large collecting area next
generation experiment and motivates the 
continued development of foreground removal algorithms.

In order to accomplish these goals, this paper will employ simple models
designed to encompass a wide range of possible scenarios.  These models
are described in Section \ref{sec:models}, wherein we describe the
models for the 
instrument (Section \ref{sec:instrument}), foregrounds (Section \ref{sec:foregrounds}),
and reionization history (Section \ref{sec:eor}).  In Section \ref{sec:results}, we
present a synthesis of these models and the resultant range of potential
power spectrum constraints, including a detailed examination
of how well one can recover physical parameters describing the EoR in 
Section \ref{sec:adrian}.
In Section \ref{sec:conclusions}, we conclude with several generic messages
about the kind of science the community can expect from 21\,cm experiments
in the next $\sim5$ years.

\section{The Models}
\label{sec:models}

In this section we present the various models for the 
instrument (Section \ref{sec:instrument}), foreground removal 
(Section \ref{sec:foregrounds}),
and reionization history (Section \ref{sec:eor}) used to explore
the range of potential EoR measurements.  
In general, these models are chosen not because they
necessarily mirror
specific measurements or scenarios, but rather because of their
simplicity while still encompassing a wide range of uncertainty about many 
parameters.
We choose several different parameterizations of the foreground removal
algorithms, and use simple simulations to probe a wide variety of reionization
histories.
Our model telescope (described below in Section \ref{sec:instrument}) 
is based off the proposed Hydrogen Epoch of Reionization Array (HERA);
we present sensitivity calculations and astrophysical constraints for
other 21\,cm experiments in the appendix.

\subsection{The Telescope Model}
\label{sec:instrument}

\begin{figure}[]
\centering
\includegraphics[width=.8\textwidth]{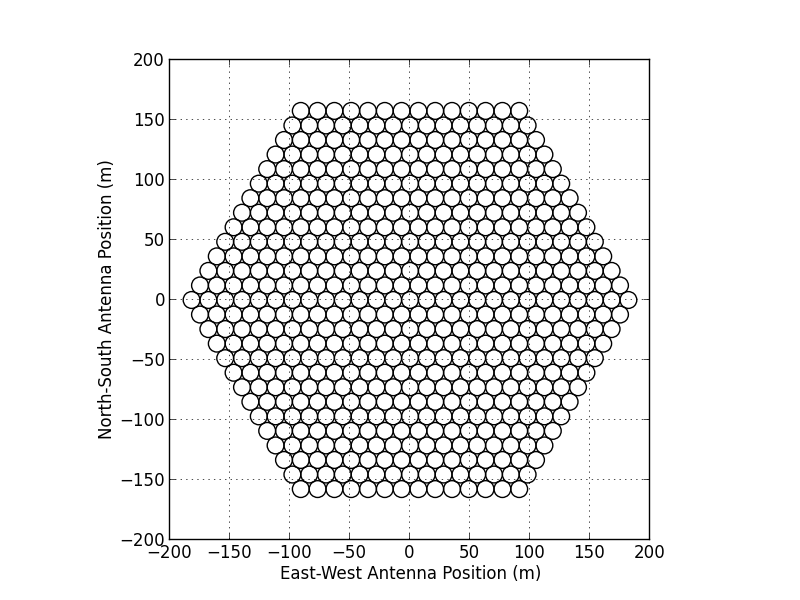}
\caption[HERA-547 layout.]{The 547-element, hexagonally packed HERA concept design, with 14~m
reflector elements.  Outrigger antennas may be included in the final design
for the purposes of foreground imaging, but they are not treated here, since
they add little to power spectrum sensitivity.}
\label{fig:hex547}
\end{figure}

The most significant difference between the current and next generations 
of 21\,cm instruments
will be a substantial increase in collecting area and, therefore, sensitivity. 
In the main body of this work, we use an instrument modeled after a concept design for the
Hydrogen Epoch of Reionization Array (HERA)\footnote{http://reionization.org/}.
This array consists of 547 zenith-pointing
14~m diameter reflecting-parabolic elements in a close-packed hexagon, as shown in Figure
\ref{fig:hex547}. 
The total collecting area of this array is $84,000~\rm{m}^2$, or approximately
one tenth of a square kilometer.  The goal of this work is not to justify
this particular design choice, but rather to show that this scale instrument
enables a next level of EoR science beyond the first generation experiments.
In the appendix, we present the resultant sensitivities and achievable
constraints on the astrophysical parameters of interest for
several other 21\,cm telescopes: PAPER, the MWA, LOFAR, and a concept
design for the SKA-Low Phase 1.  Generically, we find that
power spectrum sensitivities are a strong function of array configuration,
especially compactness and redundancy.  However, once the
power spectrum sensitivity of an array is known, constraints
on reionization physics appear to be roughly independent of other
paramters.

In many ways, the HERA concept array design
is quite representative of 21\,cm EoR
experiments over the next $\sim5\mbox{--}10$ years.  As mentioned, it has
a collecting area of order a tenth of a square kilometer --- 
significantly larger than any current instrument,
but smaller than Phase 1 of the low-frequency Square
Kilometre Array (SKA1-low)\footnote{http://www.skatelescope.org/}.
(See Table \ref{tab:instruments} for a summary of different
EoR telescopes.)
In terms of power spectrum sensitivity,
\cite{AaronSensitivity} demonstrated the power of array
redundancy for reducing thermal noise uncertainty, and showed
that a hexagonal configuration has the greatest instantaneous redundancy.
In this sense, the HERA concept design is optimized for power spectrum
measurements.
Other configurations in the literature have been optimized for foreground
imaging or other additional science; the purpose of this work is not to 
argue for or against these designs.  
Rather, we concentrate primarily on science with the 21\,cm power spectrum, and
use the HERA concept design as representative of power spectrum-focused
experiments.
Obviously,
arrays with more (less) collecting area will have correspondingly greater
(poorer) sensitivity.  
The key parameters of our fiducial concept array
are given in Table \ref{tab:array}, and constraints achievable
with other arrays are presented in the appendix.

\begin{table}
\centering
\begin{tabular}{|c|c|}
\hline
Observing Frequency & $50 \mbox{--} 225$ MHz \\ \hline
$T_{\rm receiver}$ & 100 K \\ \hline
Parabolic Element Size & 14~m \\ \hline
Number of Elements & 547 \\ \hline
Primary beam size & $8.7^\circ$ FWHM at 150~MHz \\ \hline
Configuration & Close-packed hexagonal \\ \hline
Observing mode & Drift-scanning at zenith \\ \hline
\end{tabular}
\caption{Fiducial System Parameters.}
\label{tab:array}
\end{table}

\subsubsection{Calculating Power Spectrum Sensitivity}
\label{sec:sense_calc}

To calculate the power spectrum sensitivity of our fiducial array, we use
the method presented in \cite{BAOBAB}, which is briefly summarized
here.  This method begins by creating the $uv$ coverage of the observation
by gridding each baseline into the $uv$ plane,
including the effects of earth-rotation synthesis over the course
of the observation.  
We choose $uv$ pixels the size of the antenna element in wavelengths,
and assume that any baseline samples only one pixel at a time.
Each pixel is treated as an
independent sample of one $k_{\perp}$-mode, along which the instrument
samples a wide range of $k_{\parallel}$-modes 
specified by the observing bandwidth.
The sensitivity to any one mode of the dimensionless power spectrum 
is given by Equation (4) in \cite{BAOBAB}, which is in turn
derived from Equation (16) of \cite{AaronSensitivity}:
\begin{equation}
\label{eq:sensitivity}
\Delta^2_{\rm N}(k) \approx X^2Y\frac{k^3}{2\pi^2}\frac{\Omega'}{2t}T_{\rm sys}^2,
\end{equation}
where $X^2Y$ is a cosmological scalar converting observed bandwidths and solid
angles to $h{\rm Mpc}^{-1}$, $\Omega ' \equiv \Omega_{\rm p}^2/\Omega_{\rm pp}$
is the solid angle of the power primary beam ($\Omega_{\rm p}$) squared, divided by the solid
angle of the square of the power primary beam ($\Omega_{\rm pp}$),\footnote{Although \cite{AaronSensitivity}
and \cite{BAOBAB}
originally derived this relation with the standard power primary beam
$\Omega$, it was shown in \cite{PAPER32Limits} that the power-squared
beam enters into the correct normalizing factor.}
$t$ is the integration time on that particular $k$-mode, and $T_{\rm sys}$ is
the system temperature.  It should also be noted that this equation
is dual-polarization, i.e., it assumes both linear polarizations are measured
simultaneously and then combined to make a power spectrum estimate. 
Similar forms of this equation appear in \cite{MiguelNoise}
and \cite{Matt3}
which differ only by the polarization factor and power-squared primary
beam correction. 

In our formalism, each measured mode is attributed a noise value
calculated from Equation \ref{eq:sensitivity} (see 
Section \ref{sec:observationParams}
for specifics on the values of each parameter). 
Independent modes can be combined in quadrature to form spherical or
cylindrical power spectra as desired.  
One has a choice of how to combine non-instantaneously 
redundant baselines which do in fact sample the
same $k_{\perp}/uv$ pixel.  Such a situation can arise either through
the effect of the gridding kernel creating overlapping $uv$ footprints on
similar length baselines (``partial coherence''; 
\citet{Hazelton2013}), or through the effect of earth-rotation
bringing a baseline into a $uv$ pixel previously sampled by another baseline.
Na\"{i}vely, this formalism treats these samples as perfectly coherent, i.e.,
we add the integration time of each baseline within a $uv$ pixel.  As suggested
by \cite{Hazelton2013}, however, it is possible that
this kind of simple treatment could
lead to foreground contamination in a large number of Fourier modes.
To explore the ramifications of this effect, 
we will also consider a case where only baselines which are 
instantaneously redundant are added coherently, and all other measurements
are added in quadrature when binning.  We discuss this model more in 
Section \ref{sec:foregrounds}.  

Since this method of calculating power spectrum sensitivities naturally
tracks the number of independent modes measured, sample variance is easily
included when combining modes by adding the value
of the cosmological power spectrum to each $(u,v,\eta)$-voxel
(where $\eta$ is the line-of-sight Fourier mode)
before doing any binning.  (Note that in the case where
only instantaneously redundant baselines are added coherently, partially
coherent baselines do not count as independent samples for the purpose of 
calculating sample variance.)   
Unlike \cite{BAOBAB}, we do not include the
effects of redshift-space distortions in boosting the line of sight
signal, since they will not boost the power spectrum of ionization 
fluctuations,
which is likely to dominate the 21\,cm power spectrum at these redshifts.
We also ignore other second order effects on the power spectrum,
such as the ``light-cone'' effect
\citep{datta_et_al_2012,la_plante_et_al_2013}.

\subsubsection{Telescope and Observational Parameters}
\label{sec:observationParams}

For the instrument value of $T_{\rm sys}$ we sum a frequency independent
100~K receiver temperature with a frequency dependent sky temperature,
$T_{\rm sky} = 60\rm{K}~(\lambda/1~{\rm m})^{2.55}$
\citep{ThompsonMoranSwenson}, giving a sky temperature of
351~K at 150~MHz.  Although this model is
$\sim100$~K lower than the system measured by \cite{PAPER32Limits},
it is consistent with recent LOFAR measurements \citep{InitialLOFAR1,LOFAR2013}.  Since the smaller field of view of HERA will lead
to better isolation of a Galactic cold patch, we choose this
empirical relation for our model.

For the primary beam, we use a simple Gaussian model with a Full-Width Half-Max
(FWHM) of $1.06\lambda/D = 8.7^{\circ}$ at 150 MHz.  We assume the beam
linearly evolves in shape as a function of frequency.  In the actual
HERA instrument design, the PAPER dipole serves as a feed to the larger parabolic
element.  Computational E\&M modeling suggests this setup will have a beam
with FWHM of $9.8^{\circ}$.  Furthermore, the PAPER dipole response is 
specifically designed to evolve more slowly with frequency than our linear
model.  Although the frequency dependence of the primary beam enters into
our sensitivity calculations in several places (including the pixel size in
the $uv$ plane), the dominant effect is to change the normalization of the
noise level in Equation \ref{eq:sensitivity}.  For an extreme case with
no frequency evolution in the primary beam size (relative to 150~MHz), 
we find that the resultant
sensitivities increase by up to 40\% at 100 MHz (due to a
smaller primary beam than the linear evolution model), and decrease by up
to 30\% at 200 MHz (due to larger beam).  While all
instruments will have some degree of primary beam evolution as a function
of frequency, this extreme model demonstrates that some of the poor
low-frequency (high-redshift) sensitivities reported below can be partially 
mitigated by a more frequency-independent instrument design (although at the
expense of sensitivity at higher frequencies).

It should be pointed out that for snap-shot observations,
the large-sized HERA dishes prevent
measurements of the largest transverse scales.  
At 150 MHz ($z = 8.5$), the minimum
baseline length of 14~m corresponds to a transverse $k$-mode of
$k_{\perp} = 0.0068 h\rm{Mpc}^{-1}$.  
This array will be unable to observe transverse modes on larger scales,
without mosaicing or otherwise integrating over longer
than one drift through the primary beam.
The sensitivity calculation used in this work does not account for
such an analysis, and therefore will limit the sensitivity
of the array to larger-scale modes.
For an experiment targeting unique cosmological information on the largest
cosmic scales (e.g. primordial non-Gaussianity), this effect may prove
problematic.  For studies of the EoR power spectrum, the
limitation on measurements at low $k_{\perp}$ has little effect on the end result, especially given
the near ubiquitous presence of
foreground contamination on large-scales in our models
(Section \ref{sec:foregrounds}).

The integration time $t$ on a given $k$ mode,
is determined by the length of time any baseline in the array samples each
$uv$ pixel over the course of the observation.  Since we assume a drift-scanning
telescope, the length of the observation is set by the
size of the primary beam.  The time
it takes a patch of sky to drift through the beam is the duration over which
we can average coherently.  For the $\sim 10^{\circ}$ primary beam model above, 
this time is $\sim40$ minutes.

We assume that there exists one Galactic ``cold patch'' spanning 6 hours in 
right ascension suitable for EoR observations, an assumption
which is based on measurements from both PAPER and the MWA and on previous
models (e.g. \citet{GSM}).  There are thus 9
independent fields of 40 minutes in right ascension (corresponding
to the primary beam size calculated above) 
which are observed per day.  We also assume EoR-quality
observations can only be conducted at night, yielding $\sim180$ days per year
of good observing.  Therefore, our thermal noise uncertainty 
(i.e. the $1\sigma$ error bar on the power spectrum) is reduced
by a factor of $\sqrt{9}\times 180$ over that calculated from one field,
whereas the contribution to the errors from sample variance
is only reduced by $\sqrt{9}$.

\subsection{Foregrounds}
\label{sec:foregrounds}

\begin{figure}[]
\centering
\includegraphics[width=.6\textwidth]{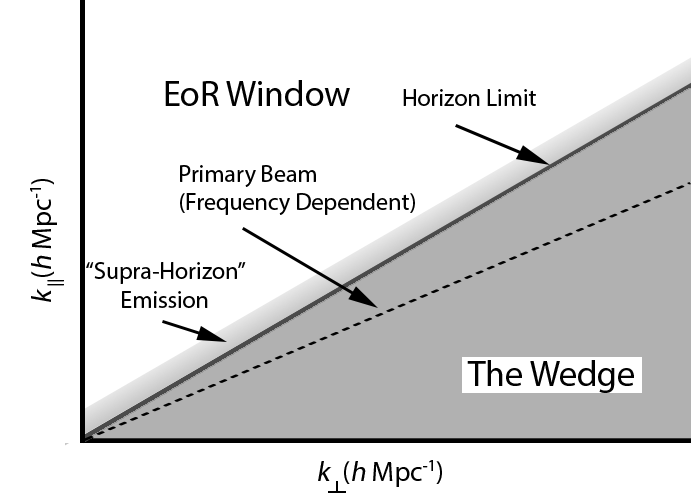}
\caption[Schematic diagram of the wedge and the EoR window.]{A schematic diagram of the wedge and EoR window in 2D $k$-space.
See Section \ref{sec:foregrounds} for explanations of the terms.}
\label{fig:wedge}
\end{figure}

Because of its spectral smoothness, foreground emission is expected to
contaminate low order line-of-sight Fourier modes in the power spectrum.
Of great concern, though, are chromatic effects in an interferometer's
response, which can introduce spectral structure into foreground emission.
However, recent work has shown that these chromatic mode-mixing effects do not
indiscriminately corrupt all the modes of the power spectrum. 
Rather, 
foregrounds are confined to a ``wedge''-shaped region in the 
2D $(k_{\perp},k_{\parallel})$ plane, with more $k_{\parallel}$ modes free
from foreground contamination on the shortest baselines (i.e. at the smallest
$k_{\perp}$ values)
 \citep{Dattapowerspec,VedanthamWedge,MoralesPSShapes,AaronDelay,CathWedge}, 
as schematically diagrammed in Figure \ref{fig:wedge}.
Power spectrum analysis in both \cite{X13} and
\cite{PoberWedge} reveal the presence of the wedge in actual 
observations.  The single-baseline approach 
\citep{AaronDelay} used in \cite{PoberWedge} yields a cleaner
EoR window, although at the loss of some sensitivity that comes from
combining non-redundant baselines.

However, there is still considerable debate about where to define the ``edge''
of the wedge.  Our three foreground models --- summarized in Table 
\ref{tab:fg}
--- differ in their choice of ``wedge edge.''  Our pessimistic model
also explores the possibility that systematic effects discussed in
\cite{Hazelton2013} could prevent to coherent addition of
partially redundant baselines.  
It should be noted that although we use the shorthand ``foreground model'' to
describe these three scenarios, in many ways these represent 
\emph{foreground removal models}, since they pertain to improvements
over current analysis techniques that may better separate foreground emission
from the 21\,cm signal.

\subsubsection{Foreground Removal Models}

At present, 
observational limits on the ``edge'' to the foreground wedge
in cylindrical $(k_{\perp},k_{\parallel})$-space are still somewhat
unclear.
\cite{PoberWedge} find the wedge to extend as much
as $\Delta k_{\parallel} = 0.05 \mbox{--} 0.1~h{\rm Mpc}^{-1}$
beyond the ``horizon limit,'' i.e., the $k_{\parallel}$ mode on a given baseline
that corresponds to the chromatic sine wave created by a 
flat-spectrum source of emission
located at the horizon.  (This mode in many ways represents a fundamental
limit, as the interference pattern cannot oscillate any faster for a 
flat-spectrum source of celestial emission; see \citet{AaronDelay}
for a full discussion of the wedge in the language of geometric delay space.)
Mathematically, the horizon limit is:
\begin{equation}
k_{\parallel,\rm{hor}} = \frac{2\pi}{Y}\frac{|\vec{b}|}{c} = \left(\frac{1}{\nu}\frac{X}{Y}\right) k_{\perp},
\end{equation}
where $|\vec{b}|$ is the baseline length in meters, $c$ is the speed of light,
$\nu$ is observing frequency, and $X$ and $Y$ are the previously described
cosmological scalars for converting observed bandwidths and solid
angles to $h{\rm Mpc}^{-1}$, respectively, defined in 
\cite{AaronSensitivity} and \cite{FurlanettoReview}.
\cite{PoberWedge} attribute the presence of ``supra-horizon'' emission 
--- emission at $k_{\parallel}$ values greater than the horizon limit --- 
to spectral structure in the
foregrounds themselves, which creates a convolving kernel in $k$-space.
\cite{AaronDelay} predict that the wedge could extend as much as 
$\Delta k_{\parallel} = 0.15~h{\rm Mpc}^{-1}$ beyond the horizon limit at the level of the 21\,cm
EoR signal.  This supra-horizon emission has a dramatic effect on the size
of the EoR window, increasing the $k_{\parallel}$ extent of the wedge
by nearly a factor of 4 on the 
$16\lambda$-baselines used by PAPER in \cite{PAPER32Limits}.

Others have argued that the wedge will extend not to the horizon limit,
but only to the edges of the field-of-view, outside of which emission is
too attenuated to corrupt the 21\,cm signal.  If achievable, 
this smaller wedge has a dramatic effect on sensitivity, since
theoretical considerations suggest that signal-to-noise
decreases quickly with increasing $k_{\perp}$ and $k_{\parallel}$. 
If one compares
the sensitivity predictions in \cite{AaronDelay} for PAPER-132 and 
\cite{MWAsensitivity} for MWA-128 (two comparably sized arrays), one
finds that these two different wedge definitions account for a large portion
of the difference between a marginal $2\sigma$ EoR 
detection and a 14$\sigma$ one.

While clearly inconsistent with the current results in
\cite{PoberWedge}, such a small wedge may be achievable
with new advances in foreground subtraction.
A large literature of work has gone into studying the removal of foreground
emission from 21\,cm data (e.g. \citet{Miguelstatistical}, 
\citet{Judd08}, \citet{paper2}, 
\citet{LT11}, \citet{Chapman1},
\citet{DillonFast}, \citet{Chapman2}).  
If successful, these techniques offer the
promise of working within the wedge.  However, despite the huge sensitivity boost,
working within the wedge clearly presents additional challenges beyond 
simply working
within the EoR window.  Working within the EoR window requires only keeping
foreground leakage from within the wedge to a level below the 21\,cm signal;
the calibration challenge for this task can be significantly reduced by
techniques which are allowed to remove EoR signal from within the wedge
\citep{PAPER32Limits}.  Working within the wedge requires
foreground removal with up to 1 part in $10^{10}$
accuracy (in mK$^2$) while leaving the 21\,cm signal unaffected.  Ensuring
that calibration errors do not introduce covariance between modes is
potentially an even more difficult task.  Therefore, 
given the additional effort it will take to be convinced that a residual 
excess of power post-foreground subtraction can be attributed to the EoR,
it seems
plausible that the first robust detection and measurements of the 21\,cm EoR
signal will come from modes outside the wedge.  

To further complicate the issue, several effects have been identified
which can scatter power from the wedge into the EoR window. 
\cite{MoorePolarization} 
demonstrate how combining redundant visibilities without image plane correction (as done by PAPER) can corrupt the EoR signal outside the wedge, due to the
effects of instrumental polarization leakage.
\cite{MoorePolarization} predict a level of contamination based on simulations
of the polarized point source population at low frequencies.
Although this predicted level of contamination
may already be ruled out by measurements from \cite{bernardi_et_al_2013},
these effects are a real concern for 21\,cm EoR experiments.
In the present analysis, however, we do not consider this contamination;
rather, we
assume that the dense $uv$ coverage of our concept array will allow
for precision calibration and image-based primary beam correction not possible
with the sparse PAPER array.  Through careful and concerted effort this
systematic should be able to be reduced to below the EoR level.

As discussed in Section \ref{sec:sense_calc}, we do consider the ``multi-baseline
mode mixing'' effects presented in \cite{Hazelton2013}.  These
effects may result when partially coherent baselines are combined
to improve power spectrum sensitivity, introducing
additional spectral structure in the foregrounds and thus 
complicating their mitigation. Conversely, the fact that only instantaneously
redundant baselines were combined in \cite{PoberWedge} and
\cite{PAPER32Limits} was partially responsible for the clear
separation between the wedge and EoR window.  
Since recent, competitive upper limits were set using this 
conservative approach, we include it as our ``pessimistic'' 
foreground strategy, noting that recent progress in accounting for 
the subtleties in  partially coherent analyses \citep{Hazelton2013} 
make it likely that better schemes will be available soon.

To encompass all these uncertainties in the foreground emission and foreground
removal techniques, we use three models for our foregrounds, which we refer to
in shorthand as ``pessimistic,'' ``moderate,'' and ``optimistic''.
These models are summarized in Table \ref{tab:fg}.
\begin{table}
\centering
\begin{tabular}{|l|p{4.4in}|}
\hline \bf{Model} & \bf{Parameters} \\
\hline
Moderate & Foreground wedge extends $0.1\ h{\rm Mpc}^{-1}$ beyond horizon limit \\ \hline
Pessimistic & Foreground wedge extends $0.1\ h{\rm Mpc}^{-1}$ beyond horizon limit, and only instantaneously redundant baselines can be combined coherently \\ \hline
Optimistic & Foreground wedge extends to FWHM of primary beam \\ \hline
\end{tabular}
\caption{Summary of the three foreground removal models.}
\label{tab:fg}
\end{table}
 
The ``moderate'' model is chosen to
closely mirror the predictions and data from
PAPER.  In this model the wedge is considered to extend 
$\Delta k_{\parallel} = 0.1\ h{\rm Mpc}^{-1}$ beyond the horizon limit.
The exact scale of the ``horizon+.1'' limit to the wedge is motivated
by the predictions of 
\cite{AaronDelay} and the measurements of 
\cite{PoberWedge} and
\cite{PAPER32Limits}.   Although the exact extent of the ``supra-horizon''
emission (i.e. the ``+.1'') at the level of the EoR signal remains to be 
determined, all of these constraints point to a range of 0.05 to 0.15
$h\rm{Mpc}^{-1}$.  The uncertainty in this value does not have a large
effect on the ultimate power spectrum sensitivity of next generation 
measurements.   
For shorthand, we will sometimes 
refer this model as having a
``horizon wedge.''

The ``pessimistic'' model uses the same horizon wedge as the moderate
model, but assumes that only 
instantaneously redundant baselines are coherently
combined. Any non-redundant baselines which sample the same $uv$ pixel
as another baseline --- either through being similar in length
and orientation or through the effects of earth rotation --- 
are added incoherently.
In effect, this model
covers the case where the multi-baseline mode-mixing of
\cite{Hazelton2013} cannot be corrected for.
Significant efforts are underway to develop pipelines
which correct for this effect and recover the sensitivity boost of partial
coherence;
since these algorithms have yet to be demonstrated on actual observations,
however, we consider this our pessimistic scenario.  

The final ``optimistic'' model, assumes the EoR window remains workable
down to $k_{\parallel}$ modes bounded by the FWHM of the primary beam,
as opposed to the horizon: $k_{\parallel,\rm{pb}} = \sin(\rm{FWHM}/2)\times k_{\parallel,\rm{hor}}$.
The specific choice of the FWHM is somewhat arbitrary; one could also
consider a wedge extending the first-null in the primary beam (although
this is ill-defined for a Gaussian beam model).  
Alternatively, one might envision a ``no wedge'' model
meant to mirror the case where
foreground removal techniques work optimally, removing all foreground
contamination down to the intrinsic spectral structure of the foreground
emission.
In practice, the small 
$\sim 10^{\circ}$ size of the HERA primary beam renders these
different choices effectively indistinguishable.
Therefore, our choice of the primary beam FWHM can also 
be considered representative
of nearly all cases where foreground removal proves highly effective.
As of the writing of this paper, no foreground removal algorithms have proven
successful to these levels, although this is admittedly a
somewhat tautological statement, since no published measurements have reached
the sensitivity level of an EoR detection.
Furthermore, the sampling point-spread function (PSF) 
in $k$-space at low $k$'s is expected to make
clean, unambiguous retrieval of these modes exceedingly difficult
\citep{LT11,AaronDelay}, although
the small size of the HERA primary beam ameliorates this problem
by limiting the scale of this PSF.  We find this
effect to represent a small ($\lesssim 5\%$)
correction to the low-$k$ sensitivities reported in this work.
In effect, the optimistic model is included to both show the
effects of foregrounds on the recovery of the 21\,cm power spectrum, and to
give an impression of what could be achievable.  For shorthand, this model will
be referred to as having a ``primary beam wedge.''

Incorporating these foreground models into the sensitivity calculations
described in Section \ref{sec:instrument} is quite straightforward.  Modes deemed
``corrupted'' by foregrounds according to a model are simply excluded
from the 3D $k$-space cube, and therefore contribute no sensitivity
to the resultant power spectrum measurements.

\subsection{Reionization}
\label{sec:eor}

In order to encompass the large theoretical uncertainties in the
cosmic reionization history, we use the publicly available 
\texttt{21cmFAST}\footnote{http://homepage.sns.it/mesinger/DexM\_\_\_21cmFAST.html/}
code v1.01 \citep{Mesinger:2007hl,21CMFAST}.
This semi-numerical code allows us to quickly generate large-scale simulations
of the ionization field (400 Mpc on a side) 
while varying key parameters to examine the possible
variations in the 21\,cm signal.  Following \cite{mesinger_et_al_2012}, we choose
three key parameters to encompass the maximum variation in the signal: 
\begin{enumerate}

\item \emph{$\zeta$, the ionizing efficiency}: $\zeta$ is a conglomeration of
a number of parameters relating to the amount of ionizing photons escaping
from high-redshift galaxies: $f_{\rm esc}$, the fraction of ionizing
photons which escape into the IGM, $f_*$, the star formation efficiency,
$N_\gamma$, the number of ionizing photons produced per baryon in stars,
and $n_{\rm rec}$ the average number of recombinations per baryon.
Rather than parameterize the uncertainty in these quantities individually,
it is common to define $\zeta = f_{\rm esc}f_*N_\gamma/(1+n_{\rm rec})$
\citep{furlanetto2}.  We explore a range
of $\zeta = 10-50$ in this work, which is generally consistent with current
CMB and Ly$\alpha$ constraints on reionization \citep{mesinger_et_al_2012}.

\item \emph{$T_{\rm vir}$, the minimum virial temperature of halos
producing ionizing photons}: $T_{\rm vir}$ parameterizes the mass
of the halos responsible for reionization.  Typically, $T_{\rm vir}$ is
chosen to be $10^4$~K, which corresponds to a halo mass of $10^8~\rm{M}_\odot$
at $z = 10.$  This value is chosen because it represents the temperature at
which atomic cooling becomes efficient.  In this work, we explore $T_{\rm vir}$
ranging from $10^3 \mbox{--} 3 \times 10^5$~K to span the uncertainty in high-redshift
galaxy formation physics as to which halos host significant stellar
populations (see e.g. \citet{haiman_et_al_1996},
\citet{abel_et_al_2002} and \citet{bromm_et_al_2002} for lower mass limits
on star-forming halos,
and e.g. \citet{mesinger_and_dijkstra_2008} and \citet{okamoto_et_al_2008}
for feedback effects which can suppress low mass halo star formation).

\item \emph{$R_{\rm mfp}$, the mean free path of ionizing photons 
through the intergalactic medium (IGM)}: $R_{\rm mfp}$ sets the
maximum size of HII regions that can form during reionization.  Physically,
it is set by the space density of Lyman limit systems, which act as sinks
of ionizing photons.  In this work, we explore a range of mean free paths
from 3 to 80 Mpc, spanning the uncertainties in current measurements
of the mean free path at $z\sim 6$ \citep{songaila_and_cowie_2010}.

\end{enumerate}
We note there are many other tunable parameters that could affect
the reionization history.  In particular,
the largest 21\,cm signals can be produced in models where
the IGM is quite cold during
reionization (cf. \citet{PAPER32Limits}).  We do not
include such a model here, and rather focus on the potential
uncertainties within ``vanilla'' reionization; for an analysis of
the detectability of early epochs of X-ray heating, see 
\cite{Christian:2013} and \cite{Mesinger:2013c}.
Also note that \texttt{21cmFAST} assumes the values of the EoR
parameters are constant over all redshifts considered.
With the exception of our three EoR variables, we use the fiducial parameters
of the \texttt{21cmFAST} code; see \cite{21CMFAST} for more details.

Note we do assume that $T_{\rm spin} \gg T_{\rm CMB}$ at all epochs, which
could
potentially create a brighter signal at the highest redshifts.  Given
that thermal noise generally dominates the signal at the highest redshifts
regardless, we choose to ignore this effect, noting that it will only
increase the difficulties of $z > 10$ observations we describe below.
(Although this situation may be changed by the 
alternate X-ray heating scenarios considered in 
\citet{Mesinger:2013c}.)

\subsubsection{``Vanilla'' Model}

\begin{figure}[]
\centering
\includegraphics[width=.6\textwidth]{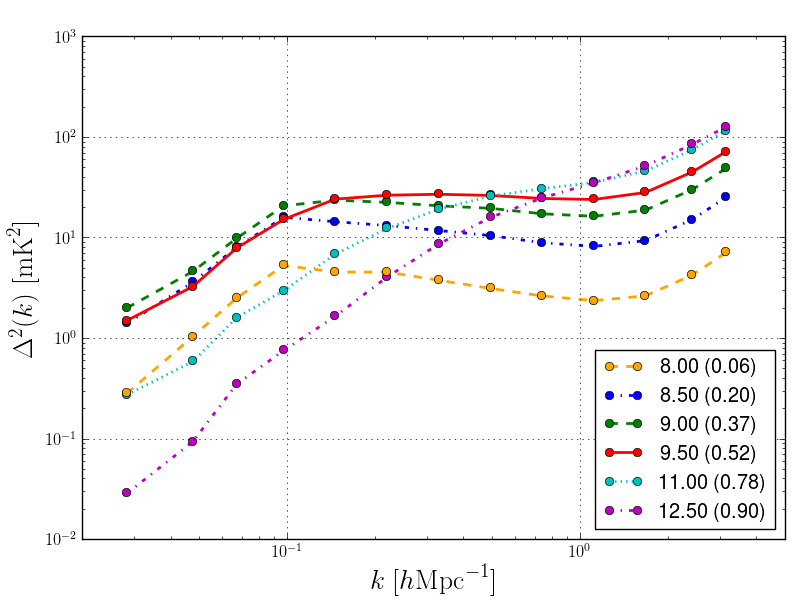}
\caption[Fiducial reionization model power spectra.]{Power spectra at several redshifts for our vanilla
reionization model with $\zeta = 31.5,\ T_{\rm vir} = 1.5 \times 10^4~\rm{K},$ 
and $R_{\rm mfp} = 30~\rm{Mpc}$.
Numbers in parentheses give the neutral fraction at that redshift.}
\label{fig:pspec_vanilla}
\end{figure}

For the sake of comparison, it is worthwhile to have one fiducial model
with ``middle-ground'' values for all the parameters in question.  
We refer to this model as our ``vanilla'' model.  Note that this
model was not chosen because we believe it most faithfully
represents the true reionization history of the universe
(though it is consistent with current observations).  Rather,
it is simply a useful point of comparison for all the other realizations
of the reionization history.
In this model, the values of the three parameters being studied are
$\zeta = 31.5,\ T_{\rm vir} = 1.5 \times 10^4~\rm{K}$ and $R_{\rm mfp} = 30~\rm{Mpc}$.
This model achieves 50\% ionization at $z\sim 9.5$, and complete ionization
at $z\sim7$.
The redshift evolution of the power spectrum in this model is shown in Figure
\ref{fig:pspec_vanilla}.

\subsubsection{The Effect of the Varying the EoR Parameters}
\label{sec:varying}

\begin{figure*}[]
\centering
\includegraphics[width=\textwidth]{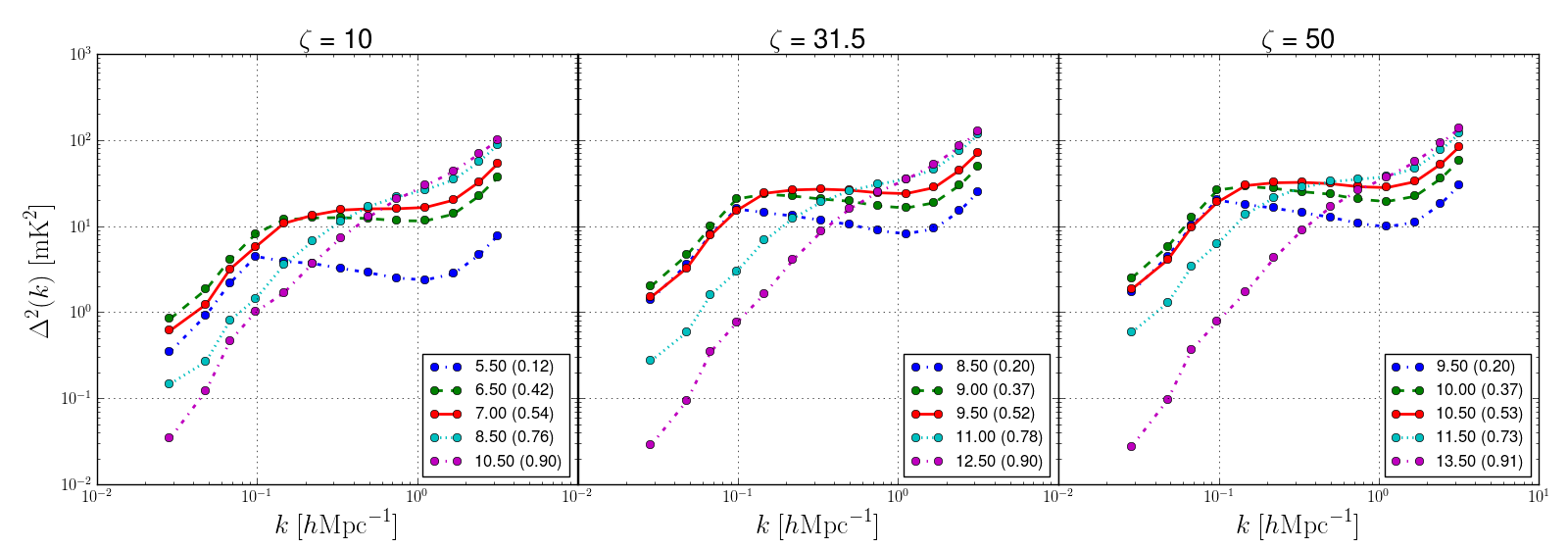}
\includegraphics[width=\textwidth]{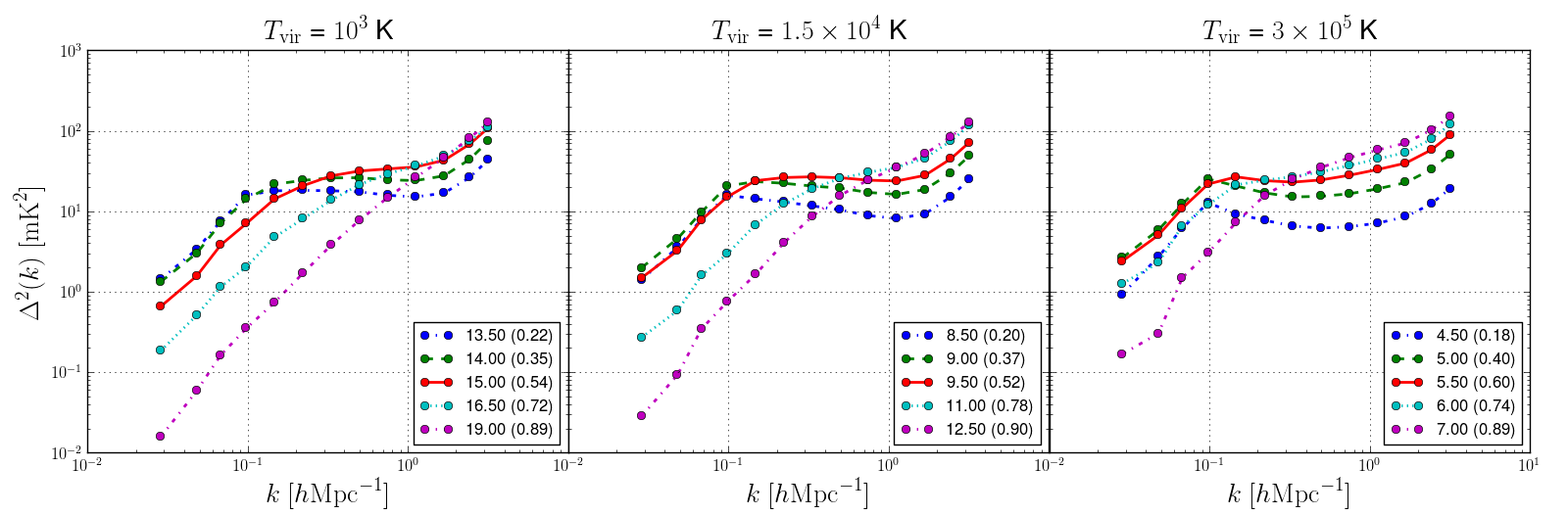} 
\includegraphics[width=\textwidth]{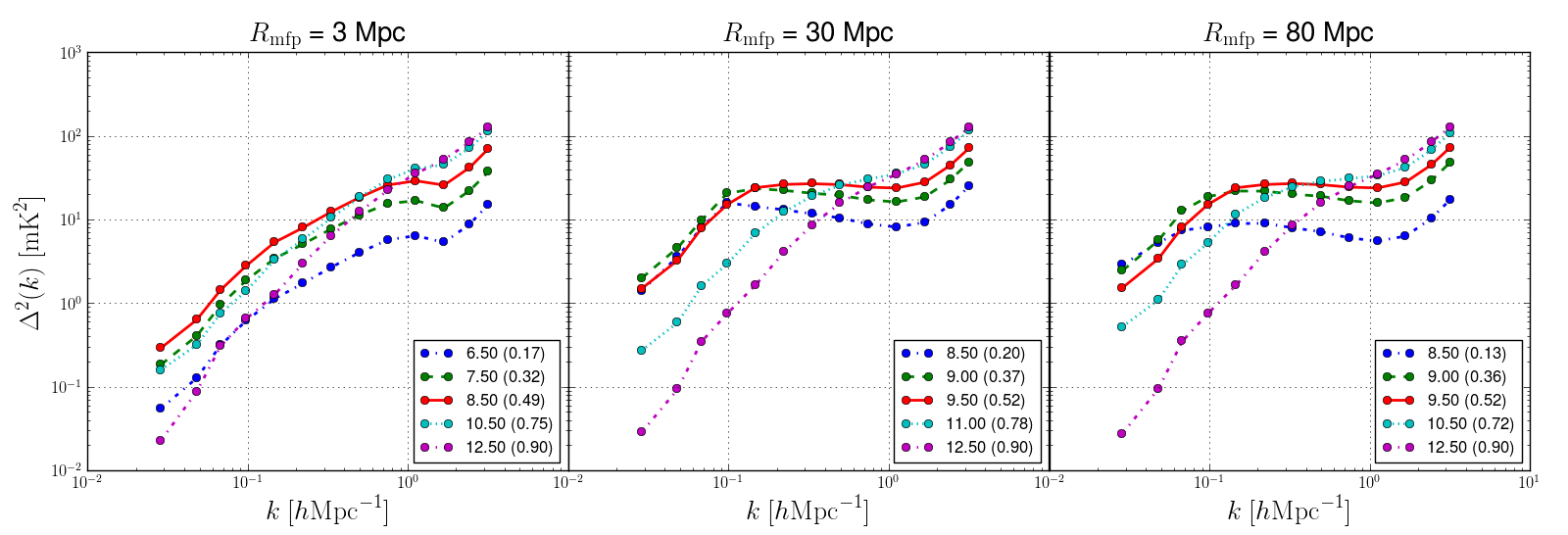} 
\caption[Power spectra varying a single reionization parameter.]{Power spectra as a function of redshift for the low, high,
and fiducial
values of the ionizing efficiency, $\zeta$ (top row), $T_{\rm vir}$ (middle
row) and $R_{\rm mfp}$ (bottom row).  Exact values of each
parameter are given in the panel title.
Numbers in parentheses give the neutral fraction at that redshift.
The central panel is the vanilla model and is identical to Figure
\ref{fig:pspec_vanilla} (although the $z=8$ curve is not included for
clarity).  Colors in each panel map to roughly the same
neutral fraction.  Qualitative effects of varying each parameter are
apparent: $\zeta$ changes the timing of reionization but not the shape
of the power spectrum; $T_{\rm vir}$ drastically alters the timing
of reionization with smaller effects on the power spectrum shape; and
small values of $R_{\rm mfp}$ reduce the amount of low $k$ power.}
\label{fig:pspecs}
\end{figure*}

The effects of varying $\zeta$, $T_{\rm vir}$ and $R_{\rm mfp}$ are
illustrated in Figure \ref{fig:pspecs}.
Each row shows the effect of varying one of the three parameters while
holding the other two fixed.
The middle panel in each
row is for our vanilla model, and thus is the same as Figure 
\ref{fig:pspec_vanilla} (although the $z=8$ curve is not included for
clarity).
Several qualitative observations can immediately be made.  Firstly, we can see
from the top row that
$\zeta$ does not significantly change the shape of the power spectrum, but only the duration
and timing of reionization.  This is expected, since the same sources
are responsible for driving reionization regardless of $\zeta$.  Rather,
it is only the number of ionizing photons that these sources produce that
varies.

Secondly, we can see from the middle row that the most dramatic
effect of $T_{\rm vir}$ is to substantially change the timing of reionization.
Our high and low values of $T_{\rm vir}$ create reionization histories
that are inconsistent with current constraints from the CMB and Ly$\alpha$
forest \citep{fan_et_al_2006,hinshaw_et_al_2012}.  This alone does not rule out these values of $T_{\rm vir}$ for the
minimum mass of reionization galaxies, but it does mean that some additional 
parameter would have to be adjusted within our vanilla model to create
reasonable reionization histories.  We can also see 
that the halo virial temperature affects the shape of the power spectrum.  When
the most massive halos are responsible for reionization, we see significantly
more power on very large scales than in the case where low-mass galaxies
reionize the universe.

Finally, the bottom row shows that the mean free path of ionizing
photons also affects the amount of large scale power in the 21\,cm power spectrum.
$R_{\rm mfp}$ values of 30 and 80~Mpc produce essentially
indistinguishable power spectra,
except at the very largest scales at late times.  However, the very small
value of $R_{\rm mfp}$ completely changes the shape of the power spectrum, 
resulting in a steep slope versus $k$, even at
50\% ionization, where most models show a fairly flat power spectrum up to
a ``knee'' feature on larger scales.  In section \ref{sec:distinguishing}, we
consider using some of these characteristic features to qualitatively assess
properties from reionization in 21\,cm measurements.

\section{Results}
\label{sec:results}

In this section, we will present the predicted sensitivities
that result from combinations of EoR and foreground models.  
We will focus predominantly on the moderate model
where one can take advantage of partially-redundant baselines,
but the wedge still contaminates $k_{\parallel}$
modes below $0.1~h\rm{Mpc}^{-1}$ above the horizon limit.`
In presenting the sensitivity levels 
achievable under the other two foreground models, we focus on the
additional science that will be prevented/allowed if these models represent
the state of foreground removal analysis.

We will take several fairly orthogonal approaches towards understanding
the science that will be achievable.
First, in Section \ref{sec:sensitivity},
our approach is to attempt to cover the broadest range
of possible power spectrum shapes and amplitudes in order to make
generic conclusions about the detectability of the 21\,cm power spectrum.
In Section \ref{sec:timing}, Section \ref{sec:imaging}, and 
Section \ref{sec:distinguishing}, we focus on our vanilla reionization model
and semi-quantitatively explore the physical lessons the predicted sensitivities
will permit.
Finally, in Section \ref{sec:adrian}, we undertake a Fisher matrix analysis
and focus on specific determinations of EoR parameters with respect to the
fiducial evolution of our vanilla model, exploring the degeneracies between
parameters and providing lessons in how to break them.
The end result of these various analyses is a holistic picture about the
kinds of information we can derive from next generation EoR measurements.

\subsection{Sensitivity Limits}
\label{sec:sensitivity}

In this section, we consider the signal-to-noise ratio of
power spectrum measurements 
achievable under our various foreground removal models.
The main results are presented in Figures \ref{fig:pspec-errs},
\ref{fig:hor-snr-matrix}, \ref{fig:inst-red-snr-matrix},
and \ref{fig:pb-snr-matrix}.
Figure \ref{fig:pspec-errs}
shows the constraints on the 50\% ionization power spectrum
in our vanilla model
for each of the three foreground models, as well as the measurement
significances of alternate ionization histories using the vanilla model.
Figures \ref{fig:hor-snr-matrix}, \ref{fig:inst-red-snr-matrix},
and \ref{fig:pb-snr-matrix} show the power spectrum measurement
signifances that result when the EoR parameters are varied
for the moderate, pessimistic, and optimistic foreground models 
respectively.
 
\subsubsection{Methodology}
In order to explore the largest range of possible power spectrum shapes and amplitudes, it is important to keep in mind the small but non-negligible spread between various theoretical predictions in the literature.  To avoid having to run excessive numbers of simulations, we make use of the observation that much of the differences between simulations is due to discrepancies in their predictions for the ionization history $x_\textrm{HI} (z)$, in the sense that the differences decrease if neutral fraction (rather than redshift) is used as the time coordinate for cross-simulation comparisons.  We thus make the ansatz that given a single set of parameters $(\zeta,T_{\rm vir},R_{\rm mfp})$, the \texttt{21cmFAST} power spectrum outputs can (modulo an overall normalization factor) be translated in redshift to mimic the outputs of alternative models that predict a different ionization history.
In practice, the \texttt{21cmFAST} simulation provides a suite
of power spectra in either (a) fixed steps in $z$ or (b) approximately
fixed steps in $x_{\rm HI}$, but constrained to appear at a single $z$. 
We utilize the latter set,
and ``extrapolate'' each neutral fraction to a variety
of redshifts by scaling the amplitude with the square of the 
mean temperature of the IGM
as $(1+z)$, as anticipated when ionization fractions
dominate the power spectrum \citep{Matt3,LidzRiseFall}.
While not completely motivated by the physics 
of the problem (since within \texttt{21cmFAST} a given set of EoR parameters
does produce only one reionization history), this approach allows
us to explore an extremely wide range of power spectrum amplitudes
while running a reasonable number of simulations.

\subsubsection{Moderate Foreground Model}

Figure \ref{fig:pspec-errs} shows forecasts for constraints on our
fiducial reionization model under the three foreground scenarios.

\begin{center}
\includegraphics[width=.48\textwidth]{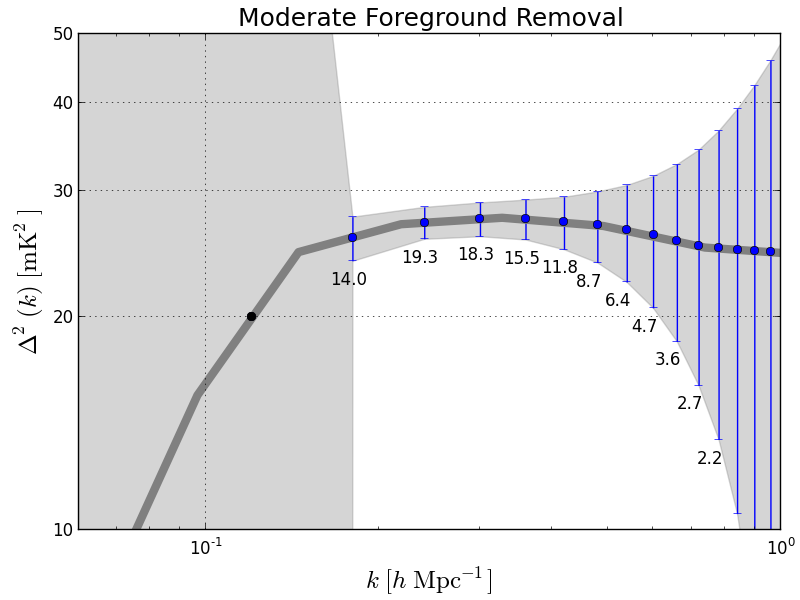}\includegraphics[width=.5\textwidth]{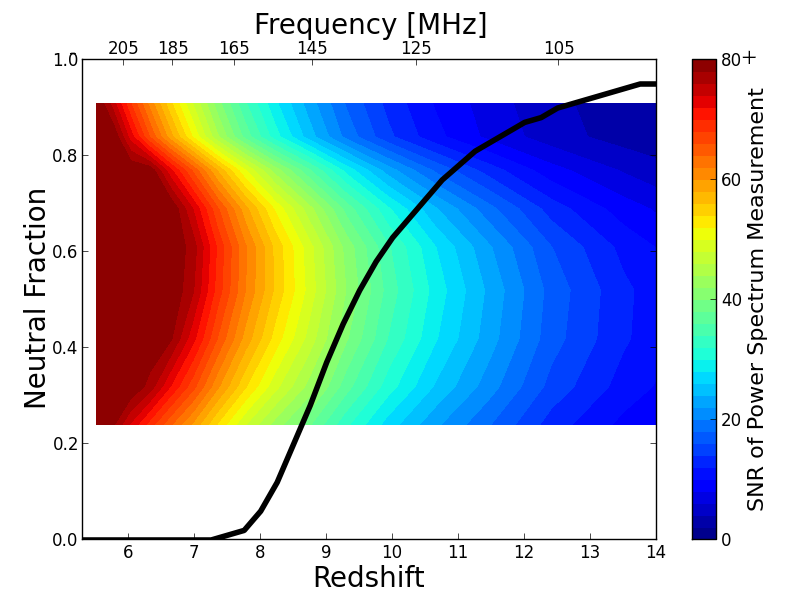}

\includegraphics[width=.48\textwidth]{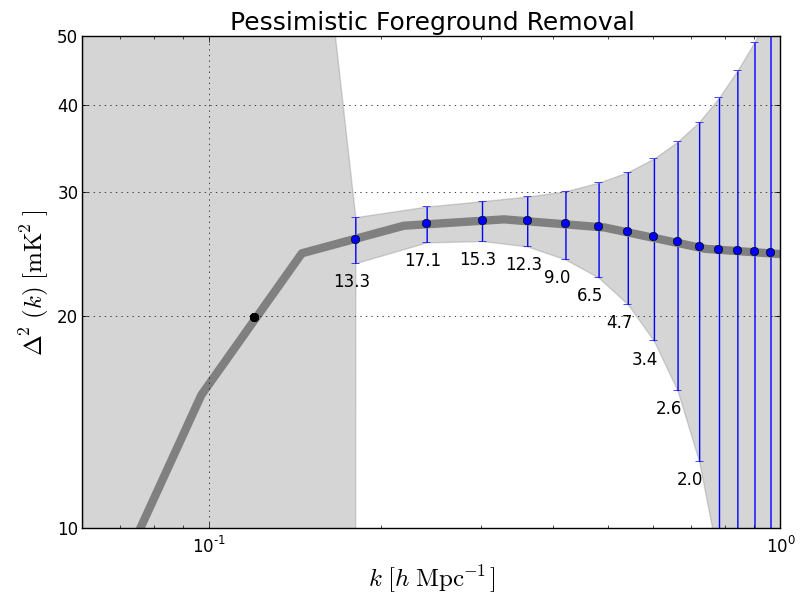}
\includegraphics[width=.5\textwidth]{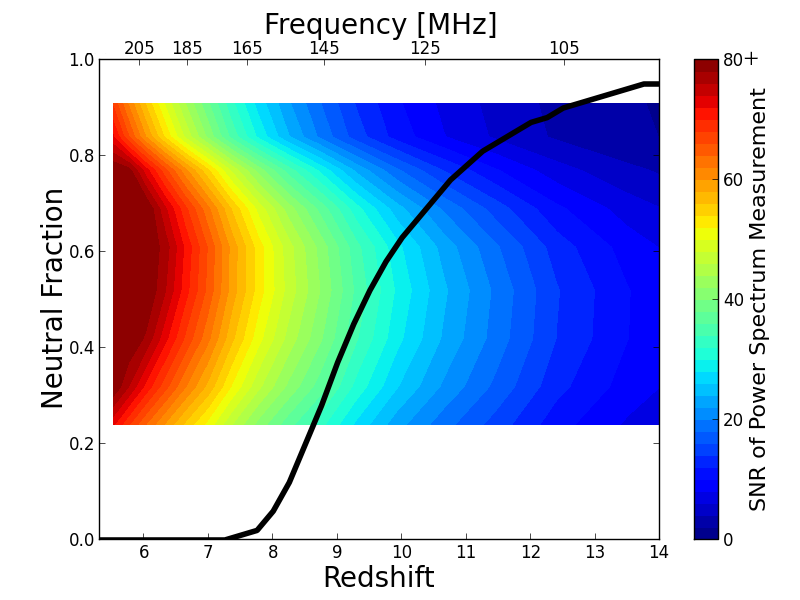}

\includegraphics[width=.48\textwidth]{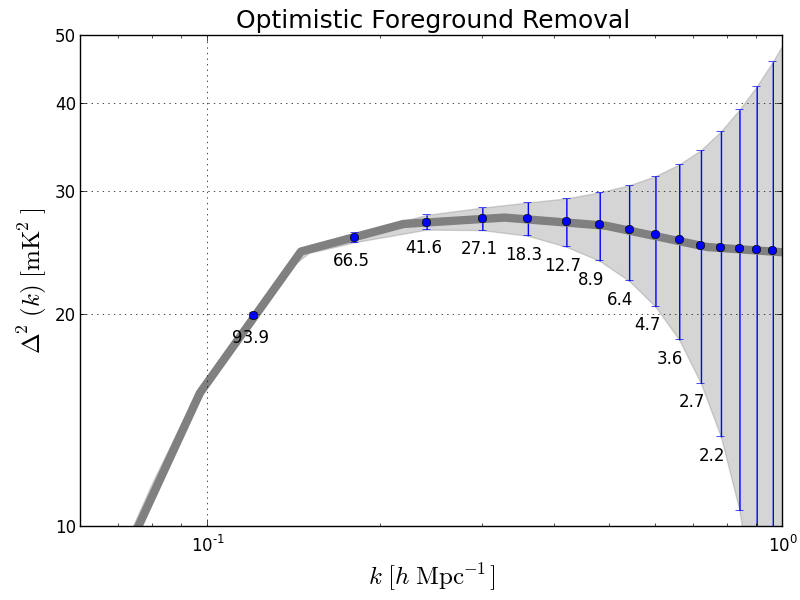}\includegraphics[width=.5\textwidth]{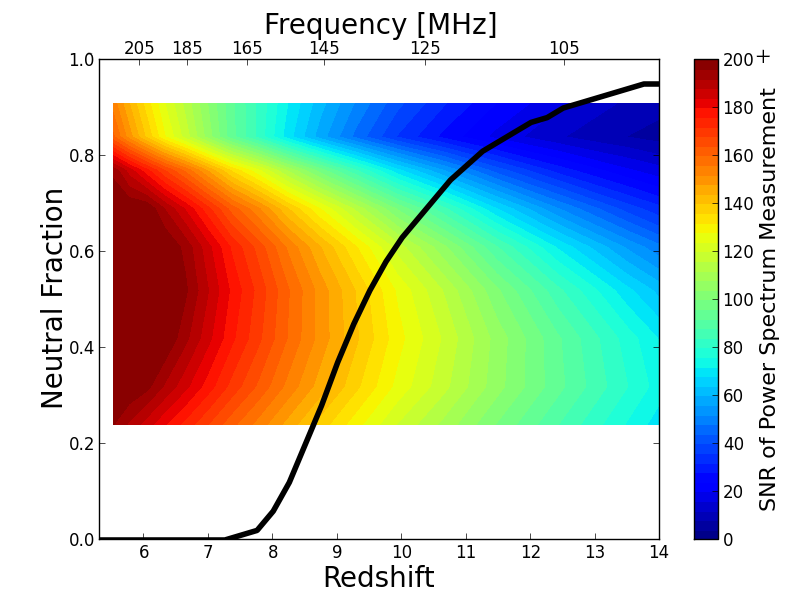}
\begin{spacing}{1}
\captionof{figure}[Sensitivity and SNR for different foreground removal models.]{\emph{Left:} Power spectrum constraints on the fiducial EoR model
at $z = 9.5\ (53\%$ ionization) for each of the three foreground removal models:
moderate (top), pessimistic (middle) and optimistic (bottom).
The shaded gray region shows the $1\sigma$ error range, whereas the
location of the blue error bars indicate the binned sampling pattern; the binning
is set by the bandwidth of 8 MHz.
Black points without error bars
indicate measurements allowed by instrumental parameters, but
rendered unusable according to the foreground model.  
The net result of
these measurements are $38\sigma$, $32\sigma$, and $133\sigma$
detections of the fiducial power spectrum for the moderate, pessimistic
and optimistic models, respectively.
Individual numbers below each error bar indicate the significance of the
measurement in that bin.
\emph{Right:} Colored contours show the total SNR of a power spectrum detection 
as a function of redshift and
neutral fraction for the three foreground removal models:
moderate (top), pessimistic (middle) and optimistic (bottom).
The black curve shows the fiducial evolution of the 
vanilla model; contour values off of the black curve are obtained by translating the
fiducial model in redshift.  This figure therefore allows one to examine
the SNR for a far broader range of reionization histories than only those
predicted by simulations with vanilla model parameters.  Alternative evolution
histories are less physically motivated, since a given set of EoR parameters
does only predict one evolution history.  
The plotted sensitivities assume 8 MHz bandwidths are used to measure
the power spectra, so not all points in the horizontal direction are 
independent.  The incomplete coverage versus $x_{\rm HI}$ does not indicate
that measurements cannot be made at these neutral fractions; rather, it
is a feature of the \texttt{21cmFAST} code, and is explained in 
Section \ref{sec:sensitivity}.
}
\end{spacing}
\label{fig:pspec-errs}

\end{center}

\noindent\hrulefill

The left-hand panels of the figure show the constraints on the
spherically averaged power
spectrum at $z = 9.5$, the point of 50\% ionization in this model, for 
the three foreground removal models.
(The 50\% ionization point generally corresponds to the peak power
spectrum brightness at the scales of interest --- as can be seen
in Figure \ref{fig:pspecs} --- making its detection
a key goal of reionization experiments \citep{LidzRiseFall,
BittnerLoeb}.) For the moderate model (top row), the
errors amount to a $38\sigma$ detection of the 21\,cm power spectrum at 
50\% ionization. 

\begin{figure*}[]
\centering
\includegraphics[width=.8\textwidth]{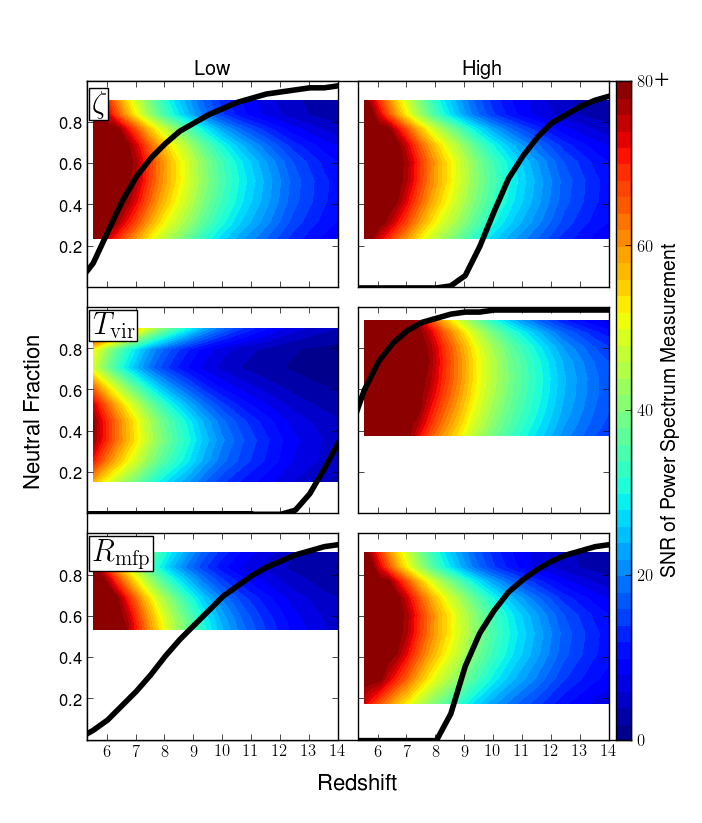}
\caption[Moderate foreground scenario SNR contours.]{Signal-to-noise ratio of 21\,cm power spectrum detections 
under the moderate foreground scenario for the high and low
values of the parameters in our EoR models as functions of neutral
fraction and redshift.  In each panel,
one parameter is varied, while the other two are held fixed
at the ``vanilla'' values.  The black curve shows the
fiducial evolution for that set of model parameters.
The incomplete coverage versus $x_{\rm HI}$ does not indicate
that measurements cannot be made at these neutral fractions; rather, it
is a feature of the \texttt{21cmFAST} code, and is explained in 
Section \ref{sec:sensitivity}.
\emph{Top}: the ionizing efficiency, $\zeta$.  
Values are $\zeta = 10$ (left) and $\zeta = 50$ (right).
\emph{Middle}: the minimum virial temperature of ionizing haloes, $T_{\rm vir}$.
Values are $T_{\rm vir} = 1\times 10^3$~K (left) and $T_{\rm vir} = 3 \times 10^5$~K (right).
\emph{Bottom}: the mean free path for ionizing photons through the IGM, $R_{\rm mfp}$.
Values are $R_{\rm mfp} = 3$~Mpc (left) and $R_{\rm mfp} = 80$~Mpc (right).
The moderate foreground removal scenario generically
allows for a high significance measurement for nearly any reasonable 
reionization history.
}
\label{fig:hor-snr-matrix}
\end{figure*}

The right-hand panels of Figure \ref{fig:pspec-errs}
warrant somewhat detailed explanation.  The three rows
again correspond to the three foreground removal models.  In each panel,
the horizontal axis
shows redshift and the vertical axis shows neutral fraction; thus this space
spans a wide range of possible reionization histories.  The black
curve is the evolution of the vanilla model through this space. 
The colored contours show
the signal-to-noise ratio of a HERA measurement of the 21\,cm power spectrum
at that point in redshift/neutral fraction space, where the power
spectrum of a given $x_{\rm HI}$ is extrapolated in redshift space
as described in the beginning of Section \ref{sec:sensitivity}.
The colorscale is set to saturate at different values in each row:
$80\sigma$ (moderate and pessimistic) and $200\sigma$ (optimistic).
These sensitivities assume 8 MHz bandwidths are used to measure
the power spectra, so not every value on the redshift-axis can be taken as
an independent measurement.
The non-uniform coverage versus ionization fraction (i.e. the white
space at high and low values of $x_{\rm HI}$) --- which appears with different
values in the panels of Figures \ref{fig:pspec-errs}, \ref{fig:hor-snr-matrix}, 
\ref{fig:inst-red-snr-matrix}, and \ref{fig:pb-snr-matrix} --- 
is a feature of the \texttt{21cmFAST} code when
attempting to produce power spectra for a set of input parameters at relatively
even spaced values of ionization fraction.  
The black line is able to extend into the white region because it 
was generated to have uniform spacing in $z$ as opposed to $x_{\rm HI}$.
The fact that these values are
missing has minimal impact on the conclusions drawn in this work.

In the moderate model, the 50\% ionization point of the fiducial power spectrum
evolution is detected at $\sim40\sigma$.  However, we see that
nearly every ionization
fraction below $z \sim 9$ is detected with equally high significance.
In general, the contours follow nearly vertical lines through this space.
This implies that the evolution of sensitivity as a function of redshift
(which is primarily driven by the system temperature) is much stronger
than the evolution of power spectrum amplitude as a function of neutral fraction
(which is primarily driven by reionization physics).

Figure \ref{fig:hor-snr-matrix} shows signal-to-noise contour plots
for six different variations of our EoR parameters, using only the
moderate foreground scenario.  (The pessimistic and optimistic equivalents
of this figure are shown in Figures \ref{fig:inst-red-snr-matrix}
and \ref{fig:pb-snr-matrix}, respectively.)
In each panel,
we have varied one parameter from the fiducial vanilla model.  In particular,
we choose the lowest and highest values of each parameter considered in
Section \ref{sec:eor}.  Since we
extrapolate each power spectrum to a wide variety of redshifts, choosing
only the minimum and maximum values leads to little loss of generality.
Rather, we are picking extreme shapes and amplitudes for the power spectrum,
and asking whether they can be detected if such a power spectrum were
to correspond to a particular redshift.  And, as with the vanilla model
shown in Figure \ref{fig:pspec-errs},
it is clear that the moderate foreground removal scenario
allows for the 21\,cm power spectrum
to be detected with very high significance
below $z\sim 8-10$, depending on the EoR model.
Relative to the
effects of system temperature, then, the actual brightness of the power
spectrum as a function of neutral fraction plays a small role in determining
the detectability of the cosmic signal.  Of course, there is still a wide
variety of power spectrum brightnesses; for a given
EoR model, however, the relative power spectrum amplitude evolution as a 
function of redshift is fairly small.

There are also several more specific points about Figure 
\ref{fig:hor-snr-matrix} that warrant comment.  Firstly, as stated in
Section \ref{sec:varying}, the ionizing efficiency $\zeta$ has
little effect on the shape of the power spectrum, but only on the timing and
duration of reionization.  This is clear from the identical sensitivity
levels for both values of $\zeta$, as well as for the vanilla model shown
in Figure \ref{fig:pspec-errs}.  Secondly, we reiterate that by tuning
values of $T_{\rm vir}$ alone, we can produce ionization histories that
are inconsistent with observations of the CMB
and Lyman-$\alpha$ forest.  In our analysis here, we extrapolate
the power spectrum shapes produced by these extreme histories to more
reasonable redshifts to show the widest range of possible scenarios.
The fact that the fiducial evolution histories (black lines)
of the $T_{\rm vir}$ row are wholly unreasonable is understood, and does
not constitute an argument against this type of analysis.  

\subsubsection{Other Foreground Models}

It is clear then, that the moderate foreground removal scenario
will permit high sensitivity measurements of the 21\,cm power with
the next generation of experiments.  Before considering what types of science
these sensitivities will enable, it is worthwhile to consider the effects
of the other foreground removal scenarios.

\begin{figure*}[]
\centering
\includegraphics[width=.8\textwidth]{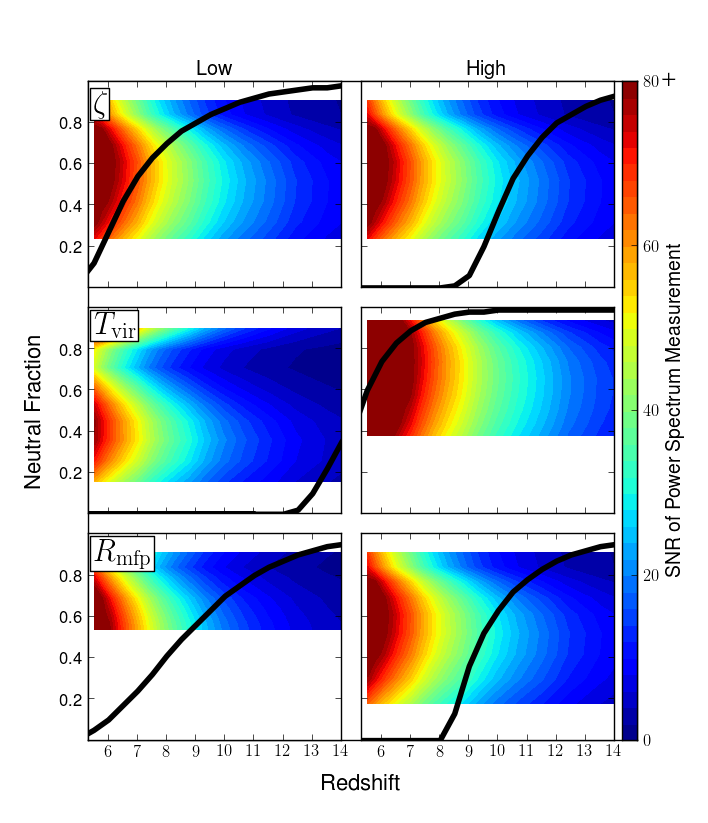}
\caption[Pessimistic foreground scenario SNR contours.]{
Same as Figure \ref{fig:hor-snr-matrix}, but for the pessimistic
foreground model.
Note that the color-scale is the same as Figure
\ref{fig:hor-snr-matrix} and saturates at $80\sigma$.}
\label{fig:inst-red-snr-matrix}
\end{figure*}

Our pessimistic scenario assumes
--- like the moderate scenario --- that
foregrounds irreparably corrupt $k_{\parallel}$ modes within the horizon
limit plus $0.1~h{\rm Mpc}^{-1}$, but also conservatively omits the 
coherent addition of partially redundant baselines in an effort to avoid
multi-baseline systematics.
As stated, this is the most
conservative foreground case we consider.
The achievable constraints on
our fiducial vanilla power spectrum under this model
were shown in the second row of Figure \ref{fig:pspec-errs}; Figure
\ref{fig:inst-red-snr-matrix} shows the sensitivities for other EoR models.
\label{fig:pspec-inst-red}
The sensitivity loss associated with coherently adding only instantaneously
redundant baselines is fairly small, $\sim 20\%$.  It should be noted
that this pessimistic model affects only the thermal noise error bars
relative to the moderate model; sample variance contributes the same
amount of uncertainty in each bin.  In an extreme sample variance limited
case, the pessimistic and moderate models would yield the same power
spectrum sensitivities.  We will further explore 
the contribution of sample variance
to these measurements in Section \ref{sec:imaging}.  Here we note that the
pessimistic model generally increases the \emph{thermal noise} uncertainties by
$30\mbox{--}40\%$ over the moderate model.  This effect will be greater
for an array with less instantaneous redundancy than the HERA concept design.

\begin{figure*}[]
\centering
\includegraphics[width=.8\textwidth]{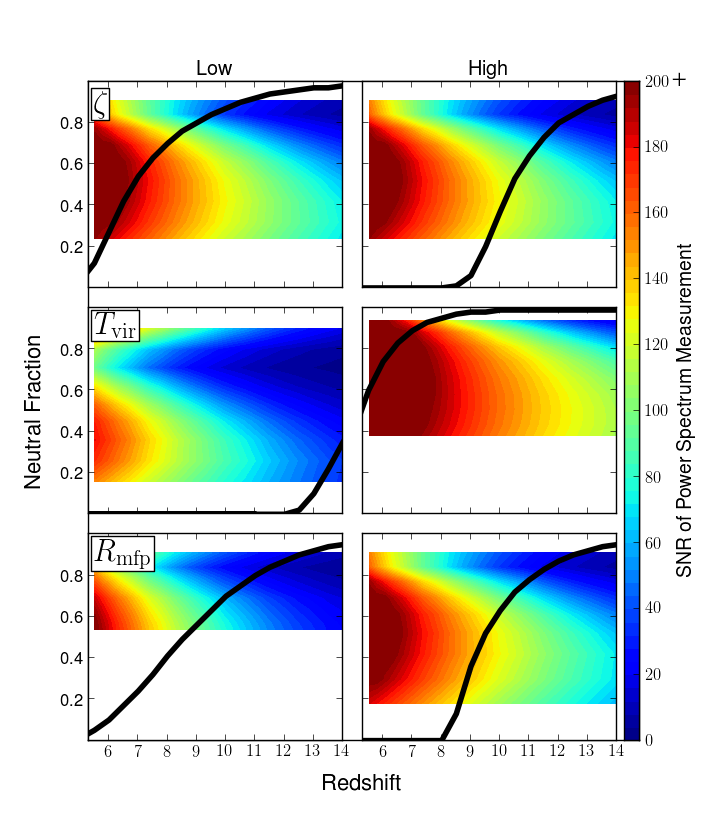}
\caption[Optimistic foreground scenario SNR contours.]{
Same as Figures \ref{fig:hor-snr-matrix} and \ref{fig:inst-red-snr-matrix}, 
but for the optimistic
foreground model.
Note that the color-scale has changed to saturate at $200\sigma$.}
\label{fig:pb-snr-matrix}
\end{figure*}

Finally, the sensitivity to the vanilla EoR model under the optimistic
foreground removal scenario is shown
in the bottom row of Figure \ref{fig:pspec-errs}.
Figure \ref{fig:pb-snr-matrix} shows the
sensitivity results for the other EoR scenarios.
The sensitivities for the optimistic model are exceedingly high.  
Comparison of the top and bottom rows of
Figure \ref{fig:pspec-errs} shows that this model does not uniformly
increase sensitivity across $k$-space, but rather the gains are entirely
at low $k$s.  This behavior is expected, since the effect of the optimistic
model is to recover large scale modes that are treated as foreground contaminated
in the other models.  The sensitivity boosts come from the fact that thermal 
noise is very low at these large scales, since noise scales as $k^3$ 
while the cosmological signal remains relatively flat in
$\Delta^2(k)$ space.  We consider the effect of sample variance in these
modes in Section \ref{sec:imaging}.

\subsection{The Timing and Duration of Reionization}
\label{sec:timing}

\begin{figure*}[]
\centering
\includegraphics[width=\textwidth]{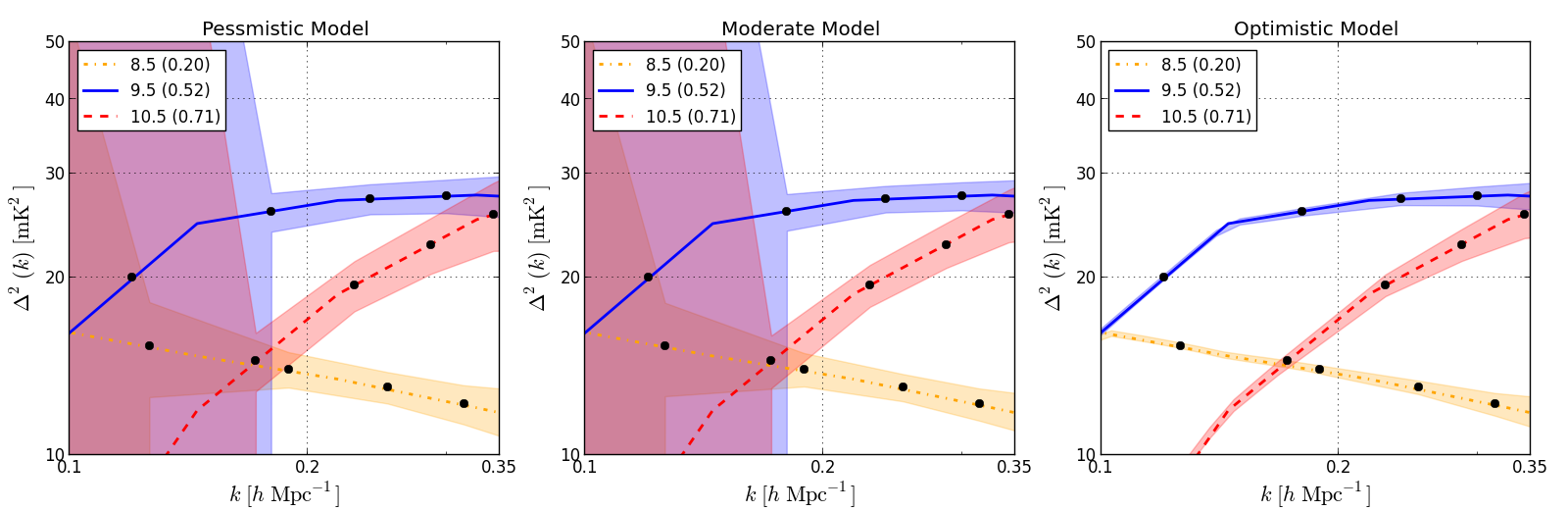}
\caption[$1\sigma$ power spectrum errors in different foreground scenarios.]{
$1\sigma$ uncertainties in the measurements of the fiducial EoR power 
spectrum at redshifts 8.5, 9.5 and
10.5, corresponding to neutral fractions of 0.20, 0.52 and 0.71, 
respectively.
Different panels show the results for the different foreground models:
pessimistic (\emph{left}), moderate (\emph{middle}), and optimistic 
(\emph{right}).  The pessimistic and moderate scenarios should both permit 
measurements of $\Delta z\sim1.0$.  The optimistic scenario
will allow for detailed characterization of the power spectrum evolution.
}
\label{fig:zpeak}
\end{figure*}

One of the first key parameters that is expected from 21\,cm measurement of the
EoR power spectrum is the redshift at which the universe was 50\% ionized,
sometimes referred to as ``peak reionization.''  The rationale behind this expectation
is evident from Figure \ref{fig:pspecs}, 
where the power spectrum generically achieves peak 
brightness at $k\sim0.1~h{\rm Mpc}^{-1}$ for $x_{\rm HI} = 0.5$.  However,
given the steep increase of $T_{\rm sys}$, one must ask if an experiment
will truly have the sensitivity to distinguish the power spectrum 
at $z_{\rm peak}$ from those on either side.  
Figure \ref{fig:zpeak}
shows the error bars on our fiducial power spectrum model at 50\% ionization
($z = 9.5$), as well as those on the neighboring redshifts $z = 8.5$ and
$z = 10.5$, under each of our three foreground models.
In both the pessimistic and moderate models (left and middle panel),
the $z=8.5$ (20\% neutral), $z=9.5$ (52\% neutral)
and $z=10.5$ (71\% neutral) are distinguishable at the few sigma level.
This analysis therefore suggests that it should
be possible to identify peak reionization to within a $\Delta z \sim 1$,
with a strong dependence on the actual redshift of reionization (since
noise is significantly lower at lower redshifts).  
 
It is worth noting, however, that even relatively high significance
detections of the power spectrum ($\gtrsim 5\mbox{--}10\sigma$) 
may not permit one to distinguish power spectrum of peak reionization from
those at nearby redshifts --- especially as one looks to higher $z$.
For our vanilla EoR model, we find a $\sim10\sigma$ detection is necessary
to distinguish the $z = 8.5,\ 9.5,$ and 10.5 power spectra at the $>1\sigma$
level.  In fact, for this level of significance, nearly all of the power spectra at redshifts higher
than peak reionization at $z = 9.5$ are indistinguishable given the steep
rise in thermal noise.  Even if the current generation of 21\,cm telescopes does yield
a detection of the 21\,cm power spectrum, these first measurements
do not guarantee stringent constraints on the peak
redshift of reionization.

Finally, one can see that the high sensitivities permitted 
by the optimistic foreground model
will allow a detailed characterization of the power spectrum amplitude and slope
as a function of redshift.  We discuss exactly what kind
of science this will enable
(beyond detecting the timing and duration of reionization)
in Section \ref{sec:adrian}.

\subsection{Sample Variance and Imaging}
\label{sec:imaging}

\begin{figure}[]
\centering
\includegraphics[width=.8\textwidth]{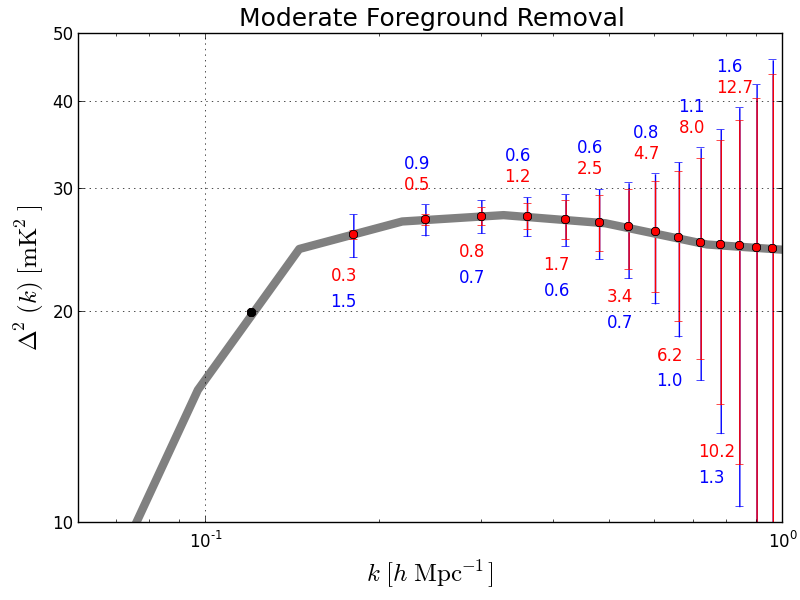}
\caption[Comparison of thermal noise and sample variance.]{The breakdown of the error bars in the top-left panel of Figure \ref{fig:pspec-errs} (vanilla EoR model and moderate foreground removal scenario).
Red shows the contribution of thermal noise, blue the contribution of sample
variance.  The text shows the value of each contribution in mK$^2$ --- 
\emph{not} the significance of the detection, as in previous plots.
Regions where the sample variance error dominates the thermal noise error
are in the imaging regime.  The placement of the numerical values above
or below the error bar has no significance; it is only for clarity.
Sample variance dominates the errors in the moderate foreground scenario
on scales $k < 0.25~h{\rm Mpc}^{-1}$.}
\label{fig:samp-var-hor}
\end{figure}

Given the high power spectrum sensitivities achievable under all of our
foreground removal models, one
must investigate the contributions of sample variance to the overall errors.
For the moderate foreground model, 
Figure \ref{fig:samp-var-hor} shows the relative contribution of sample variance
and thermal noise to the errors shown in 
the top-left panel of Figure \ref{fig:pspec-errs}.
From this plot, it is clear that sample variance contributes over half of
the total power spectrum uncertainty on scales $k < 0.3 h{\rm Mpc}^{-1}$.  If
the power spectrum constituted the ultimate measurement of reionization, this
would be an argument for a survey covering a larger area of sky.  For our
HERA concept array, which drift-scans, this is not possible, but may be
for phased-array designs.  However, the sample variance dominated regime
is very nearly equivalent to the imaging regime: thermal noise is reduced
to the point where individual modes have an SNR $\gtrsim$ 1.  Therefore, using a
filter to remove the wedge region (e.g. \citet{PoberWedge}),
a collecting area of $0.1~{\rm km}^2$ should provide sufficient sensitivity
to image the Epoch of Reionization over
$\sim800$ sq. deg. (6~hours of right ascension $\times\ 
8.7^{\circ}~\rm{FHWM})$ on scales of $0.1 \mbox{--} 0.25~h{\rm Mpc}^{-1}$.
We note that the HERA concept design is not necessarily optimized for imaging;
other configurations may be better suited if imaging the EoR is the primary
science goal.

The effect of the other foreground models on imaging is relatively
small.  The
poorer sensitivities of the pessimistic model push up thermal noise,
lowering the highest $k$ that can be imaged to $k \sim 0.2 h{\rm Mpc}^{-1}$.
The optimistic
foreground model recovers significant SNR on the largest scales, to the
point where sample variance dominates all scales up to 
$0.3~h{\rm Mpc}^{-1}$.  The effects of foregrounds and the wedge on imaging
with a HERA-like array will be explored in future work. 

\subsection{Characteristic Features of EoR Power Spectrum}
\label{sec:distinguishing}

Past literature has discussed two simple features of the 21\,cm power
spectra to help distinguish between models: the slope of the power
spectrum and the sharp drop in power (the ``knee'') on the largest
scales \citep{Matt3}. 
In particular, the mass of the halos driving reionization (parametrized
in this analysis by the minimum virial temperature of ionizing halos)
should affect the slope of the power spectrum.  Since more massive
halos are more highly clustered, they should concentrate power on
larger scales, yielding a flatter slope.
The second row of Figure \ref{fig:pspecs} suggests this effect is
small, although not implausible.
The knee of the power spectrum at large scales should correspond to the
largest ionized bubble size, since there will be little power on scales
larger than these bubbles \citep{furlanetto_et_al_2006a}.  The position
of the knee should be highly sensitive to the mean free path for ionizing
photons through the IGM, since this sets how large bubbles can grow.
This argument is indeed confirmed by the third row of Figure \ref{fig:pspecs}, where
the smaller values of $R_{\rm mfp}$ lack significant power on large
scales compared to those models with larger values.
Unfortunately, since our third parameter $\zeta$ does not change the shape
of the power spectrum, constraining different values of $\zeta$ will not be
possible with only a shape-based analysis.
In this section we first extend these qualitative arguments based on salient
features in the power spectra, and then present a more quantitative
analysis on distinguishing models in Section \ref{sec:adrian}.

To quantify the slope of the power spectrum, we fit a straight line
to the predicted power spectrum values between $k = 0.1 \mbox{--} 1.0~h{\rm Mpc}^{-1}$.
When we refer to measuring the slope, we refer to measuring the slope
of this line, given the error bars in the $k$-bins between 
$0.1 \mbox{--} 1.0~h{\rm Mpc}^{-1}$.
This choice of fit is not designed to encompass the full range
of information contained in measurements of the power spectrum shape.
Rather, the goal of this section is to find simple features of the
power spectrum that can potentially teach us about the EoR without
resorting to more sophisticated modeling.

\begin{figure*}[]
\centering
\includegraphics[width=\textwidth]{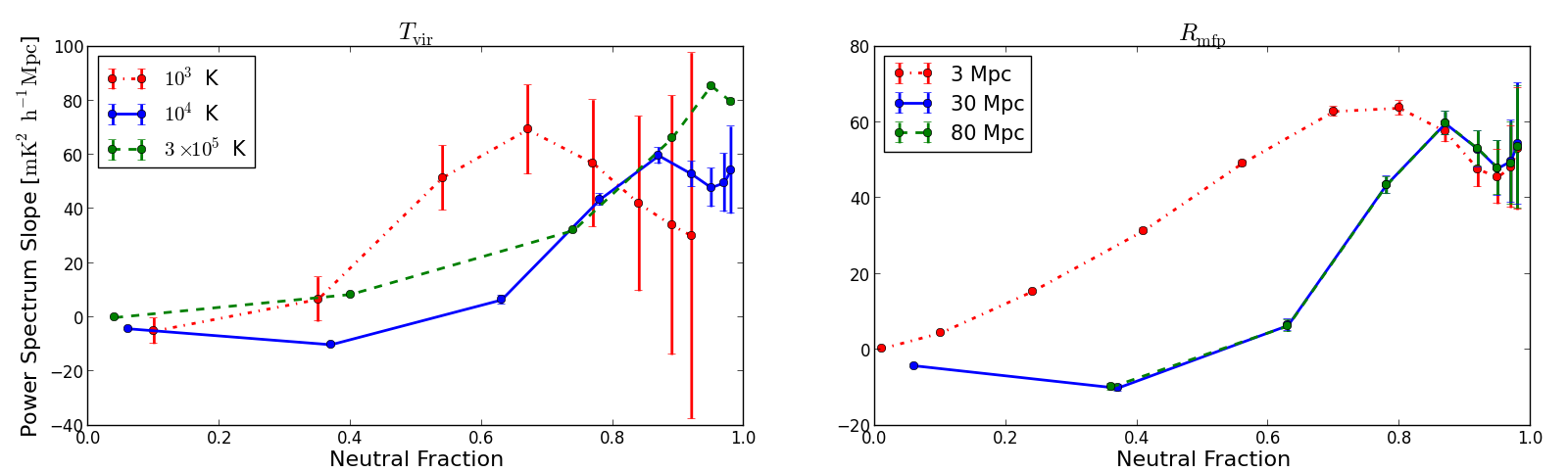}
\caption[Sensitivity to power spectrum slope.]{The power spectrum slope in units
of ${\rm mK}^2 h^{-1}\rm{Mpc}$ between $k = 0.1$ and $1.0
h{\rm Mpc}^{-1}$ as a function of neutral fraction for various EoR models.
Note that error bars are plotted on all points and correspond to the redshift
of a given neutral fraction for that model.  \emph{Left}: Different values
of $T_{\rm vir}$.  \emph{Right}: Different values of $R_{\rm mfp}$.
While different values of $T_{\rm vir}$ and $R_{\rm mfp}$ produce considerable
changes in the power spectrum slope, it will be difficult to unambiguously
interpret its physical significance.}
\label{fig:slope-comp}
\end{figure*}

Figure \ref{fig:slope-comp} shows the evolution of the slope of the
linear fit to the power spectrum
over the range $k = 0.1 \mbox{--} 1.0~h{\rm Mpc}^{-1}$, as
a function of neutral fraction for several EoR models.  
Error bars in both panels correspond
to the error measured under the moderate foreground model
for a given neutral fraction in the fiducial history
of a model.  This means that, e.g., the high neutral fractions in the
$T_{\rm vir} = 10^3$ K curve have thermal noise errors corresponding to
$z \sim 20$, outside the range of many proposed experiments.  Given that
caveat, it does appear that the low-mass ionizing galaxies produce
power spectra with significantly steeper slopes at moderate neutral fractions
than those models where
only high mass galaxies produce ionizing photons.
However, it is also clear that small-bubble (i.e. low mean free path) models
can yield steep slopes.  Therefore, while there may be some physical insight
to be gleaned from measuring the slope of the power spectrum and its evolution,
even the higher sensitivity measurements permitted by the optimistic
foreground model may not be enough to break these degeneracies.
In Section \ref{sec:adrian}, we specifically focus on the kind of information
necessary to disentangle these effects.

A comparison of the error bars in the moderate and optimistic foreground
scenario measurements of the vanilla power spectrum
(rows one and three in Figure \ref{fig:pspec-errs})
reveals the difficulty in recovering the position of the knee without
foreground subtraction: foreground contamination predominantly
affects low $k$ modes, rendering large scale features like the knee
inaccessible.
In particular, the
additive component of the horizon wedge severely restricts the large scale
information available to the array.  Without probing large scales, confirming
the presence (or absence) of a knee feature is likely to be impossible.
However, Figure \ref{fig:pspec-errs} does show that if foreground removal
allows for the recovery of these modes, the knee can be detected with
very high significance, even the presence of sample variance.

\subsection{Quantitative Constraints on Model Parameters}
\label{sec:adrian}

In previous sections, we considered rather large changes to the input parameters of the \texttt{21cmFAST} model.  These gave rise to theoretical power spectra that exhibited large qualitative variations, and encouragingly, we saw that such variations should be easily detectable using next-generation instruments such as HERA.  We now turn to what would be a natural next step in data analysis following a broad-brush, qualitative discrimination: a determination of best-fit values for astrophysical parameters.  In this section, we forecast the accuracy with which $T_\textrm{vir}$, $R_\textrm{mfp}$, and $\zeta$ can be measured by a HERA-like instrument, paying special attention to degeneracies.  In many of the plots, we will omit the pessimistic foreground scenario, for often the results from it are identical to (and visually indistinguishable from) those from moderate foregrounds.  Ultimately, our results will suggest parameter constraints that are smaller than one can justifiably expect given the reasonable, but non-negligible uncertainty surrounding simulations of reionization \citep{Zahn:2011}.  Our final error bar predictions (which can be found in Table \ref{finalErrors}) should therefore be interpreted cautiously, but we do expect the qualitative trends in our analysis to continue to hold as theoretical models improve.

\begin{figure*}
\centering
\includegraphics[width=1.0\textwidth,trim=2.5cm 3cm 3cm 4cm,clip]{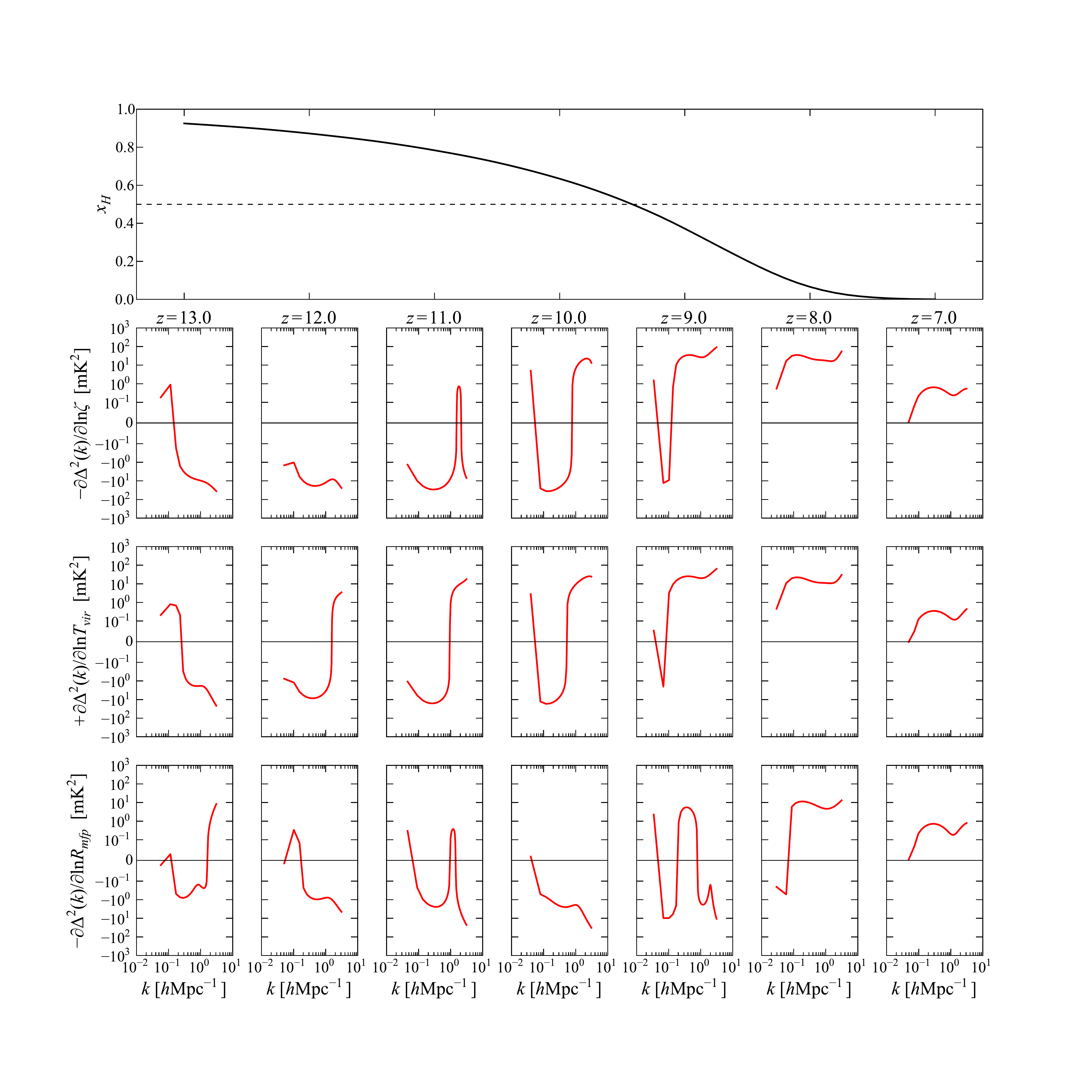}
\caption[Power spectrum derivatives for each reionization parameter.]{Power spectrum derivatives as a function of wavenumber $k$ and redshift $z$.  Each of the lower three rows shows derivatives with respect to a different parameter in our three-parameter model, and the top panel (aligned in redshift with the bottom panels) shows the corresponding neutral fraction.  Because our parameter vector $\boldsymbol \theta$ (Equation [\ref{eq:FisherDef}]) contains non-dimensionalized parameters, the derivatives $\partial \Delta^2(k,z) / \partial \theta_i$ are equivalent (if evaluated at the fiducial parameter values) to the logarithmic derivatives shown here.  Note that the derivatives with respect to $\zeta$ and $R_\textrm{mfp}$ have been multiplied by $-1$ to facililate later comparisons.  The vertical axes for the derivatives are linear between $-10^{-1}$ and $10^1$, and are logarithmic outside that range.  From this figure, we see that while the lowest redshifts are easy to access observationally, the model parameters are highly degenerate.  The higher redshifts are less degenerate, but thermal noise and foregrounds make observations difficult.}
\label{fig:unweightedPk}
\end{figure*}

\subsection{Fisher matrix formalism for errors on model parameters}

To make our forecasts, we use the Fisher information matrix $\mathbf{F}$, which takes the form
\begin{equation}
\label{eq:FisherDef}
\mathbf{F}_{ij} \equiv - \Big{\langle} \frac{\partial^2 \ln \mathcal{L}}{\partial \theta_i \partial \theta_j} \Big{\rangle} = \sum_{k,z} \frac{1}{\varepsilon^2 (k,z)} \frac{\partial \Delta^2 (k,z)}{\partial \theta_i} \frac{\partial \Delta^2 (k,z)}{\partial \theta_j},
\end{equation}
where $\mathcal{L}$ is the likelihood function (i.e. the probability distribution for the measured data as a function of model parameters), $\varepsilon (k,z)$ is the error on $\Delta^2(k,z)$ measurements as a function of wavenumber $k$ and redshift $z$, and $\boldsymbol{\theta} = ( T_\textrm{vir} / T_\textrm{vir}^\textrm{fid}, R_\textrm{mfp} / R_\textrm{mfp}^\textrm{fid}, \zeta / \zeta^\textrm{fid} )$ is a vector of the parameters that we wish to measure, divided by their fiducial values\footnote{Scaling out the fiducial values of course represents no loss of generality, and is done purely for numerical convenience.}.  The second equality in Equation \eqref{eq:FisherDef} follows from assuming Gaussian errors, and  picking $k$-space and redshift bins in a way that ensures that different bins are statistically independent \citep{MaxKL}, as we have done throughout this paper.  Implicit in our notation is the understanding that all expectation values and partial derivatives are evaluated at fiducial parameter values.  Having computed the Fisher matrix, one can obtain the error bars on the $i^{th}$ parameter by computing $1/\sqrt{ \mathbf{F}_{ii}}$ (when all other parameters are known already) or $(\mathbf{F}^{-1})_{ii}$ (when all parameters are jointly estimated from the data).  The Fisher matrix thus serves as a useful guide to the error properties of a measurement, albeit one that has been performed optimally.  Moreover, because Fisher information is additive (as demonstrated explicitly in Equation [\ref{eq:FisherDef}]), one can conveniently examine which wavenumbers and redshifts contribute the most to the parameter constraints, and we will do so later in the section.

From Equation \eqref{eq:FisherDef}, we see that it is the derivatives of the power spectrum with respect to the parameters that provide the crucial link between measurement and theory.  If a large change in the power spectrum results from a small change in a parameter --- if the amplitude of a power spectrum derivative is large --- a measurement of the power spectrum would clearly place strong constraints on the parameter in question.  This is a property that is manifest in $\mathbf{F}$.  Also important are the shapes of the power spectrum derivatives in $(k,z)$ space.  If two power spectrum derivatives have similar shapes, changes in one parameter can be mostly compensated for by a change in the second parameter, leading to a large degeneracy between the two parameters.  Mathematically, the power spectrum derivatives can be geometrically interpreted as vectors in $(k,z)$ space, and each element of the Fisher matrix is a weighted dot product between a pair of such vectors \citep{MaxKL}.  Explicitly, $\mathbf{F}_{ij} = \mathbf{w}_i \cdot \mathbf{w}_j$, where
\begin{equation}
\label{eq:vecDef}
\mathbf{w}_i (k,z) \equiv \frac{1}{\varepsilon (k,z) } \frac{\partial \Delta^2 (k,z)}{\partial \theta_i}, 
\end{equation}
with the different elements of the vector corresponding to different values of $k$ and $z$.  If two $\mathbf{w}$ vectors have a large dot product (i.e. similar shapes), the Fisher matrix will be near-singular, and the joint parameter constraints given by $\mathbf{F}^{-1}$ will be poor.

\subsubsection{Single-Redshift Constraints}

We begin by examining how well each reionization parameter can be constrained by observations at several redshifts spanning our fiducial reionization model. In Figure \ref{fig:unweightedPk}, we show some example power spectrum derivatives\footnote{Because \texttt{21cmFAST} produces output at $k$-values that differ from those naturally arising from our sensitivity calculations, it was necessary to interpolate the outputs when computing the derivatives (which were approximated using finite-difference methods).  For this paper, we chose to fit the \texttt{21cmFAST} power spectra to sixth-order polynomials in $\ln k$, finding such a scheme to be a good balance between capturing all the essential features of the power spectrum derivatives while not over-fitting any ``noise'' in the theoretical simulations.  Alternate approaches such as performing cubic splines, or fitting to fifth- or seventh-order polynomials were tested, and do not change our results in any meaningful ways.  Finally, to safeguard against generating numerical derivatives that are dominated by the intrinsic numerical errors of the simulations, we took care to choose finite-difference step sizes that were not overly fine.} as a function of $k$ and $z$.  Note that the last two rows of the figure show the \emph{negative} derivatives.  For reference, the top panel of the figure shows the corresponding evolution of the neutral fraction.  At the lowest redshifts, the power spectrum derivatives essentially have the same shape as the power spectrum itself.  To understand why this is so, note that at late times, a small perturbation in parameter values mostly shifts the time at which reionization reaches completion.  As reionization nears completion, the power spectrum decreases proportionally in amplitude at all $k$ due to a reduction in the overall neutral fraction, so a parameter shift simply causes an overall amplitude shift.  The power spectrum derivatives are therefore roughly proportional to the power spectrum.  In contrast, at high redshifts the derivatives have more complicated shapes, since changes in the parameters affect the detailed properties of the ionization field.

Importantly, we emphasize that for parameter estimation, the ``sweet spot'' in redshift can be somewhat different from that for a mere detection.  As mentioned in earlier sections, the half-neutral point of $x_H = 0.5$ is often considered the most promising for a detection, since most theoretical calculations yield peak power spectrum brightness there.  This ``detection sweet spot'' may shift slightly towards lower redshifts because thermal noise and foregrounds decrease with increasing frequency, but is still expected to be close to the half-neutral point.  For parameter estimation, however, the most informative redshifts may be significantly lower.  Consider, for instance, the signal at $z=8$, where $x_H = 0.066$ in our fiducial model.  Figure \ref{fig:pspec_vanilla} reveals that the power spectrum here is an order of magnitude dimmer than at $z = 9.5$ or $z = 9$, where $x_H = 0.5$ and $x_H = 0.37$ respectively.  However, from Figure \ref{fig:unweightedPk} we see that the power spectrum derivatives at $z =8$ are comparable in amplitude to those at higher redshifts/neutral fraction.  Intuitively, at $z=8$ the dim power spectrum is compensated for by its rapid evolution (due to the sharp fall in power towards the end of reionization).  Small perturbations in model parameters and the resultant changes in the timing of the end of reionization therefore cause large changes in the theoretical power spectrum.  There is thus a large information content in a $z=8$ power spectrum measurement.

\begin{figure*}[]
\centering
\includegraphics[width=1.0\textwidth,trim=3cm 3cm 3cm 4cm,clip]{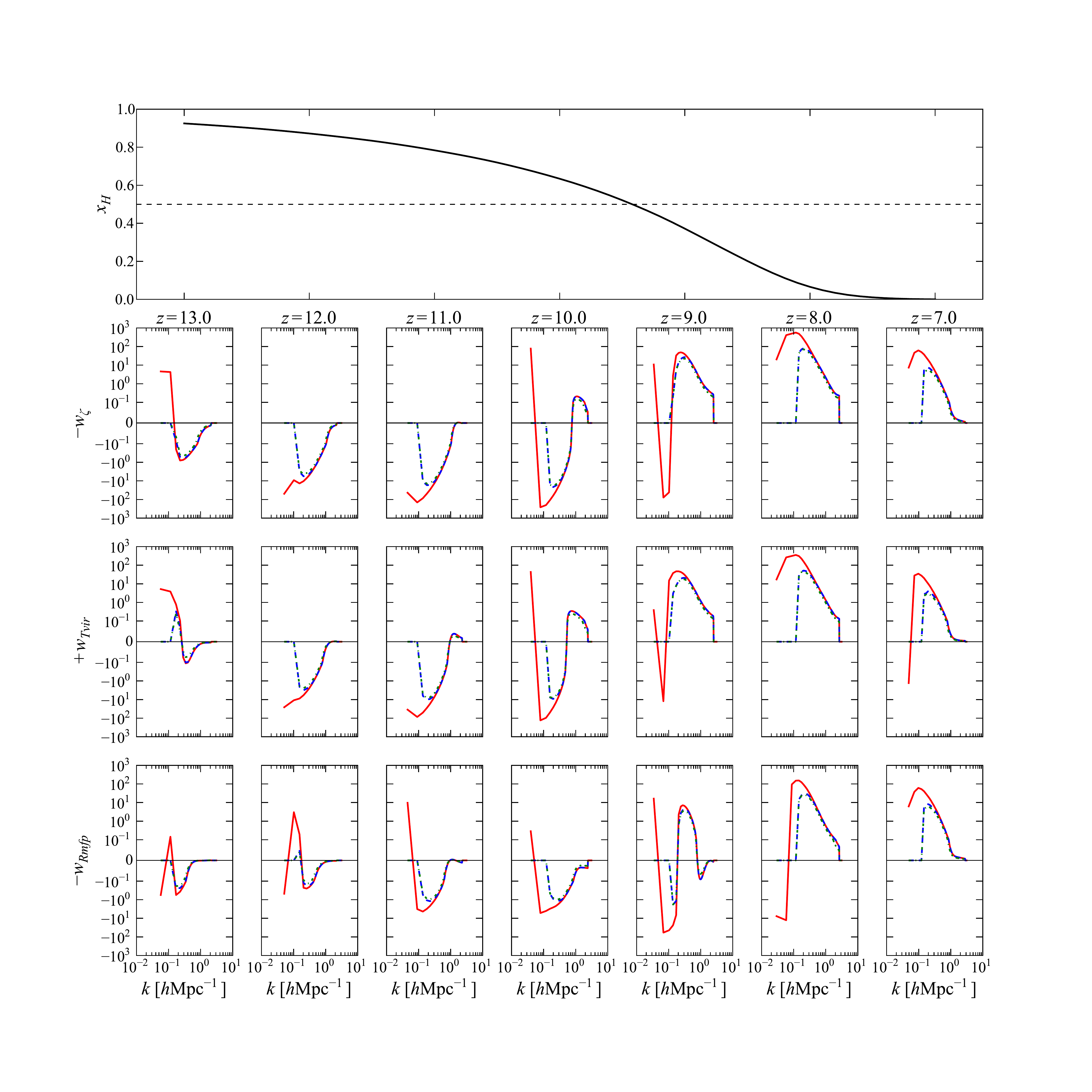}
\caption[Noise-weighted power spectrum derivatives.]{Similar to Figure \ref{fig:unweightedPk}, but inversely weighted by the error on the power spectrum measurement, i.e. plots of $\mathbf{w}_i$ from Equation \eqref{eq:vecDef}.  These weighted derivatives are computed for HERA for the optimistic (solid red curves), moderate (dashed blue curves), and pessimistic (dot-dashed green curves) foreground models.  The pessimistic curves are essentially indistinguishable from the moderate curves.  The top panel shows the corresponding evolution of the neutral fraction.  Just as with Figure \ref{fig:unweightedPk}, the vertical axes are linear from $-10^{-1}$ to $10^1$ and logarithmic elsewhere, and $\mathbf{w}_\textrm{Rmfp}$ and $\mathbf{w}_\zeta$ have been multiplied by $-1$ to facilitate comparison with $\mathbf{w}_\textrm{Tvir}$.  With foregrounds and thermal noise, power spectrum measurements become difficult at low and high $k$ values, and constraints on model parameters become more degenerate.}
\label{fig:weightedPk}
\end{figure*}

\begin{figure*}[]
\centering
\includegraphics[width=1.0\textwidth,trim=3cm 0cm 4cm 0cm,clip]{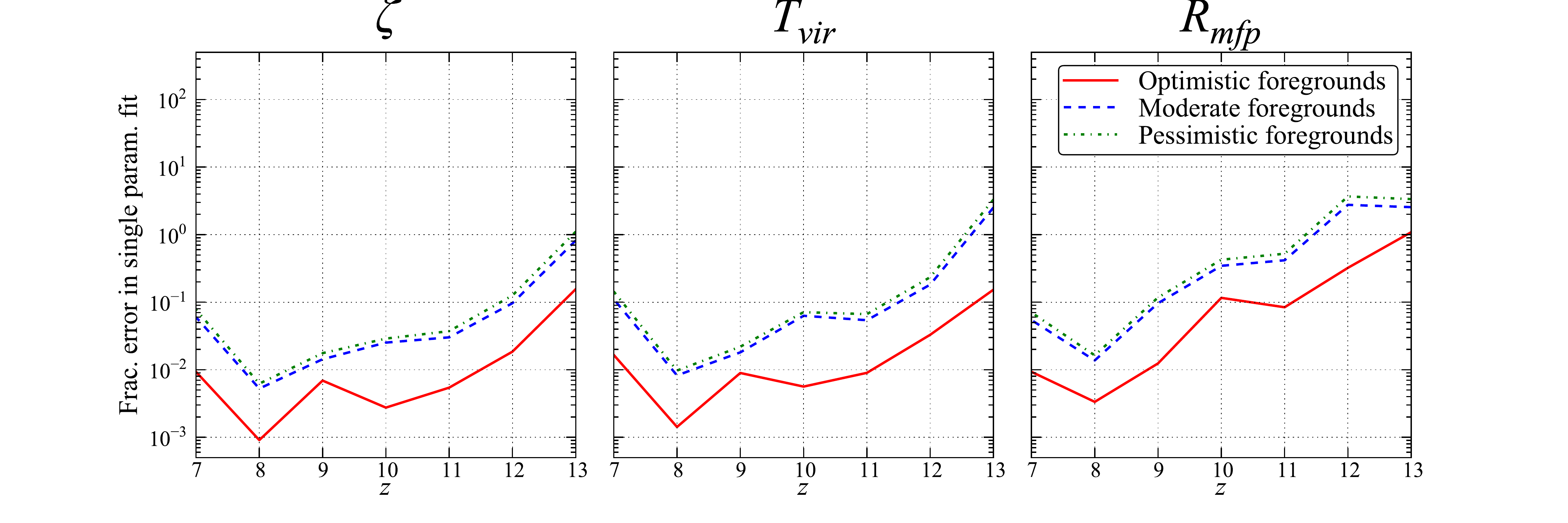}
\caption[Parameter errors as a function of redshift.]{Fractional $1\sigma$ errors ($1\sigma$ errors divided by fiducial values) as a function of redshift, with measurements at each redshift fit independently.  The errors on each parameter assume (unrealistically) that all other parameters are already known in the fit.  Solid red curves give optimistic foreground model predictions; dashed blue curves give moderate foreground model predictions; dot-dashed green curves give pessimistic foreground model predictions.  In all models, and for all three parameters, the best errors are obtained at $z=8$.  At $z=7$, the power spectrum has too small of an amplitude to yield good signal-to-noise, and at higher redshifts thermal noise poses a serious problem.}
\label{fig:singleParamErrors}
\end{figure*}

When thermal noise and foregrounds are taken into account, a $z=8$ measurement becomes even more valuable for parameter constraints than those at higher redshifts/neutral fractions.  This can be seen in Figure \ref{fig:weightedPk}, where we weight the power spectrum derivatives by the inverse measurement error\footnote{Whereas in previous sections the power spectrum sensitivities were always computed assuming a bandwidth of 8~MHz, in this section we vary the bandwidth with redshift so that a measurement centered at redshift $z$ uses all information from $z \pm 0.5$.} for HERA, producing the quantity $\mathbf{w}_i$ as defined in Equation \eqref{eq:vecDef}.  In solid red are the weighted derivatives for the optimistic foreground model, while the dashed blue curves are for the moderate foreground model.  The pessimistic case is shown using dot-dashed green curves, but these curves are barely visible because they are essentially indistinguishable from those for the moderate foregrounds.  In all cases, the derivatives peak --- and therefore contribute the most information --- at $z=8$.  Squaring and summing these curves over $k$, one can compute the diagonal elements of the Fisher matrix on a per-redshift basis.  Taking the reciprocal square root of these elements gives the error bars on each parameter assuming (unrealistically) that all other parameters are known.  The results are shown in Figure \ref{fig:singleParamErrors}.  For all three parameters, these single-parameter, per-redshift fits give the best errors at $z=8$.  At $z=7$ the neutral fraction is simply too low for there to be any appreciable signal (with even the rapid evolution unable to sufficiently compensate), and at higher redshifts, thermal noise and foregrounds become more of an influence.

Of course, what is not captured by Figure \ref{fig:singleParamErrors} is the reality that one must fit for all parameters simultaneously (since none of our three parameters are currently strongly constrained by other observational probes).  In general, our ability to constrain model parameters is determined not just by the amplitudes of the power spectrum derivatives and our instrument's sensitivity, but also by parameter degeneracies.  As an example, notice that at $z=7$, all the power spectrum derivatives shown in Figure \ref{fig:unweightedPk} have essentially identical shapes up to a sign.  This means that shifts in one parameter can be (almost) perfectly nullified by a shift in a different parameter; the parameters are degenerate.  These degeneracies are inherent to the theoretical model, since they are clearly visible even in Figure \ref{fig:unweightedPk}, where the power spectrum derivatives are shown without the instrumental weighting.  With this in mind, we see that even though Figure \ref{fig:singleParamErrors} predicts that observing the power spectrum at $z=7$ alone would give reasonable errors if there were somehow no degeneracies (making a single parameter fit the same as a simultaneous fit), such a measurement would be unlikely to yield any useful parameter constraints in practice.  To only a slightly lesser extent, the same is true for $z=8$, where the degeneracy between $R_\textrm{mfp}$ and the other parameters is broken slightly, but $T_\textrm{vir}$ and $\zeta$ remain almost perfectly degenerate.

\begin{figure*}[]
\centering
\includegraphics[width=1.0\textwidth,trim=0cm 0cm 0cm 0cm,clip]{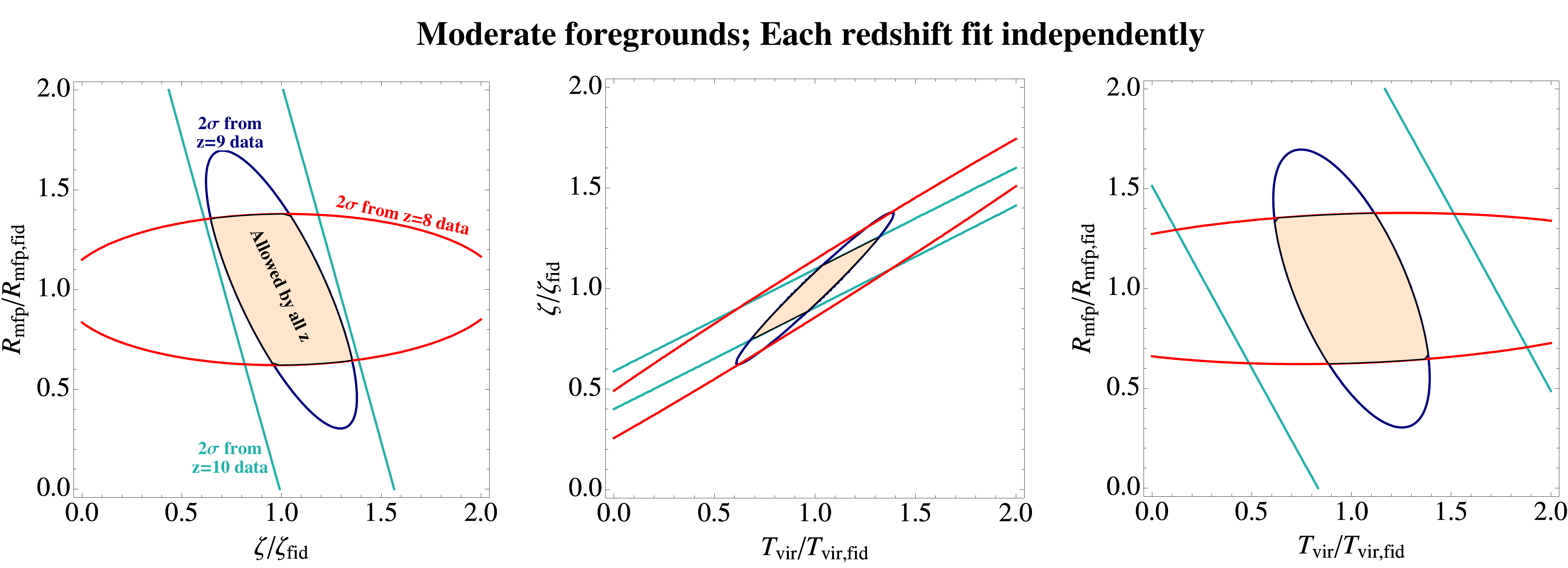}\\
\vspace{0.1in}
\includegraphics[width=1.0\textwidth,trim=0cm 0cm 0cm 0cm,clip]{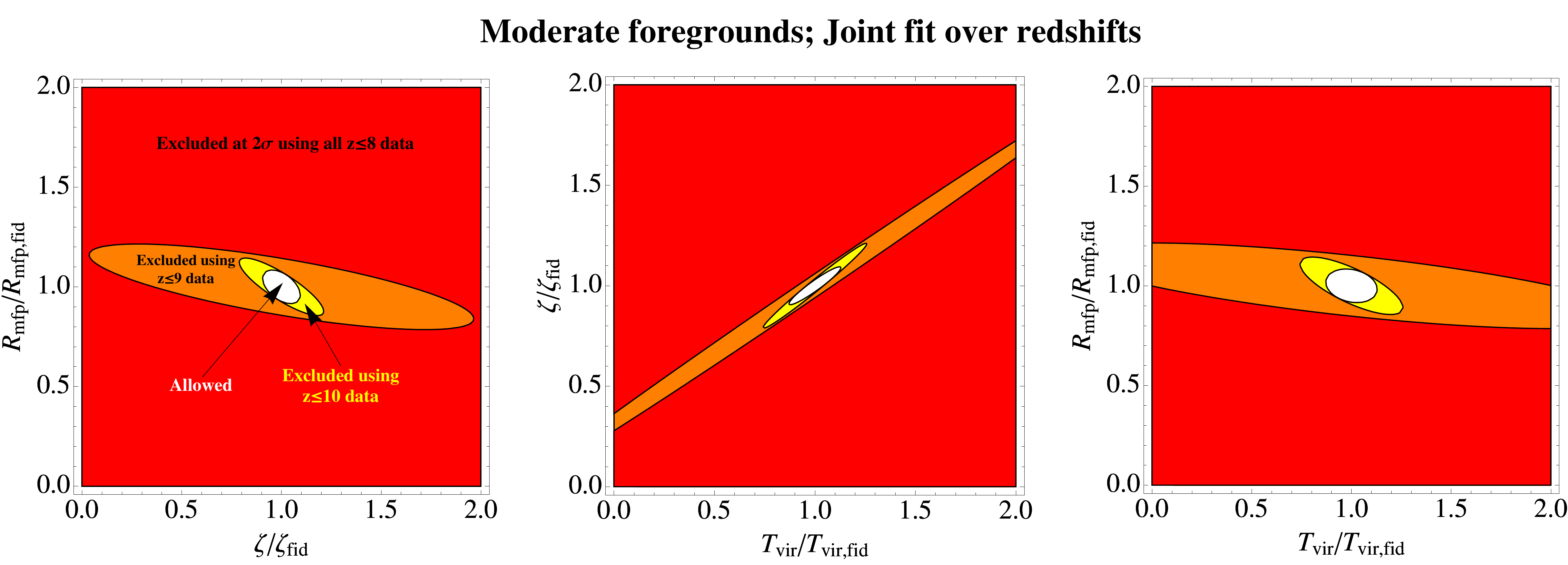}
\caption[Parameter constraints in the moderate foreground scenario.]{Pairwise parameter constraints for the moderate foreground model, shown as $2\sigma$ exclusion regions.  For each pair of parameters, the third parameter has been marginalized over.  The top row shows $2\sigma$ constraints from each redshift when fit independently; the light orange regions are not ruled out by data from any redshifts.  The bottom row shows the constraints from a joint fit over multiple redshifts.  Each color represents a portion of parameter space that can be \emph{excluded} by including data up to a certain redshift.  The white ``allowed'' region represents the final constraints from including all measured redshifts.  In both cases, a $z=7$ measurement alone does not provide any non-trivial constraints, but helps with degeneracy-breaking in the joint redshift fits (bottom row).  As one moves to higher and higher redshifts, power spectrum measurements probe different astrophysical processes, resulting in a shift in the principal directions of the exclusion regions.  Including higher redshifts tightens parameter constraints, but no longer helps beyond $z=10$ due to increasing thermal noise.}
\label{fig:MIDellipses}
\end{figure*}

The situation becomes even worse when one realizes that measurements at low and high $k$ are difficult due to foregrounds and thermal noise, respectively.  Many of the distinguishing features between the curves in Figure \ref{fig:unweightedPk} were located at the extremes of the $k$ axis, and from Figure \ref{fig:weightedPk}, we see that such features are obliterated by an instrumental sensitivity weighting (particularly for the pessimistic/moderate foregrounds).  This increases the level of degeneracy.  As an aside, note that with the lowest and highest $k$ values cut out, the bulk of the information originates from $k \sim 0.05~h\textrm{Mpc}^{-1}$ to $\sim 1~h\textrm{Mpc}^{-1}$ for the optimistic model and $k \sim 0.2~h\textrm{Mpc}^{-1}$ to $\sim 1~h\textrm{Mpc}^{-1}$ for the pessimistic and moderate models. (Recall again that since elements of the Fisher matrix are obtained by taking pairwise products of the rows of Figure \ref{fig:weightedPk} and summing over $k$ and $z$, the square of each weighted power spectrum derivative curve provides a rough estimate for where information comes from.)  Matching these ranges to the fiducial power spectra in Figure \ref{fig:pspec_vanilla} confirms the qualitative discussion presented in Section \ref{sec:distinguishing}, where we saw that slope of the power spectrum from $k \sim 0.1~h\textrm{Mpc}^{-1}$ to $\sim 1~h\textrm{Mpc}^{-1}$ is potentially a useful source of information regardless of foreground scenario, but that the ``knee'' feature at $k \lesssim 0.1~h\textrm{Mpc}^{-1}$ will likely only be accessible with optimistic foregrounds.  This is somewhat unfortunate, for a comparison of Figures \ref{fig:unweightedPk} and \ref{fig:weightedPk} reveals that measurements at low and high $k$ would potentially be powerful breakers of degeneracy, were they observationally feasible.

\subsubsection[Breaking degeneracies with multi-redshift \\observations]{Breaking degeneracies with multi-redshift observations}

Absent a situation where the lowest and highest $k$ values can be probed, the only way to break the serious degeneracies in the high signal-to-noise measurements at $z=7$ and $z=8$ is to include higher redshifts, even though thermal noise and foreground limitations dictate that such measurements will be less sensitive.  Higher redshift measurements break degeneracies in two ways.  First, one can see that at higher redshifts, the power spectrum derivatives have shapes that are both more complicated and less similar, thanks to non-trivial astrophysics during early to mid-reionization.  Second, a joint fit over multiple redshifts can alleviate degeneracies even if the parameters are perfectly degenerate with each other at every redshift when fit on a per-redshift basis.  Consider, for example, the weighted power spectrum derivatives for the moderate foreground model in Figure \ref{fig:weightedPk}.  For both $z=7$ and $z=8$, the derivatives for all three parameters are identical in shape; at both redshifts, any shift in the best-fit value of a parameter can be compensated for by an appropriate shift in the other parameters without compromising the goodness-of-fit.  At $z=8$, for instance, a given fractional increase in $\zeta$ can be compensated for by a slightly larger decrease in $R_\textrm{mfp}$, since $w_\textrm{Rmfp}$ has a slightly larger amplitude than $w_\zeta$.  However, this only works if the redshifts are treated independently.  If the data from $z=7$ and $z=8$ are jointly fit, the aforementioned parameter shifts would result in a worse overall fit, because $w_\textrm{Rmfp}$ and $w_\zeta$ have roughly equal amplitudes at $z=7$, demanding fractionally equal shifts.  In other words, we see that because the \emph{ratios} of different weighted parameter derivatives are redshift-dependent quantities, joint-redshift fits can break degeneracies even when the parameters would be degenerate if different redshifts were treated independently.  It is therefore crucial to make observations at a wide variety of redshifts, and not just at the lowest ones, where the measurements are easiest.

\begin{figure*}[]
\centering
\includegraphics[width=1.0\textwidth,trim=0cm 0cm 0cm 0cm,clip]{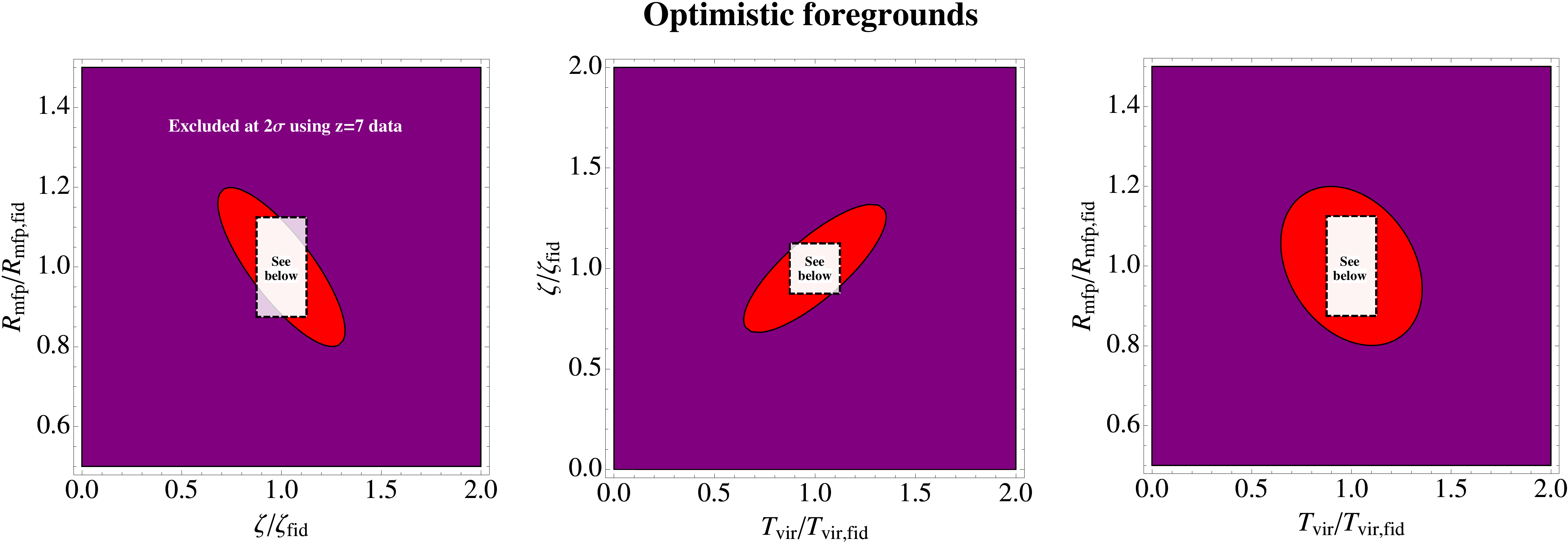}\\
\vspace{0.1in}
\includegraphics[width=1.0\textwidth,trim=0cm 0cm 0cm 0cm,clip]{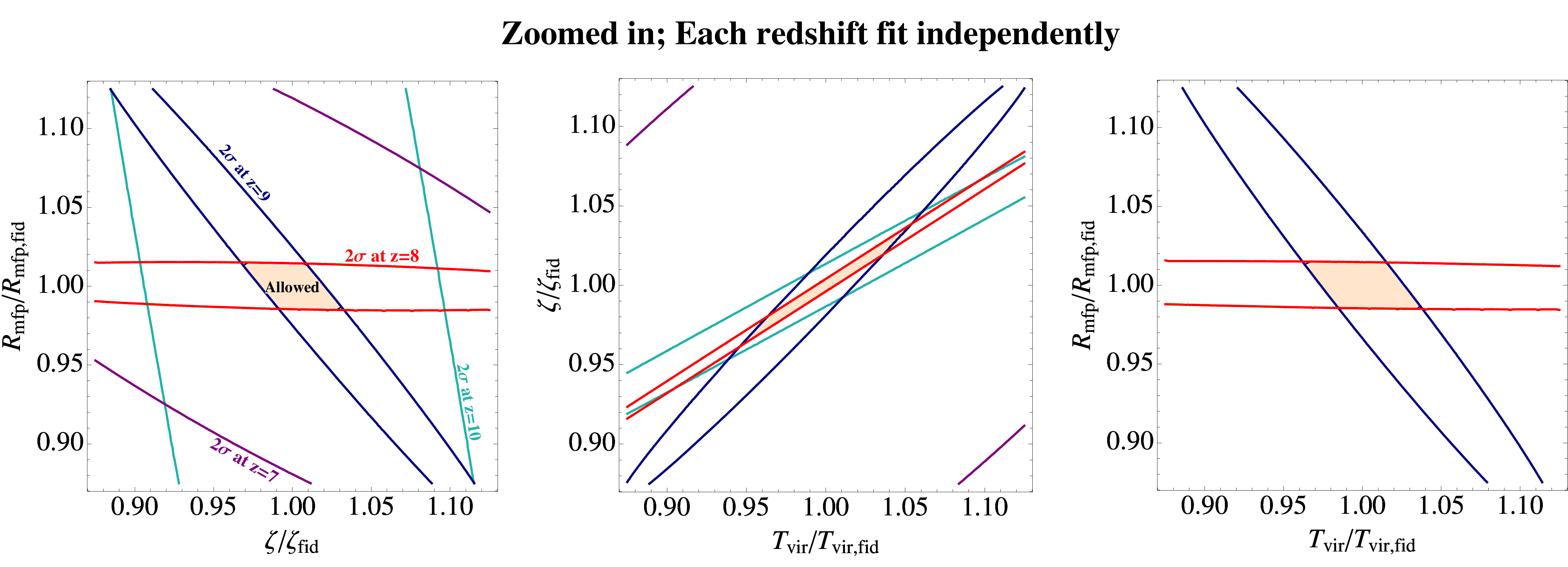}\\
\vspace{0.1in}
\includegraphics[width=1.0\textwidth,trim=0cm 0cm 0cm 0cm,clip]{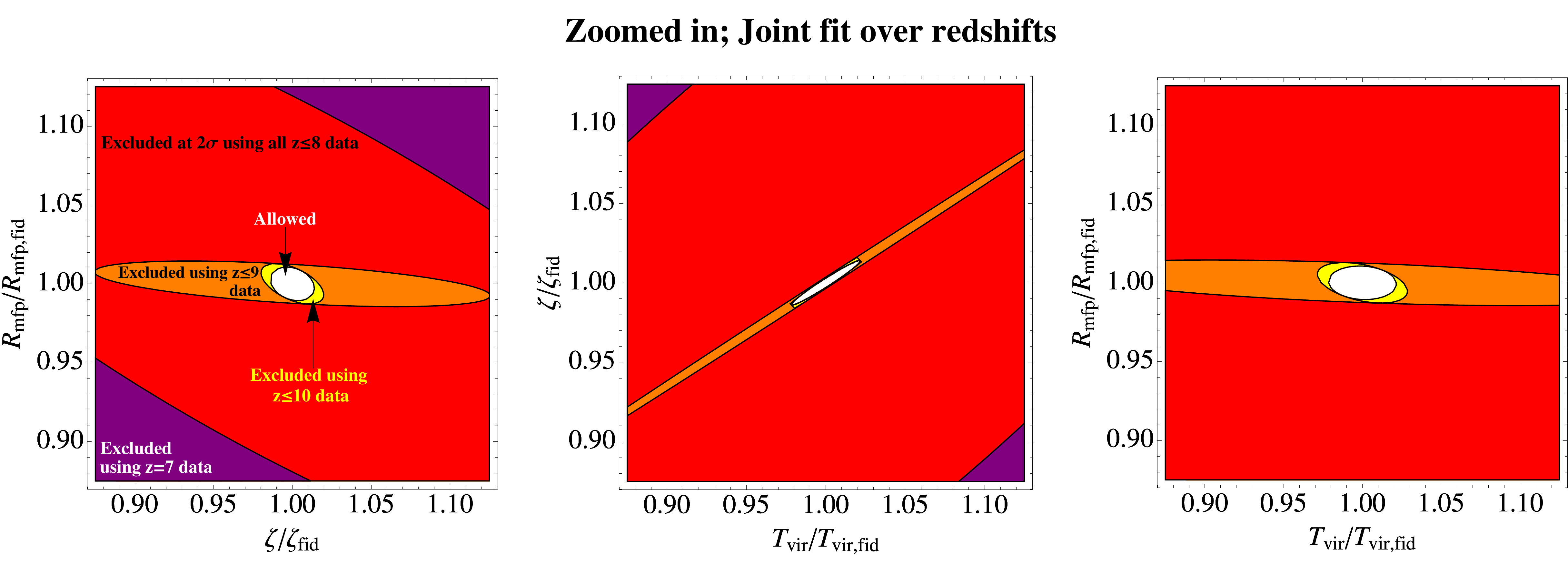}
\caption[Parameter constraints in the optimistic foreground scenario.]{Similar to Figure \ref{fig:MIDellipses}, but for the optimistic foreground model.  The top row shows the exclusion region from using $z=7$ data alone.  The middle and bottom rows show zoomed-in parameter space plots for the redshift-by-redshift and simultaneous fits respectively.  (Since $z=7$ is the lowest redshift in our model, the top panel is the same for both types of fit).  The constraints in this optimistic foreground scenario are seen to be better than those predicted for the moderate foreground model by about a factor of four.}
\label{fig:OPTellipses}
\end{figure*}

To see how degeneracies are broken by using information from multiple redshifts, imagine a thought experiment where one began with measurements at the lowest (least noisy) redshifts, and gradually added higher redshift information, one redshift at a time.  Figures \ref{fig:MIDellipses} and \ref{fig:OPTellipses} show the results for the moderate and optimistic foreground scenarios respectively.  (Here we omit the equivalent figure for the pessimistic model completely, because the results are again qualitatively similar to those for the moderate model.)  In each figure are $2\sigma$ constraints for pairs of parameters, having marginalized over the third parameter by assuming that the likelihood function is Gaussian (so that the covariance of the measured parameters is given by the inverse of the Fisher matrix).  One sees that as higher and higher redshifts are included, the principal directions of the exclusion ellipses change, reflecting the first degeneracy-breaking effect highlighted above, namely, the inclusion of different, more-complex and less-degenerate astrophysics at higher redshifts.  To see the second degeneracy-breaking effect, where per-redshift degeneracies are broken by joint redshift fits, we include in both figures the constraints that arise after combining results from redshift-by-redshift fits (shown as contours for each redshift), as well as the constraints from fitting multiple redshifts simultaneously (shown as cumulative exclusion regions).

For the moderate foreground scenario, we find that non-trivial constraints cannot be placed using $z=7$ data alone, hence the omission of a $z\le 7$ exclusion region from Figure \ref{fig:MIDellipses}.  However, we note that the constraints using $z \le 8$ data (red contours/exclusion regions) are substantially tighter in the bottom panel than in the top panel of the figure.  This means that the $z=7$ power spectrum can break degeneracies in a joint fit, even if the constraints from it alone are too degenerate to be useful.  A similar situation is seen to be true for a $z=10$ measurement, which is limited not just by degeneracy, but also by the higher thermal noise at lower frequencies.  Except for the $\zeta$-$T_\textrm{vir}$ parameter space, adding $z=10$ information in an independent fashion does not further tighten the constraints beyond those provided by $z \le 9$.  But again, when a joint fit (bottom panel of Figure \ref{fig:MIDellipses}) is performed, this information is useful even though it was noisy and degenerate on its own.  We caution, however, that this trend does \emph{not} persist beyond $z =10$, in that $z \ge 11$ measurements are so thermal-noise dominated that their inclusion has no effect on the final constraints.  Indeed, the ``allowed" regions in both Figures \ref{fig:MIDellipses} and \ref{fig:OPTellipses} include all redshifts, but are visually indistinguishable from ones calculated without $z\ge 11$ information.\footnote{We emphasize that in our analysis we have only considered the reionization epoch.  Thus, while we find that observations of power spectra at $z \ge 11$ do not add very much to measurements of reionization parameters like $T_\textrm{vir}$, $\zeta$, and $R_\textrm{mfp}$, they are expected to be extremely important for constraining X-ray physics prior to reionization, as discussed in \cite{Mesinger:2013c} and \cite{Christian:2013}.}

\begin{figure*}[!ht]
\centering
\includegraphics[width=1.0\textwidth,trim=0cm 0cm 0cm 0cm,clip]{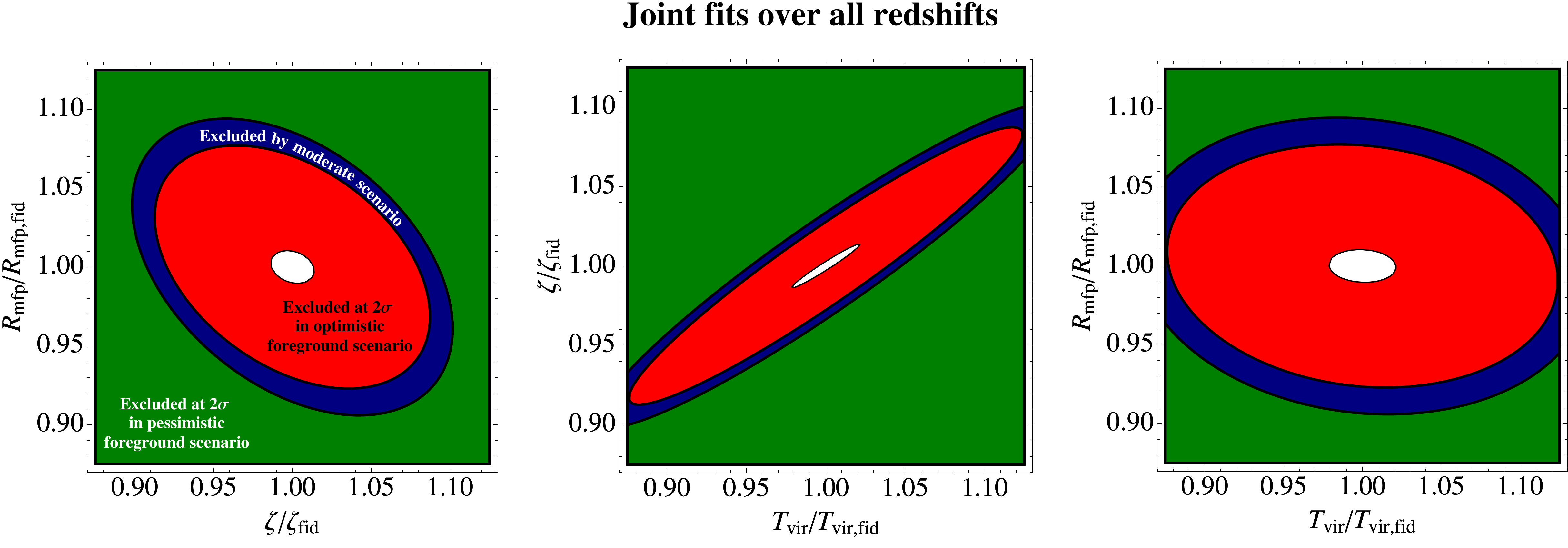}
\caption[Comparing parameter constraints between foreground scenarios.]{A comparison between the $2\sigma$ exclusion regions for pessimistic (green), moderate (blue), and optimistic (red) foregrounds, assuming that power spectra at all measured redshifts are fit simultaneously.  Going from the pessimistic foreground model to the moderate model gives only marginal improvement; going from the moderate to the optimistic model reduces errors from the $5\%$ level to the $1\%$ level.}
\label{fig:OPTMIDPESSellipses}
\end{figure*}

Comparing the predictions for the moderate foreground model to those of the optimistic foreground model (Figure \ref{fig:OPTellipses}), several differences are immediately apparent.  Whereas the $z=7$ power spectrum alone could not place non-trivial parameter constraints in the moderate scenario, in the optimistic scenario it has considerable discriminating power, similar to what can be achieved by jointly fitting all $z \le 9$ data in the moderate model.  This improvement in the effectiveness of the $z=7$ measurement is due to an increased ability to access low and high $k$ modes, which breaks degeneracies.  With low and high $k$ modes measurable, each redshift alone is already reasonably non-degenerate, and the main benefit (as far as degeneracy-breaking is concerned) in going to higher $z$ is the opportunity to access new astrophysics with a slightly different set of degeneracies, rather than the opportunity to perform joint fits.  Indeed, we see from the middle and bottom panels of Figure \ref{fig:OPTellipses} that there are only minimal differences between the joint fit and the independent fits.  In contrast, with the moderate model in Figure \ref{fig:MIDellipses} we saw about a factor of four improvement in going from the latter to the former.

\begin{table}
\centering
\begin{tabular}{|c|c|c|c|}
\hline
Foreground Model & $\Delta T_\textrm{vir}/T_\textrm{vir,fid}$ & $\Delta \zeta$/$\zeta_\textrm{fid}$ & $\Delta R_\textrm{mfp}/R_\textrm{mfp,fid}$ \\
\hline
Moderate & $0.062 $ & $0.044$ & $0.039$ \\ \hline
Pessimistic & $0.071$ & $0.051$ & $0.047$ \\ \hline
Optimistic & $0.011$ & $0.0069$ & $0.0052$ \\ \hline
\end{tabular}
\caption{Reionization Parameter Errors ($1\sigma$) for HERA.}
\label{finalErrors}
\end{table}

Figure \ref{fig:OPTMIDPESSellipses} compares the ultimate performance of HERA for the three foreground scenarios, using all measured redshifts in a joint fit.  (Note that our earlier emphasis on the differences between joint fits and independent fits was for pedagogical reasons only, since in practice there is no reason not to get the most out of one's data by performing a joint fit.)  We see that even with the most pessimistic foreground model, our three parameters can be constrained to the $5\%$ level.  The ability to combine partially-coherent baselines in the moderate model results only in a modest improvement, but being able to work within the wedge in the optimistic case can suppress errors to the $\sim 1 \%$ level.  The final results are given in Table \ref{finalErrors}.

In closing, we see that a next-generation like HERA should be capable of delivering excellent constraints on astrophysical parameters during the EoR.  These constraints will be particularly valuable, given that none of the parameters can be easily probed by other observations.  However, a few qualifications are in order.  First, the Fisher matrix analysis performed here provides an accurate forecast of the errors only if the true parameter values are somewhat close to our fiducial ones.  As an extreme example of how this could break down, suppose $T_\textrm{vir}$ were actually $1000\,\textrm{K}$, as illustrated in the middle row of Figure \ref{fig:pspecs}.  The result would be a high-redshift reionization scenario, one that would be difficult to probe to the precision demonstrated in this section, due to high thermal noise.  Secondly, one's ability to extract interesting astrophysical quantities from a measurement of the power spectrum is only as good as one's ability to model the power spectrum.  In this section, we assumed that \texttt{21cmFAST} is the ``true'' model of reionization.  At the few-percent-level uncertainties given in Table \ref{finalErrors}, the measurement errors are better than or comparable to the scatter seen between different theoretical simulations \citep{Zahn:2011}.  Thus, there will likely need to be much feedback between theory and observation to make sense of a power spectrum measurement with HERA-level precision.  Alternatively, given the small error bars seen here with a three-parameter model, it is likely that additional parameters can be added to one's power spectrum fits without sacrificing the ability to place constraints that are theoretically interesting.  We leave the possibility of including additional parameters (many of which have smaller, subtler effects on the $21\,\textrm{cm}$ power spectrum than the parameters examined here) for future work.

\section{Conclusions}
\label{sec:conclusions}

In order to
explore the potential range of constraints that will come from the proposed
next generation of 21\,cm experiments (e.g. HERA and SKA), we used simple
models for instruments, foregrounds, and reionization histories to encompass
a broad range of possible scenarios.  For an instrument model, we used the 
$\sim 0.1~\rm{km}^2$ HERA concept array, and calculated power spectrum
sensitivities using the method of \cite{BAOBAB}.  To cover
uncertainties in the foregrounds, we used three principal models.  Both
our pessimistic and moderate model assumes foregrounds occupy the wedge-like
region in $k$-space observed by \cite{PoberWedge}, extending
$0.1~h{\rm Mpc}^{-1}$ past the analytic horizon limit. 
Thus, both cases are amenable to a strategy of foreground avoidance.
What makes our
pessimistic model pessimistic is the decision to combine partially redundant
baselines in an incoherent fashion, allowing one to completely sidestep
the systematics highlighted by \cite{Hazelton2013}. 
In the moderate model,
these baselines are allowed to be combined coherently.  Finally, in our
optimistic model, the size of the wedge is reduced to a region defined
by the FWHM of the primary beam.  Given the small field of view of the 
dishes used in the HERA concept array, this model is effectively equivalent
to one in which foreground removal techniques prove successful.  Lastly,
to cover the uncertainties in reionization history, we use \texttt{21cmFAST} to generate
power spectra for a wide range of uncertain parameters: the ionizing
efficiency $\zeta$, the minimum virial temperature of halos producing
ionizing photos, $T_{\rm vir}$, and the mean free path of ionizing photons
through the IGM $R_{\rm mfp}$.

Looking at predicted power spectrum measurements for these various scenarios
yields the following conclusions:
\begin{itemize}

\item Even with no development of analysis techniques beyond those used
in \cite{PAPER32Limits}, an experiment with
$\sim 0.1~\rm{km}^2$ of collecting area can yield very high significance
$\gtrsim 30\sigma$ detections of nearly 
any reionization power spectrum (cf. Figure
\ref{fig:inst-red-snr-matrix}).

\item Developing techniques that allow for the coherent addition of partially
redundant baselines can result in a small increase of additional
power spectrum sensitivity.  In this work, we find our moderate foreground
removal model to increase sensitivities by $\sim20\%$ over our most pessimistic
scenario.  Generally, we find that coherent combination of partially
redundant baselines reduces thermal noise errors by $\sim40\%$, so addressing
this issue will be somewhat more important for smaller arrays that have
not yet reached the sample variance dominated regime.

\item With the sensitivities achievable with our moderate foreground model,
the next generation of arrays will yield high significance detections of
the EoR power spectra, and provide detailed characterization of the power
spectrum shape over an order-of-magnitude in $k$ 
($k\sim0.1\mbox{--}1.0 h{\rm Mpc}^{-1}$). 
These sensitivity levels may even allow for direct imaging of the EoR
on these scales.

\item If successful, foreground removal algorithms can dramatically
boost the sensitivity of 21\,cm measurements.  They are also the only
way to open up the largest scales of the power spectrum, which can
lead to new physical insight through observations of the generic ``knee'' 
feature.

\item Although it will represent a major breakthrough for the 21\,cm cosmology
community, a low to moderate ($\sim 5\mbox{--}10\sigma$) detection of the EoR power
spectrum may not be able to conclusively identify the redshift of 50\%
ionization.  One might expect otherwise, since the peak brightness of the power
spectrum occurs near this ionization fraction.  However, accounting for the steep
rise in $T_{\rm sys}$ at low frequencies, 
shows that the rise-and-fall
of the power spectrum versus redshift may not be conclusively measurable
without a higher significance measurement, such as those possible
with the HERA design.

\item Going beyond power spectrum measurements to astrophysical parameter constraints, lower redshifts observations are particularly prone to parameter degeneracies.  These can be partially broken by foreground removal from within the wedge region (allowing access to the lowest $k$ modes).  Alternatively, degeneracies can be broken by performing parameter fits over multiple redshifts simultaneously (which is equivalent to making use of information about the power spectrum's \emph{evolution}).  Higher redshifts ($z \ge 11$) are typically limited not by intrinsic degeneracies, but by high thermal noise (at least for a HERA-like array), and add relatively little to constraints on reionization.

\item Assuming a fiducial 21cmFAST reionization model, a HERA-like array will be capable of constraining reionization parameters to $\sim 1\%$ uncertainty if foreground removal within the wedge proves possible, and to $\sim 5\%$ otherwise.  The current generation of interferometers will struggle to provide precise constraints on reionization models; the sensitivity of a HERA-like array is necessary for this kind of science (for a quantitative comparison, see the appendix).

\end{itemize} 

From this analysis, it is clear that for 21\,cm studies to deliver the 
first conclusive scientific constraints on the Epoch of Reionization, 
arrays much larger than those currently operational must be constructed.
Advancements in analysis techniques to keep
the EoR window free from contamination can contribute additional sensitivity,
but the most dramatic gains on the analysis front will come from
techniques that remove foreground emission and allow retrieval of modes
from inside the wedge.
This is not meant to disparage the wide range of foreground removal
techniques already in the literature; rather, the impetus is on adapting
these techniques for application to real data from the current and next
generation of 21\,cm experiments.  The vast range of EoR science 
achievable under our optimistic, moderate, and even pessimistic foreground removal scenarios 
provides ample motivation for continuing these efforts.    


\begin{subappendices}

\section{Appendix: Power Spectrum Sensitivities of Other 21\,cm Experiments} \label{app:others}
 
In this appendix, we compare the power spectrum sensitivities and 
EoR parameter constraints of several 21\,cm experiments. In particular, we consider the current generation experiments of PAPER \citep{PAPER},
the MWA \citep{TingaySummary}, and 
LOFAR \citep{LOFAR2013}, as well as a concept array for Phase
1 of the SKA based on the SKA System Baseline Design document
(SKA-TEL-SKO-DD-001\footnote{http://www.skatelescope.org/wp-content/uploads/2012/07/SKA-TEL-SKO-DD-001-1\_BaselineDesign1.pdf}). 
The instrument designs are summarized
in Table \ref{tab:instruments}, and the principal results are presented
in Tables \ref{tab:signif} and \ref{tab:astrophys}, which show
the significance of the power spectrum measurements and constraints
on EoR astrophysical parameters, respectively.  
Both calculations assume the fiducial EoR history shown
in Figure \ref{fig:pspec_vanilla}.  The significances in 
Table \ref{tab:signif} assume only an 8~MHz band centered on the 50\%
ionization redshift of $z=9.5$.  The astrophysical constraints, however,
assume information is collected over a wider band from $z = 7\mbox{--}13$;
for instruments with smaller instantaneous bandwidths, the observing
times will need to be adjusted accordingly.\footnote{For the particulars
of our fiducial EoR model, a significant fraction of the information comes
from $z = 7\mbox{--}9$ (142\mbox{--}178~MHz) 
(see Figure \ref{fig:OPTellipses}), meaning that an experiment
like the MWA with an instantaneous bandwidth of 30~MHz could nearly produce
the results described here without a signficant correction
for observing time.  
Of course, this assumes that the redshift of reionization is known
a priori, and that the optimal band for constraints is actually the band
observed.}

\begin{table}
\centering
\begin{tabular}{|C{.8in}|C{.8in}|C{.65in}|C{.75in}|L{2.1in}|}
\hline
Instrument & Number of Elements & Element Size~$(\rm{m}^2)$ & Collecting Area~$(\rm{m}^2)$ & Configuration \\ \hline
PAPER & 132 & 9 & 1188 & $11 \times 12$ sparse grid \\ \hline
MWA & 128 & 28 & 3584 & Dense 100~m core with $r^{-2}$ distribution beyond \\ \hline
LOFAR NL~Core & 48\tmark[a] & 745 & 35,762 & Dense 2~km core \\
\hline
HERA & 547 & 154 & 84,238 & Filled 200~m hexagon \\ \hline
SKA1 Low~Core & 866 & 962 & 833,190 & Filled 270~m core with Gaussian distribution beyond \\
\hline
\end{tabular}
\caption{Properties of Other 21\,cm Experiments.}
\label{tab:instruments}
\end{table}
 
\begin{table}
\centering
\begin{tabular}{|c|ccc|} 
\hline
Instrument & Pessimistic & Moderate & Optimistic \\
\hline
PAPER & 1.17 & 2.02 & 4.82 \\ \hline
MWA & 0.60 & 2.46 & 6.40 \\ \hline
LOFAR NL Core & 1.35 & 2.76 & 17.37 \\ \hline
HERA & 32.09 & 38.20 & 133.15 \\ \hline
SKA1 Low Core & 10.01 & 35.95 & 218.27 \\ \hline
\end{tabular}
\caption[Power spectrum measurement significant for other experiments.]{Power spectrum measurement signifiance (number of $\sigma$s) of other 21\,cm experiments
for each of the three foreground removal models.}
\label{tab:signif}
\end{table}

\begin{table}
\centering
\begin{tabular}{|c|ccc|ccc|ccc|} 
\hline
\multicolumn{1}{|c}{} & \multicolumn{3}{c}{Pessimistic} & \multicolumn{3}{c}{Moderate} & \multicolumn{3}{c|}{Optimistic} \\
Instrument & $\frac{\Delta T_{\rm vir}}{T_{\rm vir,fid}}$ & $\frac{\Delta \zeta}{\zeta_{\rm fid}}$ & $\frac{\Delta R_{\rm mfp}}{R_{\rm mfp,fid}}$ & $\frac{\Delta T_{\rm vir}}{T_{\rm vir,fid}}$ & $\frac{\Delta \zeta}{\zeta_{\rm fid}}$ & $\frac{\Delta R_{\rm mfp}}{R_{\rm mfp,fid}}$ & $\frac{\Delta T_{\rm vir}}{T_{\rm vir,fid}}$ & $\frac{\Delta \zeta}{\zeta_{\rm fid}}$ & $\frac{\Delta R_{\rm mfp}}{R_{\rm mfp,fid}}$ \\
\hline
PAPER & 1.444 & 1.168 & 1.507 & 1.260 & 1.013 & 1.294 & 0.272 & 0.179 & 0.140 \\
MWA & 4.419 & 3.479 & 4.555 & 0.757 & 0.568 & 0.731 & 0.231 & 0.152 & 0.119 \\
LOFAR & 1.538 & 1.251 & 1.515 & 0.719 & 0.565 & 0.675 & 0.069 & 0.046 & 0.039 \\
HERA & 0.072 & 0.051 & 0.047 & 0.062 & 0.044 & 0.039 & 0.011 & 0.007 & 0.005 \\
SKA1 & 0.235 & 0.169 & 0.179 & 0.076 & 0.054 & 0.044 & 0.009 & 0.006 & 0.004 \\
\hline
\end{tabular}
\caption[Reionization parameter errors by telescope.]{Fractional errors on the reionization parameters achieveable with each instrument under the three foreground removal models,
assuming all redshifts are analyzed jointly.}
\label{tab:astrophys}
\end{table}

In order to compute the constraints achievable with other experiments,
we apply the sensitivity calculation described in Section \ref{sec:sense_calc}
to each of the five instruments under study.
We note that this sensitivity calculation assumes a drift-scanning
observing mode, with the limit of coherent sampling set by the
size of the element primary beam.  The MWA, LOFAR, and likely the SKA 
all have the capability of conducting a tracked scan to increase the
coherent integration on a single patch of sky.  Similarly, tracking can
be used to move to declinations away from zenith if sample variance
becomes the dominant source of error.  A full study of the benefits
of tracking versus draft scanning for power spectrum measurements is 
beyond the scope of this present work; rather, we assume all instruments
operate in a drift-scanning mode for the clearest comparison
with the fiducial results calculated for the HERA experiment.
We therefore also assume that each telescope observes for the fiducial
6~hours~per~day for 180~days (1080~hours).  
Finally, we also assume that 
each array has a receiver temperature of 100~K.
We discuss the important features of each instrument and the resultant
constraints in turn; see the main text for a discussion of the HERA
experiment.

\begin{enumerate}

\item \textbf{PAPER}: Our fiducial PAPER instrument is an $11\times12$
grid of PAPER dipoles modeled after the maximum redundancy arrays
presented in \cite{AaronSensitivity}.  In this configuration,
the $3\times3~\rm{m}$ dipoles are spaced in 12 north-south columns
separated by 16~m; within a column, the dipoles are spaced 4~m apart. 
In both our pessimistic and moderate scenarios, PAPER yields a
non-detection of the fiducial 21\,cm power spectrum.  In the optimistic
scenario, the array could yield a significant detection; however,
the poor PSF of the maximum redundancy array is expected to present
challenges to any foreground-removal strategy that would allow
recovery from information inside the wedge \citep{PAPER32Limits}.
Therefore, achieving the results of the optimistic scenario will be especially
difficult for the PAPER experiment.

\item \textbf{MWA}: Our model MWA array uses the 128 antenna positions
presented in \cite{TingaySummary}.  Despite having nearly three times
the collecting area of the PAPER array, we find the MWA yields a 
less significant detection in the pessimistic scenario.  
Poor sensitivity when partially redundant samples are combined
incoherently is to be expected for the MWA.  
The pseudo-random configuration of the array produces
essentially no instantaneously redundant samples, and so all
redundancy comes from partial coherence.  Therefore, one might expect
the MWA to under-perform compared to the highly redundant PAPER array in
this scenario. In the moderate and optimistic scenarios
where partial redundancy yields sensitivity boosts
the MWA outperforms the PAPER array.

\item \textbf{LOFAR}: To model the LOFAR array, we use the antenna
positions presented in \cite{LOFAR2013}.  For the purposes
of EoR power spectrum studies, we focus on the Netherlands core
of the instrument, since baselines much longer than a few km contribute
very little sensitivity.  We also assume that LOFAR is operated in a mode
where each sub-station of the HBA is correlated separately to increase
the number of short baselines.  However, the resultant sensitivities
still show that LOFAR suffers from a lack of short baselines.  Despite
having a collecting area $\gtrsim10$ times larger than PAPER and the MWA,
LOFAR still yields a non-detection of the EoR power spectrum
in the pessimistic and moderate foreground removal scenarios.  Only
in the optimistic scenario where longer baselines contribute
to the power spectrum measurements does LOFAR's collecting area
result in a high-significance measurement.  Preliminary results from the
LOFAR experiment show significant progress in subtracting
foregrounds to access modes inside the wedge 
\citep{Chapman1,InitialLOFAR1}.

\item \textbf{SKA1-Low}: We model our SKA-Low Phase 1 instrument
after the design parameters set out in the SKA System Baseline
Design document, although the final design of the SKA is still subject
to change.  This document specifies that the array will consist of 911
35~m stations, with 866 stations in a core with a Gaussian distribution
versus radius.  This distribution is normalized to have 650 stations
within a radius of 1~km.  This density in fact yields a completely
filled aperture out to $\sim300~\rm{m}$, which we model as a close
packed hexagon.
This core gives the design some degree of instantaneous redundancy,
a configuration
that is still being explored for the final design of the instrument.
We do not consider the 45 outriggers in our power
spectrum sensitivity.  Much like the case with PAPER and the MWA, the
lower instantaneous redundancy of the SKA concept array results in a poorer
performance than the highly redundant HERA array in the pessimistic
scenario.  However, in the moderate and optimistic scenarios this
SKA concept design yields high sensitivity measurements, although
not as high as might be expected from collecting area alone.  This
fact is once again due to the relatively small number of short spacings
compared to the HERA array, resulting in similar performances for the two
configurations in the moderate scenario.  As with LOFAR, the SKA design
shines in the optimistic scenario, producing a very high SNR measurement.

\end{enumerate}
  
\begin{figure*}[t]
\centering 
\includegraphics[width=1.0\textwidth]{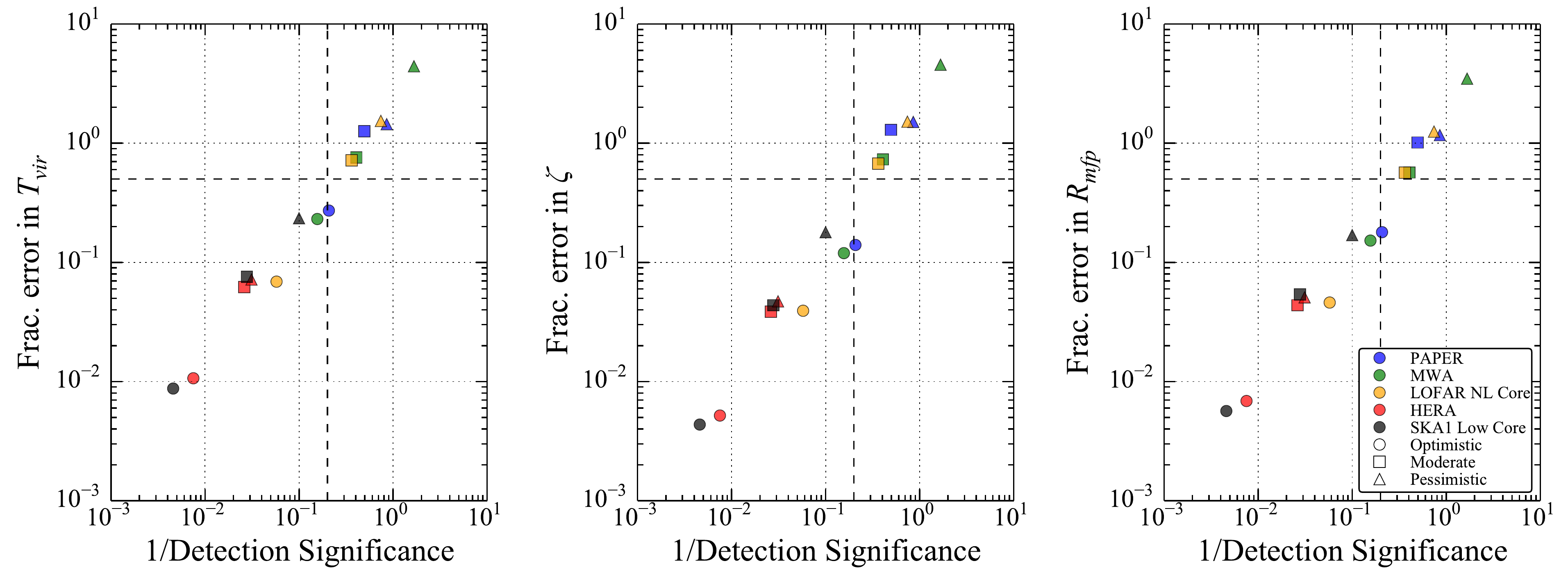}
\caption[Relationship between detection significance and parameter errors.]{Fractional errors in astrophysical parameters shown as a function of $(\textrm{detection significance})^{-1}$.  Different instruments are shown in different colors (PAPER in blue; MWA in green; 
LOFAR in yellow; HERA in red; SKA in black), and different foreground scenarios are shown using
different shapes (optimistic as circles; moderate as squares; pessimistic as triangles).  The vertical
dashed line delineates a $5\sigma$ detection of the power spectrum, while the horizontal
dashed line delineates a parameter error of $50\%$.  The tight correlations shown here suggest that the significance of a power spectrum detection can be used as a proxy for an instrument's ability to constrain
astrophysical parameters.}
\label{fig:sigmaParamCorrAllz}
\end{figure*}
\vspace{.2in}
In all cases, we find that the fractional errors on the reionization parameters
(Table \ref{tab:astrophys}) scale very closely with the overall
significance of the power spectrum measurement (Table \ref{tab:signif}).
This is shown in Figure \ref{fig:sigmaParamCorrAllz}, where we plot the fractional
errors on the reionization parameters against the reciprocal of the power spectrum
detection significance.  These two quantities are seen to be directly proportional
to an excellent approximation, regardless of foreground scenario.\footnote{We note that this is true only when all redshifts are analyzed jointly, 
where the errors are driven mainly by thermal noise.  If the errors are instead
dominated by parameter degeneracies (as is the case, for example, when
only one redshift slice is measured), the tight linear correlation breaks down.}  
Therefore, while the power spectrum sensitivity of an array can 
be a strong function of an instrument's configuration,
the resultant astrophysical constraints are fairly generalizable once
the instrument sensitivity is known.  This is strongly suggestive
that the results in the main body of the paper can be easily extended
to other instruments.  Also noteworthy is the fact that current-generation
instruments encroach on the lower-left regions (high detection significance;
 small parameter errors) of the plots in Figure \ref{fig:sigmaParamCorrAllz}
 only for the optimistic foreground model.  In contrast, the next-generation
 instruments (HERA and SKA) are clearly capable of delivering excellent
 scientific results even in the most pessimistic foreground scenario.
%



\end{subappendices}

\chapter{Conclusion}

This thesis focused on the development of and early results from the field of 21\,cm cosmology, a new and potentially transformative probe of the universe during the first billion years after the big bang, a period of dramatic change we call the ``Cosmic Dawn.'' Using the 21\,cm transition of neutral hydrogen, large volumes of the early intergalactic medium can be mapped tomographically. This will open up an enormous and largely unexplored volume to precise tests of our astrophysical and cosmological models. 

Realizing the promise of 21\,cm cosmology will be extraordinarily difficult. We need very large radio telescopes observing for hundreds or thousands of hours just to achieve the necessary sensitivity to see the faint signal from neutral hydrogen. We also need to separate out the cosmological signal from astrophysical foregrounds, which are four or more orders of magnitude brighter, using our understanding of the physical processes that create them and the way they appear in our instruments.

I began this thesis in Chapter \ref{ch:Introduction} explaining the current state of our understanding of the physics that underlies the Cosmic Dawn and the first stars, galaxies, and black holes that drove it. I then explained why the 21\,cm signal can be detected and the physical processes that affect it. I also reviewed the observational challenges---especially that of bright foregrounds as seen through an interferometer---and surveyed the current and near future efforts to detect and characterize the cosmological signal.

The rest of the thesis was split into three parts. In Part I, \emph{Novel Data Analysis Tools}, I reproduced two theoretical papers that detailed analysis techniques designed for the rigorous and robust detection the power spectrum of 21\,cm brightness temperature fluctuations. Chapter \ref{ch:FastPowerSpectrum} focused on the acceleration of the previously published quadratic estimator method of \citet{LT11} to meet the challenges of large data volumes typically encountering in 21\,cm power spectrum estimation. Chapter \ref{ch:Mapmaking} relaxed one of the key assumptions in Chapter \ref{ch:FastPowerSpectrum} to incorporate realistic chromatic effects on the point spread function---effects that create the characteristic ``wedge'' feature in the cylindrically averaged power spectrum. Like the previous chapter, Chapter \ref{ch:Mapmaking} focused on showing how the statistical relationship between interferometric maps and the true sky can be modeled and propagated through the rest of the analysis in a computationally feasible way.

In Part II, \emph{Early Results from New Telescopes}, I included three papers discussing scientific and technological progress toward eventually making a detection of the 21\,cm signal. Chapters \ref{ch:MWAX13} and \ref{ch:EmpiricalCovariance} presented upper limits on the 21\,cm brightness temperature power spectrum using the Murchison Widefield Array, in its 32-tile prototype phase and then in its full 128-tile configuration. Both papers developed various methods for adapting the techniques from Part I to the challenges presented by real-world data, including calibration errors, incomplete Fourier sampling, radio frequency interference, and foreground residual modeling. Then, in Chapter \ref{ch:MITEOR}, I reproduced a paper reporting on MITEoR, a technology demonstration array designed to test techniques---especially the redundant calibration of antenna gains and phases---for building highly scalable interferometers that may one day realize the full potential of 21\,cm cosmology.

Finally in Part III, \emph{The Cosmic Dawn on the Horizon}, I explored that potential in greater depth by reproducing two papers examining the scientific progress near future telescopes can make toward constraining the astrophysics behind the Cosmic Dawn. Chapter \ref{ch:Forest} looked at the possible effect of high redshift radio-loud active galactic nuclei on the 21\,cm power spectrum. Depending on the population of those objects and the thermal history of the intergalactic medium, they could have a significant observable effect, especially at high redshift. Chapter \ref{ch:NextGen} looked at the constraints the next generation of 21\,cm interferometers, especially the Hydrogen Epoch of Reionization Array (HERA), can place on the physics underlying reionization, by detecting and beginning to characterize the 21\,cm power spectrum from the epoch of reionization.

Together this thesis represents six years of work toward the development of 21\,cm cosmology as the next great cosmological probe---not to mention the tremendous contributions of collaborators and coauthors across three different telescope teams. And yet, to recall the words of Hubble, the thesis ``ends on a note of uncertainty.'' We still haven't found the signal we're looking for. Even as we push to ``the utmost limits of our telescopes'' we find ourselves ``measuring shadows'' and pouring though vast quantities of data, beset by overwhelming foregrounds, and ``searching among ghostly errors of measurement for landmarks that are scarcely more substantial.'' We've come a long way and we've still got a long way to go.

``The search will continue.''

\begin{singlespace}
\addcontentsline{toc}{chapter}{Bibliography}
\bibliography{main}
\bibliographystyle{plainnat}
\end{singlespace}

\end{document}